\documentclass[12pt,twoside,openright]{report} 

\title{Proton structure at the LHC}
\author{Nathan Hartland}
\date{2014}

\usepackage[phd,fancyhdr]{edmaths}
\usepackage{nnpdf}
\usepackage{cite, multirow}

\newcommand{\be}{\begin{equation}}
\newcommand{\ee}{\end{equation}}
\newcommand{\ba}{\begin{eqnarray}}
\newcommand{\ea}{\end{eqnarray}}

\graphicspath{{./}{figs/}}

\newcommand{\pt}{$p_\perp$}

\begin{document}
\maketitle
\setcounter{page}{1}
%If you want a dedication..
\declaration

\acknowledgements

%Spacing of 1.5 to abide by regulations. Don't write outside this... Change for quotations and notes (if any)
\begin{spacing}{1.5}

\begin{abstract}
A determination of Parton Distribution Functions (PDFs) from a global fit to a dataset including measurements from the LHC has been performed for the first time. The determinations
have been performed according to the NNPDF methodology, leading to a fit relatively free of parametrisation bias and with an accurate account of PDF uncertainty.

In this thesis the importance of QCD measurements at the LHC to PDF extraction are discussed, and we summarise some of the technical difficulties in their inclusion into PDF fits. A number of methods are presented that permit the efficient inclusion of these observables
into PDF determinations. 

Firstly a Bayesian reweighting procedure taking advantage of the Monte Carlo representation of PDF uncertainties in NNPDF sets is discussed. The utility of the Bayesian reweighting method is demonstrated by a study of the impact of early $W$ production asymmetry measurements from ATLAS, CMS and LHCb upon an earlier PDF set.

A package for the fast computation of observables in an automated NLO framework is presented, providing an interface between Monte Carlo event generators and NLO interpolation tools.

Finally, a new method of combining PDF evolution with interpolating codes for hadronic observable computation is also described. This method largely overcomes the computational difficulties in performing fast perturbative QCD predictions for collider observables. The method has been applied to the determination of PDFs from a global dataset including electroweak vector boson production data from LHCb, ATLAS and CMS along with inclusive jet data from ATLAS. The resulting set, NNPDF2.3 provides the most accurate determination of parton distributions via the NNPDF methodology to date.

Finally, the method of closure testing is introduced, and the method is applied to the study of the NNPDF methodology. A number of improvements are found in the minimisation and stopping procedures, which are adopted for the development of the next NNPDF release, NNPDF3.0. Alongside the improved methodology, the NNPDF3.0 PDF set will provide a determination based upon an expanded dataset in order to produce a comprehensive upgrade to the NNPDF2.3 family of fits.
\vspace{10mm}
\normalsize

\end{abstract}

%%%%%%%%%%%%%%%%%%%%%%%%%%%%%%%%%%%%%%%%%%%%%%%%%%%%%%%%%%%%%%%%%%%%%

{\pagestyle{plain}
\tableofcontents
\cleardoublepage}
 \cleardoublepage
    \addcontentsline{toc}{chapter}{List of Figures}
    \listoffigures
 \cleardoublepage
    \addcontentsline{toc}{chapter}{List of Tables}
    \listoftables

%%%%%%%%%%%%%%%%%%%%%%%%%%%%%%%%%%%%%%%%%%%%%%%%%%%%%%%%%%%%%%%%%%%%%

\pagenumbering{arabic}
\chapter*{Introduction}
\label{ch:intro}
\addcontentsline{toc}{chapter}{Introduction}

The study of elementary particles and their behaviour relies on a great many sources of experimental information.
In order to verify the predictions of the Standard Model (SM) of particle physics or indeed extensions beyond, precise and accurate measurements
must be made of the fundamental properties of matter. Building upon decades of advances in the study of elementary particles,
today the foremost source of cutting edge measurements in particle physics is the Large Hadron Collider (LHC) based at CERN in Switzerland.
The LHC, through colossal scientific and human effort has opened up the study of the properties of nature to scales that were previously inaccessible. 

The LHC probes the building blocks of nature by the collision of high energy protons. Maximising the physics potential of the LHC therefore requires a deep understanding
of the composite nature of the proton. The short range dynamics of a proton's constituent particles can be described by perturbative Quantum Chromo-Dynamics (QCD),
however an understanding of the low energy behaviour is impossible to obtain through perturbative methods, therefore making its determination by a calculation from first principles challenging. In practice the structure of the proton is understood through a comprehensive analysis of experimental data, and described in terms of Parton Distribution Functions (PDFs). These functions parametrise the unknown non-perturbative dynamics of the proton. As a universal property of protons, the PDFs may be determined from available experimental data and then applied in the calculation of predictions for other experiments, therefore making the application of QCD in hadron collisions into a predictive theory which may be tested via comparison to data.

The accurate determination of parton densities in the proton is an ongoing effort, with several groups providing competing sets of parton distributions. The NNPDF collaboration provides a set of parton distributions determined through a rather different methodology than the standard procedures, resulting in a PDF set suffering from little of the parametrisation bias possible in competing approaches. Furthermore the NNPDF methodology has a unique treatment of the experimental uncertainty propagation, leading to a statistically sound estimation of the uncertainties in the resulting PDFs.

While a precise knowledge of the dynamics of the proton is vital for LHC studies of physics in the standard model and beyond, LHC data also has the potential to provide the most in depth information on parton densities to date. This thesis is based upon work conducted in the study of early LHC standard model measurements of particular sensitivity to parton distributions. The inclusion of such an experimental dataset into a fit in the NNPDF framework has necessitated the development of a number of tools for the efficient calculation of collider observables. These tools and their applications shall be discussed alongside the development of the NNPDF methodology to better handle the ever-enlarging LHC dataset.

This thesis is arranged as so. In Chapter One we shall provide a brief discussion of the theoretical structure of parton distributions, where they arise in the calculation of deep-inelastic scattering cross-sections and further theoretical background relevant to the reliable determination of PDFs from experimental data. Chapter Two is concerned with the practical extraction of PDFs and shall describe experimental observables of interest along with the different approaches used by major PDF collaborations to fit the data. In Chapter Three, the tools that have been developed to enable the inclusion of a large LHC dataset into a computationally intensive fit such as the NNPDF procedure are introduced and described. A brief summary of experimental measurements at the LHC of interest to the determination of PDFs is provided in Chapter Four. In Chapter Five, we shall examine the impact of some of these measurements made by LHC collaborations upon PDF determinations, enabled by the tools developed in Chapter Three. The data impact will be assessed in the context of the two most recent public releases of the NNPDF collaboration; providing a summary of their datasets and the tools used in their extraction. Finally in Chapter Six we examine some of the methodological improvements that have been made in the NNPDF procedure in order to ensure the maximal efficiency in extracting new information on PDFs from future LHC measurements, and demonstrate their application in early prototypes of the NNPDF3.0 PDF set.

\chapter{\label{chapter1} Parton distribution functions} 
Parton distributions are one of the central pillars of perturbative QCD, factorising as they do the perturbatively incalculable long distance dynamics present in calculations involving hadronic initial states. Combined with
the perturbative description of the short-distance cross-section what could seem at first a hopeless situation is alleviated, and QCD becomes a predictive and useful theory when applied to hadronic scattering.

In this chapter a brief overview of how parton distribution functions arise in QCD calculations will be presented.
We shall explore the prototypical example of the deep inelastic scattering (DIS) of leptons off a hadronic target, first in the \emph{naive parton model} arising before the advent of QCD and then with the QCD-improved parton model which allows for an excellent description of DIS measurements across a wide range of hard scales.

The treatment of heavy quarks in parton distributions is a particularly delicate issue and therefore will
also be discussed in this introductory theory section. Finally there will be some exploration of the general properties of parton distributions in order to provide a summary of the available theoretical constraints upon PDFs.

\section{Partons in deep inelastic scattering}
We shall begin by introducing parton distribution functions as they arise in the early parton model. The model was originally introduced by Feynman and Bjorken~\cite{feynman1,Feynmanparton,feynmanparton2, Bjorken:1968dy} in the late 1960's in an effort to understand the scattering behaviour of hadronic states and successfully
describes many properties observed in early deep inelastic scattering experiments.

\begin{figure}[ht]
\centering
\includegraphics[scale=0.5]{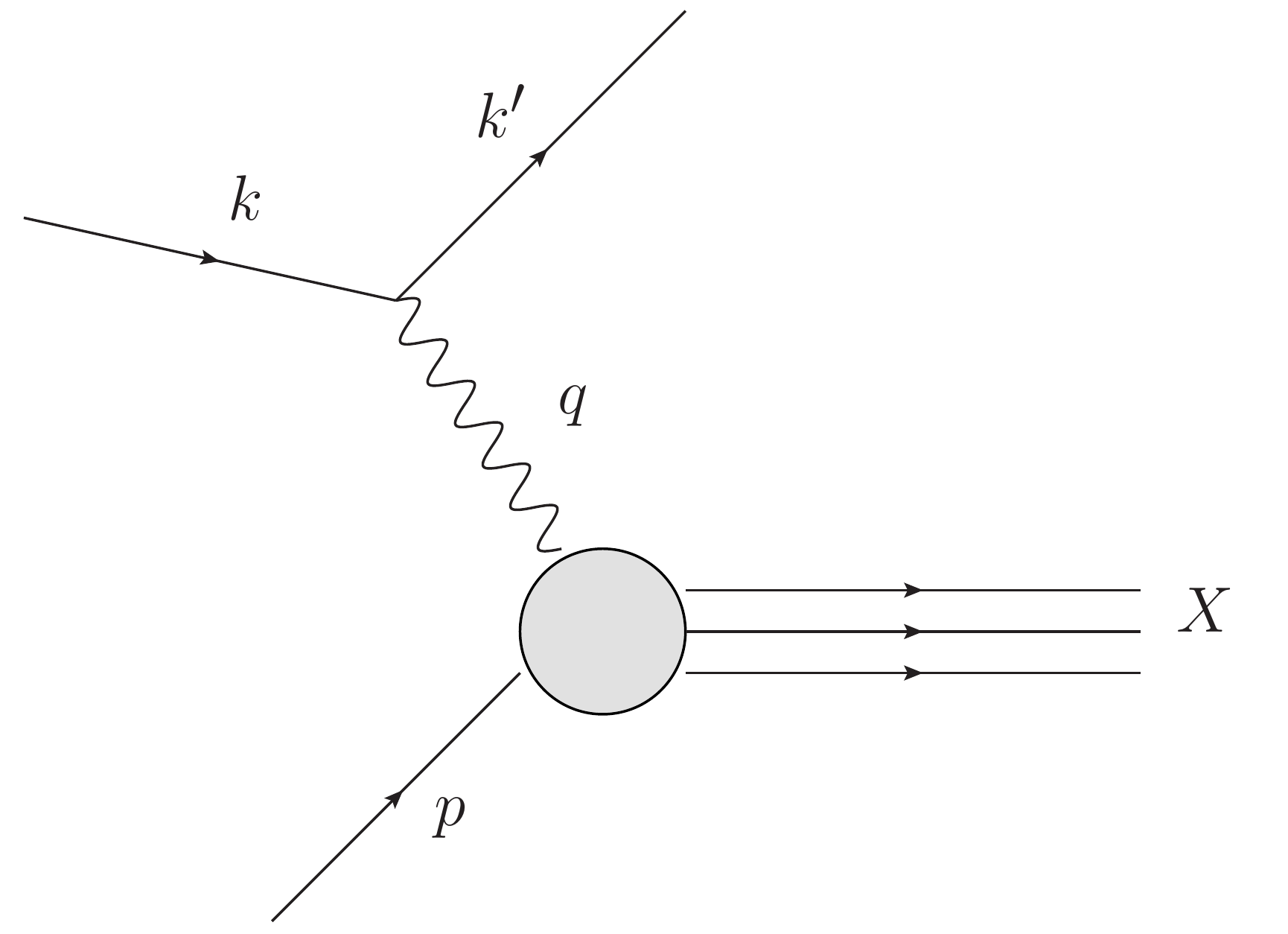}
\caption{Deep inelastic scattering of a charged lepton with a proton target.}
\label{fig:DIS}
\end{figure}

In this process, a charged lepton $l$ probes a proton $P$ by the exchange of a gauge boson. For simplicity we shall describe here the neutral current process where a photon is exchanged. In the inelastic regime where the momentum transfer to the target proton is large, the proton does
not survive the scattering process and fragments into an arbitrary hadronic final state $X$. The process $l(k) + P(p) \to l(k^\prime) + X$ is illustrated at tree level in Figure \ref{fig:DIS}. 

In this system we can define the standard DIS kinematic variables;  $Q^2$ denotes the momentum transfer from the electron to the target proton, $\nu$ the energy transfer and $y$ the measure of the reaction's inelasticity, or fractional energy transfer. In the rest frame of the proton these are given by
\begin{eqnarray}
 Q^2 &=& -q^2 = -(k - k^{\prime})^2, \\
 \nu &=& M(E- E^\prime), \\
 y &=& (q \cdot p)/(k \cdot p),
\end{eqnarray}
where $M$ refers to the mass of the proton, and the inelasticity ranges between 0 (elastic scattering) and 1. $E$ and $E^\prime$ denote the energies associated with the four-momenta $k$ and $k^\prime$ respectively. Additionally, we may introduce the Bjorken scaling parameter $x$, central to the parton model,
\be x = \frac{Q^2}{2\nu}. \ee
Neglecting spin labels, the amplitude for this diagram in the Feynman gauge is given by
\be\mathcal{M} = ie^2\bar{u}(k^\prime)\gamma^\mu u(k)\left( i\frac{g_{\mu\nu}}{Q^2} \right)\left<X\right|J_h^\nu\left|P\right>,  \label{eq:DISme}\ee
where $J_h^\nu$ represents the hadronic current. 
The fundamental difficulty in attempting to compute the cross section for this process is our ignorance of the wavefunction for the hadronic states $\left|X\right>$ and $\left|P\right>$. To isolate the problem, we are able to factorise the spin averaged square of the amplitude in Equation \ref{eq:DISme} into a leptonic ($L_{\mu\nu}$) and a hadronic ($W^{\mu\nu}$) part
\be |\overline{\mathcal M}|^2 = \frac{1}{Q^2} L_{\mu\nu}W^{\mu\nu}, \ee
where the leptonic tensor is straightforwardly calculable:
\ba L_{\mu\nu} &=& e^2\sum_{spin} \bar{u}(k^\prime)\gamma_\mu u(k) \bar{u}(k)\gamma_\nu u(k^\prime), \\
&=& e^2 \mathrm{ tr}\left[ \slashed{k}^\prime\gamma_\mu\slashed{k}\gamma_\nu \right], \\
&=& 4e^2[k_\mu k^\prime_\nu + k_\nu k^\prime_\mu - g_{\mu\nu}k\cdot k^\prime],  \ea
where here we have neglected the fermion masses. The hadronic part of the calculation is considerably more difficult to evaluate, and indeed impossible to compute from first principles in perturbation theory as it is sensitive to the low-scale, and therefore strongly coupled dynamics of the proton target:
\ba  
W^{\mu\nu} &\sim& \sum_X \left<P(p)\right| {J_h^\mu}^{\dagger} \left|X\right>\left<X\right| J_h^\nu \left| P(p)\right>, \\
 &\sim& \left<P(p)\right| {J_h^\mu}^{\dagger} J_h^\nu \left| P(p)\right>. \ea

However, we can gain some insight into its structure by noting that the tensor must obey the conservation requirements of the hadronic current $q_\mu W^{\mu\nu}=0$ and $q_\nu W^{\mu\nu}=0$. The tensor may therefore be parametrised without loss of generality by the following structure:
\be W_{\mu\nu} = -\left( g_{\mu\nu} - \frac{q_\mu q_\nu}{q^2}\right) F_1(x,Q^2) +\left(p_\mu -q_\mu \frac{p \cdot q}{q^2}\right)\left(p_\nu -q_\nu \frac{p \cdot q}{q^2}\right)\frac{1}{\nu}F_2(x,Q^2).\label{eq:htensor}\ee
Here we have introduced the parameters in our tensor $F_i$ which are known as the electromagnetic structure functions. For interactions involving parity-violating currents, there is a third contribution to the hadronic tensor arising through the $F_3$ structure function. Here the only possible functional dependence for the structure functions is upon the quantities $Q^2$ and $x$.

It is convenient now to define a projection vector $n$ with the properties $ p \cdot n = 1$, $ n \cdot q = 0$, and $n^2 = p^2 = 0$, where the assumption of negligible proton mass has been made. Any vector can now be written as a combination of $n$, $p$ and a component transverse to the proton momentum as a \emph{Sudakov decomposition}. Using this projection vector we may obtain the structure functions from the hadronic tensor as so:
\begin{eqnarray}
 F_2 &=& \nu n^\mu n^\nu W_{\mu\nu},  \label{eq:proj1} \\
 F_L &=& F_2 - 2xF_1 = \frac{Q^4}{\nu^3}  p^\mu p^\nu W_{\mu\nu}, \label{eq:proj2}
\end{eqnarray}
where the quantity in the second equation is known as the longitudinal structure function. So far, few assumptions have been made about the form of the EM hadronic tensor $W_{\mu\nu}$, we have simply parametrised it in terms of a Lorentz invariant tensor structure and structure functions. Feynman's parton model allows us to describe more of the hadronic tensor with perturbation theory by proposing a composite proton formed as a bound state of fundamental, spin-$1/2$ constituents: the \emph{partons}. 

The parton model approximation states that for a sufficiently hard interaction, the virtual photon only interacts with a single point-like parton inside the target proton and we can treat the partons as approximately free particles. The hadronic tensor then admits a probabilistic expansion in terms of Parton Distributions which encode the probability of the hard photon interacting with a constituent parton carrying a faction $\xi$ of the parent proton's momentum. The probability of interacting with a parton carrying between $\xi$ and $\xi+\delta\xi$ of the proton's momentum being given by $f(\xi)\delta\xi$ where $f(\xi)$ is the interaction probability for a parton with momentum $\xi p$. Diagrammatically we may therefore construct the photon-hadron interaction as a weighted sum of partonic diagrams:
\be \left| \vcenter{\hbox{\includegraphics[height=2cm]{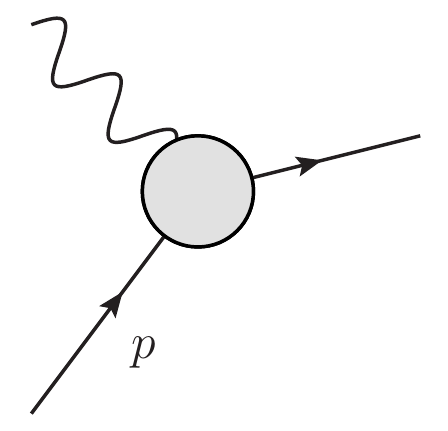}}} \right|^2= \sum_i^{N_{part}}f_i(\xi,Q^2) \otimes \left| \vcenter{\hbox{\includegraphics[height=2cm]{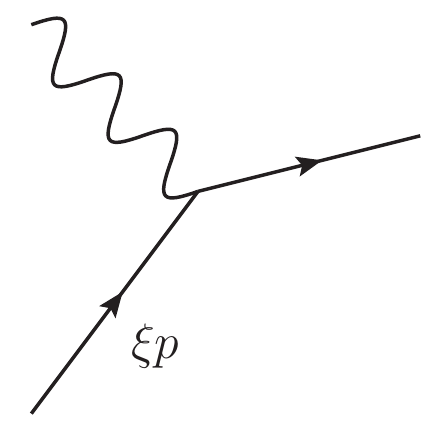}}}\right|^2(\xi), \nonumber\ee
where we have introduced the multiplicative convolution
\be (f \otimes g )(x) = \int_0^1\; \frac{d\xi}{\xi}\; f\left(\frac{\xi}{x}\right) g(\xi).\ee
The hadronic tensor is then given in terms of a sum of individual hard scattering partonic tensors, denoted  $\widetilde{W}^i_{\mu\nu}(\xi)$ for a target parton of type $i$. Writing the hadronic tensor as the probabilistic sum over all constituent parton types we obtain
\be W_{\mu\nu} =\int_0^1 \frac{d\xi}{\xi} \sum_i f_i(\xi,Q^2)\; \widetilde{W}^i_{\mu\nu}(\xi,Q^2). \label{eq:htensorexpan}\ee
As the parton level tensors must obey the same conservation relations as the full hadronic tensor, we can once again form a general parameterization of $\widetilde{W}^i_{\mu\nu}$:
\be \widetilde{W}^i_{\mu\nu} = -\left( g_{\mu\nu} - \frac{q_\mu q_\nu}{q^2}\right) \widetilde{F_1^i}(\xi,Q^2) +\xi^2 \left(p_\mu -q_\mu \frac{p \cdot q}{q^2}\right)\left(p_\nu -q_\nu \frac{p \cdot q}{q^2}\right)\widetilde{F_2^i}(\xi,Q^2),\ee
where the factors of $\xi^2$ arise from taking $p^\mu \to \xi p^\mu$. Substituting this form for $\widetilde{W}^i_{\mu\nu}(\xi)$ into Eqn \ref{eq:htensorexpan} and comparing with the form in Eqn \ref{eq:htensor}, we find two expressions for the proton EM structure functions,
\be F_1(x,Q^2) = \int_0^1 \frac{d\xi}{\xi} \sum_i f_i(\xi) \widetilde{F_1^i}(\xi,Q^2),  \label{eqn:f1}\ee
\be F_2(x,Q^2) = \int_0^1 \xi d\xi \sum_i f_i(\xi) \widetilde{F_2^i}(\xi,Q^2). \label{eqn:f2}\ee
The naive parton level structure functions $\widetilde{F_1^i}(\xi,Q^2)$ describe the hard scattering subprocess involving a parton of species $i$ and may be computed by considering the parton level squared amplitude for the subprocess, $\gamma^*(q) + q(\xi p) \to q(l)$ and projecting out the desired quantities with the operators defined previously. At leading order, using the parton level version of the projector Eqn \ref{eq:proj1}:
\be { \mathcal M }_\mu = -i e_{q^i}\bar{u}(l)\gamma^\mu u(\xi p),\ee
\be \frac{n^\mu n^\nu}{\xi^2} \widetilde{W}^i_{\mu\nu}  = \frac{n^\mu n^\nu}{\xi^2}\overline{\sum} \left| \mathcal{M} \right|^2_{\mu\nu} = 4e_{q^i}^2,\ee
where we have made the approximation that momenta transverse to the beam axis vanish. Including the phase space for the final state quark in the CM frame we obtain:
\be \widetilde{F^i_2} =  2 e_{q^i}^2 \delta(l^2),\ee
where the delta function can be rewritten in terms of $\xi p$ and $q$:
\be \delta (l^2) = \delta ((\xi p + q )^2 ) = \delta (2\xi \nu - Q^2) = \delta (2\nu (\xi - x)).\ee 
This is an interesting result of the analysis at leading order, the kinematical variable $x$ actually describes the momentum fraction of the interacting parton. The parton level structure function
$\widetilde{F^i_2}$ is therefore given by:
\be \widetilde{F^i_2} = 2 e_{q^i}^2 \delta (\xi-x).\ee
The parton level longitudinal structure function is also straightforwardly projected out of the same amplitude,
\be \widetilde{F}^i_L = \frac{Q^4}{\xi \nu^3}p^\mu p^\nu \widetilde{W}^i_{\mu\nu} =  \widetilde{F}^i_2 - \frac{2x}{\xi^2} \widetilde{F}^i_1. \ee
At leading order this projection, and therefore the longitudinal structure function, are exactly zero, consequently
\be \widetilde{F}^i_1 = \frac{\xi^2}{2x} \widetilde{F}^i_2 = e_{q^i}^2\frac{ \xi^2 }{x} \delta (\xi - x).\ee
We may therefore write the full EM proton structure functions in the naive parton model as
\be F_1(x,Q^2) =  \int_0^1 d\xi \sum_i f_i(\xi) e_{q^i}^2\frac{ \xi }{x} \delta (\xi - x) =  \sum_if_i(x)e^2_{q^i},\ee
\be F_2(x,Q^2) = 2 \int_0^1 \xi d\xi \sum_i  f_i(\xi) e_{q^i}^2 \delta (\xi-x) = 2 x \sum_if_i(x)e^2_{q^i}.\ee
These results have a number of important features. Firstly in this model the structure functions have no dependence upon the resolution parameter $Q^2$, a phenomenon known as Bjorken scaling~\cite{Bjorken:1968dy}. This scaling effect was an important achievement of the original parton model, as it was able to describe contemporary experimental results rather well. The lack of any scale dependence in the structure functions is a consequence of the model's assumptions treating interactions with the proton's constituent partons as point like, and consequently having no characteristic length scale.

Secondly we note that $F_2(x) = 2xF_1(x)$, which is known as the Callan-Gross relation~\cite{callangross}. It illustrates a fundamental property of spin-1/2 particles, that they are unable to absorb a longitudinally polarised photon~\cite{pQCDhandbook}.
\section{QCD and the parton model}
The naive parton model was able to provide a good phenomenological description of early DIS measurements. Its success also provided great support for QCD as the correct description of the strong interaction. The phenomenon of Bjorken scaling placed substantial constraints upon the theory governing the internal dynamics of the proton. The asymptotic freedom of QCD allows for a consistent description of Bjorken-scaling, where the constituents of the hadron can be viewed as independent, non-interacting point like particles at high  values of the resolution parameter $Q^2$. The partons in Feynman's model were therefore quickly associated with the quarks and gluons of QCD.
\begin{figure}[t]
\centering
\includegraphics[scale=0.5]{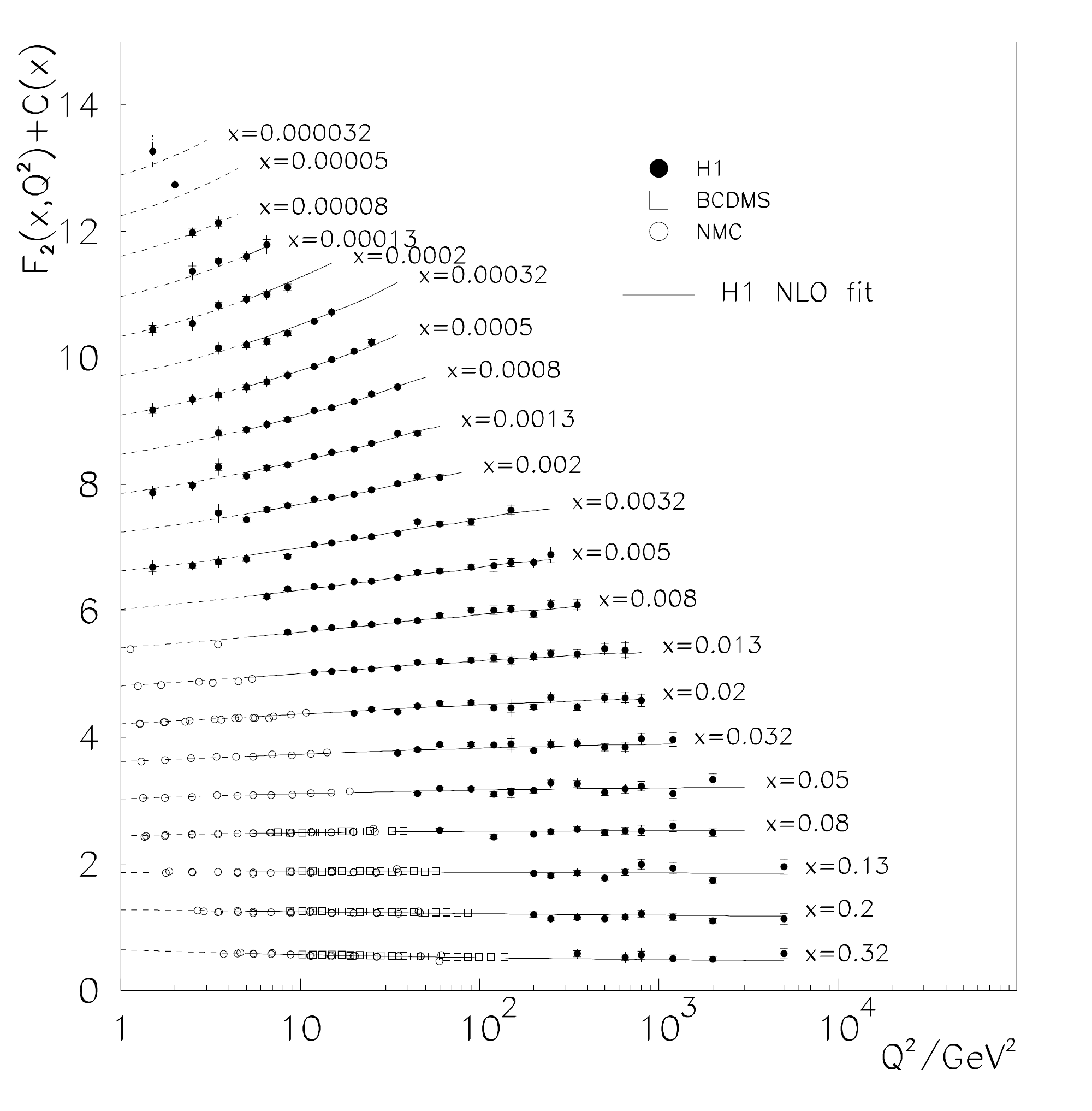}
\caption[Scaling violations in the proton structure function $F_2$]{Scaling violations in the proton structure function $F_2$. Here each curve in $x$ is scaled by a function $C(x)= 0.6(i- 0.4)$ for presentation purposes, where $i$ denotes the bin in $x$. Figure from~\cite{Aid:1996au}.}
\label{fig:F2H1}
\end{figure}

Despite the `snapshot' picture of non-interacting partons at leading order in QCD, we cannot neglect the higher order corrections to the point vertex calculated in the previous section. These corrections introduce logarithms of $Q^2$ which break the naive Bjorken scaling of the structure functions. Indeed, the measurement of such scaling violations provided one of the most powerful experimental verifications of QCD. Such violations are demonstrated in measurements of $F_2$ in Figure \ref{fig:F2H1}. In this section we shall perform an overview of the extension of the parton model to $\mathcal{O}(\alpha_s)$ in QCD.

At one loop order, there are three diagrams that contribute to the $qq\gamma$ vertex studied in the previous section; the real emission of a gluon from the initial (a) or final state (b) quarks, and the virtual correction diagram (c). Additionally at one loop order in QCD there arises a diagram initiated by a gluon splitting into a $q\bar{q}$ pair (d).

\begin{figure}[ht]
\centering
\includegraphics[scale=0.6]{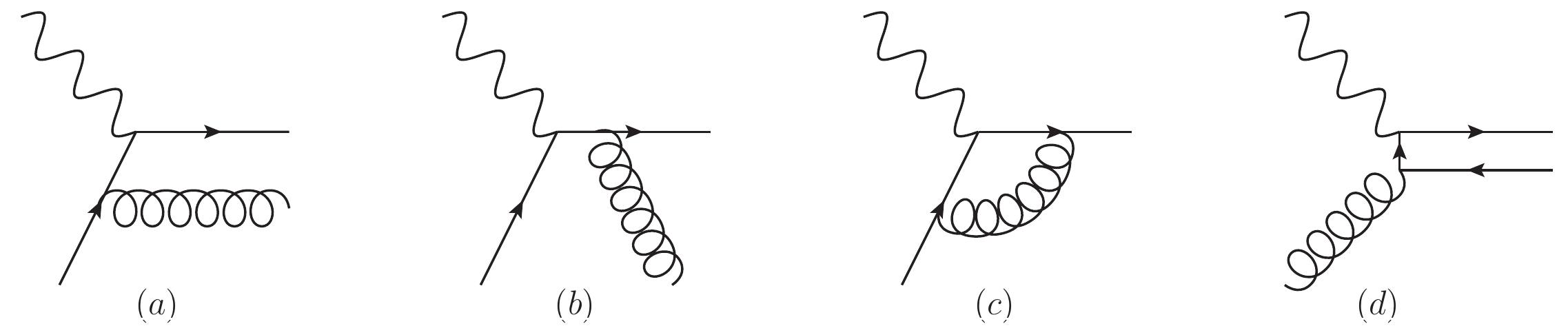}
\end{figure}

All four of these diagrams are separately divergent. When appropriately regularised however, the divergences in the final state real emission and virtual correction diagrams cancel explicitly as a consequence of the IR safety of QCD, yielding a finite contribution to the cross section. However the divergences present in the real emission diagrams from the initial state partons are not subject to the same cancellations, as they modify the momenta at the interaction vertex. 
 
Like the real emission diagram of a gluon from an initial state quark, the initial state gluon diagram (d) suffers from an equivalent divergence mediated by a perturbatively calculable $g\to q\bar{q}$ splitting function $P_{gq}$. Including all of the finite contributions from the other contributing diagrams as the coefficient $W(x)$, the parton level structure function at next to leading order in QCD is given by
\ba
 \widetilde{F_2^i}(\xi,Q^2) &=& 2 e_i^2\left[ \delta(\xi-x) \right. \nonumber\\
 				 &+& \frac{\alpha_S}{2\pi}\sum_j\left(P_{ij}(\xi)\log\frac{Q^2}{\kappa^2} + W_{ij}(\xi)\right) \nonumber\\
				 &+&  \left. \mathcal{O}(\alpha_S^2) \right]. \label{eq:f2plnlo}
\ea
Here the $i$ once again refers to the partonic species at the interaction vertex, and we have introduced an infrared cutoff $\kappa$ to regulate the parton splitting. The sum over splitting functions arises from the multiple contributions from partonic species $j$ splitting to $i$:
\begin{figure}[ht]
\centering
\includegraphics[scale=0.6]{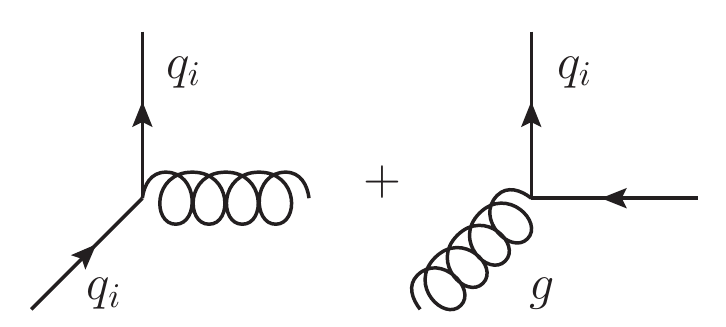}
\end{figure}

The splitting functions $P_{ij}$  were known for some time at leading and next-to-leading accuracy \cite{Gross:1973ju,Georgi:1951sr,Floratos:1977au,Altarelli:1977zs,GonzalezArroyo:1979df,Floratos:1978ny,Furmanski:1980cm,Curci:1980uw,GonzalezArroyo:1979he,Floratos:1981hs,Hamberg:1991qt}, and more recently extended to next-next-to-leading order accuracy \cite{Moch:2004pa,Vogt:2004mw}. After convoluting the parton level functions with the PDFs, we obtain the full structure function 
\ba
 F_2(x,Q^2) &=& \sum_i xe_i^2\left[\; f_i(x) \right.  \nonumber\\
 				 &+& \frac{\alpha_S}{2\pi}\int_0^1 \frac{d\xi}{\xi}\sum_j\left(P_{ij}\left(\frac{x}{\xi}\right)\log\frac{Q^2}{\kappa^2} + W_{ij}(x)\right)\; f_j(\xi) \nonumber \\
				 &+&  \left. \mathcal{O}(\alpha_S^2) \right]. \label{eq:f2nlo}
\ea
Our expression for the parton level structure function still suffers from the IR divergence when we take the limit $\kappa\to 0$. This issue may be resolved by concluding that the singularity arises from a breakdown of the ability of perturbation theory to describe physics in the strongly-coupled infrared. We may therefore attempt to factorise out the long distance behaviour of the structure functions into some bare parameters of the theory; analogously to the treatment of ultraviolet divergences by renormalisation of the strong coupling. In this instance we shall absorb the divergences present in the parton level structure functions into our parton distribution functions by replacing the bare quantities $f(x)$ with a physically accessible quantity measured at the \emph{factorisation scale} $\mu_f$. We can express these in terms of an expansion in the bare PDFs as
\be
f_i(x,\mu_F^2) = f_i(x) + \frac{\alpha_S}{2\pi}\int_0^1 \frac{d\xi}{\xi} \Delta^{(1)}_{ij}\left(\frac{x}{\xi}, \frac{\mu_F}{\kappa}\right)\; f_j(\xi) + \mathcal{O}(\alpha_S^2),
\ee
where the counter terms $\Delta^{(n)}_{ij}$ are formed as a sum of a regular part $\Delta^{(n)}_{r,ij}$ and a singular part $\Delta^{(n)}_{s,ij}$, and the sum over the dummy index $j$ is implicit. The singular part of these counterterms is uniquely specified by having to remove the divergence present in the structure functions due to the collinearly divergent parton splitting. Comparing to Eqn.~\ref{eq:f2nlo}, this divergence may be subtracted by setting
\be
\Delta^{(1)}_{s,ij} = P_{ij}\left(\frac{x}{\xi}\right)\log\frac{\mu_F^2}{\kappa^2}.
\ee
Unlike the divergent part, the regular part of the counter-term is not uniquely defined by the factorisation procedure. The choice of a specific regular counter-term is known as a \emph{factorisation scheme}; a choice consisting of shuffling terms between the regular part of the PDF definition and the coefficients present in the calculation. For example one may make a process-specific choice where all of the regular coefficients are absorbed into the PDF definition. In our example case of $F_2$ this is known as the DIS scheme~\cite{Altarelli:1978id}, $\Delta^{(1)}_{r,ij} = W_{ij}(x)$, in terms of which the form of the calculation becomes particularly simple:
\be
 F_2(x,Q^2) = 2 \int_0^1 \xi d\xi \sum_i  f^{\mathrm{DIS}}_i(\xi) e_{i}^2.
\ee
In practice this scheme choice is often rather unhelpful, as it does not permit a consistent definition of PDFs across multiple processes. With this in mind, the most common choice is the \emph{Modified Minimal Subtraction} or $\overline{\mathrm{MS}}$ scheme where the only regular counterterms are a process independent $\Delta^{(1)}_{r,ij} = \log 4\pi - \gamma_E$. In the $\overline{\mathrm{MS}}$ scheme therefore our factorised PDFs are given by
\be 
f_i(x,\mu_F^2) = f_i(x) + \frac{\alpha_S}{2\pi} \sum_j  \left[\left(P_{ij}\left(x\right)\log\frac{\mu_F^2}{\kappa^2} + \log 4\pi - \gamma_E \right) \right] \otimes f_j(x) + \mathcal{O}(\alpha_S^2), \label{eq:renormpdf}
\ee
and the expression for $F_2$ becomes
\be
F_2(x,Q^2) = x \sum_i e_i^2 \left\{ f_i(x,\mu_F^2) +  \frac{\alpha_S}{2\pi}\int_x^1 \frac{d\xi}{\xi} f_i(\xi,\mu_F^2)\;\widetilde{W}_i\left(\frac{x}{\xi},\frac{Q^2}{\mu_F^2},\alpha_S\right) \right\},
 \ee
where the $\widetilde{W}_i$ are the finite contributions remaining after factorisation. While the relationship between the PDFs at the factorisation scale and the bare distributions is now divergent, the renormalised quantities may be measured at some scale and used in subsequent calculations, thus making the theory predictive. In general, under a universal factorisation scheme such as $\overline{\mathrm{MS}}$, structure functions may be calculated as 
\be F(x,Q^2) = \sum_i \int_x^1 \frac{d\xi}{\xi} C_i\left(\frac{x}{\xi},\frac{Q^2}{\mu_F^2}, \alpha_S \right) f_i(\xi,\mu_F^2), \label{eq:DISsf} \ee
where the $C_i$ are the finite Wilson coefficients determined perturbatively and the PDFs $f_i$ encode the non-perturbative structure of the calculation. This differs from the naive parton model in that the Bjorken-scaling is now broken by logarithms of the hard scale $Q^2$, and the sum over parton species not only runs over spin-$1/2$ partons (the quarks of QCD), but also contains a contribution from an initial state gluon splitting into a quark-antiquark pair. 
\subsection{DGLAP and PDF evolution} \label{sec:DGLAP} As a measurable quantity, the structure function itself clearly must be independent of the unphysical factorisation scheme and scale choices. The requirement of scheme independence is of course met when the factorisation scheme is followed consistently for the definition of PDFs and Wilson coefficients in all subsequent calculations. The requirement of factorisation scale independence leads to a renormalisation group equation (RGE) for the structure function
\be \mu_F \frac{d}{d\mu_F} F(x,Q^2) = 0,\ee
and consequently RGEs for the parton distributions and Wilson coefficients, once again in terms of the Altarelli-Parisi splitting functions $P_{ij}$
\be \mu_F \frac{d}{d\mu_F}f_i(y,\mu_F^2) = \sum_j \int_z^1 \frac{dz}{z} P_{ij}\left(\frac{y}{z},\alpha_S \right) f_j(z,\mu_F^2), \label{eq:DGLAP}\ee
\be \mu_F \frac{d}{d\mu_F}C_i\left(x,\frac{Q^2}{\mu_F^2}, \alpha_S \right) = -\sum_i \int_z^1 \frac{dy}{y} C_j\left(y,\frac{Q^2}{\mu_F^2}, \alpha_S \right) P_{ij}\left(\frac{x}{y},\alpha_S \right).\ee
These are known as the Altarelli-Parisi equations~\cite{AP} or the Dokshitzer-Gribov-Lipatov-Altarelli-Parisi (DGLAP) equations~\cite{dokshitzer,gribovlipatov,lipatov}, and they describe how PDFs change, or \emph{evolve} with the factorisation scale. Identically as the RGE for the running of the strong coupling performs a resummation of contributions arising from self energy diagrams, the DGLAP equation resums scale logarithms arising from collinear parton splittings. 

The equations may be greatly simplified by moving to a PDF basis that largely diagonalises the matrix of splitting functions $P_{ij}$. For example we may construct a basis of \emph{non-singlet} PDFs, e.g the valence distributions
\be V_i = q_i - \bar{q_i}, \ee
and differences between quark sea distributions $q_s = q + \bar{q}$
\begin{eqnarray}
T_3 &=& u_s - d_s, \\
T_8 &=& u_s + d_s - 2s_s,  \\
T_{15} &=& u_s + d_s +s_s - 3c_s, \\
T_{24} &=&  u_s + d_s +s_s + c_s - 4b_s, \\
T_{35} &=&  u_s + d_s +s_s + c_s + b_s - 5t_s. \label{eq:evolbasis2}
\end{eqnarray}
As QCD is flavour blind, the gluon contribution to the evolution of these PDFs cancels, therefore diagonalising the matrix of splitting functions in this basis. For the nonsinglet distributions the DGLAP equation reduces to
\be \mu_F \frac{d}{d\mu_F}f^{\mathrm{NS}}_i(y,\mu_F^2) =\int_z^1 \frac{dz}{z} P^{\mathrm{NS}}_{i}\left(\frac{y}{z},\alpha_S \right) f^{\mathrm{NS}}_i(z,\mu_F^2).\label{eq:NSDGLAP}\ee
Completing this basis are the gluon and the flavour singlet $\Sigma = \sum_i (q_i +\bar{q}_i)$ PDFs. These remain coupled leading to a $2\times 2$ matrix of integro-differential equations for their evolution:
\be
 \mu_F \frac{d}{d\mu_F} 
 \begin{pmatrix} g(x,\mu_F) \\  \Sigma(x,\mu_F) \end{pmatrix}  =
\int_z^1 \frac{dz}{z} 
  \begin{pmatrix} P_{gg} & P_{g\Sigma} \\  P_{\Sigma g} & P_{\Sigma\Sigma} \end{pmatrix} 
   \begin{pmatrix} g(z,\mu_F) \\  \Sigma(z,\mu_F) \end{pmatrix}. \label{eq:gSDGLAP}\ee
These equations may be solved for a PDF at some scale $Q^2$ evolved from an initial scale $Q_0^2$. Solutions typically follow one of two procedures; arguably the most direct consists of solving the equations iteratively through numerical methods in $x$-space. This method is followed in codes such as HOPPET~\cite{Salam:2008qg}, QCDNUM~\cite{Botje:2010ay} and APFEL~\cite{Bertone:2013vaa} which employ interpolation techniques to improve the speed of the solution. Alternatively the DGLAP equations may be solved by making use of the Mellin convolution theorem
\be \mathcal{M}\left\{f \otimes g\right\} = \mathcal{M}\left\{f\right\}\cdot \mathcal{M}\left\{g\right\}, \ee
whereby the multiplicative convolution present in equations \ref{eq:NSDGLAP}, \ref{eq:gSDGLAP} is reduced to a product in Mellin space; the method employed by QCD-Pegasus~\cite{Vogt:2004ns}. In the Mellin space approach, the emphasis largely lies on a fast numerical implementation of the Mellin inversion integral.

Through either method, the solution of the DGLAP equations provides a perturbative description of the behaviour of parton distributions as they vary in scale. However we remain short of a full description of the distributions having not determined their dependence upon the momentum fraction $x$. Furthermore the precise behaviour of the PDF and structure function renormalisation may be complicated in the attempt to overcome some of the approximations we have made so far regarding the masses of quarks contributing to our parton model, which we shall address here before discussing how the $x$ behaviour of the PDFs may be elucidated.
\clearpage
\section{Treatment of heavy quarks}
So far in our discussion of the QCD parton model we have made the assumption that all the quarks contributing in the theory are massless, an approximation that becomes increasingly untenable when investigating scattering processes with a hard scale approaching a quark's physical mass. A careful treatment of terms depending on quark masses is therefore vital for making theoretical predictions to a dataset that spans heavy quark mass thresholds. 

Dealing with heavy quark mass effects is a delicate issue in that different treatments generally have different regions of applicability. The specific combination of approaches to quark masses used when confronting a dataset with a broad reach in hard scale is known as a heavy quark \emph{scheme}, although not necessarily in the spirit of factorisation or renormalisation schemes as the choice often lies in the particulars of the approximation rather than in some arbitrary shuffling of parameters. A heavy quark scheme choice can therefore potentially lead to differences with alternative calculations that do not in principle vanish in the limit of an all-orders calculation.

The space of heavy quark renormalisation schemes is bounded by two regimes where the treatment is fairly simple, the fixed flavour number scheme (FFNS) and the zero-mass variable flavour number scheme (ZM-VFNS). The remaining schemes, known as general-mass variable flavour number schemes (GM-VFNS) aim to interpolate between the FFNS and ZM-VFNS, reducing to the simpler calculations in certain kinematic limits. Motivated by observations suggesting that a more careful treatment of quark mass effects is phenomenologically relevant at the LHC~\cite{Tung:2006tb}, a number of such schemes have arisen in an attempt to better describe experimental data. These typically differ by sub-leading terms in the method of interpolation between the two limiting regimes. We shall now outline some of the available choices and their potential impact in the case of a deep-inelastic scattering analysis. For simplicity we shall discuss a theory with $n_l$ light quarks, and attempt to introduce a single massive quark $h$ with mass $m_h$.

\subsection{The FFN and ZM-VFN schemes}
We consider first the kinematical regime where the hard scale of our scattering problem is of similar order or smaller than our heavy quark mass; $Q^2 \lesssim m_h^2$. Making the assumption that the initial state proton has no intrinsic heavy quark component it is reasonable to treat the heavy quark as a purely final state particle, and the only partons in the theory are the $n_l$ light quark flavours and the gluon. In this instance, setting the factorisation and renormalisation scales $\mu_F^2=\mu_R^2 = \mu^2$; the calculation of a structure function in Eqn.~\ref{eq:DISsf} takes the form
\be F(n_l, Q^2, m_h^2) = \sum_i^{n_l}  C_i\left(n_l, \frac{Q^2}{m_h^2}, \frac{\mu^2}{m_h^2}, \frac{Q^2}{\mu^2} \right) \otimes f_i(n_l, \mu^2), \label{eq:FFN} \ee
where the sum is over light quark flavours only and the full mass dependence of the heavy quark is intact in the calculation. The structure function can be separated into a contribution where only light flavours are present, $F^{L}$, and a contribution including the heavy flavour $F^{H}$ as,
\be F(n_l, Q^2, m_h^2) = F^{L}(n_l, Q^2) + F^{H}(n_l, Q^2,m_h^2), \ee
where
\ba
F^{L}(n_l, Q^2) &=& \sum_i^{n_l}  L_i\left(n_l, \frac{Q^2}{\mu^2} \right) \otimes f_i(n_l, \mu^2),\\
F^{H}(n_l, Q^2,m_h^2) &=& \sum_o^{n_l} H_i\left(n_l, \frac{Q^2}{m_h^2}, \frac{\mu^2}{m_h^2}, \frac{Q^2}{\mu^2} \right) \otimes f_i(n_l, \mu^2).
\ea
Here $L$ denotes the Wilson coefficients that do not contain heavy quark lines, and $H$ includes only the diagrams that do. In this instance the heavy quark structure function first contributes at $\mathcal{O}(\alpha_S)$ via the splitting of an initial state gluon into a $h\bar{h}$ pair:
\begin{figure}[ht]
\centering
\includegraphics[scale=0.6]{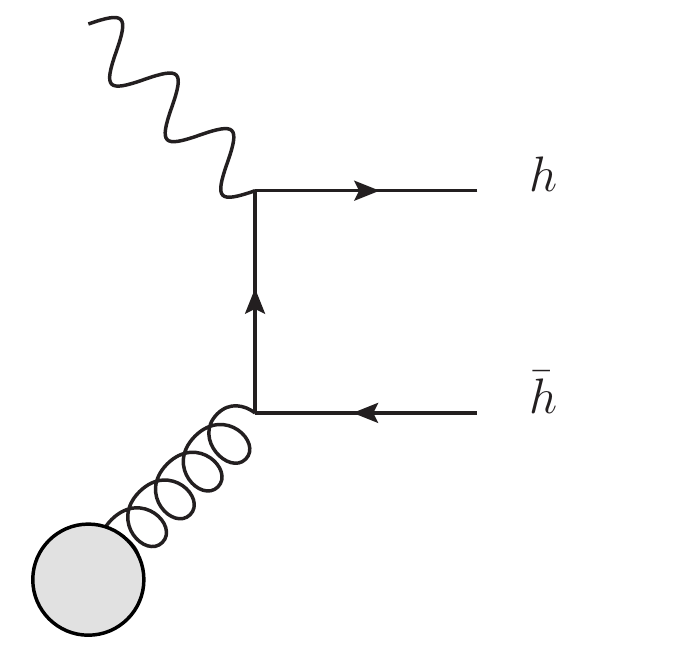}
\end{figure}

This approach is known as the \emph{decoupling} or FFN scheme where the only quarks treated as partons are the $n_l$ light quarks. The expression in Eqn.~\ref{eq:FFN} is unique up to terms of order $m_l^2/Q^2$ in the light quark masses which are typically treated as part of the factorisation level corrections of $\mathcal{O}(\Lambda^2_{\mathrm{QCD}}/Q^2)$. While accurate in the quark mass threshold region and below, this scheme suffers from unresummed logarithms of the ratio $Q^2/m_h^2$ contained in the Wilson coefficients which can become large and damage the convergence of the perturbative series at scales much larger than the heavy quark mass.

These problems may be resolved in a scheme which treats the heavy quark as a massless parton above its mass threshold with the introduction of an associated heavy quark PDF. The subsequent renormalisation of the PDF resums the logarithmic contributions due to parton splitting via solution of the DGLAP equation, removing a significant disadvantage present in the FFN treatment. As this scheme is identical to the zero mass scheme discussed previously, but with an additional partonic flavour, this procedure is known as the Zero-Mass Variable Flavour Number (ZM-VFN) scheme. In the ZM-VFN a structure function calculation is simply

\be F(n_l+1, x,Q^2) = \sum_i^{n_l+1} C_i\left(n_l+1,\frac{Q^2}{\mu^2} \right) \otimes f_i(n_l+1,\mu^2). \label{eq:ZMVFN} \ee
In this instance the heavy quark contribution to the structure function first arises now at leading order via diagrams of the type:

\begin{figure}[h]
\centering
\includegraphics[scale=0.6]{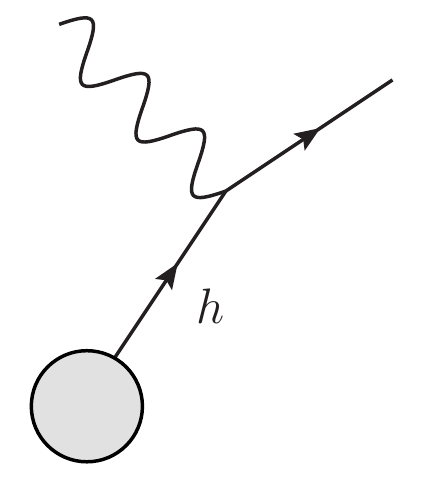}
\end{figure}

In the ZM-VFNS the heavy quark PDFs are set to zero below mass threshold and evolved as a massless parton according to the DGLAP equations for scales greater than the heavy quark mass. While this method alleviates the difficulties present in the FFN scheme at large scales, its treatment of heavy quarks only in terms of massless partons completely ignores the massive contributions to the Wilson coefficients and is therefore no longer exact. The reliability of the ZM scheme is therefore particularly reduced in the region where powers of $m_h^2/Q^2$ are significant. 
\subsection{General mass schemes}
Analyses of QCD measurements are often performed by making a choice between using a suitable FFN scheme at scales in the region of heavy quark mass thresholds or a ZM scheme at high scales where the associated powers of $m_h^2/Q^2$ can be safely neglected. In either case the treatment of heavy quarks is at least unambiguous, with the ZM approach yielding a simpler procedure as there is no requirement to calculate coefficient functions with the heavy quark masses intact.

For analyses of a large dataset, potentially spanning several heavy quark thresholds and extending to very high scales, the desire to improve the perturbative reliability of the calculations has led to the development of a number of hybrid or \emph{general mass} schemes. In such schemes the treatments generally reduce to the FFN regime at low scales and the ZM treatment at high scales, with the intermediate regime handled via some interpolation between the two. Generally in a Variable Flavour Number (VFN) scheme one requires that
\be F^{L}(n_l, Q^2) + \lim_{Q^2 \gg m_h^2} \left[ F^{H}(n_l, Q^2,m_h^2)\right] =  F(n_l+1, x,Q^2), \label{eq:VFN}\ee
i.e. that the ZM-VFN and FFN calculations coincide at large scales, where the heavy quark mass dependance of the FFN Wilson coefficients can be neglected. The constraint in Eqn. \ref{eq:VFN} means that parton distributions in the two schemes may be related by a perturbatively calculable transformation.

\be f_i(n_l+1,\mu^2) = \sum_j^{n_l}A_{ij}\left(n_l, \frac{\mu^2}{m_h^2}\right)\otimes f_j(n_l, \mu^2), \label{eq:FLVreln}\ee
where the $A$ are determined to NNLO in $\alpha_S$ in Refs.~\cite{Buza:1995ie,Buza:1996wv}. It should be noted that the $A$ are not square matrices, with $i$ running over the $n_l+1$ partons in the zero mass scheme, and the $j$ running over the $n_l$ partons in the FFN.

In general a GM-VFN operates as a tower of FFN-type schemes with increasing $n_l$ as the scale increases over each quark mass threshold. In constructing a GM-VFN, the guiding principle is that physical observables should be continuous across these thresholds and therefore continuous across the $n_l$ and $n_l+1$ regimes. 
Taking the heavy quark mass itself as the matching point between the two regimes, we demand that the GM-VFN structure function $F^{\text{GM}}$ obeys
\ba 
F^{\text{GM}}(m_h^2) &=& \sum_j^{n_l} C^{\text{GM}}_j \left(n_l, m_h^2 \right) \otimes f_j(n_l) \nonumber \\
&=&   \sum_i^{n_l+1} C^{\text{GM}}_i\left( n_l+1, m_h^2 \right) \otimes f_i(n_l+1). \label{eq:GMmatching} 
\ea 
where the dependance upon the perturbative scales has been omitted for notational simplicity, and the GM superscripts refer to the coefficients in a general mass scheme. Using the relation in Eqn. \ref{eq:FLVreln} we can express the $n_l+1$ expression in the matching Eqn. \ref{eq:GMmatching} in terms of the $n_l$ scheme PDFs, therefore obtaining the relation
\ba && \sum_j^{n_l} C^{\text{GM}}_j \left(n_l, m_h^2 \right) \otimes f_j(n_l) \\
=  && \sum_i^{n_l+1} \sum_j^{n_l } C^{\text{GM}}_i\left( n_l+1, m_h^2 \right) \otimes A_{ij}\left(n_l, m_h^2\right)\otimes f_j(n_l). \ea
Subsequently, we may make the identification
\be  C^{\text{GM}}_j \left(n_l, m_h^2 \right) 
=   \sum_i^{n_l+1} C^{\text{GM}}_i\left( n_l+1,m_h^2 \right) \otimes A_{ij}\left(n_l, m_h^2\right),\label{eq:minimalGM}\ee
which provides the minimal description for the construction of a GM-VFN scheme~\cite{Thorne:2008xf}. Ensuring that Eqn.~\ref{eq:minimalGM} is satisfied order by order in $\alpha_S$, we can construct the expression for the GM-VFN scheme coefficient functions above the heavy quark mass threshold. Taking the simplistic example case of Ref.~\cite{Kramer:2000hn} with a theory including only a gluon and a single heavy quark ($h=\bar{h}$), the GM-VFN coefficients may be constructed to order $\alpha_S$ as
\ba
C_g^{\text{LO}}(n_l+1,m_h) &=& C_g^{\text{LO}}(n_l,m_h), \\
C_g^{\text{NLO}}(n_l+1,m_h) &=& C_g^{\text{NLO}}(n_l,m_h) \nonumber \\
&-& C_h^{\text{LO}}(n_l+1,m_h)\otimes A^{\text{LO}}_{hg}(n_l, m_h^2). \label{eq:ACOT}
\ea
where the GM superscript has been omitted, the new superscript specifying the order of the term in the perturbative expansions of the quantities $C$ and $A$. Here the rightmost term in the $\mathcal{O}(\alpha_S)$ expression Eqn.~\ref{eq:ACOT} is known as the \emph{subtraction term} which ensures the cancellation of the IR-unsafe scale logs present in the FFN calculation. The ambiguity in the definition of a GM-VFNS arises upon noticing that terms proportional to powers of $m_h/Q$ may be interchanged between the Wilson coefficients in Eqn.~\ref{eq:ACOT} without changing the final value of the structure function. In this respect changing the distribution of terms between the gluon and heavy quark initiated diagrams in Eqn.~\ref{eq:ACOT} provides the opportunity to perform a \emph{scheme choice}, a freedom which has been exploited by several different GM-VFN scheme implementations. 

The earliest complete description of a GM-VFNS was provided by the ACOT method~\cite{Collins:1978wz} which ensures the continuity of physical quantities through Eqn.~\ref{eq:minimalGM}, but does not attempt to take advantage of the degeneracy in the GM-VFN procedure. An important result was achieved with the Simplified-ACOT or S-ACOT  scheme~\cite{Collins:1998rz,Kramer:2000hn} which was able to exploit this ambiguity to considerably simplify the calculation of physical observables. In the S-ACOT scheme it was noted that shifts of the Wilson coefficients by their zero-mass limits may be absorbed into a redefinition of the GM-VFNS. That is, terms such as
\be C_h(n_l+1, m_h) - C_h(n_l+1, 0),\ee
vanish in the limit $Q^2 \gg m_h^2$, and therefore do not spoil the interpolation between the FFN and ZM schemes. This leads to the option of shifting to a simpler scheme where the massive heavy quark initiated coefficients may instead be evaluated with the heavy quark mass set to zero. Other options for the scheme definition were explored by Thorne and Roberts in the TR type schemes~\cite{Thorne:1997uu,Thorne:1997ga}, with the additional constraint that scale derivatives of heavy flavour structure functions should also be continuous at the matching scale.
\subsubsection{The FONLL approach}
A more recent approach was developed by examining methods previously used to combine fixed order calculations with next-to-leading log resummation via the FONLL method~\cite{Cacciari:1998it}. The method was extended from the original application of studying the \pt spectrum in heavy flavour hadroproduction to the treatment of heavy quarks in DIS by Forte \emph{et al.}~\cite{Forte:2010ta}. The procedure begins by inverting the relationship in Eqn.~\ref{eq:FLVreln} so as to express an $n_l$ flavour structure function in terms of $n_l+1$ flavour PDFs,
\be F(n_l,Q^2) = \sum_i^{n_l} B_i\left(\frac{Q^2}{m_h^2}\right)\otimes f_i(n_l+1,Q^2), \label{eq:FONLLmassive}\ee
where it is important to note that the sum over flavours does not include the heavy flavour PDF, and the full heavy quark mass dependence is present in the coefficients $B$. To perform a matching with the massless scheme, the ZM result in Eqn.~\ref{eq:ZMVFN} can be expressed in terms of light flavour PDFs only, given the assumption that the heavy flavour PDF is generated perturbatively. In this case, Eqn.~\ref{eq:ZMVFN} can be written
\be F(n_l+1, Q^2) = \sum_i^{n_l} \widetilde{C}_i\left(n_l+1,\frac{Q^2}{m_h^2} \right) \otimes f_i(n_l+1,\mu^2). \label{eq:FONLLmassless}\ee
where once again, the sum runs over only light flavours, this time with the heavy flavour contribution being generated via DGLAP evolution included into the modified coefficient function $\widetilde{C}$. To understand which terms are common in the two descriptions, the massive coefficient functions may be decomposed into terms logarithmically dependant upon the heavy quark mass, and terms suppressed by powers of $m_h/Q$: 
\be B_i\left(\frac{Q^2}{m_h^2}\right) = \overline{B}_i\left(\frac{Q^2}{m_h^2}\right) + \mathcal{O}\left(\frac{m_h}{Q}\right).\ee  
As only the power suppressed terms vanish in the limit of $Q^2 \gg m_h^2$, the terms remaining must be common to both the ZM and massive scheme calculations. We can therefore express the massive structure function in a `\emph{massless}' limit, having dropped those terms in the coefficient functions that are suppressed by powers of $m_h/Q$:
\be \overline{F}(n_l,Q^2) = \sum_i^{n_l} \overline{B}_i\left(\frac{Q^2}{m_h^2}\right)\otimes f_i(n_l+1,Q^2).\label{eq:FONLLdoublecount}\ee
The FONLL result for the structure function is given by the sum of the massive calculation in Eqn.~\ref{eq:FONLLmassive}, and the massless calculation in Eqn.~\ref{eq:FONLLmassless} with the asymptotic limit of the massive calculation in Eqn.~\ref{eq:FONLLdoublecount} subtracted. 
\be
F^{\mathrm{FONLL}}(Q^2) =\left[ F(n_l,Q^2) + F(n_l+1, Q^2) \right] - \overline{F}(n_l,Q^2).
\ee
With the subtraction ensuring the cancellation of terms which are double counted between the massive and massless calculations. Therefore in this expression the mass-suppressed terms present in the FFN calculation are fully accounted for in the GM scheme, with the duplicate terms subtracted. The simplicity of this approach helped to elucidate many of the differences between general mass schemes.

It should be noted that while general mass schemes suffer from an ambiguity in their definition compared to the simpler fixed-flavour and zero mass schemes, the differences between them are always of higher order compared to the calculation at hand, as is the case in any true scheme choice. Indeed, a well-defined GM-VFNS will always reduce to the decoupled result at low scales and the zero-mass result at scales much higher than the quark mass, behaving effectively as a tower of fixed flavour schemes with increasing number of partonic quarks. The general-mass schemes therefore do not suffer from a significant loss of predictive power, and are able to provide considerable improvement over the simpler schemes when dealing with datasets spanning quark mass thresholds.
\section{General features of parton distributions}
While we have now described how the parton distributions functions at an experimental scale $Q^2$ may be found by evolving parton distributions from an initial scale, and discussed briefly how the renormalisation of heavy quark distributions may be accomplished, the issue of determining the functional dependence of the parton distributions upon the momentum fraction $x$ at some initial scale $f_i(x,Q_0^2)$ remains. 

The number of independent PDFs to be determined is dependent upon the choice of initial scale, as quark distributions that can be considered \emph{heavy} with respect to $Q^2_0$ may be generated perturbatively through the DGLAP procedure outlined previously. The typical choice is to determine the parton distributions at some scale $ m_s^2< Q^2_0 \le m_c^2$ such that the flavours $c$, $b$, $t$ are produced by evolution. These scale choices minimise the number of distributions to be determined while remaining perturbatively reliable.

As the remaining seven distributions\footnote{The gluon, the $u$, $d$, $s$ quarks and their antiquarks.} are fundamentally a parametrisation of the nonperturbative dynamics of the proton, they are by definition out of reach of a perturbative analysis. There are however some general statements that may be made of their $x$-dependence that are independent of the hard scale. The most important of which are the parton distribution \emph{sum rules} which constrain the relative normalisation of PDFs.

Firstly, the \emph{momentum sum rule} (MSR) ensures that the parton distributions' fractional momenta sum to the momentum of the parent proton
\be  \int_0^1 dx \left[  x\Sigma(x,Q^2) + xg(x,Q^2) \right]= 1, \label{eq:MSR} \ee
where $\Sigma$ is the singlet distribution defined previously. Following this are the quark valence sum rules. These fix the quark distributions such that the resulting proton has the appropriate quantum numbers,
\begin{subequations}
\label{eq:VSR}
\ba \text{up-valence:}& \: &\int_0^1 dx\left( f_u(x,Q^2) - f_{\bar{u}}(x,Q^2) \right) = 2, \label{eq:UVSR} \\
      \text{down-valence:}&  \: & \int_0^1 dx\left( f_d(x,Q^2) - f_{\bar{d}}(x,Q^2) \right) = 1, \label{eq:DVSR}\\
      \text{strange-valence:}&  \: & \int_0^1 dx\left( f_s(x,Q^2) - f_{\bar{s}}(x,Q^2) \right) = 0. \label{eq:SVSR} 
\ea
\end{subequations}

From these rules we may infer additional constraints upon individual PDFs. The MSR suggests a form for the large-$x$ behaviour of the distributions, in that they should parametrically tend to zero as $x\to1$. The number sum rules in Eqns. \ref{eq:VSR} require the valence-type distributions to be integrable over the whole $x$-range. While there is no requirement for the singlet and gluon distributions to be integrable, their first moments must be, as required by the MSR. Combining these three constraints we may parametrise the large and small-$x$ behaviour of both valence-like and gluon or singlet-like distributions as:
\ba f_V(x,Q_0^2) &= N_V\;x^{\alpha_V}(1-x)^{\beta_V}\,r_V(x), \nonumber \\
 f_\Sigma(x,Q_0^2) &= N_\Sigma\;x^{\alpha_\Sigma}(1-x)^{\beta_\Sigma}\,r_\Sigma(x). \label{eq:pdflimits}\ea
In these expressions, the parameters $\alpha$ and $\beta$ control the small and large-$x$ PDF behaviour respectively. The $\beta$ should be such that the PDFs tend to zero smoothly at large-$x$, and the $\alpha$ such that the valence distributions are integrable, and the first moment of the gluon and singlet are integrable. The overall PDF normalisations $N$ being constrained via the appropriate sum rules.

Finally, what remains in the determination of the distributions are the remainder terms $r(x)$ which describe the PDFs between the two $x$-limits. Their determination is considerably more complex and is a ongoing source of research. Much of this thesis will be dedicated to discussing the determination of these remainder functions.
\chapter{Review of PDF determination}
\label{ch:pdfdet}
Understanding the functional structure of parton distributions is a complex task that has been subject to a number of approaches over the years. As nonperturbative quantities describing the behaviour of QCD bound states, in principle they may be subject to analysis using Lattice QCD methods. While a great deal of effort and progress has been made in understanding PDFs through nonperturbative methods~\cite{Dolgov:2000ca,Horsley:2004uq,Gockeler:2004wp,Schroers:2005rm}, results remain short of providing distributions for practical application at hadron colliders.

The majority of PDF analyses are therefore performed analogously to the determination of many other QCD parameters; via a fit to appropriate experimental data. The fundamental difficulty in PDF fits being that they are determinations of \emph{functions} rather than single parameters and therefore one must attempt to find some optimum solution in an (in principle) infinite-dimensional functional parameter space. This is of course complicated by having only a finite set of experimental data points upon which to perform a fit. Moreover as the applications involving PDFs have become more precise, a detailed understanding of the uncertainties in the determination of PDFs has become vital. The problem of PDF fitting is therefore one of finding a reliable estimator for a probability distribution in a space of functions.

The complexity of the task, along with the inherent ambiguities in the QCD treatment of data, led to the emergence of several competing methodologies and determinations. Today there are a diverse array of fitting groups producing sets of parton distribution functions, the most important of which being the ABM~\cite{Alekhin:2013nda,Alekhin:2012ig} (formerly ABKM~\cite{Alekhin:2009ni}), CTEQ/CJ~\cite{Gao:2013xoa,Lai:2010vv,Nadolsky:2008zw,Owens:2012bv}, JR/GJR~\cite{JimenezDelgado:2008hf,Gluck:2007ck}, HERAPDF~\cite{Aaron:2009aa,::2014uva}, MSTW~\cite{Martin:2009iq,Martin:2012da} (formerly MRST~\cite{Martin:1998sq,Martin:2001es,Martin:2002dr,Martin:2004dh}) and NNPDF~\cite{Ball:2012cx,Ball:2011gg,Ball:2011uy,Ball:2011mu,Ball:2010de,Ball:2008by} groups. Typically PDF sets are provided for a variety of theory input parameters such as perturbative order, and value of the strong coupling. All modern PDF sets now include a quantitative assessment of their associated uncertainties. In this chapter we shall review the ingredients and methods utilised in a modern PDF determination, primarily focusing on the methodology of the three global PDF fits recommended for LHC phenomenology by the PDF4LHC working group~\cite{Botje:2011sn}, namely the procedures of the CTEQ, MSTW and NNPDF collaborations.

These three groups produce PDF sets determined from a fit to a wide range of experimental data, including DIS, Drell-Yan and inclusive jet cross sections. The CTEQ and MSTW determinations follow a similar fitting procedure and method of uncertainty estimation, with the NNPDF group taking a rather different approach to both. We will now describe the basic fitting procedure of these groups, with an eye to detailing areas where the groups have different solutions.

\section{Experimental data on parton distributions before the LHC}
The most important ingredient in the determination of parton distributions is naturally the selection of the dataset from which to extract PDF constraints. The first step in performing a PDF fit is therefore to identify which datasets are most sensitive to input parton distributions, and offer precise and reliable data. As PDF determinations to date have relied only upon fixed-order perturbation theory, the dataset chosen should probe sufficiently inclusive observables which are therefore relatively insensitive to resummation effects. In general PDF fitting collaborations also require data to be taken at a sufficiently high scale that leading-twist factorisation remains reliable, although there are some exceptions which we shall discuss later in the section. Here we shall briefly discuss some of the most important processes in terms of PDF sensitivity, and review some of the most relevant experimental measurements. For this section we shall restrict ourselves to data available before the start of LHC operation in order to provide a background for the methodological developments made in the light of LHC data. 

\subsubsection{Fixed-Target and collider DIS}
Deep inelastic scattering data provides the backbone for much of a PDF analysis, and data is available from a wide array of sources. Precise electron-proton scattering data from HERA provides the cleanest probe of proton structure function data, while high-luminosity fixed-target experiments can provide important constraints, at the expense of potentially having to deal with additional data corrections due to nuclear and higher-twist effects. As DIS is one of the best understood scattering processes in QCD, precise theoretical predictions are available up to 3-loop order in the zero-mass scheme~\cite{Vermaseren:2005qc,Moch:2008fj} and 2-loop order with full heavy quark masses intact~\cite{Buza:1997mg,Buza:1996xr,Blumlein:2006mh,Bierenbaum:2007qe,Bierenbaum:2007dm,Bierenbaum:2009zt,Blumlein:2014fqa}.

At leading order, neutral current DIS measurements from a proton target directly probe the quark sea distributions $q_i+\bar{q}_i$, with the relative power of each flavour contribution mediated via its coupling to $\gamma,Z$. Charged current, and $Z$-mediated neutral current data can provide some constraint upon PDF flavour separation via the $F^3$ structure function. 

In addition to proton structure function measurements, data obtained from scattering off deuterium targets can be important in constraining light quark flavour separation under the assumption of isospin symmetry. Data may be presented as direct measurements of $F_d$ structure functions or as the ratio $F_d/F_p$. A simultaneous fit to deuterium and proton data may therefore provide important constraints upon the $u-d$ and $u/d$ PDF combinations. Data determined via deuteron scattering are subject to nuclear corrections e.g. shadowing effects~\cite{Badelek:1994qg} which may be estimated as part of the theoretical treatment or neglected; the corrections to be considered part of the theory uncertainty.

Alongside the direct information on quark distributions, scaling violations present in structure function data provide constraints upon the gluon. While rather indirect, the wealth of DIS measurements available at a wide range of scales provides a great deal of information on the structure of the gluon distribution.

DIS data may be presented either as experimental cross sections, or separated into structure functions. Fixed target structure function data on $F_2$ from muon scattering is available for both proton and deuteron targets from the BCDMS~\cite{Benvenuti:1989rh,Benvenuti:1989fm}, NMC~\cite{Arneodo:1996qe,Arneodo:1996kd} and Fermilab E665~\cite{Adams:1996gu} experiments. Electron scattering $F_2$ data is also available from SLAC data on both proton and deuteron targets~\cite{Whitlow:1991uw}. The longitudinal structure function $F_L$ is measured in fixed target experiments also by SLAC~\cite{Whitlow:1990gk} , BCDMS~\cite{Benvenuti:1989rh} and NMC~\cite{Arneodo:1996qe}. 

In addition to the large datasets available from fixed target experiments, HERA data provides a clean probe of DIS properties, although with HERA data the separation of cross-sections into structure functions is typically not performed. Neutral current cross-section data is provided by ZEUS~\cite{Breitweg:1998dz,Chekanov:2001qu,Chekanov:2002ej,Chekanov:2003yv} and H1~\cite{Adloff:2000qk,Adloff:2000qj,Adloff:2003uh}. Charged-current DIS data is also provided by the HERA collaborations \cite{Chekanov:2003vw,Adloff:2003uh} along with information on the longitudinal structure function $F_L$~\cite{Andreev:2013vha,Chekanov:2009na}. Information on charm hadroproduction in DIS is available via $F_2^{\mathrm{charm}}$ measurements at HERA also~\cite{Adloff:1996xq,Adloff:2001zj,Aktas:2005iw,Aktas:2004az,Breitweg:1999ad,Chekanov:2003rb,Chekanov:2007ch}. This data provides particular constraint upon the gluon PDF, and has been an important testing ground for heavy quark flavour schemes. The clean $ep$ environment means that data is unaffected by nuclear or deuteron corrections, although low energy datapoints may still suffer from substantial higher-twist corrections. These corrections are typically kept under control by kinematic cuts on the affected points, however some groups (notably the ABM/CJ groups) include the affected data and attempt to model the corrections.

HERA measurements from the two collaborations have been examined as a combined analysis and dataset, so far resulting in two studies of direct interest to PDF determination; a combination of HERA-1 inclusive DIS data~\cite{aaron:2009wt}, and of charm production cross-sections~\cite{Abramowicz:1900rp}.

\begin{figure}[ht!]
\centering
\includegraphics[width=0.9\textwidth]{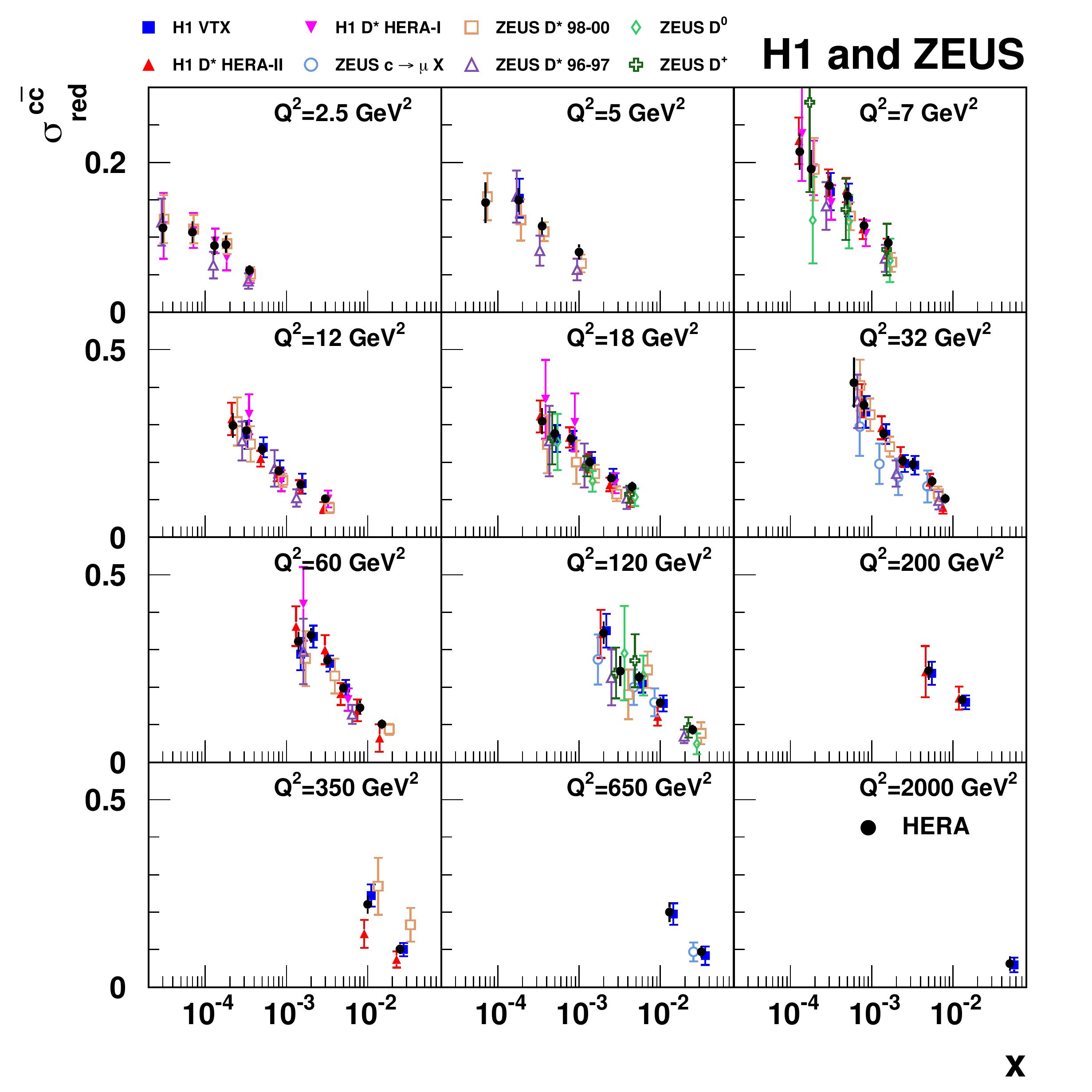}
\caption[Combined reduced charm cross-section data from HERA]{Reduced charm cross-section data from the HERA combined measurement. Data from the measurements contained in the combination analysis is shown for comparison. Figure from \cite{Abramowicz:1900rp}.}
\label{fig:HERAF2c}
\end{figure}

The very large quantity of deep-inelastic scattering measurements performed at a variety of experimental facilities means that generally DIS data forms the backbone for PDF fits, providing a substantial proportion of the experimental data points used in a fit.

\subsubsection{Neutrino DIS}
There are a number of measurements available for the scattering of neutrino beams from heavy nuclear targets. For example the NuTeV~\cite{Tzanov:2005kr} and CHORUS~\cite{Onengut:2005kv} data on neutrino $F_2$ and $F_3$. Assuming an approximately isoscalar target, and neglecting CKM factors, the PDF dependence of the neutrino structure function data at leading order is given by~\cite{Forte:2013wc}
\begin{eqnarray}
	F_2^\nu(x) &=& x\left( u^+(x) + d^+(x) + 2s(x) + 2\bar{c}(x)\right), \\
	F_2^{\bar{\nu}}(x) &=& x\left( u^+(x) + d^+(x) + 2\bar{s}(x) + 2c(x)\right),
\end{eqnarray}
and for the $F_3$ structure function,
\ba
	F_3^\nu(x) &=& x\left( u^-(x) + d^-(x) + 2s - 2\bar{c}\right), \\
	F_3^{\bar{\nu}}(x) &=& x\left( u^-(x) + d^-(x) - 2\bar{s}(x) + 2c(x)\right).
\ea
A simultaneous fit of these data points therefore provides a good handle upon the valence quark distributions $q-\bar{q}$. These datasets are relatively precise; however they are subject to potentially large nuclear corrections which introduce an uncertainty that is poorly understood.

\begin{figure}[h]
\centering
\includegraphics[width=0.5\textwidth]{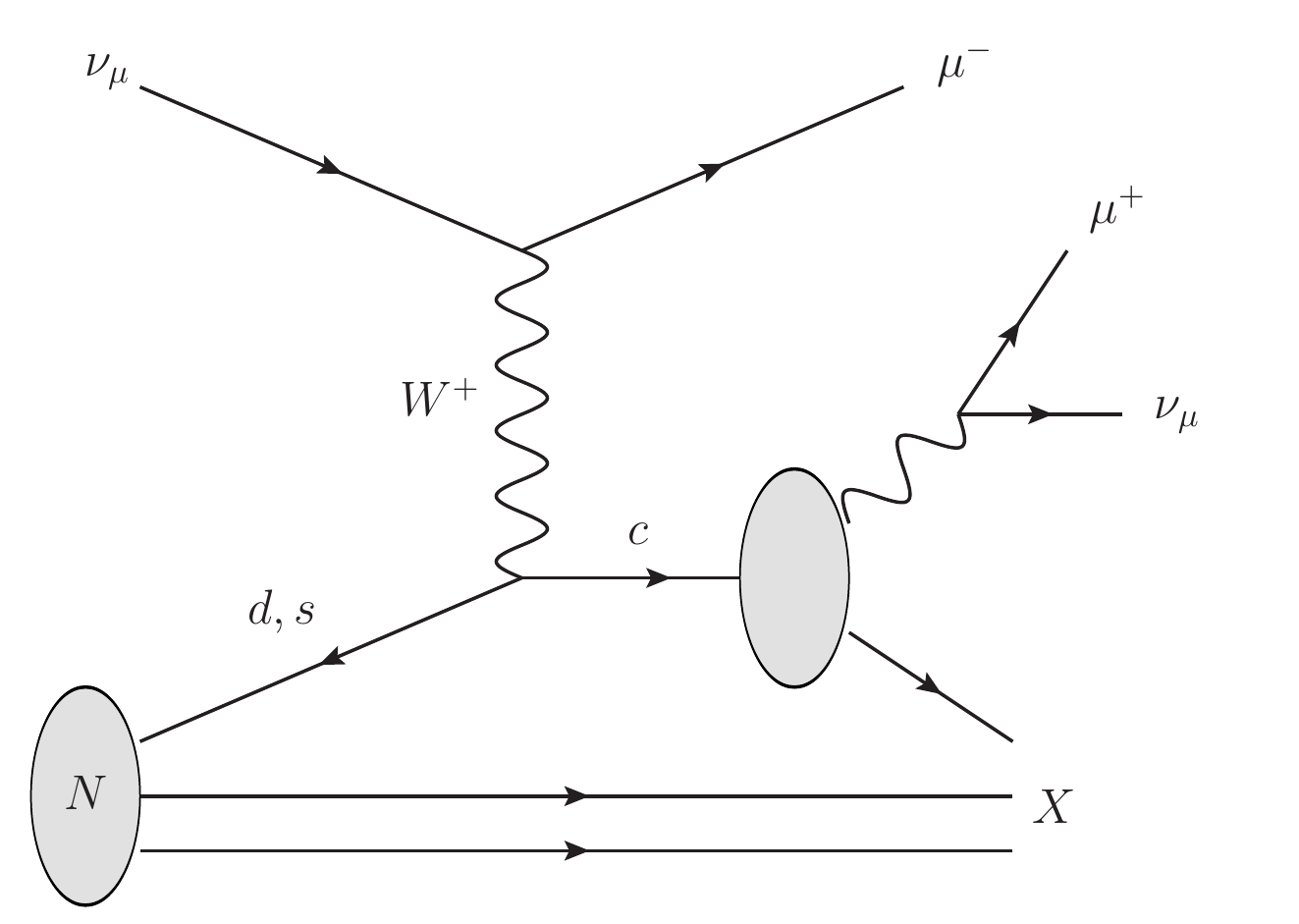}
\caption[Leading order diagram for dimuon production in neutrino DIS]{Leading order diagram for dimuon production in neutrino DIS.}
\label{fig:dimuon}
\end{figure}

Neutrino DIS becomes particularly valuable for PDF determination when considering the semi-inclusive DIS dimuon production process $\nu N \to \mu\mu X$ illustrated in Figure \ref{fig:dimuon}. In this process the contribution from initial state strangeness is Cabbibo favoured, therefore providing a direct handle on the strange distribution whose contribution is ordinarily difficult to discern from total structure function measurements. Measurements of this process are therefore commonly used as a strangeness probe, and data has been provided by the NuTeV/CCFR collaborations~\cite{Goncharov:2001qe}.
\subsubsection{Fixed-target and collider Drell-Yan}
After DIS measurements, the production of electroweak vector bosons in hadronic collisions provides the next most important contribution to the constraint of parton densities, with precise predictions available at NNLO in QCD~\cite{Anastasiou:2003ds,Catani:2009sm,Catani:2010en}.  At leading order the neutral current Drell-Yan process is moderated by the PDF combination
		\be q(x_1)\bar{q}(x_2) +  \bar{q}(x_1)q(x_2),\ee
and provides a direct probe of various partonic combinations depending upon the experimental configuration. In the Drell-Yan process the relevant kinematic variables are the invariant mass of the lepton pair
\be M_{ll}^2 = (E_1 + E_2)^2 - (\mathbf{p}_1 + \mathbf{p}_2)^2,\ee
and the intermediate boson's rapidity, given in the detector frame by
\be y = \frac{1}{2}\log \frac{E+ p_L}{E-p_L},\ee
where $E$ is the detector frame energy of the intermediate boson, and $p_L$ its longitudinal momentum.  in terms of which the parton-$x$ is given by;
\be x_\pm = M_{ll} e^{\pm y} / \sqrt{s}, \ee
where $s$ is the centre-of-mass energy squared of the reaction and the $\pm$ denotes the parton direction with respect to the beam frame. High rapidity measurements therefore constrain PDFs at both high and low-$x$. 

Additionally the charged-current process $qq^\prime \to l^\pm \nu_l$ provides information on quark flavour separation in the initial state hadrons. While the rapidity of the lepton pair resulting from $Z/\gamma$ decay in neutral current Drell-Yan is experimentally straightforward to distinguish, the presence of a neutrino in the final state of $W$ production processes complicates the direct resolution of the $W$ rapidity. Therefore data is often presented in the pseudorapidity of the detected lepton,
\be \eta = -\log \tan \theta,\ee
defined in terms of the angle $\theta$ between the final state lepton and the beam axis. It can therefore be measured without knowledge of the particle mass and momentum. The pseudorapidity coincides with the standard rapidity in the case of massless particles where $E = |\mathbf{\bar{p}}|$.

Lepton asymmetries are another common form for experimental results in Drell-Yan, defined in terms of $W^{\pm}\to l^\pm\nu_l $ differential cross-sections $d\sigma_{l^\pm}/d\eta_l$ as
\be 
  A^l_W=\frac{d\sigma_{l^{+}}/d\eta_{l}-d\sigma_{l^{-}}/d\eta_{l}}
  {d\sigma_{l^{+}}/d\eta_{l}+d\sigma_{l^{-}}/d\eta_{l}}, 
\ee
such measurements also benefit from the cancellation of shared systematic uncertainties. Measurements of lepton pair production from proton beams incident upon heavy nuclear targets, such as the E605\cite{Moreno:1990sf} experiment determining dimuon production from a copper target are useful for the constraint of the light quark sea $q+\bar{q}$. These measurements are typically very precise but suffer from poorly determined nuclear corrections. Several approaches have been performed to study the extent of these corrections~\cite{deFlorian:2003qf,Hirai:2007sx,Kulagin:2007ju,Eskola:2009uj}, although the effects are typically small and may sometimes be discounted in comparison to experimental uncertainties~\cite{Ball:2009mk}. Contributions from initial state heavy quarks and strangeness are typically suppressed in these measurements due to the relatively low scales.

\begin{figure}[ht]
\centering
\includegraphics[width=0.45\textwidth]{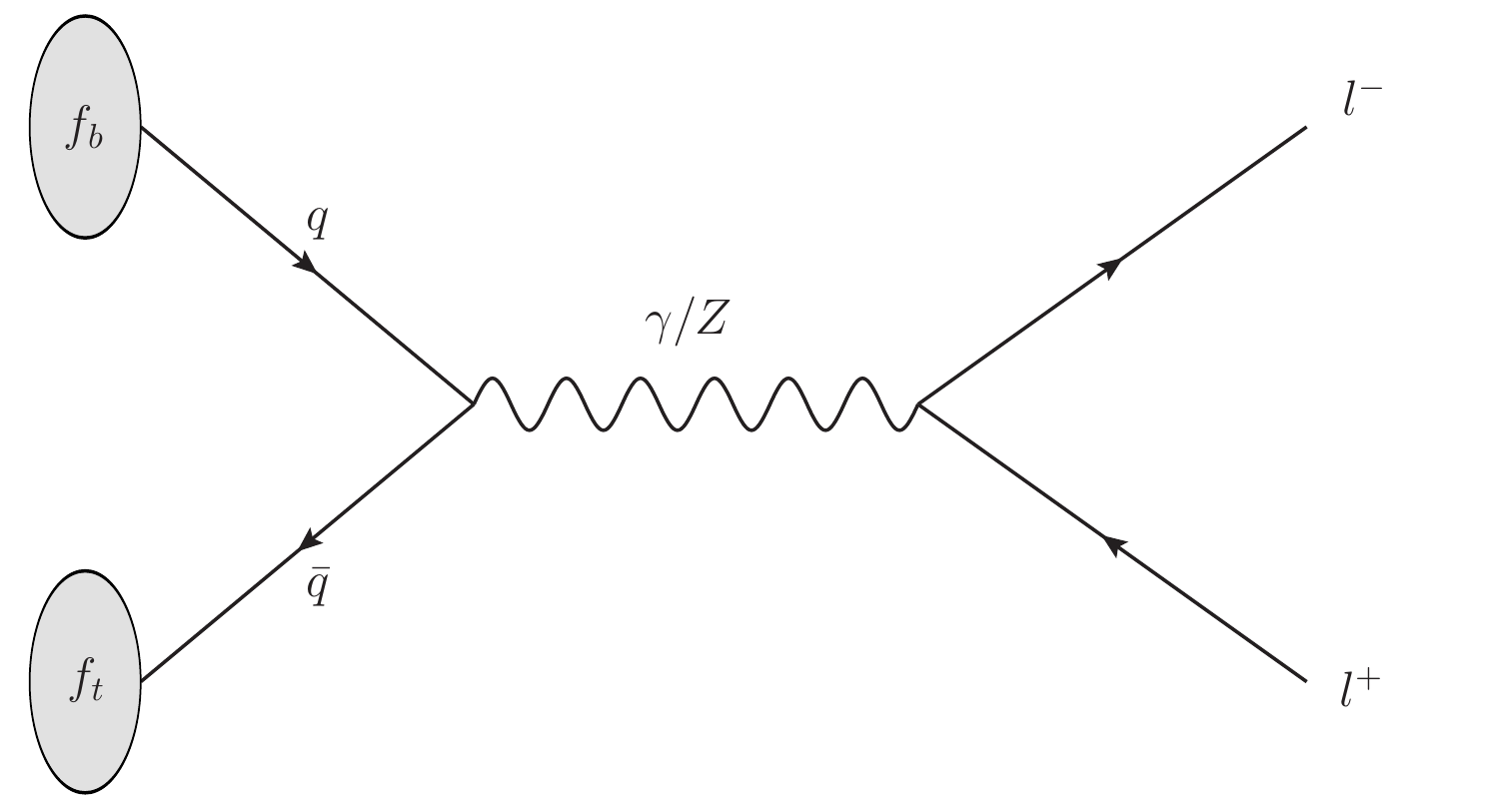}
\includegraphics[width=0.45\textwidth]{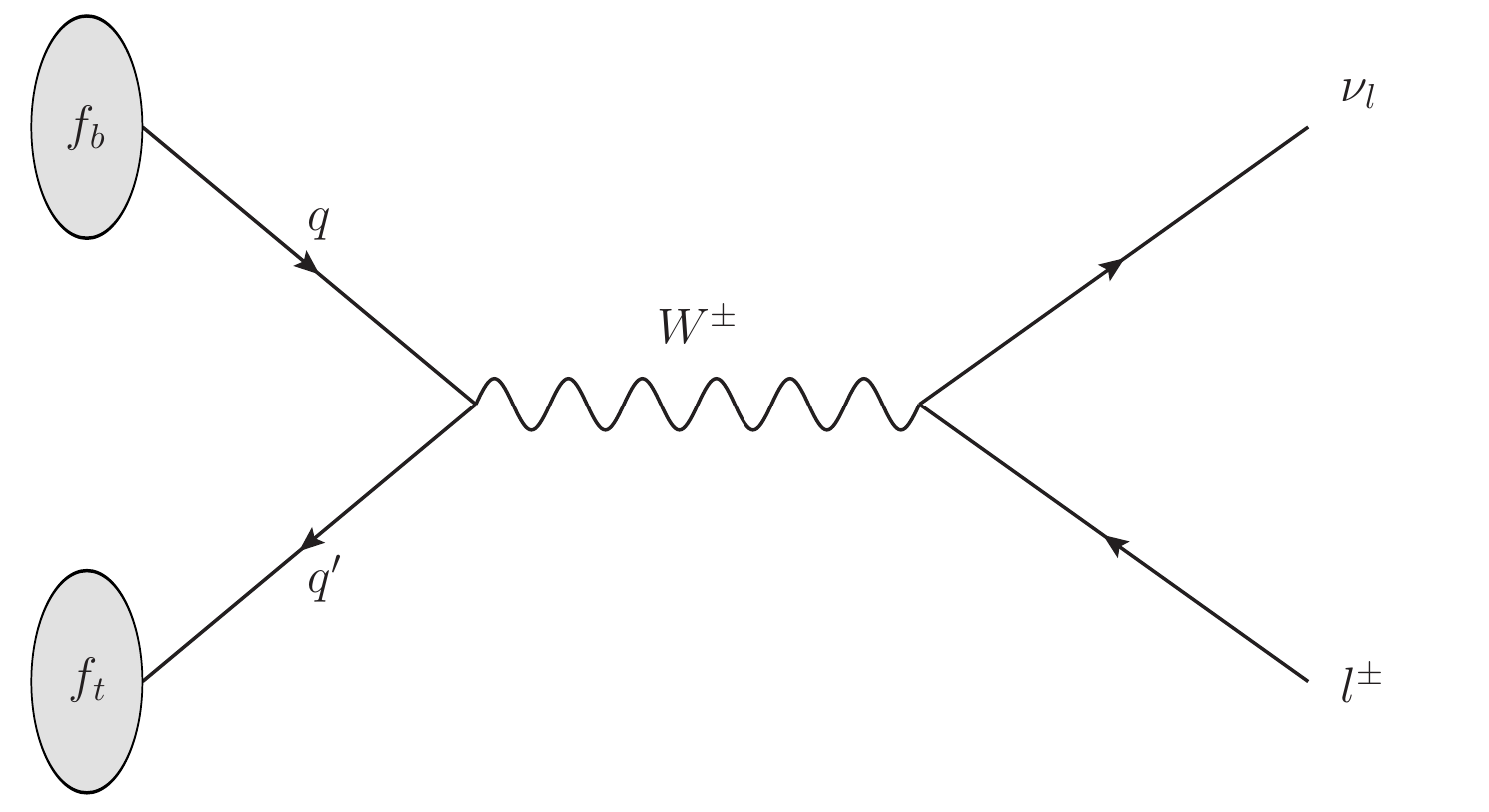}
\caption[Leading order diagram for the Drell-Yan process]{Drell-Yan process at leading order, initiated by beam protons with PDF $f_b$ and target protons with PDF $f_t$. The neutral current process is shown on the left, and the charged current process on the right.}
\label{fig:ncdy}
\end{figure}

Fixed target experiments upon hydrogen or deuterium targets provide a relatively clean probe and the ratio of Drell-Yan cross sections in proton to deuteron targets can provide crucial information on the $u/d$ PDF combination. While relatively free of nuclear effects, deuteron data still suffers from poorly understood corrections, which have been the subject of extensive study\cite{Martin:2012da, Badelek:1994qg,Accardi:2011fa,Brady:2011hb}.
Experimental measurements from the Fermilab NuSea/E866 collaboration are commonly used, providing data from $pp$~\cite{Webb:2003bj} and $pd/pp$~\cite{Towell:2001nh} experiments.  

The theoretically cleanest environment to examine the Drell-Yan process is at high scales at a collider. Several measurements are available from the Tevatron collaborations which provide information free of nuclear or deuteron corrections. As a $p\bar{p}$ collider, neutral-current Drell-Yan at the Tevatron targets the quark valence contribution and asymmetry data provides information on the $u/d$ ratio. A measurement of the $Z$ rapidity distribution  is available from D0~\cite{Abazov:2007jy}, and several measurements are available for $W$ lepton asymmetries from both Tevatron collaborations \cite{Acosta:2005ud,Abazov:2007pm,Abazov:2008qv,Abe:1998rv}.

In order to obtain a handle on the contribution of initial state strange quarks to the Drell-Yan process it is once again necessary to examine less inclusive processes. Of particular interest are measurements of $W$ production in association with a charm jet, analogous to the usefulness of dimuon measurements in neutrino DIS where the strange contribution is favoured in terms of CKM elements. Measurements of this process were initially made at the Tevatron by both CDF~\cite{Aaltonen:2007dm} and D0~\cite{Abazov:2008qz}. More precise determinations can be obtained by normalisation with respect to the total $W+$ jets rate~\cite{Stirling:2012vh}. 

\subsubsection{Jet production data}
While DIS data provides constraints upon the gluon distribution via scaling violations and contribution to heavy quark and longitudinal structure functions, DIS and Drell-Yan data do not provide a substantial direct constraint upon gluon densities. The most constraining datasets for the gluon, particularly in the uncertain large-$x$ region, are those of jet production measurements. The large strong coupling of the gluon combined with a high gluon luminosity in the proton at high scales results in $gg$ initiated diagrams being the dominant sub channels for the production of inclusive jet and dijet events.

Cross-section calculations for inclusive jet and dijet data in hadron-hadron collisions are available at NLO in QCD~\cite{Ellis:1992en,Giele:1994gf,Nagy:2001fj,Nagy:2003tz}, however a great deal of progress has been made in the determination of the NNLO corrections~\cite{Currie:2013dwa,Glover:2001af,Glover:2001rd}, with the exact gluon-gluon sub channel calculation recently determined~\cite{Currie:2013dza}. For the full calculation however, only approximate NNLO results are available via threshold resummation techniques~\cite{deFlorian:2013qia,Kidonakis:2000gi,Kumar:2013hia}. Jet data may therefore only be included into an NNLO PDF fit through an approximate treatment if at all.

Jet data must be included via some clustering algorithm which takes a QCD final state and identifies suitable jet-like structures. Earlier measurements were performed with so-called cone algorithms, although these are potentially very sensitive to infrared and collinear effects. More recent experiments typically utilise sequential-combination algorithms such as the Cambridge-Aachen~\cite{Dokshitzer:1997in,Wobisch:1998wt}, $k_T$~\cite{Ellis:1993tq} or anti$-k_T$~\cite{Cacciari:2008gp} algorithms, often used as implemented in the efficient {\tt FastJet}~\cite{Cacciari:2011ma} package.

\begin{figure}[ht]
\centering
\includegraphics[width=0.6\textwidth]{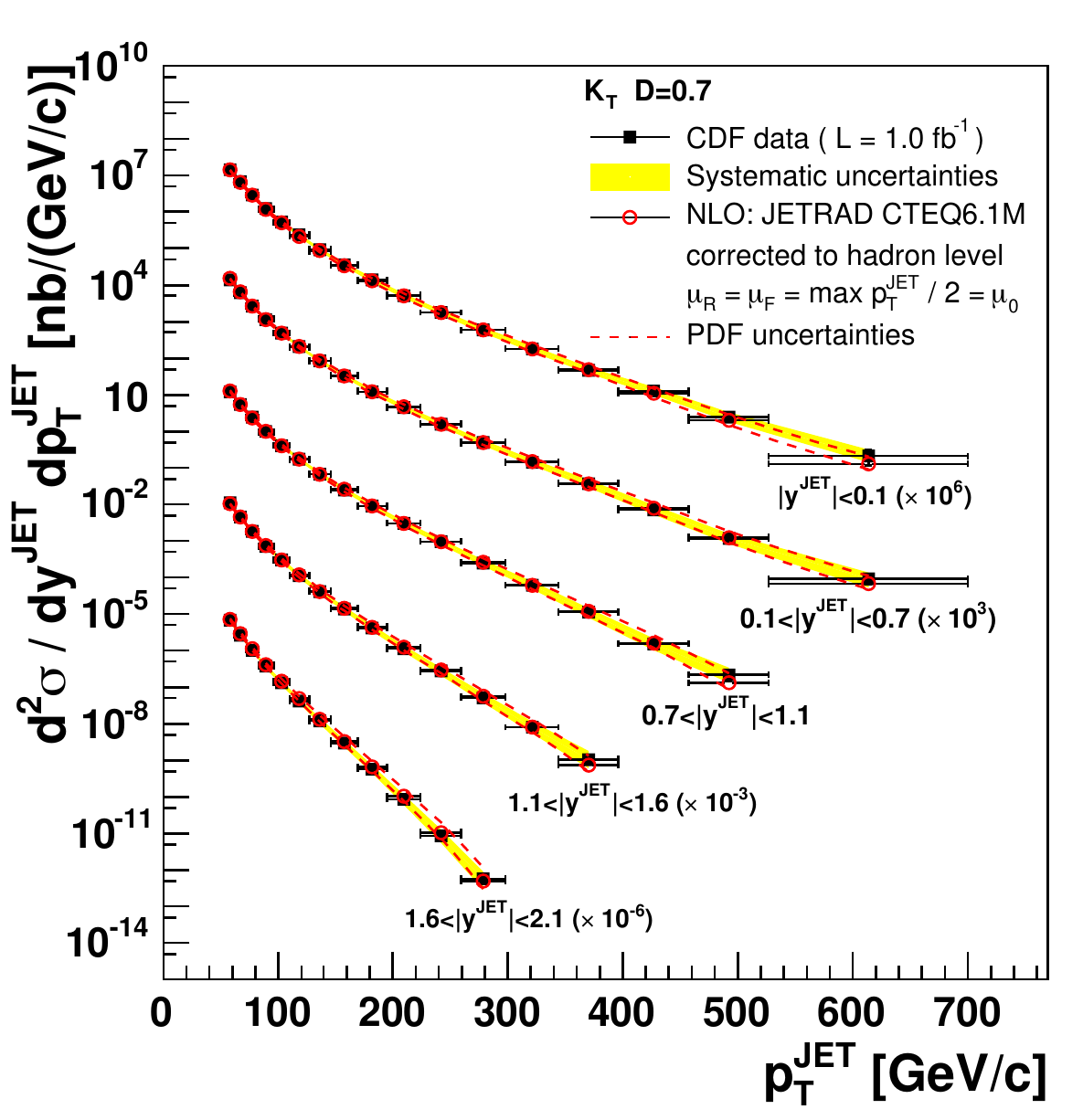}
\caption[Inclusive Jet data from the CDF experiment]{Inclusive jet data from CDF using the $k_T$ jet clustering algorithm, compared to predictions from the CTEQ6.1M PDF set. Figure from~\cite{Abulencia:2007ez}. }
\label{fig:CDFkTJet}
\end{figure}

The CDF collaboration has published precise measurements of inclusive jet~\cite{Abulencia:2007ez,Aaltonen:2008eq} and dijet~\cite{Aaltonen:2008dn} cross sections. Data is also available from the D0 experiment, once again for inclusive~\cite{Abazov:2008ae} and dijet~\cite{Abazov:2010fr} quantities.
Figure \ref{fig:CDFkTJet} shows the results of an inclusive jet measurement at CDF using the $k_T$ clustering algorithm.

\subsubsection{Prompt photon measurements}
\begin{figure}[ht]
\centering
\includegraphics[width=0.48\textwidth]{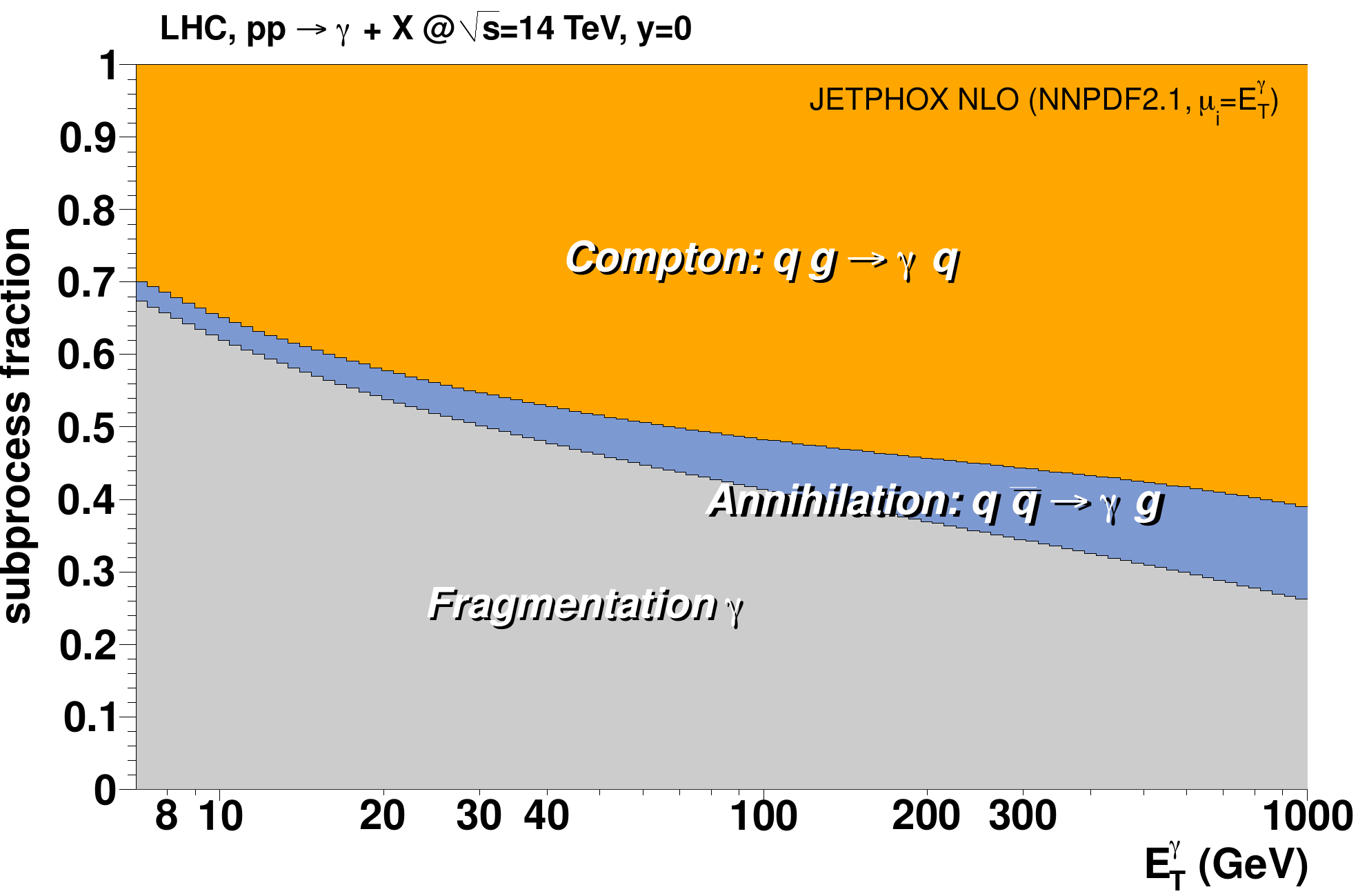}
\includegraphics[width=0.48\textwidth]{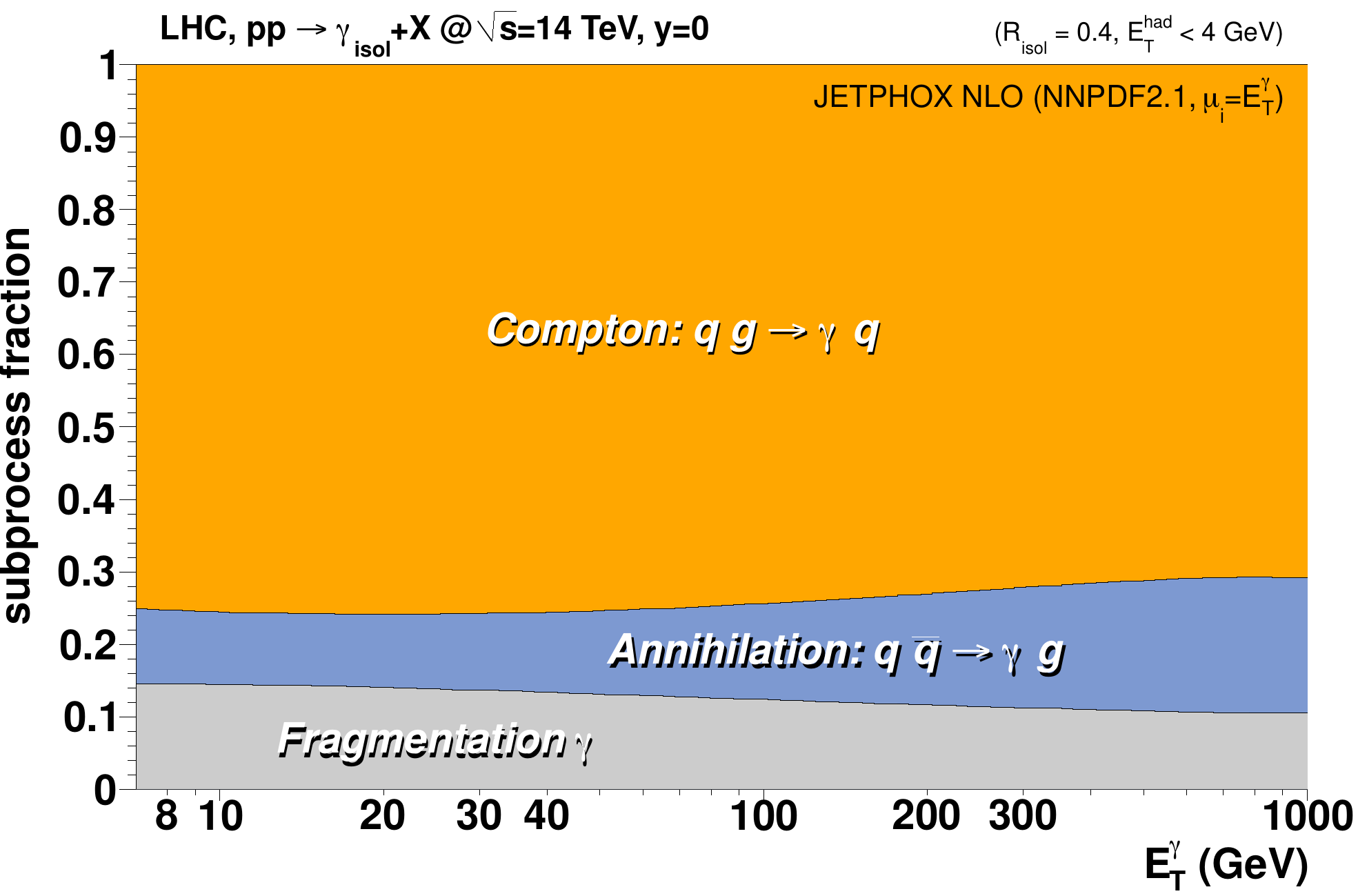}\\
\includegraphics[width=0.48\textwidth]{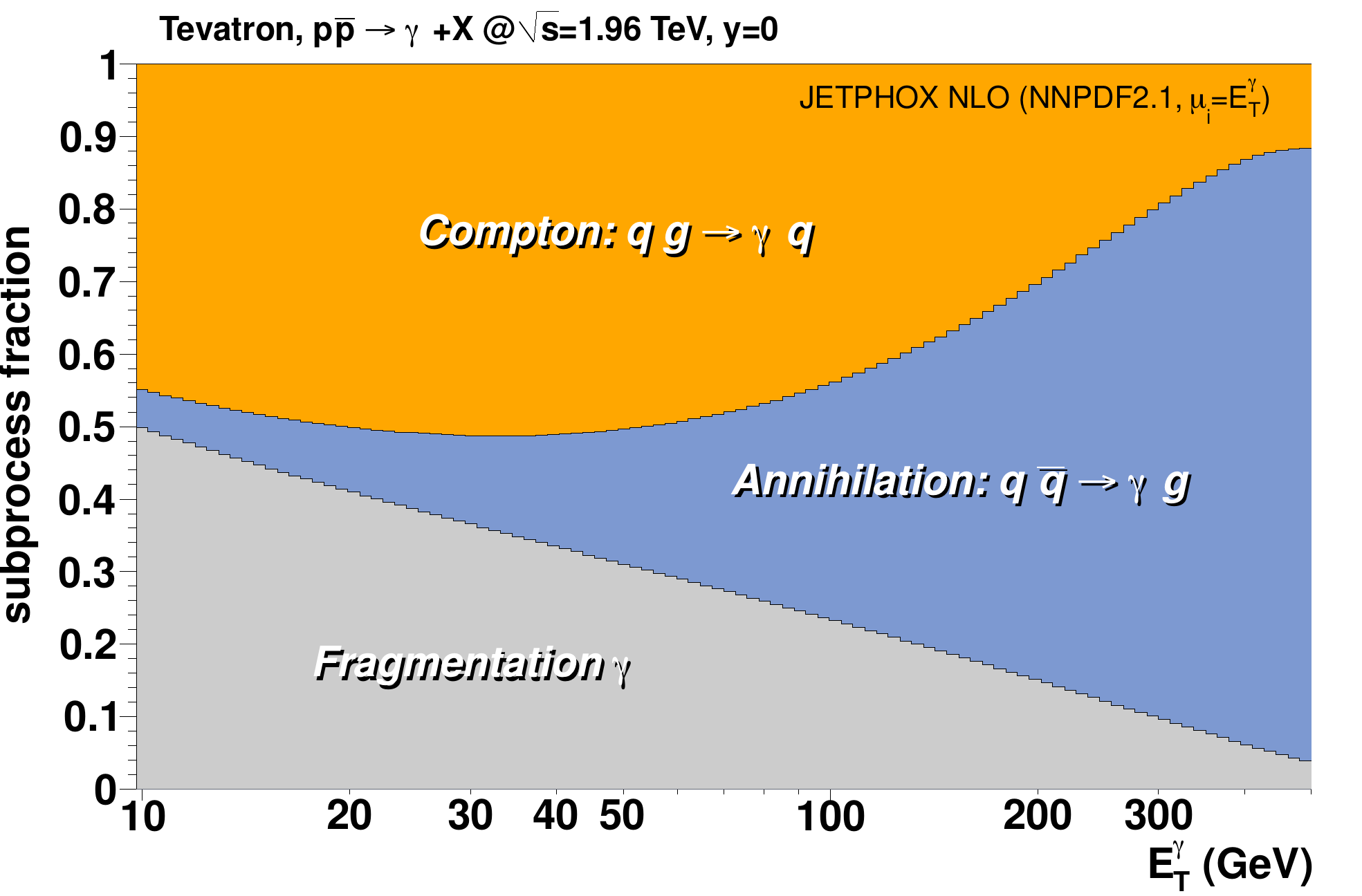}
\includegraphics[width=0.48\textwidth]{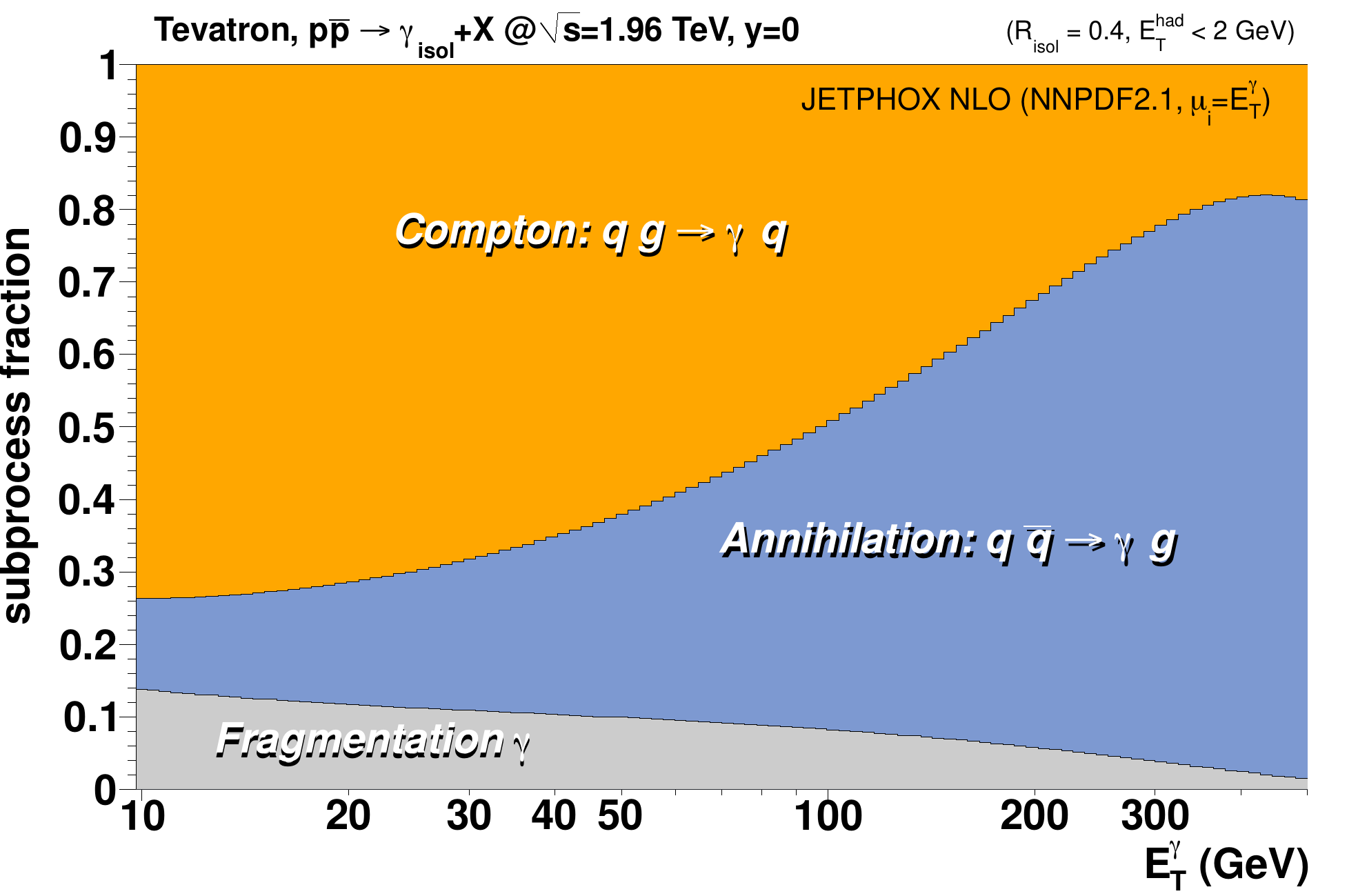}
\caption[Relative contribution of partonic subprocesses to $pp/p\bar{p}\to \gamma X$]{Relative contribution of partonic subprocesses to $pp/p\bar{p}\to \gamma X$. Figures on the left refer to the inclusive case, and on the right to the observable after isolation cuts on the final state photon. Figure from~\cite{dEnterria:2012yj}. }
\label{fig:jrpromptphoton}
\end{figure}

Complementary to the data on jet production, measurements of prompt photon processes $pp/p\bar{p} \to \gamma X$ can also provide an important handle on the gluon. The term \emph{prompt} photon refers the production of a photon in the hard scatter rather than in subsequent emissions. Prompt photons in the final state can originate either from Compton scattering processes $gq \to \gamma q$ or annihilation events $q\bar{q} \to \gamma g$, processes denoted \emph{direct} photon production. Alternatively prompt photons may be produced via the fragmentation of final state hadrons into photons via so-called fragmentation functions\cite{Bourhis:1997yu,Gluck:1992zx}. In $pp$ collisions the Compton scatter is typically the dominant process, particularly at higher scales where the fragmentation contribution is suppressed. For $p\bar{p}$ events the annihilation contribution becomes more important due to the enhanced $q\bar{q}$ PDF luminosity. Figure \ref{fig:jrpromptphoton} demonstrates the relative fraction of these contributions to the cross-section for a range of photon transverse energy $E_T$.

For the purposes of PDF determination direct photon measurements which are free of the additional uncertainties introduced when performing calculations with photon fragmentation functions are the ideal measurement. While performing selection cuts to measure only the direct photon contribution is experimentally challenging, the relative contribution of fragmentation photons may be suppressed by making isolation cuts upon the final state photon. These cuts admit only photons with no hadronic material in close proximity. Smooth-cone cuts such as the Frixione isolation criterion~\cite{Frixione:1998jh} in principle can remove entirely the fragmentation contribution. However these cuts remain challenging to implement experimentally, with experimental data usually obtained with simpler isolation cuts which aim to suppress rather than eliminate fragmentation photons.

Theoretical predictions are available at NLO for the Compton process~\cite{Owens:1987qy,Aurenche:1988vi} and commonly used as implemented in the {\tt JETPHOX} program~\cite{Catani:2002ny,Aurenche:2006vj,Belghobsi:2009hx}. While inclusive data is challenging to include in a PDF determination due to contamination by fragmentation photons, results are available from a wide range of isolated photon measurements. Isolated data is available from  UA1/UA2 at the Sp$\bar{\mathrm{p}}$S~\cite{Albajar:1988im,Alitti:1992hn,Ansari:1988te}, PHENIX at RHIC~\cite{Adler:2006yt}, CDF~\cite{Aaltonen:2009ty,Abazov:2005wc,Abe:1994rra,Acosta:2002ya,Acosta:2004bg} and D0~\cite{Abachi:1996qz,Abazov:2001af,Abbott:1999kd}.

\subsubsection{Top quark pair production data}
The production of top-antitop pairs is potentially a process of great interest in the determination of PDFs, with calculations available up to NNLO for the total cross-section~\cite{Czakon:2013goa,Baernreuther:2012ws,Czakon:2012zr,Czakon:2012pz}. The impact of the total top pair production cross-section upon PDFs is quite sensitive to the kinematics of the collider, with Tevatron data probing directly the quark content of the proton, while data from colliders with higher centre of mass energies being dominated by the gluon-gluon channel. Precise data from the Tevatron is available in the form of a combined D0-CDF analysis~\cite{Aaltonen:2012ttbar}.

\subsubsection{Experimental cuts}
 A simple cut is typically performed on the hard scale $Q^2$ and for DIS the final state invariant mass $W^2$ to ensure the reliability of perturbative predictions. The MSTW2008 parton fit uses an initial scale for evolution of $Q_0^2=1$ $\mathrm{ GeV}^2$, CT10 uses $Q_0^2=1.3$ $\mathrm{ GeV}^2$ and NNPDF2.3 $Q_0^2=2$ $\mathrm{ GeV}^2$. Most of the data included in global parton fits has a minimum of $Q^2\sim2$ to $5$ GeV$^2$\cite{DeRoeck:2011na}.

\section{Methodological elements}
\subsection{Parametrisation} 
Given an experimental dataset, one must choose a convenient and effective parametrisation of the parton distribution functions such that their predictions may be compared to data. Nominally there are a total of 13 PDFs, six quarks, six antiquarks and a gluon. However as mentioned in the previous section, the heavy quarks $c$, $b$, $t$ are determined perturbatively. There are therefore typically seven free PDFs remaining to be fitted. The parton parametrisation basis is chosen for ease of fitting and perturbative evolution; a basis close to the DGLAP basis in Eqn.~\ref{eq:DGLAP} is desirable for efficiency. However often a different basis is chosen to avoid fitting quantities that are poorly defined by the experimental dataset. 

For example, MSTW2008\cite{Martin:2009iq} uses the following basis for their determination:
\ba
	g,&& \nonumber\\
        q_v &\equiv & q - \bar{q},\nonumber \\
	\Delta &\equiv & \bar{d} - \bar{u},\nonumber \\
	S &\equiv & 2(\bar{u}+\bar{d})+s+\bar{s}\nonumber,\\
	s^\pm &\equiv & s \pm \bar{s},
	\ea	
where g is the gluon PDF and the $q_v$ correspond to the $u$, $d$ quark valence PDFs. These fully parameterise the degrees of freedom to be determined. A functional form in $x$ is then chosen for each of the distributions (the value of $Q^2$ is kept fixed at the input scale for fitting). While all groups include the limiting-$x$ description of Eqn.~\ref{eq:pdflimits}, the choice of parametrisation for the remainder function $r$ varies substantially between fitting groups. As an example, the valence quark PDF $q_v$ parametrisation in MSTW2008 is provided by the expression
\be xq_v(x,Q_0^2) = ax^{b}(1-x)^{c}(1+d\sqrt{x}+e x),\ee
and the equivalent parametrisation in CT10\cite{Lai:2010vv} is 
\be xq_v(x,Q_0^2) = ax^b(1-x)^b \exp{(cx + dx^2 + e\sqrt{x})},\ee
where the $(a$,...,$e)$ are the parameters to be determined in the fit. In total the MSTW08 basis has 30 free parameters (taking into account sum rule constraints), the CT10 parametrisation is a little less flexible, having 26 free parameters. The problem is now reduced to finding the optimum parameters for the $7$ PDFs that minimise some measure of fit quality, the differing versions of which we shall discuss later in the chapter.

The NNPDF procedure is markedly different from that of the other PDF fitting groups and the first major difference lies in the choice of parametrisation. Unlike in the general procedure outlined above, neural networks are used to provide the functional $x$ dependence of the PDFs. Neural networks are a typical computational tool in machine learning environments, often used in regression applications where flexibility and a lack of bias with respect to a conventional fixed parametrisation are desired. A typical neural network in a fitting context will usually have considerably more functional freedom (and therefore parameters) than a normal parametric model, with the neural network compensating for its relative generality with respect to the problem by having much greater flexibility. 

The use of neural networks as applied to the determination of the proton structure function $F_2^p$ was first suggested in Ref.~\cite{Forte:2002fg} and subsequently developed in~\cite{DelDebbio:2004qj}. The approach was later extended to the determination of quark distributions~\cite{DelDebbio:2007ee} before becoming a global analysis of PDFs as of NNPDF2.0~\cite{Ball:2010de} as part of the wider NNPDF methodology.

In the NNPDF approach the specific networks used in the parametrisation are multi-layer feed forward neural networks configured with 2-5-3-1 architecture. This architecture applied over seven PDFs results in a fit with a total of 259 free parameters, considerably more than in competing approaches. The architecture chosen in fact has considerable redundancy to minimise potential bias due to inflexibility or choice of architecture. The flexibility of the approach was demonstrated in Ref.~\cite{Ball:2011eq} where the architecture was modified considerably, with no significant change in the fit results.

Due to the redundant parametrisation provided by the neural networks, there is a great deal of freedom in the choice of the input parton distribution basis. In the more recent NNPDF analyses: sets NNPDF 2.1 and NNPDF 2.3, the basis is chosen for simplicity of evolution as:
\ba
\mathrm{gluon}\quad \quad&g,& \nonumber \\
\mathrm{singlet}\quad \quad&\Sigma &\equiv  \sum_{i=1}^{n_f} (q_i + \bar{q_i}),\nonumber \\
\mathrm{valence}\quad\quad &V &\equiv  \sum_{i=1}^{n_f} (q_i - \bar{q_i}),\nonumber \\
\mathrm{triplet}\quad\quad &T_3 &\equiv   (u + \bar{u}) - (d+\bar{d}),\nonumber \\
\mathrm{sea}\; \mathrm{asymmetry}\quad\quad &\Delta &\equiv  \bar{d}-\bar{u},\nonumber \\
\text{strange sea/valence}\quad\quad &s\pm &\equiv  s \pm \bar{s}.
\ea
\clearpage
The equivalent functional forms for the fitting in terms of the Neural Networks are;
\ba \Sigma(x,Q_0^2)&=&x^{-\alpha_\Sigma}(1-x)^{\beta_\Sigma} \mathrm{NN}_\Sigma(x), \nonumber  \\
V(x,Q_0^2)&=&A_V x^{-\alpha_V} (1-x)^{\beta_V} \mathrm{NN}_V(x),\nonumber \\
T3(x,Q_0^2)&=&x^{-\alpha_{T3}}(1-x)^{\beta_{T3}} \mathrm{NN}_{T3}(x),\nonumber \\
\Delta(x,Q_0^2)&=&A_\Delta x^{-\alpha_\Delta}(1-x)^{\beta_\Delta} \mathrm{NN}_\Delta(x),\nonumber \\
g(x,Q_0^2)&=&A_gx^{-\alpha_g}(1-x)^{\beta_g} \mathrm{NN}_g(x)\nonumber, \\
s^+(x,Q_0^2)&=&x^{-\alpha_{s^+}} (1-x)^{\beta_{s^+}}\mathrm{NN}_{s^+}(x),\nonumber \\
s^-(x,Q_0^2)&=&x^{-\alpha_{s^-}}(1-x)^{\beta_{s^-}} \mathrm{NN}_{s^-}(x) - s_{\mathrm{aux}}(x,Q_0^2), \label{eq:NNPDF23param}
\ea
where the NN denote the 2-5-3-1 neural network parametrisations and the $A$ are set by enforcing the appropriate sum rules. In the NNPDF approach the treatment of the limiting exponents $\alpha$, $\beta$ is rather different. These factors are introduced in order to speed up the convergence of the neural network fitting, with the intention of providing a rough preprocessing function as a backbone for the neural networks to deviate from, and ensuring that the functions have the correct behaviour under integration. These exponents are therefore randomised within an optimised range at the start of the fit and are not modified by the fitting procedure. The final results should therefore be reasonably independent of the preprocessing factor and of the coefficients involved. 
\\
While determinations with fixed parametrisations typically design the strange valence functional form such that the strange valence sum rule is automatically satisfied, this cannot be done with a neural net parametrisation. In the determinations up to NNPDF2.3 the strange auxiliary term $s_{\mathrm{aux}}(x,Q_0^2)$ in Eqn.~\ref{eq:NNPDF23param} is therefore introduced to ensure the strange valence sum rule is followed, and has the form~\cite{Ball:2009mk}:
\be s_{\mathrm{aux}}(x,Q_0^2) = A_{s^-}(x^{r}(1-x))^s.\ee
\subsection{Fit quality and minimisation}
With an experimental dataset selected and a choice made for the parametrisation of the PDFs, the optimal fit should be determined by varying fit parameters and attempting to minimise some measure of fit quality. Different groups make quite different choices not only in the minimisation method but also in the measure used to determine fit quality. The most general statement that can be made is that the global fit quality (generally denoted $\chi^2$) is built from the quality of fit to individual datasets as

\be \chi^2 = \sum_k^{n} \chi^2_k,\ee
for a fit with $n$ data sets, each with a consistent normalisation. In the NNPDF approach the full covariance matrix of the data is used in determining the quality of fit, including all appropriate correlations within and between datasets. The $\chi^2$ measure for a set of data with common correlations is then given by
\be \chi^2_k=\sum_{i,j=1}^{N_{\mathrm{dat}}}\frac{(D_{k,i}-T_{k,i})(D_{k,j}-T_{k,j})}{\mathrm{Cov}[i,j]}.\ee
Here the $T$ are the theoretical predictions for the experimental data points $D$ calculated from the neural network parametrisation, and $\mathrm{Cov}[i,j]$ is the covariance between data points $i$ and $j$. In practice there is a ensemble of neural networks each associated with a single Monte Carlo sample of the experimental data, for the purposes of error propagation. This point will be discussed in more detail later in the chapter.  In NNPDF determinations the full experimental correlations should be available for a dataset to be included into the determination.
Other groups take a different strategy, often with the suggestion that correlation effects are small to negligible with the exception of overall normalisations. Adopting the same practice as earlier MRST fits, the MSTW2008 fit uses an uncorrelated $\chi^2$ measure over much of its dataset~\cite{Martin:2009iq}, with the normalisation of the theory predictions set by a fitted parameter $\mathcal{N}$
\be \chi^2_k=\sum_{i=1}^{N_{\mathrm{dat}}} \frac{(D_{k,i}-T_{k,i}/\mathcal{N}_k)^2}{\mathrm{Var}[i]} + \left(\frac{1-\mathcal{N}_k}{\sigma_k^{\mathcal{N}}}\right)^4, \label{eq:MSTWchi2}\ee 
where the final quartic penalty is intended to prevent the normalisation deviating too far from the experimental normalisation uncertainty $\sigma_{\mathcal{N}}$, and the variance $\mathrm{Var}[i]$ is constructed by the sum in quadrature of the statistical and uncorrelated systematic errors. The CT series of fits utilise a $\chi^2$ measure that includes systematic uncertainties in terms of explicit shifts~\cite{Stump:2001gu,Pumplin:2002vw}. In this arrangement, the fit quality measure is given by
\begin{equation} \label{eq:CTchi2}
\chi^{2}_k =  \sum_{i=1}^{N_{\mathrm{dat}}} \frac{1}{%
\mathrm{Var}[i]} \left(D_{k,i}-T_{k,i}-\sum_{n=1}^{N_{\mathrm{corr}}}r_{n}\sigma^{\mathrm{corr}}_{k,n,i}\right)^{2}
+\sum_{n=1}^{N_{\mathrm{corr}}} r_{n}^{2},
\end{equation}
where here the $\sigma^{\mathrm{corr}}$ are the $N_{\mathrm{corr}}$ correlated systematic uncertainties. In this procedure the theory predictions $T$ are shifted parametrically by the variables $r$. The optimal shift values are found by minimising the $\chi^2$ with respect to the $r$ analytically at each stage of the fit. This procedure was introduced to accommodate for overall shifts in the CT10 distributions. A similar method which was adopted in MSTW2008 for a limited number of datasets where correlations were deemed to be important, with the normalisations also determined in the fit as per the uncorrelated case.

\subsubsection{Normalisation uncertainty}

A key point that must be addressed when constructing a measure of fit quality is the treatment of normalisation uncertainties, or multiplicative uncertainties in general. Even using the same definition of the fit quality measure, substantial deviations may be produced by defining the covariance matrix and therefore the breakdown into systematic errors, differently.

The full experimental uncertainty information is characterised by the sum of all uncorrelated errors for a datapoint $\sigma^{\mathrm{unc}}$; the set of $N_{\mathrm{add}}$ correlated additive systematics $\sigma^{\mathrm{add}}$; and the set of $N_{\mathrm{mul}}$ correlated multiplicative systematics $\sigma^{\mathrm{mul}}$.
Given this information one may naively define an \emph{`experimental'} prescription~\cite{Ball:2012wy} for constructing a covariance matrix as
\be
\label{eq:covmat}
\mathrm{Cov}[i,j]=
\delta_{ij}\; \sigma^{\mathrm{unc}}_{i}\sigma^{\mathrm{unc}}_{j} + 
\sum_{k=1}^{N_{\mathrm{add}}}\sigma^{\mathrm{add}}_{i,k}\sigma^{\mathrm{add}}_{j,k}
+ \left( \sum_{k=1}^{N_{\mathrm{mul}}} \sigma_{i,k}^{\mathrm{mul}}\sigma_{j,k}^{\mathrm{mul}}
\right) D_{i} D_{j},
\ee
where once again the $D$ represent the experimental data points. This method of constructing the covariance matrix is therefore unambiguously defined by the experimental results. While a perfectly valid definition for analysing the description of data after a PDF determination, it is unreliable for use directly within a fitting procedure. The use of the experimental definition has for some time been understood to result in a \emph{d'Agostini bias}\cite{DAgostini:1993uj}. That is, the theoretical values determined via a minimisation of a $\chi^2$ function with the experimental covariance matrix are systematically shifted lower than the true value, an effect which only worsens as the number of data points subject to a common multiplicative error increases. The bias is generated by downward statistical fluctuations of data, if these low data points are used to generate the normalisation uncertainty, the result is a smaller uncertainty for the lower points, causing the fit to systematically undershoot the data. 

\begin{figure}[ht]
\centering
\includegraphics[width=0.8\textwidth]{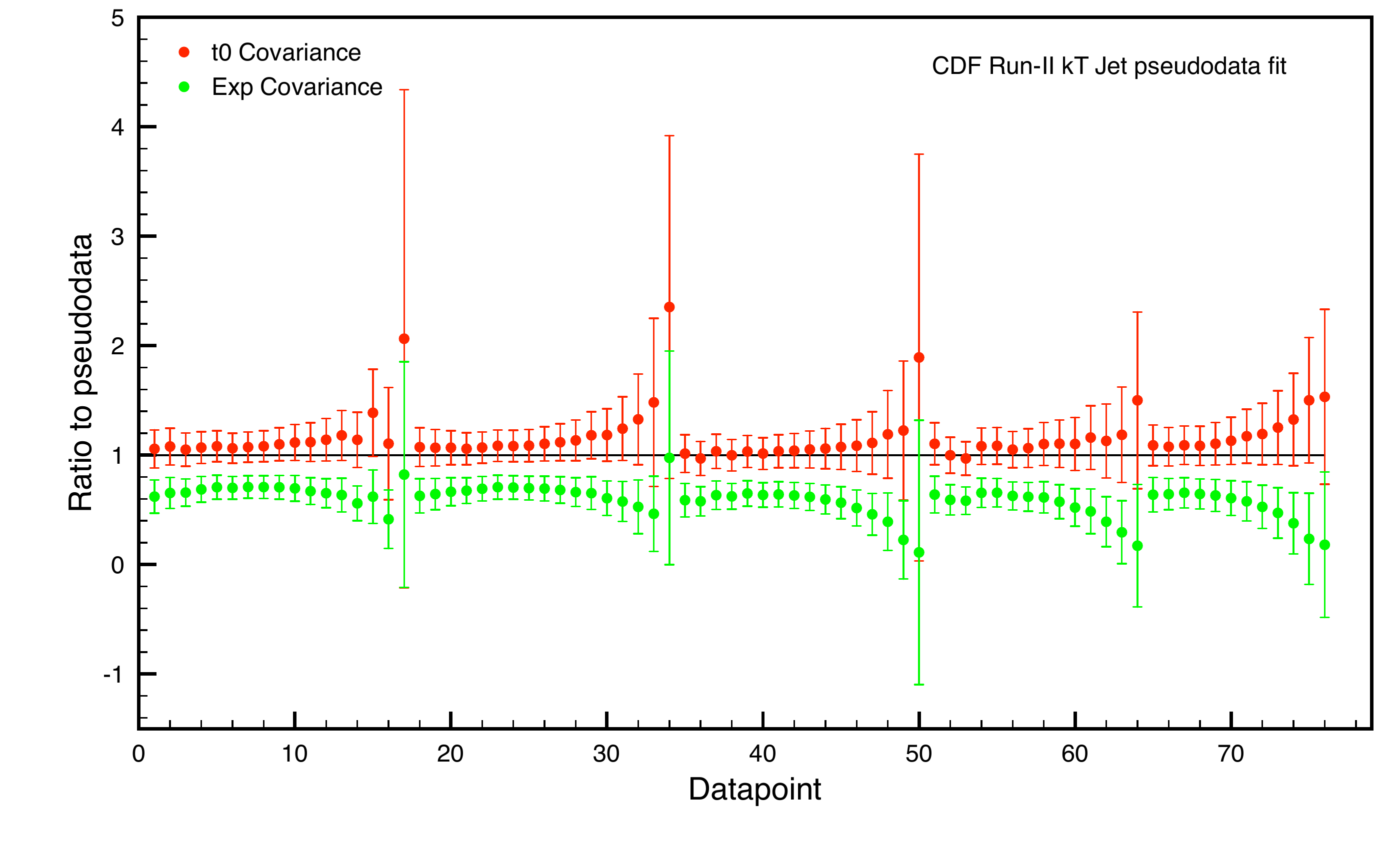}
\caption[Demonstration of d'Agostini bias in a fit to pseudodata generated according to the kinematics of CDF inclusive jet data]{Demonstration of d'Agostini bias in a fit to pseudodata generated according to the kinematics of CDF inclusive jet data. Fit results are shown as a ratio to the `true' value used to generate the pseudodata. The fit performed with the experimental definition of the covariance matrix results in predictions shifted systematically downwards with respect to the underlying law. The predictions from the fit using a $t_0$ covariance matrix do not suffer from such a bias.}
\label{fig:expbias}
\end{figure}

The typical method employed to avoid the d'Agostini bias proceeds by including the normalisation as a fitted parameter and penalising large deviations as shown in Eqn. \ref{eq:MSTWchi2}. This procedure largely corrects for the problem, although when applied to a dataset with several different normalisation uncertainties it still suffers from a bias. This effect was demonstrated by the NNPDF collaboration in Ref.~\cite{Ball:2009qv}.  The bias can be avoided by using the so-called $t_0$ prescription~\cite{Ball:2009qv} for defining the covariance matrix. In this method the covariance matrix is constructed using the predictions from a previous fit rather than the experimental data values, to multiply with the multiplicative uncertainties.
\be
\label{eq:covmat_t0}
\mathrm{Cov}^{t_0}[i,j]=
\delta_{ij}\; \sigma^{\mathrm{unc}}_{i}\sigma^{\mathrm{unc}}_{j} + 
\sum_{k=1}^{N_{\mathrm{add}}}\sigma^{\mathrm{add}}_{i,k}\sigma^{\mathrm{add}}_{j,k}
+ \left( \sum_{k=1}^{N_{\mathrm{mul}}} \sigma_{i,k}^{\mathrm{mul}}\sigma_{j,k}^{\mathrm{mul}}
\right) T_{i}\, T_{j},
\ee
where here the $T$ are theory predictions for the associated datapoint, generated by some prior (fixed) PDF set. The prior, or $t_0$ set should be determined self-consistently via an iterative procedure in which the $t_0$ set is obtained from the previous result for the full fit. As the theory predictions are not subject to the same fluctuations as the data, the fit is not subject to the aforementioned bias. This effect can be seen explicitly in a fit to artificial pseudodata, performed with the experimental and $t_0$ covariance matrix definitions in Figure \ref{fig:expbias}.

\subsubsection{Minimisation}
With a figure of merit constructed, the PDF determination now becomes a problem of varying the free parameters in the PDF basis to minimise said measure. Even for those groups utilising a fixed parametrisation, performing a minimisation of the global $\chi^2$ for a large, $n\sim\mathcal{O}(1000)$ dataset with a fairly large number of free parameters (approximately $50$ in the MSTW analysis once normalisation uncertainties are added as free parameters) is a challenging numerical task. For performing the minimisation, the MINUIT~\cite{MINUIT} package is a common choice, although other function minimisation methods are applied such as the Levenberg-Marquardt~\cite{levenberg,marquardt} method as used in the MSTW fits.

In the NNPDF case the minimisation is complicated by the very large number of parameters and highly nonlocal behaviour in the error function, making conventional methods of minimisation difficult. These difficulties are overcome in the NNPDF methodology by the use of \emph{genetic algorithms}, which are particularly efficient at exploring large parameter spaces. The implementation of the genetic algorithm is discussed in detail in Refs.~\cite{Ball:2010de,DelDebbio:2007ee}.

In addition to the basic difficulty of minimisation in a large parameter space, there is a further issue that arises when considering the fitting of a function with a great deal of redundant flexibility. Because of the flexibility of the parametrisation, it is possible that training the neural networks so that each reaches the global minimum in the error function actually results in the networks fitting to statistical noise. This effect is known as \emph{overlearning} and is a problem often encountered in the training of large neural networks~\cite{bishopnn,nnoverlearn}. In previous NNPDF determinations, the widely used \emph{cross-validation} technique~\cite{Ball:2010de,bishopnn} was employed in order to identify when overlearning occurred. 

\begin{figure}[t!]
\centering
\includegraphics[scale=0.5]{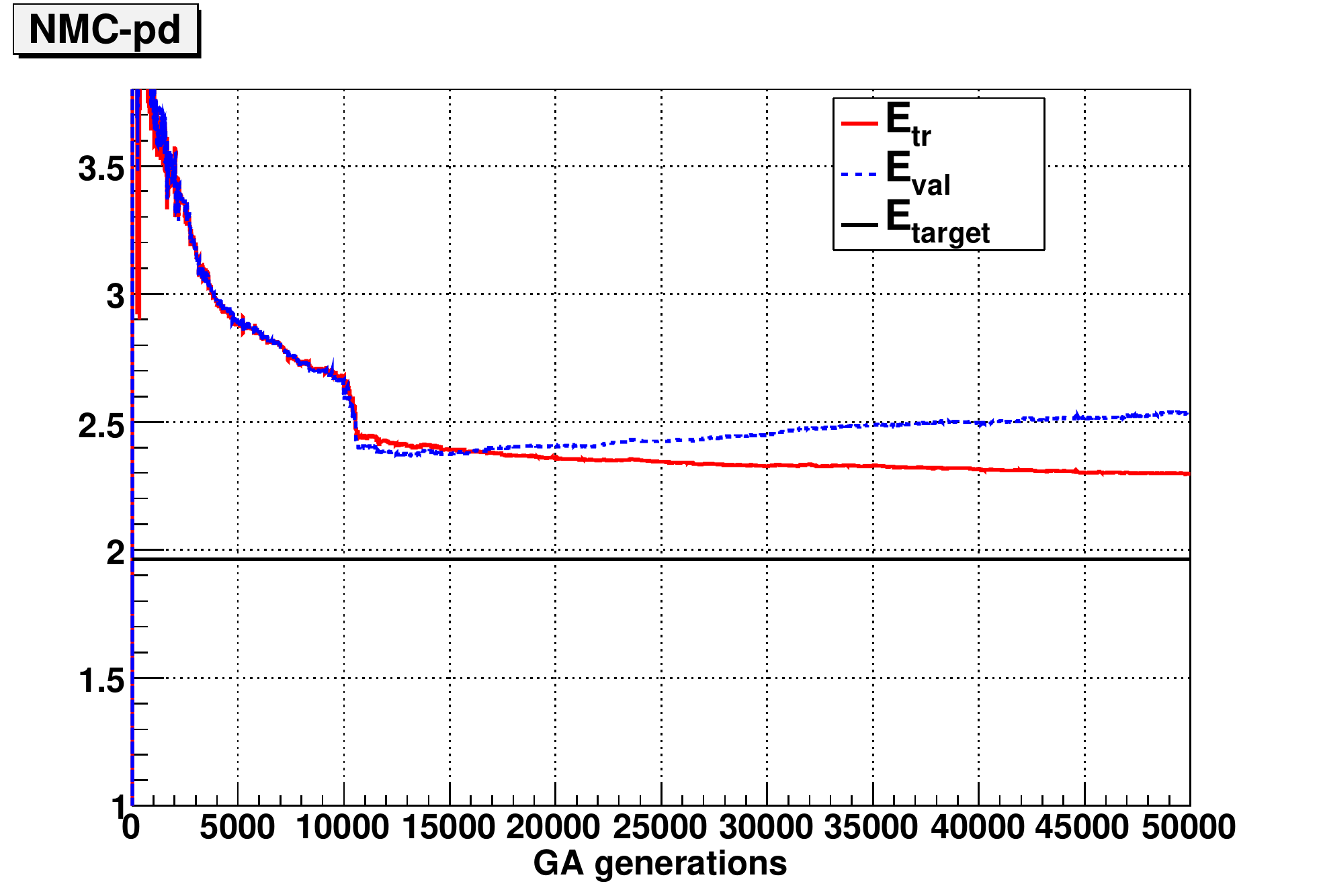}
\caption[Demonstration of overlearning in the cross-validation of a neural network fit]{A typical signal of overlearning in a neural network fit. E$_{\mathrm{tr}}$ and E$_{\mathrm{val}}$ represent the training and validation figures of merit respectively. As the number of genetic algorithm generations proceeds, eventually the network begins to fit statistical noise in the training set and the validation fit quality begins to decrease. Figure from~\cite{Ball:2010de}.}
\label{fig:crossval}
\end{figure}

In this method the experimental data set is split into two separate sets. The first, a fitting set which is used for the minimisation of the error function, and a second validation set which is not used directly in the fitting procedure. For each iteration in the genetic algorithm minimisation the error function is computed between the neural network predictions and both data sets. In the early stages of the training both error functions should decrease. However in the latter stages of the training where statistical noise begins to become an important contribution, the goodness-of-fit calculated to the fitting data set may continue to decrease while the same value calculated to the validation set has stopped decreasing or even begun to increase. This is a clear signal of overlearning, where fitting to statistical noise in the fitting set means that the fit to the validation set is no longer improving. At this stage the training of the neural networks is stopped. A typical signal of overlearning in a cross-validated fit can be seen in Figure~\ref{fig:crossval} which compares the fit quality for both the training and validated sets over a number of fit iterations.

\subsection{Error propagation}
\label{sec:errors}
In order to undertake precision QCD studies, some estimate of the uncertainty on PDFs is required for a meaningful interpretation of the measured observables. The need for PDF sets with quantified uncertainties has been long recognised, and all modern determinations provide sets with at least experimental uncertainty estimation. While performing a comprehensive quantification of the theoretical uncertainty in a PDF fit is challenging, many methods have been developed in order to propagate the uncertainty from the dataset to the fitted PDFs. Ideally, one would like to determine a representation of the probability distribution in the whole functional space. That is given a dataset ${d}$, we would like to find the probability of a certain PDF candidate $f$ such that our fitted PDF central value is given by
\be \left<f\right>\left(x\right) = \int \mathcal{D}f \;f\left(x\right)  \mathcal{P}\left(f \middle| d \right), \ee
and the uncertainty by
\be \mathrm{Var}\left[f\right]\left(x\right) = \int \mathcal{D} f \; \left[ f\left(x\right) - \left<f\right>\left(x\right) \right]^2   \mathcal{P}\left(f \middle| d \right). \ee
The probability distribution for an observable $\mathcal{O}$ is then simply $\mathcal{O}\left[f\right]\mathcal{P}\left(f \middle| d \right)$, in terms of which an observable's central
value and PDF uncertainty can be calculated by
\be \left<O\right> = \int \mathcal{D}f \; \mathcal{O}\left[f\right]  \mathcal{P}\left(f \middle| d \right), \ee
\be \mathrm{Var}\left[\mathcal{O}\right] =  \int \mathcal{D}f \; \left(\mathcal{O}\left[f\right] - \left<\mathcal{O}\right> \right)^2 \;\mathcal{P}\left(f \middle| d \right). \ee
The probability distribution $\mathcal{P}\left(f \middle| d \right)$ is however a difficult quantity to determine. In this section we shall examine a number of the methods used in the literature to provide an estimate of PDF uncertainties.
\subsubsection{The Hessian method}
The Hessian method is the most widely used method of uncertainty determination in PDFs. In essence, the method involves examining how the fit quality $\chi^2$ varies when the $n$ fit parameters $a$ are perturbed about the values which minimise the $\chi^2$, here denoted by $a^\mathrm{min}$. A tolerance in the $\chi^2$ variation is then chosen, and the error on an observable is determined geometrically from observables calculated with parameters perturbed by the selected tolerance. To examine this quantitatively, we first define the difference in $\chi^2$ from the minimum value
\be \Delta\chi^2(a) \equiv \chi^2(a) - \chi^2(a^\mathrm{min}) = \sum^n_{i,j=1}H_{ij}(a_i-a_i^\mathrm{min})(a_j - a_j^\mathrm{min}), \ee
where the $a_i$ represent the $i$th component of the parameter set $a$ (and likewise, for the minimised set $a^\mathrm{min}$). Here we assume that the variation around the $\chi^2$ minimum is approximately quadratic. The Hessian matrix $H$ has values determined by
\be H_{ij} = \frac{1}{2}  \frac{ \partial^2 \chi^2(a) }{\partial a_i\partial a_j}\bigg|_{\mathrm{min}},  \ee
where the \emph{min} subscript refers to the parameters obtained at the $\chi^2$ minimum Early Hessian uncertainty estimates~\cite{Adloff:2000qk,Alekhin:2002fv} were based upon the standard formula for linear error propagation
\be (\Delta F)^2 = T^2\sum^n_{i,j=1}\frac{\partial F}{\partial a_i}C_{ij}\frac{\partial F}{\partial a_j}, \ee
where $T^2=\Delta\chi^2$ is the tolerance in $\chi^2$ variation and $C=H^{-1}$ is the inverse Hessian matrix. This procedure is however a little inconvenient due to the requirement of the partial derivatives of the observable with respect to the fit parameters. There are also numerical issues relating to this method which give rise to peculiar uncertainty estimates~\cite{DeRoeck:2011na}.  In order to overcome these issues the geometrical method outlined above was developed by the CTEQ collaboration~\cite{Pumplin:2000vx,Pumplin:2001ct}.

For this method it is convenient to work in a rescaled orthogonal eigenbasis for the covariance matrix. The orthonormal eigenbasis is defined in the usual way

\be H v_i = \lambda_i v_i, \ee
and the rescaled eigenbasis is defined as $e_{i}=1/\sqrt{\lambda_i} v_{i}$. The difference between a parameter set $a$ and $a^\mathrm{min}$ can now be expanded as
\be a_i-a_i^\mathrm{min} = \sum^n_{k=1}e_{ik}z_k, \ee
where $e_{ik}$ is the $i$th component of the $k$th rescaled eigenvector, and the $z_k$ are the coefficients for the expansion of the parameter difference onto the rescaled eigenbasis.  Therefore the expression for $\Delta\chi^2$ reduces to
\be\Delta\chi^2(a) = \sum_{k=1}^n z_k^2 \quad\quad \mathrm{or,}\quad\quad \chi^2(a) = \chi^2(a^\mathrm{min}) + \sum^n_{k=1}z_k^2.\ee
This defines a hypersphere in the parameter space of radius $\Delta\chi^2$ centred around $a^\mathrm{min}$, which corresponds to the variation in the parameters that is consistent with the tolerance $T=\sqrt{\Delta\chi^2}$ in the quadratic approximation. It is now possible to construct an ensemble of $2n$ PDF sets corresponding to the fits on the boundaries of the volume. A PDF set $S^\pm_k$ therefore has the parameter set
\be a_i(S_k^\pm) = a_i^\mathrm{min} \pm te^{ik},\ee
i.e. each parameter is perturbed by $t$ in the direction of the $e_{k}$ eigenvector. In the quadratic approximation $t=T$, when the approximation breaks down $t$ can be determined by an iterative procedure to obtain the desired $\Delta\chi^2$. The error on an observable $F$ is then given simply by Pythagoras' theorem on the hypersphere
\be(\Delta F)^2=\frac{1}{2}\sum_i^n(F(S^+_i)-F(S^-_i))^2.\ee 

In this procedure there is something of an ambiguity in the determination of the tolerance (and hence, the volume of the sphere in parameter space). Ideally the difference in $\chi^2$ values should be exactly one for a confidence level of one-sigma\footnote{It should be noted that this is only the case when, either the data errors are uncorrelated, or when the correlations are included in the definition of the global goodness-of-fit $\chi^2$\cite{Stump:2001gu}}. In the case of PDF fits, this tolerance often leads to uncertainties far lower than expected. In practice, the CTEQ group uses a value of $\Delta \chi^2\sim 100$ and MSTW uses a value $\sim50$. The more recent MSTW PDF sets have uncertainties calculated with a dynamically determined tolerance. More specialised fits such as ABM11 or the HERAPDF series,  based upon relatively restrictive datasets may use the standard tolerance of $\Delta\chi^2=1$. Their use of a more restrictive dataset perhaps leading to fewer conflicts between experimental datasets that could require a more flexible tolerance.

The uncertainties produced via the Hessian procedure are difficult to analyse in a statistical sense due to the (occasional) inflation of the $\Delta\chi^2$ and the approximations made in the procedure. It is therefore difficult to find a representation in the Hessian approach of the full probability distribution $\mathcal{P}\left(f \middle| d \right)$. Furthermore the uncertainty in the choice of functional form, or estimation of parametrisation bias, is not typically take account of. The HERAPDF family of fits however do attempt to estimate this uncertainty by performing a series of fits with slightly modified parametrisations.

\subsubsection{Lagrange multiplier method}
Another method of error propagation that has been explored is the Lagrange multiplier method. The method has the advantage of not assuming that the $\chi^2$ function is quadratic around the global minimum. We shall briefly discuss the method applied to the PDF error determination as suggested by Pumplin~\cite{Pumplin:2000vx} and Stump~\cite{Stump:2001gu}. A description of the process can also be found in~\cite{DeRoeck:2011na,Martin:2009iq}.

Firstly, a general global fit is performed to the data as described above. This yields a set of parameters ${a^\mathrm{min}}$ which minimise the $\chi^2$ measure. Using these parameters we calculate
the best fit prediction for the observable in question $F(a^\mathrm{min})$. A new PDF fit can now be performed, where instead of minimising the $\chi^2$ the following function is minimised
\be \Psi= \chi^2(a) + \lambda (F(a)-F(a^\mathrm{min}))\ee
i.e. we introduce the observable $F$ as a parameter in the fitting procedure and constrain the fit so that the minimal $\Psi$ occurs when $F(a)=F(a^\mathrm{min})$. The value $\lambda$ in this function is the Lagrange multiplier. The fit above is performed for many values of $\lambda$, each time leading to a parameter set that depends on that particular value of $\lambda$, this parameter set will be denoted $a_\lambda$. Using these parameters, we now calculate values for $\chi^2(a_\lambda )$ and $\mathcal{O}(a_\lambda )$.

At this stage we now have a set of values for $\chi^2(a_\lambda)$ and $\mathcal{O}(a_\lambda)$ over a large range of $\lambda$ values. This allows a determination of the relationship between the goodness-of-fit and the prediction for $F$ via the parameter $\lambda$. We obtain an approximate function $\chi^2(F)$ over a range of observable values, with a minimum at $F=F(a^\mathrm{min})$ i.e $\lambda=0$ and $a_\lambda=a^\mathrm{min}$. Also we have a set of the $a$ parameters for every point on the curve which are optimised for the best fit to the observable $F$. This means that we have a set of fully optimised parameters for any arbitrary confidence level determined by the $\Delta\chi^2$ that we select as a tolerance. Uncertainties for the PDFs can therefore be given in a way that utilises the whole of the $a$ parameter space, rather than just perturbing around the global minimum as in the Hessian approach.

Of course, the disadvantage of this method is that the PDF uncertainties must be calculated for each observable in a rather computationally intensive process. The errors are naturally optimised for the particular observable, but the process is inconvenient for a PDF end-user, and so it is not widely-used in error determination. In this sense the Lagrange multiplier approach suggests a method of estimating $\mathcal{P}\left(\mathcal{O} \middle| a \right)$, or the probability density of an observable in the space of parameters. The Lagrange multiplier method also relies on the same somewhat arbitrary choice of tolerance in $\chi^2$ as the Hessian method. The method has however been applied as a cross-check to the Hessian results~\cite{Pumplin:2000vx,Martin:2002aw}.

\subsubsection{Monte Carlo method}
Another quite distinct method of PDF uncertainty determination is the Monte Carlo method, first suggested by Giele and Keller~\cite{Giele:1998gw,Giele:2001mr} where a Monte Carlo procedure in the space of fit parameters was outlined. The NNPDF collaboration uses a similar method in all of its fits, although with the Monte Carlo performed in the space of experimental data. The method is designed to faithfully represent the uncertainties present in the initial data, and to propagate the errors in a way that does not assume anything of the nature of the error propagation. The Monte Carlo approach was also analysed and compared to the results of a Hessian fit by the MSTW group in~\cite{Watt:2012tq}.

In the Monte Carlo procedure an ensemble of $N_{\text{rep}}$ artificial data replicas is produced for every data point in the fit, generated according to the probability distribution of the initial data. Typically this distribution is multi-Gaussian with central values and variances provided by experimental results, but any probability distribution may be used if required. If we use $F_{p}^{(\mathrm{art})(k)}$ to represent a single element $k$ of the pseudo-data sample (the \emph{art} superscript designates the data as an artificial sample) of the observable $F$ at the kinematical point $\left\{ x_p,Q^2_p \right\} $. Then we can generate such a pseudo-data element as in~\cite{Ball:2008by} by
\be F_{p}^{(\mathrm{art})(k)}=S_{p,N}^{(k)}F_{p}^{(exp)}\left( 1 + \sum_{l=1}^{N_c}r_{p,l}^{(k)}\sigma_{p,l} + r_{p}^{(k)}\sigma_{p,s}\right),\ee
where the $r$ are independent Gaussian random numbers centred upon the experimental central value. The $\sigma_{p,s}$ term contains the uncorrelated systematic uncertainties and the statistical uncertainty added in quadrature. The $\sigma_{p,l}$ are the correlated errors for the data provided by experiment. The normalisation of the probability distribution is fixed by the term $S_{p,N}$. Provided a large enough quantity of these artificial replicas ($N_{\mathrm{rep}}$) is generated, this form of the generating distribution for the Monte Carlo ensemble reproduces all of the statistical qualities of the original experimental data. In Ref.~\cite{Ball:2010de} it is demonstrated that $N_{\mathrm{rep}}=1000$ is sufficient to reproduce the experimental central values and variances to an accuracy of better than one percent.

Now that a good Monte Carlo sample of the experimental data is available, instead of performing just the one fit to the data, $N_{\mathrm{rep}}$ independent fits are performed, one for each of the data replicas. At the end of the fitting procedure we obtain an ensemble of $N_{\mathrm{rep}}$ equally probable PDFs which reliably describe the probability distribution of the PDFs based upon the original experimental uncertainties. The central values and uncertainties of an observable can be simply obtained by computing the average and the variance over the ensemble of PDFs.
\be \left< F\right> = \frac{1}{N_{\mathrm{rep}}}\sum_i^{N_{\mathrm{rep}}} F^{(k)}, \label{eq:MCCV}\ee
 \be \sigma^2[F] = \frac{1}{N_{\mathrm{rep}}-1} \sum_{i=1}^{N_{\mathrm{rep}}} (F^{(k)} - \left< F\right>)^2, \label{eq:MCVAR}\ee
 where $F^{(k)}$ denotes the observable $F$ computed using PDF replica $k$.

\begin{figure}[ht]
\centering
\includegraphics[width=0.48\textwidth]{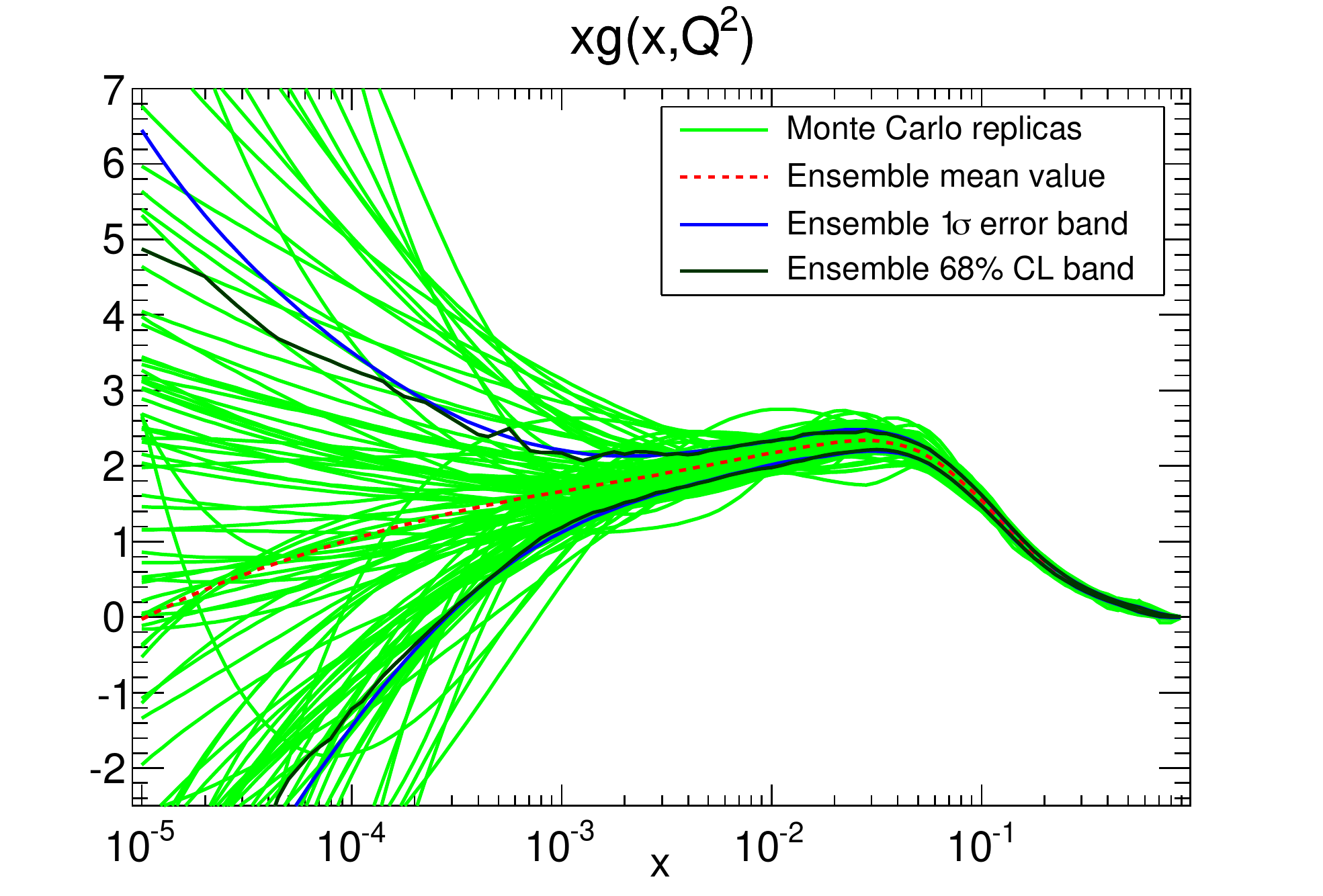}
\includegraphics[width=0.48\textwidth]{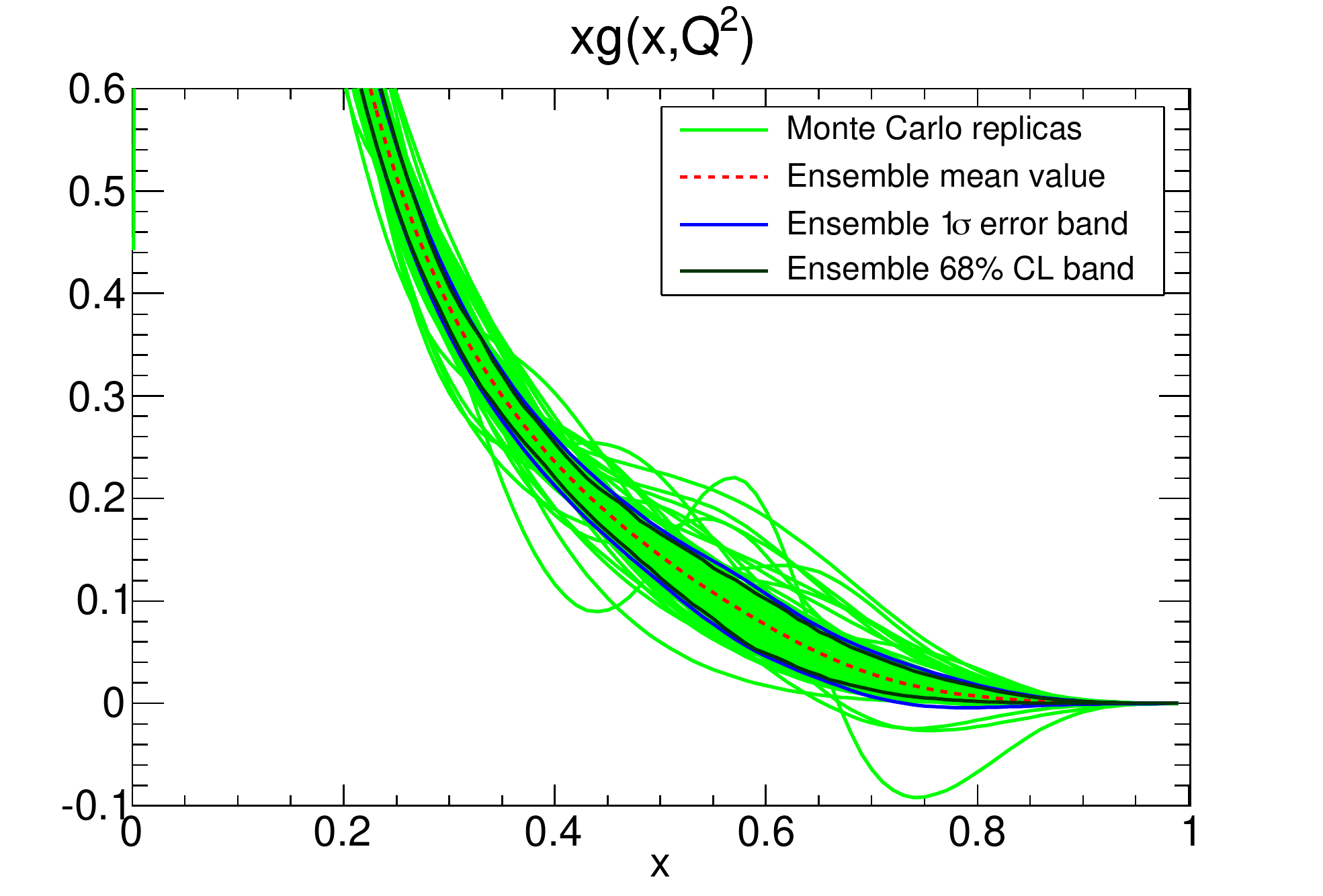}
\caption[A Monte Carlo representation of the gluon PDF probability distribution]{A Monte Carlo representation of the gluon PDF probability distribution. Individual PDF replicas are shown as green lines, and the ensemble average, standard deviation and 68\% confidence level are shown.}
\label{fig:mcerror}
\end{figure}

The Monte Carlo method therefore propagates the errors from the experimental data through to the PDFs in a natural way, without the need for a linear propagation of errors assumption, or the need for an inflated tolerance in the $\chi^2$ distribution. Figure \ref{fig:mcerror} demonstrates a Monte Carlo ensemble of PDF replicas for the gluon distribution.

%
%In the NNPDF approach the method has another benefit. Considering again the method of cross-validation used to prevent overfitting in the neural network parametrisation. At first glance, it would seem that separating the data sets into a fitting and validation set means that not all of the data is directly used in a fit. However, as the selection of data points for each set is done at randomly on a replica, by replica basis, for a large enough $N_{\mathrm{rep}}$ the whole data set is used for fitting across the ensemble.  The use of an ensemble of replicas additionally helps with estimating the functional uncertainty present when determining a function from a finite set of data points. 
%
%\subsection{Sources of theoretical uncertainty}
%
%
%Maybe have here a short summary of all the sources of theoretical uncertainty, and how they may be taken into account?
%alphas, GM scheme, deuteron/nuclear corrections, fragmentation in prompt $\gamma$, NP uncertainties e.g hadronisation/pileup.

\section{Status of PDF determination before the LHC}
\begin{figure}[t]
\centering
\includegraphics[width=0.48\textwidth]{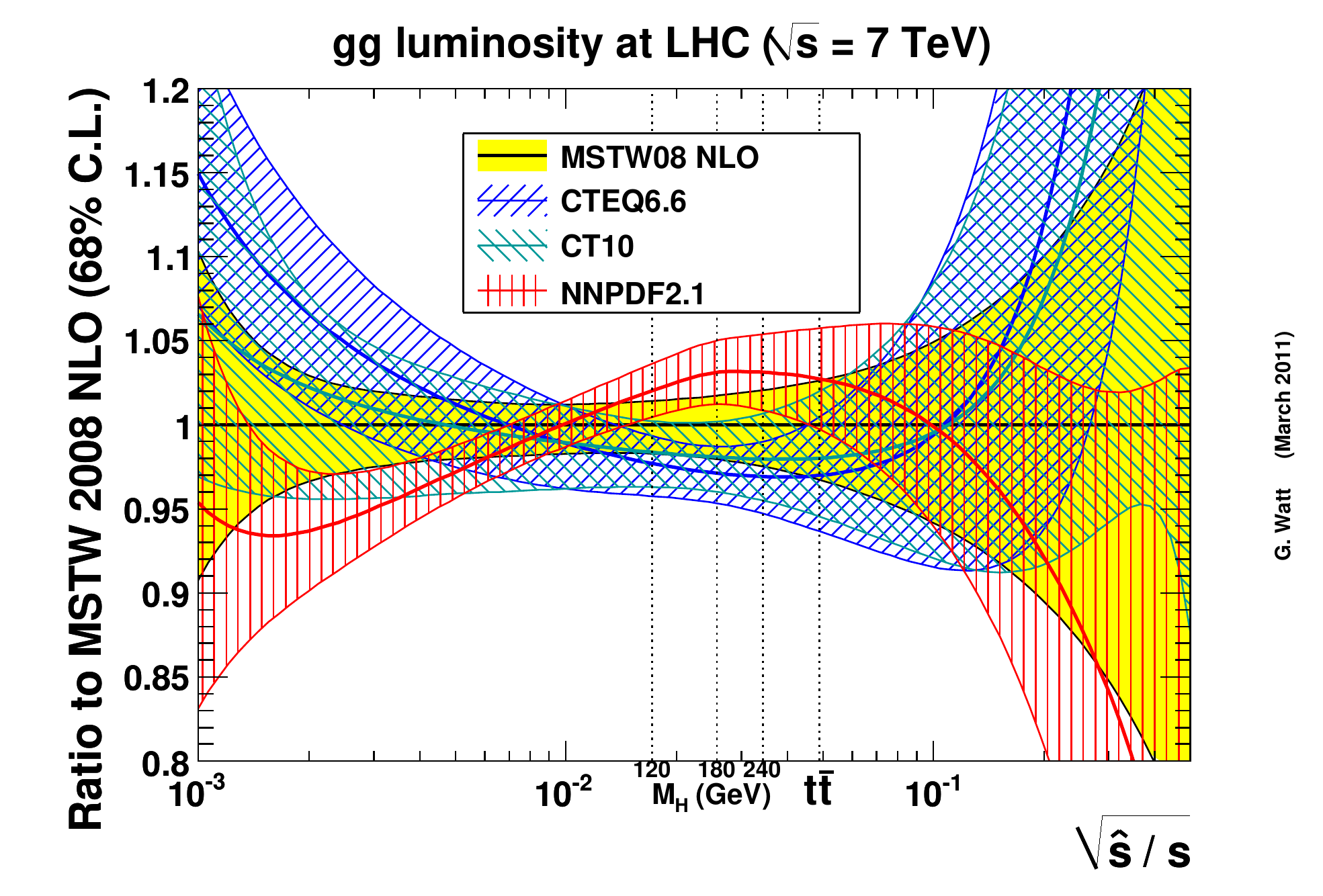}
\includegraphics[width=0.48\textwidth]{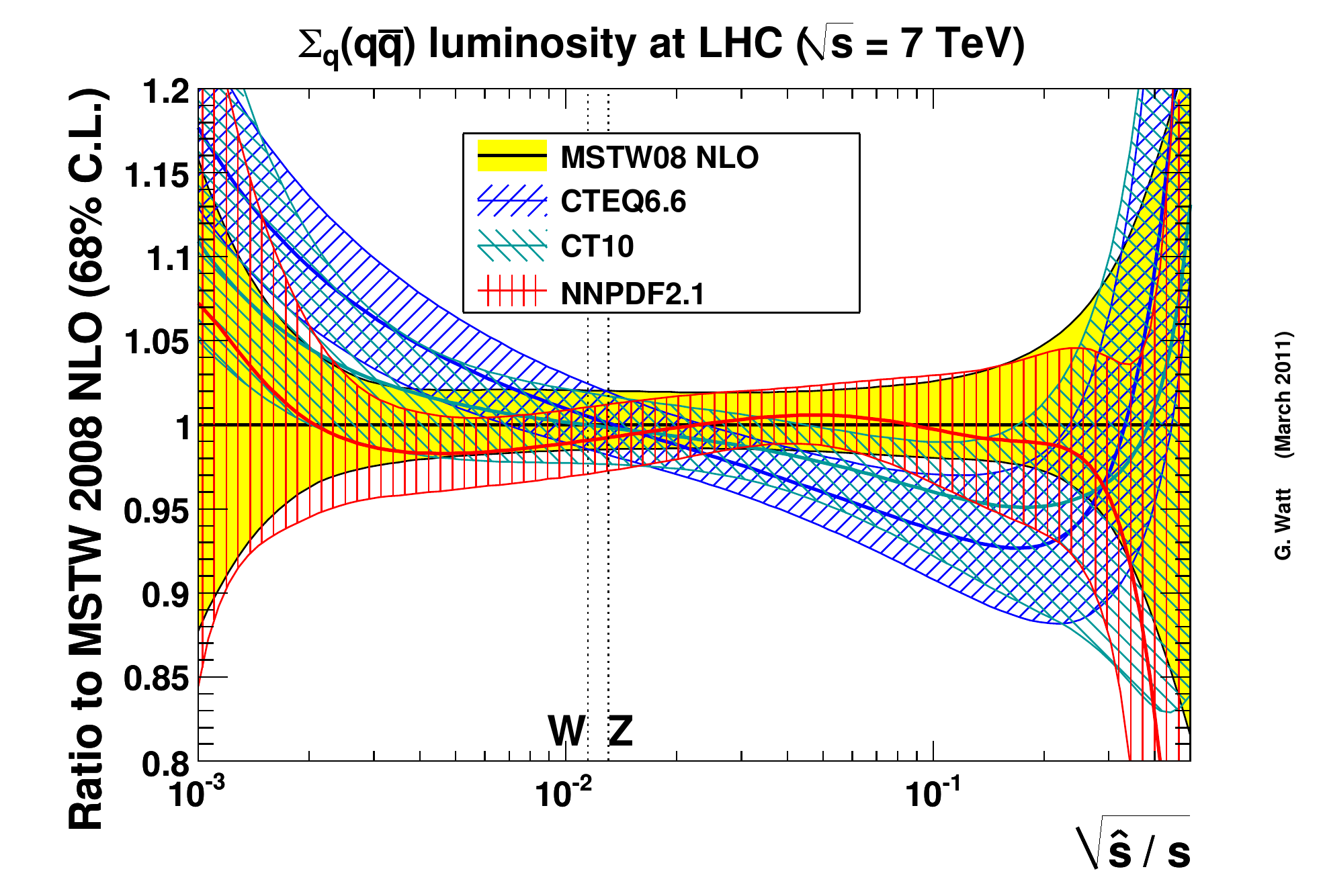}
\caption[Luminosities for $gg$ and $q\bar{q}$ PDF combinations at the $7$ TeV LHC]{Luminosities for $gg$ (left) and $q\bar{q}$ (right) PDF combinations at the $7$ TeV LHC. Figure from~\cite{Watt:2011kp}.}
\label{fig:pdflumidiff}
\end{figure}

\begin{figure}[ht]
\centering
\includegraphics[width=0.48\textwidth]{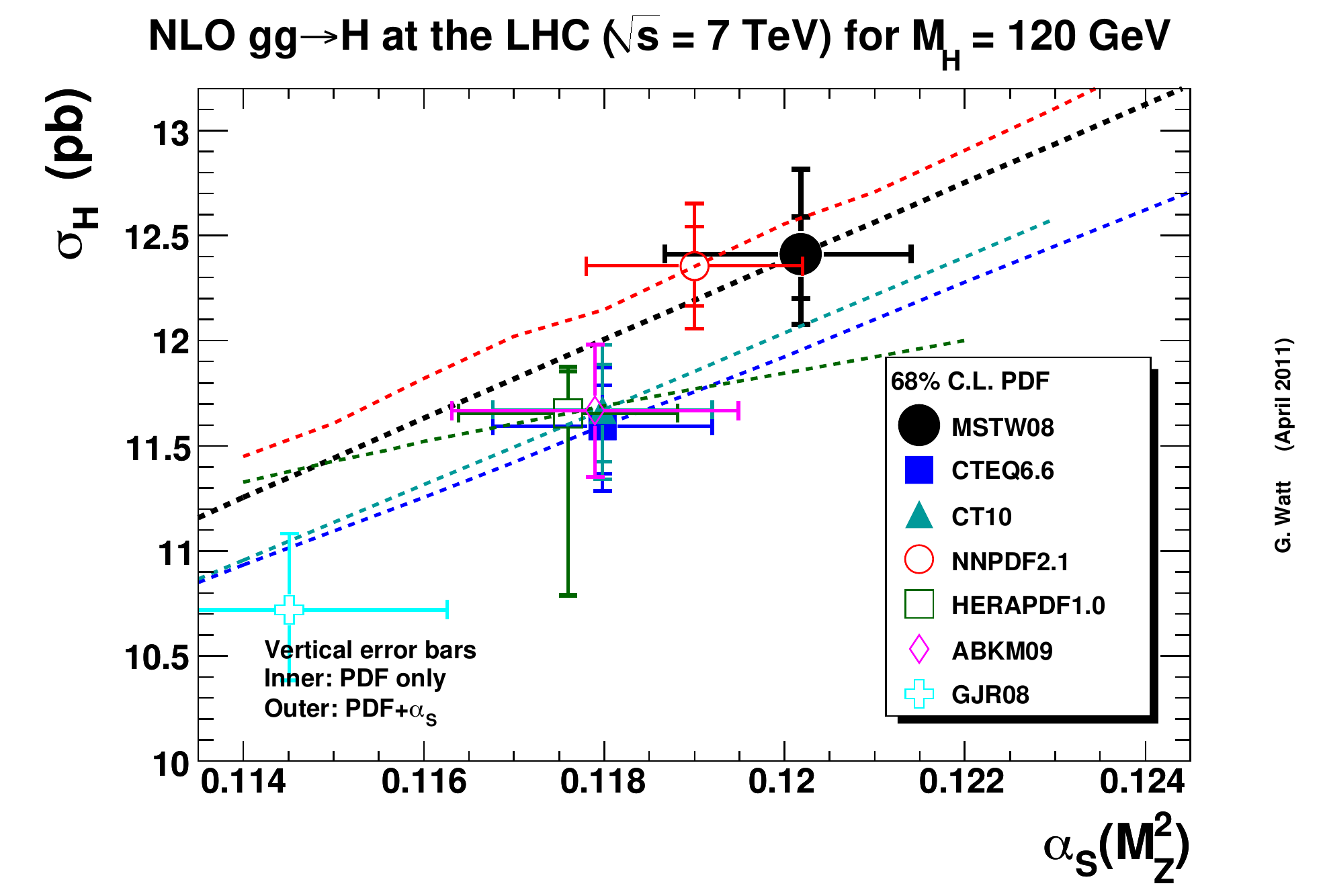}
\includegraphics[width=0.48\textwidth]{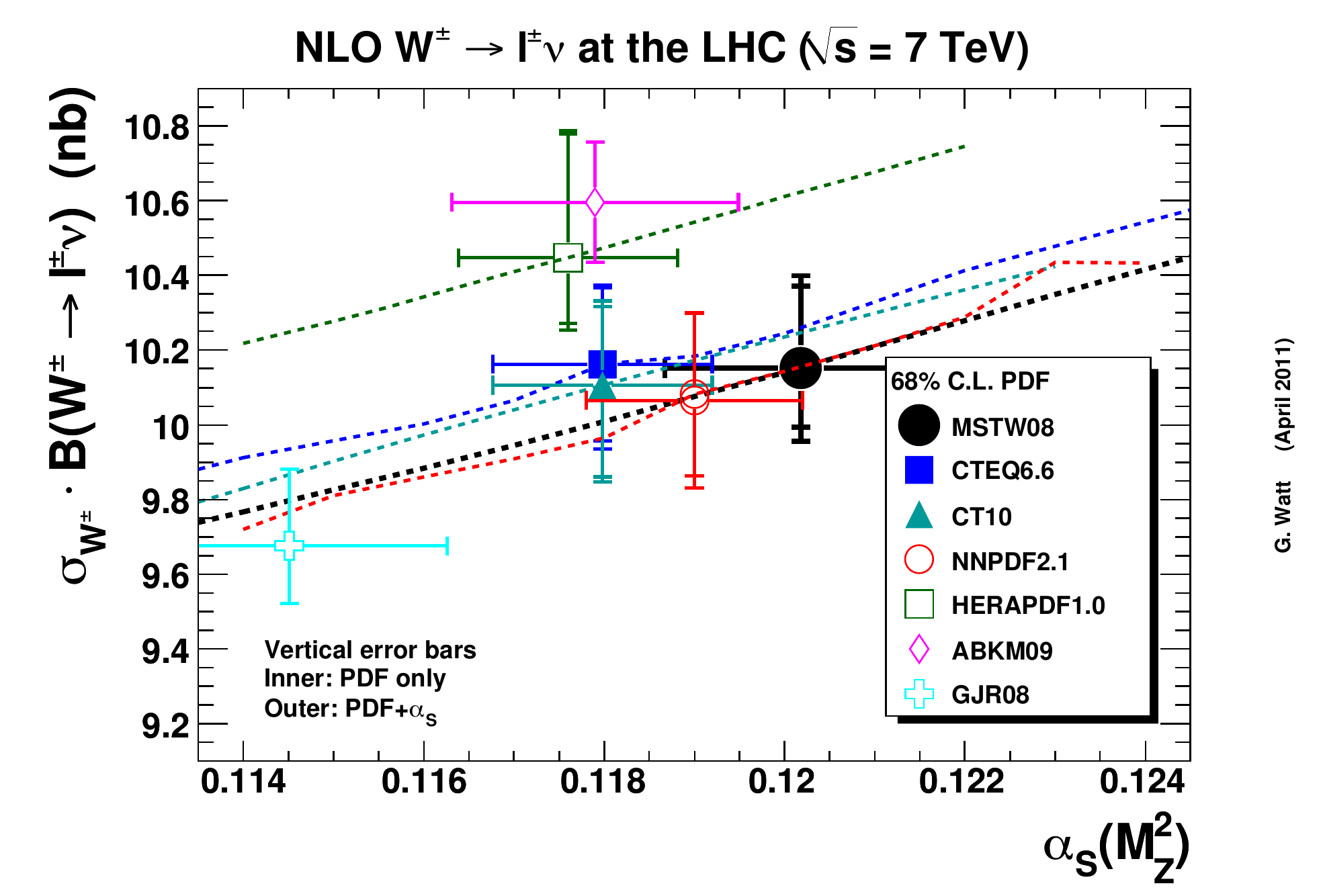}
\caption[Predictions for example LHC processes based upon a number of PDF determinations]{Predictions for LHC processes based upon a number of PDF determinations. Left figure: cross section for Higgs production in gluon fusion. Right figure: cross section for the production of $W$ bosons. Figure from~\cite{Watt:2011kp}.}
\label{fig:standardcandleerror}
\end{figure}

In preparation for the application of parton distributions at the LHC, extensive studies were performed in order to benchmark and understand areas of agreement and discrepancy across fitting collaborations~\cite{Watt:2011kp,Dittmar:2009ii}. While agreement had generally improved as the level of sophistication applied in parton fits increased, there were still notable regions where PDF fits from the widest datasets remained in disagreement at levels greater than their quoted uncertainties. Figure \ref{fig:pdflumidiff} illustrates the situation for two important PDF luminosities before the LHC. These discrepancies extended not only to so far unmeasured quantities such as Higgs production cross sections, but also to PDF standard candle observables such as $W$ boson production (c.f. Figure \ref{fig:standardcandleerror}).

The Les Houches benchmark exercise~\cite{Dittmar:2009ii} helped to elucidate the methodological source of many of these differences by testing fits from various methodologies to a standard dataset.

Many of the observed discrepancies arise due to differences in the theoretical description of data, with the choice of flavour number scheme providing the largest differences. Dataset choice and methodological choices introducing significant differences also. These differences led to the conservative PDF4LHC recommendation for observables to be calculated as the central contour of the CTEQ-MSTW-NNPDF uncertainty envelope. Despite the differences, for the LHC Run-I the range of available sets allowed for experimental collaborations to effectively explore the differences in the resulting predictions.

While providing accurate determinations for use at the LHC has been the primary concern in the years leading up to the LHC's first operation, there was substantial interest in the potential of the LHC to provide constraints upon PDFs and potentially provide discriminating power between sets. Data from the LHC provides the best opportunity for distinguishing the most effective approaches both theoretically and methodologically. Additionally LHC data provides particularly valuable input in the field of collider-only determinations, which aim to provide a cleaner description of data by avoiding the inclusion of nuclear-corrected and low energy data. The inclusion of a large LHC dataset into PDF fits is however a challenging problem, and one which has inspired a great deal of progress in the efficient calculation of collider observables. The remainder of this work will therefore deal with the both the technical inclusion of LHC data into parton distribution fits and the subsequent phenomenological results.

\chapter{Tools for the LHC}
\label{ch:LHCtools}
Including data from a wide range of LHC or collider sources into a global PDF determination provides several challenges, particularly in the context of the computationally intensive 
NNPDF methodology. In this chapter we shall discuss some of the methods that have been developed in order to study the impact of collider data, and include their constraints into
PDF fits.

Firstly we shall describe the method of Bayesian reweighting of Monte Carlo error sets, along with the associated set of tools made available by a Bayesian study of PDF sets and their uncertainties.
Secondly the FastKernel method developed by the NNPDF collaboration for the fast evolution of PDFs will be introduced, along with its extension to the fast computation of experimental observables in the {\tt FK} method.
Finally we shall discuss the application of interpolation methods such as FastKernel to the automated calculation of cross sections at next-to-leading order accuracy in QCD. To this end
we shall perform a brief overview of such calculations in the context of general purpose event generators.

\section{Bayesian reweighting}
When examining the statistical properties of PDF fits it is important to note that in the Monte Carlo approach, not only the uncertainties on PDFs are provided, but a full representation of the probability distribution.
As described in Section~\ref{sec:errors}, the integral over the PDF probability distribution is approximated by a sum over replicas,
\ba \left<f\right>(x,Q^2) &=& \int f(x,Q^2) \mathcal{P}\left(f \middle| d\right) \mathcal{D}f \nonumber \\
&\approx& \frac{1}{N_{\text{rep}}}\sum_i^{N_{\text{rep}}}f_i(x,Q^2),
\ea
where the subscript $i$ here refers to the PDF replica in the Monte Carlo ensemble. This correspondence leaves PDFs in the Monte Carlo representation open to standard statistical analysis methods. One of the most important of which is the \emph{Bayesian reweighting} technique, first proposed by Giele and Keller
alongside the original Monte Carlo procedure~\cite{Giele:1998gw} and then subsequently developed by the NNPDF collaboration~\cite{Ball:2011gg,Ball:2010gb}. The problem that reweighting seeks to address is the rapid addition of experimental data into an existing parton determination. The method is particularly useful in cases where there are no fast implementations of a calculation, and allows for the fast assessment of experimental impact upon PDFs and their uncertainties. 

Given a probability distribution for PDFs, Bayes' theorem suggests that we can update the experimental information in an existing PDF fit, here denoted $\mathcal{P}(f)$ by determining the conditional probability of the PDF given the new dataset ${y}$,  
\begin{equation}
  {\mathcal P}(f|y) \mathcal Df  =  \frac{{\mathcal P}(y|f)}{{\mathcal P}(y)} \, {\mathcal P}(f) \mathcal Df\, .   \label{eq:bayes2}
\end{equation}
However it was noted in Ref.~\cite{Ball:2010gb} that the probability of a PDF given the new data is not strictly what a fitting procedure would obtain. Rather the fitting procedure aims to find the probability distribution of the PDFs given some measure of fit quality to the new data, e.g $\chi^2$. Therefore to obtain a distribution statistically equivalent to a refit, one should attempt to determine
\begin{equation}
  \label{eq:bayes3}
  {\mathcal P}(f|\chi) \mathcal Df =  \frac{{\mathcal P}(\chi |f)}{{\mathcal P}(\chi)} \, {\mathcal P}(f) \mathcal Df\, ,
\end{equation}
where $\mathcal{P}(\chi)$ may be marginalised over to obtain the correct normalisation for ${\mathcal P}(f|\chi)$. Armed with such a distribution, we may then compute our predictions for a general observable given the information contained in the new dataset,
\begin{eqnarray}
\left<\mathcal{O}\right>_{\text{new}}&=&\int \mathcal{O}[f] \,
\mathcal{P}(f|\chi)\,Df\nn\\ &=&\int \mathcal{O}[f]
\,\frac{\mathcal{P}(\chi|f)}{\mathcal{P}(\chi)}
\mathcal{P}(f)\,Df,\nn
\end{eqnarray}
where $\left<\mathcal{O}\right>_{\text{new}}$ is the central value prediction for the observable $\mathcal{O}$ provided by a PDF distribution updated with the new experimental data. Given this probability distribution we can form a Monte Carlo representation in terms of PDF replicas once again,
\begin{eqnarray} 
\left<\mathcal{O}\right>_{\text{new}}&=&\frac{1}{N_{\text{rep}}}\sum_{i=1}^{N_{\text{rep}}}
\frac{\mathcal{P}(\chi|f_i)}{\mathcal{P}(\chi)}\,\mathcal{O}[f_i], \nn \\
&=& \frac{1}{N_{\text{rep}}}\sum_{i=1}^{N_{\text{rep}}} w_i\,\mathcal{O}[f_i]. 
\label{eq:avgnewx}
\end{eqnarray}
The weights $w_i$ for the individual replicas encoding the information from the new dataset, may be obtained from the $\chi^2$ goodness-of-fit measure to the new data
\begin{equation}
w_i = \frac{\mathcal{P}(\chi|f_i)}{\mathcal{P}(\chi)} \propto \chi_i^{n-1} e^{-\half\chi_i^2}.
\label{eq:weightscsq}
\end{equation}
Where $n$ denotes the number of new datapoints. The new data may therefore be included into an existing MC parton set by the simple calculation of a $\chi^2$ for each replica in the set. In comparison to a fitting procedure where many thousands of $\chi^2$ computations are required, this procedure is extremely fast. Furthermore, as a purely statistical exercise
this PDF reweighting does not suffer from any of the inherent vagaries of a fitting procedure.

The reweighting technique does however come at a cost in that it may reduce the overall efficiency of the Monte Carlo ensemble's representation of the underlying probability distribution. As can be seen from Eqn.~\ref{eq:weightscsq}, replicas in the prior distribution which do not provide a good description
of the new experimental data and therefore have a large $\chi^2$ value are penalised by small weights. For a sufficiently large or constraining dataset this can mean that many of the replicas are effectively switched out of the distribution, leaving a smaller number of \emph{effective} replicas. The efficiency of the
representation can be quantified by the Shannon entropy, which provides the number of effective replicas as
  \be N_{\textrm{eff}} \equiv \exp \left(\frac{1}{N_{\mathrm{rep}}}\sum_{i=1}^{N_{\mathrm{rep}}}w_i\ln(N_{\mathrm{rep}}/w_i)\right). \label{eq:shannonentropy}\ee
As the constraining power of the new dataset increases, so the Shannon entropy $N_{\text{eff}}$ decreases. Consequently a larger number of replicas sampling the prior distribution are required to maintain a fixed level of ensemble accuracy. Despite this limitation, reweighting can provide an extremely useful method
for analysis of a typical experimental dataset.

\subsection{Error rescaling parameter}
A Bayesian analysis of the Monte Carlo probability representation opens up other avenues of investigation. Of particular interest is the examination of an \emph{error rescaling parameter}.
When examining the impact of an experimental measurement, we can study how the constraints are modified under a global rescaling of the experimental error, i.e the $\chi^2$ values
\be \chi^2_k \to \chi^2_{k,\alpha} = \chi^2_k/\alpha^2, \ee
where $\alpha$ is the rescaling parameter. In our reweighting exercise the weights are subsequently given by
\be w_k(\alpha) \propto (\chi^2_{k,\alpha})^{(n-1)/2}\mathrm{e}^{-\chi^2_{k,\alpha}/2}.\ee
Our Bayesian expression for the updated probability density is now also a function of the rescaling parameter $\alpha$. A further application of Bayes' theorem inverts
this relationship, and allows us to form a probability density for the rescaling parameter itself.
\be \mathcal{P}(\alpha | \chi^2)\propto \smallfrac{1}{\alpha}\sum_{k=1}^N w_k(\alpha).\ee
This probability distribution provides an estimate as to whether the experimental errors in the new dataset may have been under or overestimated, based upon agreement with the prior distribution. An experimental result where the uncertainties have accurately estimated leads to a $\mathcal{P}(\alpha)$ distribution peaked at $\alpha=1$, whereby an over(under)-estimated set of uncertainties leads to a lower(higher) peak in the distribution.  This is a particularly
useful tool for analysing experimental uncertainties, and can provide some differentiation between inconsistent and constraining data in cases where $N_{\mathrm{eff}}$ is small.

\subsection{PDF unweighting}

While the PDF reweighing approach is a powerful method for the addition of new data to an existing set, a reweighted PDF set is unsuitable for general distribution. For use in typical calculational codes, a standard interface is required through packages such as LHAPDF. Therefore the provision of a PDF ensemble with an  associated set of weights would require the retooling of codes in which a reweighted calculation is desired. To alleviate this a method was developed in order to present a reweighted distribution as a standard MC replica ensemble~\cite{Ball:2011gg}.

\begin{figure}[ht!]
\centering
\includegraphics[width=1\textwidth]{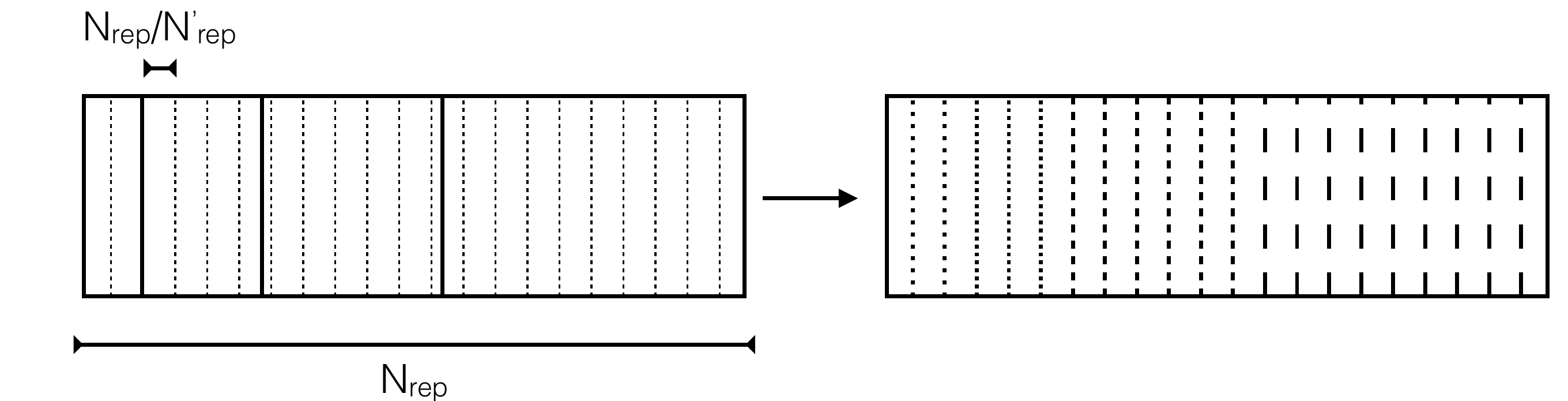}
\caption[Unweighting of a Bayesian reweighted Monte Carlo PDF set]{The unweighting of a Bayesian reweighted Monte Carlo PDF set. The left hand figure shows the weight cumulant segments for the original weighted set, with four replicas of weight $w_i = \left\{ 1, 2, 3, 4 \right\} $. The line is subdivided by $N_{\text{rep}}^\prime = 20$ lines. The right hand figure illustrates the \emph{unweighted} set in this case. Here each replica in the unweighted set has equal weight, with different line strokes denoting different replicas from the weighted distribution. }
\label{fig:unweighting}
\end{figure}

This is done by representing the reweighted set upon a cumulative line of weights as in Figure~\ref{fig:unweighting}. Each line segment corresponds to the weight of an individual replica. The total cumulant line therefore being normalised to $N_{\text{rep}}$, the number of replicas in the reweighted distribution. Replicas in an `unweighted' set are then chosen by distributing evenly $N_{\text{rep}}^\prime$ replicas across this cumulant line. When one of these replicas falls into the weight segment of a corresponding reweighted replica, that PDF is selected for inclusion in the unweighted set. Importantly, the same reweighted replica may be selected more than once to appear in the unweighted set.

As an example, consider the case where there are four replicas in an initial distribution, with weights $w_i = \left\{ 1, 2, 3, 4 \right\} $. The cumulant line formed by these weighted replicas is shown on the left side of Figure~\ref{fig:unweighting}. This line is subdivided into $N_{\text{rep}}^\prime + 1$ intervals. With $N_{\text{rep}}^\prime = 20$ as shown in the Figure, two unweighted replicas fall in the first weighted segment, three in the second, six in the third and nine in the fourth. Therefore the unweighted ensemble is formed by duplicating the original weighted replicas with a frequency dictated by how many unweighted replicas fall in their respective line segment.

The weights of the original set are therefore approximately represented as replica multiplicities in the unweighted set, with low-weight replicas selected few times (if at all), and large weight replicas selected multiple times. In this way a conventional MC ensemble can be formed with the usual LHAPDF interface, this time including duplicate replicas for those with high weights and excluding replicas with weights that fall under the unweighted set's resolution. Therefore the unweighting procedure can provide an exact representation of the reweighted ensemble in the limit $N_{\text{rep}}^\prime \to \infty$. 

However in practice a number of unweighted replicas of the order of the number of effective replicas $N_{\text{eff}}$ is typically sufficient for a good level of accuracy in the reproduction.

\subsection{Reweighting validation}
\begin{figure}[ht!]
\centering
\includegraphics[width=0.45\textwidth]{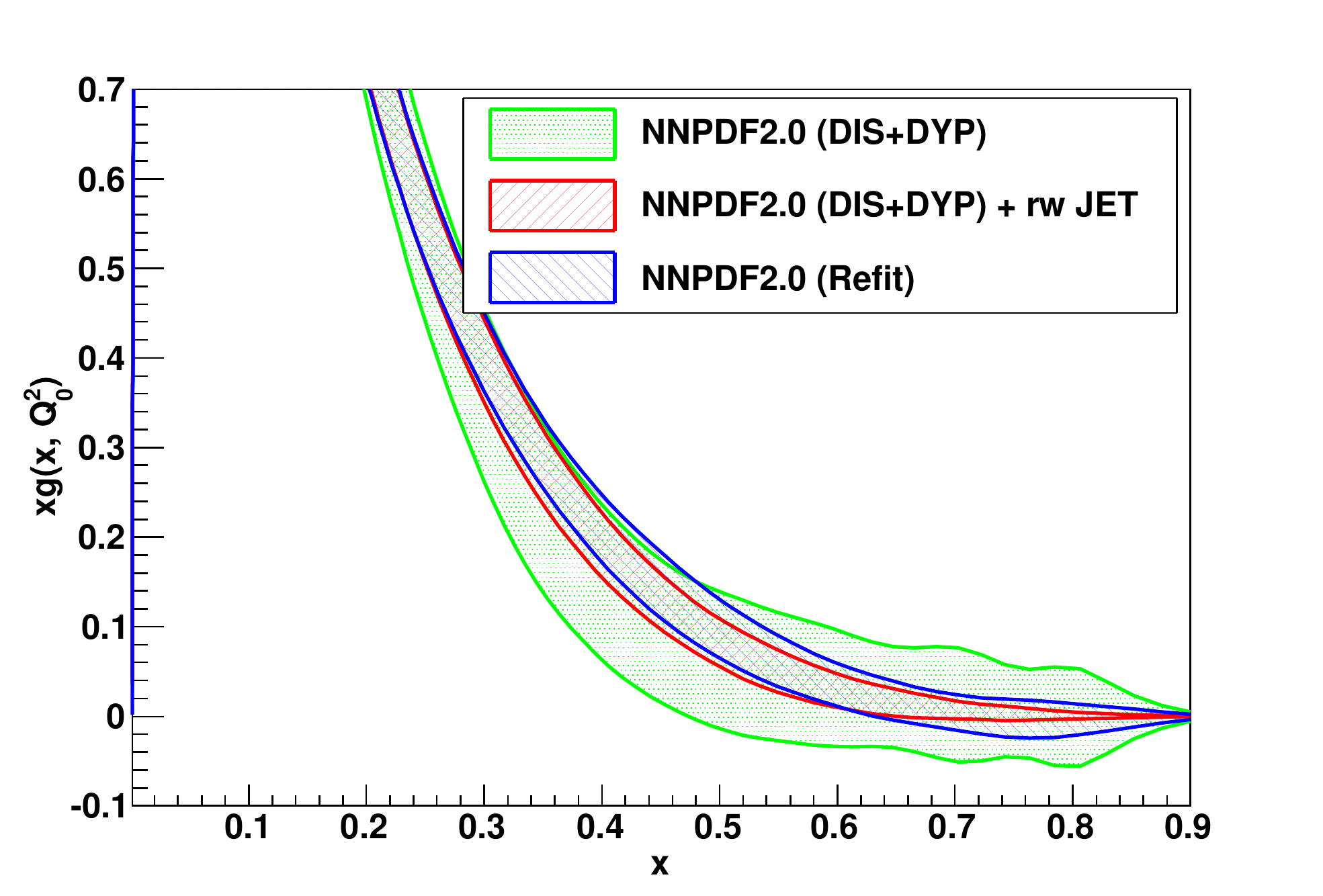}
\includegraphics[width=0.45\textwidth]{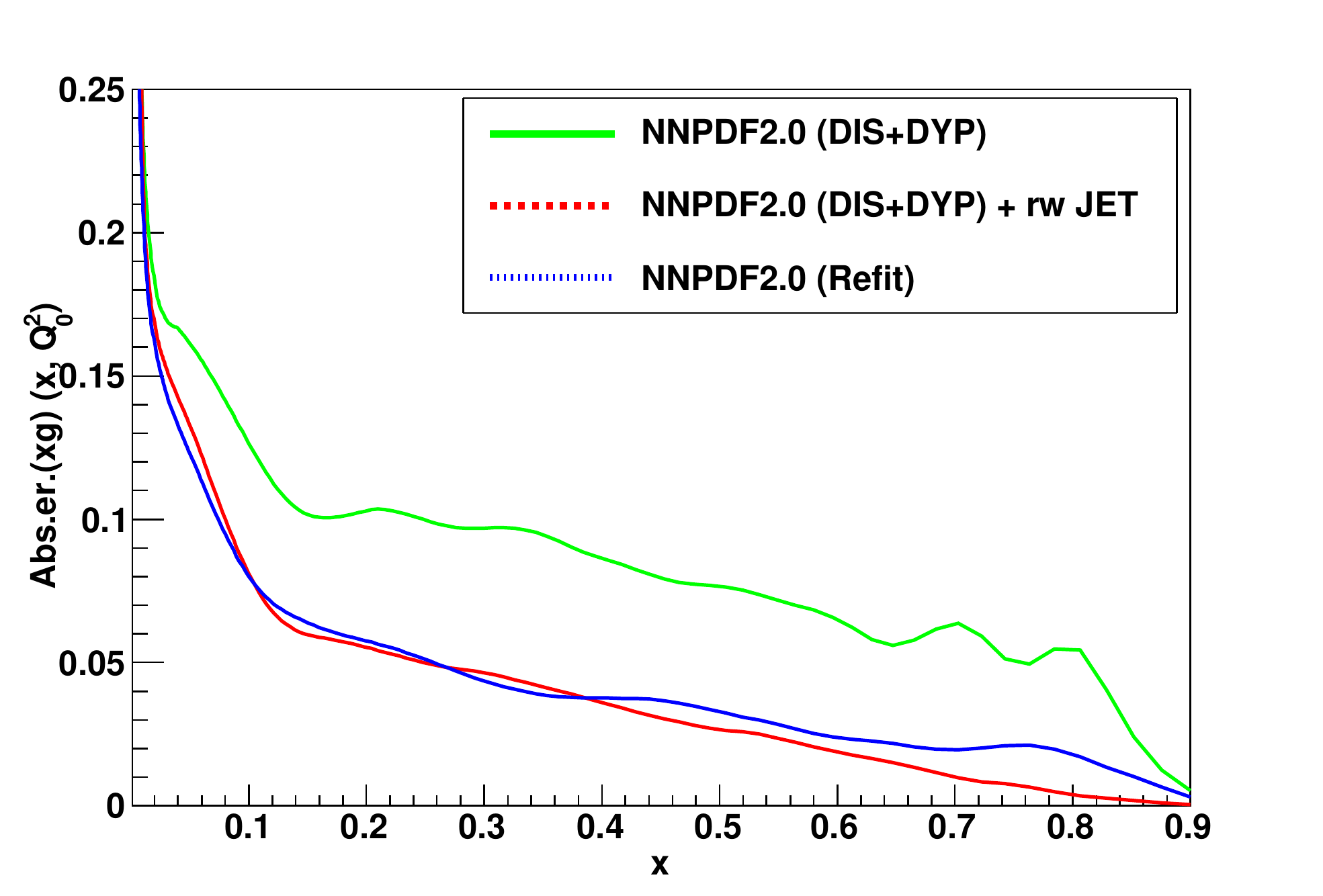}
\caption[Validation of Bayesian reweighting by the inclusion of Tevatron jet data]{The validation of Bayesian reweighting by the inclusion of Tevatron jet data. The left figure demonstrates the prior distribution along with the reweighted and refitted distributions upon the addition of Tevatron jet data. The right plot shows the absolute error upon the PDFs for the three sets. Figures are from~\cite{Ball:2010gb}.}
\label{fig:rwvalid}
\end{figure}

The Bayesian reweighting procedure has been extensively validated by the NNPDF collaboration in a number of highly non-trivial tests of the methodology. As the method has been designed to update a prior distribution with new information analogously to the approach used in an ideal fit, the first test is to ensure that a PDF set reweighted with a new dataset is statistically equivalent to a new set refitted from scratch utilising the new data. This was first performed in~\cite{Ball:2010gb} by reweighting an NNPDF 2.0 fit which included only DIS and Drell-Yan data with information from Tevatron inclusive jet measurements. The reweighted set was compared to the full NNPDF 2.0 fit including the data. As Figure~\ref{fig:rwvalid} demonstrates, the reweighted set is able to reproduce the refitted set up to the level of statistical fluctuation.
\clearpage

The development of the unweighting method as outlined in the previous section, allowed for further tests of the reweighting method. A series of tests were carried out in order to assess the behaviour of PDFs under successive reweighting operations.

When including multiple datasets into a PDF fit via reweighting, there are three possibilities. One can reweight with the combined $\chi^2$ values for the two experiments, or reweight first with one experiment, unweight the PDF ensemble, then reweight with the second. The resulting PDFs should be reasonably independent of the method chosen, and of the order in which the successive reweighting is performed. This requirement is a stringent test of the Monte Carlo PDF representation, as it determines whether or not the ensemble truly behaves as a probability distribution. More pragmatically, the test verifies whether the loss of ensemble efficiency in one reweighting operation is not so great as to prevent a further reweighing. This investigation was carried out in Ref.~\cite{Ball:2011gg} with a DIS only prior. The E605 Drell-Yan experiment and CDF/D0 inclusive jet measurements were included into this set by reweighting. As the E605 experiment provides global fits with rather stringent constraints compared to the moderate effect of the jet data, this is a rather asymmetrical and therefore effective test.

\begin{figure}[ht]
\centering
\includegraphics[width=0.48\textwidth]{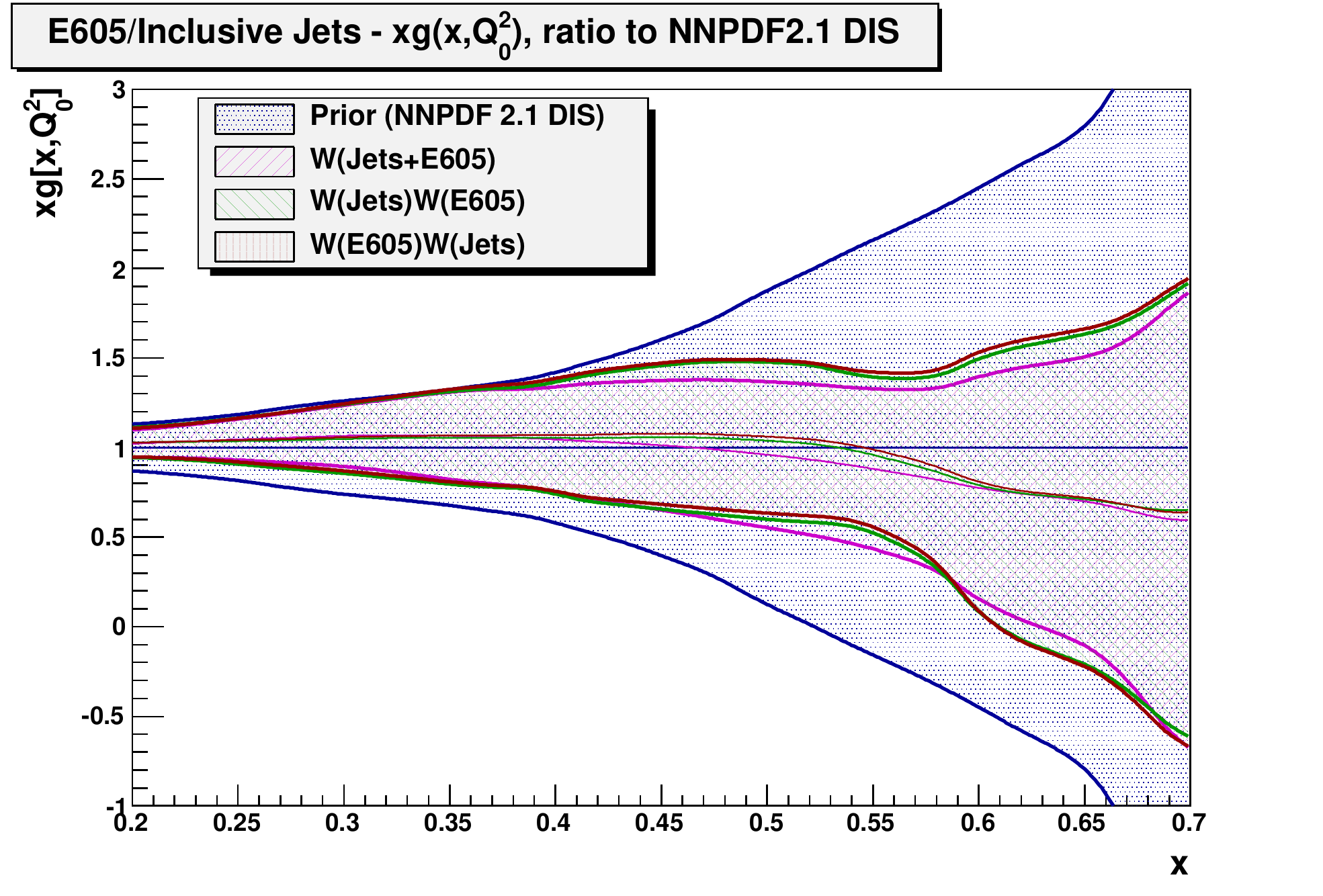}
\includegraphics[width=0.48\textwidth]{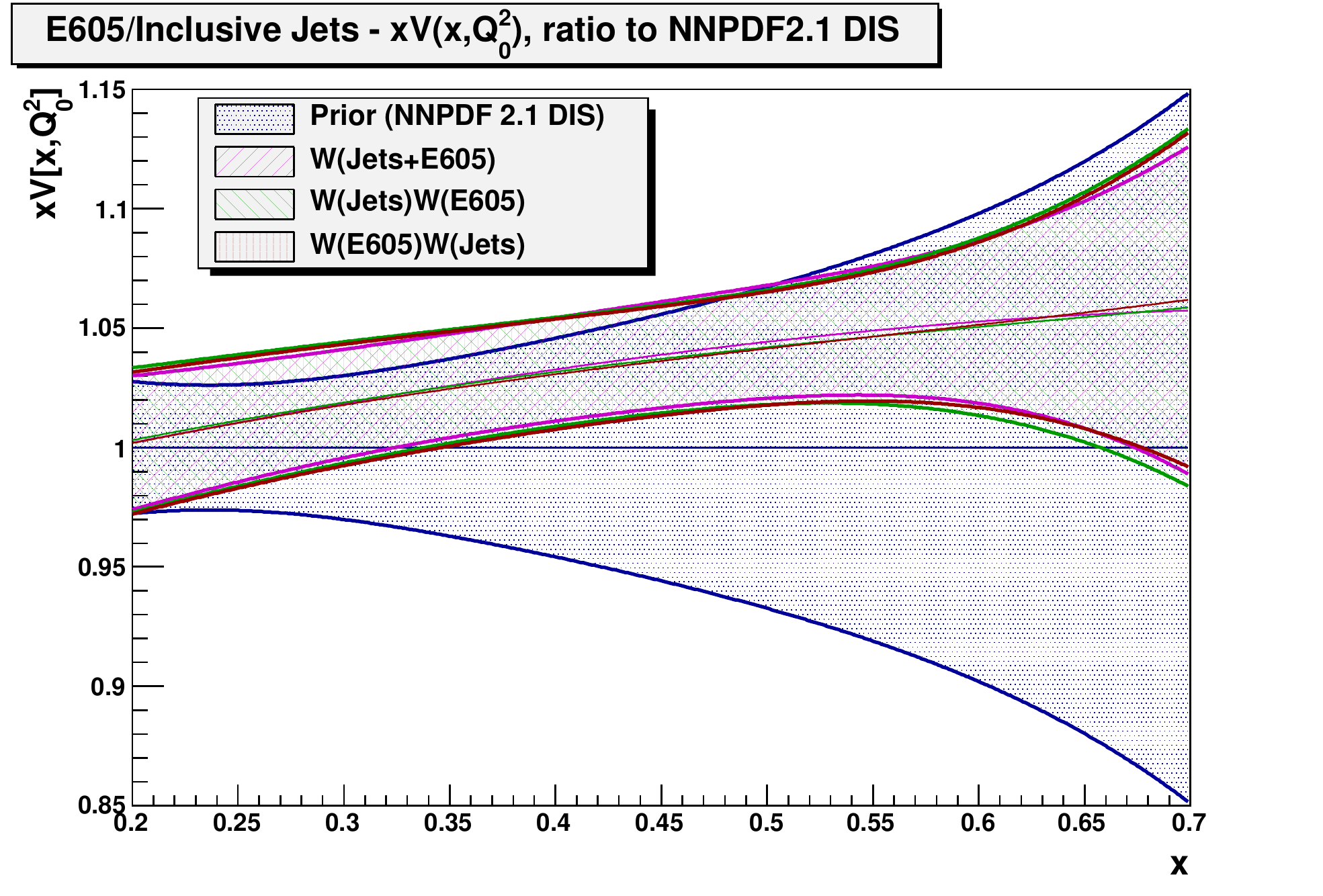}
\caption[Test of Bayesian reweighting under a successive reweighting operation]{Test of Bayesian reweighting under a successive reweighting operation. Inclusive Jet and Drell-Yan data are added as a combined dataset, and as individual reweightings separated by an unweighting operation. The resulting distributions, the gluon PDF on the left and the valence distribution on the right, show excellent agreement between the different procedures. All curves are normalised to the prior, NNPDF 2.1 DIS result. The figures are taken from~\cite{Ball:2011gg}.}
\label{fig:SRWvalid}
\end{figure}

In Figure~\ref{fig:SRWvalid} these reweighting procedures are compared for the case of the gluon and valence distributions of the NNPDF2.0 DIS only fit. It is clear that while the impact of the data upon the prior is substantial, the three reweighting methods hardly differ in their results. There is therefore a strong confirmation of the statistical properties of both the Monte Carlo representation of PDFs, and of the reweighting method.

\section{The FastKernel method}
The method of Bayesian reweighting provides an extremely fast and efficient method of including new data into a determination. However as described previously, the method is
ill-suited to the addition of a large or very constraining dataset as the required size of the prior distribution in replicas rapidly becomes unmanageable. Therefore the standard fitting methodology
remains the most important procedure in the determination of parton distributions.

The primary issue in the standard methodology upon the addition of a large LHC dataset is the computational time required to perform the theoretical predictions for experimental data.
Not only must the standard double convolution over the two parton densities be performed, but also each PDF must be evolved from some initial fitting scale to the scale of the experimental data by yet another set of convolutions. We shall first describe the methods used for fast PDF evolution, before going on to discuss the extension to the calculation of physical observables at colliders.

\subsection{Fast PDF evolution}
While there are many methods for performing the evolution of parton distributions, the technique used in NNPDF fits must be particularly efficient due to the computational complexity of the NNPDF procedure. The evolution of a flavour basis PDF of flavour $i$ from an initial scale $Q^2_0$ to a target scale $Q^2_\tau$ can be expressed as 
\be f_i(x_{\alpha}, Q^2_\tau) = \sum_j^{N_{f}} \int_{x_\alpha}^1 d\xi \; \Gamma_{ij}\left(\frac{x_\alpha}{\xi},\frac{Q^2_\tau}{Q_0^2}\right) f_j(\xi, Q_0^2), \label{eq:DGLAPconv} \ee
where the $\Gamma$ are found by solution of the DGLAP equation as shown in Eqn.~\ref{eq:DGLAP}. In order to take advantage of the sparse nature of the DGLAP evolution kernels, we work in the evolution basis defined in Section~\ref{sec:DGLAP},
\be N_i(x_{\alpha}, Q^2_\tau) = \sum_j^{N_{f}} \int_{x_\alpha}^1 d\xi \; \widetilde{\Gamma}_{ij}\left(\frac{x_\alpha}{\xi},\frac{Q^2_\tau}{Q_0^2}\right) N_j(\xi, Q_0^2), \ee
where here we have introduced the notation $N$ for the evolution basis PDFs; related to the flavour basis by a simple rotation
\be f_i(x,Q^2_\tau) =  \sum_j^{N_f}R_{ij}N_j(x,Q_\tau^2). \label{eq:LHA2EVLN}\ee 
Having to perform many instances of the convolution integral in Eqn.~\ref{eq:DGLAPconv} would be prohibitively expensive in most fitting applications, and so an alternative approach must be used. In the NNPDF framework this is based upon the {\tt FastKernel} interpolation method introduced in Ref.~\cite{Ball:2010de}, and shares the general approach with other interpolating methods, while maintaining a hybrid $x$ and Mellin space solution. Here we shall outline the general method used in all interpolating tools.

The first step is to expand the initial-state PDFs upon some set of interpolating basis functions $\mathcal{I}$,
\be f_i(x,Q^2_0) \approx \sum_\beta^{N_{\mathrm{fn}}} c_i^{(\beta)} \;\mathcal{I}^{(\beta)}(x),\ee
with the coefficients of this expansion calculable through the usual overlap integral
\be c^{(\beta)}_i = \int_0^1 dx\;f_i(x,Q^2_0) \; \mathcal{I}^{(\beta)}(x). \ee
Substituting the interpolated version of the initial state PDF into the evolution equation and applying the inverse transformation of Eqn.~\ref{eq:LHA2EVLN} to work in the evolution basis we obtain
\be N_i(x_{\alpha}, Q^2_\tau) = \sum_{j,k}^{N_{f}} \sum_\beta^{N_{\mathrm{fn}}} \int_{x_\alpha}^1 d\xi \; \widetilde{\Gamma}_{ij}\left(\frac{x_\alpha}{\xi},\frac{Q^2_\tau}{Q_0^2}\right) R^{-1}_{jk}\;c_k^{(\beta)} \;\mathcal{I}^{(\beta)}(x). \ee
In this expression, we can actually factorise the PDF-dependent expansion coefficients $c$ from the integral, and perform the convolution over the interpolating functions
\be N_i(x_{\alpha}, Q^2_\tau) = \sum_{k}^{N_{f}} \sum_\beta^{N_{\mathrm{fn}}} E_{ik\alpha\beta}^\tau\;c_k^{(\beta)}, \ee
where the \emph{evolution tables} $E$ are given by 
\be E_{ik\alpha\beta}^\tau =  \sum_{j}^{N_{f}} \int_{x_\alpha}^1 d\xi \; \widetilde{\Gamma}_{ij}\left(\frac{x_\alpha}{\xi},\frac{Q^2_\tau}{Q_0^2}\right) R^{-1}_{jk} \;\mathcal{I}^{(\beta)}(x). \label{eq:evolfactors} \ee
While it may seem as if we have simply moved the problem from the convolution with the DGLAP kernel to the overlap integral required to compute the coefficients $c$, this can be
avoided via a careful choice in the interpolating functions. A suitable choice of interpolating function yields the following identification for the coefficients
\be c_k^{(\beta)} = f_k(x_\beta,Q^2_0),\ee
that is, the interpolants effectively pick out the value of the PDF at some point $\beta$ in an $x$-grid. Providing the grid in $\beta$ is dense enough the interpolation accuracy can still be very high.
With such a choice of functional basis, the full evolution product becomes particularly simple
\ba 
f_i(x_{\alpha},Q^2_\tau) &=&  \sum_{j,k}^{n_f}\sum_{\beta}^{N_x} R_{ij}E^\tau_{\alpha\beta j k}\; f_k(x_\beta,Q^2_0), \\
&=& \sum_{k}^{n_f} \sum_\beta^{N_x} A^\tau_{\alpha\beta ik}\; f_k(x_\beta,Q^2_0). \label{eq:fastPDFfinal}
\ea 
The convolution required by the initial solution to the DGLAP equation has now been reduced via interpolation methods to a simple product over a rotated evolution table $A$.

\subsection{Fast calculation of collider observables}
Similar methods to what we have discussed for fast PDF evolution have also been applied to the calculation of collider observables. For a typical observable with two partons in the initial state,
a full calculation is given by a double convolution over two parton densities,
\be  \sigma_{pp\to X} = \left( \frac{\alpha_s(Q^2)}{2\pi} \right)^{p} \int dx_1\, dx_2\,  f_i(x_1,Q^2)\; d\hat{\sigma}_{ij\to X}\; f_j(x_2,Q^2) \, . \label{eq:hadconv} \ee
The double convolution can once again be avoided by inserting interpolated versions of the PDFs, and performing the convolution over the interpolating functions.
\be  \sigma_{pp\to X} = \left( \frac{\alpha_s(Q^2)}{2\pi} \right)^{p} \sum_{\alpha,\beta}^{N_X}  f_i(x_\alpha,Q^2)\; W_{\alpha\beta,ij}\; f_j(x_\beta,Q^2) \, . \label{eq:hadconv2} \ee
The weight grid $W$ is calculated analogously to the evolution tables in Eqn.~\ref{eq:evolfactors},
\be W_{\alpha\beta,ij} =  \int dx_1\, dx_2\,  \mathcal{I}^{(\alpha)}(x_1)\; d\hat{\sigma}_{ij\to X}\; \mathcal{I}^{(\beta)}(x_2) \, . \ee
Identical methods can be used to interpolate over the hard scale $Q^2$ in multi-scale processes. These techniques are used in publicly available tools such as { \tt APPLgrid}~\cite{Carli:2010rw} and { \tt FastNLO}~\cite{Kluge:2006xs}. In the { \tt APPLgrid} framework, the full product used to calculate a hadronic observable is
\be
\label{eq:applconv}
\sigma = \sum_p \sum_{s}^{N_{\mathrm{sub}}} \sum_{\alpha,\beta}^{N_x} \sum_{\tau}^{N_{Q}}
W_{\alpha\beta\tau}^{(p)(s)} \, \left( \frac{\alpha_s\left(Q^2_{\tau}\right)}{2\pi}\right)^{p}
F^{(s)}\left(x_{\alpha}, x_{\beta},  Q^2_{\tau}\right),
\ee
where the interpolation over a grid of points in hard scale runs over the index $\tau$, and the perturbative order of the contributions is separated by the index $p$. The initial state parton combinations have been grouped into the appropriate QCD subprocesses $s$, according to a table of coefficients $C$,

\be F^{(s)}\left(x_{\alpha}, x_{\beta},  Q^2_{\tau}\right) =  \sum_{i,j}^{13} C^{(s)}_{ij}  \left( f_i(x_{\alpha},Q^2_\tau)f_j(x_{\beta},Q^2_\tau) \right).  \label{eq:APPLsubproc}\ee
The resulting product in Eqn~\ref{eq:applconv} allows for the simple variation of PDFs, strong coupling and perturbative scales in a fast calculation; the product taking typically of order milliseconds rather than the hours to days required to obtain reliable statistics in an NLO code.

Despite the dramatic speed improvement, the { \tt APPLgrid}/{ \tt FastNLO} products represent a considerable computational expense when introducing a large dataset. The NNPDF methodology in particular is extremely sensitive to the convolution speed due to the nature of the genetic algorithm minimisation, orders of magnitude more convolutions are required than in competing approaches. Therefore in order to practically include a large collider dataset into an NNPDF fit more work must be done on improving the convolution algorithm.

\subsection{Combined evolution and observable calculation}
The { \tt APPLgrid}/{ \tt FastNLO} approach maintains a great deal of flexibility, in that scale, $\alpha_S$ and PDF variations are all possible within the same framework. In a PDF fit the only requirement is an efficient variation of input parton distributions. We can therefore try to improve the efficiency of the calculation at the cost of some of the flexibility available in the fast convolution tools. The {\tt FK} procedure and toolchain has therefore been developed, implementing a combined PDF evolution and collider observable calculation.

Recalling the fast PDF evolution method in Eqn.~\ref{eq:fastPDFfinal} with the suitable grids precomputed, PDF evolution can be performed simply as  
\be
f_i(x_{\alpha},Q^2_\tau) = \sum_{k}^{n_f} \sum_\beta^{N_x} A^\tau_{\alpha\beta ik}\; f_k(x_\beta,Q^2_0). \label{eq:fastPDFfinal_recalled}
\ee 
The evolution of the { \tt APPLgrid} subprocess in Eqn.~\ref{eq:APPLsubproc} from an initial state distribution is therefore
\ba F^{(s)}\left(x_{\alpha}, x_{\beta},  Q^2_{\tau}\right) &=&  \sum_{i,j}^{13} \sum_{k,l}^{n_f}  \sum_{\delta,\gamma}^{N_x} C^{(s)}_{ij}  \left[  A^\tau_{\alpha\delta ik}\; f_k(x_\delta,Q^2_0) A^\tau_{\beta\gamma jl}\; f_l(x_\gamma,Q^2_0) \right]\;\;\; \\
&=&   \sum_{k,l}^{n_f}\sum_{\delta,\gamma}^{N_x} \widetilde{C}^{(s),\tau}_{kl,\alpha\beta\gamma\delta} f_k(x_\delta,Q^2_0) f_l(x_\gamma,Q^2_0), \label{eq:FK1}
\ea
where the evolved subprocess coefficients are
\be \widetilde{C}^{(s),\tau}_{kl,\alpha\beta\gamma\delta} = \sum_{i,j}^{13} C^{(s)}_{ij} A^\tau_{\alpha\delta ik} A^\tau_{\beta\gamma jl}.\ee
Substituting the expression for the subprocess in terms of initial state PDFs, Eqn.~\ref{eq:FK1}, into the { \tt APPLgrid} expression for the full convolution shown in Eqn.~\ref{eq:applconv} we obtain
\be
\sigma = \sum_p \sum_{s}^{N_{\mathrm{sub}}} \sum_{\alpha,\beta}^{N_x^\prime} \sum_{\tau}^{N_{Q}}
W_{\alpha\beta\tau}^{(p)(s)} \, \left( \frac{\alpha_s\left(Q^2_{\tau}\right)}{2\pi}\right)^{p}
 \sum_{k,l}^{n_f}\sum_{\delta,\gamma}^{N_x} \widetilde{C}^{(s),\tau}_{kl,\alpha\beta\gamma\delta} f_k(x_\delta,Q^2_0) f_l(x_\gamma,Q^2_0).
\ee
where the number of points in the { \tt APPLgrid} $x$-grid is denoted $N_x^\prime$ to indicate that the grid is different to the input parton $x$-grid which runs over $\gamma,\delta$ up to $N_x$ points. Now that the PDF evolution has been factorised into the coefficients $\widetilde{C}$, much more of this sum may now be precomputed. Specifically we are now able to sum over the indices for subprocess $s$, perturbative order $p$, hard scale $\tau$, and the { \tt APPLgrid} $x-$grids $\alpha$ and $\beta$. The resulting expression for the combined evolution and observable calculation is therefore
\be
\label{eq:FK}
\sigma = \sum_{k,l}^{n_f}\sum_{\delta,\gamma}^{N_x} 
\widetilde{W}_{kl\delta\gamma} \,f_k(x_\delta,Q^2_0) f_l(x_\gamma,Q^2_0),
\ee
with the combined grid, which may be precomputed and stored, given by
\be
\label{eq:FKTable}
\widetilde{W}_{kl\delta\gamma} = \sum_p \sum_{s}^{N_{\mathrm{sub}}} \sum_{\alpha,\beta}^{N_x^\prime} \sum_{\tau}^{N_{Q}}
W_{\alpha\beta\tau}^{(p)(s)} \, \left( \frac{\alpha_s\left(Q^2_{\tau}\right)}{2\pi}\right)^{p}\widetilde{C}^{(s),\tau}_{kl,\alpha\beta\gamma\delta} .
\ee

The quantity $\widetilde{W}_{kl\delta\gamma}$ is the {\tt FK} table for the observable $\sigma$ and encodes all of the theoretical treatment of the observable. The product in Eqn.~\ref{eq:FK} is therefore completely
agnostic with regards to all theory parameters such as process, scales, perturbative order and strong coupling value. This makes the {\tt FK} table particularly simple to implement in a fitting procedure, and allows a clean separation
of theory concerns from the calculation.

The {\tt FK} convolution also benefits from requiring considerably fewer floating point operations than a typical { \tt APPLgrid} convolution. This is particularly evident when studying multi-scale processes, where the sum over the scale grid is precomputed. The product over PDF flavours is now also limited to the $n_f$, typically seven, light partons rather than the general 13 parton basis. Of course in the {\tt FK} procedure the ability to vary scales and the strong coupling with a single grid is lost, and new {\tt FK} tables $\widetilde{W}$ must be generated for different theoretical treatments.

The procedure outlined above for generating {\tt FK} tables from { \tt APPLgrid} or { \tt FastNLO} files has been implemented in a {\tt C++} framework, alongside a comprehensive toolchain for performing {\tt FK} table {\tt I/O} and optimisation. The convolution in Eqn.~\ref{eq:FK} has been implemented for a general PDF input (for example Neural Network or LHAPDF) and extensively optimised. The optimisation ensures only the relevant parton sub channels and $x$-grid entries enter the product, which is performed as a memory-aligned scalar product with the use of SSE intrinsics~\cite{SSE}.   Table~\ref{tab:FKtimings} compares the relative speed improvement compared to the { \tt APPLgrid} calculation of the basic  {\tt FK} convolution and the optimised version, using PDFs obtained through the LHAPDF library.
\begin{table}[htdp]
\begin{center}
\begin{tabular}{|c|c|c|c|c|}
\hline Observable &APPLgrid & FK & optimised FK \\
\hline Total $W^+$ xsec &1.03 ms & 0.41 ms (2.5X) & 0.32 ms (3.2X) \\
\hline Jet distribution &2.45 ms & 20.1 $\mu$s (120X) & 6.57 $\mu$s (370X) \\ 
\hline
\end{tabular}
\caption[Comparison of { \tt APPLgrid} and {\tt FK} convolution timings]{Typical timings per observable for several convolution methods. Two observables are presented, the total cross-section for $W^+$ production and the inclusive jet \pt  distribution. Values are given per datapoint. In brackets the relative speed-up compared to the native { \tt APPLgrid} convolution is shown. For this test, the timings were calculated with a 2.9 GHz Intel Core i7 processor.}
\end{center}
\label{tab:FKtimings}
\end{table}
\clearpage
In Table~\ref{tab:FKtimings} a reasonable speed improvement is evident for the example single-scaled process of $W^+$ production, and a very significant improvement on the multi-scale jet production observable. It is important to note that these figures were obtained via convolutions with LHAPDF parton densities, and in a PDF fit a considerably greater speed advantage is gained via the {\tt FK} procedure as no additional operation is required to evolve the PDFs.

While for most applications, the original { \tt APPLgrid} convolution speed is more than sufficient, these speed improvements make the inclusion of a large LHC dataset possible, rather than prohibitively expensive in the NNPDF methodology. For example, in a typical NNPDF fit of 20,000 genetic algorithm generations, including a 100 datapoint jet dataset via the { \tt APPLgrid} interface would add several days of additional computer time to each individual replica fit. With the {\tt FK} procedure this additional cost is reduced to minutes.

The speed improvement is achieved without any loss of accuracy, as the interpolation procedure used to perform the PDF evolution is required in both the { \tt APPLgrid} and {\tt FK} convolutions. The two methods were benchmarked in Ref.~\cite{Ball:2012cx}, with the results shown in Table~\ref{tab:fkbenchatlasjet}. The relative discrepancy $\epsilon$ noted in the table is largely due to the additional interpolation in hard scale $Q^2$ from LHAPDF required in the { \tt APPLgrid} convolution that is not present in the {\tt} FK method, as evolution is performed directly to the required scale.

\begin{table}[ht!]
\begin{center}
\footnotesize
\begin{tabular}{c|c|c|c||c|c|c|}
\cline{2-7}
   & \multicolumn{3}{|c||}{${\rm W}^{+}$ distribution [pb]} & \multicolumn{3}{|c|}{${\rm W}^{-}$ distribution [pb]}  \\
\hline
\multicolumn{1}{|c||}{$| \eta_{l} |$} & {\tt FK}	 & {\tt APPLgrid} &  $\epsilon_{\rm rel}$ & {\tt FK}  &  {\tt APPLgrid} &  $\epsilon_{\rm rel}$\\
\hline
\multicolumn{1}{|c||}{0.00--0.21} & 617.287 & 617.345 & 0.01\% & 456.540 & 456.819 & 0.06\% \\
\multicolumn{1}{|c||}{0.21--0.42} & 616.988 & 617.062 & 0.01\% & 453.045 & 453.315 & 0.06\%\\
\multicolumn{1}{|c||}{0.42--0.63} & 620.237 & 620.290 & 0.01\% & 448.902 & 449.172 & 0.06\%\\
\multicolumn{1}{|c||}{0.63--0.84} & 624.192 & 624.235 & 0.01\% & 441.789 & 442.045 & 0.06\%\\
\multicolumn{1}{|c||}{0.84--1.05} & 630.235 & 630.286 & 0.01\% & 432.206 & 432.435 & 0.05\%\\
\multicolumn{1}{|c||}{1.05--1.37} & 636.835 & 636.886 & 0.01\% & 419.027 & 419.222 & 0.05\%\\
\multicolumn{1}{|c||}{1.37--1.52} & 642.800 & 642.861 & 0.01\% & 403.908 & 404.084 & 0.04\%\\
\multicolumn{1}{|c||}{1.52--1.74} & 642.499 & 642.569 & 0.01\% & 390.564 & 390.724 & 0.04\%\\
\multicolumn{1}{|c||}{1.74--1.95} & 642.351 & 642.437 & 0.01\% & 377.328 & 377.473 & 0.04\%\\ 
\multicolumn{1}{|c||}{1.95--2.18} & 628.592 & 628.693 & 0.02\% & 359.373 & 359.498 & 0.03\%\\
\multicolumn{1}{|c||}{2.18--2.50} & 590.961 & 591.079 & 0.02\% & 337.255 & 337.366 & 0.03\% \\
\hline
 \end{tabular}
 
 \vspace{2mm}
 
 \begin{tabular}{c|c|c|c|}
 \cline{2-4}
   & \multicolumn{3}{|c|}{Z distribution [pb]}  \\
    \hline
\multicolumn{1}{|c||}{$|y|$} & {\tt FK}	  &    {\tt APPLgrid}  &   $\epsilon_{\rm rel}$\\
\hline
\multicolumn{1}{|c||}{0.0--0.4} & 124.634 & 124.633 & 0.001\%\\
\multicolumn{1}{|c||}{0.4--0.8} & 123.478 & 123.488 & 0.01\%\\
\multicolumn{1}{|c||}{0.8--1.2} & 121.079 & 121.108 & 0.02\%\\
\multicolumn{1}{|c||}{1.2--1.6} & 118.057 & 118.108 & 0.04\%\\
\multicolumn{1}{|c||}{1.6--2.0} & 113.512 & 113.549 & 0.03\%\\
\multicolumn{1}{|c||}{2.0--2.4} & 106.552 & 106.562 & 0.01\%\\
\multicolumn{1}{|c||}{2.4--2.8} & 93.7637 & 937.838 & 0.02\%\\
\multicolumn{1}{|c||}{2.8--3.6} & 55.8421 & 558.538 & 0.02\%\\
\hline
 \end{tabular}
 
  \vspace{2mm}
  
\begin{tabular}{c|c|c|c|}
\cline{2-4}
   & \multicolumn{3}{|c|}{ATLAS 2010 jets [pb]}  \\
    \hline
\multicolumn{1}{|c||}{$p_T$ (GeV)} & {\tt FK}	  &    {\tt APPLgrid}  &    $\epsilon_{\rm rel}$ \\
\hline
\multicolumn{1}{|c||}{20--30}   & $6.1078 \times 10^6$ & $6.1090 \times 10^6$ &	0.02\% \\
\multicolumn{1}{|c||}{30--45}   & 986285 	                     & 98654      & 0.03\% \\
\multicolumn{1}{|c||}{45--60}   & 190487 	                     & 190556    & 0.04\% \\
\multicolumn{1}{|c||}{60--80}   & 48008.7 	                     & 48029.7   & 0.04\% \\
\multicolumn{1}{|c||}{80--110} & 10706.6 	                     & 10710.4   & 0.03\% \\
\multicolumn{1}{|c||}{110--160} & 1822.62 			  & 1822.87   & 0.01\% \\
\multicolumn{1}{|c||}{160--210} & 303.34 			  & 303.443   & 0.03\% \\
\multicolumn{1}{|c||}{210--260} & 76.1127 			  & 76.1338   & 0.03\% \\
\hline
 \end{tabular}

 \end{center}
\caption[Benchmark of the {\tt FK} result for datasets with different underlying processes]{ Benchmark of the {\tt FK} result for datasets with different underlying processes, all generated according to ATLAS experimental kinematics and acceptances. The { \tt APPLgrid} and {\tt FK} results are presented along with the relative discrepancy between the two. Table from \cite{Ball:2012cx}.\label{tab:fkbenchatlasjet} }
\end{table}
%%%%%%%%%%%%%%%%%%%

%

%
%\be f_i(x_{\alpha},Q^2_\tau) =  \sum_j^{13}R_{ij}N_j(x_{\alpha},Q_\tau^2) =\sum_{j}^{13} \sum_{\gamma}^{N_x}  \sum_{k}^{N_{\mathrm{pdf}}} R_{ij}E^\tau_{\alpha\gamma j k}N^0_k(x_\gamma).\ee 
%\be F^{(l)}\left(x_{\alpha}, x_{\beta},  Q^2_{\tau}\right) =   \sum_{i,j}^{13} \sum_{k,l}^{N_{\mathrm{pdf}}}  C^{(l)}_{ij}A^\tau_{\alpha\gamma i k}A^\tau_{\beta\delta j l}N^0_k(x_\gamma)N^0_l(x_\delta),
%\quad\quad A^\tau_{\alpha\gamma i k} =  \sum_j^{13} R_{ij}E^\tau_{\alpha\gamma j k}.\ee
%\be \sigma =  \sum_{i,j}^{N_{\mathrm{pdf}}} \sum_{\alpha,\beta}^{N_x} \widetilde{W}_{\alpha\beta i j} N_i^0(x_\alpha)N_j^0(x_\beta),\ee
%\be \widetilde{W}_{\alpha\beta i j} = \sum_p \sum_{l=0}^{N_{\mathrm{sub}}}\sum_{k,l}^{13} \sum_{\gamma,\delta}^{N_x} \sum_{\tau}^{N_{Q}}
%W_{\gamma\delta\tau}^{(p)(l)} \, \left( \frac{\alpha_s\left(Q^2_{\tau}\right)}{2\pi}\right)^{p} C^{(l)}_{kl}A^\tau_{\gamma\alpha k i}A^\tau_{\delta\beta l j}, \ee
%

\section{Interpolating tools for automated NLO}
Tools such as { \tt FastNLO}/{ \tt APPLgrid} and their extension for fast PDF fitting in the $\tt FK$ method, are invaluable in the analysis of collider data. Their usefulness is not limited to applications such as fitting, but can also be used to perform thorough QCD analysis with rigorous theory uncertainty estimation in situations where obtaining sufficient statistics with an NLO code or event generator would be extremely expensive computationally. Despite this, at the outset of LHC data taking the amount of codes interfaced to such interpolating tools was extremely limited. Additionally the need for separate interfaces to existing codes meant a great deal of duplication in terms of analysis tools and software.  The { \tt APPLgrid} group provided a direct interface to the NLO codes {\tt MCFM}~\cite{Campbell:2000bg} and {\tt nlojet++}~\cite{Nagy:2001fj,Nagy:2003tz}. { \tt FastNLO} provided a set of precomputed scenarios generated through a private interface to NLO codes. More recently, a public toolkit was released to allow for the interfacing of { \tt FastNLO} to external calculations. 

A conspicuous absence was an interface to tools providing automated NLO calculations via computer algebra suitable one-loop methods\cite{Anastasiou:2006jv,Berger:2008sj,Denner:2005nn,Ellis:2007br,Giele:2008ve,Ossola:2006us} and their implementations in parton level Monte Carlo codes such as {\tt MadGraph}~\cite{Alwall:2011uj}, {\tt HELAC}~\cite{Bevilacqua:2011xh} and {\tt SHERPA}~\cite{Gleisberg:2003xi,Gleisberg:2008ta}. In this section we shall discuss the implementation of a fast interface to such codes, the {\tt MCgrid}~\cite{DelDebbio:2013kxa} package; developed with the aid of funding from the MCnet initial training network.

\subsection{Reweighting Monte Carlo calculations}
Recalling Eqn.~\ref{eq:hadconv}, a hadronic observable calculation proceeds via
\be  \sigma_{pp\to X} = \sum_p \left( \frac{\alpha_s(Q^2)}{2\pi} \right)^{p} \int dx_1\, dx_2\,  F_l(x_1,x_2,Q^2)\; d\hat{\sigma}_{l\to X}^{(p)}\; , \label{eq:doubleconv2} \ee
where the initial state PDFs have been grouped according to Eqn.~\ref{eq:APPLsubproc} and the sum over subprocesses is implicit. In an event generator this integral is performed via Monte Carlo integration. At leading order this is a relatively straightforward procedure,
\begin{align}
  \label{eq:basicMC}
  \sigma_{pp\to X}^{\text{LO}} &= \sum_{e=1} \tilde{w}_e(k_e) = \sum_{e=1}^{N_\evt} \left(  \frac{\alpha_s\left(k_e\right)}{2\pi} \right)^{p_{\text{LO}}} w_e(k_e) F_{l_e}(k_e)\, ,
\end{align}
where $\tilde{w}$ is the full event weight and the $w$ are the matrix element weights generated via importance sampling of the integrand of Eqn.~\ref{eq:doubleconv2}. The $F_{l_e}$ refer to the parton density of the event's subprocess $l_e$. Each event is generated according to a set of kinematics
\begin{align}
  k_e &= \left\{p_1, ..., p_n,\,x_1,\,x_2,\,\frac{\mu_F^2}{Q^2},\,\frac{\mu_R^2}{Q^2} \right\}.
\end{align}

As a full re-run of the event generator for every parameter variation is extremely expensive, the variation is typically performed via an event-by-event reweighting procedure. The full set of events is stored
in a common format such as { \tt HepMC}~\cite{Dobbs:2001ck} and re-processed by dividing out the appropriate factors of the old PDFs and $\alpha_S$ and multiplying in the desired new values. 

\be \tilde{w}_e (k_e)\to \left(  \frac{\alpha_s^\prime\left(k_e\right)}{\alpha_s\left(k_e\right) } \right)^{p_{\text{LO}}}  \frac{F^\prime_{l_e}(k_e)}{F_{l_e}(k_e)}\; \tilde{w}_e(k_e), \ee
where the primed quantities denote the new, reweighted strong coupling and PDF choices. Having the full generated event sample stored also has the advantage of being able to rerun analysis software with varying parameters/selections without the need to rerun the potentially expensive event generation.

The reweighting situation in an NLO calculation is considerably more complicated. In order to be able to solve the integral numerically, a divergence-subtraction scheme e.g Catani-Seymour~\cite{Catani:1996vz}
or Frixione-Kunst-Signer (FKS)~\cite{Frixione:1995ms,Frixione:1997np} must be employed. These subtraction algorithms separate the calculation into distinct sections which are to be numerically evaluated individually. Here we shall discuss the implementation in terms of a Catani-Seymour dipole scheme. The four contributions to the total NLO cross section are 

\begin{equation}
  \label{eq:NLOxsect}
  \sigma_{pp\to X}^{\text{NLO}}= \int d\hat{\sigma}^\mathrm{B}
  + \int d\hat{\sigma}^\mathrm{V}
  + \int d\hat{\sigma}^\mathrm{I}
  + \int d\hat{\sigma}^\mathrm{RS}
  \, .
\end{equation}
The terms $B$, $V$, $I$ and $RS$ refer to the Born (B), Virtual (V), Integrated subtraction (I) and Real Subtracted (RS) cross section elements respectively. Neglecting terms used in the variation of perturbative scales, the $B$, $V$ and $RS$ terms may be integrated via a Monte Carlo procedure equivalently to Eqn.~\ref{eq:basicMC}. The integrated dipole term however has a rather more complicated dependance upon the initial state PDFs, originating as it does through the splitting of an initial state parton. The Monte Carlo solution to the $I$ integral is given by 

\def\varref#1{{\tt #1}}
\begin{eqnarray}
 \int d\hat{\sigma}^\mathrm{I}  &=&  \sum_{e=1}^{N_\evt}   \left(  \frac{\alpha_s(\mu_R^2)}{2\pi} \right)^{p_{\text{NLO}}} \Bigg\{  F_{l_e}(x_1,x_2,\mu^2_F)\,  w^{(0)}_e 
\nonumber\\
&&+\sum_{s}^{N_{\text{sub}}} F_s(x_1/x'_1, x_2, \mu^2_F)\, \tilde{w}^{(1)}_{e,s} \label{eq:NTupleBreakdown} \\
&&+\sum_{s}^{N_{\text{sub}}} F_s(x_1, x_2/x'_2, \mu^2_F)\, \tilde{w}^{(2)}_{e,s}
 \Bigg\},
  \, \nonumber 
\end{eqnarray}
in which the weight $w^{(0)}$ arises through the usual Born-like PDF dependance, and the weights $w^{(1/2)}$ arise from integration over parton-$x$ from the first or second parton in the initial state splitting. To reweight such events these weights must therefore be properly distinguished
in the event record. While is is not the case in the standard {\tt HepMC} layout, a format based upon {\tt ROOT NTuples} was designed by the {\tt BlackHat-Sherpa} group for the reweighting of NLO event weights~\cite{Bern:2013zja}. In the {\tt BlackHat NTuple} format the weights that must be distinguished for the accurate treatment of scale variations are also stored.  \\
While the event reweighting approach is considerably faster than an entire rerun of the Monte Carlo, a reweight of a full event sample can still take a considerable amount of computer time. The key issue being that the statistical accuracy of the calculation is limited by the number of events in the sample, and therefore for a more accurate calculation, more computational expense is incurred. This dependence on the event loop is not removed by the event reweighting procedure. The dependance can however be removed by applying the interpolation methods of the previous section, by providing an interface for event generators to interpolating packages such as { \tt APPLgrid}.
\subsection{An interpolation interface for automated NLO} 
\label{sec:interp}
The {\tt MCgrid} project began as a direct interface for the {\tt SHERPA} event generator framework to the {\tt APPLgrid} interpolation package. The development of a new interface between an event generator and the {\tt APPLgrid} framework in principle requires the implementation of an analysis suite to provide the categorisation of event final states into appropriate observable bins. However the {\tt MCgrid} interface is built upon standard analysis tools and formats to provide a more general interface between standards-compliant Catani-Seymour event generators to the {\tt APPLgrid} package.

{\tt MCgrid} is written as a set of additional tools for the {\tt Rivet} MC analysis system. The {\tt Rivet} system implements a wide range of experimental analysis tools and provides the flexibility for the user to define their own selection criteria and processing tools to operate on an event final state. Writing the {\tt APPLgrid} interface as a {\tt Rivet} extension therefore removes the need to implement a separate toolchain, and allows a degree of generator agnosticism. As {\tt} Rivet operated upon events in the standard {\tt HepMC} format, any generator equipped to output events in this format may potentially be interfaced to {\tt APPLgrid} through {\tt MCgrid}.

The interface requires additional information over the standard data available in {\tt HepMC}, as the information on the weight breakdown as per Eqn.~\ref{eq:NTupleBreakdown} must be available. However this can be straightforwardly appended in the {\tt HepMC} user defined weights fields. The interface then provides the correct handling of initial state parton mappings from the PDF basis used in the Catani-Seymour process to the APPLgrid flavour basis.

With the appropriate mapping to initial state parton flavours performed, the weights must be converted to the appropriate subprocess basis. The minimal initial state PDF basis can be automatically determined by a set of packaged scripts.
General purpose Monte Carlo codes such as {\tt SHERPA} will typically generate events will the full initial parton flavours explicit rather then generating events based upon QCD subprocesses. Weights originating from different flavour basis channels are generated via importance sampling of the distribution in order to ensure an efficient description of the most important channels.  Accordingly there is a selection weight present in each event weight, given by 
\begin{equation}  
w_e(i,j,k_e) =\mathcal{N}_{ij}\;d \hat{\sigma}_{l_e \to X}(k_e) \Pi_{\text{ps}}(k_e) \Theta (k_e-k_{\text{cuts}}), \label{eq:mcweight}
\end{equation}
where the factor $\mathcal{N}_{ij}$ is approximately given by
\begin{equation}
\mathcal{N}_{ij} \sim \frac{N_{\text{tot}}}{N_{ij}}, \label{eq:selectionweight}
\end{equation}
where $N_{\text{tot}}$ denotes the total number of events in the sample, and $N_{ij}$ is the number of events initiated by partons of flavour $i$ and $j$. In Eqn.~\ref{eq:mcweight} we use $\Pi$ to represent the phase space weight associated with the kinematics $k_e$, and the $\Theta$ as a step function implementing the desired kinematic cuts in the analysis. 
The selection weights $\mathcal{N}$ must be converted into the appropriate subprocess selection weight to prevent the statistical uncertainty in poorly-sampled distributions from overwhelming the subprocess. {\tt MCgrid} monitors the relative population of the channels and subprocesses in order to provide a statistically sound subprocess combination. The selection weight in Eqn.~\ref{eq:selectionweight} must be converted into the appropriate subprocess selection weight as
\be \mathcal{N}_{ij} \to \mathcal{N}_{l} = \frac{N_{\mathrm{tot}}}{N_{l}}, \ee
where $N_{l}$ is the number of events falling into the initial state subprocess $l$. Converting the selection weights to the appropriate subprocess selection weight is therefore a matter of multiplying each event weight by a factor $N_{ij}/N_{l}$.

\begin{figure}[h!]
\centering
\includegraphics[width=0.48\textwidth]{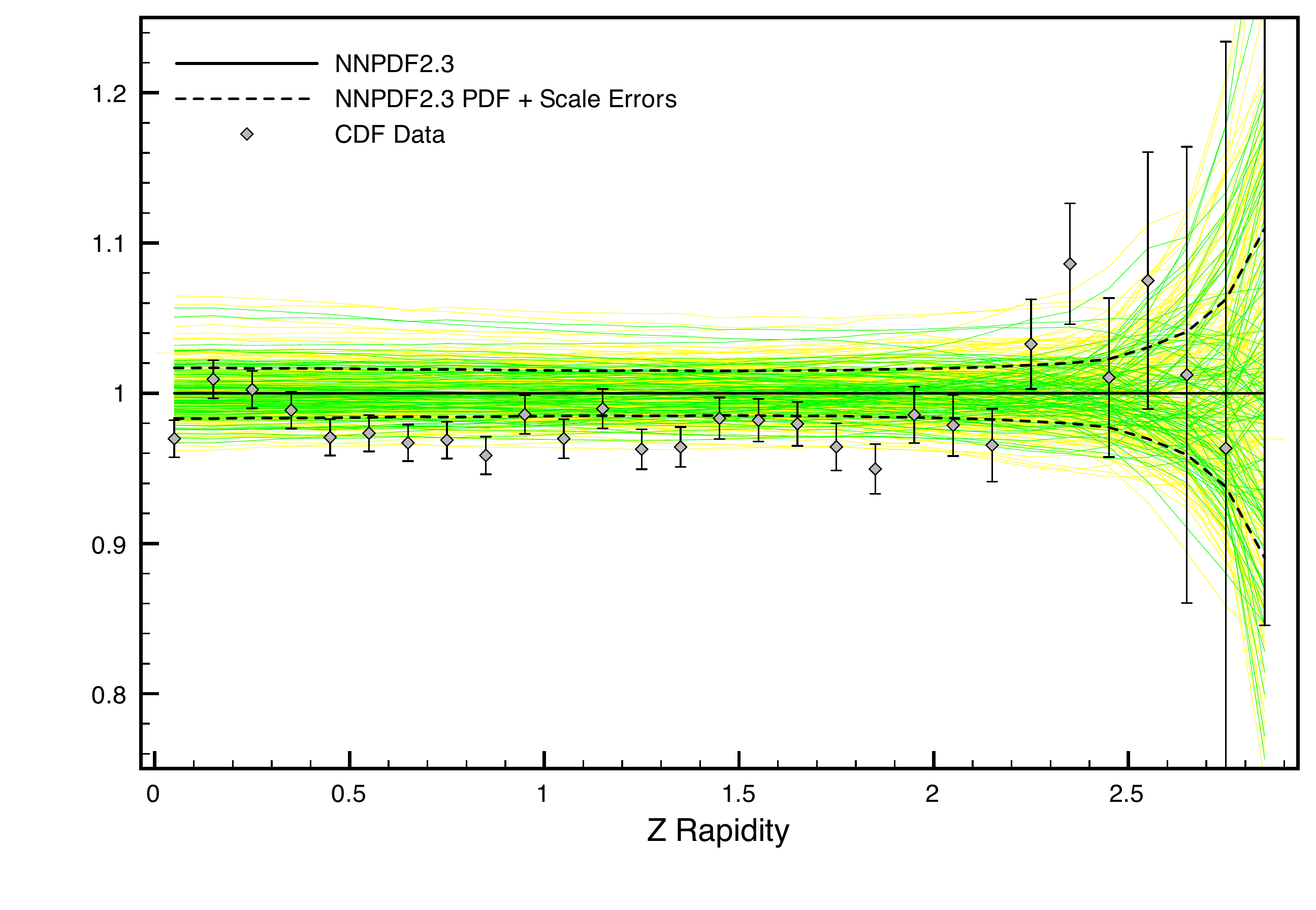}
\includegraphics[width=0.48\textwidth]{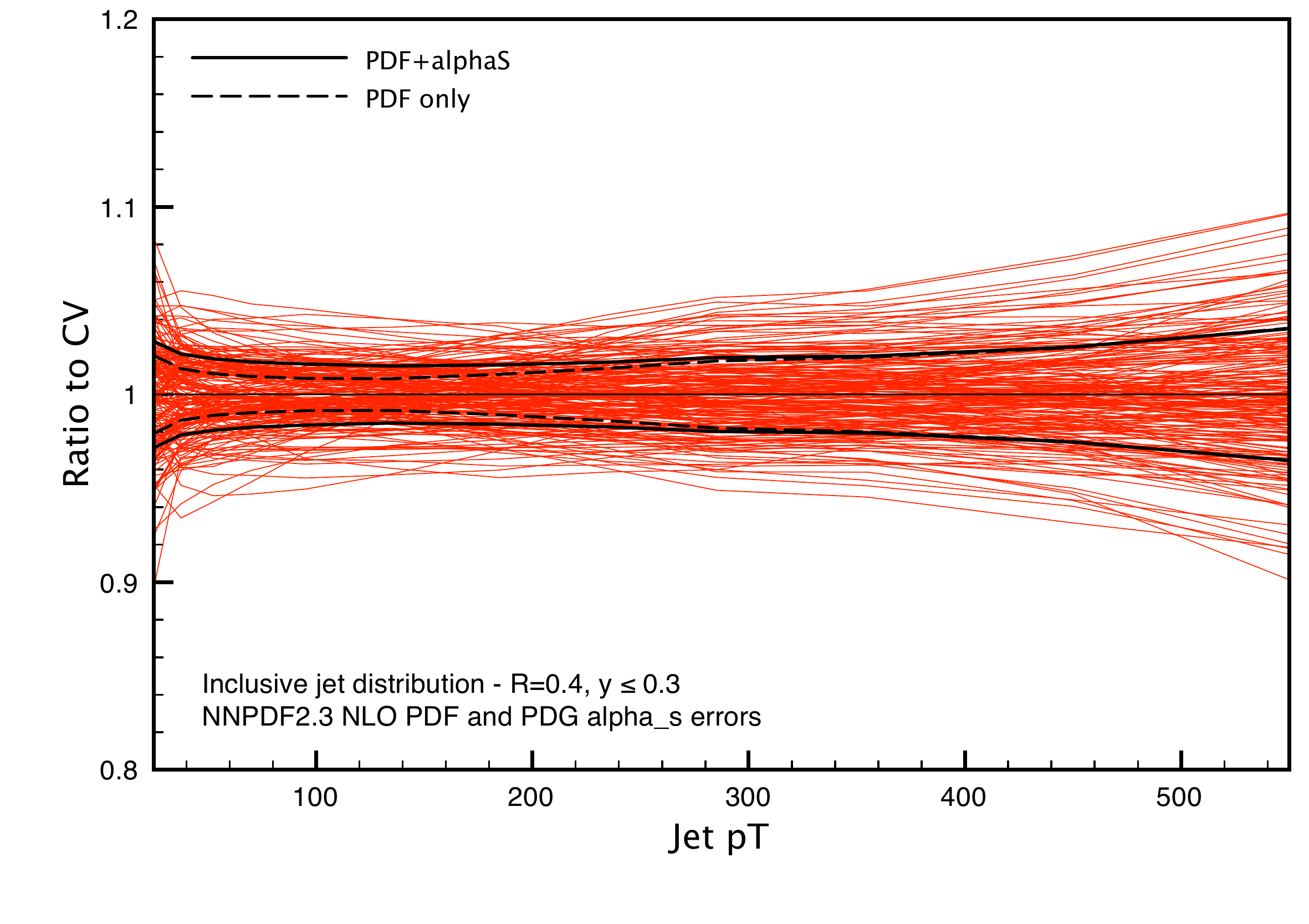}
\caption[An example of the output of the {\tt MCgrid} package]{An example of the output of the {\tt MCgrid} package. A $Z$ boson rapidity distribution plot is shown on the left, with scale error estimation. The plot on the right demonstrates the grids applied to inclusive jet data, with $\alpha_S$ error estimation. Both plots are normalised to their central values, to demonstrate the level of uncertainty.}
\label{fig:MCgridreps}
\end{figure}

In this way the fully exclusive predictions given in a typical Monte Carlo event generator may be effectively converted into the relevant subprocess basis. With this accomplished, and the correct weight conversion performed according to the exact PDF dependance of the Catani-Seymour counterterms, the weights may be filled directly into an APPLgrid type weight grid. Figure~\ref{fig:MCgridreps} demonstrates the application of the {\tt MCgrid} package when used in conjunction with {\tt Sherpa} and {\tt BlackHat} as a one-loop generator. The tools are applied to the test cases of Drell-Yan and inclusive jet production, with the resulting { \tt APPLgrid} applied to the estimation of scale and $\alpha_S$ uncertainties alongside standard PDF error, requiring a very large number of replicas.

The {\tt MCgrid} project is publicly available\footnote{The software is available at \emph{http://mcgrid.hepforge.org}.}, and allows for the first time calculations from automated NLO event generators to be interfaced to interpolation tools, for potential application in PDF fits, or indeed fast parameter variation studies in phenomenological applications.

\chapter{LHC Data for Parton Determinations}
\label{ch:LHCdata}
The Large Hadron Collider has the ability to provide a comprehensive examination of QCD and electroweak physics at a wide range of scales. The requirement of precise and reliable
determinations of proton structure is clear in order to fully exploit the LHC's potential. LHC data also has the potential to provide deep new insights into parton distributions, examining hitherto
poorly determined flavours and kinematic regimes. A great deal of effort has therefore been expended in providing and validating tools for the inclusion of LHC data in an efficient manner into
NNPDF fits.

In this section the Standard Model measurements of relevance to PDF determination so far performed by the LHC shall be briefly summarised. While the general processes have been described previously,
here we shall look directly at the experimental data along with a brief examination of the areas of agreement or discrepancy with regard to PDF sets made available before the first data runs of the LHC.

\section{Jet measurements}
At the LHC, data on the production of collimated jets of particles originating from partonic final states provides valuable information on proton structure and additional constraints for $\alpha_S$ determinations. The LHC's centre-of-mass energies
mean that jets with transverse momenta in the TeV range are observable for the first time. Forward jets probing the very large-$x$ gluon that has suffered from poor constraints prior to the LHC. As the prototypical QCD measurement, data
is available from both of the general purpose LHC experiments, and preliminary data on jets in the forward region is available from LHCb~\cite{LHCb:2011xqa}. LHC measurements are based upon modern infrared and collinear safe jet-finding algorithms such as anti-$k_T$~\cite{Cacciari:2008gp}. In PDF fits the jet quantity of interest is typically the inclusive measurement rather than dijet data. In principle dijet measurements offer more discriminating power over the parton distributions, however they typically suffer from larger scale uncertainties and often must be corrected for higher order effects, typically modelled through parton showers.

 Here we shall summarise the relevant jet measurements at the LHC with a focus on the data most relevant to PDF determination.

The first ATLAS inclusive jet and dijet measurements were based upon a partial analysis of $17$ nb$^{-1}$ of data available from the 2010 data run at a centre of mass energy of $7$ TeV~\cite{Aad:2010ad}. This result was then updated to the full 2010 dataset of $37$ pb$^{-1}$~\cite{Aad:2011fc}.
The full 2010 measurement presents the inclusive jet cross section differentially in both the jet $p_T$ and rapidity. Data is available for the $20 \le p_T < 1500$ GeV range for jets with rapidity $|y|<4.4$, and is available for two choices of the anti-$k_T$ cone size, $R=0.4$ and $R=0.6$.
\begin{figure}[ht]
\centering
\includegraphics[width=0.48\textwidth]{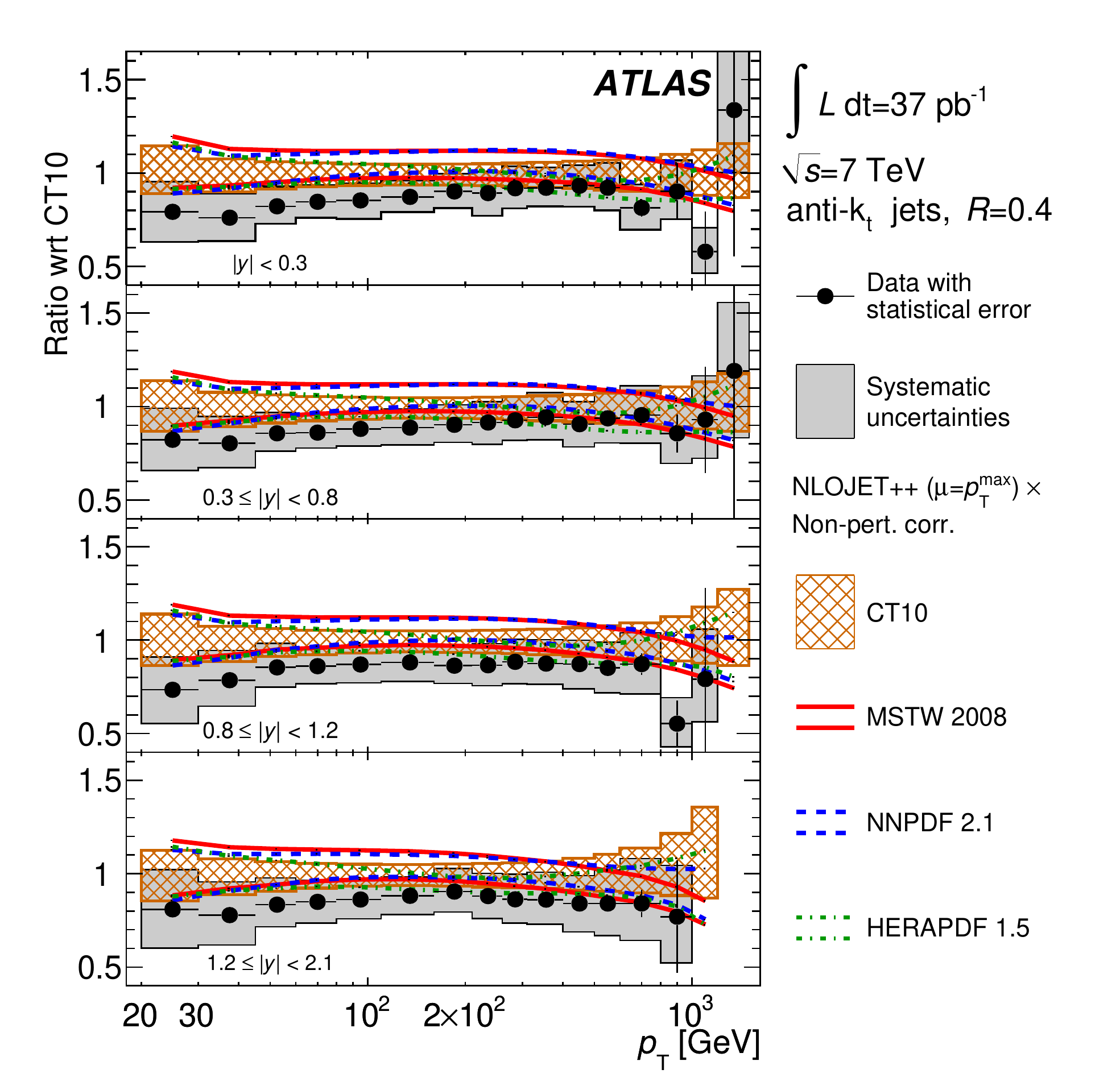}
\includegraphics[width=0.48\textwidth]{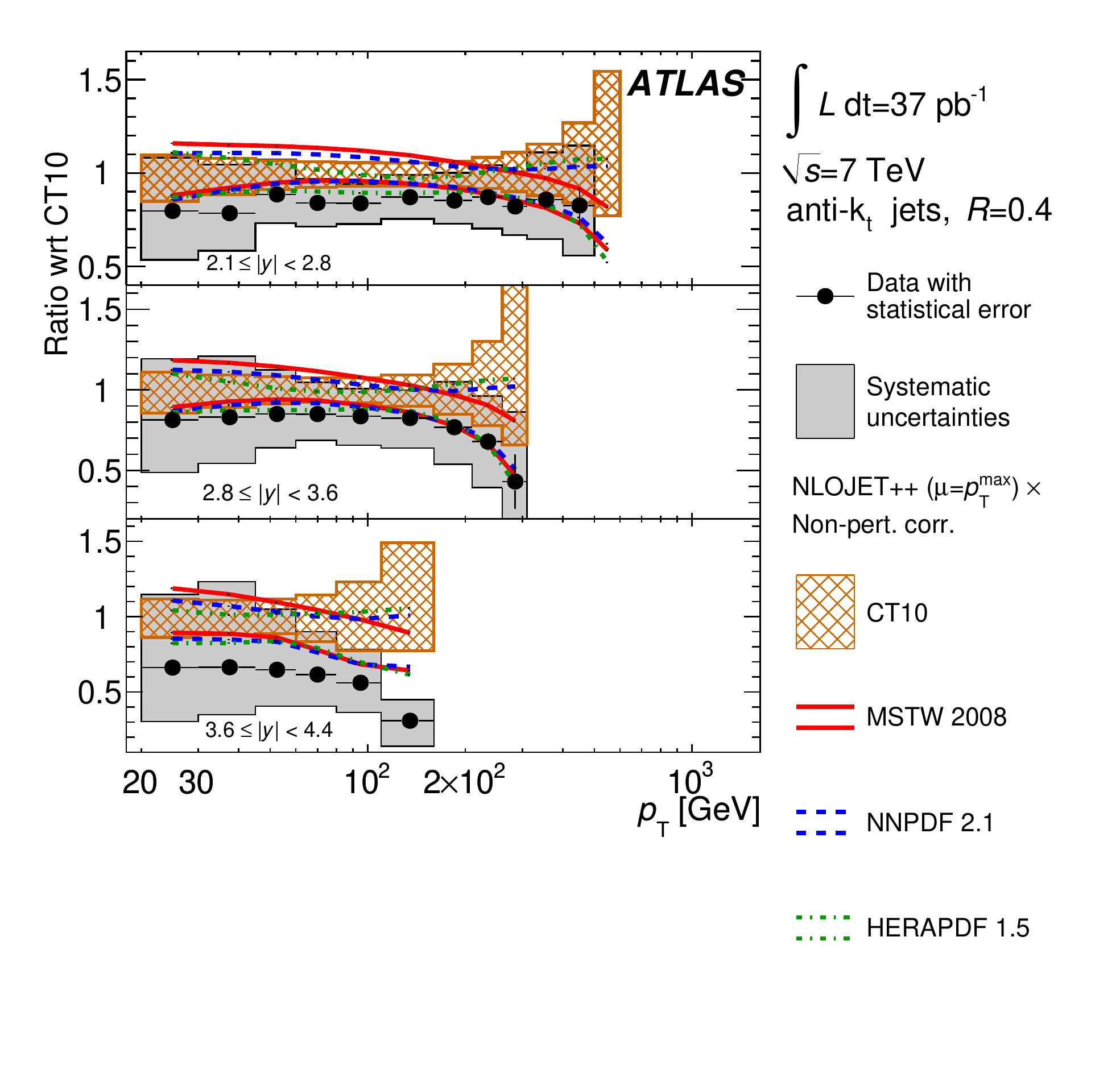}
\caption[ATLAS inclusive jet data with anti-$k_T$ algorithm $R=0.4$]{ATLAS inclusive jet data with anti-$k_T$ algorithm $R=0.4$ from the 2010 dataset. Figures from~\cite{Aad:2011fc}. Predictions are shown based upon MSTW2008, NNPDF2.1 and HERAPDF1.5 PDFs, with all data and theory normalised to the CT10 central value.}
\label{fig:ATLASR04Jets}
\end{figure}

Figure~\ref{fig:ATLASR04Jets} from the ATLAS $37$ pb$^{-1}$ result demonstrates the level of agreement of the fixed-order NLO inclusive jet computation present in {\tt NLOJet++} with the experimental data given four choices of PDFs: CT10, MSTW2008, NNPDF2.1 and HERAPDF1.5. Predictions from the four sets largely agree within their PDF uncertainties, and the experimental data also shows good agreement for most of the data range. Some evidence of a systematic discrepancy is visible at large $p_T$, an effect that becomes more noticeable in the larger rapidity bins (and therefore more extreme values of parton-$x$).  

\begin{figure}[ht]
\centering
\includegraphics[width=0.90\textwidth]{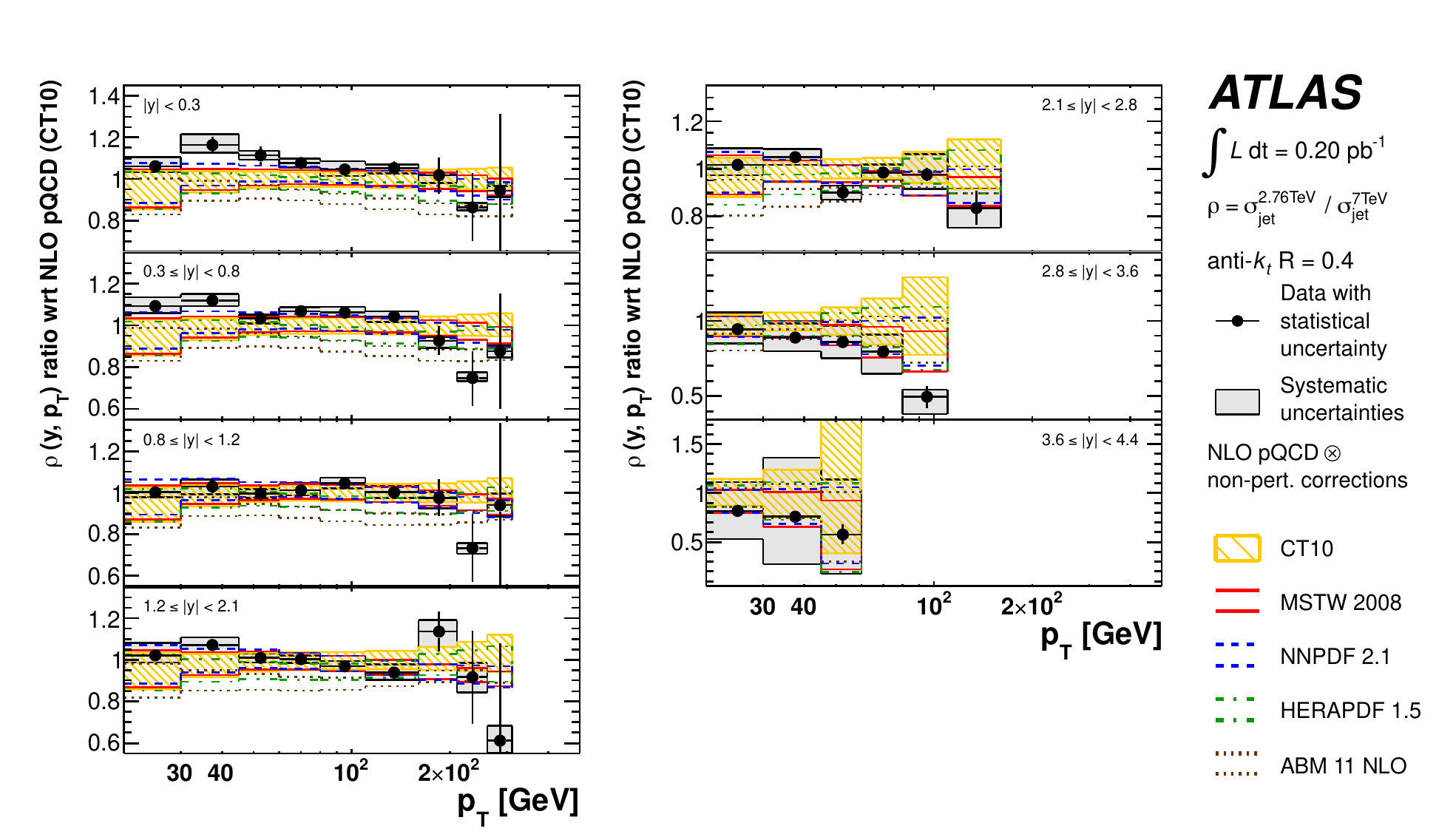}
\caption[ATLAS inclusive jet ratio between $\sqrt{s} = 2.76$ GeV and $7$ GeV data] {ATLAS inclusive jet ratio between $\sqrt{s} = 2.76$ GeV and $7$ GeV data, anti-$k_T$ $R=0.4$. Figures from ~\cite{Aad:2013lpa}.}
\label{fig:ATLASJETSRAT}
\end{figure}

ATLAS has also published data on the inclusive jet cross-sections at $\sqrt{s} = 2.76$ GeV measured during the 2011 run~\cite{Aad:2013lpa}. The data provides an important link between jet measurements at lower centre-of-mass energies at the Tevatron and the higher scale measurements previously published. In addition, the ratio of the $\sqrt{s} = 2.76$ GeV data to the 2010 $\sqrt{s}=7$ GeV measurement is presented. The ratio offers additional important constraints in that the dominant uncertainties upon the jet measurements are systematic across both datasets, and therefore largely cancel in the ratio. Figure~\ref{fig:ATLASJETSRAT} demonstrates the reduced uncertainty in the measurement, and therefore the additional constraint that the data may provide parton fits.

CMS has published three measurements of inclusive and dijet observables to date. The first provided data in the $18 < p_T < 1100$ GeV interval for jets with $|y|<3$ based upon $34$pb$^{-1}$ of 2010 data~\cite{CMS:2011ab}. This was followed up
by a study of jets in the forward region~\cite{Chatrchyan:2012gwa}, examining inclusive jets with pseudorapidities $3.2<|\eta|<4.7$, and dijets with one forward jet and one central $|\eta|<2.8$ jet. A study of 2011 data totalling $5.0$fb$^{-1}$ was also
performed of jets in the central $|y|<2.5$ region up to very high jet transverse momenta $p_T < 2$ TeV~\cite{Chatrchyan:2012bja}. CMS also utilises the anti-$k_T$ clustering algorithm, with cone sizes $R=0.5$ and $R=0.7$.

\begin{figure}[ht]
\centering
\includegraphics[width=0.48\textwidth]{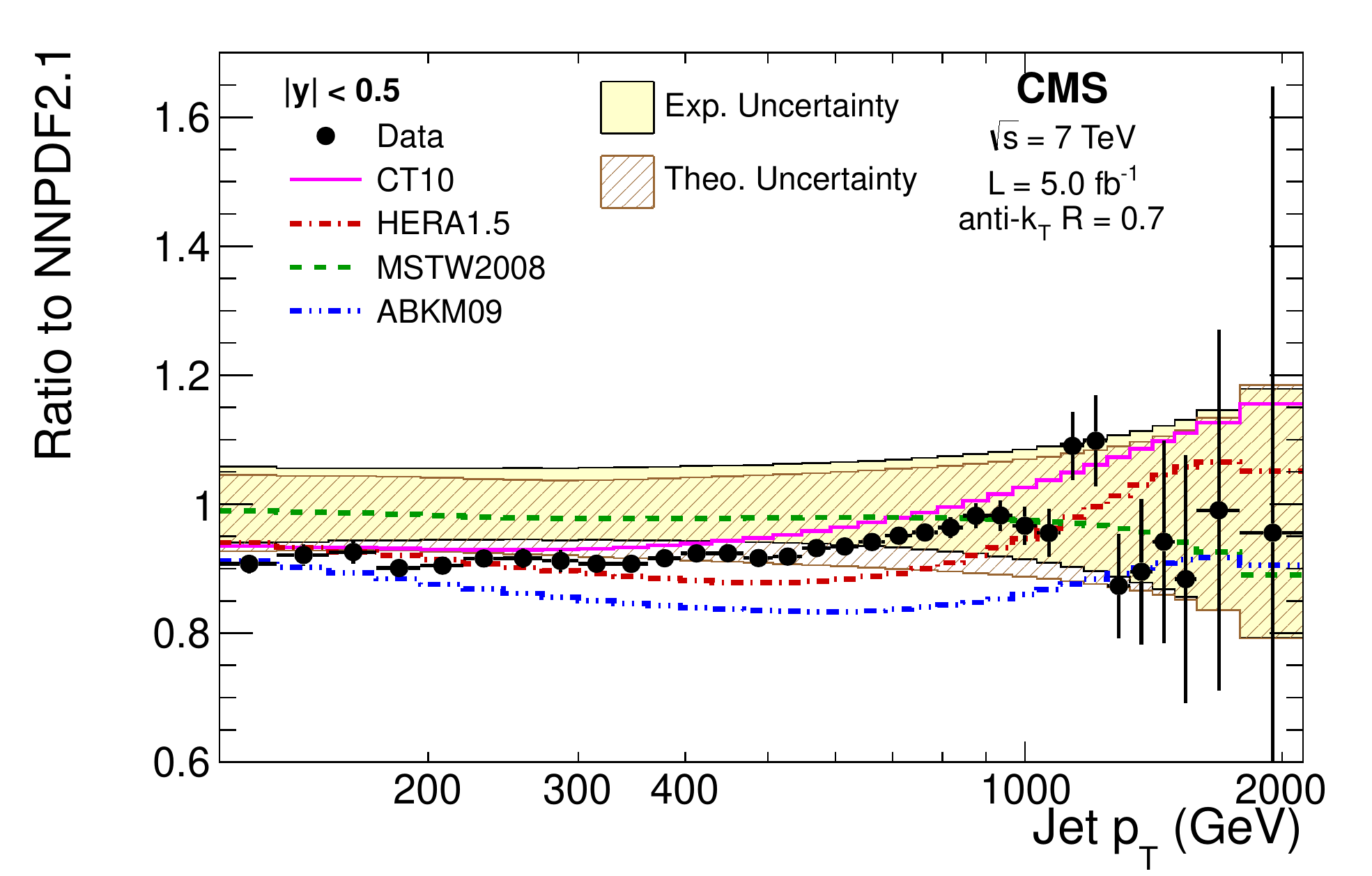}
\includegraphics[width=0.48\textwidth]{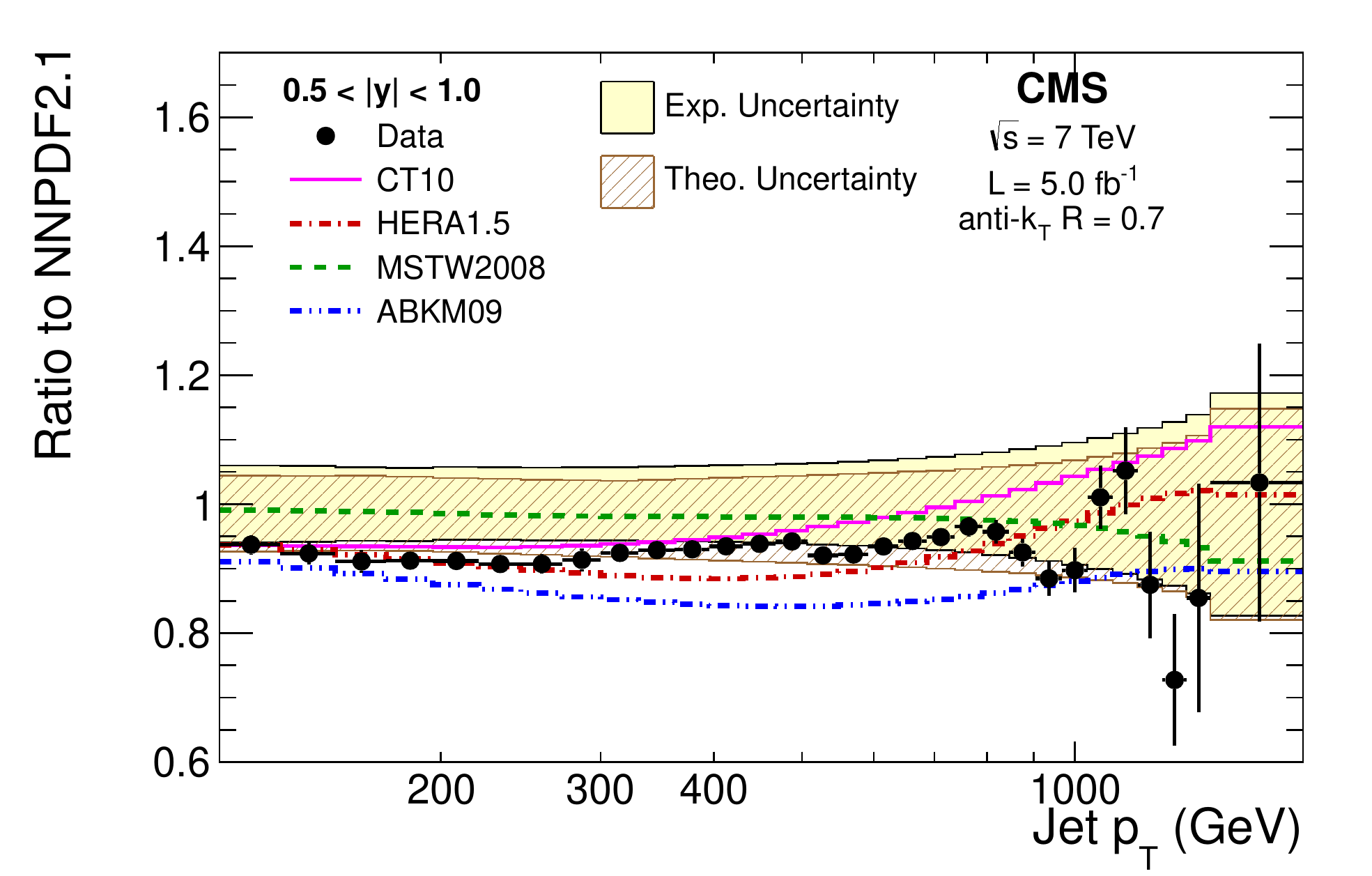}
\includegraphics[width=0.48\textwidth]{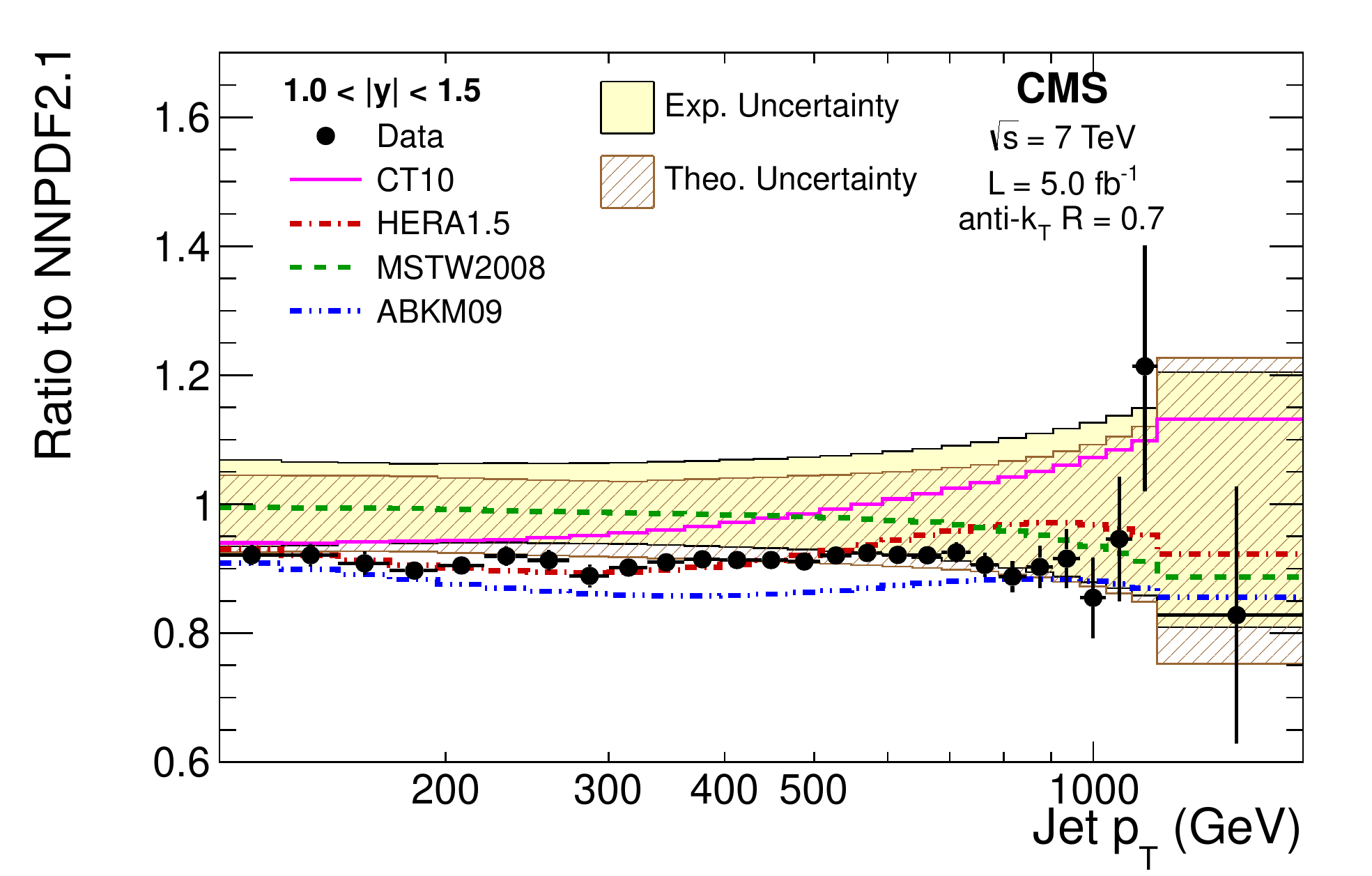}
\includegraphics[width=0.48\textwidth]{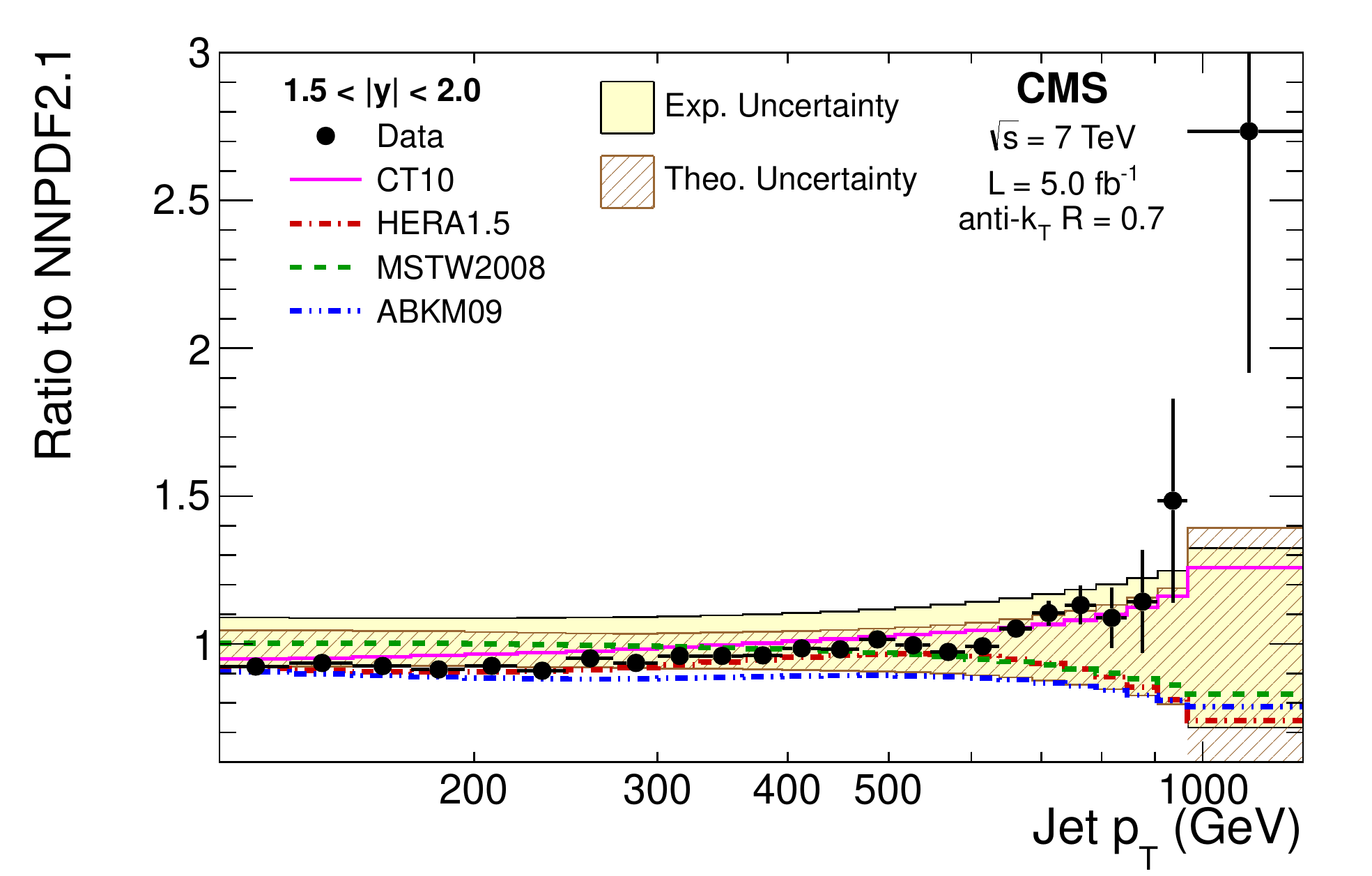}
\caption[CMS inclusive jet data with anti-$k_T$ algorithm $R=0.7$]{CMS inclusive jet data with anti-$k_T$ algorithm $R=0.7$ from the 2011 dataset. Figures from~\cite{Chatrchyan:2012bja}. Predictions are shown based upon MSTW2008, NNPDF2.1 and HERAPDF1.5 PDFs, with all data and theory normalised to the NNPDF2.1 central value.}
\label{fig:CMSR07Jets}
\end{figure}

Figure~\ref{fig:CMSR07Jets} shows the inclusive data from the CMS central region jet measurement normalised to the NNPDF2.1 central value. Results are once again largely consistent with PDFs determined with pre-LHC data. 

%
%\begin{figure}[ht]
%\centering
%\includegraphics[width=0.48\textwidth]{5-LHCdata/figs/fig_16a.pdf}
%\includegraphics[width=0.48\textwidth]{5-LHCdata/figs/fig_16b.pdf}
%\caption[ATLAS dijet data with anti-$k_T$ algorithm $R=0.4$]{ATLAS dijet data with anti-$k_T$ algorithm $R=0.6$. Figures from ~\cite{Aad:2011fc}.}
%\label{fig:ATLASR06JETS}
%\end{figure}

\section{$W$/$Z$ boson production}

The measurement of electroweak vector boson production and Drell-Yan cross sections are standard candle measurements for the LHC, and have been widely studied by ATLAS, CMS and LHCb in the first run.

\begin{figure}[ht]
\centering
\includegraphics[width=0.80\textwidth]{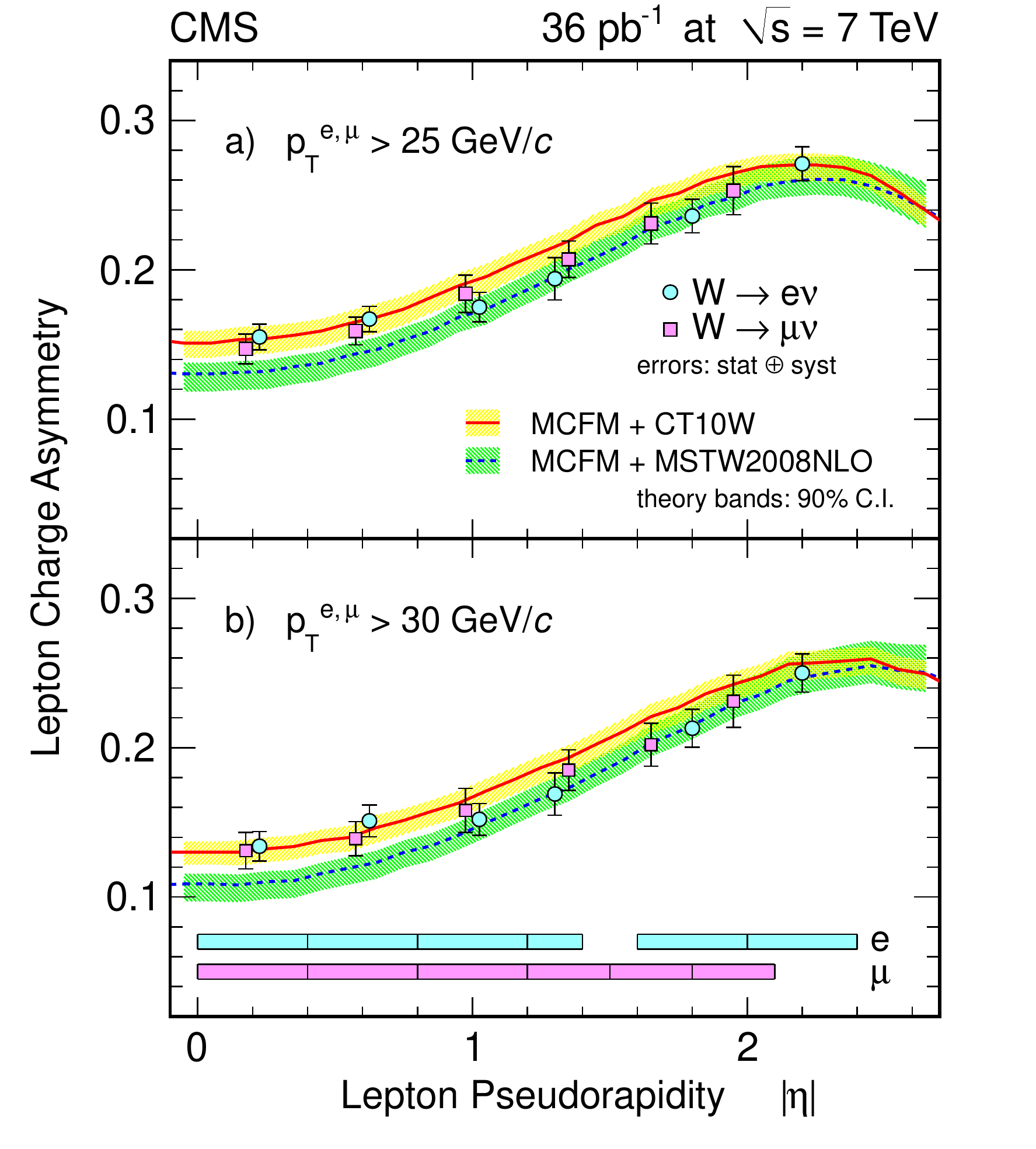}
\caption[CMS 2010 $W$ asymmetry data]{CMS 2010 $W$ asymmetry data from the electron and muon decay channels. Figure from ~\cite{Chatrchyan:2011jz}. The figure demonstrates the good agreement of the CT10 PDF set with the experimental data, and the somewhat poorer agreement of the MSTW2008 set, a typical feature of the MSTW fit with LHC electroweak data.}
\label{fig:CMS2010WASY}
\end{figure}

CMS has presented measurements of the $Z$ boson $p_T$ and rapidity distributions, initially upon $36$pb$^{-1}$ of $7$ TeV 2010 data~\cite{Chatrchyan:2011wt}, and more recently a preliminary
study of $8$ TeV data on Z decay to dimuons~\cite{CMS-PAS-SMP-12-025,CMS-PAS-SMP-13-013}. The first differential measurements of $W$ boson production at CMS were lepton charge asymmetry measurements
based upon 2010 data~\cite{Chatrchyan:2011jz}, which were superseded by the muon asymmetry measurement based upon $840$pb$^{-1}$, and then $4.6$pb$^{-1}$ of 2011 data~\cite{Chatrchyan:2012xt,Chatrchyan:2013mza}. In Figure~\ref{fig:CMS2010WASY}
the 2010 data $W$ asymmetry measurement of CMS is shown, demonstrating the constraining power of the earlier CMS result, where agreement is generally good with the pre-LHC parton distributions with the exception of the MSTW 2008 description.

ATLAS initially published a study of the $W$ muon asymmetry distribution with $31$ pb$^{-1}$ of $7$ TeV data~\cite{Aad:2011yna}. This was followed by studies of the $Z$~\cite{Aad:2011gj} and $W$~\cite{Aad:2011fp}  $p_T$ distributions. The most recent data is provided by a combined study of the $W$ and $Z$ $p_T$ distributions based upon the full 2010 dataset~\cite{Aad:2011dm}.

\begin{figure}[ht]
\centering
\includegraphics[width=0.48\textwidth]{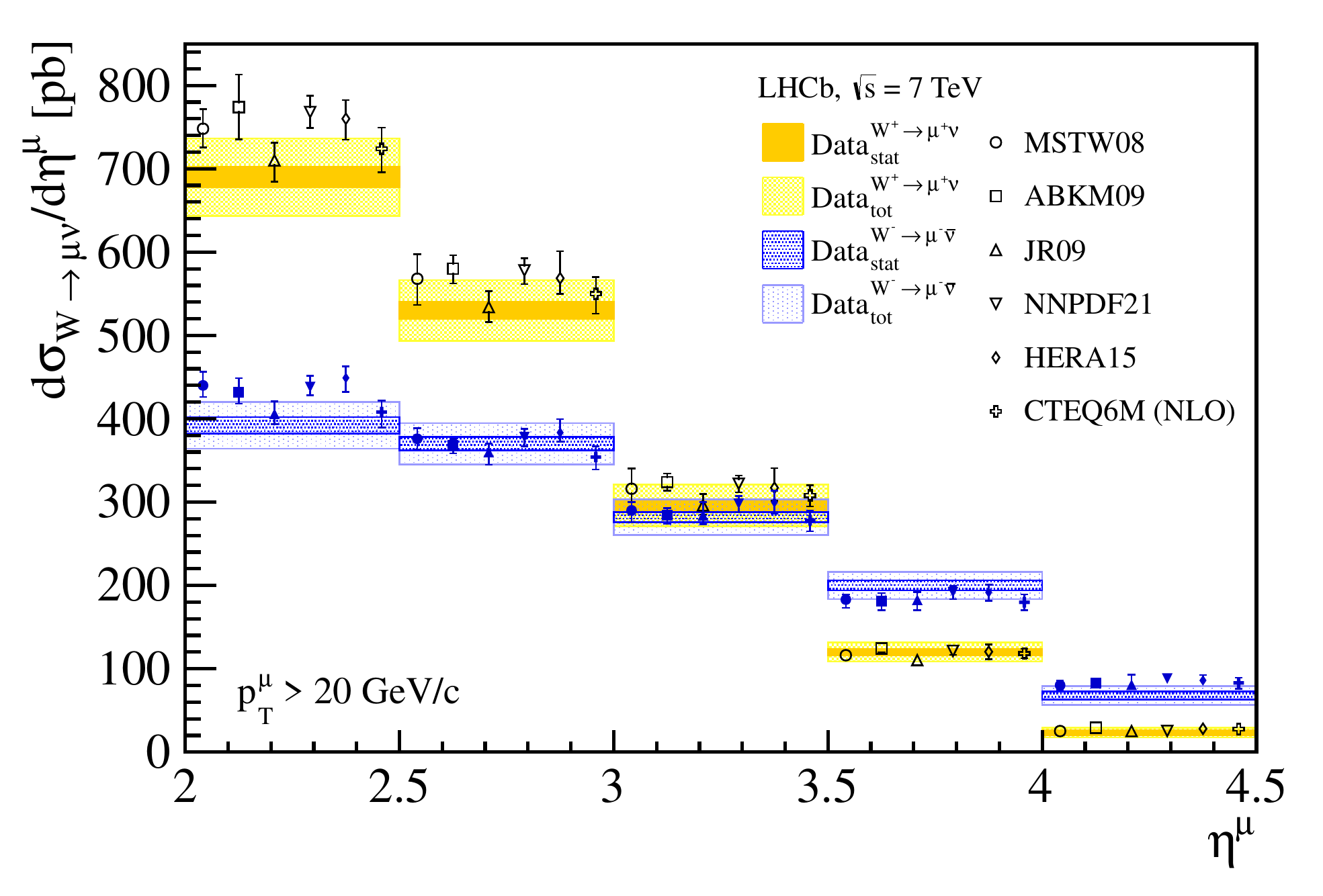}
\includegraphics[width=0.48\textwidth]{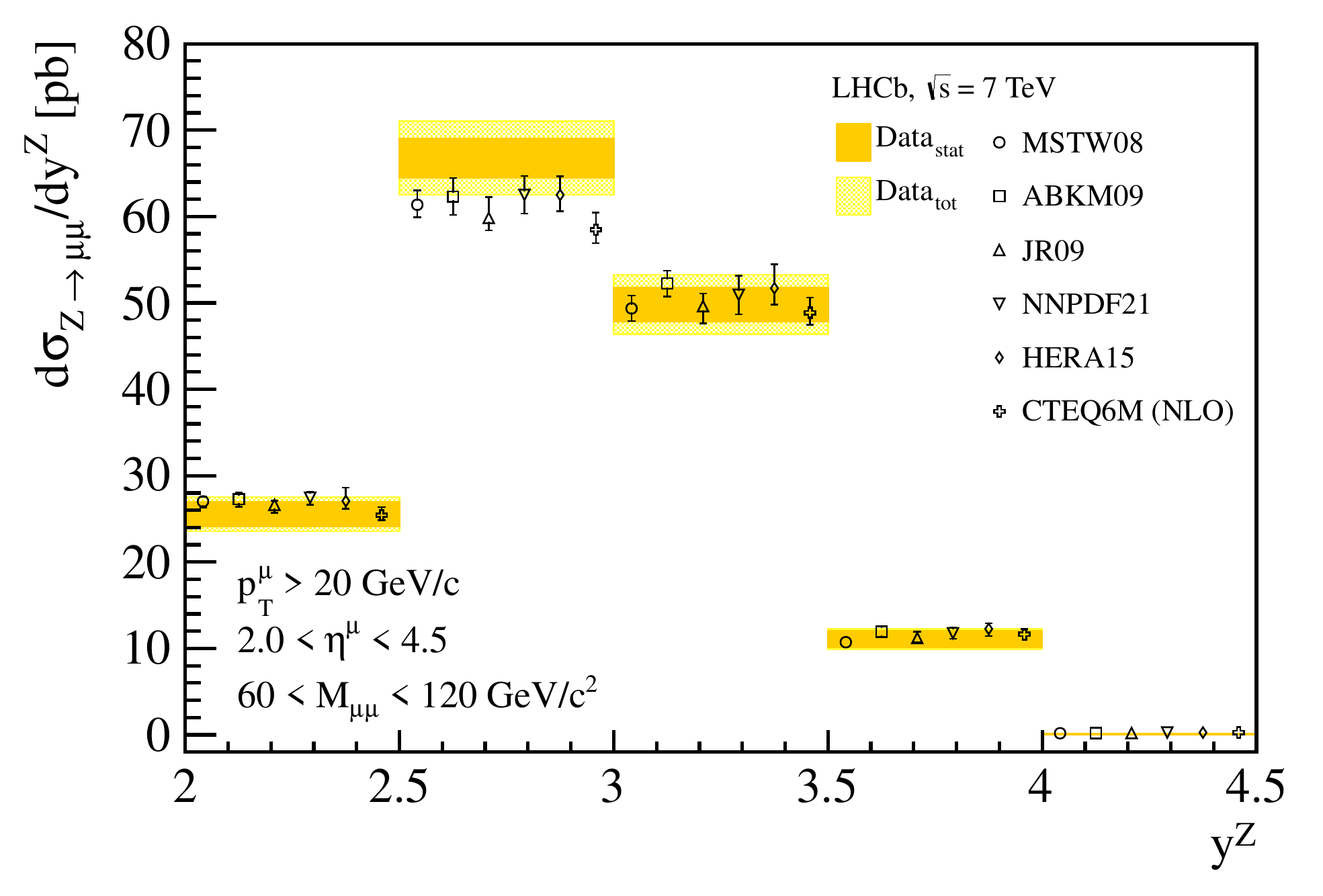}
\caption[LHCb $W$/$Z$ boson (pseudo)rapidity data]{LHCb $W$/$Z$ boson (pseudo)rapidity data. The left panel shows the $W^{\pm}$ distributions in the pseudorapidity of the detected lepton compared to leading PDF sets. The right panel shows the $Z$ in the rapidity of the resulting lepton pair. Figures from~\cite{Aaij:2012vn}.}
\label{fig:LHCBWZ}
\end{figure}

The LHCb detector has a window upon electroweak vector boson production in the very forward region, a kinematic regime that cannot be explored by the general-purpose detectors. $W$ and $Z$ to muon production data based upon an integrated luminosity sample $37$pb$^{-1}$ was published in Ref.~\cite{Aaij:2012vn}, where data was taken in the pseudorapidity range $2.0 < |\eta| < 4.5$ and presented differentially in the (pseudo)rapidity of the detected lepton (pair). Figure~\ref{fig:LHCBWZ} shows the main result of the LHCb $W/Z$ study and demonstrates the good agreement of the theoretical predictions, within the limited statistical precision available in the forward data sample.

%\subsubsection{Drell-Yan}
%2011 CMS differential in invariant mass Drell-Yan \cite{CMS:2011dxa}
%2011 CMS double differential Y M DY \cite{CMS:2012xxs}
%

\section{Prompt photon data}

Constraints upon the gluon distribution are possible through measurements made of direct photon production at the LHC. Both CMS and ATLAS have published prompt photon data. ATLAS provides inclusive data in photon pseudorapidity intervals of $|\eta| <1.37$ and $1.52 \le |\eta| < 2.37$, for transverse energies $45 \le E_T < 400$ GeV~\cite{Aad:2011tw}, the data showing excellent agreement with predictions from CTEQ6.6 and JETPHOX. Additionally data is available for isolated prompt photon data in association with a jet, based upon the same dataset~\cite{ATLAS:2012ar}, where once again NLO predictions provide a good description of the data, albeit with a small discrepancy arising for photons with $E_T < 45$ GeV.

\begin{figure}[ht]
\centering
\includegraphics[width=0.48\textwidth]{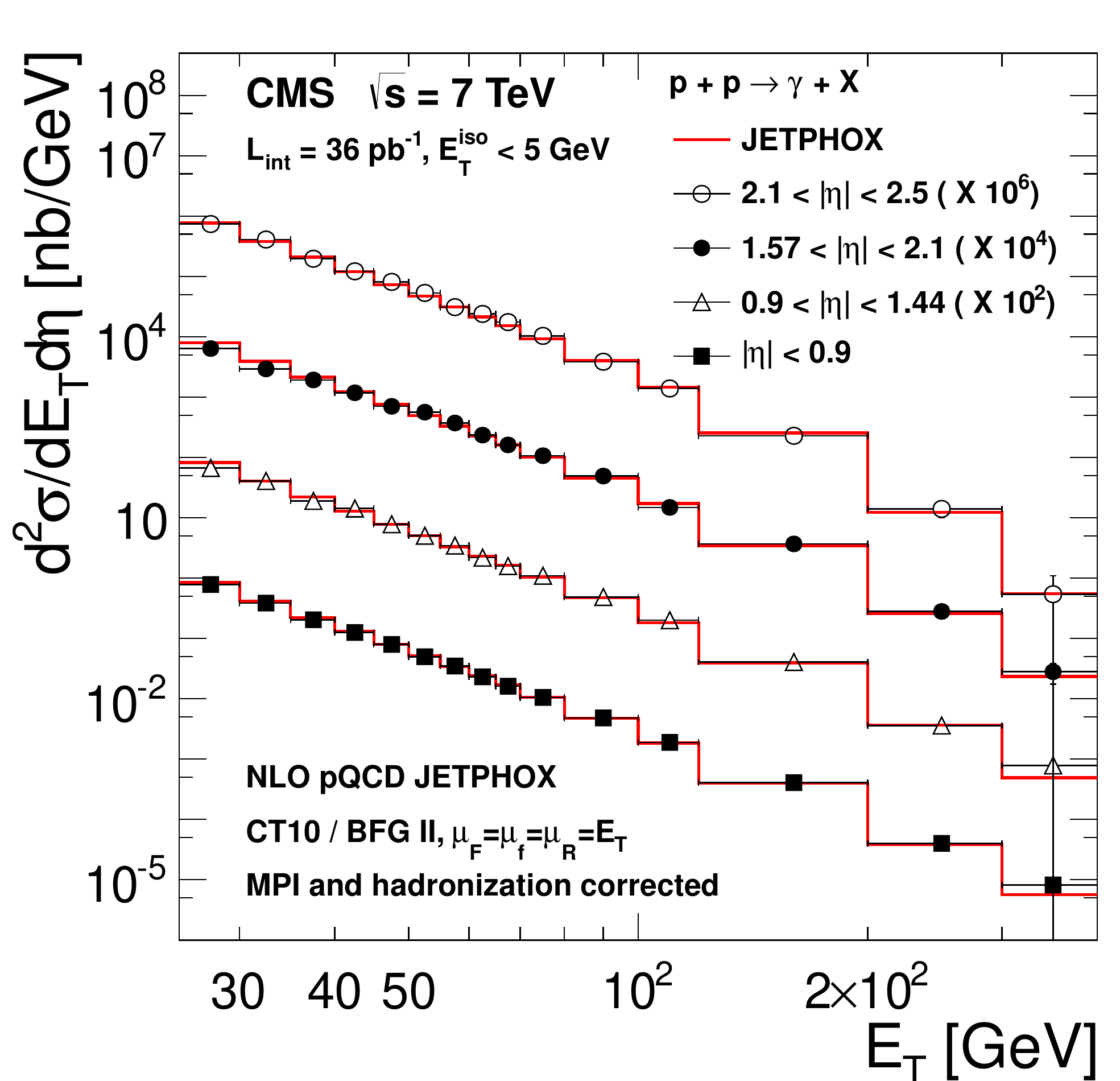}
\includegraphics[width=0.48\textwidth]{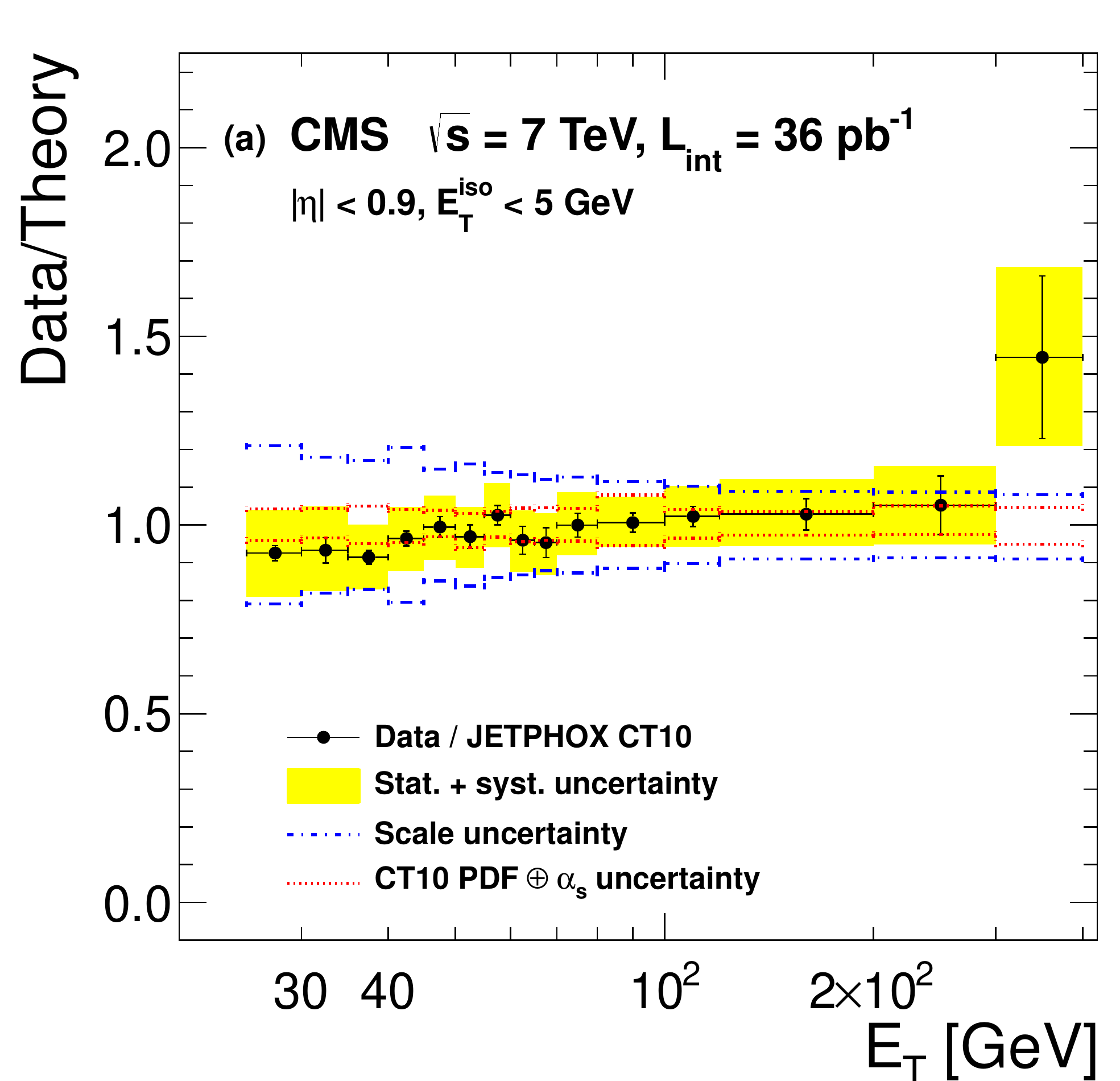}
\caption[CMS isolated prompt photon measurement]{Figures from the CMS isolated prompt photon measurement~\cite{Chatrchyan:2011ue}. The left panel shows the full dataset compared to the CT10/JETPHOX predictions. The right panel shows the result for the lowest pseudorapidity bin as a ratio to the CT10 central value.}
\label{fig:CMSpromptphoton}
\end{figure}

CMS has performed an isolated photon measurement based upon the same 2010 data run, in the pseudorapidity range $|\eta| < 2.3$ for photons with $25 < E_T < 400$ GeV~\cite{Chatrchyan:2011ue}. The CMS result is plotted in Figure~\ref{fig:CMSpromptphoton}, which shows the agreement between the NLO calculation and the experimental data. The figure demonstrates clearly the precision available of the experimental measurement, however the theoretical predictions clearly suffer from relatively large scale uncertainties. The inclusion of such data into PDF determinations is therefore likely to be challenging without further theoretical progress.

\section{Top pair production data}
\begin{figure}[!]
\centering
\includegraphics[width=0.8\textwidth]{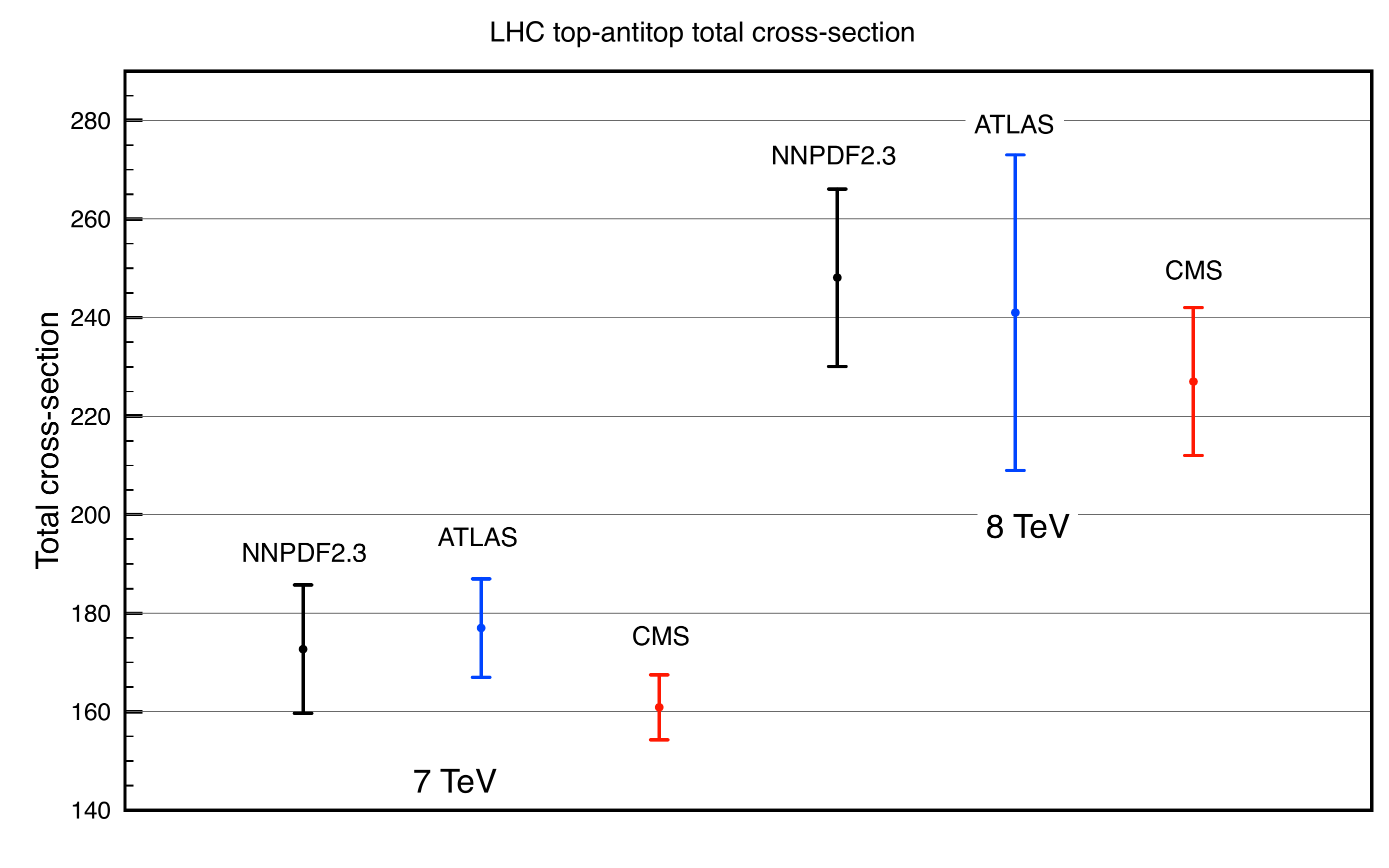}
\caption[LHC 7 and 8 TeV $t\bar{t}$ total cross-section data and predictions from NNPDF2.3]{LHC 7 and 8 TeV $t\bar{t}$ total cross-section data and predictions from NNPDF2.3 at NNLO+NNLL precision. Values in the figure are taken from~\cite{Czakon:2013tha}.}
\label{fig:LHCttbar}
\end{figure}

LHC collaborations have made extensive measurements of the top pair production cross-section, building upon the combined Tevatron analysis of~\cite{Aaltonen:2012ttbar}. Unlike at the Tevatron where the $qq$ initiated channel is favoured, $t\bar{t}$ data at the LHC is primarily a probe of the gluon content of the proton through the $gg \to t\bar{t}$ subprocess.
The ATLAS collaboration has published measurements of the $t\bar{t}$ cross section in a number of channels, with combination results available at both $7$ TeV~\cite{ATLAS:2012jyc} and $8$ TeV ~\cite{ATLAS:2012fja} centre of mass energies. Likewise CMS have published combined $t\bar{t}$ analyses at $7$~\cite{Chatrchyan:2012bra} and $8$~\cite{CMS:2012iba} TeV. These results are compared to the theoretical prediction obtained from NNPDF2.3 at NNLO+NNLL with { \tt top++ v2.0}~\cite{Czakon:2011xx} in Figure~\ref{fig:LHCttbar}.

\chapter{The impact of LHC data on PDFs}
\label{ch:LHCimpact}
The series of measurements made to date during the first runs of the LHC have been studied to assess the impact upon PDFs and their uncertainties, and where appropriate, have been included into PDF fits through the NNPDF methodology. Early LHC measurements serve not only as useful constraints in their own right, but also as a testing ground for tools developed to include such data into PDF determinations, ready for future datasets with even higher precision. 

In this section we shall provide an overview of the work performed in the inclusion of LHC data and some of the results obtained. The methods introduced in Chapter~\ref{ch:LHCtools} are applied to some of the datasets in Chapter~\ref{ch:LHCdata} and the resulting PDFs discussed and compared to results obtained before the LHC era.

\section{The NNPDF2.2 parton set}
The NNPDF2.2 parton set~\cite{Ball:2011gg} was the first practical demonstration of the Bayesian reweighing and unweighting methods. These methods were applied to the inclusion of a series of $W$ boson charge asymmetry measurements made by the ATLAS, D0 and CMS experiments. In this way LHC data was included for the first time in a public parton set.
\subsection{NNPDF2.2 dataset}

\begin{figure}[h!]
  \centering
  \includegraphics[width=0.48\textwidth]{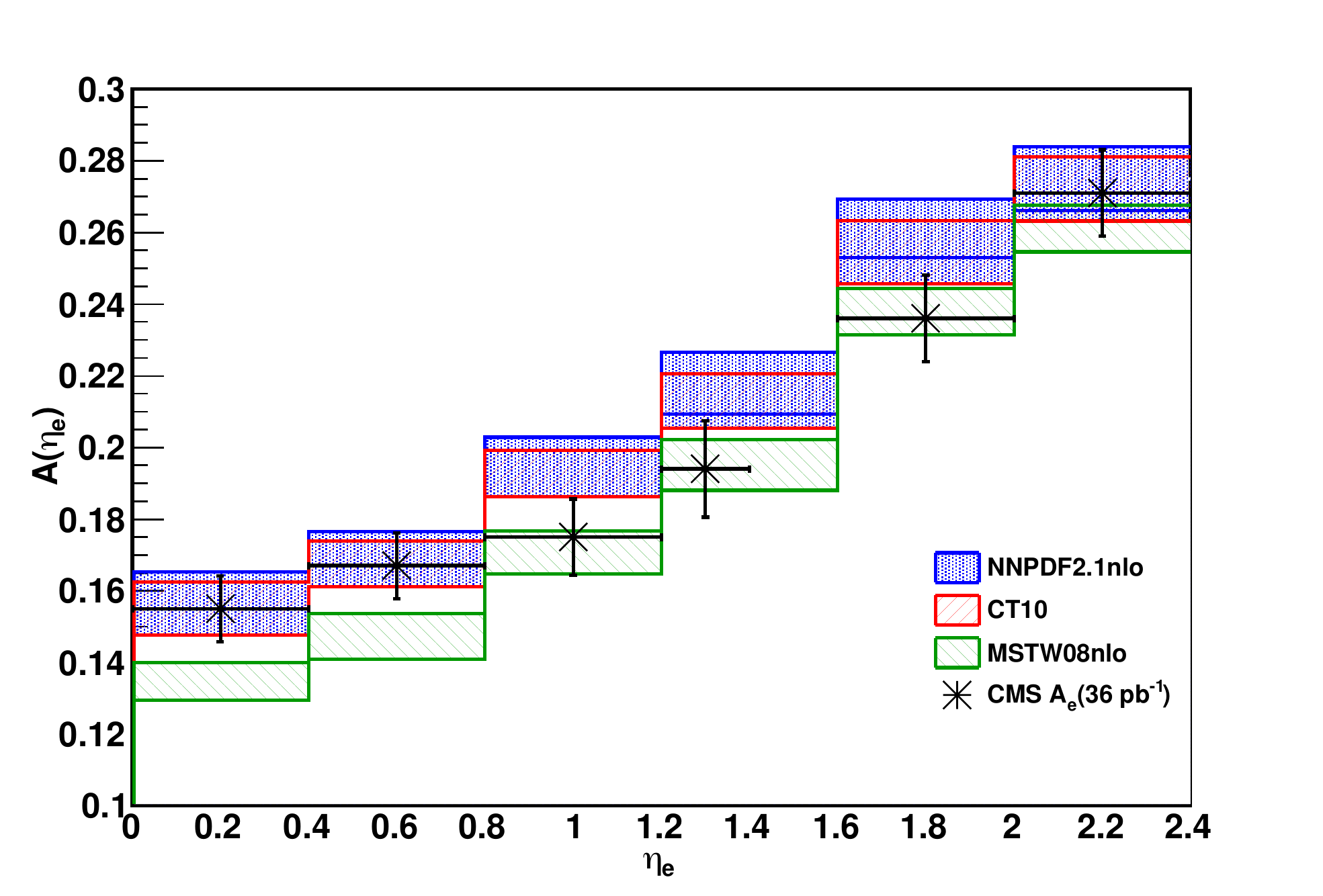}
  \includegraphics[width=0.48\textwidth]{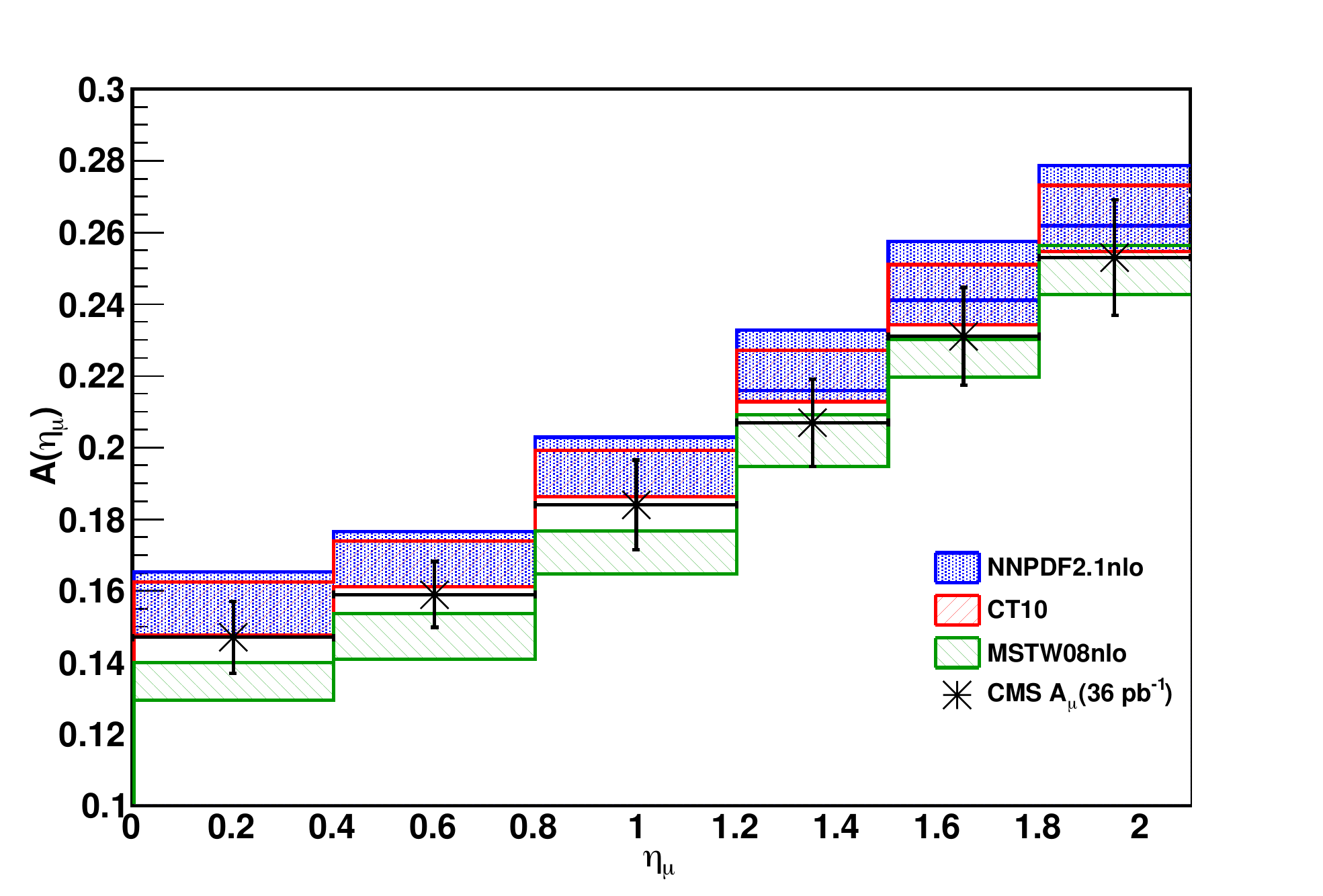}\\
    \includegraphics[width=0.48\textwidth]{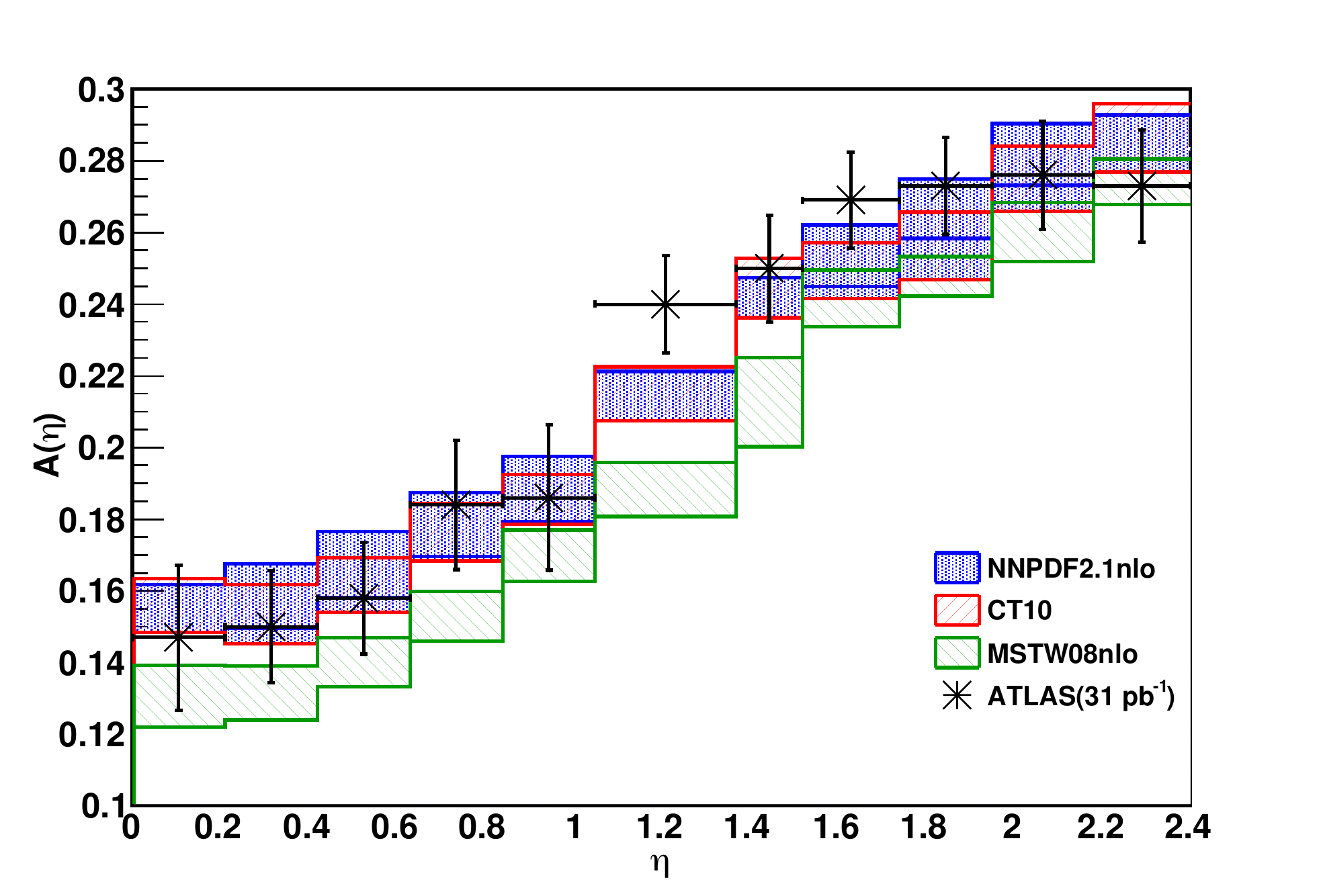}

  \caption[Plot of LHC data to be included in the NNPDF2.2 determination. ]{Plot of LHC data to be included in the NNPDF2.2 determination. Data from the CMS electron (top-left) and muon (top-right) data is given alongside the ATLAS muon asymmetry data (bottom). Figure from~\cite{Ball:2011gg}.}
  \label{fig:22LHCdat}
\end{figure}

The dataset studied included the $1.96$ TeV $p\bar{p}$ data from D0 on both the $W$ electron~\cite{Abazov:2008qv} and muon asymmetries~\cite{Abazov:2007pm}. The LHC dataset consisted of the 2010 run $W$ lepton asymmetry measurement of CMS~\cite{Chatrchyan:2011jz} and ATLAS~\cite{Aad:2011yna}. LHCb asymmetry data with a full covariance matrix was not available at the time and so was not included in the dataset. Agreement for the LHC data points is generally reasonable for PDF sets obtained without LHC data, as shown in Figure~\ref{fig:22LHCdat}. While NNPDF2.1 and CT10 obtain good overall agreement, the MSTW2008 prediction tends to be systematically lower than the data. 

The level of agreement taking into account systematic uncertainties is of course most clearly quantified with a $\chi^2$ measure. In Table~\ref{tab:atlas-cms-chi2} we compare the fit quality of NNPDF2.1, CT10 and MSTW2008 NLO sets to the new data. The result show that while generally the consistency with the new datasets is good in NNPDF2.1, there is certainly room for improvement.

\begin{table}[ht]
  \begin{center}
    \begin{tabular}{|c|c|c|c|c|}
      \hline
       & $N_{\mathrm{dat}}$& NNPDF2.1 & CT10 & MSTW08 \\
      \hline
      ATLAS(31pb$^{-1}$)                       & 11 & 0.77 & 0.77 & 3.32 \\
      \hline
      CMS(36pb$^{-1}$) electron $p_{T}>25$ GeV &  6 & 1.83 & 1.19  & 1.70 \\
      \hline
      CMS(36pb$^{-1}$) muon $p_{T}>25$ GeV     &  6 & 1.24 & 0.73  & 0.77 \\
      \hline
      \hline
      D0(0.3fb$^{-1}$) muon $p_{T}>20$ GeV     &  10 & 1.48  &  -  & -  \\
            \hline
      D0(0.75fb$^{-1}$) electron $E_{T}>25$ GeV     &  12 & 4.39 & -   & -  \\
      \hline
    \end{tabular}
  \end{center}
  \caption[Table of $\chi^2$ values for new data included in NNPDF2.2]{Table of $\chi^2$ values for new data included in NNPDF2.2. }
  \label{tab:atlas-cms-chi2}
\end{table}

The combined goodness of fit value for the LHC and Tevatron datasets for NNPDF2.1 is $\chi^2/N_{\text{data}} = 2.22$ which suggests a less than ideal description of the data, largely due to the precise D0 electron asymmetry measurement. 

For this dataset the reweighting technique presented an ideal method for the data inclusion. With a total of 45 points the dataset is relatively small, and together with the fair agreement of the prior PDF set (NNPDF 2.1) the reweighting can be accomplished with a reasonable number of prior replicas. Also the lack of a fast method of determining these asymmetries within the NNPDF framework at the time meant that the data could not be included via a conventional fit, necessitating the reweighting approach.

\subsection{NNPDF2.2 results}

The LHC and Tevatron $W$ boson asymmetry datasets were included into the NNPDF 2.1 determination by a reweighting both individually and upon the combined $\chi^2$ figure for the whole dataset. To ensure maximal final ensemble efficiency, an NNPDF 2.1 prior with $N_{\text{rep}} = 1000$ replicas was used for the reweighting. Theoretical predictions for the various datasets were computed at NLO using DYNNLO~\cite{Catani:2007vq}. After computing the $\chi^2$ values using the $t_0$ method to ensure consistency with the fit procedure, the number of effective replicas remaining in the ensemble is given by the Shannon entropy (Eqn.~\ref{eq:shannonentropy}). For the different reweighting combinations attempted, the number of effective replicas is given in Table~\ref{tab:22Neff}.

\begin{table}[ht]
  \begin{center}
    \begin{tabular}{|c|c|c|c|c|}
      \hline
       & ATLAS & CMS & LHC & LHC + TeV\\
      \hline
      $N_{\text{eff}}$ & 928 & 531 & 619 & 181 \\
      \hline
    \end{tabular}
  \end{center}
  \caption[Number of effective replicas for each dataset reweighting in NNPDF 2.2]{Number of effective replicas for each dataset reweighting in NNPDF 2.2. Figures are given for the ATLAS and CMS experiments, along with their combination (LHC) and their combination with the Tevatron data (LHC+TeV).}
  \label{tab:22Neff}
\end{table}

Both ATLAS and CMS show good consistency with the prior in the reweighting, with the CMS data providing the greater constraint and resulting in a lower number of replicas surviving the reweighting process. The reweighting with ATLAS data only leading to 928 effective replicas and the CMS reweighting resulting in 531. The D0 data goes further to provide a great deal of extra constraint. In the final combined reweighting, roughly one fifth of the prior replicas remain active, a figure which demonstrates that the $W$ asymmetry data available at the time was able to provide a great deal of additional information on parton distributions.

The PDF set resulting from the reweighting with the combined dataset was then unweighted to 100 replicas via the mechanism described in Chapter~\ref{ch:LHCtools}. The unweighted set forms the NNPDF 2.2 determination, available as part of the LHAPDF platform. In Table~\ref{tab:chisq22} the full $\chi^2$ breakdown for every experiment
in the NNPDF2.2 dataset is shown. It is clear from the table that a great deal of improvement in the new $W$ asymmetry data is achieved by the addition of the new data, and there is no associated cost to the $\chi^2$ values for the rest of the dataset, suggesting that the new data maintains a good consistency with the measurements already utilised in NNPDF 2.1. The global fit quality therefore has a modest improvement from $\chi^2/N_{\text{data}}=1.165$ to $1.157$.

\begin{table}[hp!]
  \begin{center}
    \begin{tabular}{|c|c|c|c|}
      \hline 
      Experiment & $N_{\mathrm{dat}} $ & NNPDF2.1 & NNPDF2.2 \\
      \hline
      \hline
      NMC-pd     &  132 &  0.97 &   0.97 \\
      \hline
      NMC        &  221 &  1.73 &    1.72 \\
      \hline
      SLAC       &   74 &  1.33 &    1.28 \\
      \hline
      BCDMS      &  581 &  1.24 &   1.23 \\
      \hline
      HERAI-AV   &  592 &  1.07 &   1.07 \\
      \hline
      CHORUS     &  862 &  1.15 &    1.15 \\
      \hline
      FLH108     &    8 &  1.37 &    1.37 \\
      \hline
      NTVDMN     &   79 &  0.79 &    0.70 \\
      \hline
      ZEUS-H2    &  127 &  1.29 &    1.28 \\
      \hline
      ZEUSF2C    &   50 &  0.78 &   0.78 \\
      \hline
      H1F2C      &   38 &  1.51 &    1.51 \\
      \hline
      DYE605     &  119 &  0.84 &    0.86 \\
      \hline
      DYE886     &  199 &  1.25 &    1.27 \\
      \hline
      CDFWASY    &   13 &  1.85 &    1.81 \\
      \hline
      CDFZRAP    &   29 &  1.66 &    1.70 \\
      \hline
      D0ZRAP     &   28 &  0.60 &    0.58 \\
      \hline
      CDFR2KT    &   76 &  0.98 &    0.96 \\
      \hline
      D0R2CON    &  110 &  0.84 &   0.83 \\
      \hline
      \hline
      ATLASmuASY &   11 & [0.77]  &   1.07  \\
      \hline
      CMSeASY    &   6 &  [1.83]  &    1.08  \\
      \hline
      CMSmuASY   &   6 &  [1.24]  &    0.56  \\
      \hline
      D0eASY     &   12 & [4.39]  &    1.38  \\
      \hline
      D0muASY    &   10 & [1.48]  &    0.35  \\
      \hline
      \hline
      Total      &      &  1.165 &    1.157 \\
      \hline
    \end{tabular}
  \end{center}
  \caption[The global fit quality to all experiments included in the NNPDF 2.2 fit]{The global $\chi^2/N_{\text{dat}}$ values to all experiments included in the NNPDF 2.2 fit. Values presented within square brackets were not included in the associated fit, and do not contribute to the total at the end of the table. Values from~\cite{Ball:2011gg}.}
  \label{tab:chisq22}
\end{table}

\begin{figure}[hp!]
  \centering
  \includegraphics[width=0.44\textwidth]{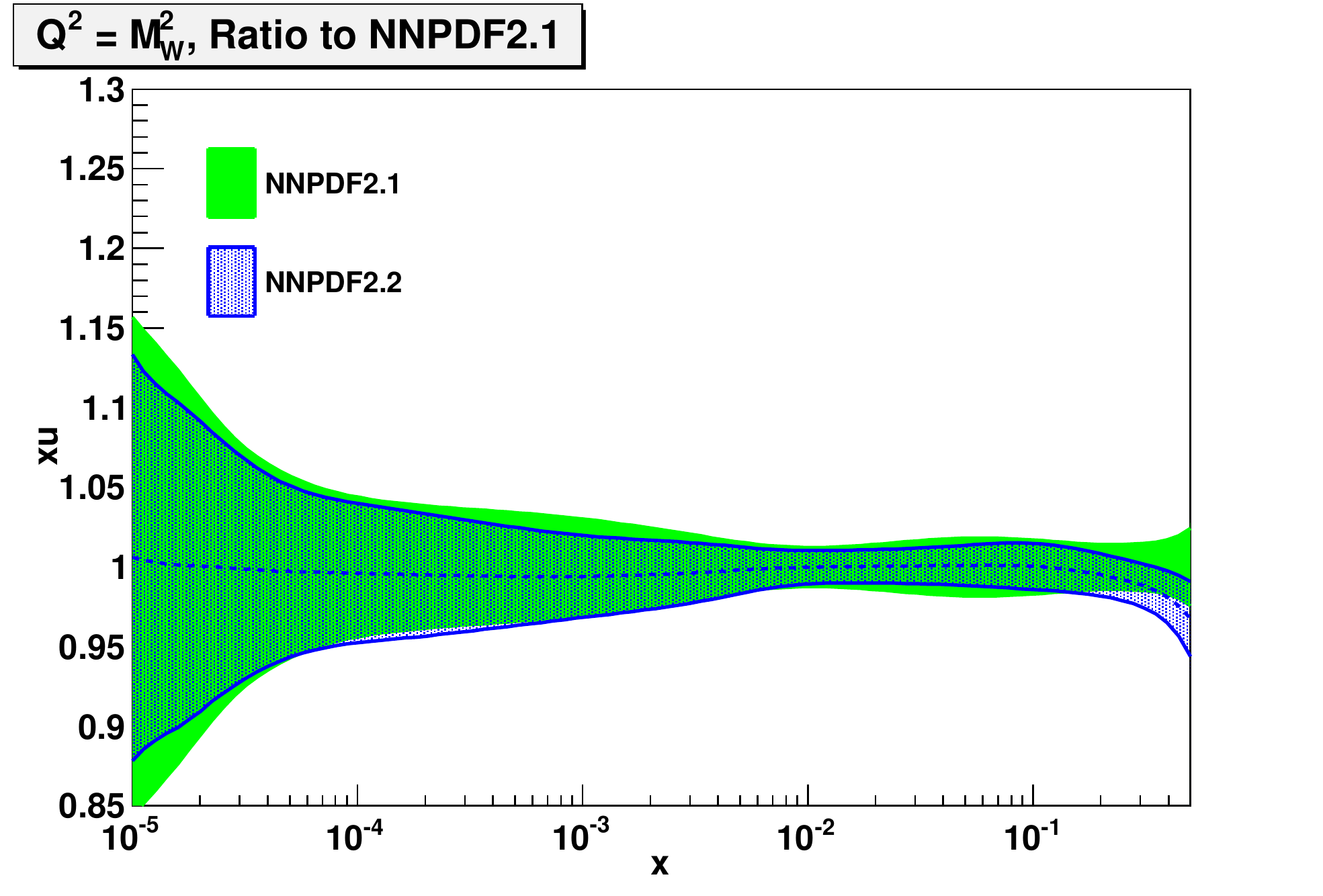}
    \includegraphics[width=0.44\textwidth]{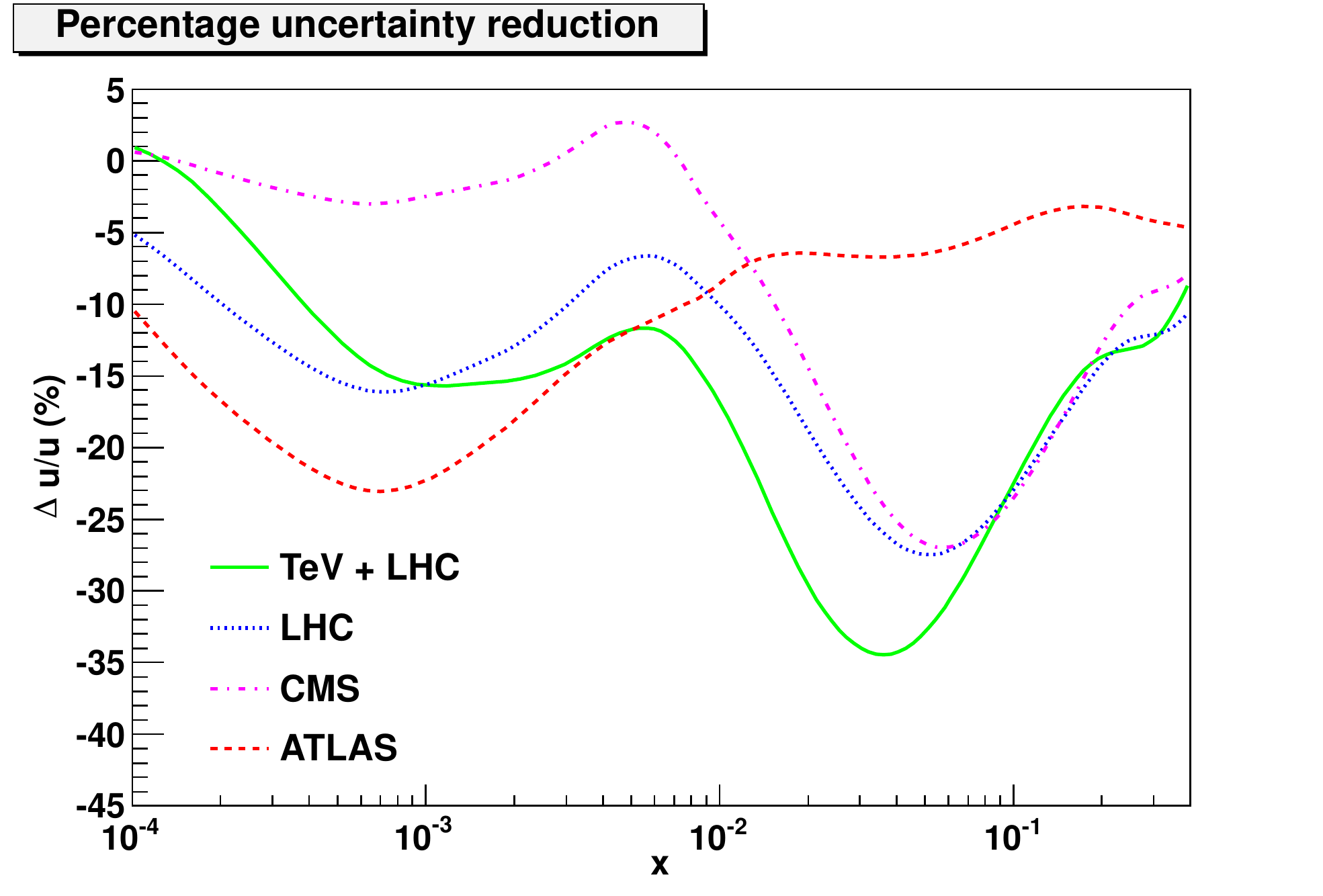}

  \includegraphics[width=0.44\textwidth]{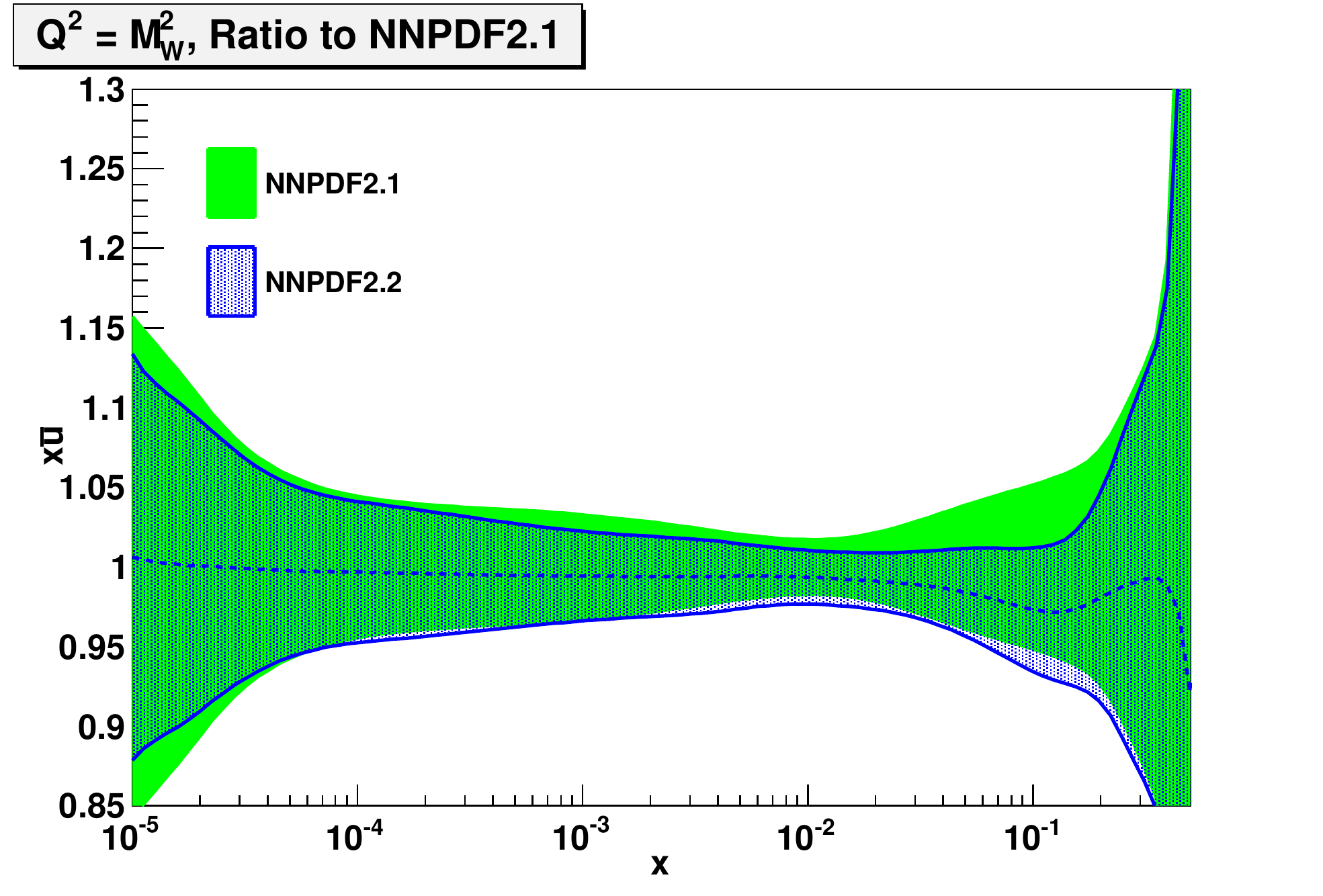}
    \includegraphics[width=0.44\textwidth]{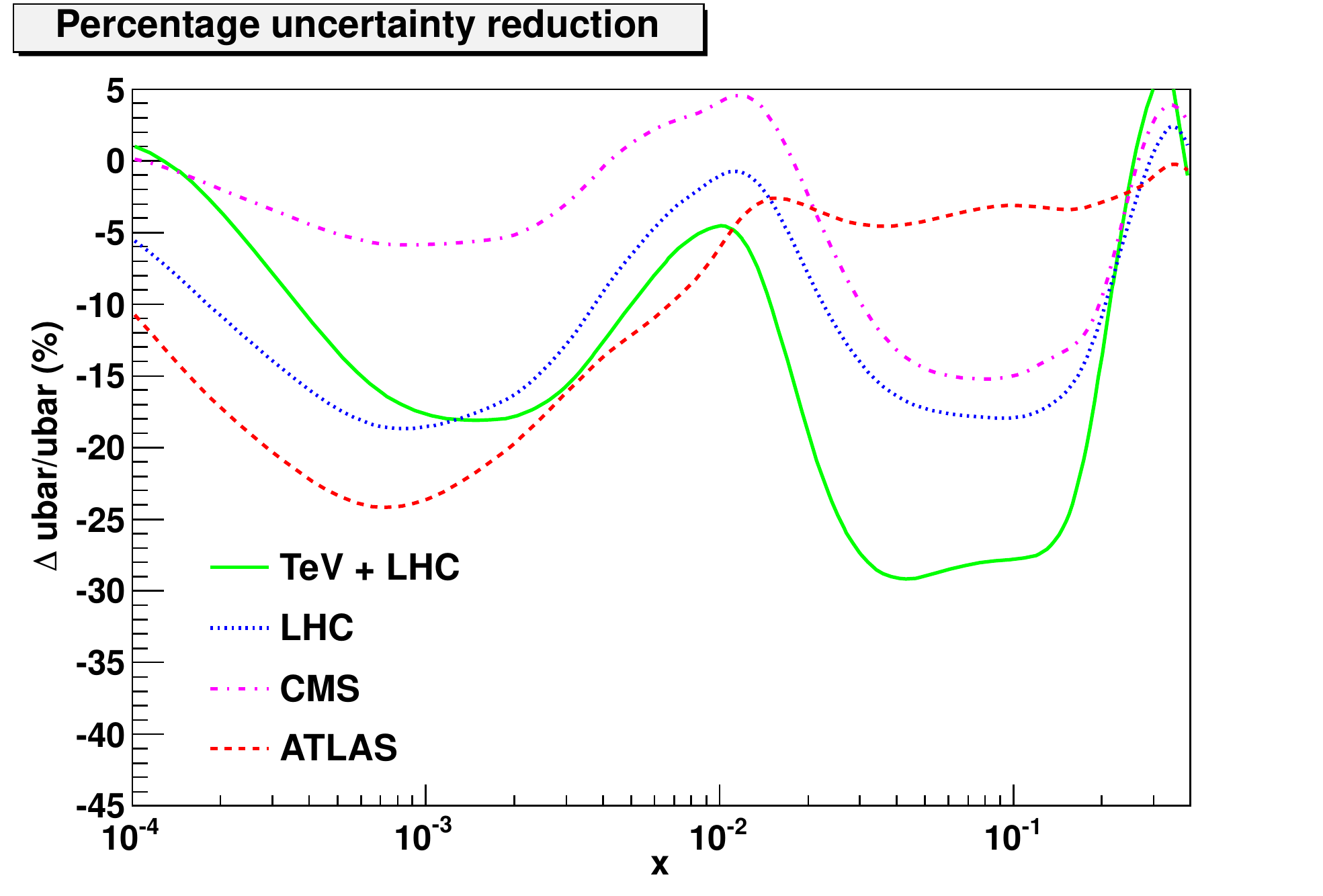}

  \includegraphics[width=0.44\textwidth]{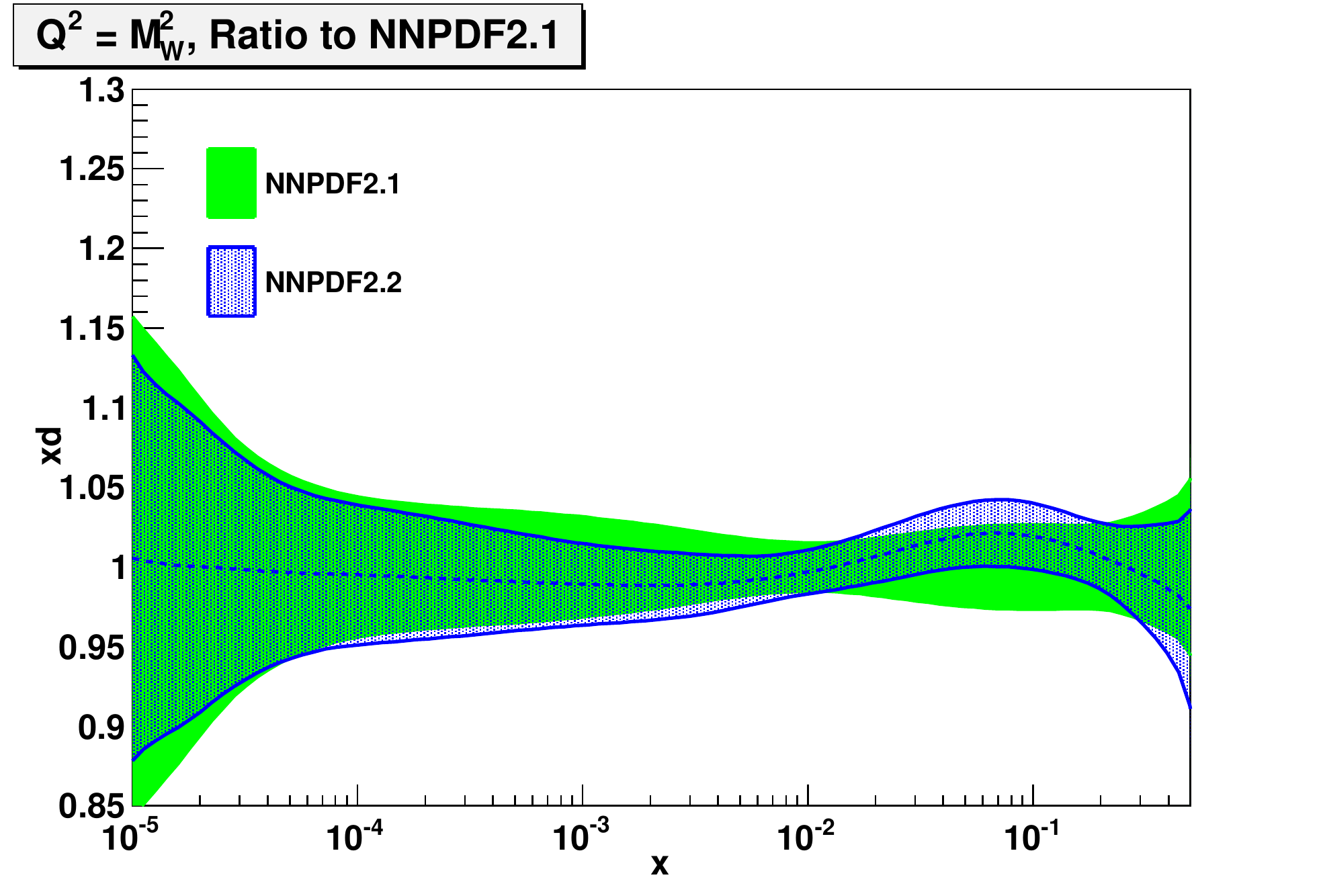}
    \includegraphics[width=0.44\textwidth]{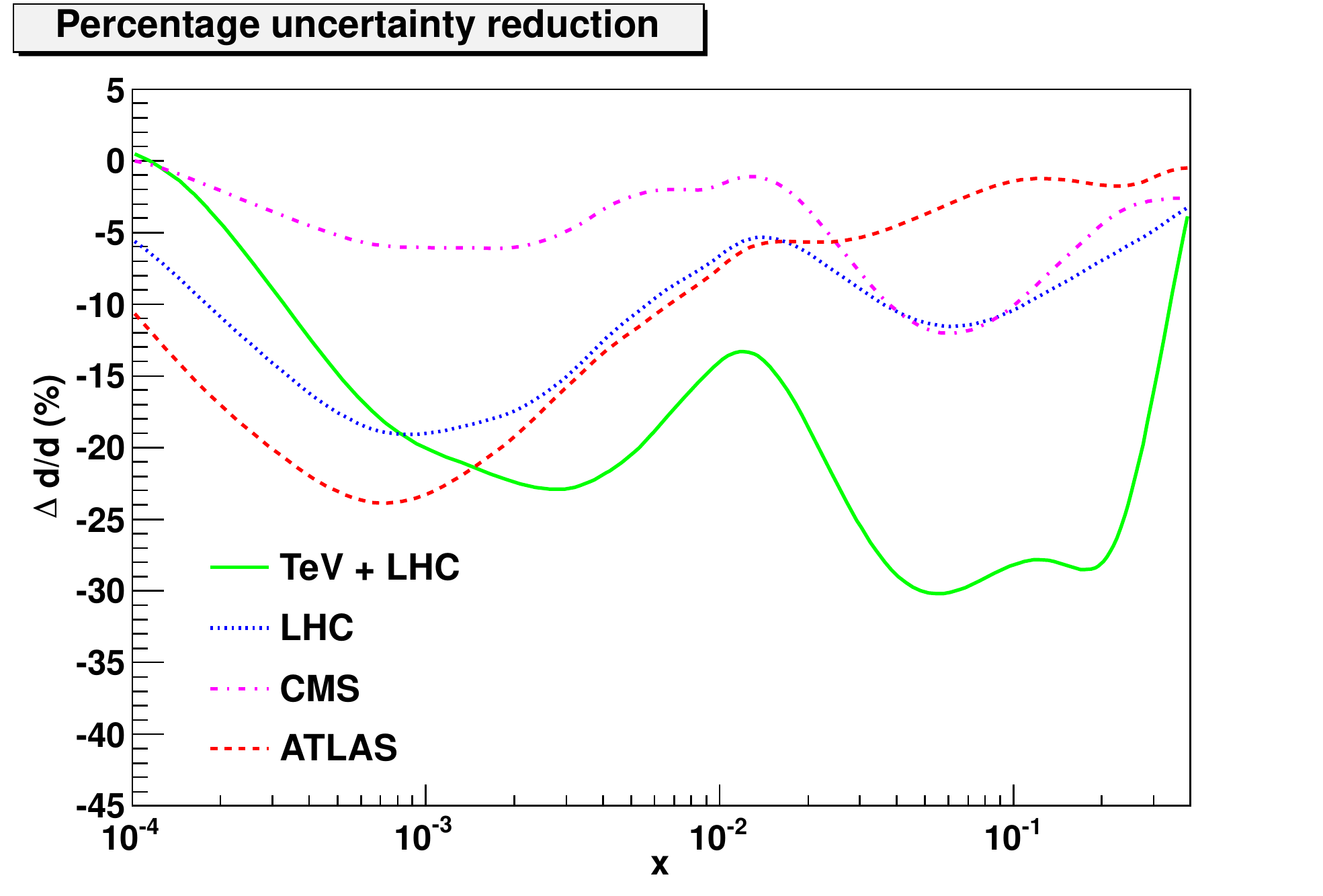}

  \includegraphics[width=0.44\textwidth]{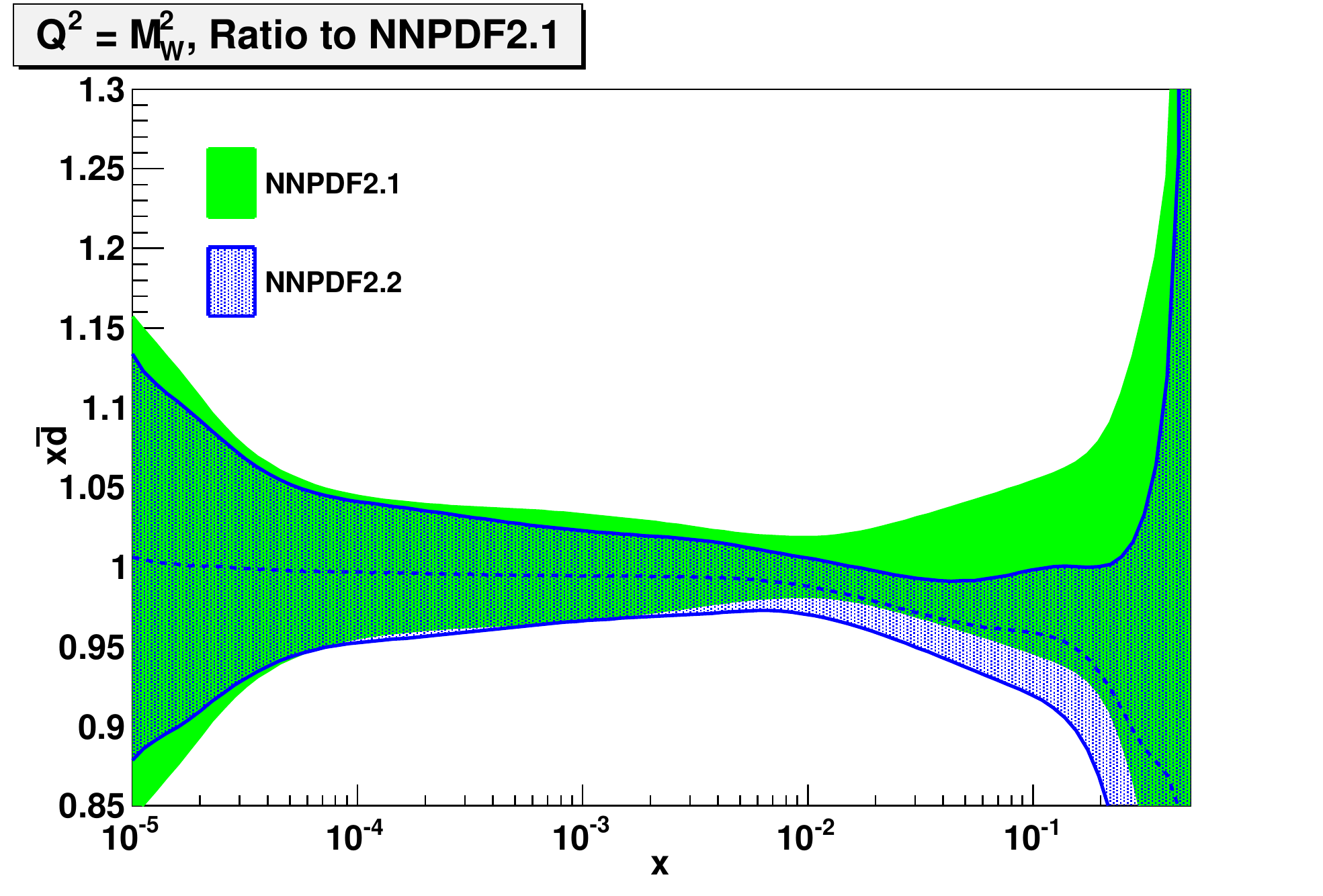}
    \includegraphics[width=0.44\textwidth]{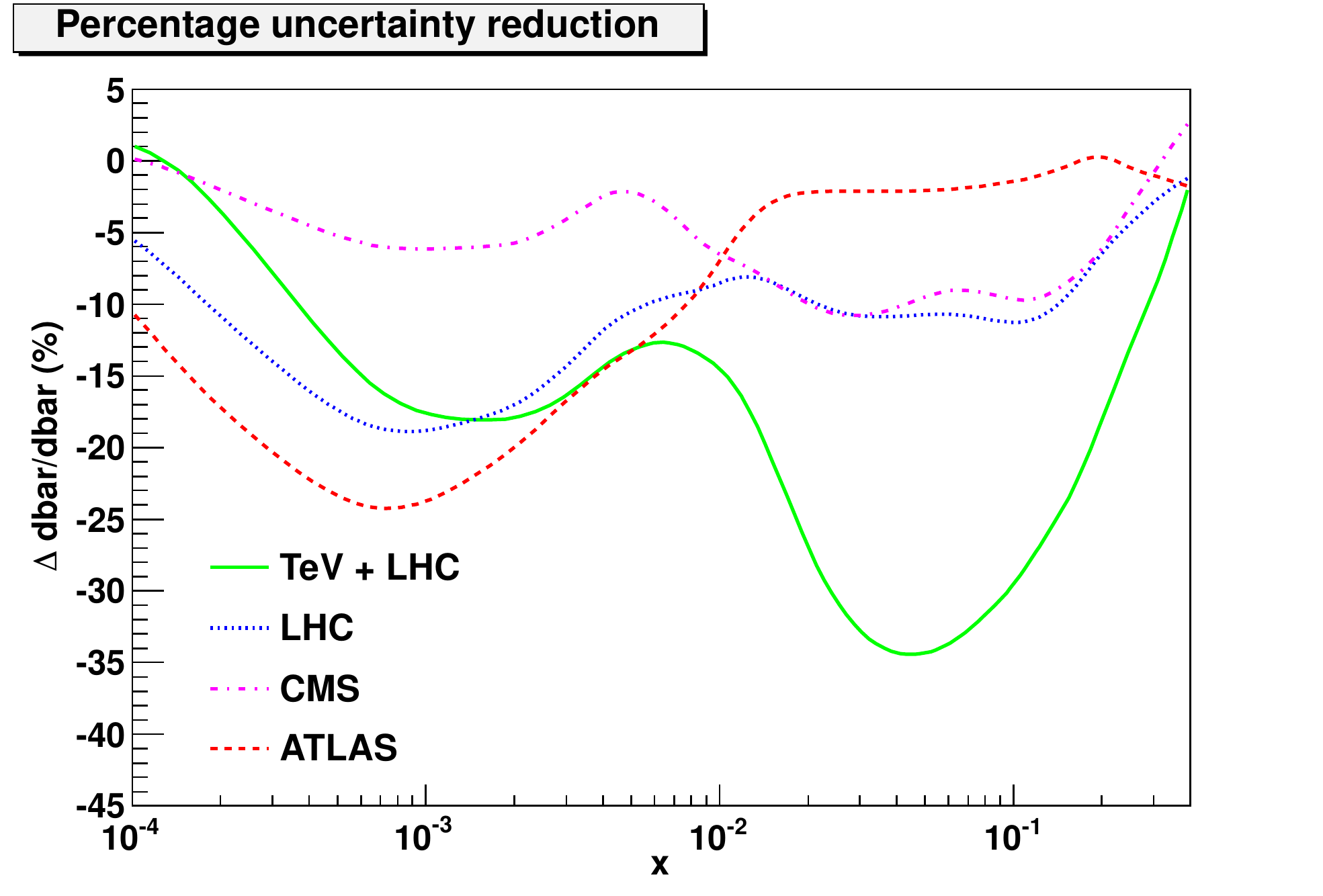}

  \caption[Impact of the LHC and Tevatron $W$ asymmetry data upon PDFs]{Impact of the LHC and Tevatron $W$ asymmetry data upon PDFs. On the left, the NNPDF 2.1 (prior) and NNPDF 2.2 (reweighted) distributions are shown for the light quarks $u,\bar{u},d,\bar{d}$. On the right are the relative uncertainty changes in the equivalent PDFs under a reweighting with the various dataset options, with the green lines indicating the final NNPDF2.2 result. The plots on the right therefore demonstrate the impact of the new data upon light quark PDF uncertainties. Figures from~\cite{Ball:2011gg}.}
 \label{fig:22pdfimpact}
\end{figure}

Examining the NNPDF 2.2 PDFs directly, the largest differences with respect to the prior arise as expected in the light quark PDFs, the most relevant initial states for the $W$ asymmetry. Figure~\ref{fig:22pdfimpact} demonstrates the effect that the new data has upon the PDFs. For all of the light quark distributions a substantial reduction of the uncertainties can be observed, with a typical reduction of around 25\%. The PDF central values also undergo a slight shift in the large$-x$ region, typically demonstrating a preference for softer light quarks. Phenomenologically these improvements will manifest in reduced uncertainties for observables sensitive to light/valence quarks over a large kinematic range, and a slightly tweaked distribution for those observables probing high-$x$ physics, such as the high rapidity observable region.

\begin{figure}[h!]
\centering
\includegraphics[width=0.9\textwidth]{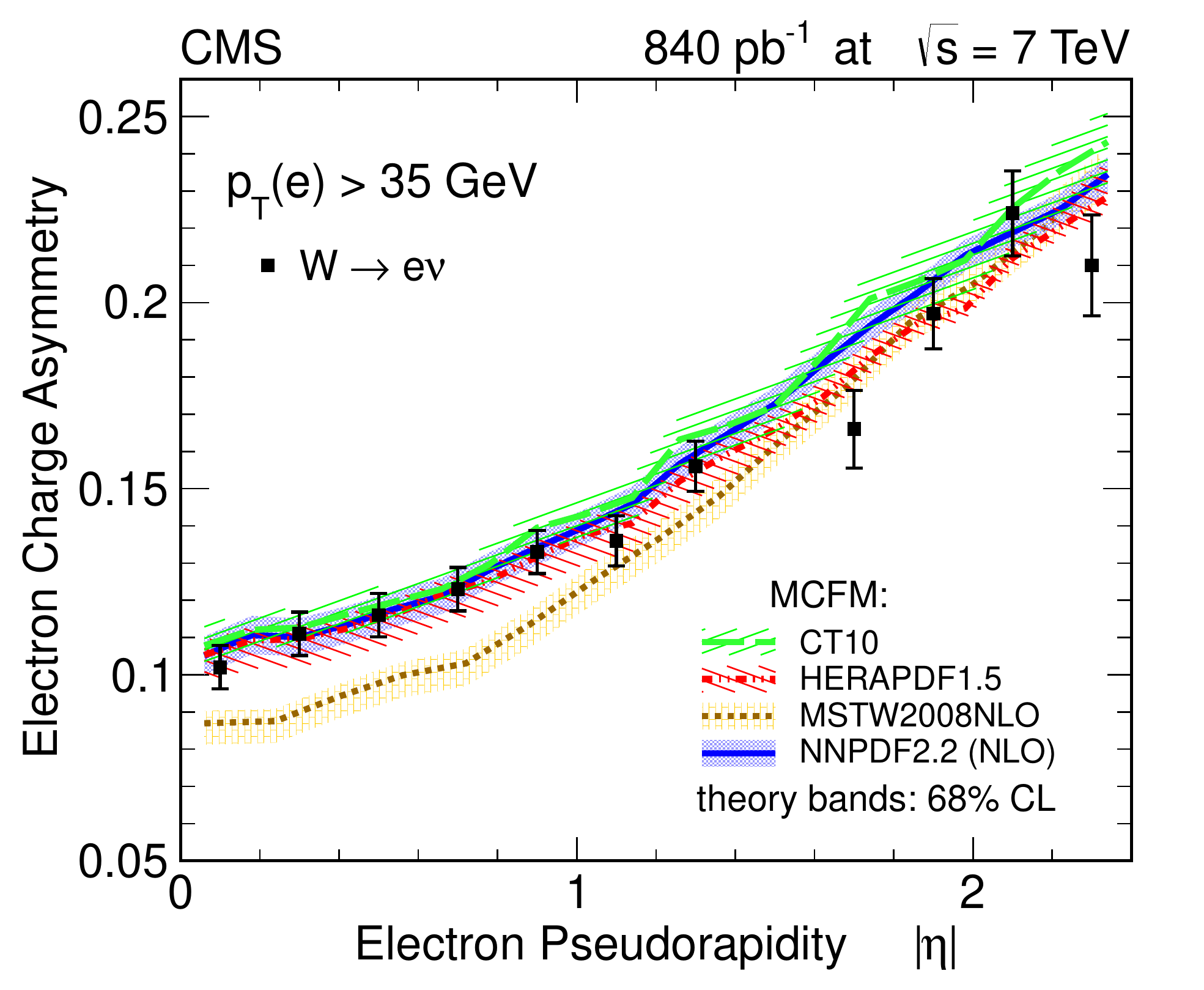}
\caption[Comparison of NNPDF2.2 predictions with CMS $W$ asymmetry measurement]{Comparison of NNPDF2.2 predictions with updated CMS $W$ asymmetry measurement at 840 pb$^{-1}$.  The comparison also includes the theory predictions from CT10, MSTW 2008 and HERAPDF 1.5. Agreement is generally very good for the PDF sets, although the MSTW2008 set demonstrates a significant discrepancy. Figure from~\cite{Chatrchyan:2012xt}.}
\label{fig:Wasy22}
\end{figure}

The NNPDF2.2 parton set was used in the CMS 840pb$^{-1}$ $W$ electron asymmetry measurement~\cite{Chatrchyan:2012xt}, where excellent agreement was demonstrated alongside the high precision available for electroweak observables with the 2.2 set. Figure~\ref{fig:Wasy22} taken from the CMS paper illustrates the level of agreement in comparison to the CT10, MSTW 2008 and HERAPDF 1.5 predictions. 

\section{The NNPDF2.3 parton set}
The NNPDF2.2 fit demonstrated the constraining power of early LHC measurements, and provided a showcase for the reweighting technique as a method of analysing the impact of new data and indeed producing a new PDF set including the data's constraints. Nevertheless, the rapid pace of new experimental measurements meant that the data included in the set was soon superseded with higher integrated luminosity samples, and datasets sampling other processes of interest were being explored at the LHC. As the reweighting exercise in NNPDF2.2 had demonstrated, the inclusion of much more data into the fit would require priors with a rather unwieldy number of replicas, needing in excess of a thousand to include even a modest additional dataset. Therefore to include a large set of up to date measurements from the LHC into a parton fit, the conventional fitting methodology must still be applied.

The development of the {\tt FK} method and associated toolchain enabled these fits to be performed without the requirements of extremely long fitting times, potentially requiring weeks of computer time per replica for a standard fit on a typical 2.4GHz Intel Xeon processor with the earlier technology. In this section we shall discuss the NNPDF2.3 fit, the successor to the NNPDF2.2 fit in that an updated and enlarged LHC dataset is included in a full NNPDF fit. We shall outline the datasets included in the determination, along with a discussion of methodological improvements made, as several optimisations were enabled by the faster fitting framework.

\subsection{NNPDF2.3 dataset and methodology}
\label{sec:23meth}
For the NNPDF2.3 determination, the electroweak data included in NNPDF2.2 has been upgraded. From CMS the 840pb$^{-1}$ $W$ electron asymmetry data~\cite{Chatrchyan:2012xt} replaces the previous measurement. The full $W/Z$ (pseudo)rapidity distributions replace the asymmetry measurements for ATLAS, based upon $35$ pb$^{-1}$ of 2010 data~\cite{Aad:2011dm}. From LHCb, the $W^{\pm}$ distributions in the forward region were included~\cite{Aaij:2012vn}. Beyond the electroweak sector, the ATLAS 2010 inclusive jet data was also included to obtain an additional handle upon the gluon. At the time of publication, the NNPDF2.3 dataset included all relevant published LHC data with publicly available covariance matrices. Theoretical predictions for these observables were implemented as {\tt FK} tables obtained via APPLgrid files from {\tt MCFM} for the electroweak processes, and  {\tt nlojet++} for the jet data.

\begin{figure}[h!]
\centering
\includegraphics[width=0.9\textwidth]{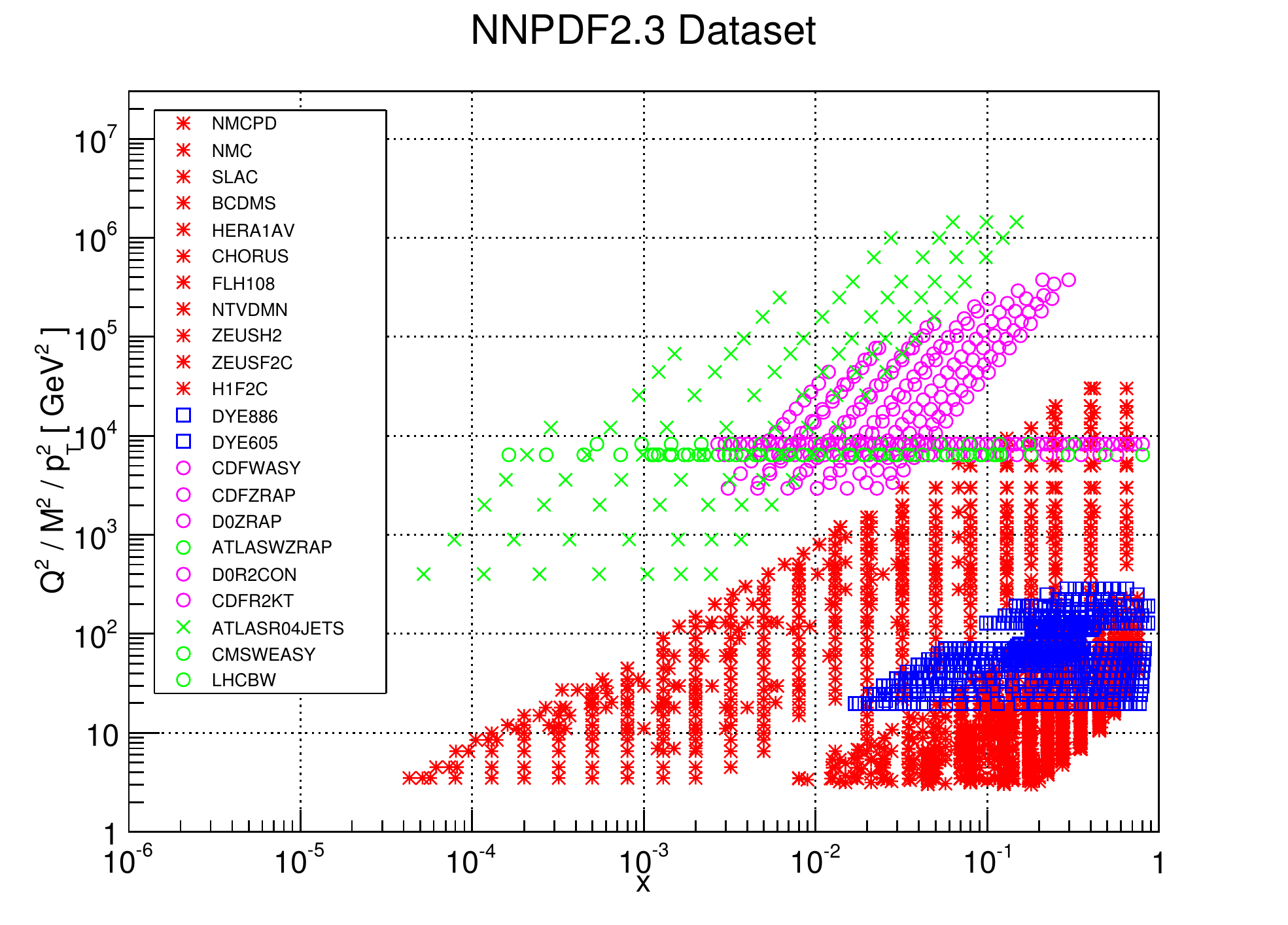}
\caption[Kinematic distribution of data points in the NNPDF2.3 analysis]{Kinematic distribution of points in the NNPDF2.3 analysis. The green points show the LHC data which was added to the analysis over the NNPDF2.1 dataset, and demonstrates the additional kinematic reach of the dataset. DIS datapoints are represented in red, the fixed-target Drell-Yan data in blue, and Tevatron data in pink. In the case of data with two hadrons in the initial state, the smaller parton-$x$ value is plotted.}
\label{fig:kin23}
\end{figure}

Figure~\ref{fig:kin23} demonstrates the additional reach of the NNPDF2.3 dataset upon the addition of the LHC data. The electroweak measurements extend those performed at the Tevatron to considerably lower values of parton-$x$. The inclusive jet data spans a large range in kinematics, providing points at large and small$-x$  across a wide range of scales. Examining the description provided by earlier PDF sets, Table~\ref{tab:23datagreement} demonstrates the agreement at NLO and NNLO of the previous 2.1 PDF set to the new experimental data. While fair agreement is reached for most sets the description is often sub-optimal therefore the data can provide useful additional constraints. This is particularly evident for the ATLAS electroweak data at NNLO ($\chi^2/N_{\text{dat}} = 2.21$) and the CMS $W$ electron asymmetry data at NLO ($\chi^2/N_{\text{dat}} = 2.02$).

\begin{table}[htdp]
\begin{center}
\begin{tabular}{c|c|c}

\hline
  & \multicolumn{2}{c}{NNPDF2.1}\\ \hline
    & NLO& NNLO\\
 \hline \hline 
 ATLAS $W/Z$ & 1.57 & 2.21 \\
 \hline
 LHCb $W$ &0.89 &1.13 \\
 \hline
 CMS $We$ Asy & 2.02 & 1.27 \\
 \hline\hline
 ATLAS Jets & 1.06& 0.95\\
 \hline
\end{tabular}
\caption[Description of the NNPDF2.3 LHC dataset by the NNPDF2.1 PDF set]{Description of the NNPDF2.3 LHC dataset provided by the NNPDF2.1 PDF set, provided as $\chi^2$ per degree of freedom, $\chi^2/N_{\text{dat}}$. The new data shows good consistency with the previous data available in NNPDF2.1 however there is room for improvement upon the inclusion of the data.}
\label{tab:23datagreement}
\end{center}
\end{table}

In the 2.3 fit, the theoretical prediction mechanism for all previously included observables was converted to the {\tt FK} procedure, leading to a substantial decrease in fitting times. These speed improvements were exploited in order to perform a more aggressive fitting procedure. The NNPDF minimisation procedure involves a genetic algorithm where the best fit network per iteration undergoes a set of random adjustments or `mutations' , the best of which is selected for the next iteration. In the NNPDF2.1 NLO fits two genetic algorithm epochs are used. The first, or \emph{`a'} phase with $N_{\text{mut}}^a=80$ mutants and the second \emph{`b'} phase with $N_{\text{mut}}^b=10$ mutants per generation. This was upgraded to the more explorative settings of $N_{\text{mut}}^b=30$ mutants in the second epoch. The maximum number of training generations was extended to $N_{\text{gen}}^{\text{max}}=$ 50,000 generations from the 30,000 used in the NNPDF2.1 series.  For mutation rates, the number of mutations $N_{\text{mut}}$ were increased for a number of PDF combinations in order to better explore the fit quality minima, and the mutation sizes $\eta$ optimised on a PDF by PDF basis. In Table~\ref{tab:gapars} we summarise the modifications made in the genetic algorithm minimisation in terms of the parameters that have been modified. Additionally the parameters controlling the dynamical stopping criterion were tightened, requiring a clearer overlearning signal from the cross-validation.

%%%%%%%%%%%%%%%%%%%%%%%%%%%%%%%%%%%%%%%%%%%%%%%%
\begin{table}
\begin{center}
  \begin{tabular}{|c||c|c|c|c|c|c|}
    \hline 
 &  $N_{\rm gen}^{\rm mut}$
&   $N_{\rm gen}^{\rm max}$ & $N_{\rm mut}^a$ 
&  $N_{\rm mut}^b $\\
    \hline
2.1 NLO &   2500 & 30000 &  80 & 10\\
    \hline
2.1 NNLO &    2500 & 30000  & 80 & 30\\
    \hline
    \hline
2.3 NLO &    2500 & 50000 & 80 & 30\\
    \hline
2.3 NNLO &   2500 & 50000 & 80 & 30\\
    \hline
  \end{tabular}\\
%%%%%%%%%%%%%%%%%%%%%%%%%%%%%%%%%%%%%%%%%%%%%%%%%%%

\bigskip
%\bigskip

%%%%%%%%%%%%%%%%%%%%%%%%%%%%
  \begin{tabular}{|c||c|c|c|c|}
\hline
&  \multicolumn{2}{|c|}{2.1 NLO} &
   \multicolumn{2}{|c|}{2.1 NNLO and 2.3}    \\
    \hline 
PDF &   $N_{\rm mut}$ &  $\eta^{\rm k}$ &  $N_{\rm mut}$ &  $\eta^{k}$  \\
    \hline
\hline 
$\Sigma(x)$   &2 & 10, 1 & 2 & 10, 1 \\
$g(x)$  & 2& 10, 1& 3 & 10, 3, 0.4 \\
$T_3(x)$  &2 & 1, 0.1 & 2 &  1, 0.1 \\
$V(x)$  &2 & 1, 0.1 & 3 &  8, 1, 0.1\\
$\Delta_S(x)$  &2 & 1, 0.1 &3 & 5, 1, 0.1 \\
$s^+(x)$  & 2&  5, 0.5 & 2 & 5, 0.5 \\
$s^-(x)$  & 2&  1, 0.1& 2 & 1, 0.1\\
\hline 
  \end{tabular}
  \end{center}
  \caption[Summary of modifications to the genetic algorithm minimisation between NNPDF2.1 and NNPDF2.3]{Summary of modifications to the genetic algorithm minimisation between NNPDF2.1 and NNPDF2.3. The table on top describes the number of mutations  ($N_{\rm mut}$) and the number of generations ($N_{\rm gen}$) in the different training epochs, while the lower table shows the number of mutations per PDF and the corresponding mutation sizes. Table from \cite{Ball:2012cx}.}
  \label{tab:gapars}
\end{table}
%%%%%%%%%%%%%%%%%%%%%%%%%%%%%%%%%%%%%%%%%

Other small methodological changes included the addition of a maximum $\chi^2$ criterion, whereby replicas with a fit quality outside a $4\sigma$ band in $\chi^2$ are vetoed from the ensemble as outliers. The training/validation split used in the cross-validation was also modified for experiments with smaller than 30 data points, where as of NNPDF2.3 all of these points enter in the training set to prevent them from \emph{underlearning} or being ignored in the fit in favour of larger datasets.

With these methodological modifications, a number of determinations were performed to different datasets. Firstly the global fit was performed to the entire 2.1 dataset with the addition of the LHC data. This was followed by a `noLHC' fit which applied the methodological improvements to the same dataset as NNPDF2.1, both in order to understand the impact of the methodological modifications upon the fit and to provide a set for applications where the inclusion of an LHC dataset is undesirable. Finally a collider-only determination was performed, which excluded the older low scale fixed-target data in an attempt to reduce the effect of nuclear, higher twist and non-perturbative corrections.

\subsection{NNPDF2.3 results}

Here we shall discuss some of the results obtained in the NNPDF2.3 family of PDF determinations. Assisted by the developments in the fitting methodology, all of the 2.3 determinations were able to provide an excellent description of their included datasets. Table~\ref{tab:23chi2} details the agreement through the $\chi^2$ measure to each experiment in the analysis, for every variation of the dataset.

\begin{table}[hp]
\scriptsize
\begin{tabular}{c||c|c||c|c||c|c||c|c||c|c}
\hline 
& \multicolumn{2}{c||}{\bf NNPDF2.1} & \multicolumn{8}{|c}{\bf NNPDF2.3}  \\
\hline 
&  \multicolumn{2}{c||}{Global} &  \multicolumn{2}{|c||}{Global Fit} &
 \multicolumn{2}{|c||}{Global RW} &  \multicolumn{2}{|c||}{noLHC} &
 \multicolumn{2}{|c}{Collider}   \\
\hline 
\hline 
Experiment  & NLO & NNLO  & NLO  & NNLO  & NLO  & NNLO  & NLO  & NNLO  &
NLO  & NNLO   \\
\hline 

Total  &   1.145&   1.162 &    1.101 &    1.139 &    1.105 &    1.139 &    1.101 &    1.142 &    0.971 &    0.993 \\  
 \hline 
NMC-pd              & $          0.97      $ & $          0.93      $  &  $          0.95      $  &  $          0.95      $  &  $         0.93      $ &  $         0.93      $ &  $          0.93      $  &  $          0.94      $  &  $  \lc     5.33 \rc  $  &  $  \lc     5.13 \rc  $  \\  
NMC                 & $          1.68      $ & $          1.58      $  &  $          1.61      $  &  $          1.59      $  &  $         1.62      $ &  $         1.57      $ &  $          1.59      $  &  $          1.56      $  &  $  \lc     1.89 \rc  $  &  $  \lc     1.83 \rc  $  \\  
SLAC                & $          1.34      $ & $          1.04      $  &  $          1.24      $  &  $          1.00      $  &  $         1.27      $ &  $         1.01      $ &  $          1.28      $  &  $          1.04      $  &  $  \lc     1.72 \rc  $  &  $  \lc     1.41 \rc  $  \\  
BCDMS               & $          1.21      $ & $          1.29      $  &  $          1.20      $  &  $          1.28      $  &  $         1.20      $ &  $         1.28      $ &  $          1.20      $  &  $          1.28      $  &  $  \lc     1.85 \rc  $  &  $  \lc     2.15 \rc  $  \\  
CHORUS              & $          1.10      $ & $          1.08      $  &  $          1.10      $  &  $          1.07      $  &  $         1.10      $ &  $         1.06      $ &  $          1.09      $  &  $          1.07      $  &  $  \lc     1.73 \rc  $  &  $  \lc     1.70 \rc  $  \\  
NTVDMN              & $          0.70      $ & $          0.50      $  &  $          0.43      $  &  $          0.56      $  &  $         0.42      $ &  $         0.51      $ &  $          0.42      $  &  $          0.48      $  &  $  \lc    26.69 \rc  $  &  $  \lc    21.13 \rc  $  \\  
 \hline
HERAI-AV            & $          1.04      $ & $          1.04      $  &  $          1.00      $  &  $          1.01      $  &  $         1.00      $ &  $         1.02      $ &  $          1.01      $  &  $          1.03      $  &  $          0.97      $  &  $          0.99      $  \\  
FLH108              & $          1.34      $ & $          1.23      $  &  $          1.29      $  &  $          1.20      $  &  $         1.29      $ &  $         1.20      $ &  $          1.29      $  &  $          1.21      $  &  $          1.35      $  &  $          1.25      $  \\  
ZEUS-H2             & $          1.21      $ & $          1.21      $  &  $          1.20      $  &  $          1.22      $  &  $         1.20      $ &  $         1.22      $ &  $          1.20      $  &  $          1.22      $  &  $          1.29      $  &  $          1.32      $  \\  
ZEUS $F_2^c$        & $          0.75      $ & $          0.81      $  &  $          0.82      $  &  $          0.90      $  &  $         0.80      $ &  $         0.90      $ &  $          0.81      $  &  $          0.86      $  &  $          0.71      $  &  $          0.77      $  \\  
H1 $F_2^c$          & $          1.50      $ & $          1.44      $  &  $          1.59      $  &  $          1.53      $  &  $         1.57      $ &  $         1.52      $ &  $          1.58      $  &  $          1.49      $  &  $          1.33      $  &  $          1.30      $  \\  
 \hline
DYE605              & $          0.94      $ & $          1.08      $  &  $          0.86      $  &  $          1.04      $  &  $         0.88      $ &  $         1.04      $ &  $          0.85      $  &  $          1.06      $  &  $  \lc     3.58 \rc  $  &  $  \lc     1.02 \rc  $  \\  
DYE886              & $          1.42      $ & $          1.69      $  &  $          1.27      $  &  $          1.58      $  &  $         1.27      $ &  $         1.55      $ &  $          1.24      $  &  $          1.55      $  &  $  \lc     5.65 \rc  $  &  $  \lc     5.14 \rc  $  \\  
 \hline
CDF $W$ asy         & $          1.88      $ & $          1.63      $  &  $          1.57      $  &  $          1.64      $  &  $         1.57      $ &  $         1.72      $ &  $          1.45      $  &  $          1.67      $  &  $          1.05      $  &  $          1.21      $  \\  
CDF $Z$ rap         & $          1.77      $ & $          2.38      $  &  $          1.80      $  &  $          2.03      $  &  $         1.77      $ &  $         2.17      $ &  $          1.76      $  &  $          2.13      $  &  $          1.32      $  &  $          1.37      $  \\  
D0 $Z$ rap          & $          0.57      $ & $          0.67      $  &  $          0.56      $  &  $          0.61      $  &  $         0.57      $ &  $         0.63      $ &  $          0.57      $  &  $          0.63      $  &  $          0.56      $  &  $          0.58      $  \\  
ATLAS $W,Z$         & $  \lc     1.57 \rc  $ & $  \lc     2.21 \rc  $  &  $          1.26      $  &  $          1.43      $  &  $         1.31      $ &  $         1.65      $ &  $  \lc     1.37 \rc  $  &  $  \lc     1.94 \rc  $  &  $          1.02      $  &  $          1.05      $  \\  
CMS $W$ el asy      & $  \lc     2.02 \rc  $ & $  \lc     1.27 \rc  $  &  $          0.82      $  &  $          0.81      $  &  $         1.09      $ &  $         0.99      $ &  $  \lc     1.32 \rc  $  &  $  \lc     1.20 \rc  $  &  $          0.87      $  &  $          0.85      $  \\  
LHCb $W$            & $  \lc     0.89 \rc  $ & $  \lc     1.13 \rc  $  &  $          0.67      $  &  $          0.83      $  &  $         0.77      $ &  $         0.98      $ &  $  \lc     0.76 \rc  $  &  $  \lc     1.03 \rc  $  &  $          0.74      $  &  $          0.72      $  \\  
 \hline
CDF RII $k_T$       & $          0.68      $ & $          0.65      $  &  $          0.60      $  &  $          0.68      $  &  $         0.61      $ &  $         0.67      $ &  $          0.60      $  &  $          0.67      $  &  $          0.60      $  &  $          0.59      $  \\  
D0 RII cone         & $          0.90      $ & $          0.98      $  &  $          0.84      $  &  $          0.94      $  &  $         0.84      $ &  $         0.93      $ &  $          0.84      $  &  $          0.94      $  &  $          0.85      $  &  $          0.92      $  \\  
ATLAS jets          & $  \lc     1.06 \rc  $ & $  \lc     0.95 \rc  $  &  $          1.00      $  &  $          0.94      $  &  $         1.00      $ &  $         0.92      $ &  $  \lc     1.01 \rc  $  &  $  \lc     0.94 \rc  $  &  $          0.98      $  &  $          0.93      $  \\

\hline
\end{tabular}

\caption[Fit quality in the NNPDF2.3 family of PDF determinations]{\small \label{tab:estfit2dataset}  The fit quality to each individual dataset in the global NNPDF2.3 determination provided by various NNPDF sets. The global, noLHC and collider only 2.3 determinations are shown along with the NNPDF2.1 values for comparison. Additionally the values for a reweighting of 2.1 with LHC data is shown in order to test the efficacy of the fitting procedure. The figures in square brackets are for datasets that were not included in the associated PDF set. }
\label{tab:23chi2}
\end{table}

The total $\chi^2$ values achieved by the global fits were $1.101$ at NLO and $1.139$ at NNLO, both indicating fine agreement with the experimental data and demonstrating improvement over the fit quality obtained in the NNPDF2.1 series. The noLHC fits obtained similar levels of fit quality, while the collider only determinations demonstrated the excellent consistency in the dataset with $\chi^2$ values of 0.971 and 0.993 for the NLO and NNLO fits respectively.

Notably the collider only dataset fails to describe the older fixed-target data, particularly the NuTeV dimuon measurements, by a large margin. A $\chi^2$ value of $26.69$ at NLO to the NuTeV data suggests that the collider-only dataset may be in some tension with the older, low scale measurements. Despite this the global fit is able to provide a good description of both the collider only and fixed target data simultaneously, therefore any tension present between the datasets remains at the moment compatible within experimental errors.

\begin{figure}[h!]
\centering
\includegraphics[width=0.48\textwidth]{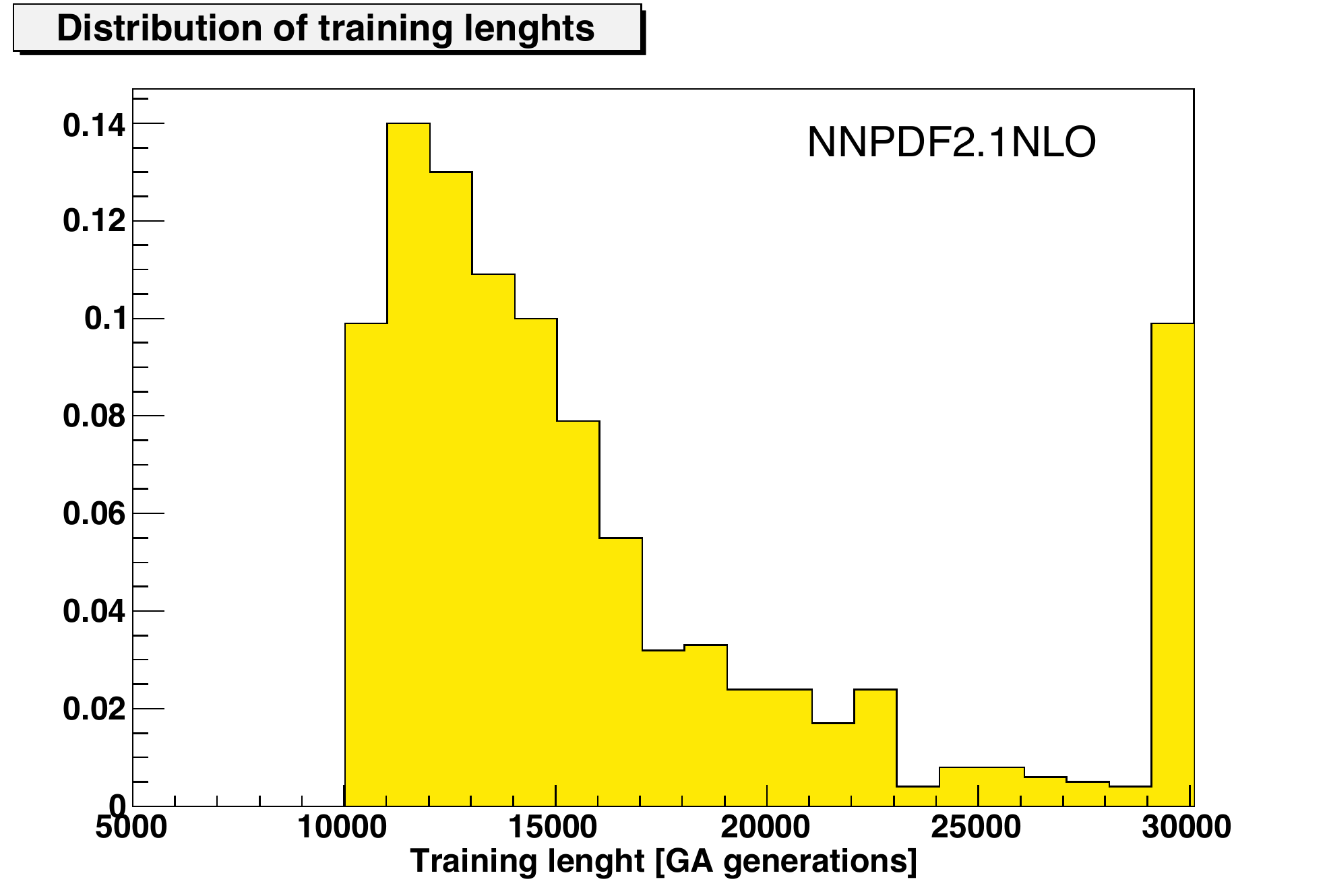}
\includegraphics[width=0.48\textwidth]{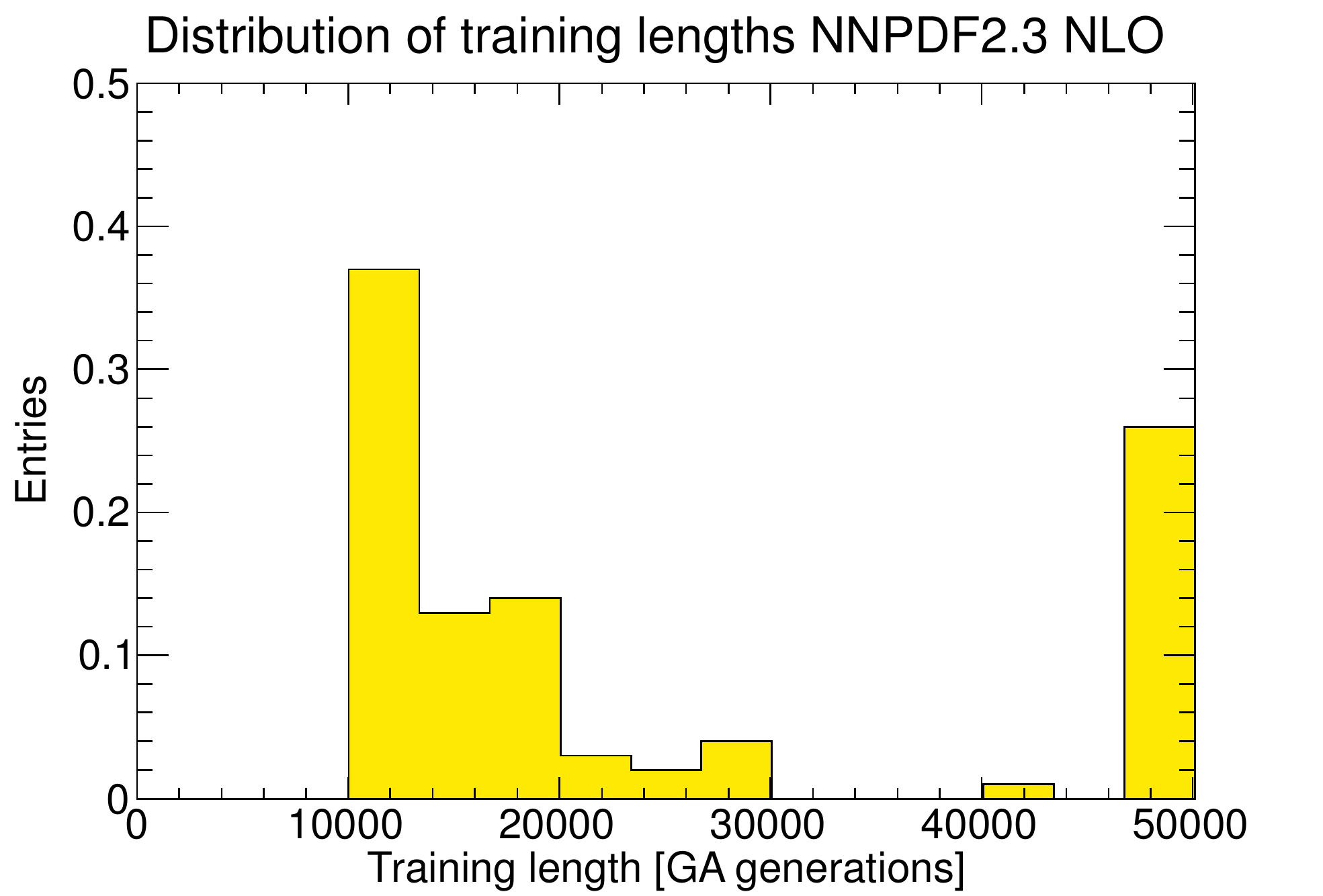}
\caption[Replica training lengths in NNPDF2.1 and NNPDF2.3]{Replica training lengths in NNPDF2.1 NLO (left) and NNPDF2.3 NLO (right). These histograms display the relative frequency of replicas stopping in bins along the full training length. In NNPDF2.3 both the maximum number of generations was increased to 50,000, and the criteria governing the replica stopping was tightened, causing more replicas to stop later. } 
\label{fig:tlcomp}
\end{figure}

The average training length at NLO is predictably extended in 2.3 over 2.1. The more stringent stopping condition leading to more replicas running for the extended maximum $N_{\text{gen}} = 50,000$. The training length comparison is shown in Figure~\ref{fig:tlcomp}.

We shall now examine the changes between the NNPDF2.1 and NNPDF2.3 distributions at the level of PDFs. Firstly discussing the impact of the methodological changes to the NNPDF determination by examining the NNPDF2.3 noLHC fits, before moving on to look at the direct impact of the LHC dataset by performing comparisons of the noLHC and full 2.3 datasets. Finally we shall discuss the impact of the LHC data upon a collider only determination. The issue of the strange content of the proton is a particularly delicate one, and therefore will be discussed separately from the other five light quark distributions.

\subsubsection{NNPDF2.3 noLHC}
\begin{figure}[hp!]
\centering
\includegraphics[width=0.48\textwidth]{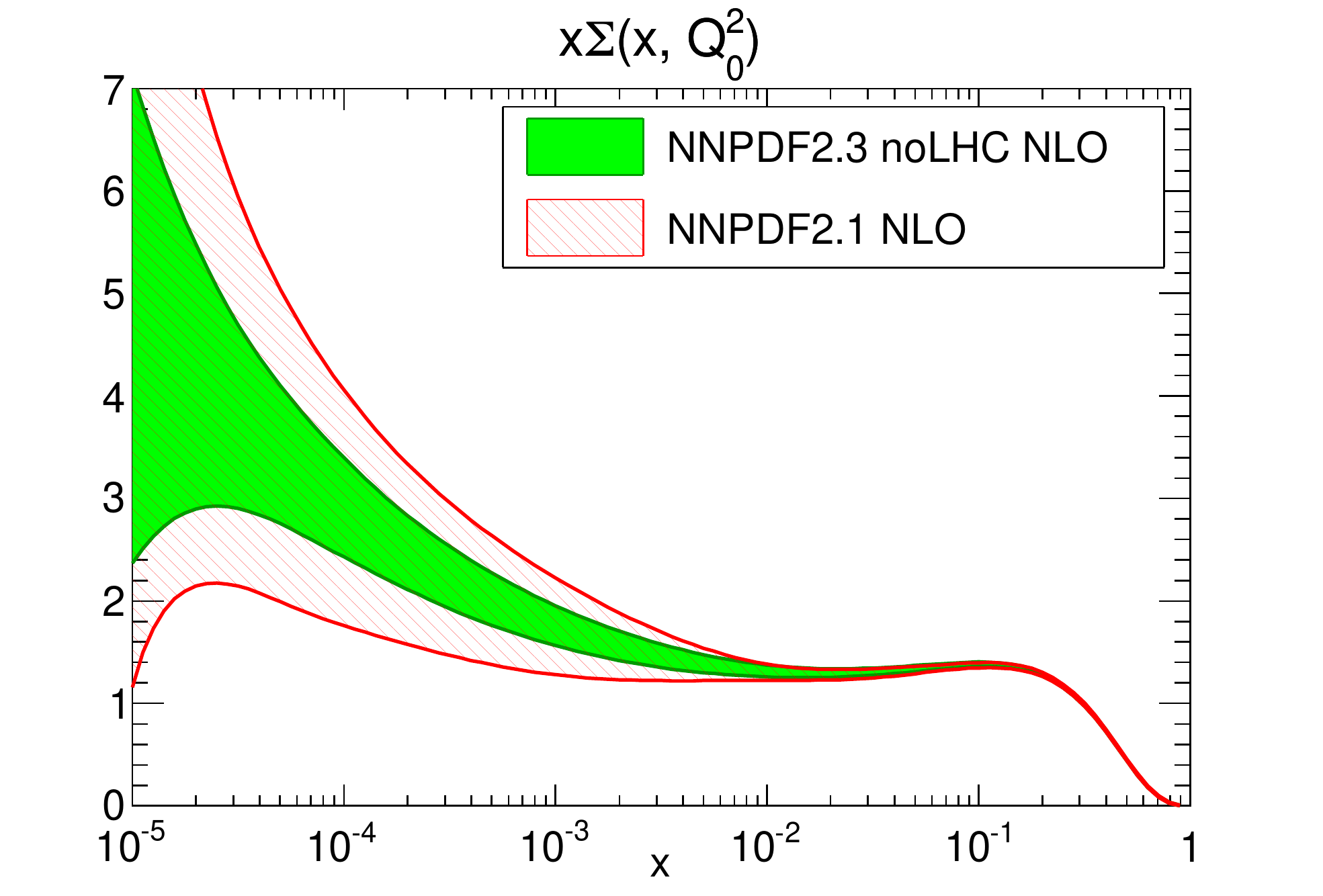}
\includegraphics[width=0.48\textwidth]{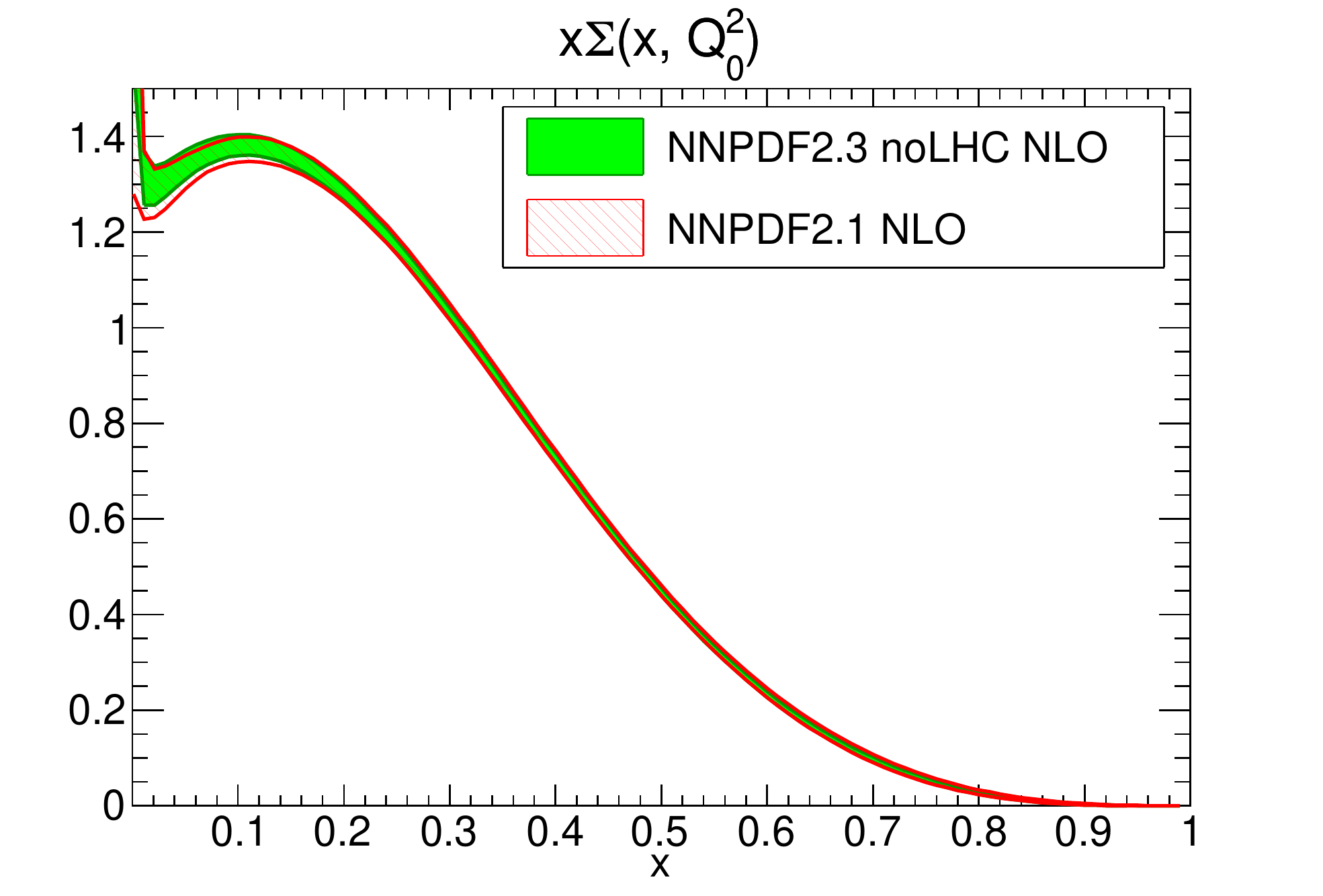}\\
\includegraphics[width=0.48\textwidth]{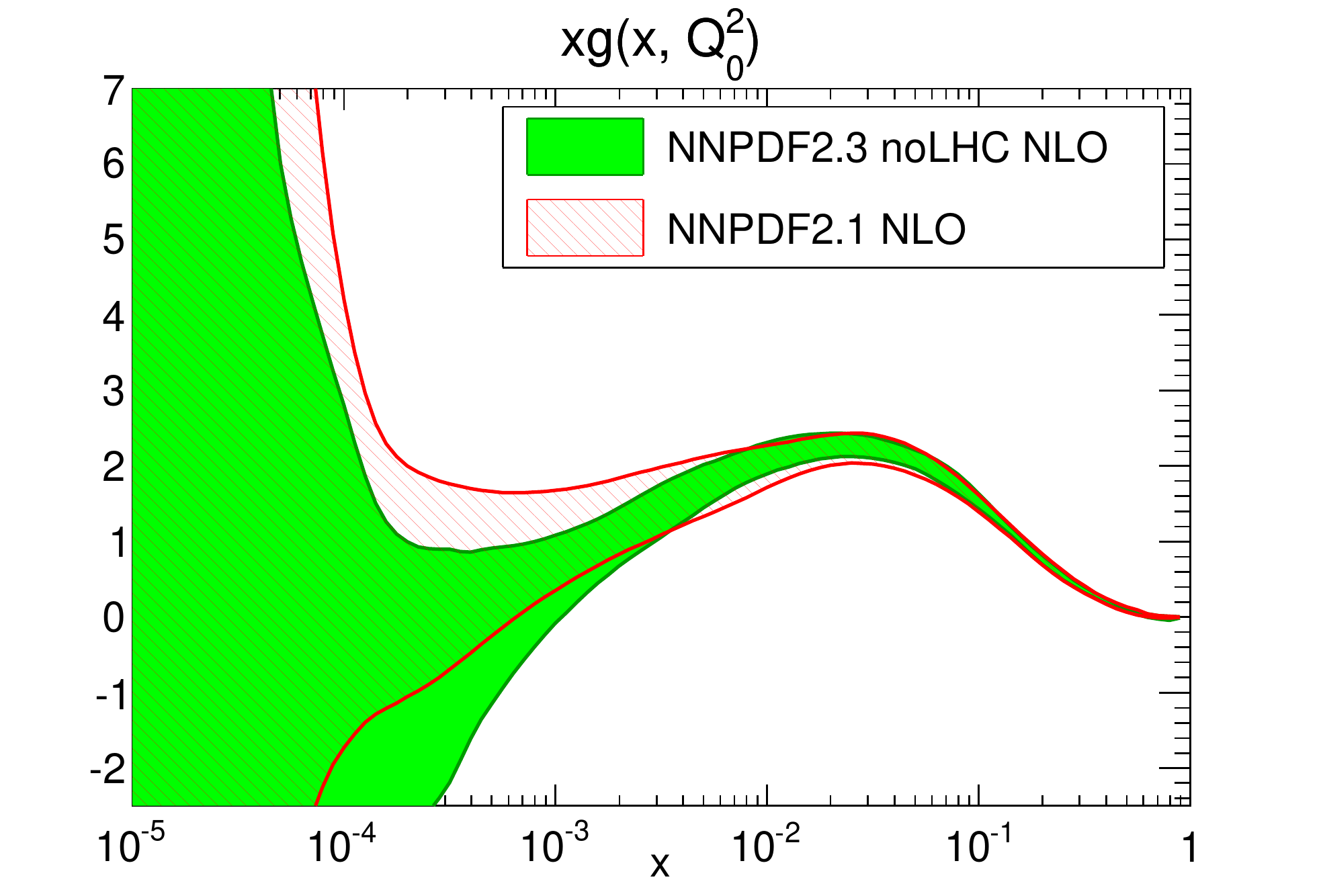}
\includegraphics[width=0.48\textwidth]{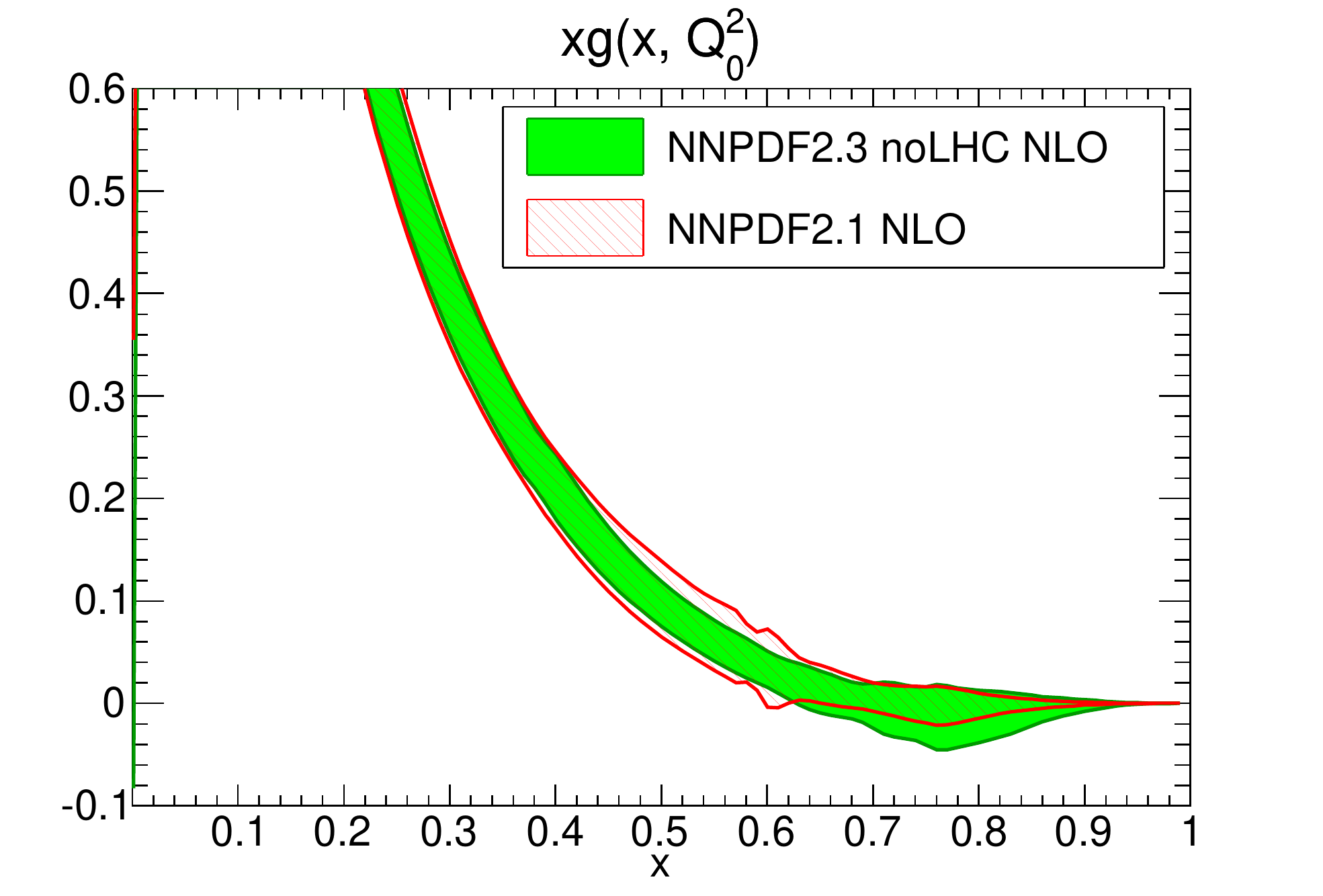}

\noindent\rule[0.5ex]{\linewidth}{1pt}

\includegraphics[width=0.48\textwidth]{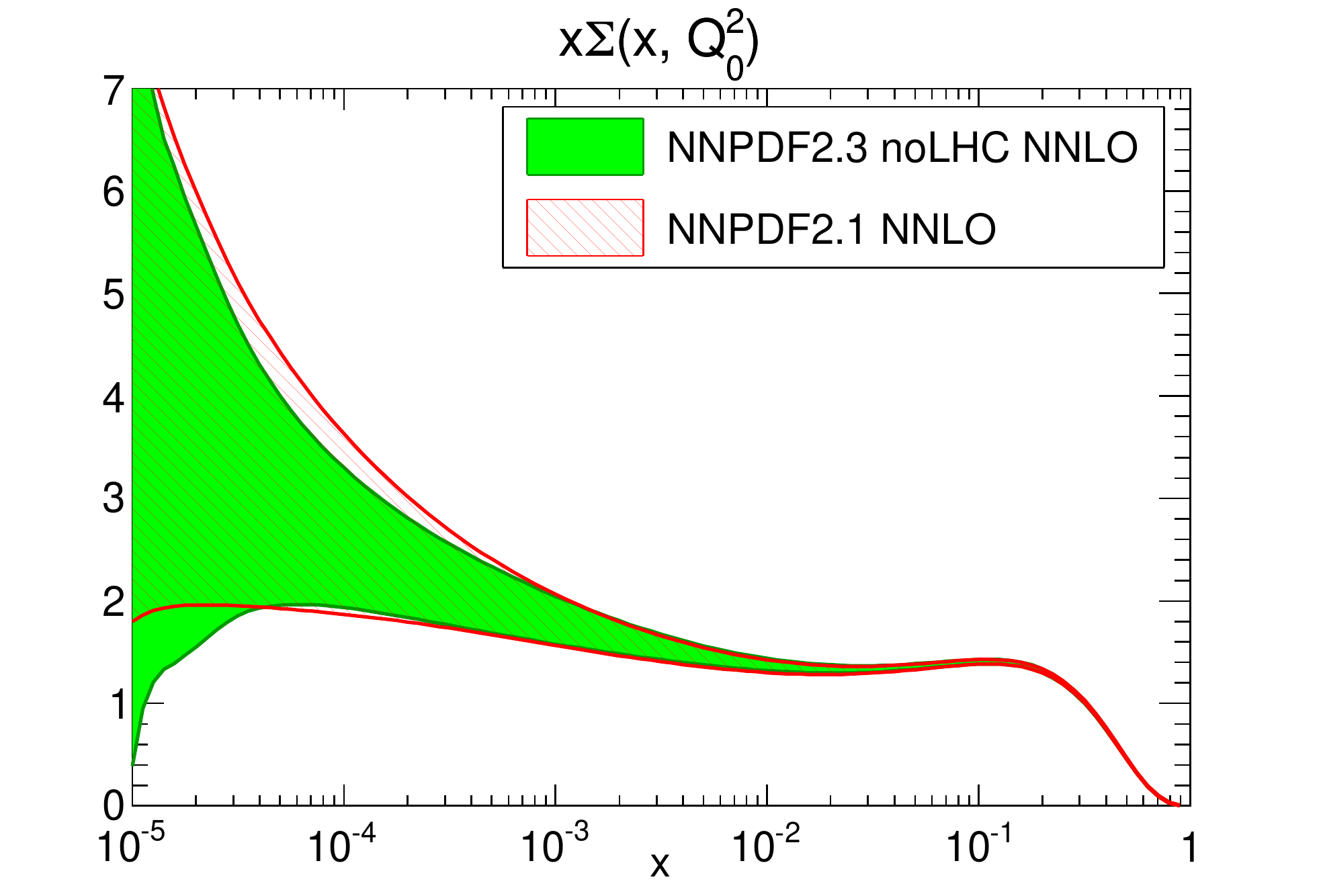}
\includegraphics[width=0.48\textwidth]{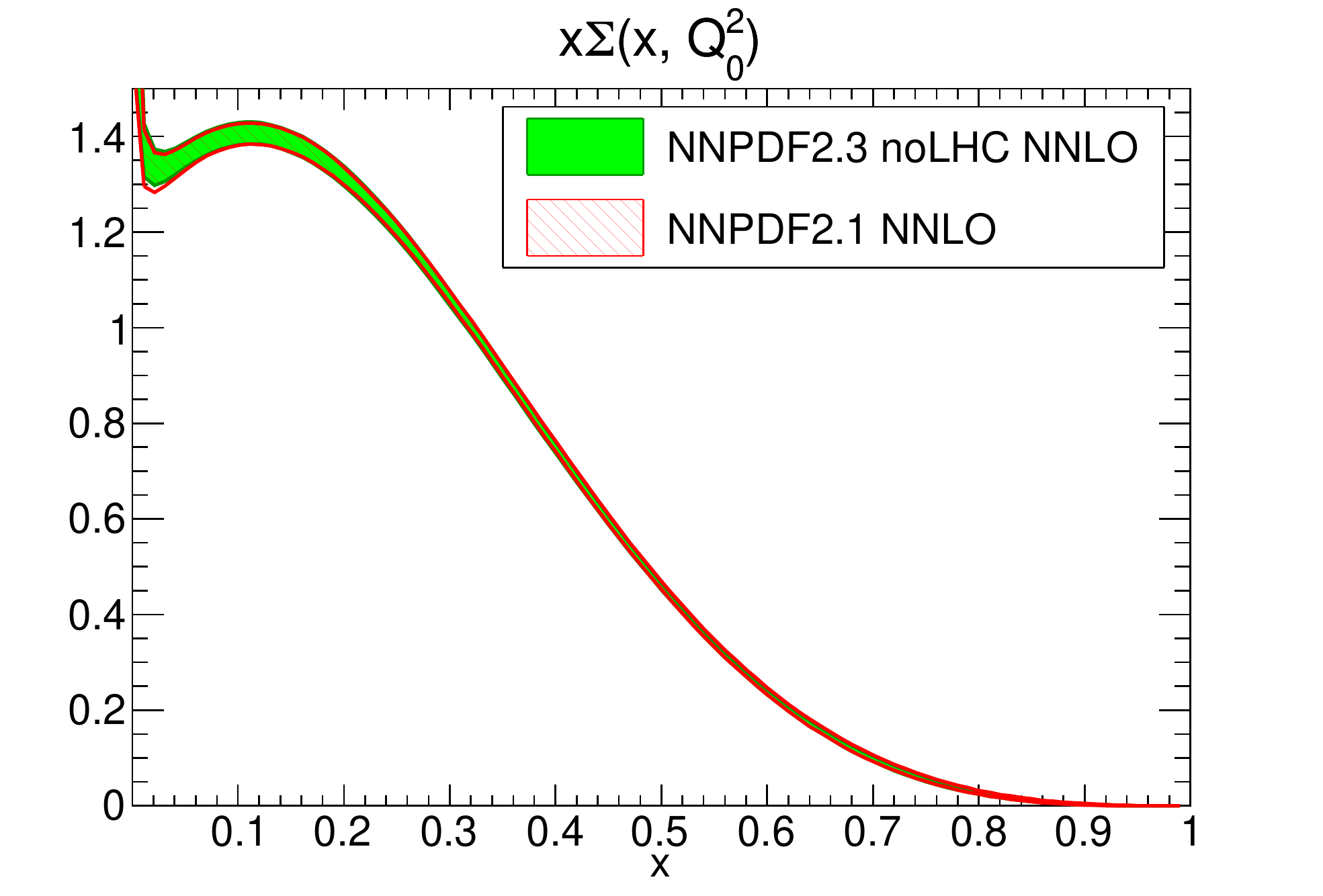}\\
\includegraphics[width=0.48\textwidth]{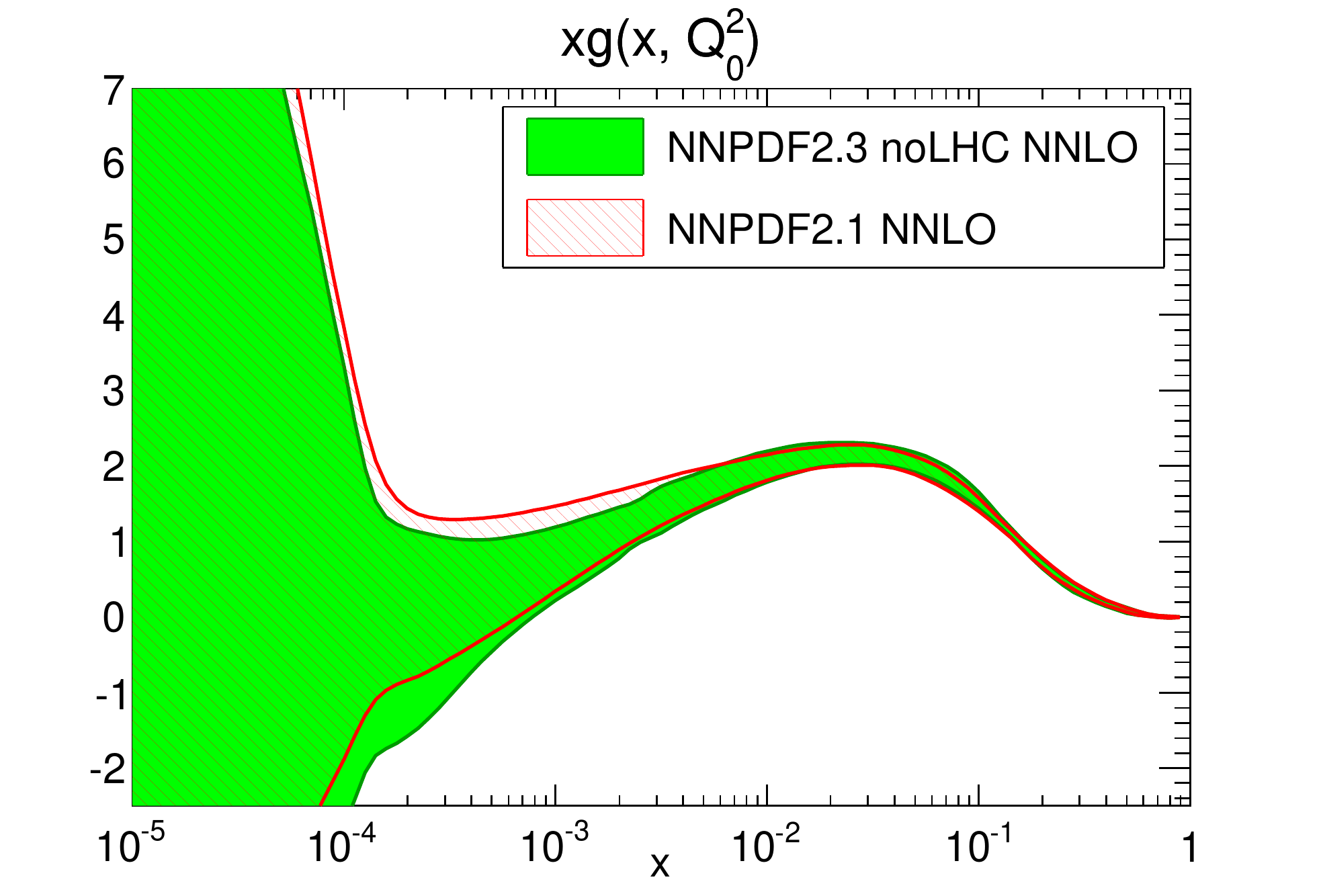}
\includegraphics[width=0.48\textwidth]{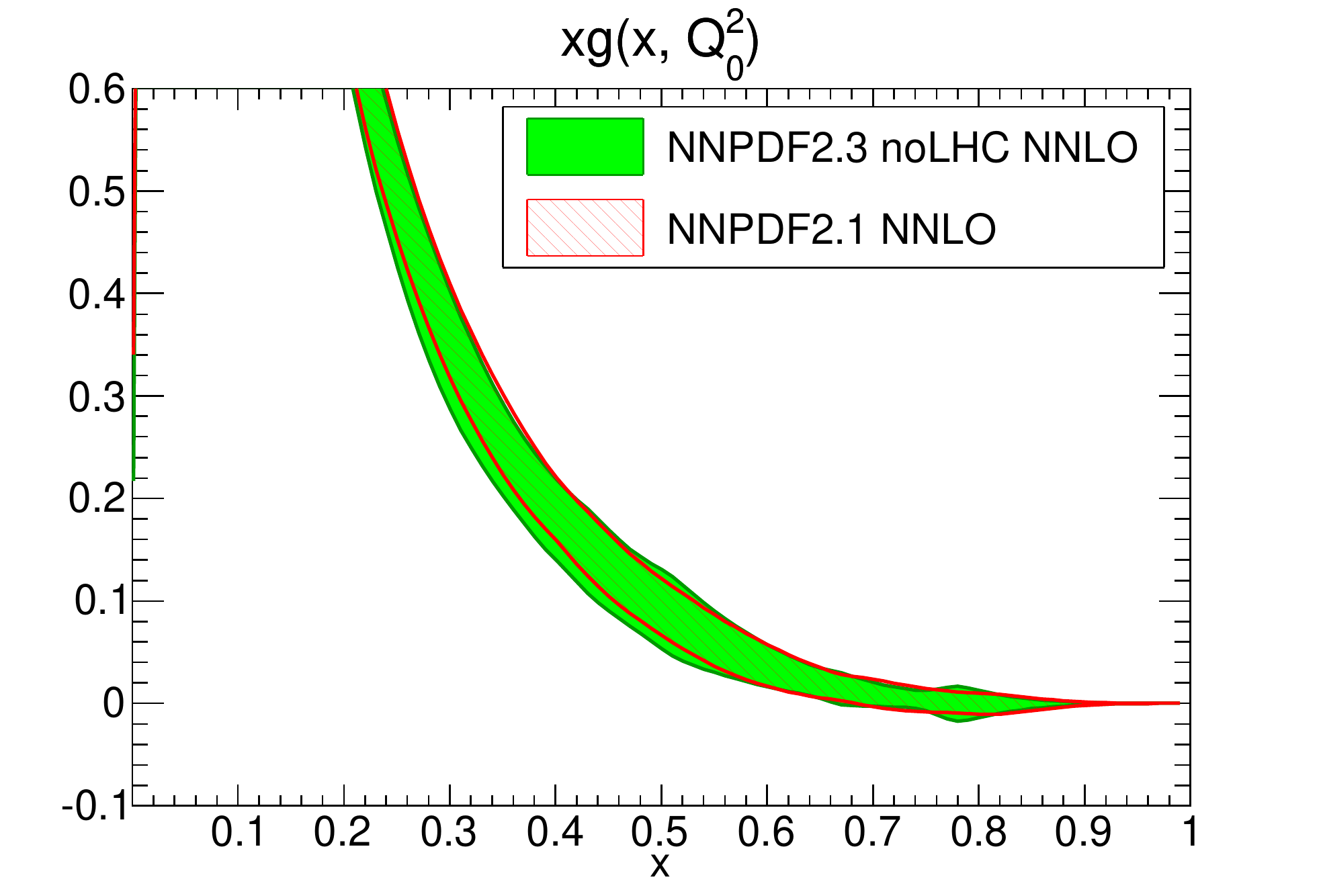}
\caption[The impact of the improved methodology in NNPDF2.3 against NNPDF2.1 in the gluon and singlet sectors]{The impact of the improved methodology in NNPDF2.3 against NNPDF2.1 in the gluon and singlet sectors for the NLO (top) and NNLO (bottom) distributions. The red curves show the results of NNPDF2.1 while the green curves show NNPDF2.3 noLHC. Figures on the left are shown in a logarithmic scale in $x$.}
\label{fig:23noLHC}
\end{figure}

The NNPDF2.3 noLHC set has two primary uses. To understand the improvements made in the NNPDF methodology by applying the updated procedure to the older dataset, and for applications such as BSM searches at the LHC where a dataset without the influence of LHC data may be desirable. Here we shall directly compare the 2.3 noLHC results with NNPDF2.1 to see the methodological improvement. These improvements were expected to be clearer in the NLO PDF sets, as for NNPDF2.1 NNLO several of the improvements in the minimisation were already implemented. Aside from the strange sector (which will be discussed later), the methodological changes largely only impact the gluon and singlet distributions, with other distributions undergoing small changes, so we shall restrict ourselves here to comparisons of the gluon and singlet PDFs. The upper section of Figure~\ref{fig:23noLHC} compares NNPDF2.1 and NNPDF2.3 noLHC at NLO, for those PDFs most affected by the improvements; the singlet and gluon. The clearest improvements are in the low-$x$ region, where the combination of more aggressive minimisation and tighter stopping criteria lead to substantially smaller uncertainty in the singlet, and a moderate shift in the gluon. These improvements suggests that there was potentially a degree of underlearning present in the small-$x$ region of NNPDF2.1 generated by stopping too early.

The lower part of Figure~\ref{fig:23noLHC} demonstrates the same comparison for the NNLO determination. From this figure it is clear that the degree of underlearning present in the NLO fit was avoided by the use of the updated fit settings, leading to slightly narrower uncertainty bands. The relatively insignificant differences remaining due to the presence of more data in the training sets, although the difference remains statistically insignificant at the level of PDFs.

\subsubsection{NNPDF2.3 global}

\begin{figure}[hp!]
\centering
\includegraphics[width=0.48\textwidth]{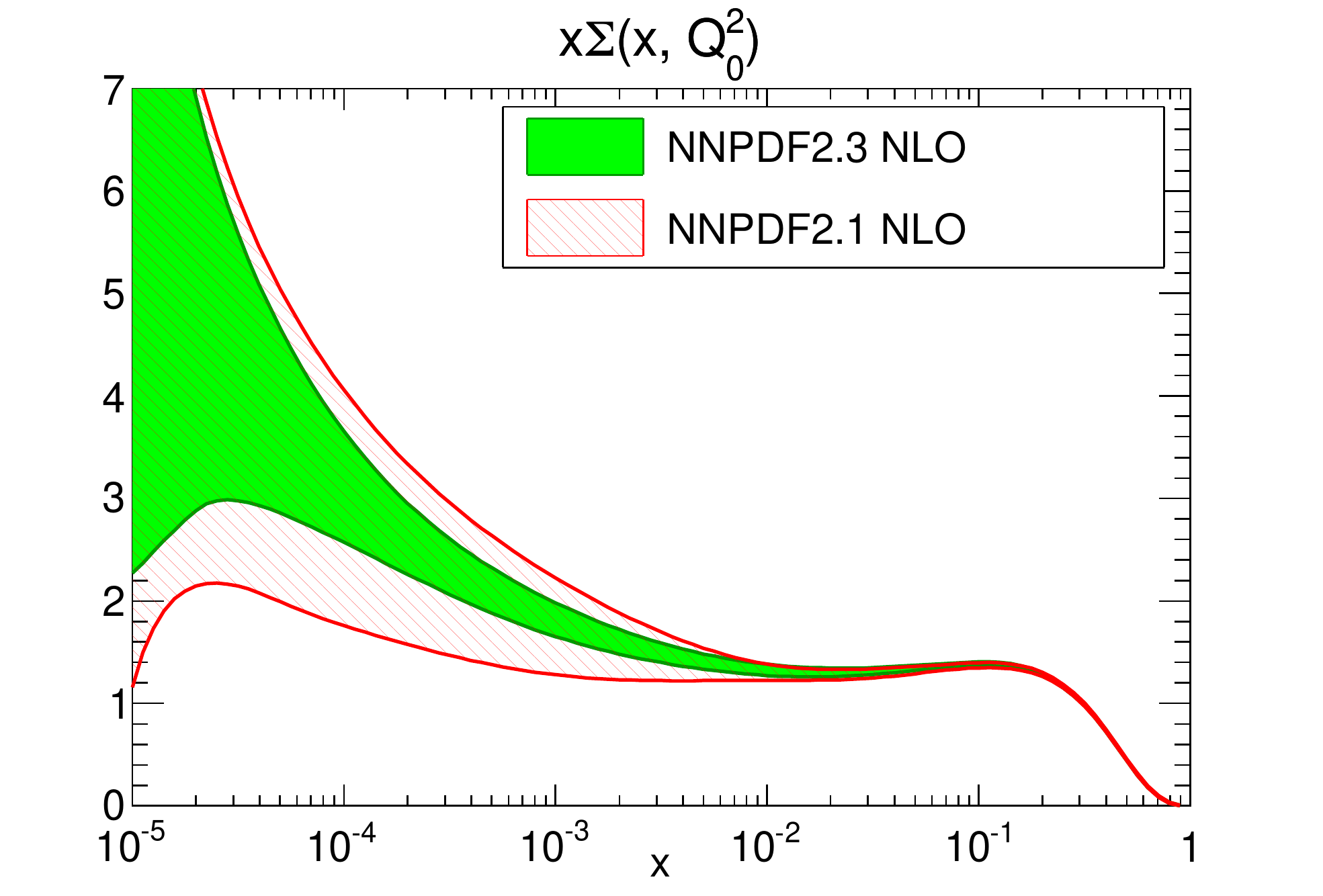}
\includegraphics[width=0.48\textwidth]{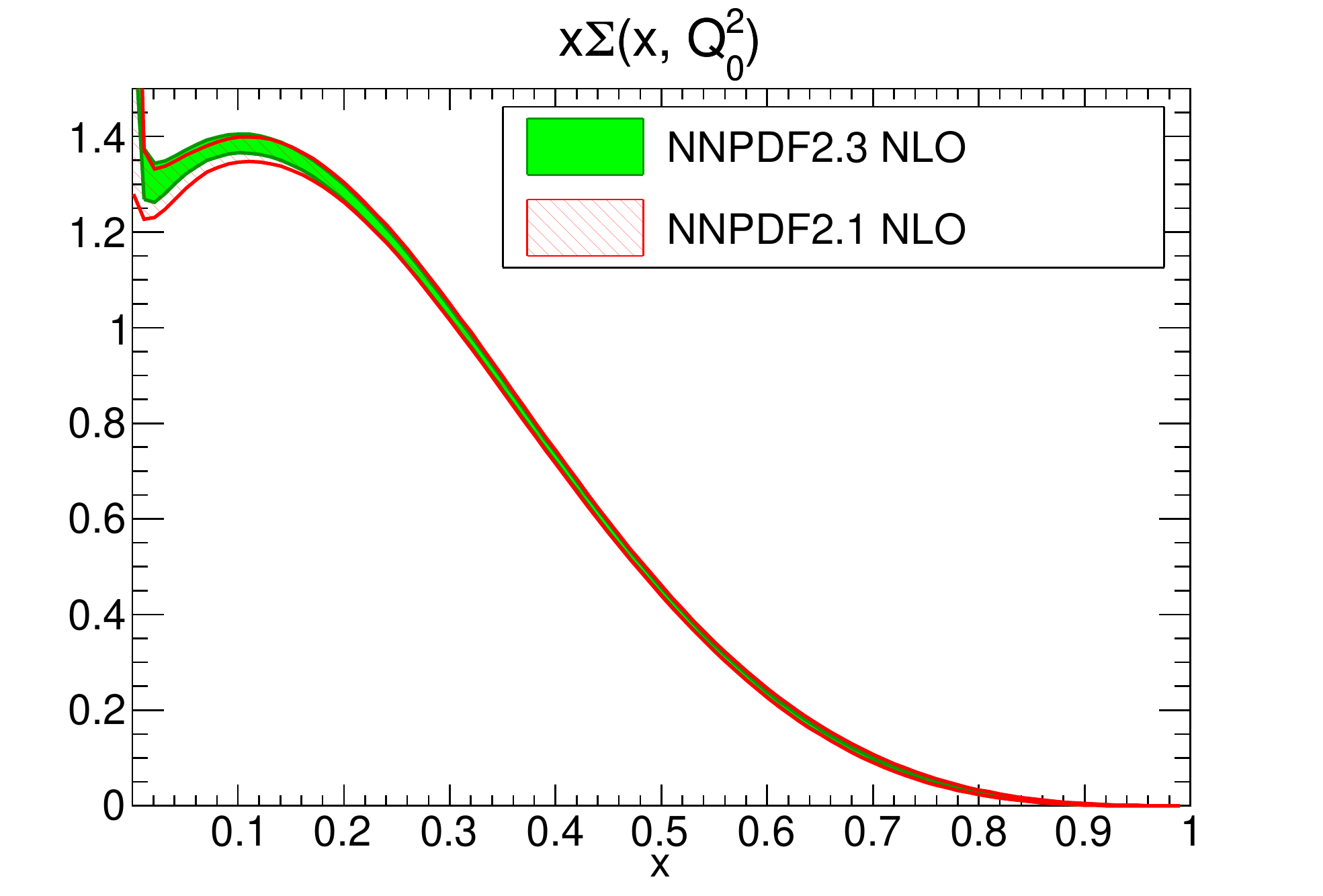}\\
\includegraphics[width=0.48\textwidth]{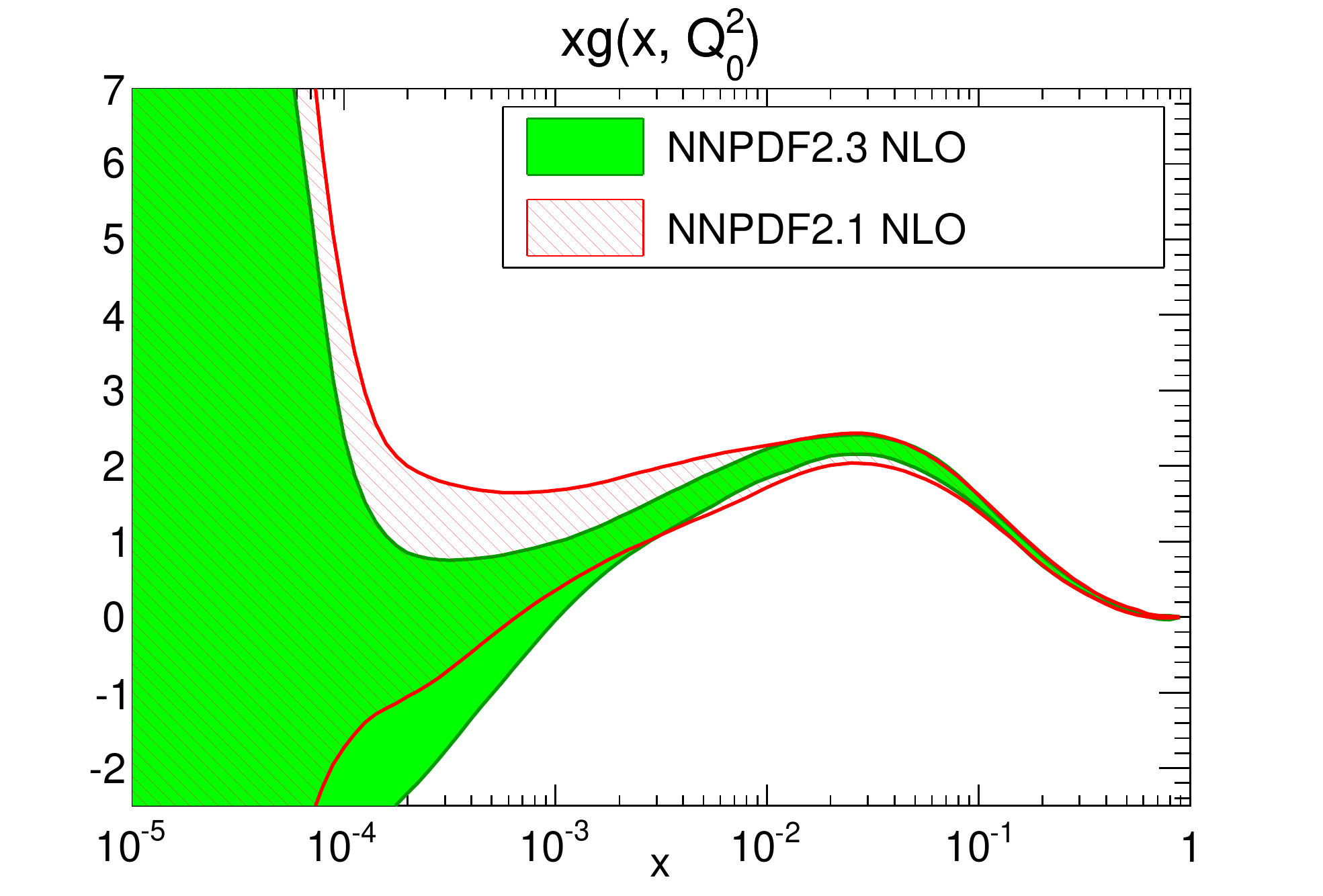}
\includegraphics[width=0.48\textwidth]{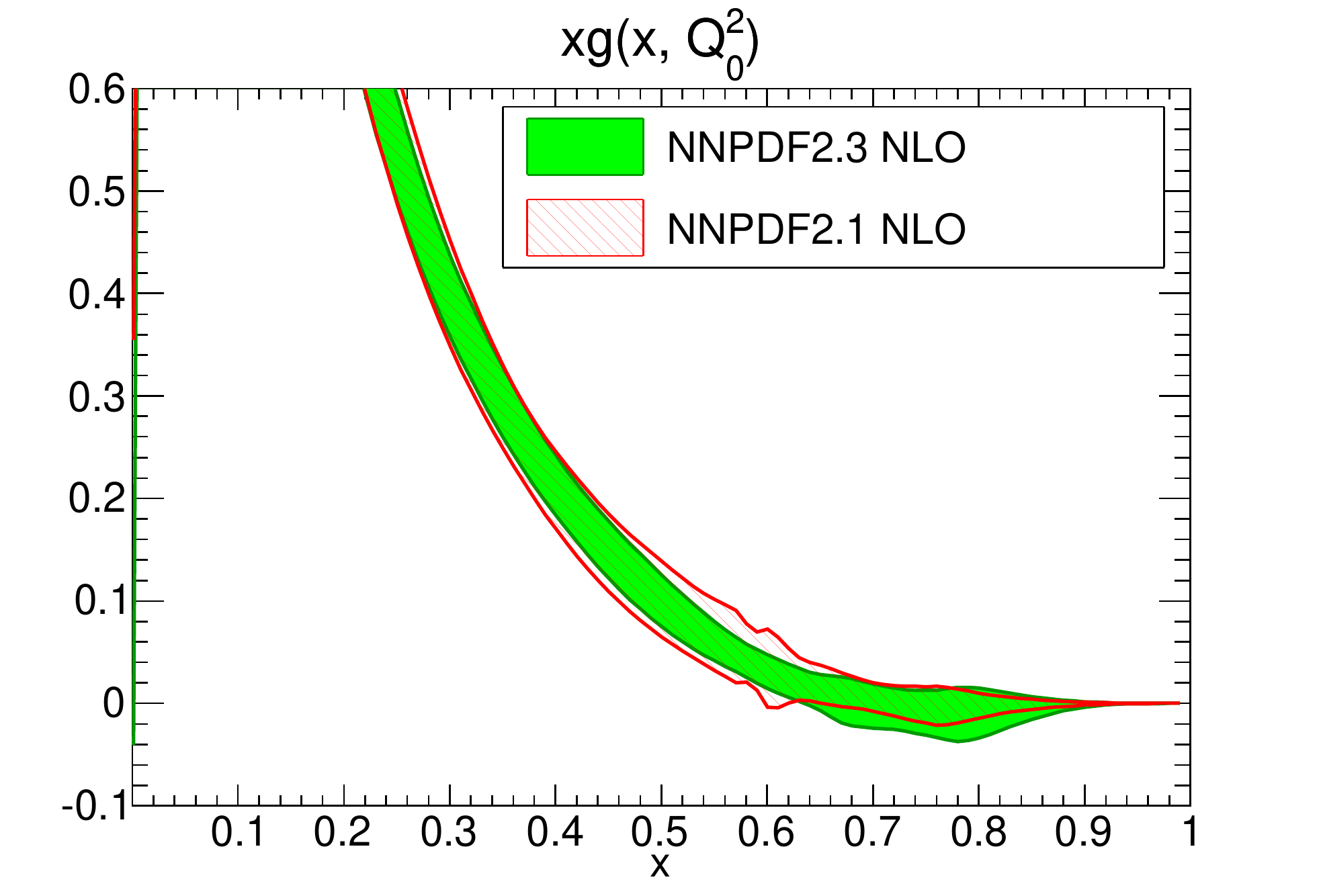}

\noindent\rule[0.5ex]{\linewidth}{1pt}

\includegraphics[width=0.48\textwidth]{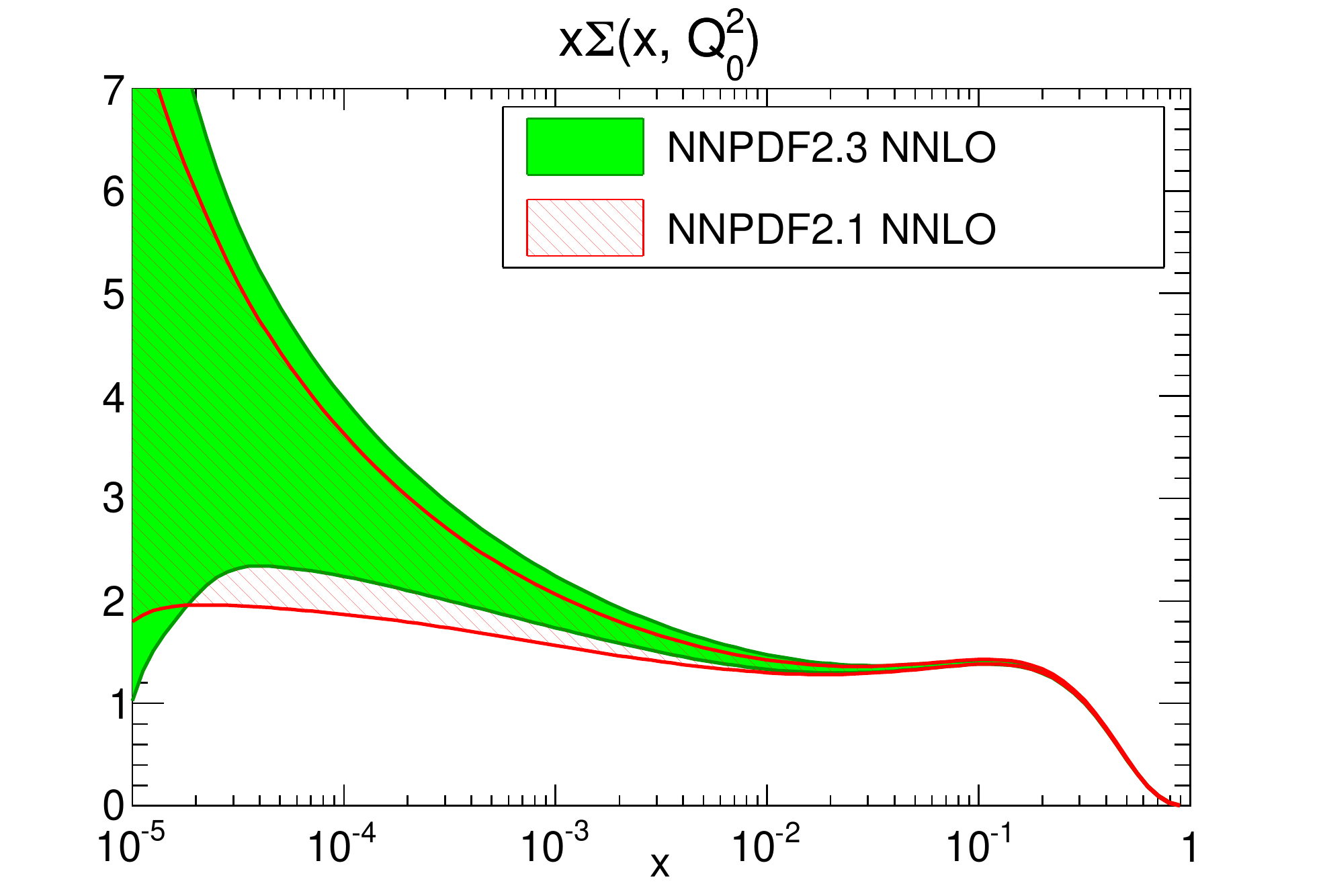}
\includegraphics[width=0.48\textwidth]{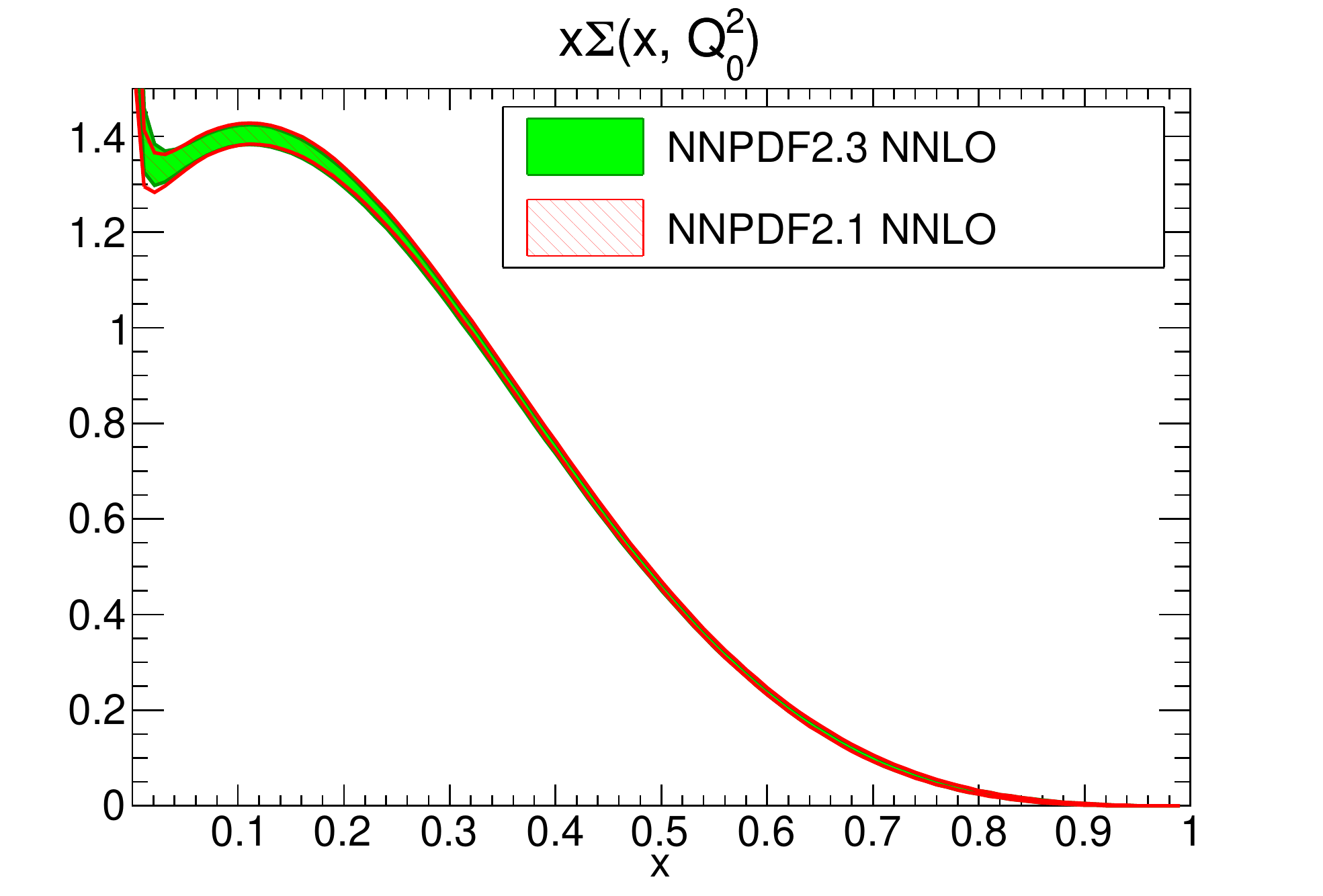}\\
\includegraphics[width=0.48\textwidth]{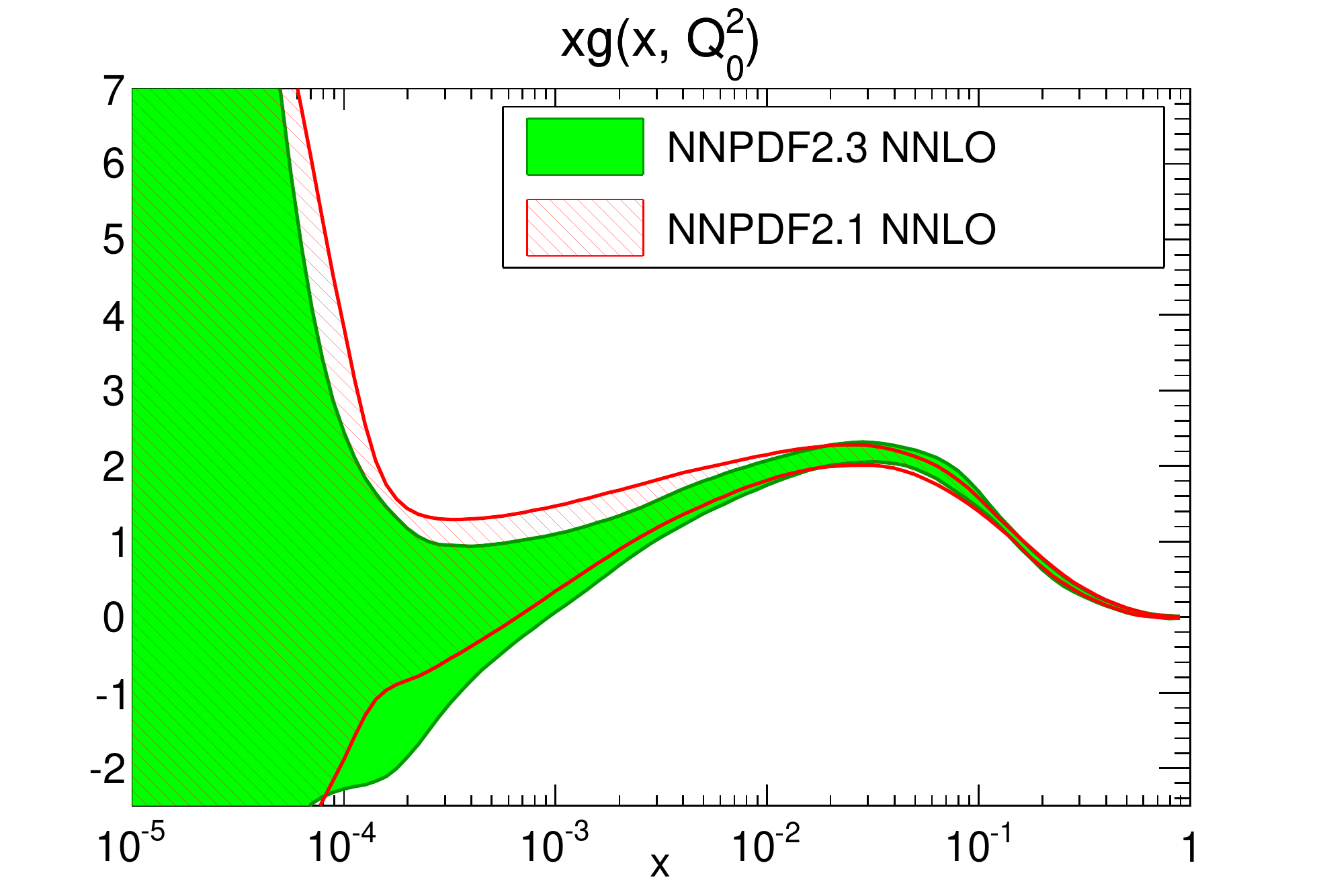}
\includegraphics[width=0.48\textwidth]{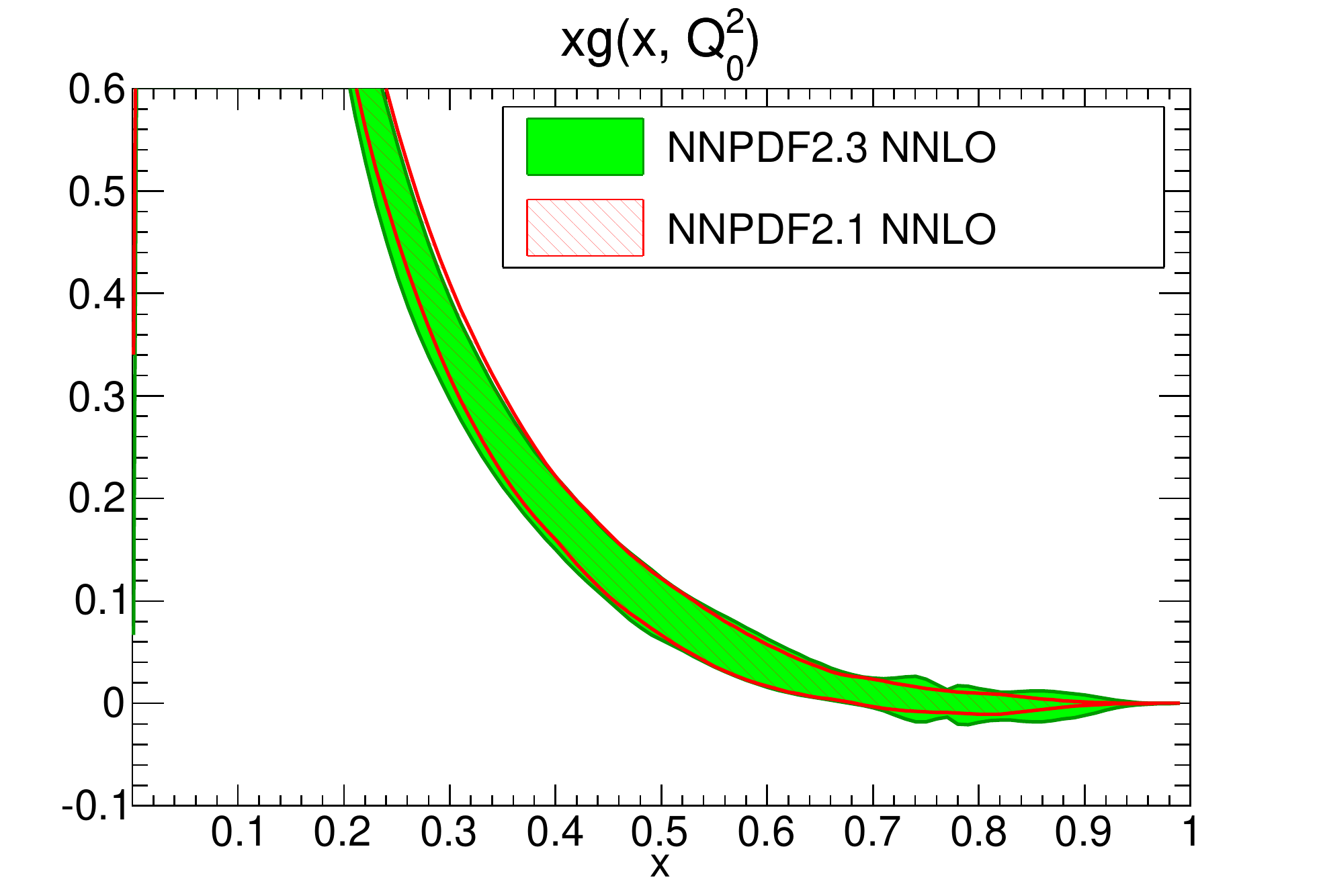}
\caption[A comparison of the NNPDF2.1 and NNPDF2.3 gluon and singlet global determinations at NLO and NNLO]{A comparison of the NNPDF2.1 and NNPDF2.3 global determinations at NLO (top) and NNLO (bottom). The gluon and singlet distributions are shown, with a logarithmic $x$ scale on the left, and linear on the right. Red distributions are those given by the NNPDF2.1 set, while green represent NNPDF2.3.}
\label{fig:23vs21}
\end{figure}

The NNPDF2.3 global set includes all of the methodological improvements along with the constraints from the new LHC dataset. There are therefore comprehensive improvements available in the 2.3 set over 2.1, both at NLO and NNLO in QCD. Figure~\ref{fig:23vs21} shows the same comparison as in Figure~\ref{fig:23noLHC}, but including the impact of the LHC data by comparing NNPDF2.1 to the full global NNPDF2.3 set. As much of the improvements are driven by methodology, the largest modifications in the global comparison can also be found in the gluon and singlet distributions. To obtain a clearer view of the impact of the LHC data upon the PDFs we can compare the 2.3 noLHC fit with the global determination, with the only differences in the two sets due to the LHC data. 

\begin{figure}[hp!]
\centering
\includegraphics[width=0.48\textwidth]{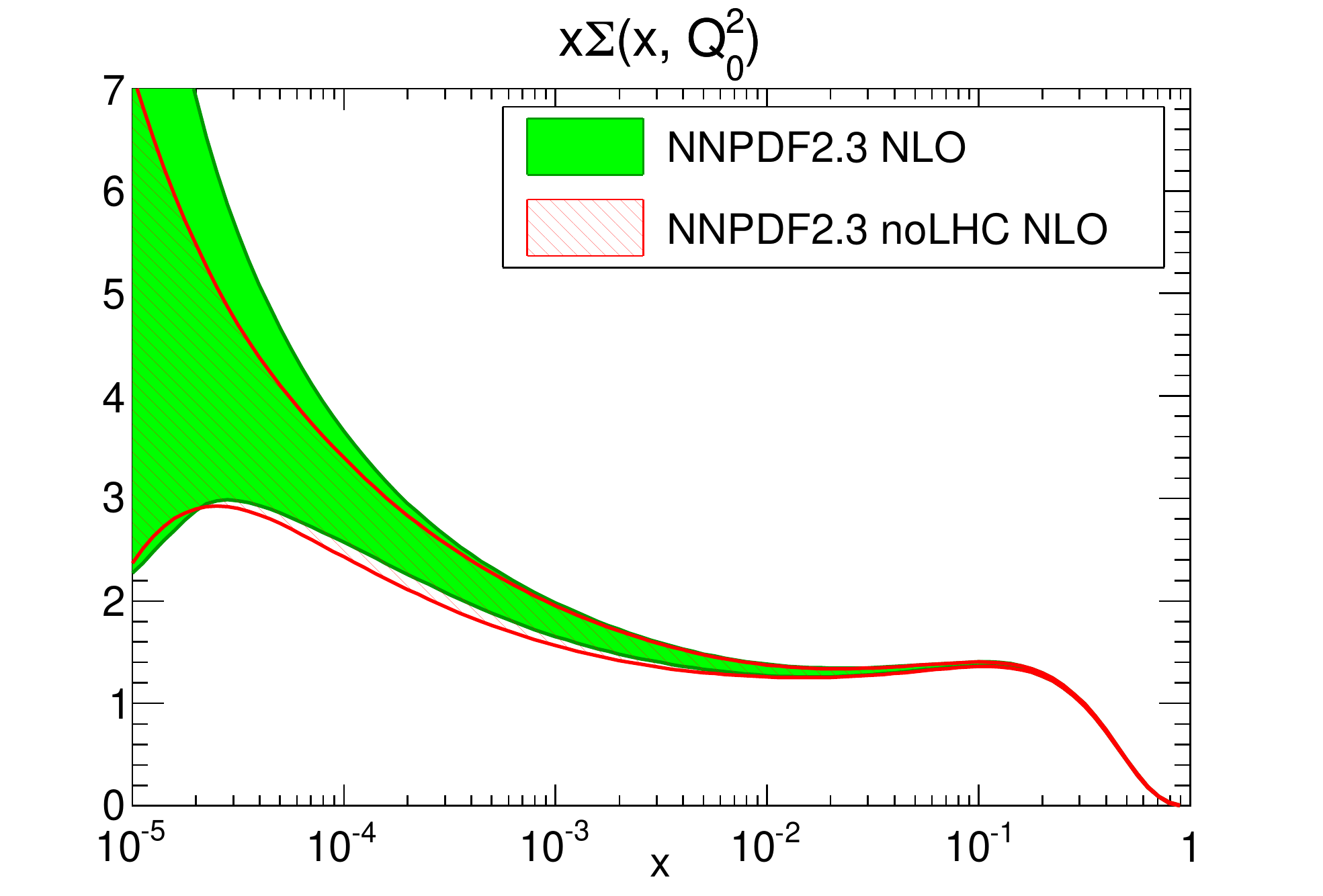}
\includegraphics[width=0.48\textwidth]{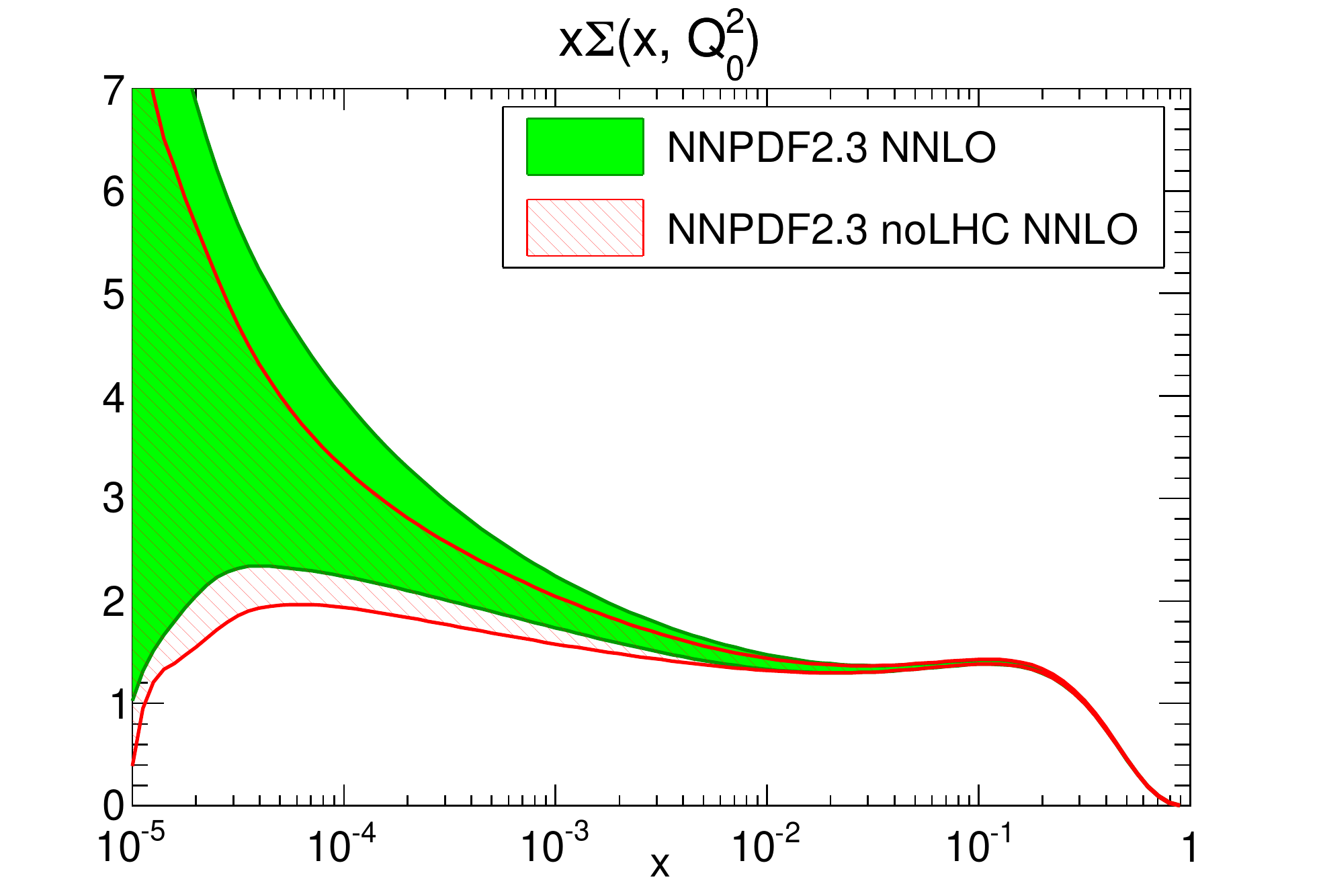}\\
\includegraphics[width=0.48\textwidth]{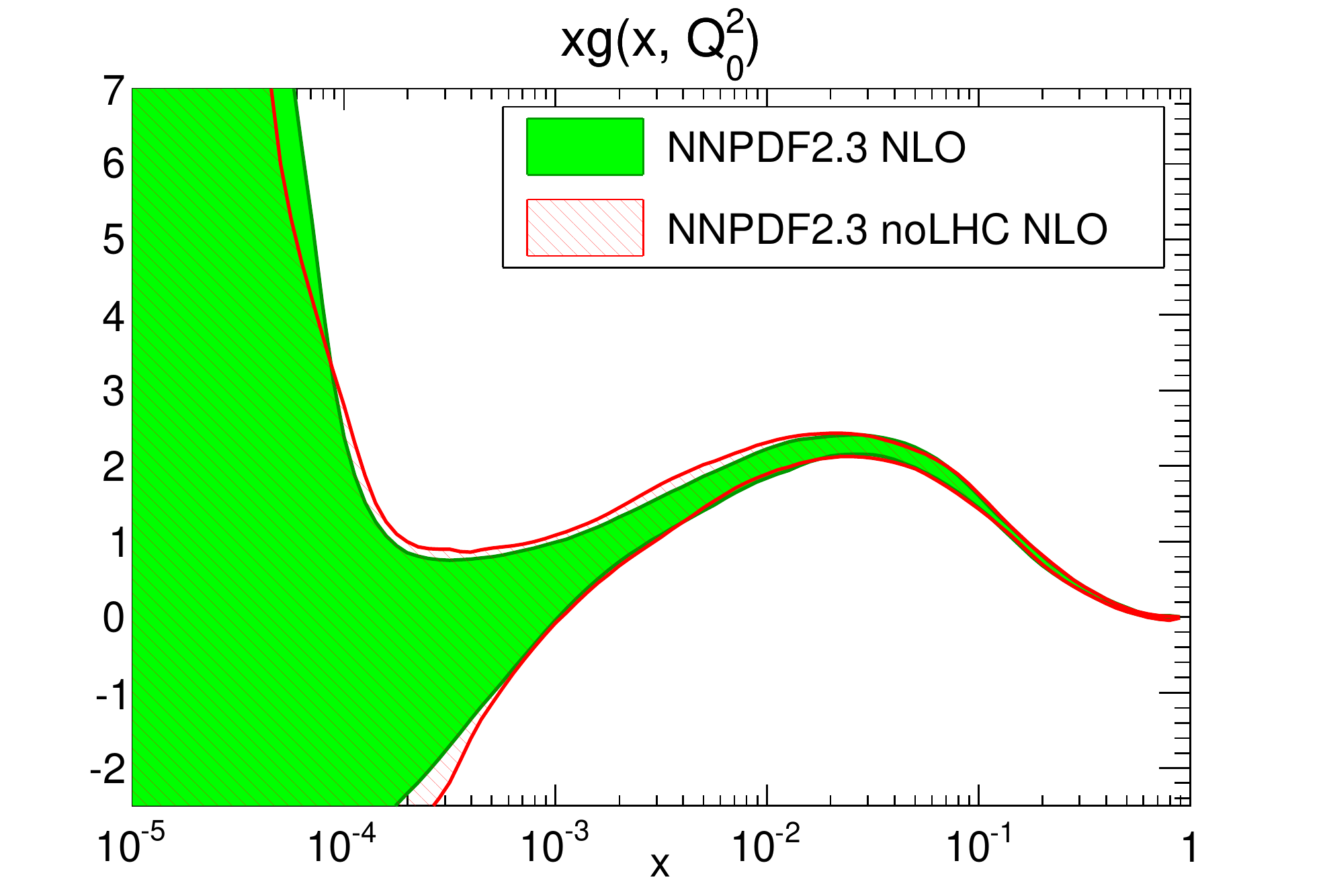}
\includegraphics[width=0.48\textwidth]{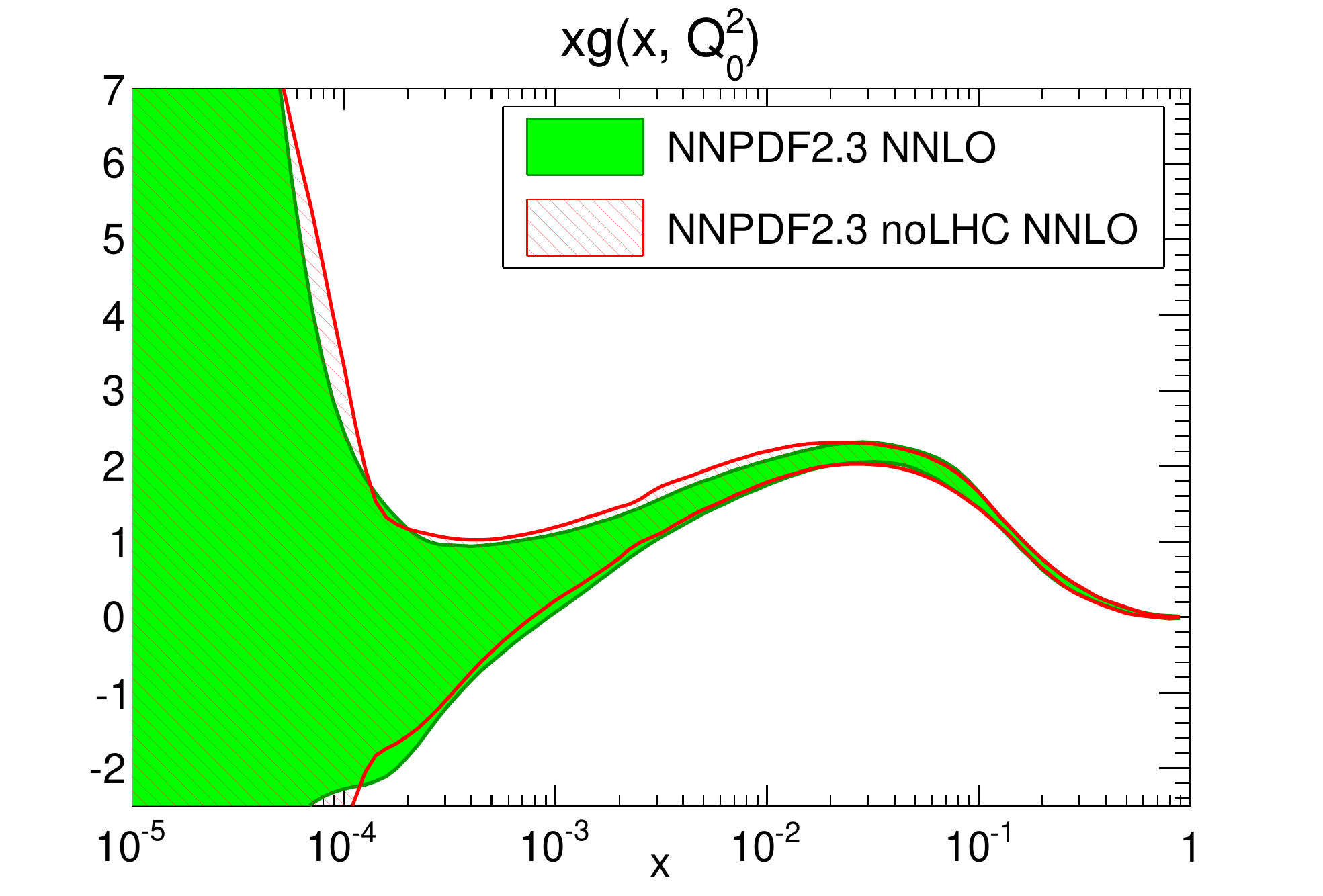} \\
\includegraphics[width=0.48\textwidth]{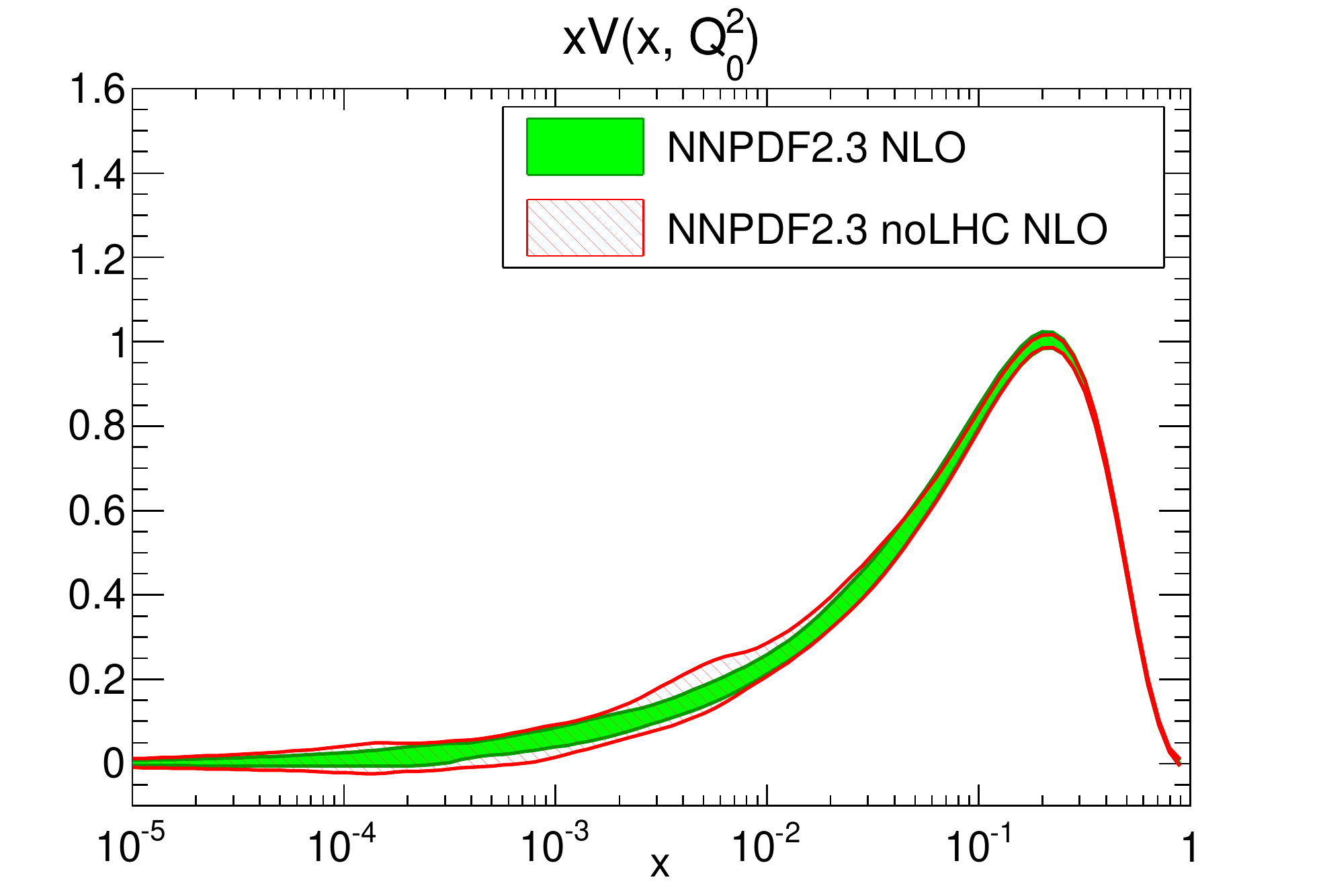}
\includegraphics[width=0.48\textwidth]{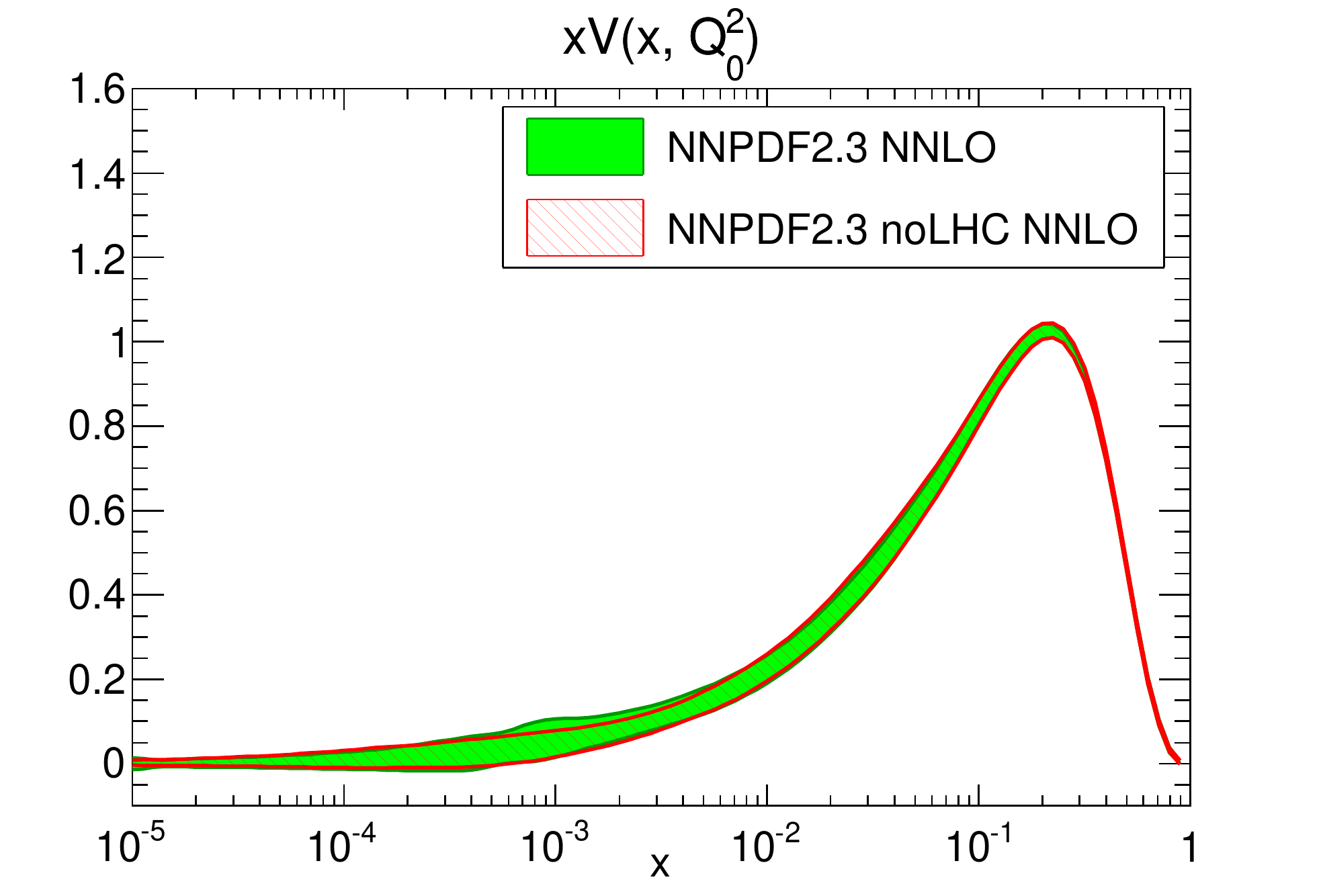}
\caption[Comparison of NNPDF2.3 and NNPDF2.3 noLHC at NLO and NLO for the singlet, gluon and valence distributions]{Comparison of NNPDF2.3 and NNPDF2.3 noLHC at NLO (left) and NLO (right) for the singlet (top), gluon (middle) and valence (bottom) distributions. The figures therefore show directly the influence of the LHC data in the fit.}
\label{fig:23vs23noLHC}
\end{figure}

In comparing the 2.3 global and noLHC sets, the clearest improvements can be found in the singlet, gluon and valence sectors as would be expected from the expanded dataset. Figure~\ref{fig:23vs23noLHC} compares these distributions at NLO and NNLO to study the influence of the new data. In the singlet sector, the LHC data prefers a rather higher value for the PDF in the small-$x$ region, with the central value being systematically higher below $x\sim 0.1$, an effect which is clearer at NNLO. In the NLO fit there is a broadening of uncertainties for the extrapolation region $x < 10^{-4}$, but a moderate degree of uncertainty reduction in the data region. For the NNLO singlet the uncertainties are larger over a broad kinematic range, generated by the larger upwards shift preferred by the LHC data at NNLO.

The gluon distribution at NLO enjoys a great deal of consistency between the 2.3 noLHC and 2.3 global fits. With the additional LHC data contributing to a broad reduction of uncertainties in the data region. The NNLO fit, while demonstrating a good deal of consistency, does not make any significant reduction in uncertainty outside the region of $x\sim10^{-2}$.

The PDF benefiting the most from the inclusion of the LHC data is the NLO valence distribution, where significant reductions in uncertainty are achieved across a wide kinematic range. Despite this improvement, the NNLO fit is not able to make such significant gains on the basis of the new data, with improvements constrained to the moderate to large-$x$ region.

\begin{figure}[h!]
\centering
\includegraphics[width=0.48\textwidth]{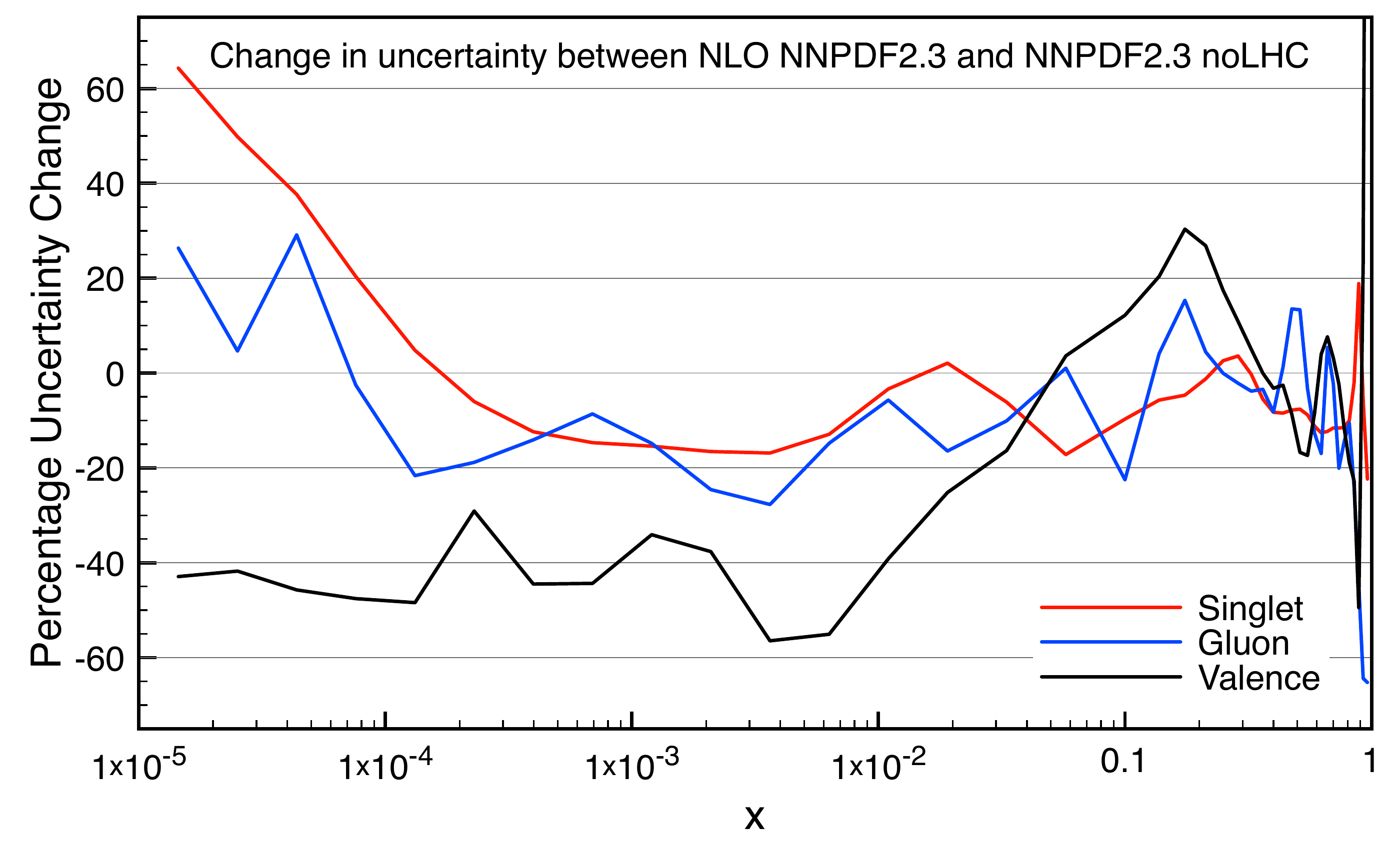}
\includegraphics[width=0.48\textwidth]{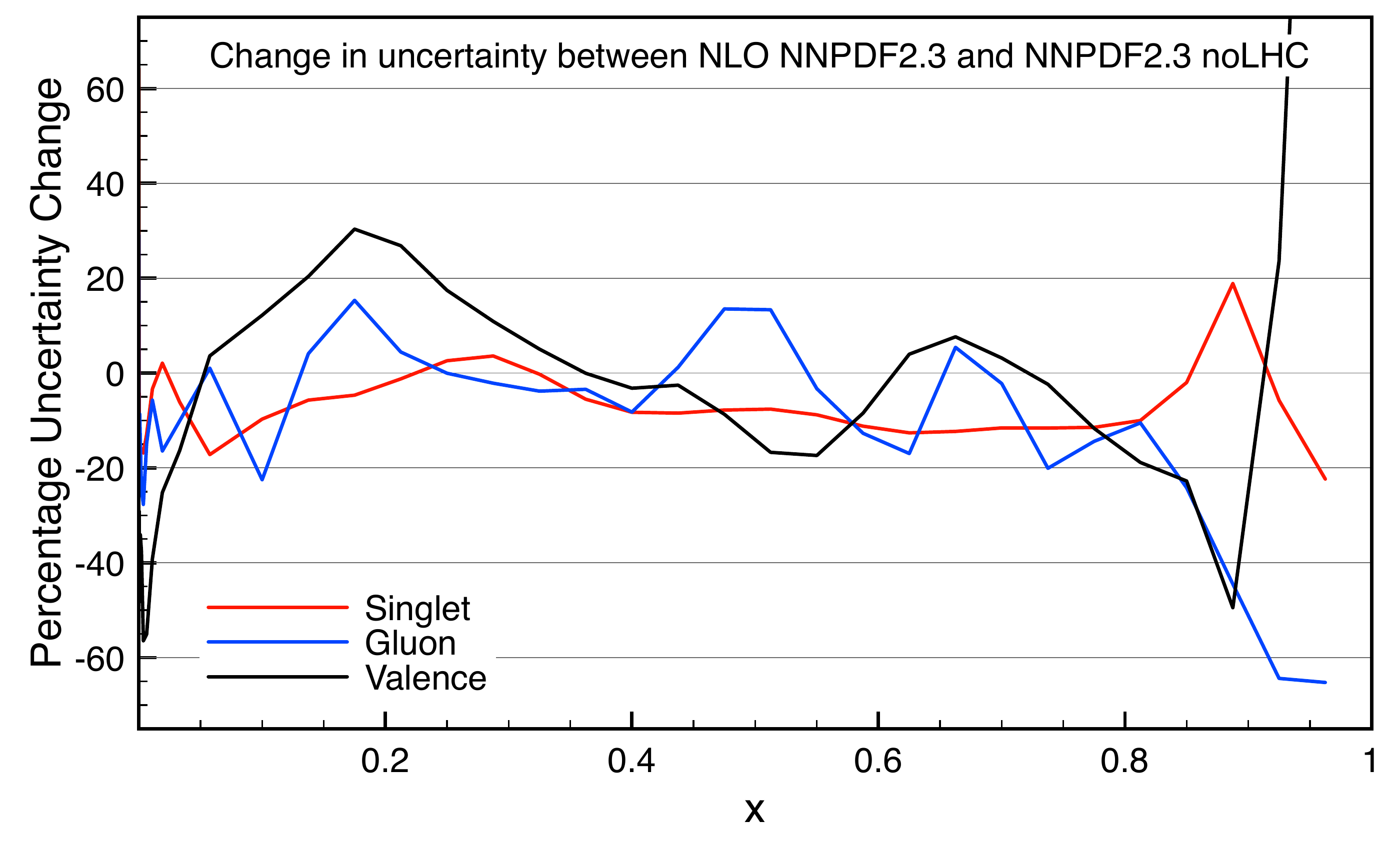}\\
\includegraphics[width=0.48\textwidth]{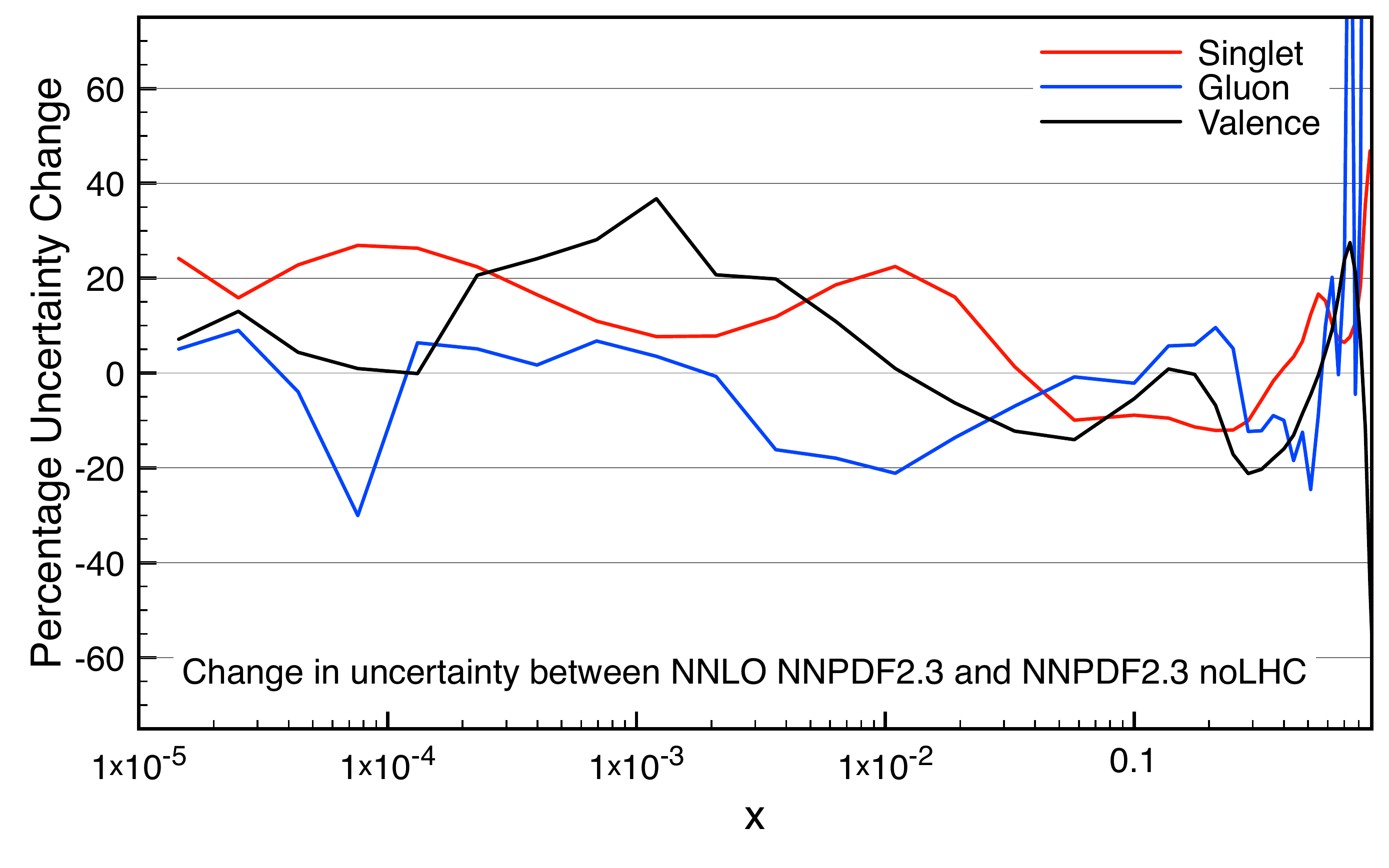}
\includegraphics[width=0.48\textwidth]{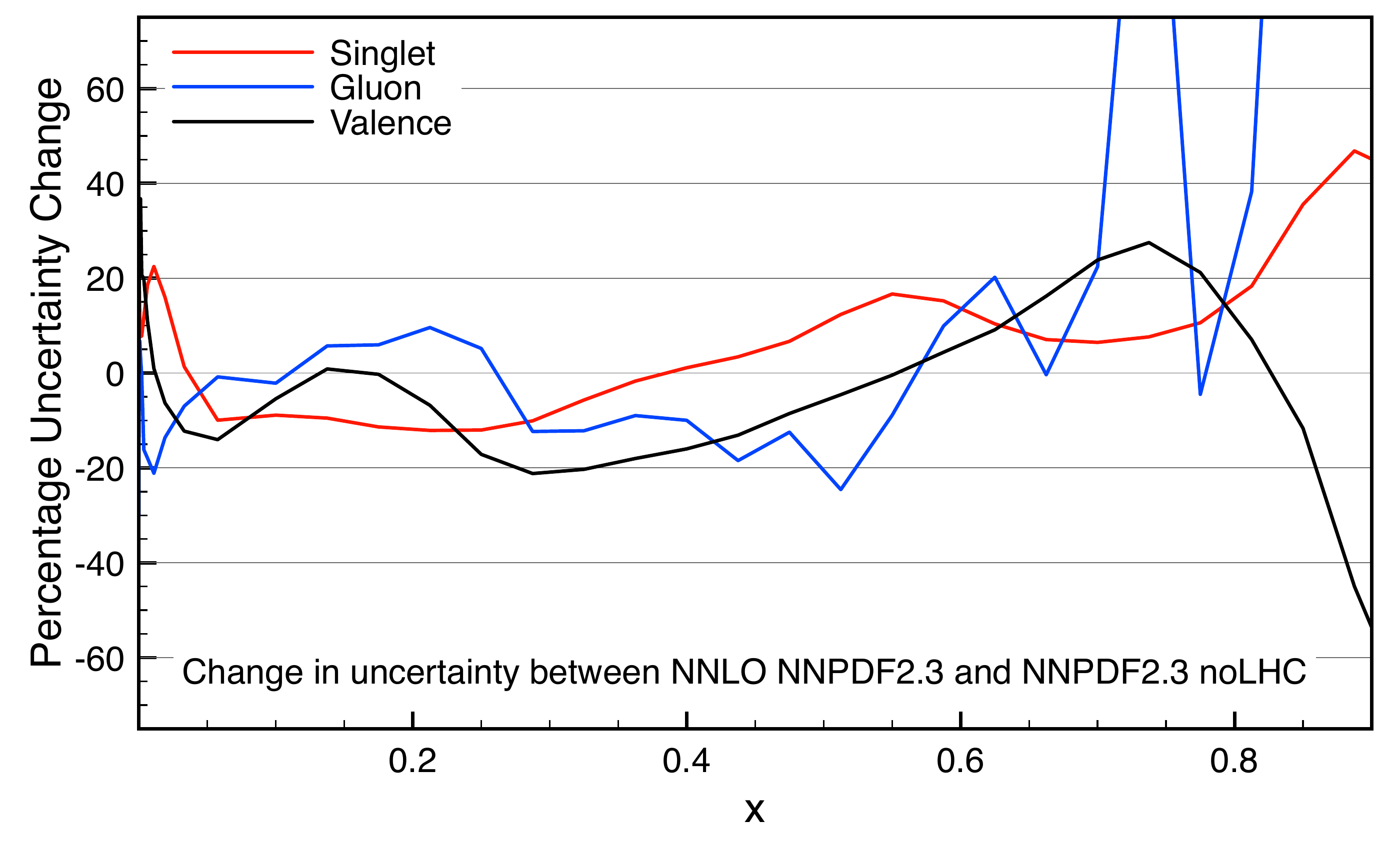}
\caption[Uncertainty change in NNPDF2.3 under the addition of LHC data]{Uncertainty change in NNPDF2.3 under the addition of LHC data. The top figures represent the percentage change in uncertainty between NNPDF2.3 global and NNPDF2.3 noLHC at NLO, while the bottom plots show the equivalent comparison at NNLO.}
\label{fig:23vs23noLHCunc}
\end{figure}

Figure~\ref{fig:23vs23noLHCunc} specifically demonstrates the changes in uncertainties upon the addition of the new data. While uncertainty reduction has been achieved for some PDF combinations, several areas undergo an increase in their uncertainties due to central value shifts.

\subsubsection{NNPDF2.3 collider only}

\begin{figure}[ht]
\centering
\includegraphics[width=0.48\textwidth]{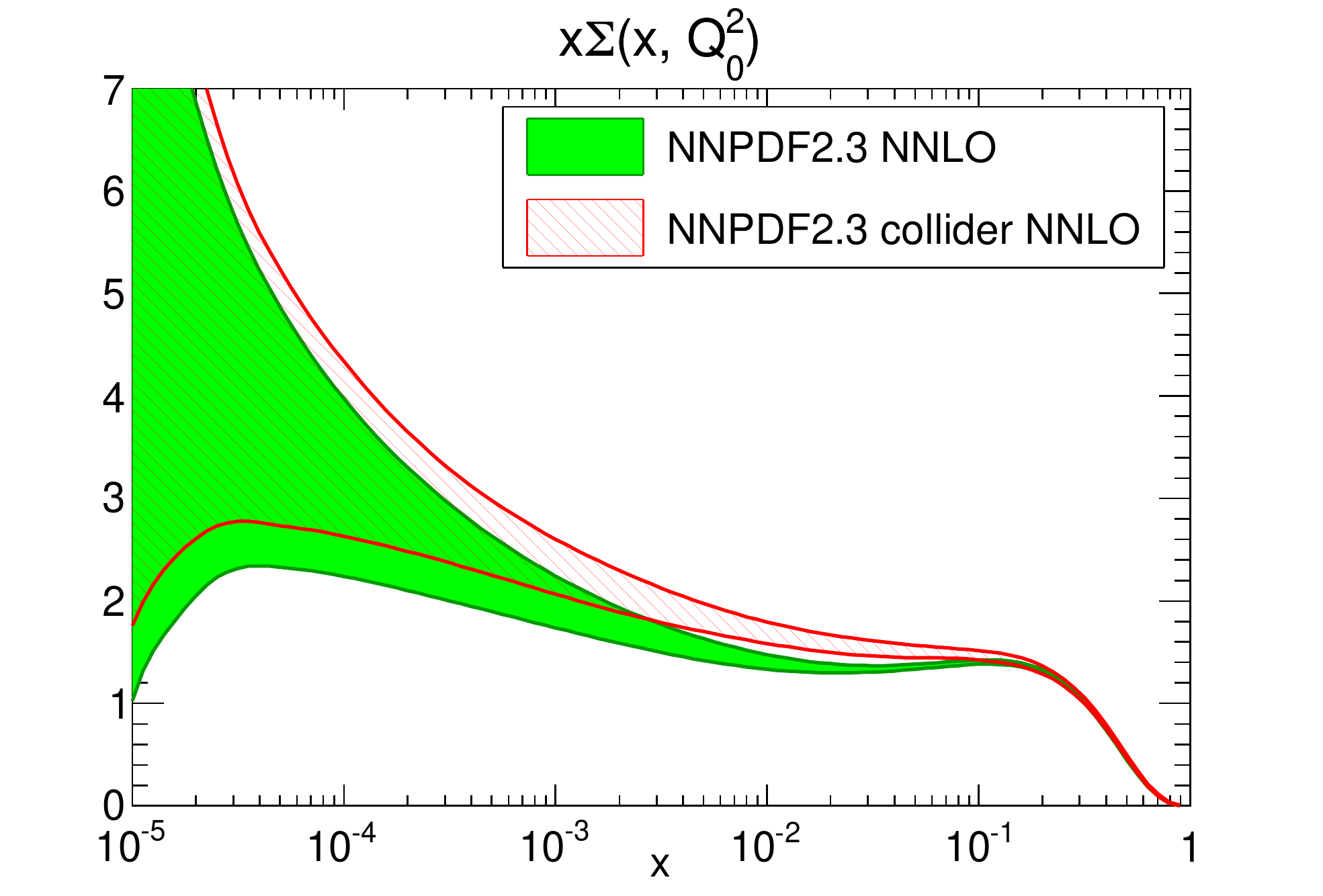}
\includegraphics[width=0.48\textwidth]{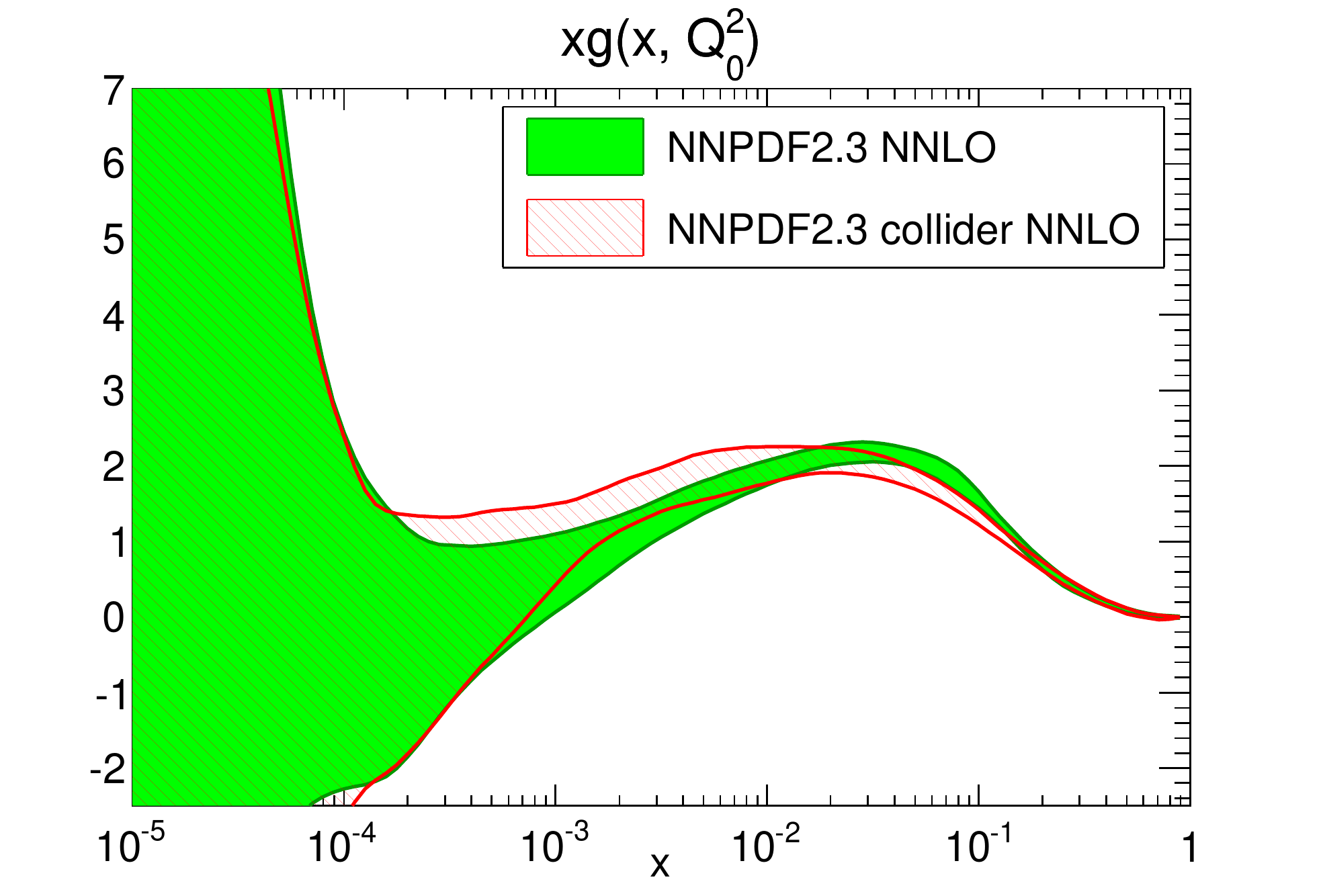}\\
\includegraphics[width=0.48\textwidth]{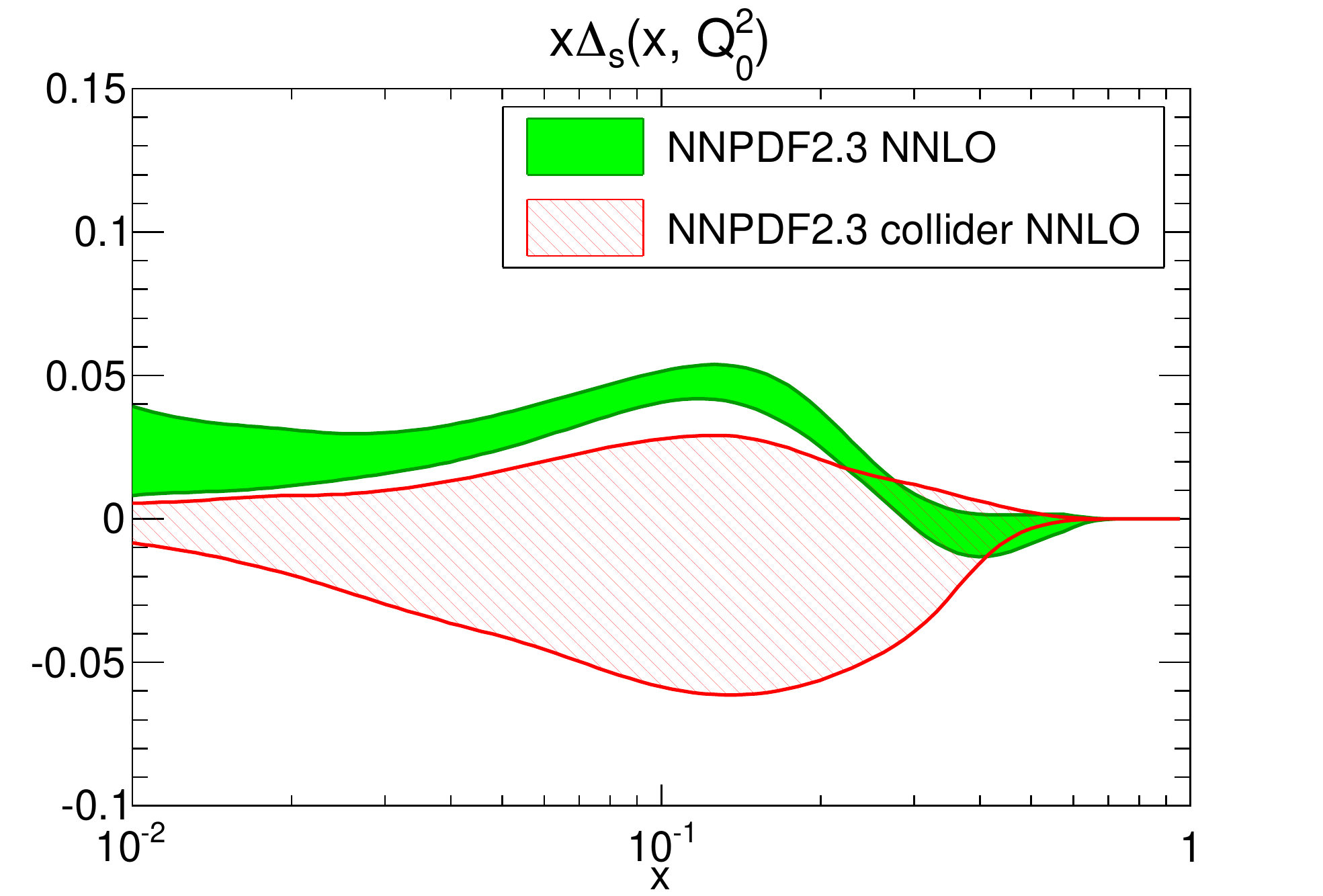}
\includegraphics[width=0.48\textwidth]{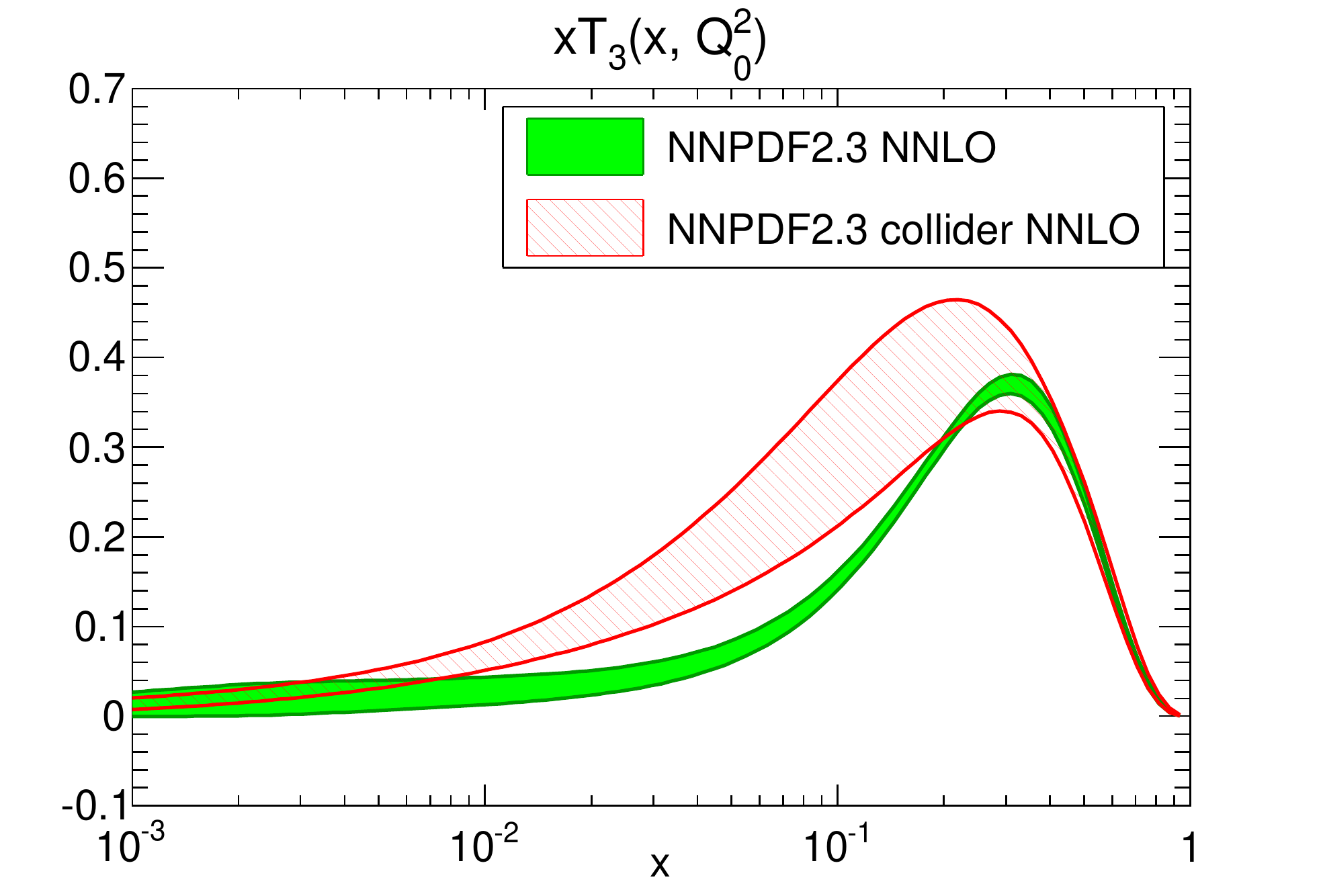}
\caption[NNPDF2.3 collider only compared to the global determination at NNLO]{NNPDF2.3 collider only compared to the global determination at NNLO. The red distributions shown are those determined via a fit to a collider-only dataset, while the green curves show the results of the NNPDF2.3 global fit.}
\label{fig:23vs23coll}
\end{figure}

In order to investigate the viability of an NNPDF collider only determination with the available dataset, we now compare the resulting distributions from the NNPDF2.3 global and collider-only fits. In Figure~\ref{fig:23vs23coll} the distributions for the singlet, gluon, sea-asymmetry and triplet are shown for the two fits at NNLO. The combination of HERA DIS data along with Tevatron and LHC inclusive jet data ensure that the gluon and singlet, although deviating not insignificantly from the global fit, are well constrained by data. The preference for a higher singlet distribution by the LHC data seen in the comparison between the global and noLHC fits is very clear in this comparison, with the collider only dataset preferring a significantly higher singlet also. The gluon distribution demonstrates also rather different shape in the medium-$x$ region. One may therefore at first be be tempted to prefer the collider only determination for phenomenology, given its theoretically cleaner underpinnings. However the ability of the fit to obtain a good handle on PDF combinations involving flavour separation is substantially reduced in the collider only fit. The lower two plots in Figure~\ref{fig:23vs23coll} demonstrate that the collider dataset is not able to provide sufficient constraints for these combinations even after the addition of the LHC dataset.

\begin{figure}[h]
\centering
\includegraphics[width=0.48\textwidth]{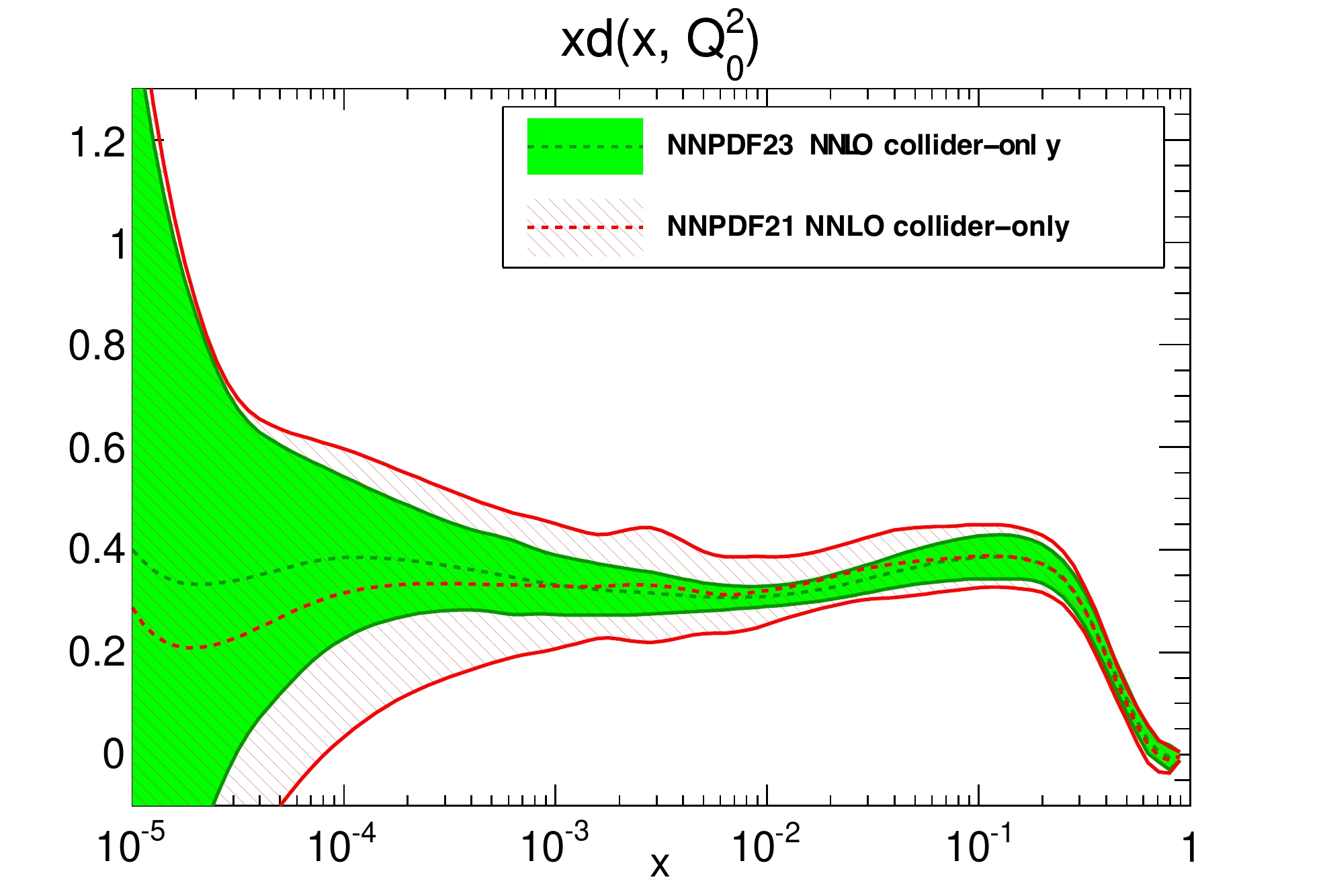}
\includegraphics[width=0.48\textwidth]{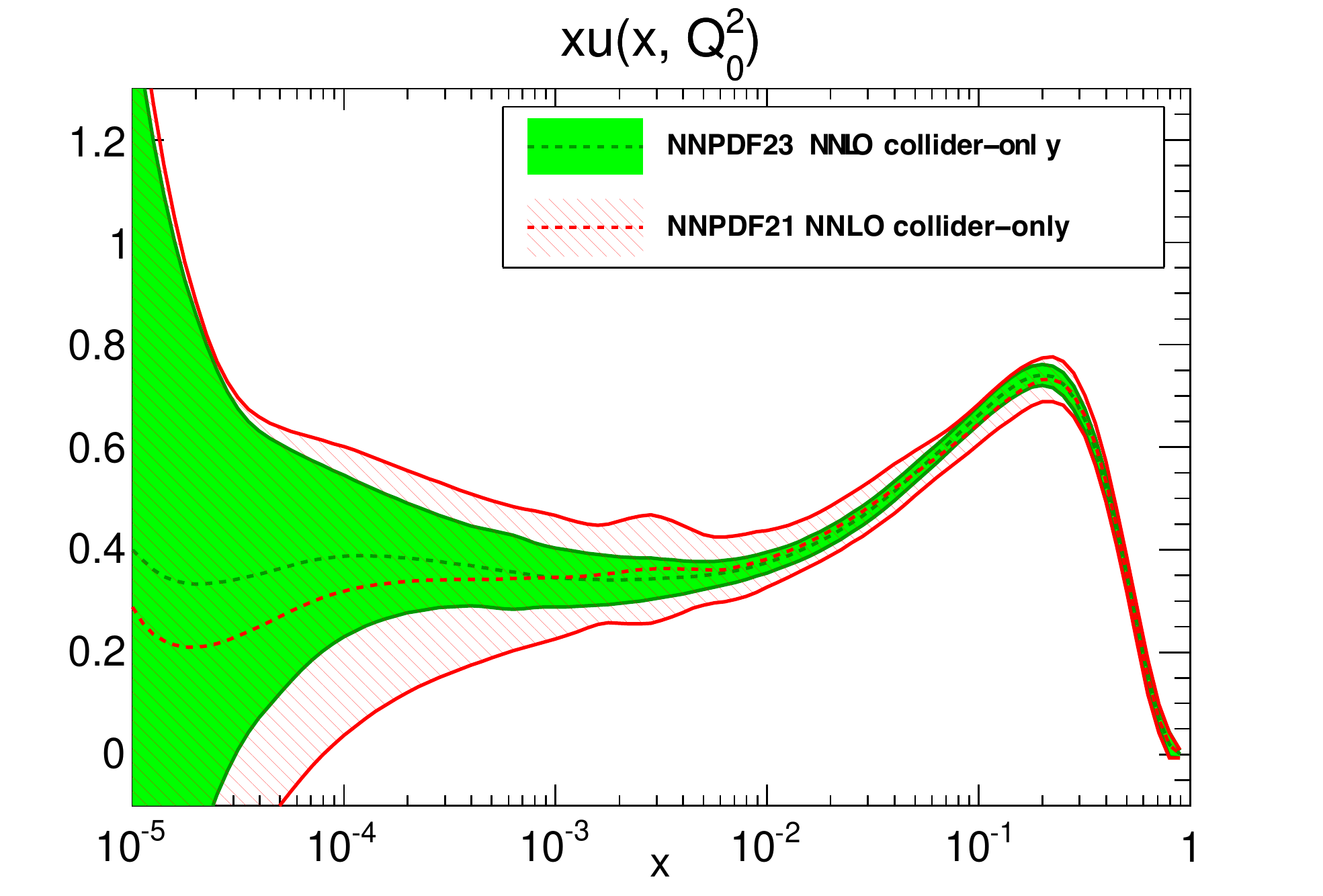}
\caption[Impact of NNPDF2.3 LHC data upon collider only determinations for light flavour PDFs]{Impact of NNPDF2.3 LHC data upon collider only determinations for light flavour PDFs. The green solid curves show the collider only results including the LHC dataset and the red dashed curves show the results of the NNPDF2.1 collider only fits.}
\label{fig:23vs21coll_lqs}
\end{figure}

The NNPDF2.3 collider only dataset therefore remains too imprecise to provide an accurate determination of flavour-separation, and is therefore unsuitable for applications sensitive to such parton combinations. Nevertheless, a great deal of progress is evident when we compare PDFs obtained via a collider-only fit to the pre-LHC NNPDF2.1 dataset, and those obtained with the new LHC data. Examining the NNLO PDFs where methodological differences are slight, Figure~\ref{fig:23vs21coll_lqs} compares the light quark distributions $u$ and $d$ between NNPDF2.1 collider only and NNPDF2.3 collider only, with the only significant difference being the presence of the LHC dataset in the 2.3 determination.

\begin{figure}[h]
\centering
\includegraphics[width=0.48\textwidth]{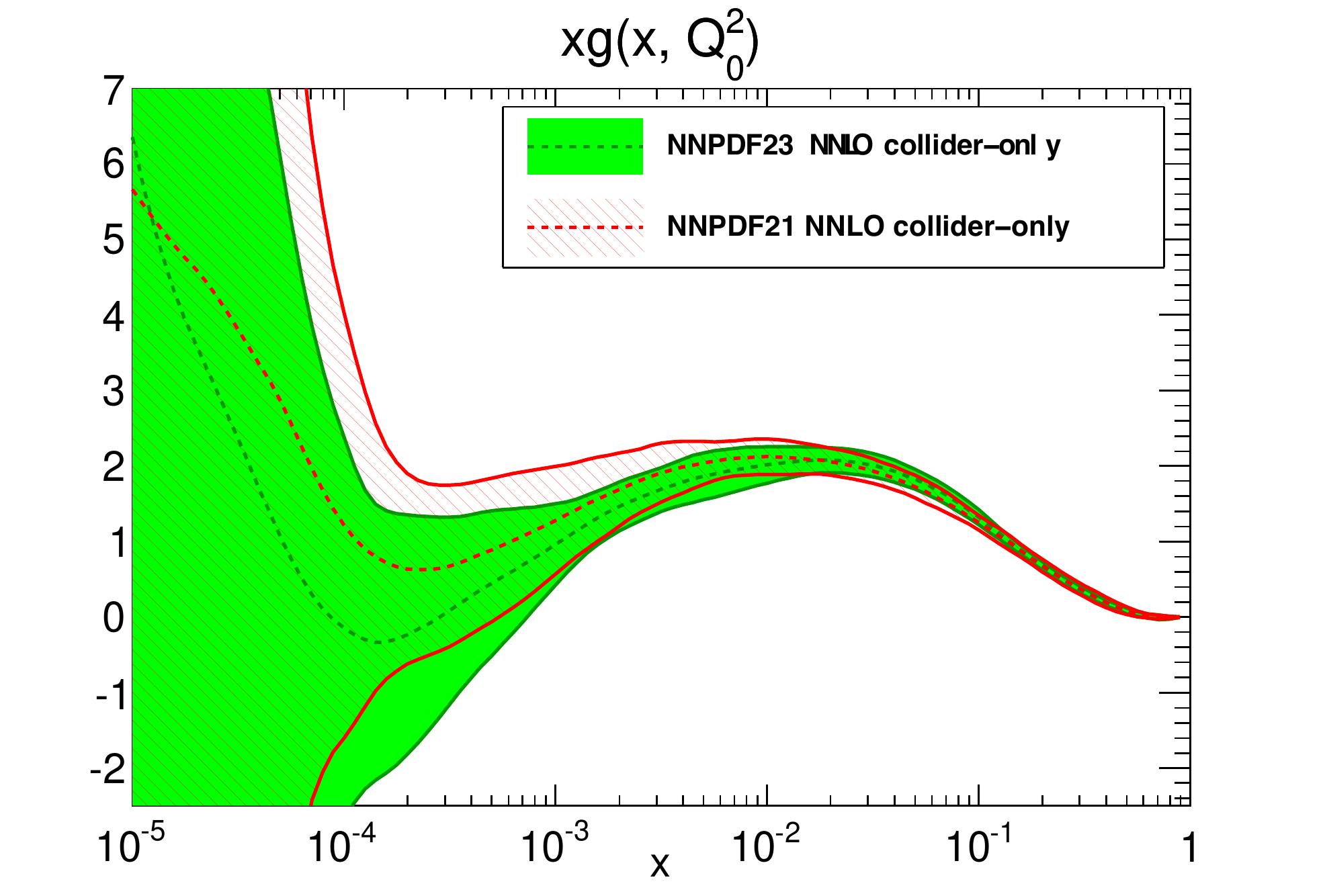}
\includegraphics[width=0.48\textwidth]{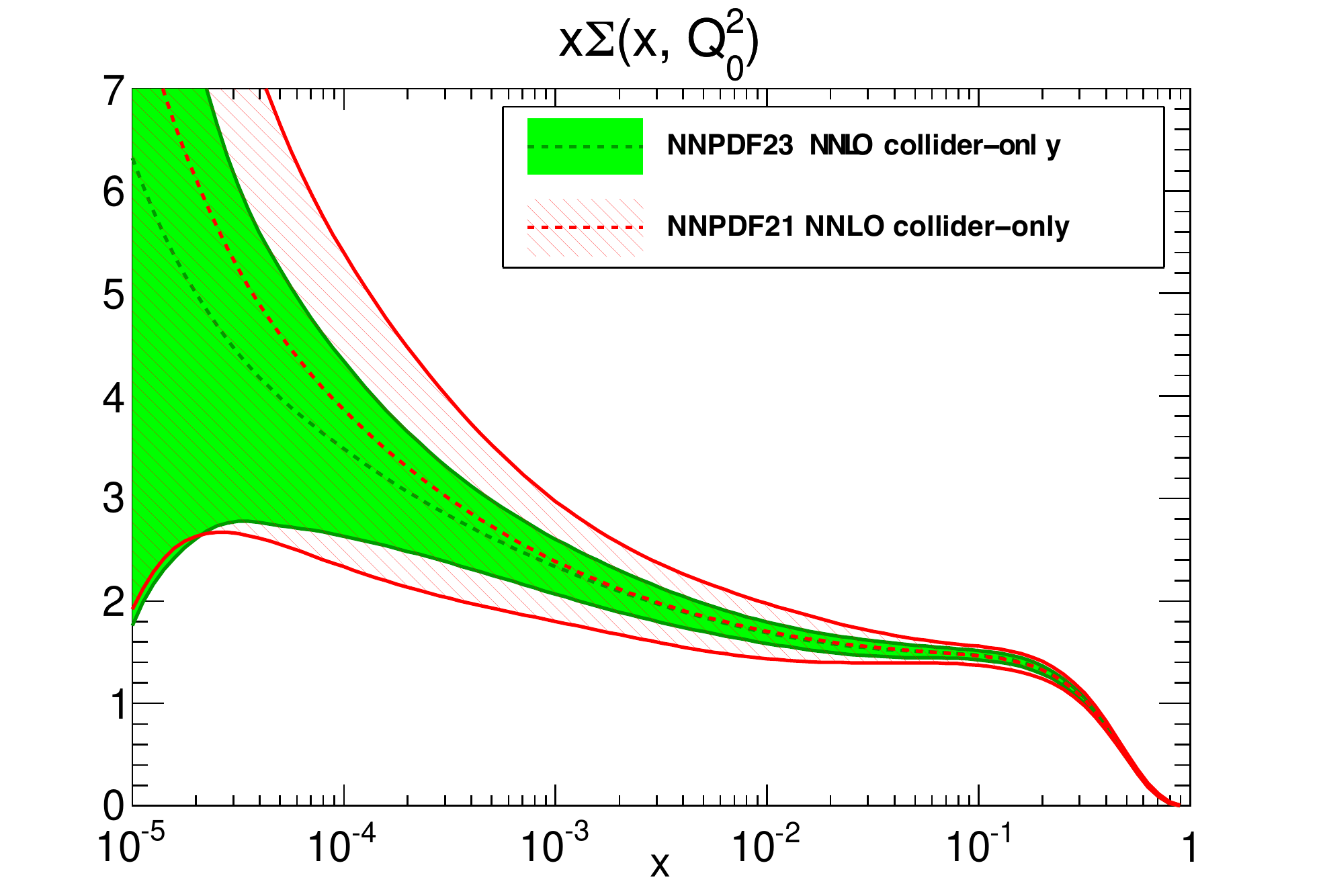}
\caption[Impact of NNPDF2.3 LHC data upon collider only determinations for singlet and gluon PDFs]{Impact of NNPDF2.3 LHC data upon collider only determinations for singlet and gluon PDFs. The green curves show the collider only results including the LHC dataset and the red curves show the results of the NNPDF2.1 collider only fits.}
\label{fig:23vs21coll_gs}
\end{figure}

\clearpage
From the figure it is clear that the LHC data provides very large constraints upon the collider only distributions when we compare to the version available in the NNPDF2.1 series. Examining the gluon and singlet PDFs in Figure~\ref{fig:23vs21coll_gs} we see that the improvements made in the light quarks carry through to the quark singlet. The gluon distribution was already relatively well determined in the 2.1 series due to the Tevatron jet data, therefore it does not experience such a large improvement. 

\begin{figure}[h!]
\centering
\includegraphics[width=0.48\textwidth]{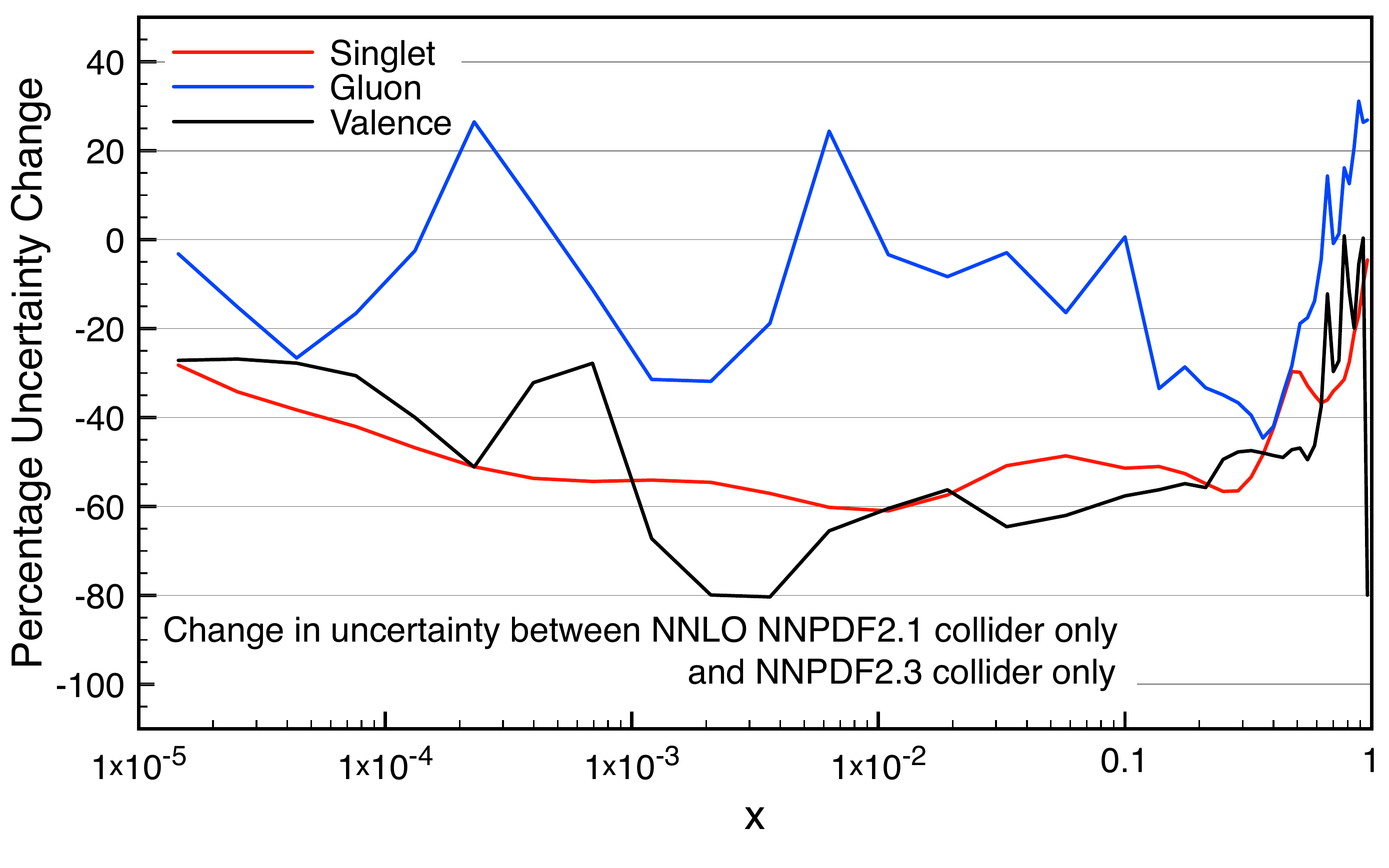}
\includegraphics[width=0.48\textwidth]{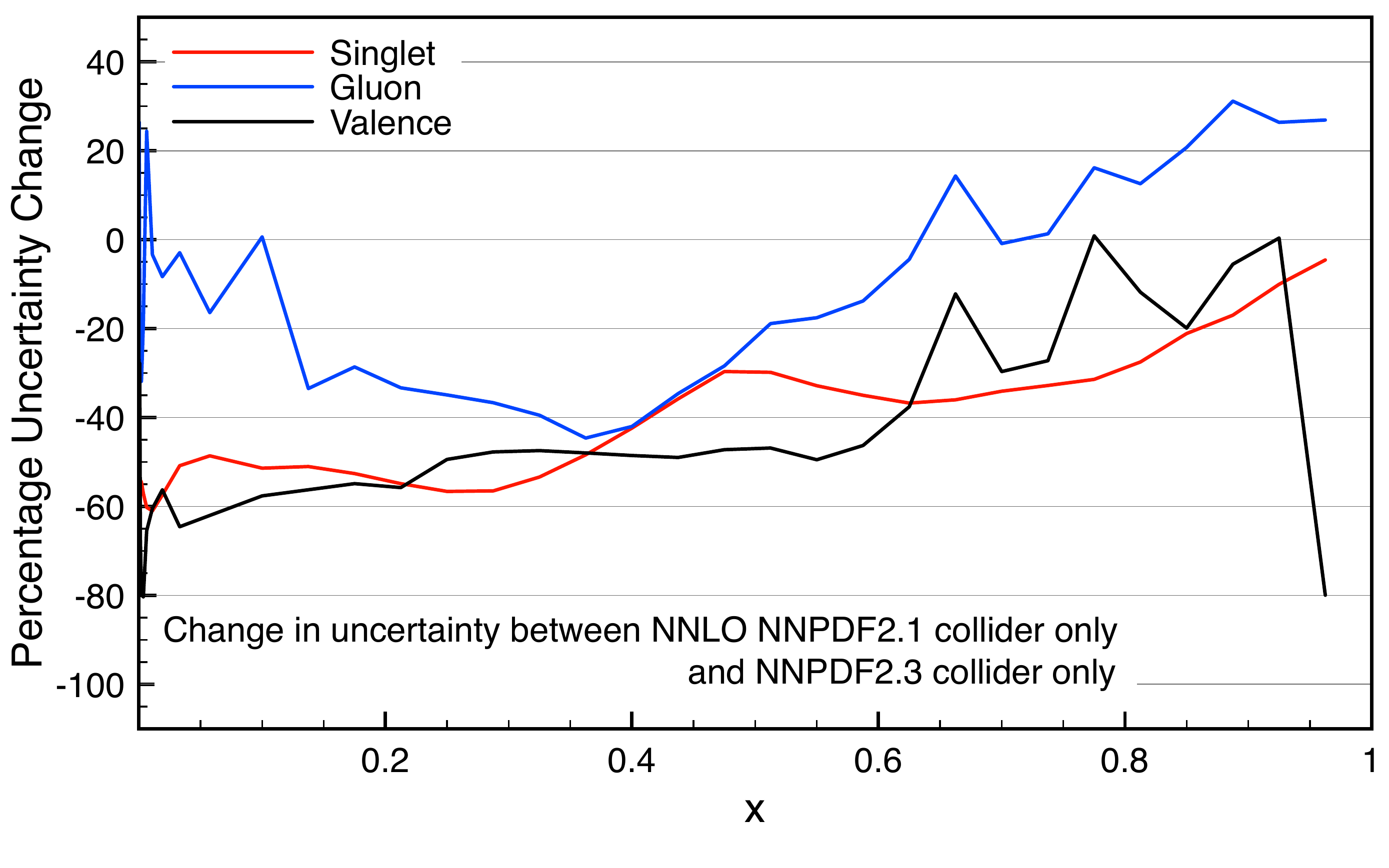}
\caption[Impact of LHC data upon collider only fit uncertainties]{Impact of LHC data upon collider only fit uncertainties. Figures show the percentage improvement in the NNPDF2.3 collider only uncertainty compared to the NNPDF2.1 collider only fit, for the singlet, gluon and valence distributions at NNLO.}
\label{fig:23vs21coll_unc}
\end{figure}

In Figure~\ref{fig:23vs21coll_unc} we can see explicitly the impact the new data has upon collider only uncertainties. Across nearly the whole kinematic range, very substantial improvements can be seen in both the singlet and valence distributions. The results from the LHC are therefore vital in providing a handle on the collider only distributions, and updated measurements may be able to bring the accuracy of such determinations to near the level of the global fits.

\subsection{Proton strangeness}
The issue of the strange content of the proton is a particularly interesting one, and has been the source of discussion due to new results and analyses arising from the LHC experiments. A particular complication lies in the treatment of the NuTeV dimuon data. 
In the NNPDF2.1 series of fits the expression used for the dimuon data suffered from an error originating from the heavy quark mass handling. Specifically, Eqn.~34 of Ref.~\cite{Ball:2011mu} presented an incorrect expression for the charm production reduced cross-section in neutrino charged current DIS, where the correct expression reads

\begin{eqnarray}
  \label{eq:nuxsecdimuon}
  &&\tilde{\sigma}^{\nu (\bar{\nu}),c}(x,y,Q^2)\equiv 
  \frac{1}{E_{\nu}}\frac{d^2\sigma^{\nu(\bar{\nu}),c}}{dx\,dy}
  (x,y,Q^2)\nonumber\\
  &&\qquad =\frac{G_F^2M_N}{2\pi(1+Q^2/M_W^2)^2}
  \Bigg[ 
    \left( \left( Y_+ - \frac{2M^2_Nx^2y^2}{Q^2} -y^2\right) +y^2\right)
    F_{2,c}^{\nu(\bar{\nu})}(x,Q^2) \nonumber\\ 
    &&\qquad\qquad\qquad\qquad\qquad\qquad
    -y^2F_{L,c}^{\nu(\bar{\nu})}(x,Q^2)\pm 
    \,Y_-\,xF_{3,c}^{\nu(\bar{\nu})}(x,Q^2)
    \Bigg], \,
\end{eqnarray}
where here $x$ and $y$ are the usual DIS kinematic variables, $ Y_\pm = 1\pm(1-y)^2$ and the momentum transfer is given by $Q^2=2M_NE_{\nu}xy$. The expression in Ref.~\cite{Ball:2011mu} differs from this by a spurious additional $\left( 1+ \frac{m_c^2}{Q^2}\right)$ term which was corrected prior to the NNPDF2.3 determination.

\begin{figure}[h!]
\centering
\includegraphics[width=0.48\textwidth]{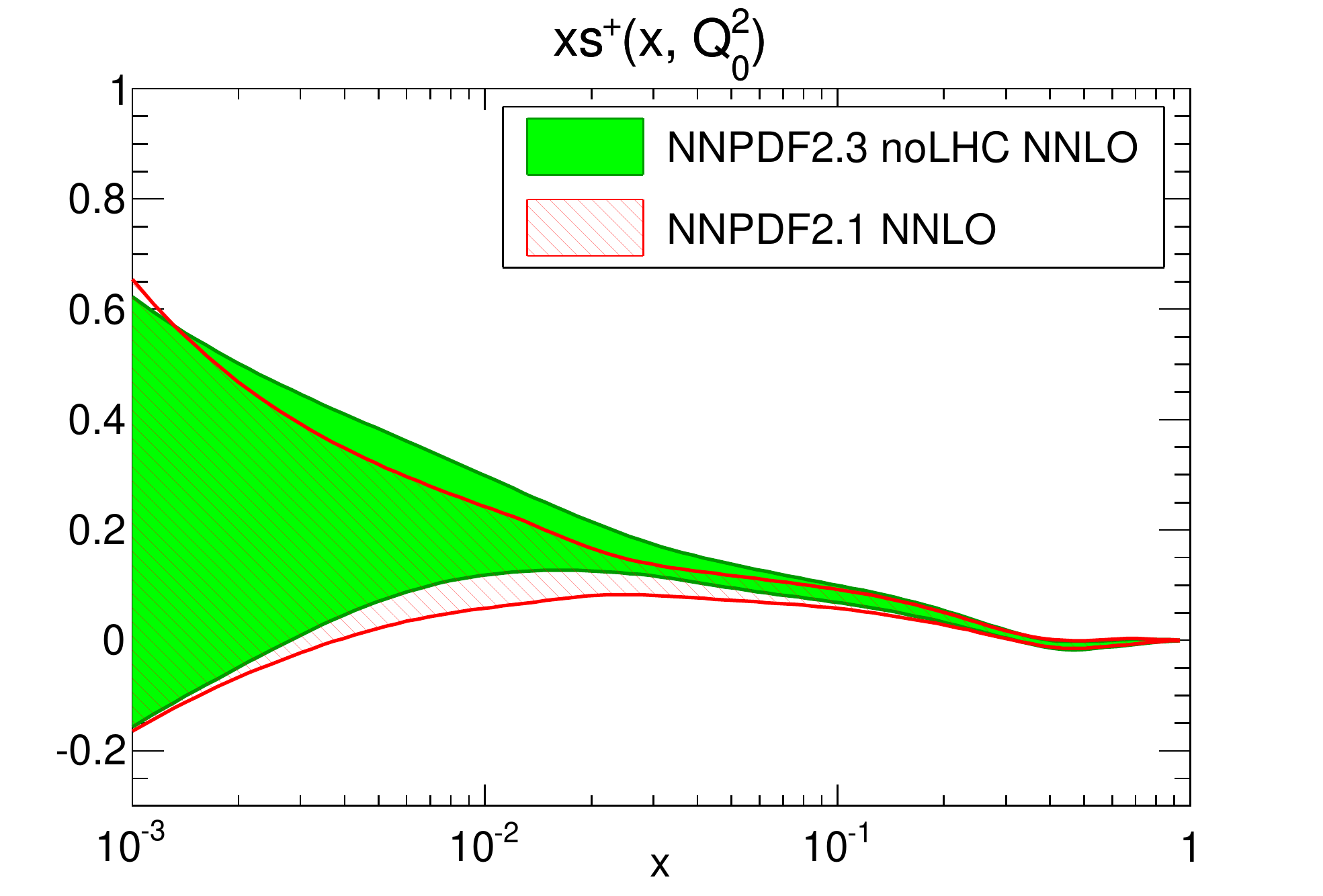}
\includegraphics[width=0.48\textwidth]{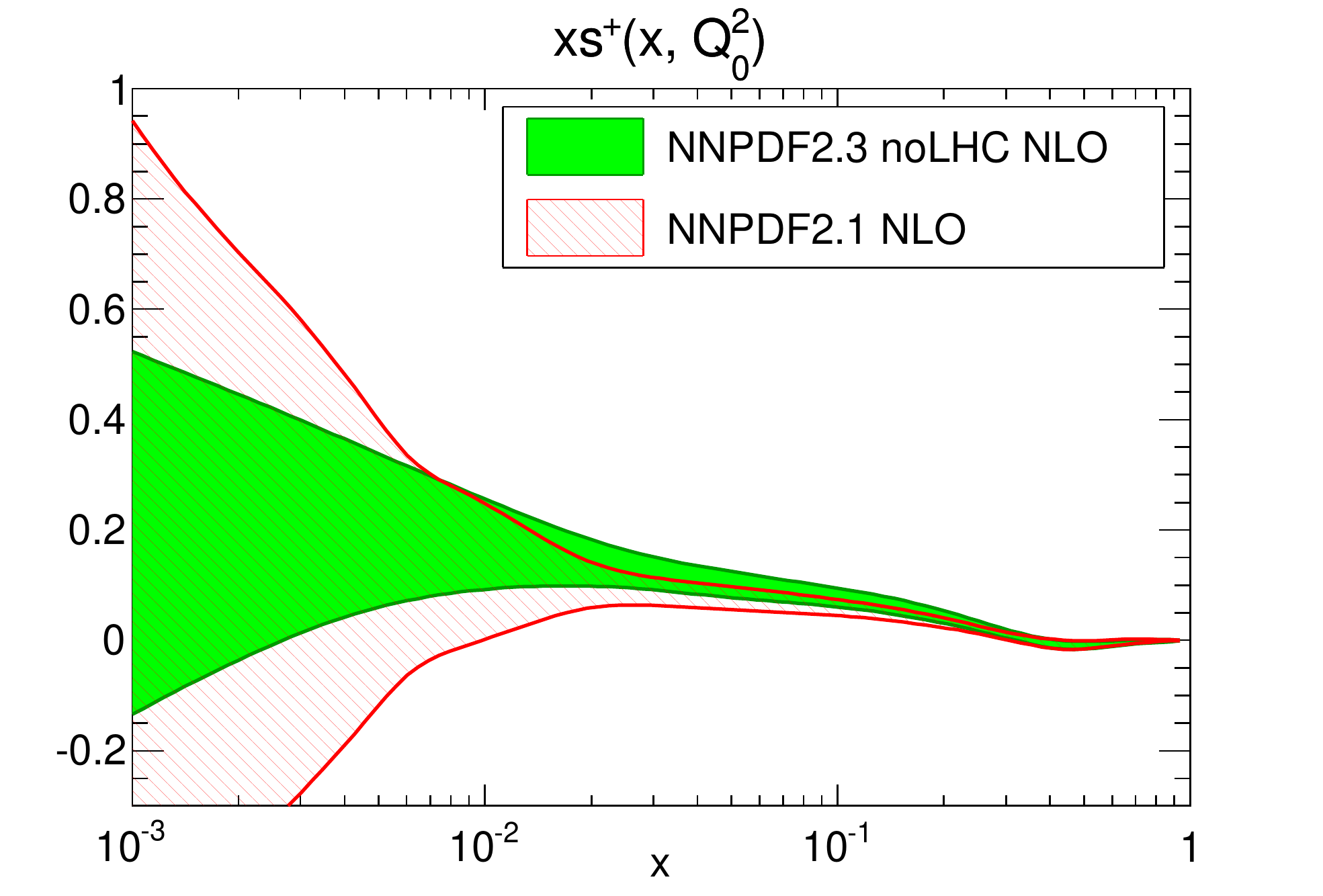}\\
\includegraphics[width=0.48\textwidth]{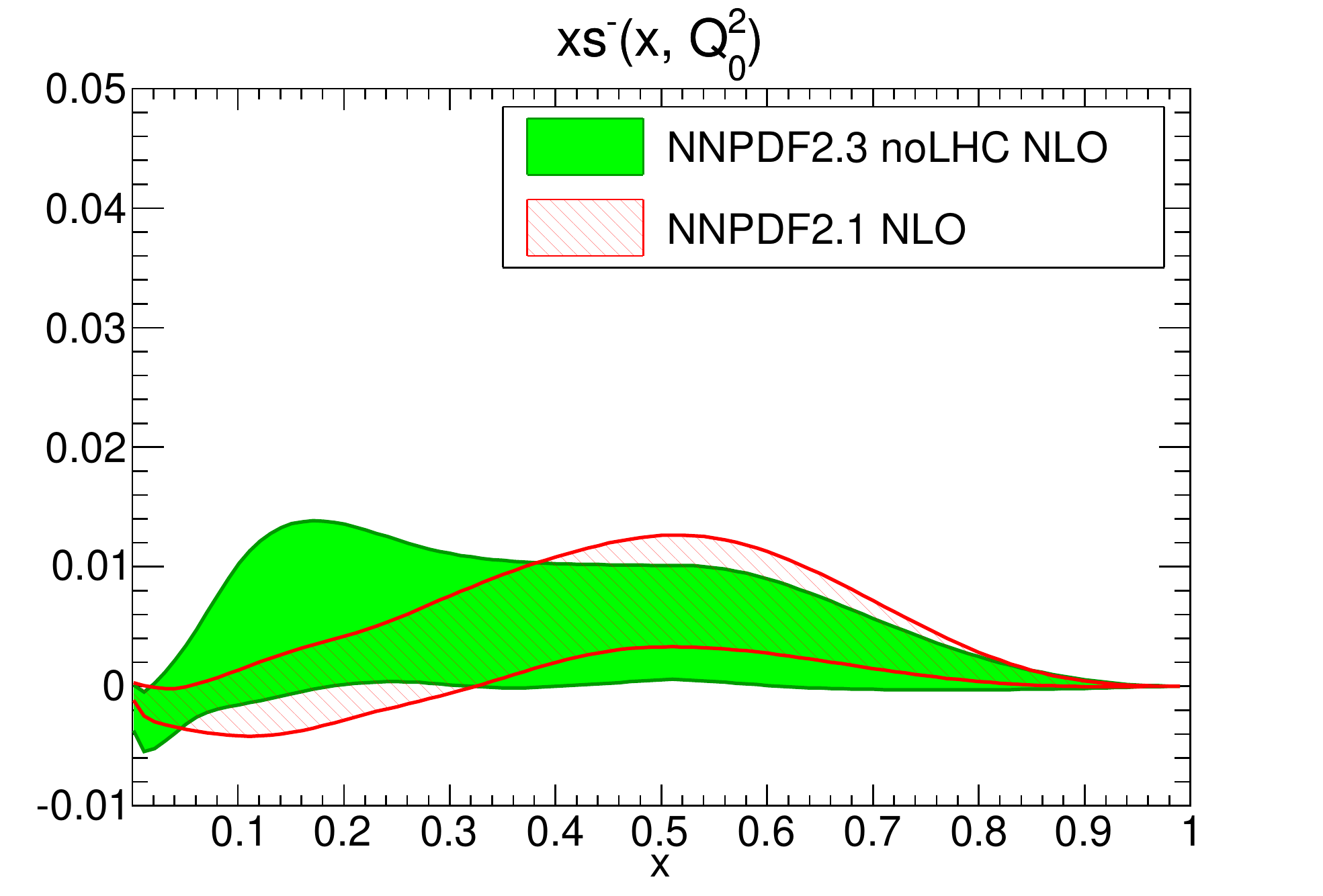}
\includegraphics[width=0.48\textwidth]{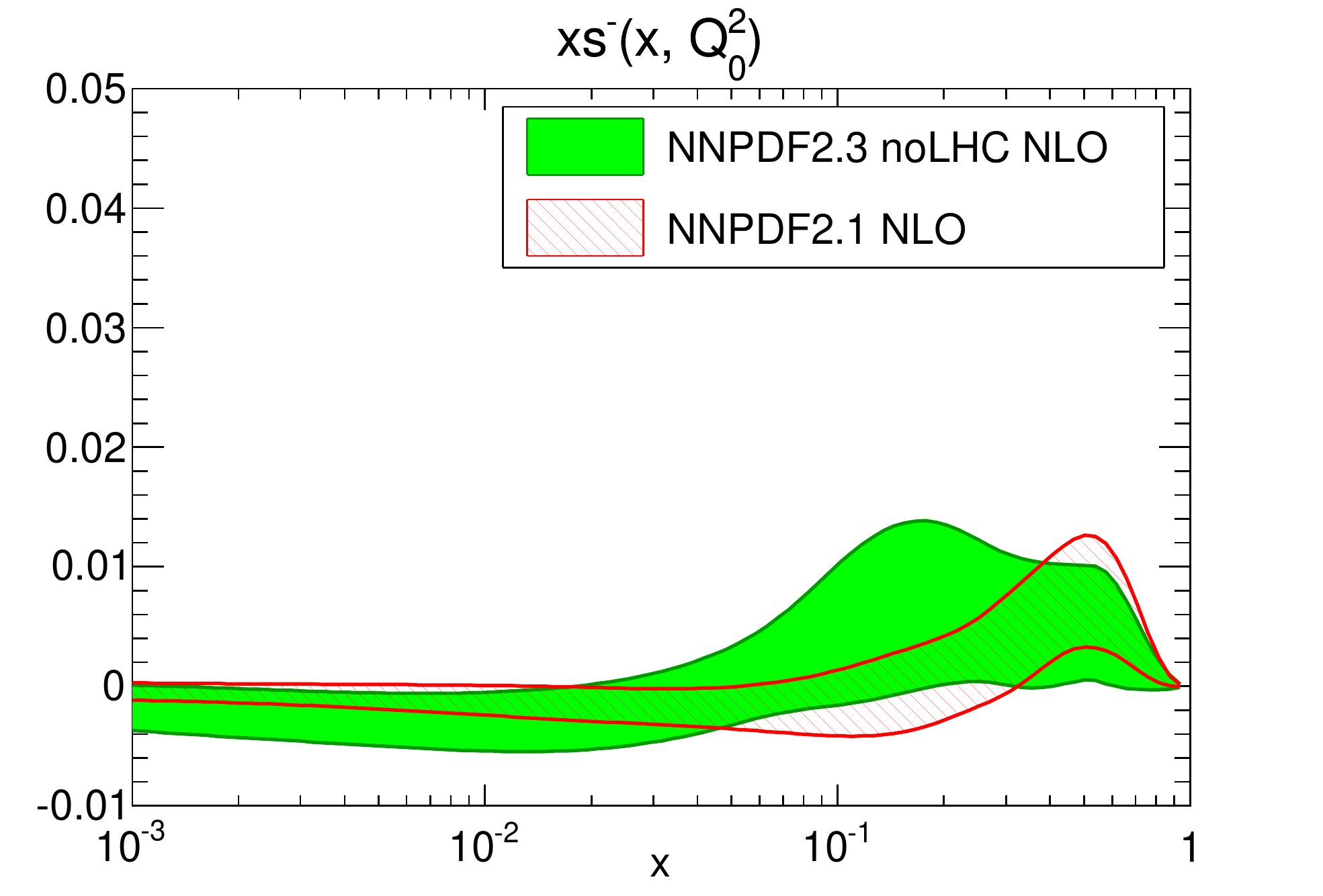}
\caption[Total strangeness and strange valence distributions compared between NNPDF2.1 and NNPDF2.3 noLHC]{Total strangeness and strange valence distributions compared between NNPDF2.1 and NNPDF2.3 noLHC. The NLO bands demonstrate also the improvements due to the more aggressive minimisation, particularly evident at low-$x$.} \label{fig:23noLHCvs21_strangeness}
\end{figure}

This error affected only the predictions for the NuTeV data, and consequently after the error was corrected the impact upon PDFs themselves was largely restricted to the strange sector. The impact of the error is shown in Figure~\ref{fig:23noLHCvs21_strangeness}, where it can be clearly seen that the error led to a small suppression of the total strange distribution across most of the kinematic range, peaking at around half a standard deviation.

\begin{figure}[h!]
\centering
\includegraphics[width=0.48\textwidth]{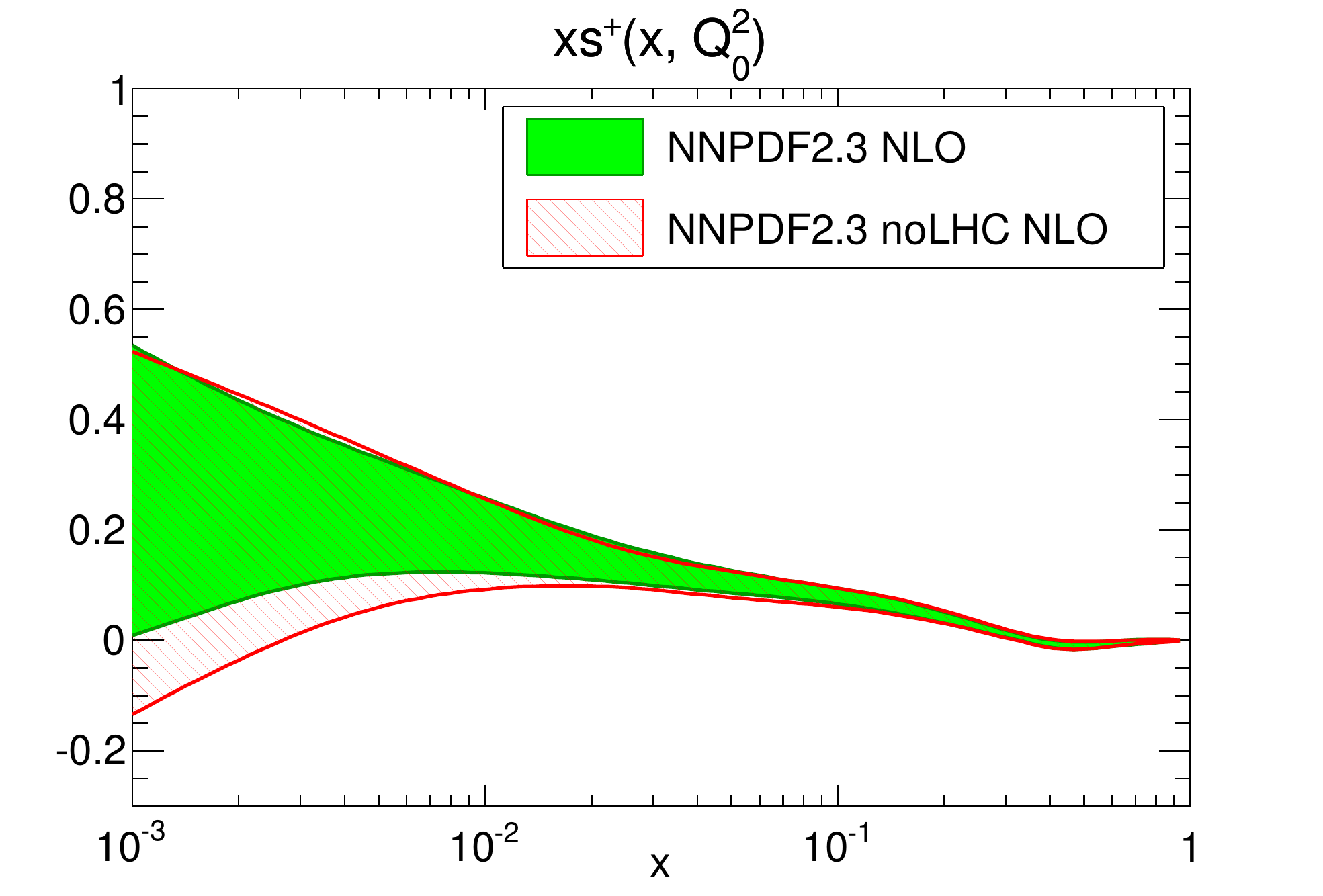}
\includegraphics[width=0.48\textwidth]{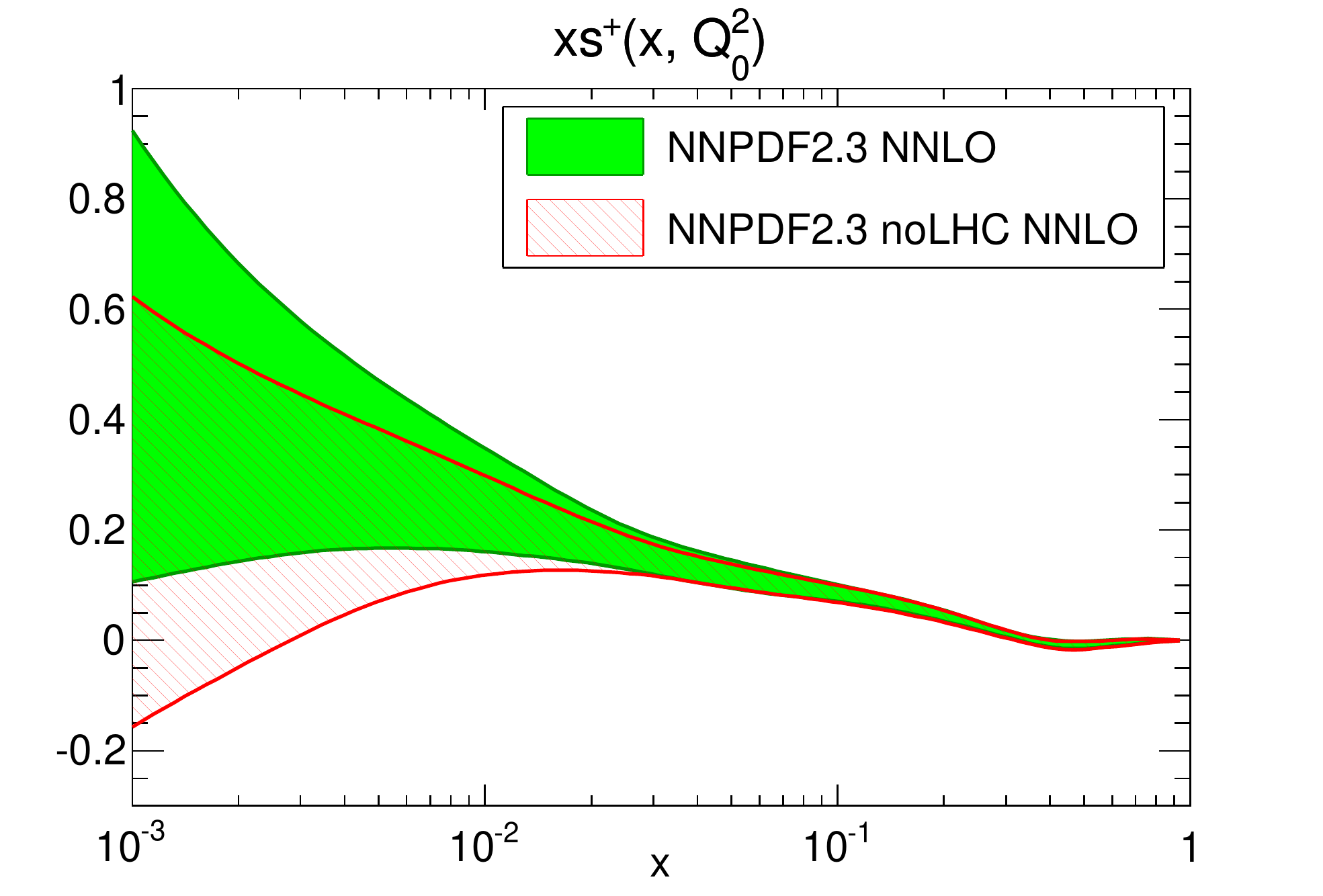}
\caption[Strange sea distributions in NNPDF2.3 and NNPDF2.3 noLHC]{Strange sea distributions in NNPDF2.3 and NNPDF2.3 noLHC, for the NLO (left) and NNLO(right) PDF sets. The NNPDF2.3 global set, differing from the noLHC set by the inclusion of LHC measurements, prefers a marginally larger strange distribution.}
\label{fig:23noLHCvs23_strangeness}
\end{figure}

The shift towards slightly higher total strangeness is continued upon the addition of the LHC dataset. Figure~\ref{fig:23noLHCvs23_strangeness} shows how the strange sea distribution changes under the addition of the new data. The electroweak measurements present in the LHC dataset seem to marginally prefer a slightly larger strange sea at small-$x$ for both the NLO and NNLO distributions.

To investigate the relative contribution of the strange sea with respect to the light quark sea, a commonly used measure~\cite{Lai:2007dq,Alekhin:2008mb,Martin:2009iq,Ball:2009mk}
 is the integrated ratio of the two PDF combinations,
\be
K_s = \frac{\int_0^1 dx \, x\left(s(x,Q^2) + \bar{s}(x,Q^2)\right)}{\int_0^1 dx \, x\left(\bar{u}(x,Q^2) + \bar{d}(x,Q^2)\right)}.
\ee
In most global determinations a significant suppression of the strange sea is typically observed at low scales, with $K_s < 1$. In Table~\ref{tab:strangesupp} we see the results for $K_s$ obtained through the NNPDF2.1, NNPDF2.3 and NNPDF2.3 noLHC sets demonstrating such a suppression at NNLO. The impact of the incorrect dimuon treatment in NNPDF2.1 is manifest in an exaggerated level of suppression, although it remains consistent with the newer determinations within uncertainties. The preference for a larger strange sea by the LHC measurements is also demonstrated in the difference between NNPDF2.3 and the noLHC dataset fit.

\clearpage

\clearpage

\begin{table}[htb]
\begin{center}
\begin{tabular}{|c|c|c|}
\hline
PDF	& $K_s(2$ GeV$^2)$ & $K_s(M_{\text{Z}}^2)$ \\
\hline
NNPDF2.1 & $0.26^{+0.08}_{-0.08}$ & $0.63^{+0.04}_{-0.05}$ \\
NNPDF2.3 noLHC & $0.30^{+0.09}_{-0.08}$ & $0.65^{+0.05}_{-0.05}$ \\
NNPDF2.3 & $0.35^{+0.10}_{-0.08}$ & $0.68^{+0.05}_{-0.05}$ \\
\hline
\end{tabular}
\caption[Strange sea suppression in NNPDF2.3 and NNPDF2.1 NNLO]{Strange sea suppression in NNPDF2.3 and NNPDF2.1, with the uncertainties given by the 68\% confidence interval. }
\label{tab:strangesupp}
\end{center}
\end{table}%

\begin{figure}[ht]
\centering
\includegraphics[width=0.60\textwidth]{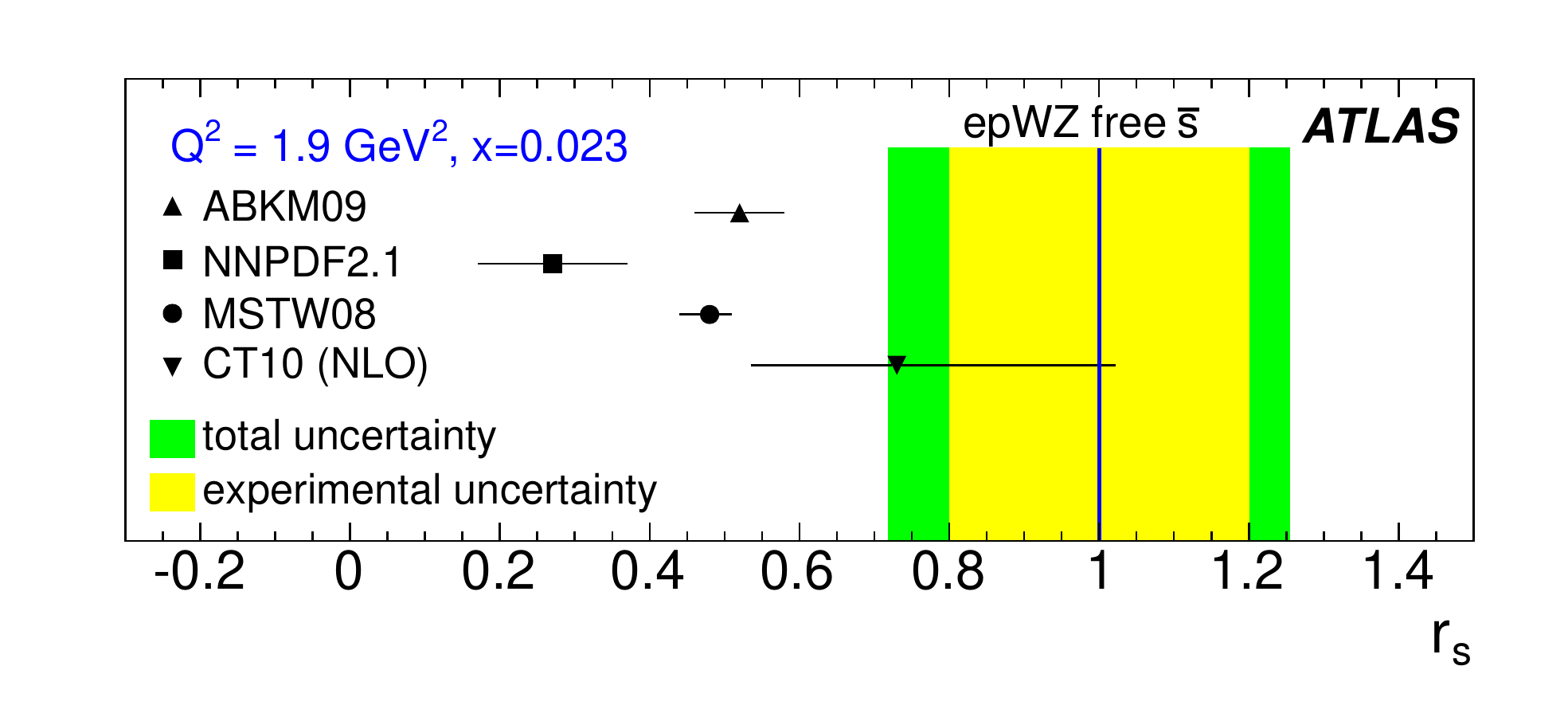}
\caption[ATLAS determination of strange sea suppression]{ATLAS determination of strange sea suppression at $x=0.023$ for a number of PDF sets. The ATLAS result is consistent with no suppression for the strange distributions. Figure from~\cite{Aad:2012sb}.}
\label{fig:ATLASepWZplot}
\end{figure}

Such a strange sea suppression was challenged by an ATLAS determination of the strange content of the proton\cite{Aad:2012sb} based upon a fit to a combined HERA DIS and ATLAS $W$ and $Z$ production dataset. Defining a more exclusive measure, the ratio of the strange sea to twice the $\bar{d}$ distribution at specific points of $x$ and $Q^2$,

\be r_s(x,Q^2) = \frac{s(x,Q^2) + \bar{s}(x,Q^2)}{2\bar{d}(x,Q^2)},\ee
the ATLAS study reported values that significantly differed from the typical results of global fits, with the most extreme disagreement with the NNPDF2.1 set where the two values are separated by more than two sigma. The disagreement is particularly large in the region $x=0.023$, at the initial scales, as shown in the ATLAS plot in Figure~\ref{fig:ATLASepWZplot}, where the ATLAS result is consistent with no suppression of the strange sea, $r_s \sim 1$.

\begin{figure}[ht]
\centering
\includegraphics[width=0.48\textwidth]{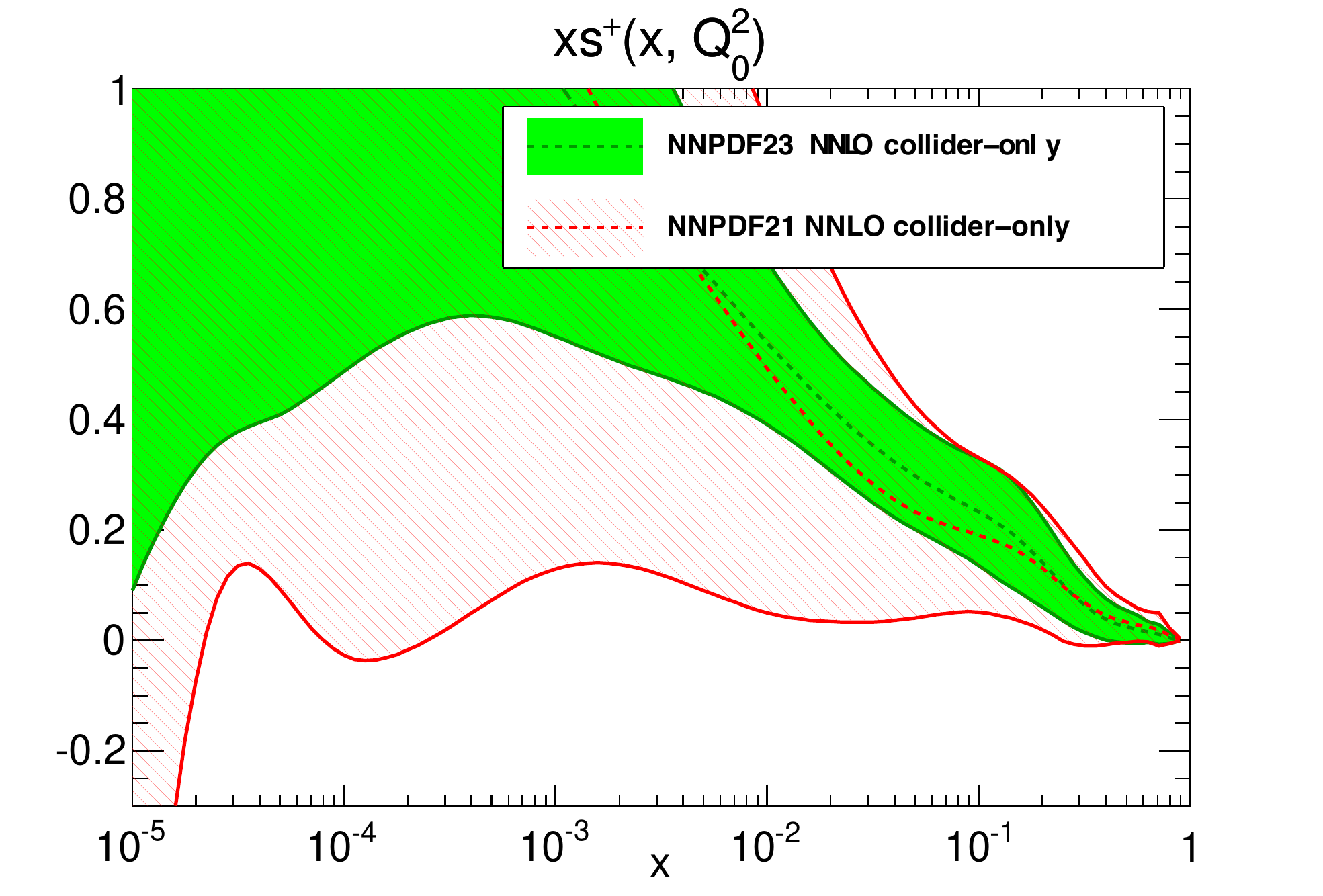}
\includegraphics[width=0.48\textwidth]{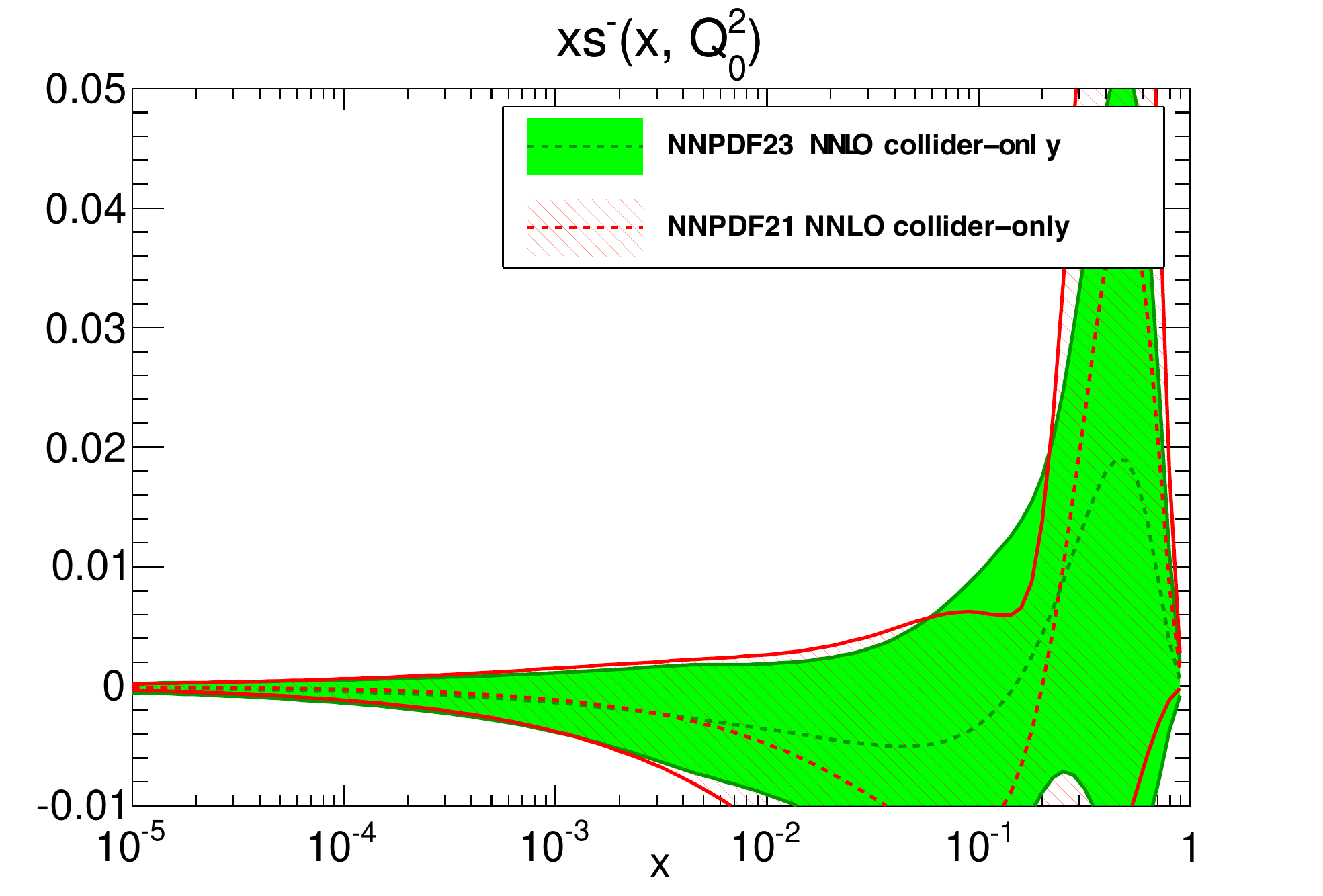}
\caption[Impact of LHC data on collider only strangeness]{Impact of LHC data on collider only strangeness. The strange sea (left) and strange valence (right) distributions are plotted at NNLO, comparing the NNPDF2.1 and NNPDF2.3 collider only results. A very significant impact upon the total strangeness uncertainties can be observed, however little constraint is afforded to the valence distribution.}
\label{fig:21vs23coll_strange}
\end{figure}

The impact of the NNPDF2.3 LHC dataset clearly has a preference for a more strange-symmetric sea, as is particularly demonstrated upon the inclusion of the LHC dataset (including the ATLAS data used in their strangeness analysis) to the NNPDF2.1 collider only strange distribution. Figure~\ref{fig:21vs23coll_strange} demonstrates the extensive constraint placed upon the NNPDF2.1 collider only set by the LHC electroweak data in the strange sector, and a clear preference for a larger strange sea. Despite this preference, the results of the global fit remain consistent within the larger uncertainties of the NNPDF collider only determination.

\begin{figure}[ht]
\centering
\includegraphics[width=0.48\textwidth]{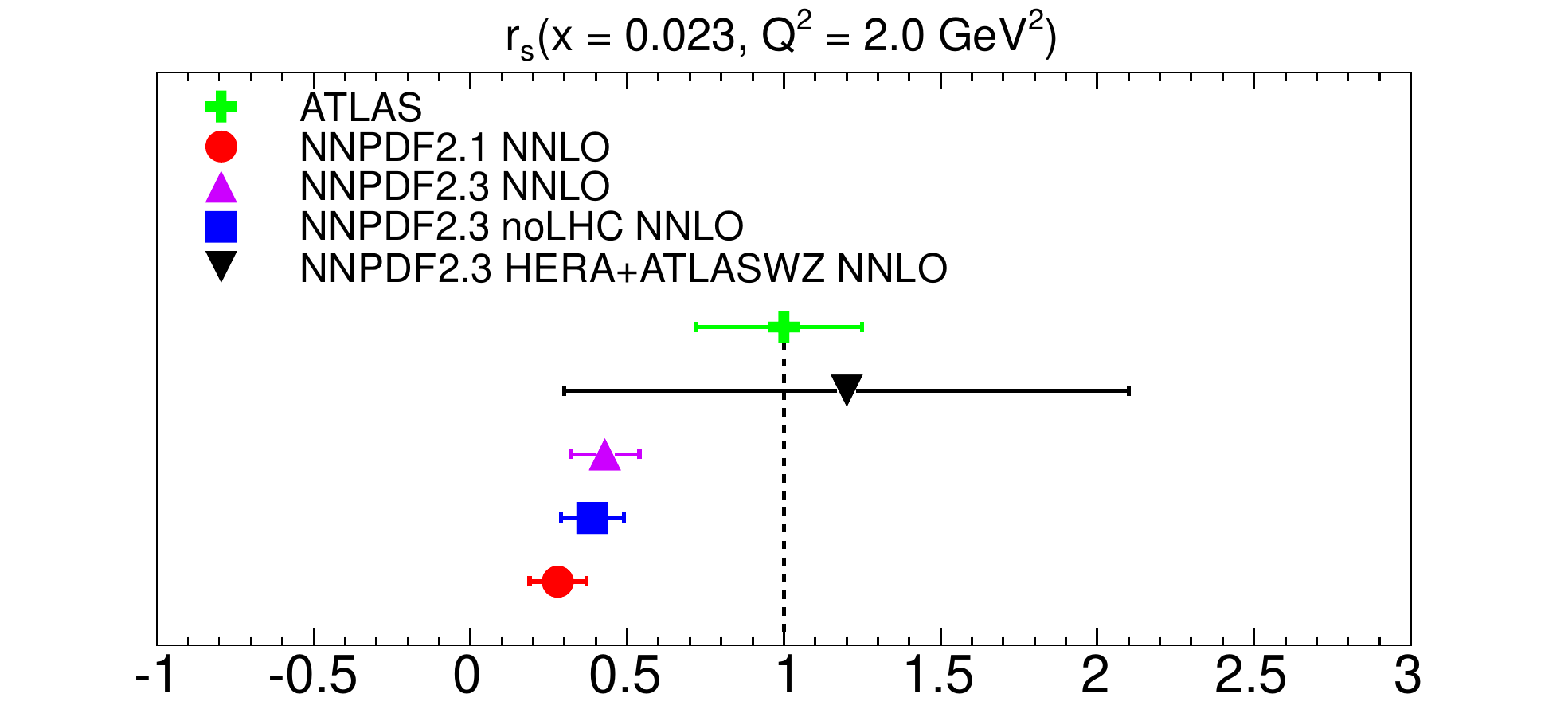}
\includegraphics[width=0.48\textwidth]{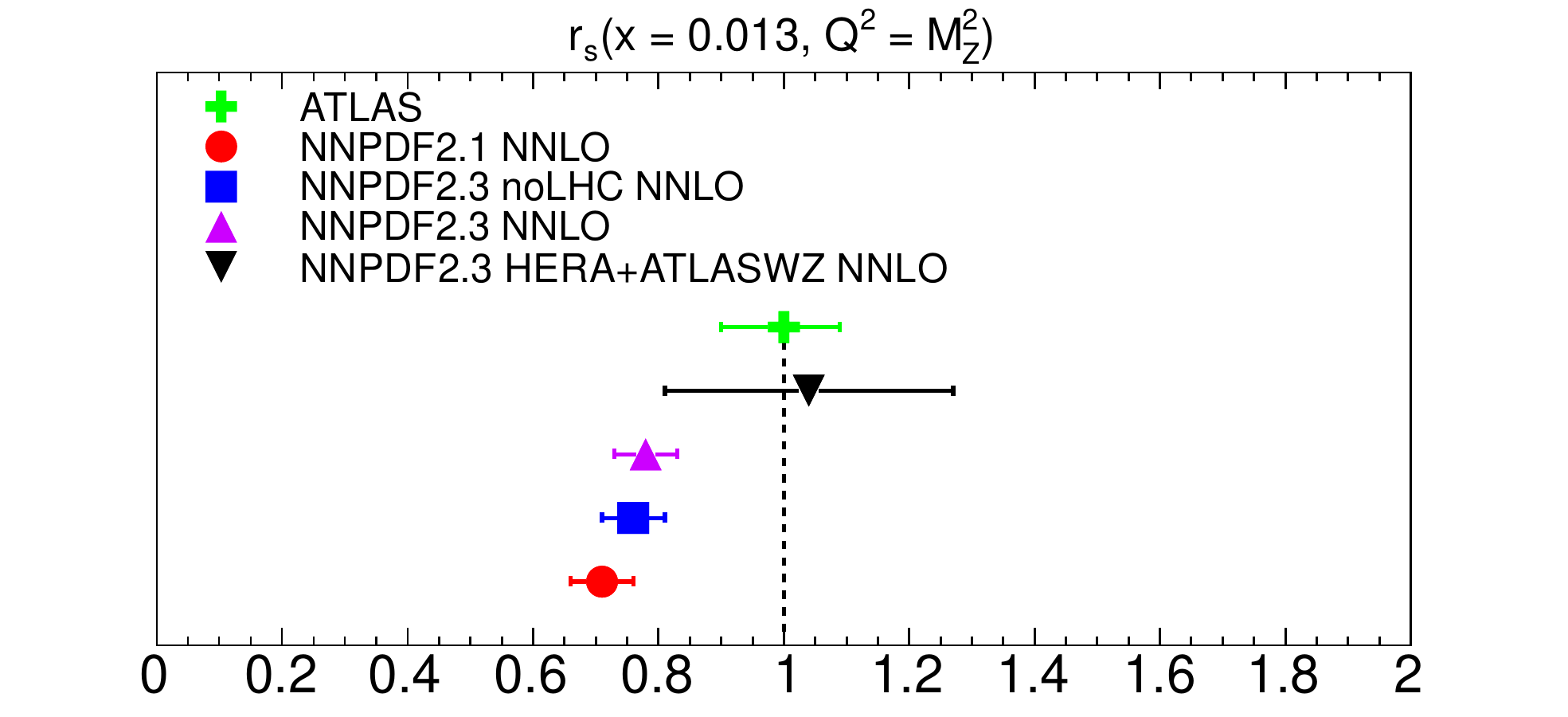}\\
\includegraphics[width=0.48\textwidth]{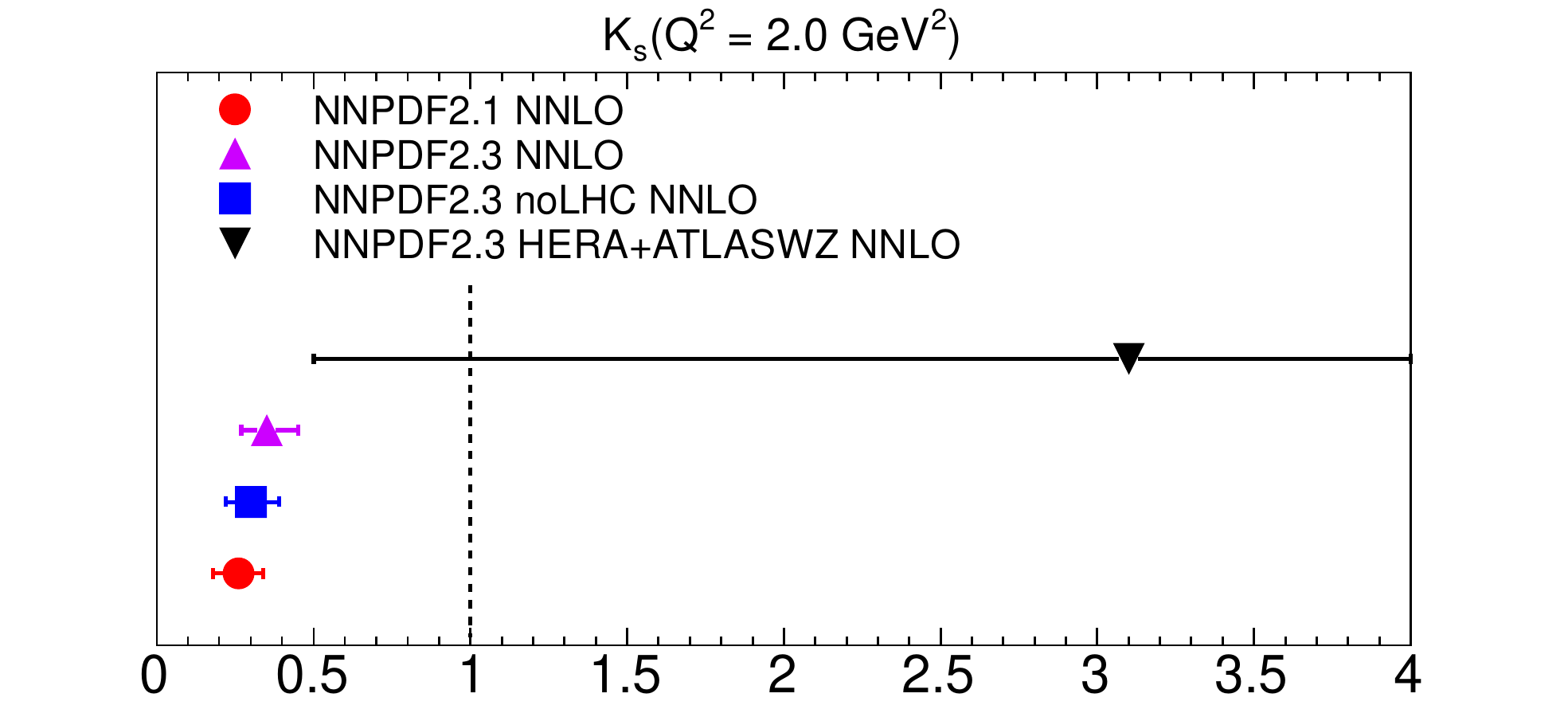}
\includegraphics[width=0.48\textwidth]{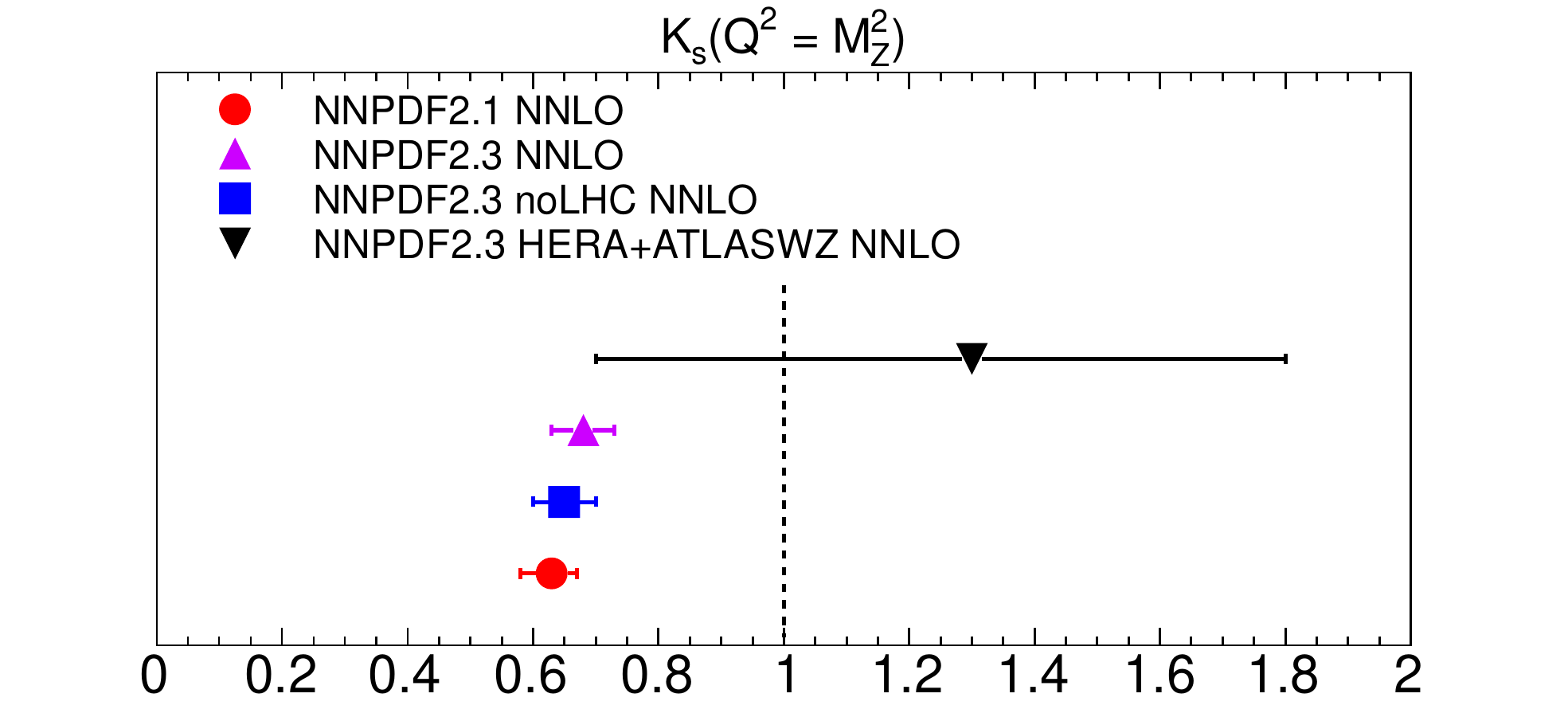}\\
\caption[ Results on the strangeness fraction of the proton from restricted dataset fits ]{Results on the strangeness fraction of the proton from restricted dataset fits. Results are shown for $r_s$ with the ATLAS kinematics (top plots) and for the integrated strangeness fraction $K_s$ (below). Values are given for NNPDF2.1, NNPDF2.3, NNPDF2.3 noLHC and the NNPDF2.3 HERA + ATLAS $W/Z$ dataset.}
\label{fig:NNPDFrs}
\end{figure}

To investigate the ATLAS result, an NNPDF2.3 fit was performed to the same dataset as in the ATLAS `epWZ' fit. The results of this fit for both the $r_s$ values quoted by the ATLAS collaboration, and the integrated $K_s$ values are shown in Figure~\ref{fig:NNPDFrs}. While the results of the NNPDF2.3 series fits to global datasets remain incompatible with the ATLAS result, the results of all of the fits are perfectly compatible within the very large uncertainties of the NNPDF fit to the restricted ATLAS and HERA dataset used for the ATLAS result.

The much greater uncertainty present in the NNPDF fit to the HERA and ATLAS $W/Z$ dataset suggests that the uncertainty in the ATLAS result was underestimated significantly, a conclusion also reached by similar analyses by the MSTW and ABM groups\cite{Watt:2012tq,Alekhin:2014sya}.

\begin{figure}[h!]
\centering
\includegraphics[width=0.48\textwidth]{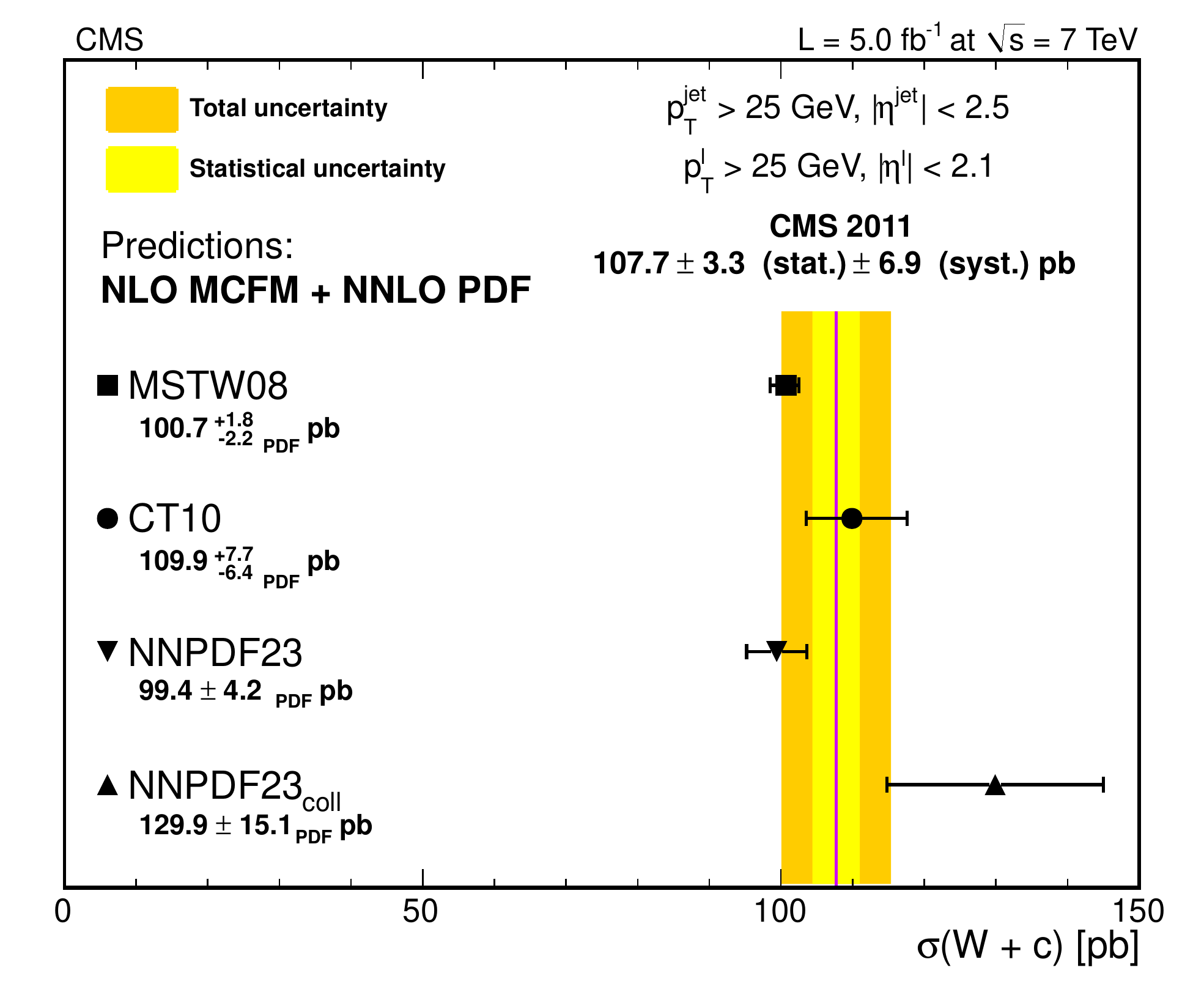}
\includegraphics[width=0.48\textwidth]{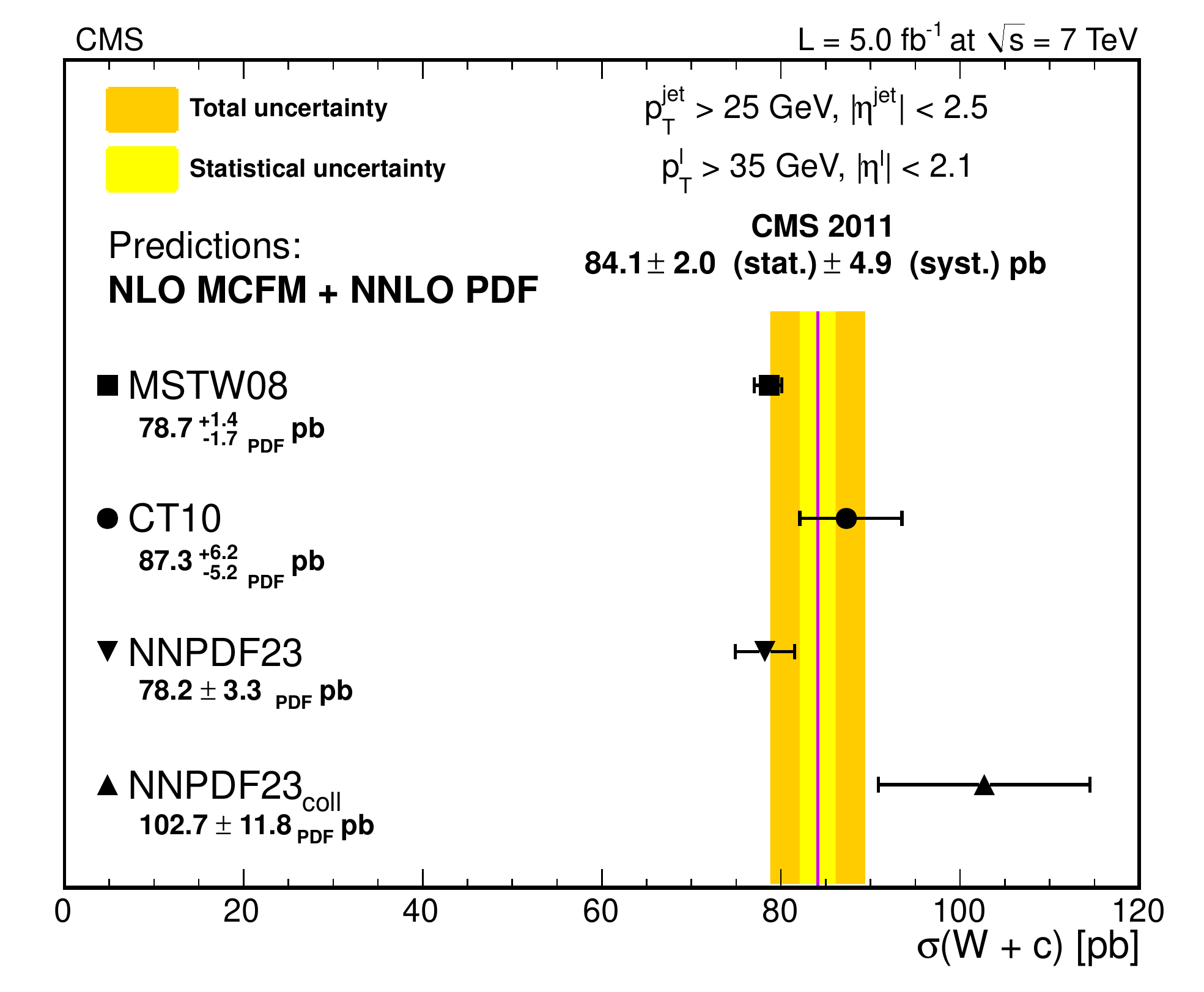}
\caption[CMS $W+c$ production data]{CMS $W+c$ production data, figure from~\cite{Chatrchyan:2013uja}.}
\label{fig:WplusC}
\end{figure}

 Measurements of $W+c$ production, particularly sensitive to the strange fraction can provide additional information for future fits. As an example, the CMS $W+c$ measurement based upon $5.0 $fb$^{-1}$ of $7$ TeV data~\cite{Chatrchyan:2013uja} demonstrates good agreement with the results of global PDF sets, as shown in Figure~\ref{fig:WplusC}.

\subsection{NNPDF2.3 phenomenology}

We will begin the study of the phenomenological applications of the NNPDF2.3 set and comparisons to previous determinations, by comparing computations of the LHC measurements included in the 2.3 fit. In this way the improvements in precision available for LHC standard candle predictions can be assessed. We shall follow by looking at some
typical total cross-sections of interest.

\begin{figure}[hp]
\centering
\includegraphics[width=0.48\textwidth]{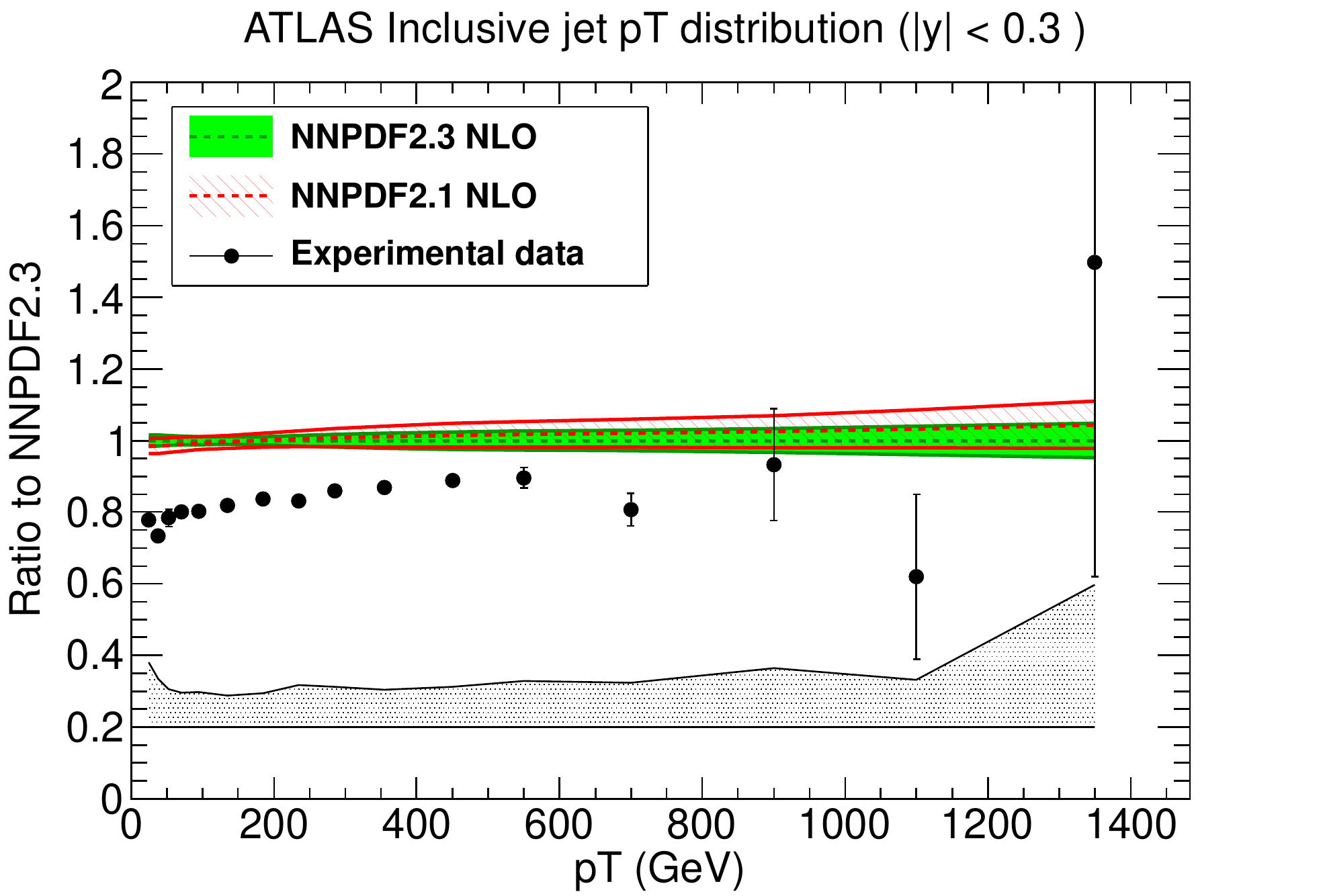}
\includegraphics[width=0.48\textwidth]{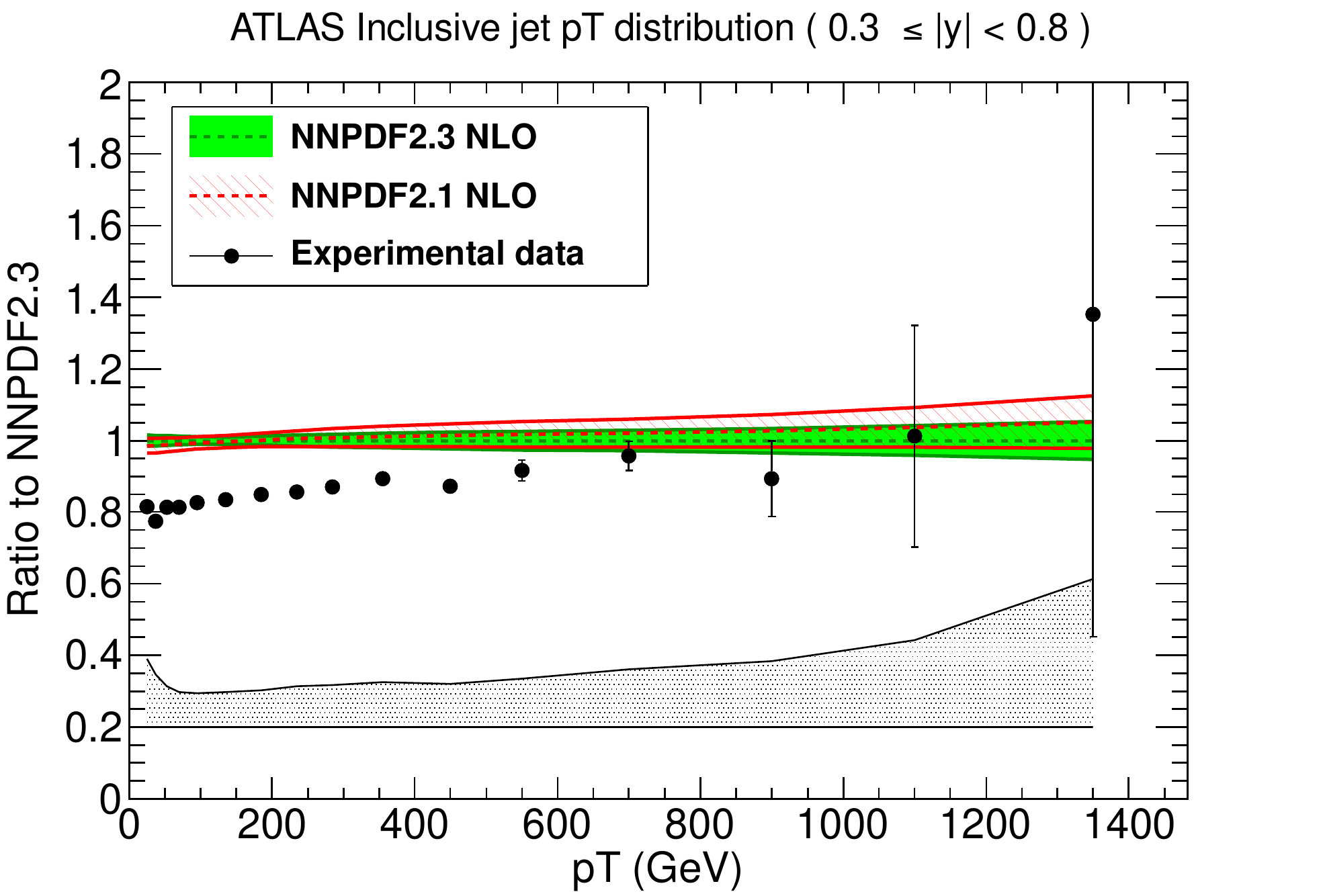}
\includegraphics[width=0.48\textwidth]{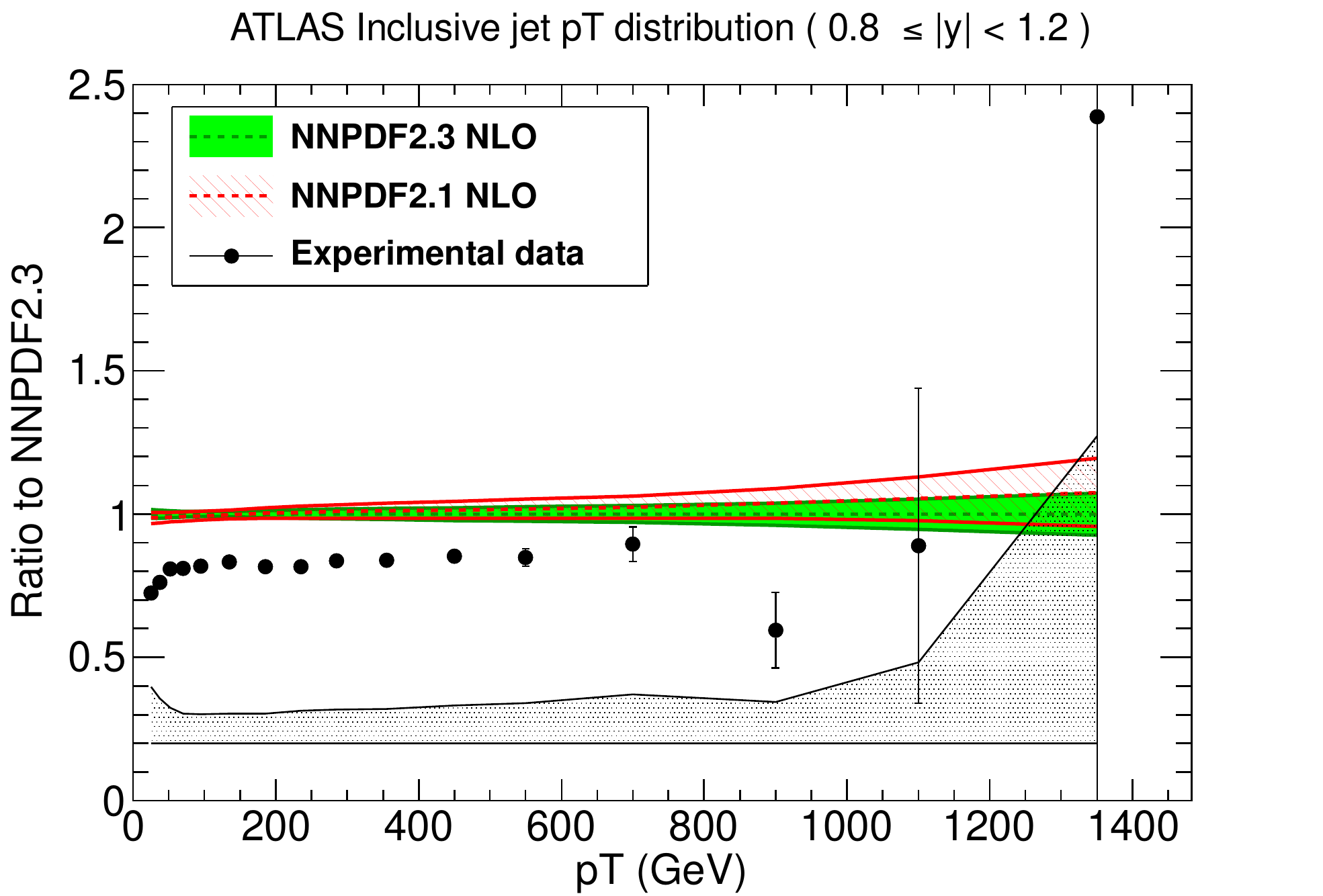}
\includegraphics[width=0.48\textwidth]{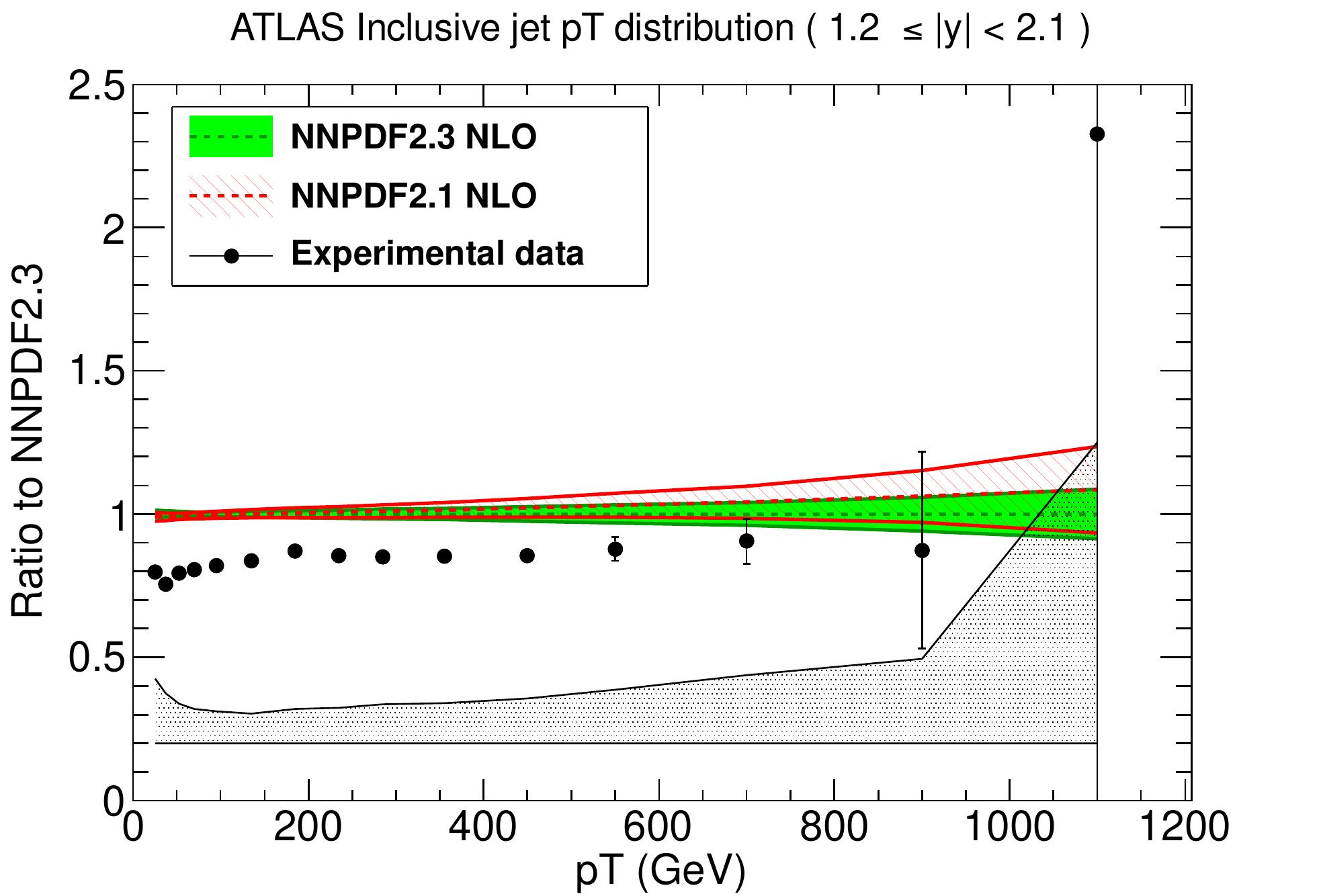}
\includegraphics[width=0.48\textwidth]{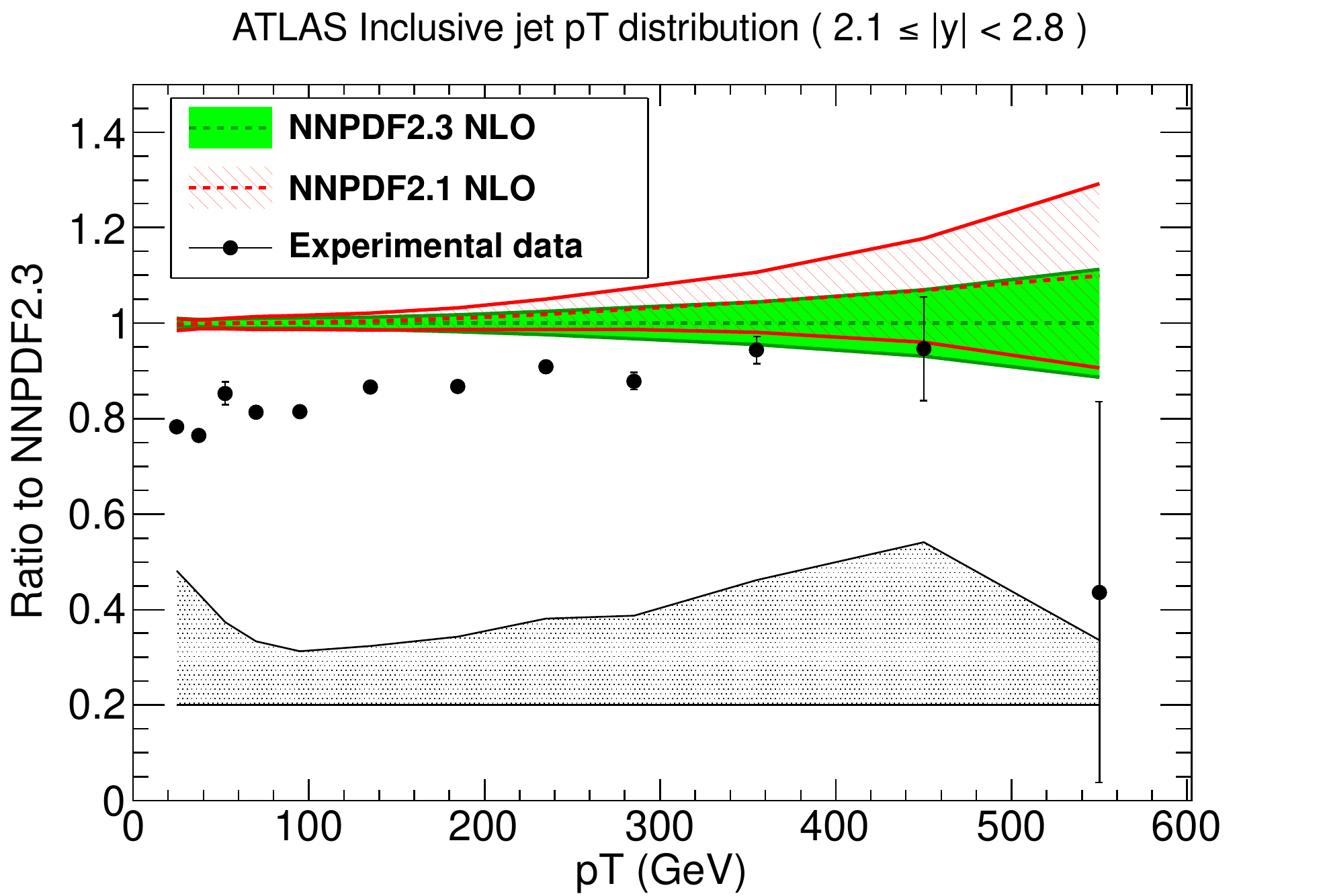}
\includegraphics[width=0.48\textwidth]{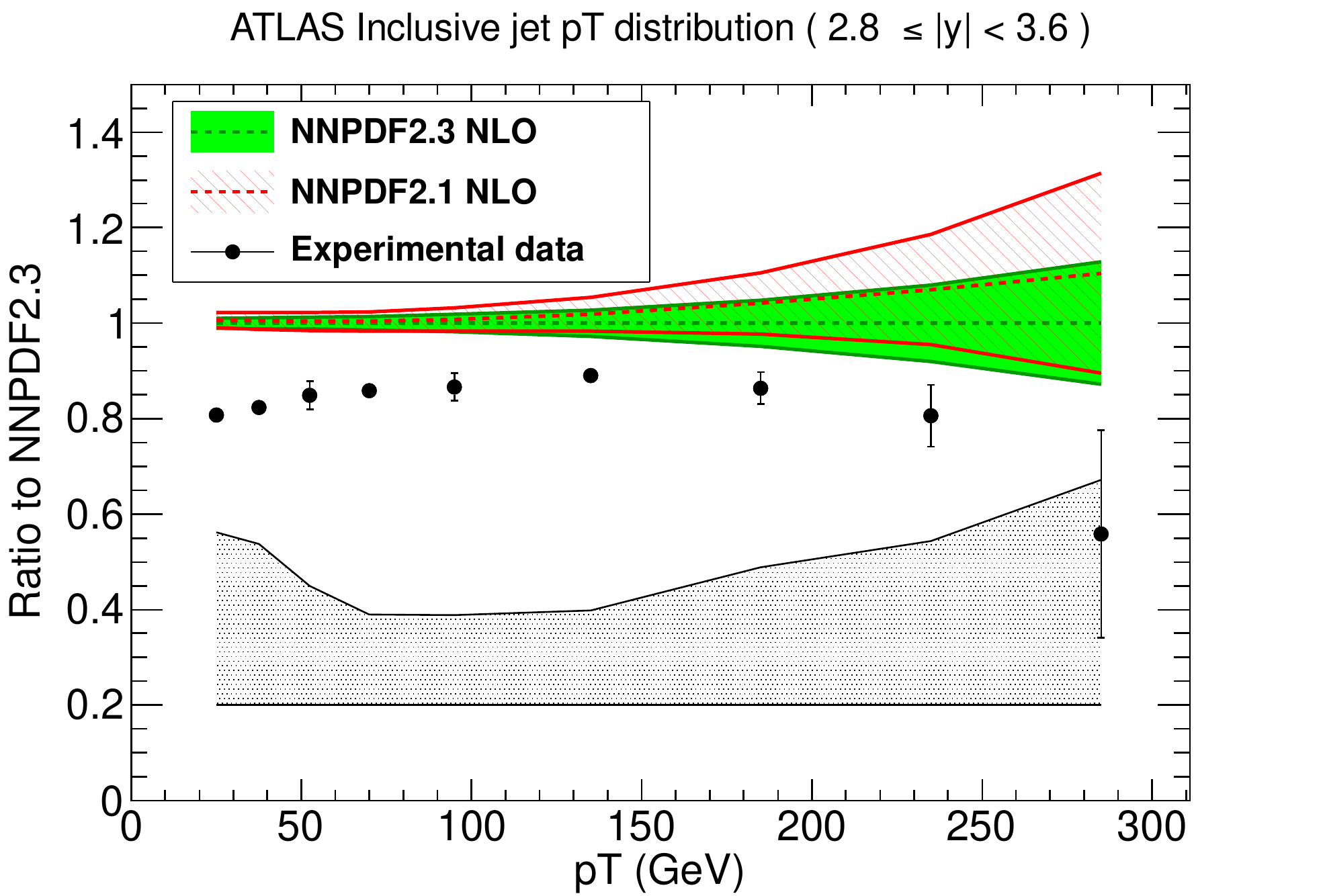}
\caption[Predictions for the ATLAS 2010 inclusive jet data, using NNPDF2.1 and NNPDF2.3]{Predictions for the ATLAS 2010 inclusive jet data, using NNPDF2.1 (green) and NNPDF2.3 (red). The grey band at the bottom of each figure represents the systematic uncertainty in the data, while the error bars are the statistical error only. Predictions are given for all rapidity bins for the $R=0.4$ data as included in NNPDF2.3}
\label{fig:ATLASjetspred}
\end{figure}

The impact of the ATLAS inclusive jet measurements upon NNPDF is made clear in Figure~\ref{fig:ATLASjetspred} where the data is compared to the predictions from the NNPDF2.1 and NNPDF2.3 sets. Uncertainties on the predictions are reduced across all datapoints, and there is a general shift to lower values of the differential cross-section. Despite the shift downwards, the theory remains systematically above the experimental datapoints. However the dataset suffers from relatively large systematic uncertainties, within which both NNPDF2.1 and NNPDF2.3 are consistent as demonstrated by the excellent agreement at the level of $\chi^2$ shown in Table~\ref{tab:23chi2}.

\begin{figure}[h!]
\centering
\includegraphics[width=0.48\textwidth]{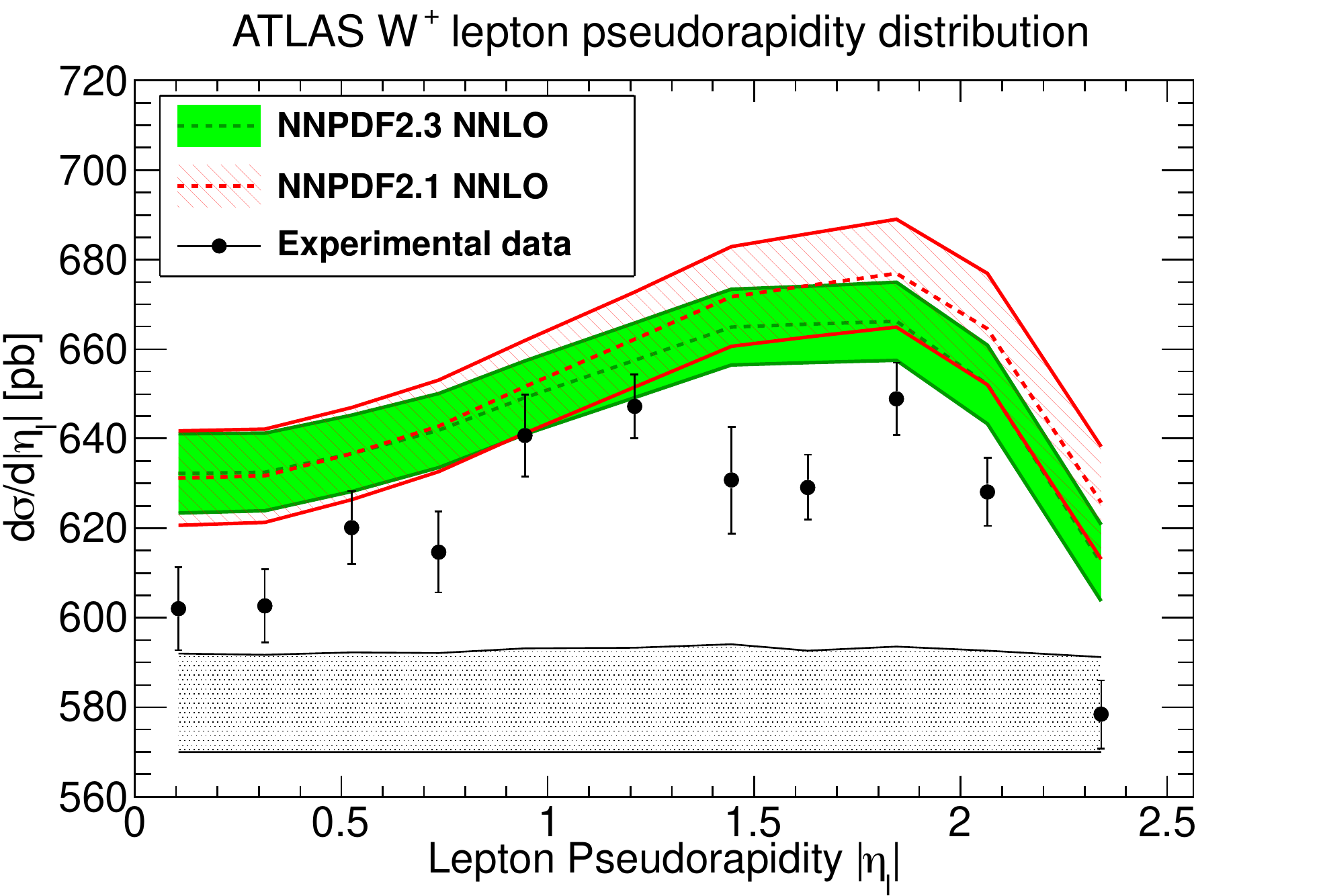}
\includegraphics[width=0.48\textwidth]{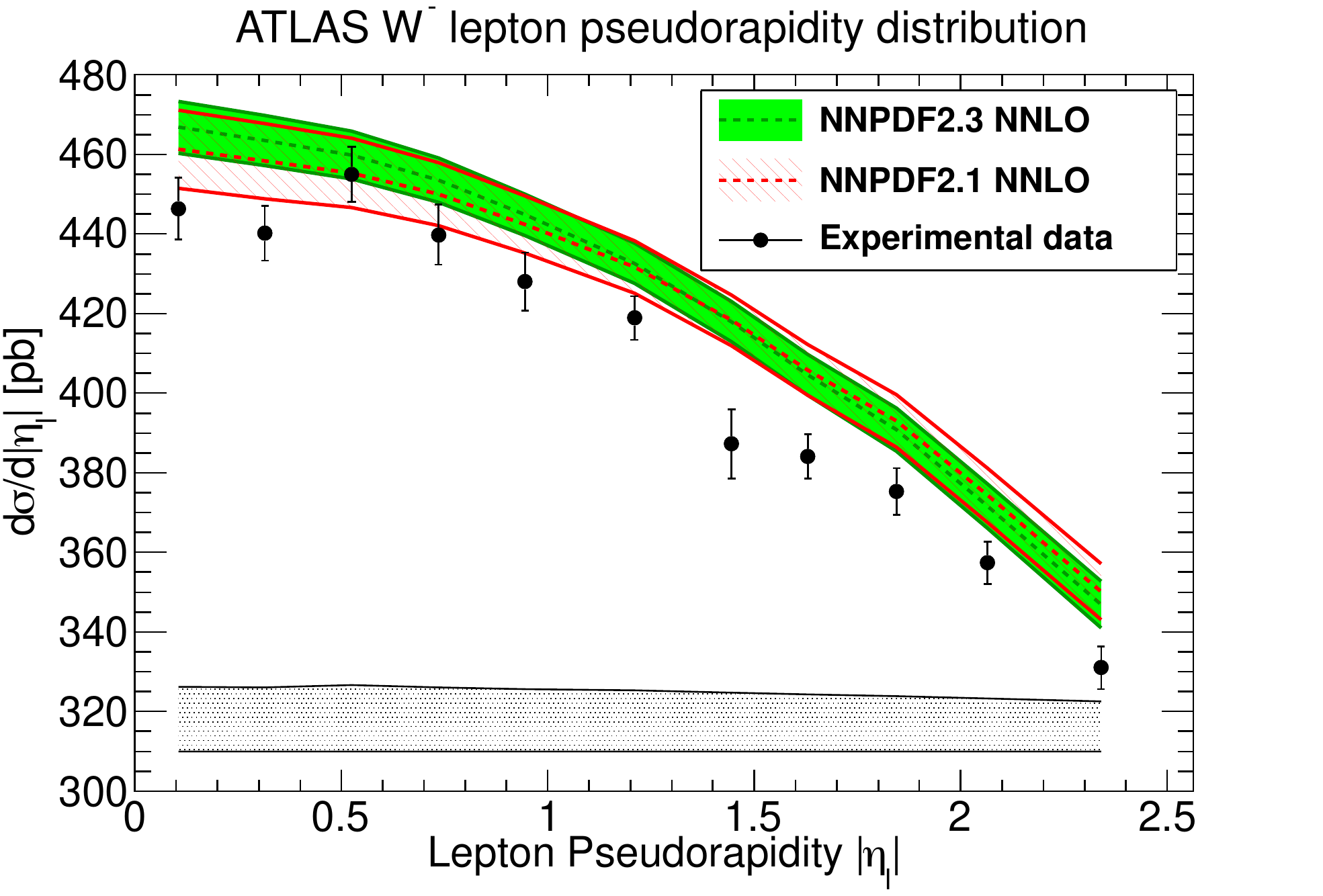}
\includegraphics[width=0.48\textwidth]{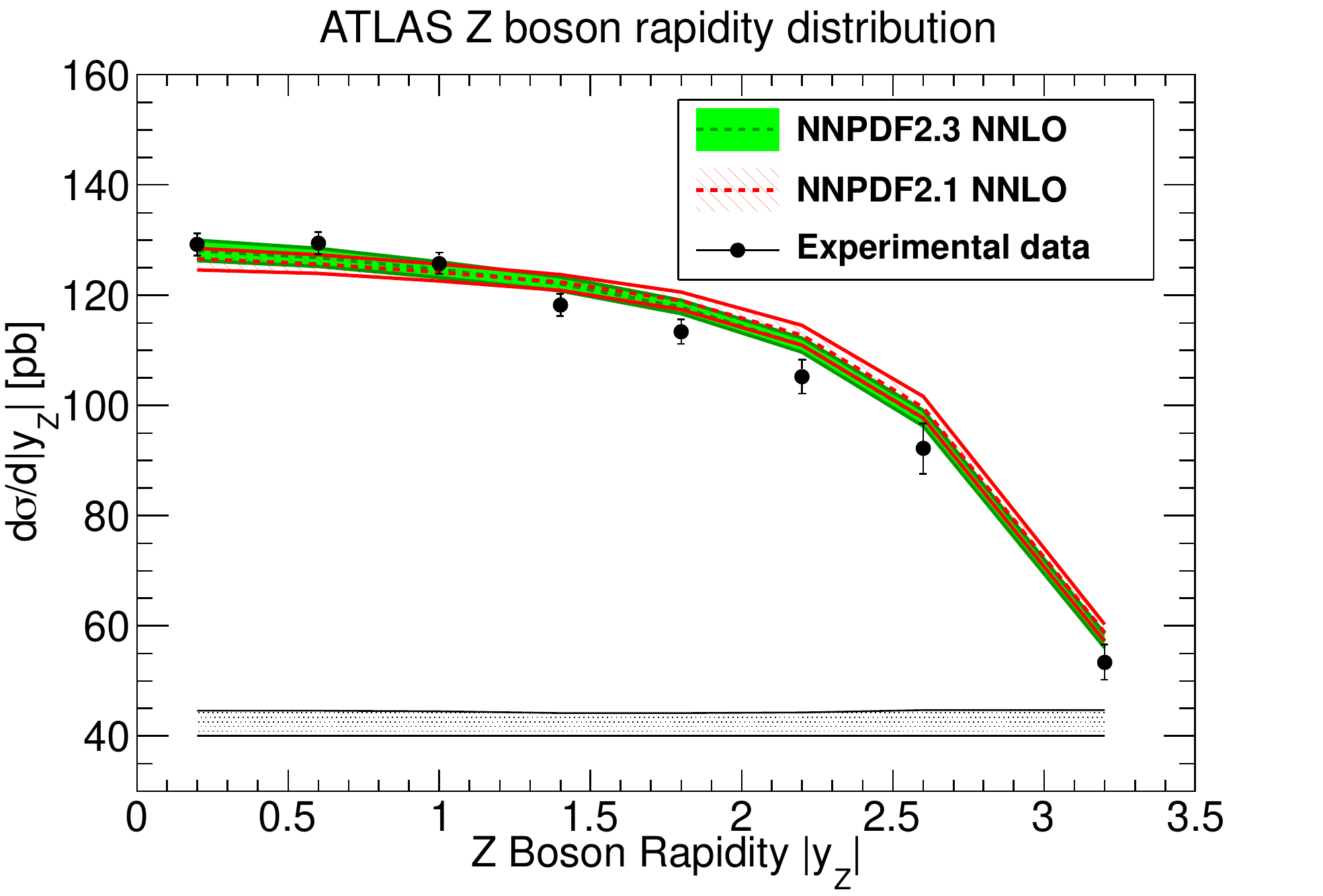}
\includegraphics[width=0.48\textwidth]{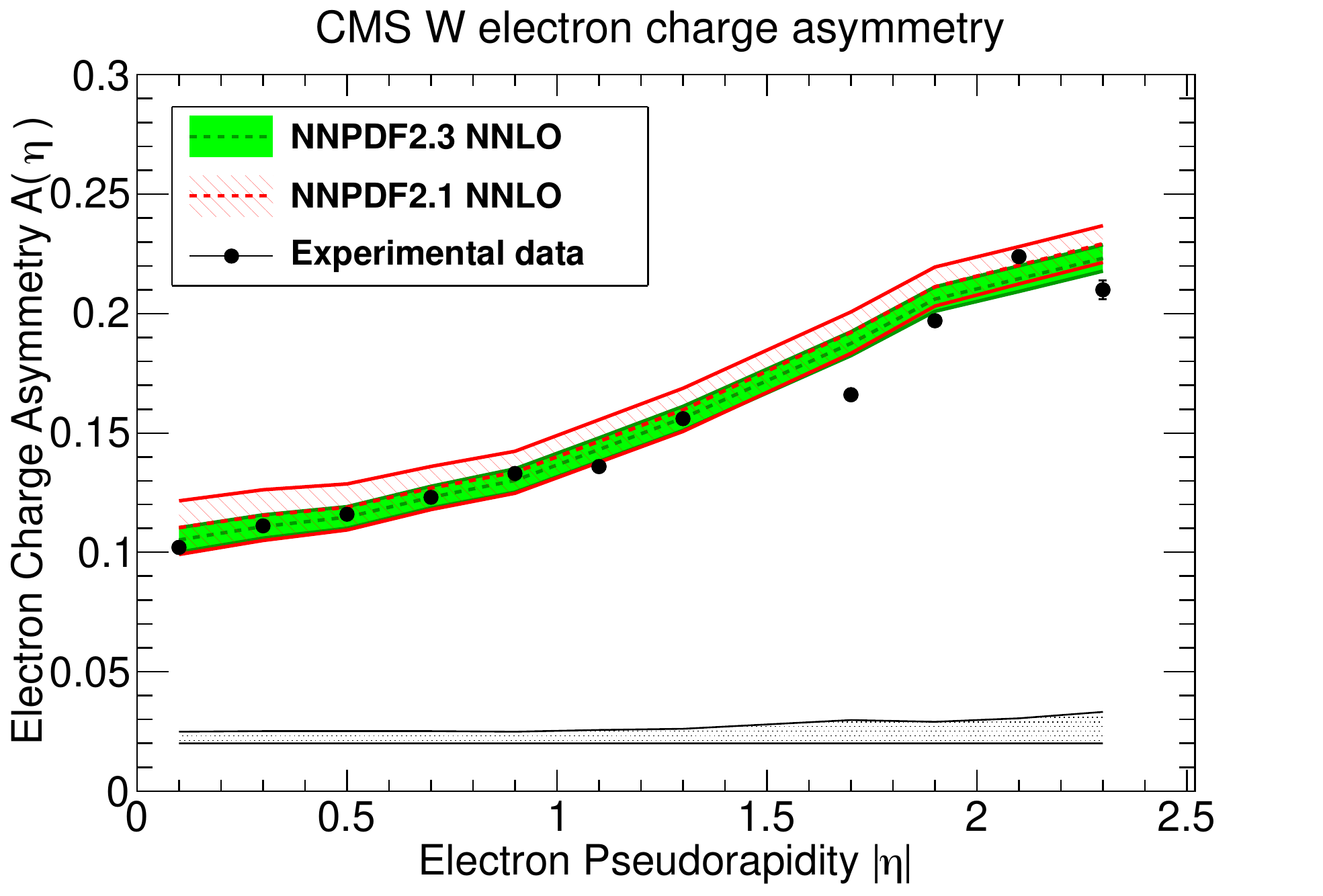}
\caption[Predictions for the ATLAS 2010 electroweak vector boson production and CMS 2011 $W$ electron asymmetry data, using NNPDF2.1 and NNPDF2.3]{Predictions for the ATLAS 2010 electroweak vector boson production and CMS 2011 $W$ electron asymmetry data, using NNPDF2.1 (green) and NNPDF2.3 (red). The grey band at the bottom of each figure represents the systematic uncertainty in the data, while the error bars are the statistical error only.}
\label{fig:ATLASplusCMSwzpred}
\end{figure}

\begin{figure}[h!]
\centering
\includegraphics[width=0.48\textwidth]{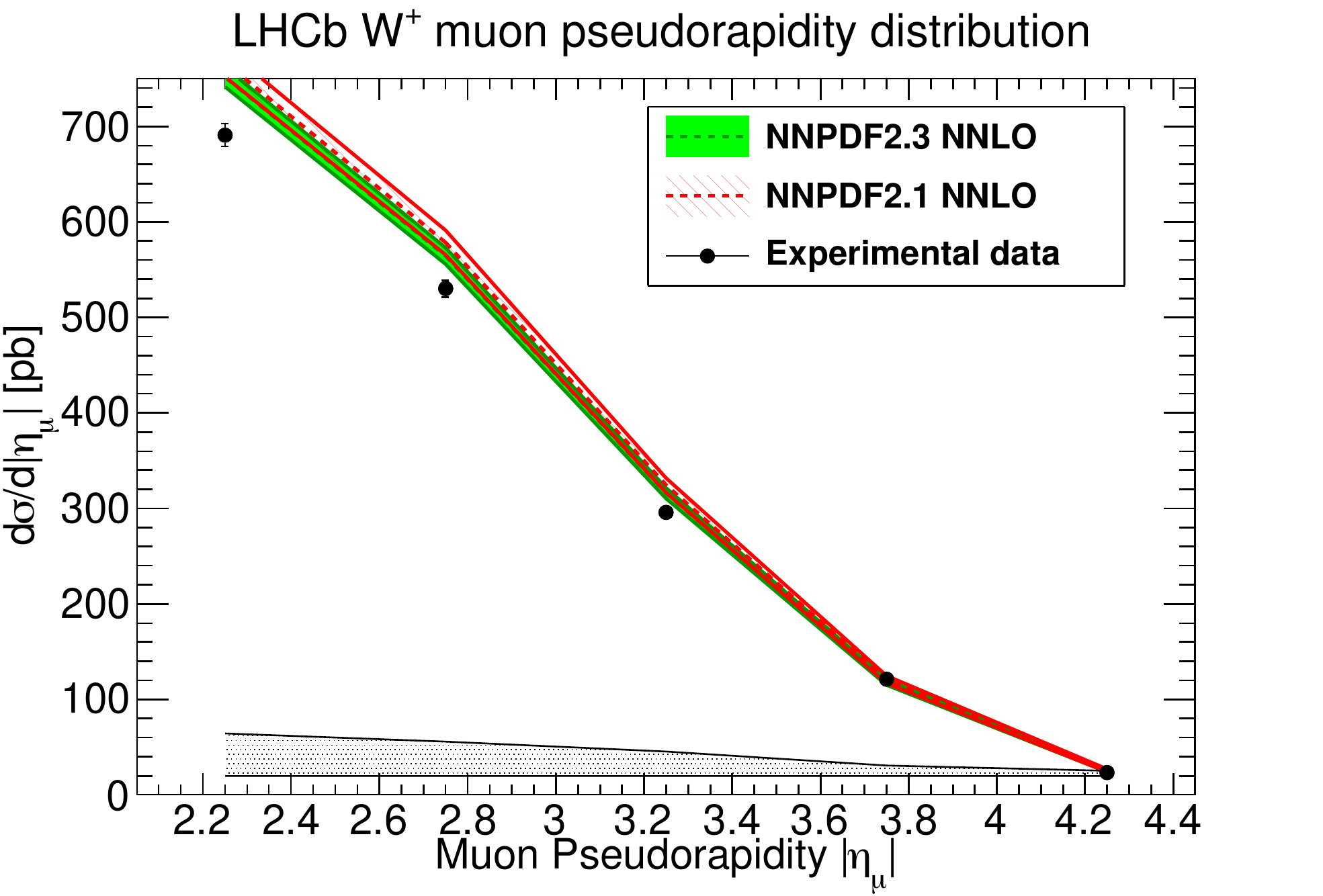}
\includegraphics[width=0.48\textwidth]{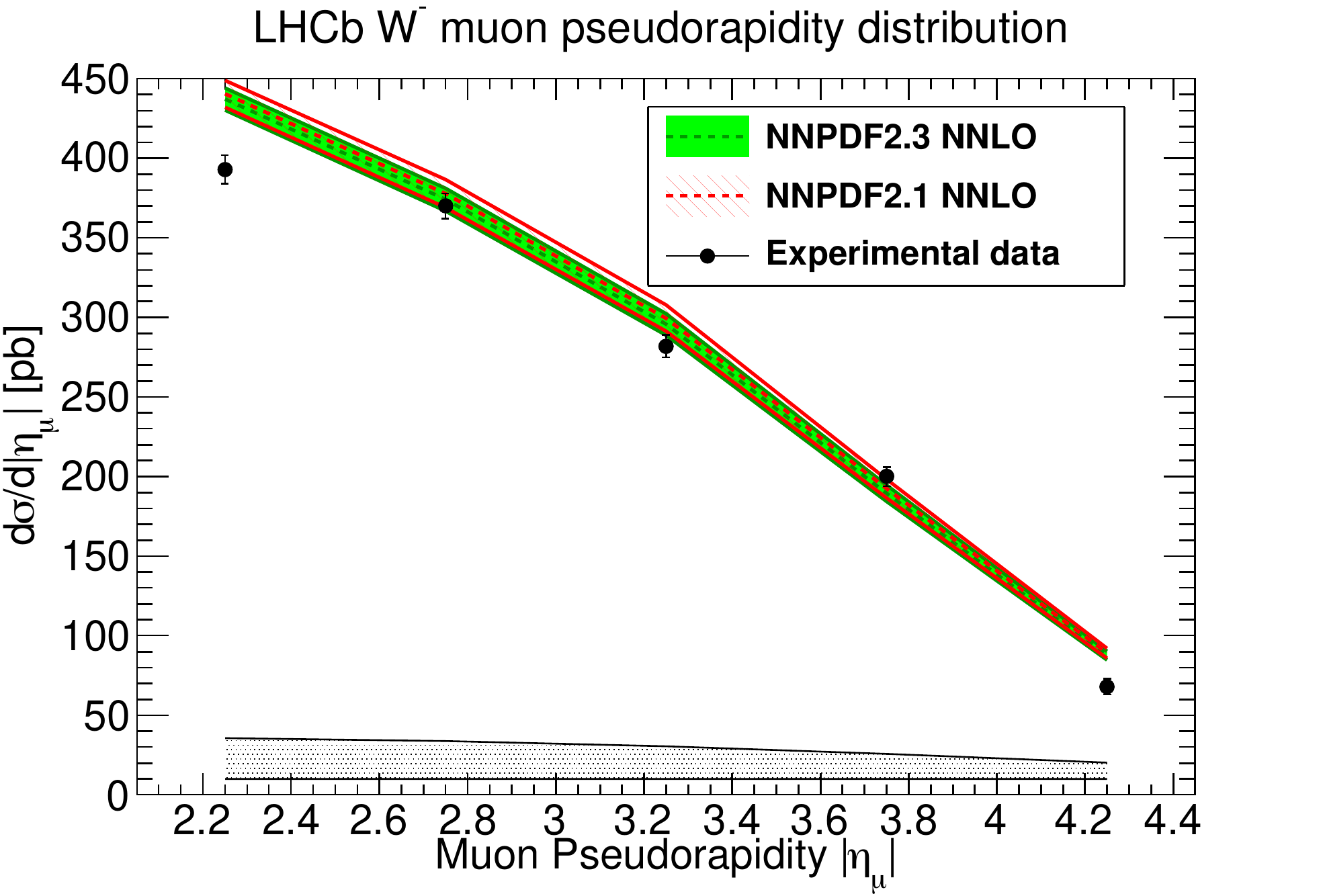}
\caption[Predictions for the LHCb 2010 $W$ boson production data, using NNPDF2.1 and NNPDF2.3]{Predictions for the LHCb 2010 $W$ boson production data, using NNPDF2.1 (green) and NNPDF2.3 (red). The grey band at the bottom of each figure represents the systematic uncertainty in the data, while the error bars are the statistical error only.}
\label{fig:LHCbWpred}
\end{figure}

\begin{figure}[ht]
\centering
\includegraphics[width=0.48\textwidth]{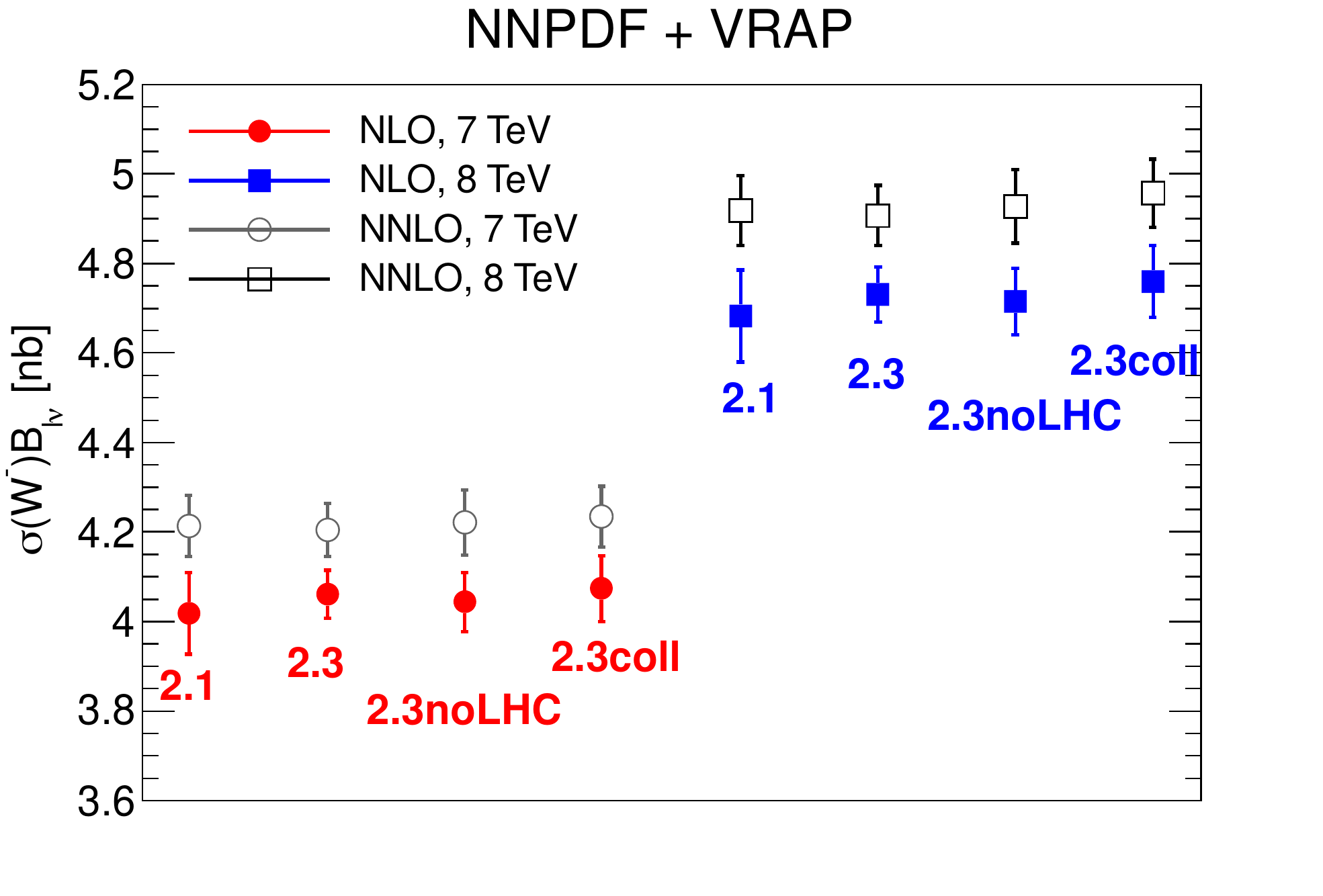}
\includegraphics[width=0.48\textwidth]{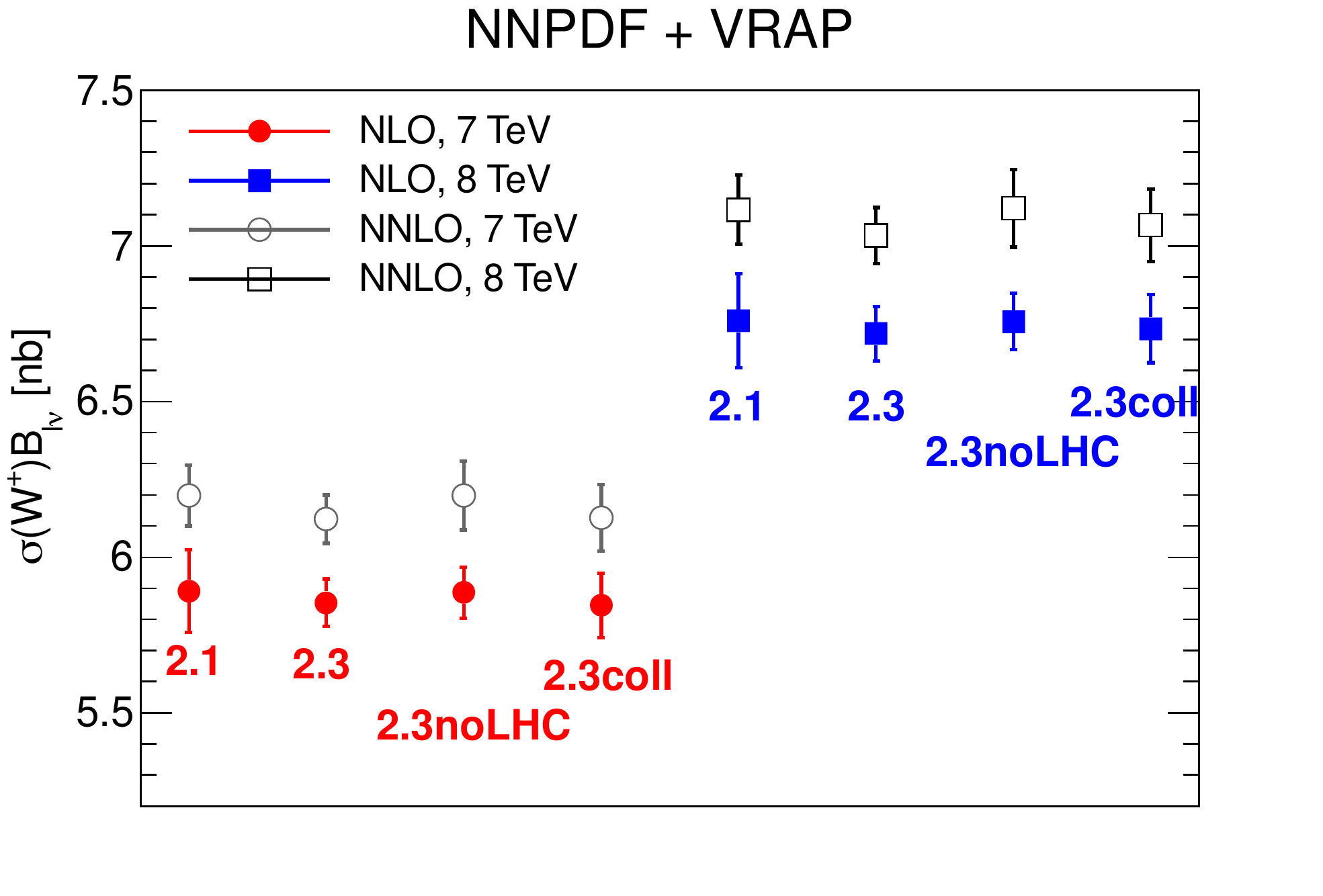}
\includegraphics[width=0.48\textwidth]{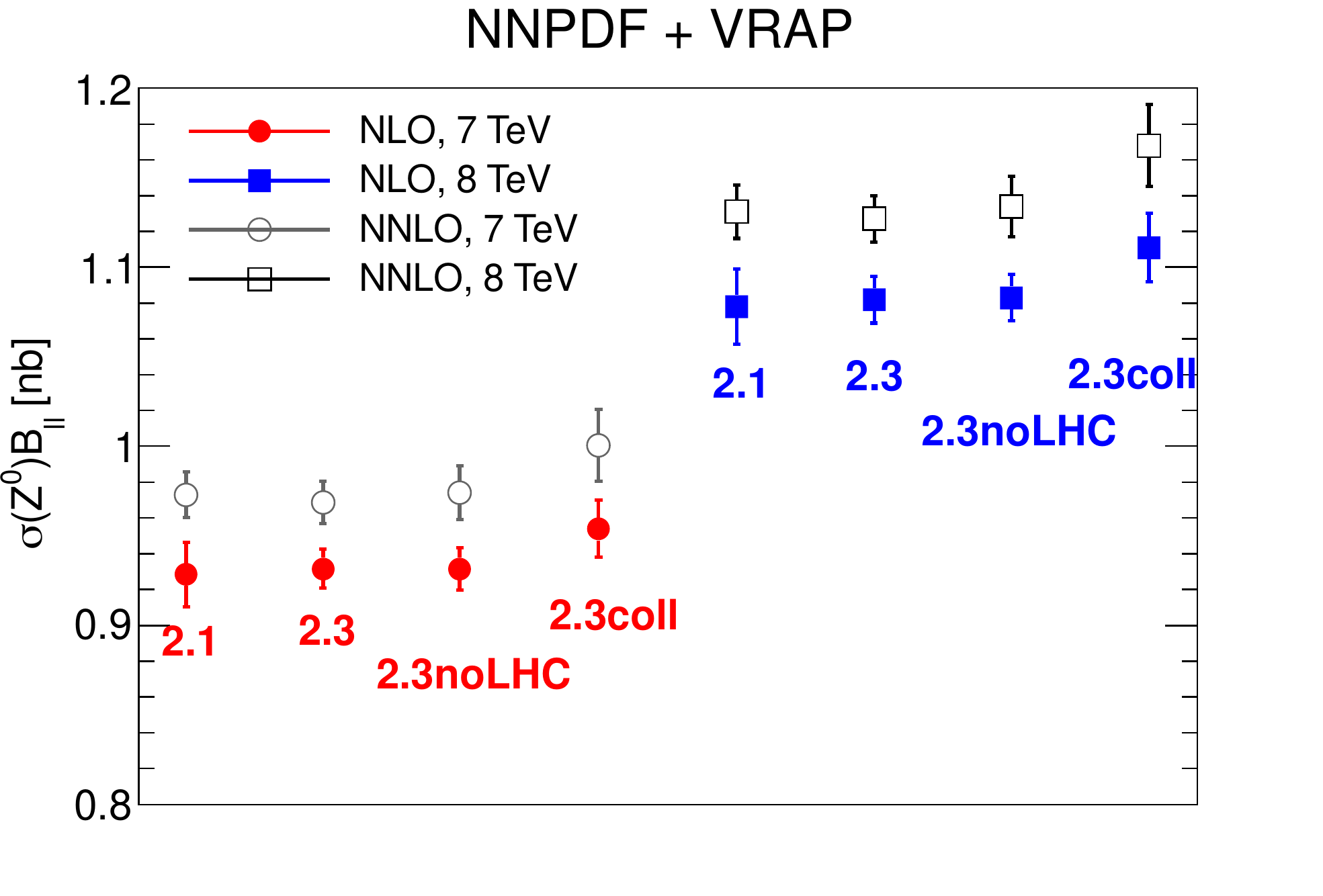}
\includegraphics[width=0.48\textwidth]{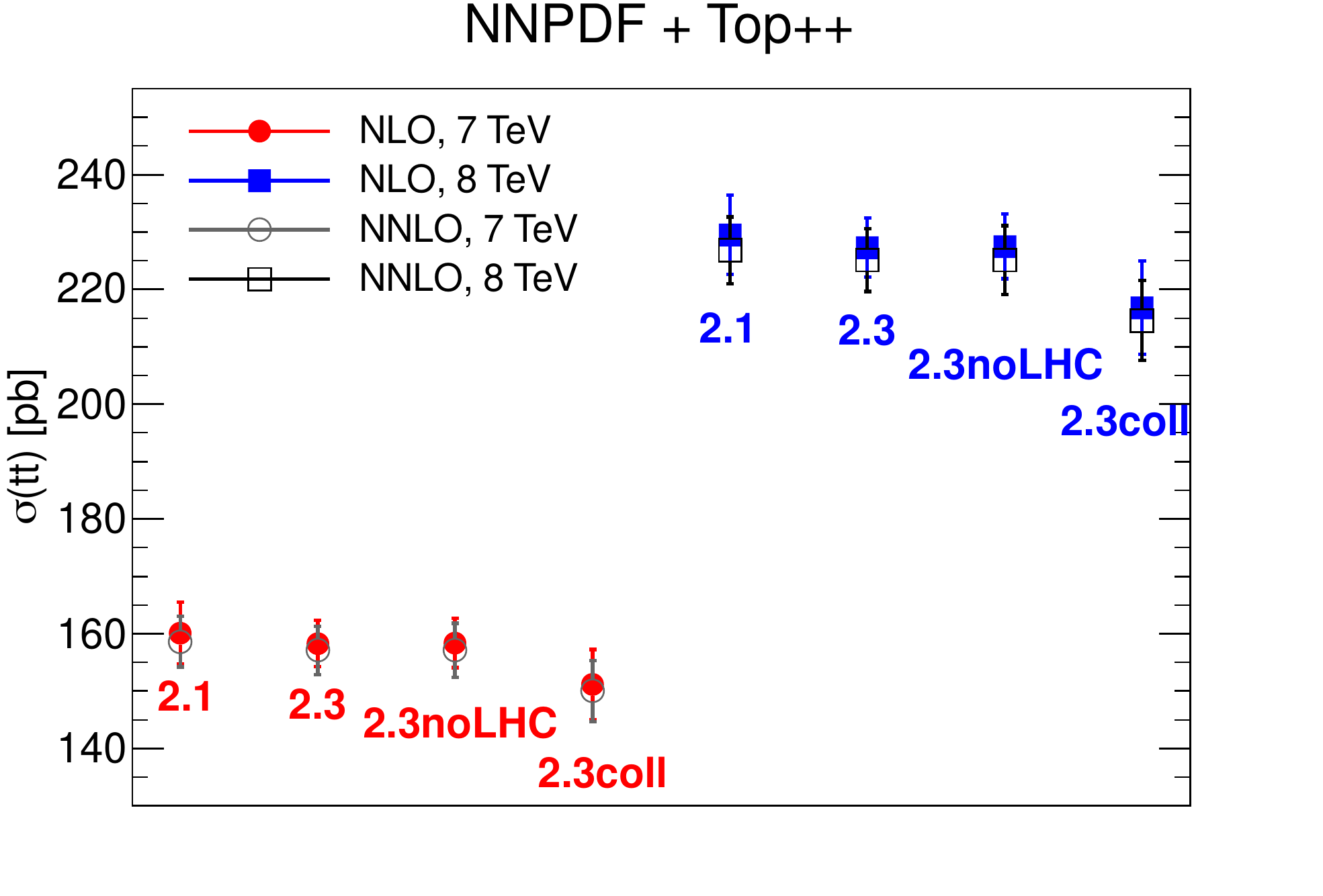}
\caption[Predictions for total  $W/Z$ and top production cross-sections at the 7 and 8 TeV LHC]{Predictions for total cross-sections at the 7 and 8 TeV LHC from NNPDF2.1 and the NNPDF2.3 series at NLO and NNLO. 7 TeV points are shown with circular markers, and 8 TeV with square markers. Theoretical predictions are given for the total $W^+$ (top left), $W^-$ (top right), $Z$ (bottom left) and $t\bar{t}$ (bottom right) production cross-sections.}
\label{fig:totalxsecWZt}
\end{figure}

\begin{figure}[ht]
\centering
\includegraphics[width=0.48\textwidth]{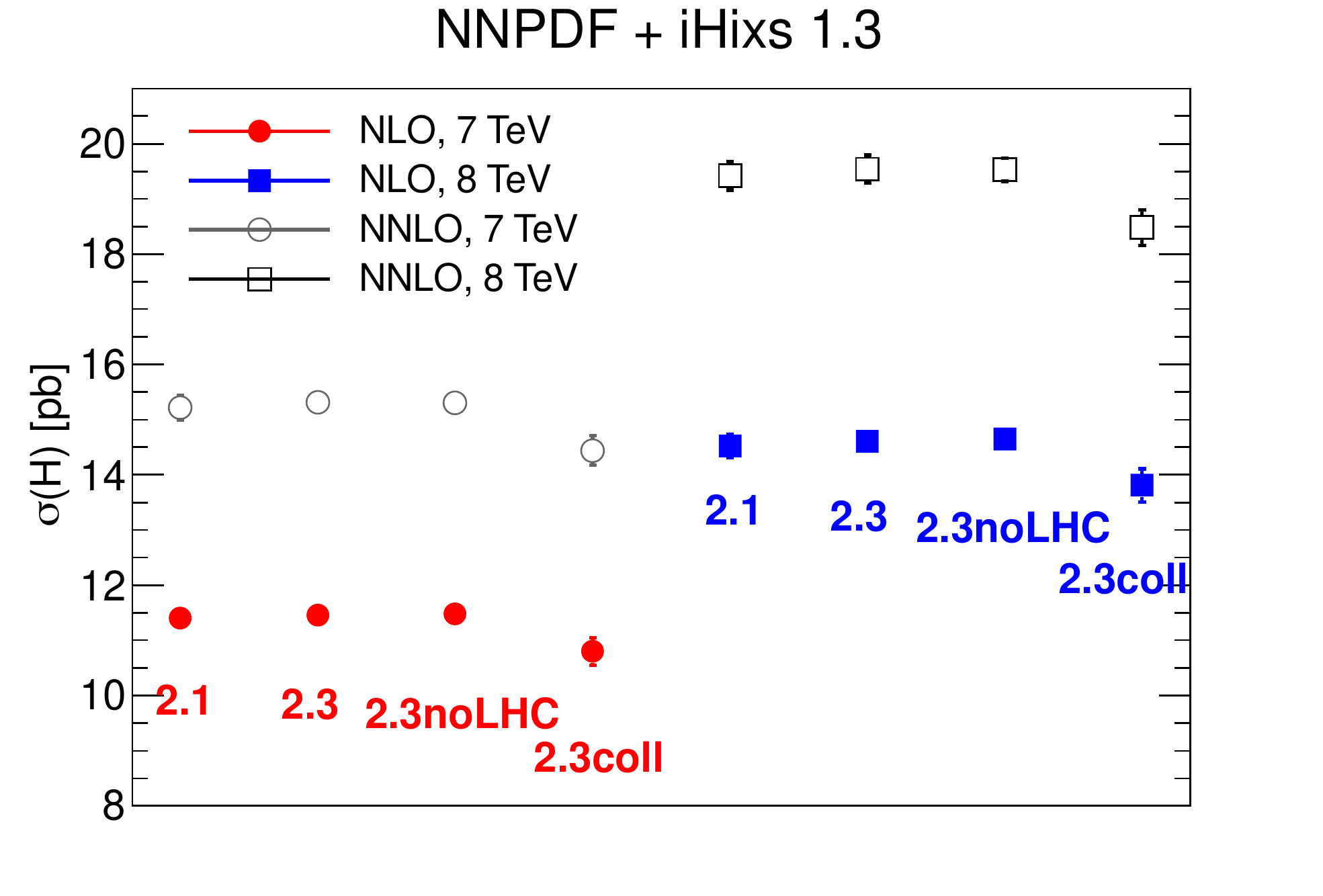}
\includegraphics[width=0.48\textwidth]{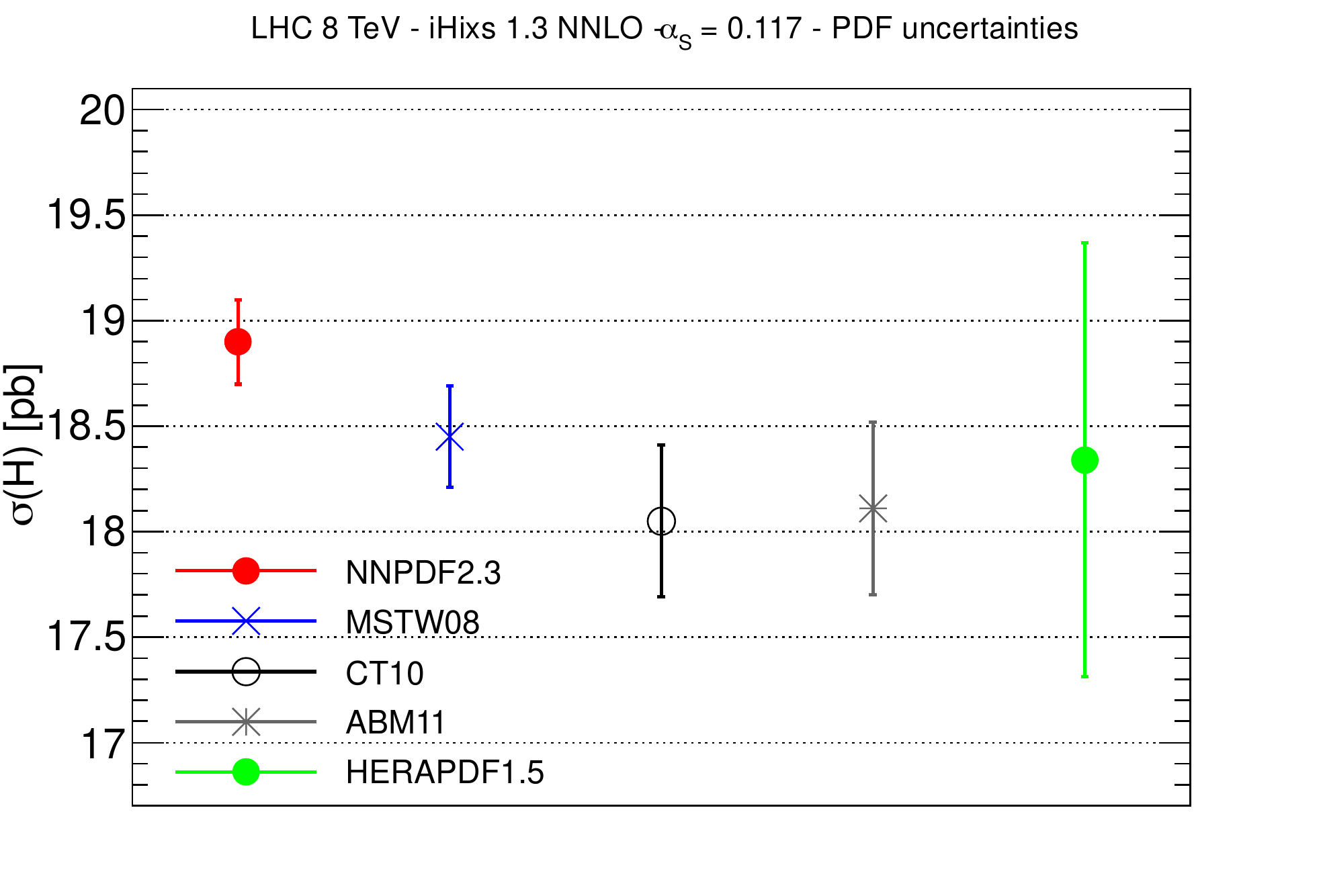}
\caption[Predictions for the total Higgs production cross-section in the gluon fusion channel at the 7 and 8 TeV LHC]{Total Higgs production cross-section in the gluon fusion channel at the 7 and 8 TeV LHC. Predictions are given at NLO and NNLO.}
\label{fig:totalxsecHiggs}
\end{figure}

In the electroweak sector, significant improvements are made across all observables included in the fit. Figure~\ref{fig:ATLASplusCMSwzpred} compares the predictions of NNPDF2.1 and NNPDF2.3 to the experimental data for the LHC electroweak measurements, demonstrating the improved agreement between theory and data in the ATLAS and CMS results, while Figure~\ref{fig:LHCbWpred} shows the same comparison for the LHCb data, demonstrating the improvements made in the very forward region measured by LHCb. The precise and consistent CMS data provide the clearest reduction of uncertainty of all the datasets, while the ATLAS and LHCb measurements suggest that the previous determination overestimated the electroweak cross-sections, leading to lower distributions with much improved agreement in the new fit.
\clearpage
Moving to inclusive cross-sections, predictions for total $W^\pm$ and $Z$ boson production, along with the total $t\bar{t}$ cross-section are shown in Figure~\ref{fig:totalxsecWZt}. Predictions for the electroweak observables were calculated using the {\tt VRAP}~\cite{Anastasiou:2003ds} code, and for the top predictions, {\tt top++}~\cite{Czakon:2011xx,Baernreuther:2012ws} was used. Predictions are provided for the $7$ TeV and $8$ TeV LHC with $\alpha_s(M_{\rm Z})=0.119$. In Figure~\ref{fig:totalxsecHiggs} the total cross-section for Higgs production in gluon fusion is shown with the same settings, predictions provided by {\tt iHixs}~\cite{Anastasiou:2011pi}. Results across the NNPDF2.1 and NNPDF2.3 sets demonstrate generally good consistency within their errors, with the NNPDF2.3 set providing the most precise predictions. The collider only determination is shown to be reasonably competitive when applied to the electroweak observables, where improved constraint is available from the LHC dataset. A similar pattern can be observed in the top and Higgs production observables, however errors remain systematically larger than for the global set.

The benefits of the NNPDF2.3 PDF set in phenomenological applications to LHC measurements are then clear, with the 2.3 set being the most precise and accurate determination in the NNPDF family.

\clearpage

\chapter{Fitting in the light of LHC data}
\label{ch:LHClight}
The addition of the first LHC datasets into an NNPDF fit allowed for important gains to be made in the precision of the resulting sets. However the potential dataset available for PDF determination from the LHC is increasing at a considerable rate, and datasets are being rapidly updated with more precise measurements. There is therefore still much more potential in the LHC to provide PDF constraint, especially in collider only fits.

With the ever enlarging dataset comes an important question: whether the fitting methodology applied to the pre-LHC dataset is still the best procedure for the extraction of precise parton densities in the LHC era. In order to accommodate the growing LHC dataset and to be able to efficiently explore methodological options, the toolchain used by the NNPDF collaboration had to be updated. The need for an updated fitting apparatus was recognised near the end of development of the NNPDF2.3 PDF set. Previous NNPDF sets were generated by a {\tt FORTRAN} codebase which grew out of the earliest NNPDF determinations. Consequently the codebase suffered from a great deal of inflexibility with regard to the treatment of data. In particular, performing varying cuts and fits to reduced or special datasets were complicated procedures. Additionally as the fits to the pre-LHC dataset were considerably less computationally intensive the core fitting apparatus was not designed with computational efficiency as the first priority, meaning that fits with the LHC dataset were rather sluggish. Beyond being a mere technicality, such slow fits actually meant that detailed studies of the methodology applied to an LHC dataset were prohibitively expensive in computer time.

With these issues in mind, the {\tt nnpdf++} project was initiated, whereby the full NNPDF toolchain has been implemented from scratch in {\tt C++}. The core of the project was built around the efficient $\tt FK$ method described previously, allowing for a much clearer separation in code between theoretical predictions and experimental data along with a much greater efficiency in the convolution. The {\tt FK} products themselves are accelerated via explicit use of {\tt SIMD} vectorisation, and {\tt OpenMP}~\cite{openmp08} provides multiprocessor options.

The framework was designed to be as modular as possible, to allow for the simple and safe modification of sections of the NNPDF methodology without requiring major modifications to the remaining codebase. The re-implementation of the whole NNPDF toolchain also provided an extremely thorough cross-check of the two implementations, and allowed for the step-by-step evaluation of several methodological elements. The results of this re-evaluation and investigation of alternative procedures shall be described in this chapter along with the consequences for future determinations.

\section{Closure testing}
\label{sec:closuretest}
The central element in the methodological review conducted with the {\tt nnpdf++} code after NNPDF2.3 is the closure testing procedure.

In a closure test, a PDF fitter takes their tools and applies them to a set of pseudo-experimental data generated from a known prior parton distribution set. Provided that the theory used to generate the pseudodata is identical to that used in the fitting procedure, the results of the fit should reproduce the generating function to within the estimation of PDF error. The test is an extremely sensitive check of a fitting procedure, in that it tests the ability of a methodology to resolve the underlying law when said law is known exactly. The method can also be used to study the effect of data inconsistencies by artificially modifying data uncertainties as is examined in Ref.~\cite{Watt:2012tq}, however here we shall restrict ourselves to examining the quality of reproduction of the underlying law.

Closure tests in the NNPDF methodology can be performed in a number of ways. One possible method is a direct fit to theoretical predictions generated from a known distribution, in this way the pseudo-dataset is free from the statistical noise that would be present in experimental data. $N_{\text{rep}}$ PDF replicas are then fitted to the theory predictions, without performing the generation of a Monte Carlo artificial data sample. In this type of fit one aims to reproduce as well as possible the generating function at the end of the fit. As no statistical noise is inserted at any point the final fit quality should approach $\chi^2=0$, we shall therefore denote such a fit a \emph{level zero} closure test.

Alternatively one may perform a fit where statistical noise is introduced to the pseudo-dataset according to the experimental uncertainty present in the real dataset. This can be done in two ways; either the noise is introduced directly to the pseudo-data itself whereby all Monte Carlo replicas fit to the same noisy sample, or noise is introduced on a replica-by-replica basis as in the normal Monte Carlo procedure. These types of fit we denote \emph{level one} closure tests.

\begin{figure}[hp]
\centering
\includegraphics[width=0.9\textwidth]{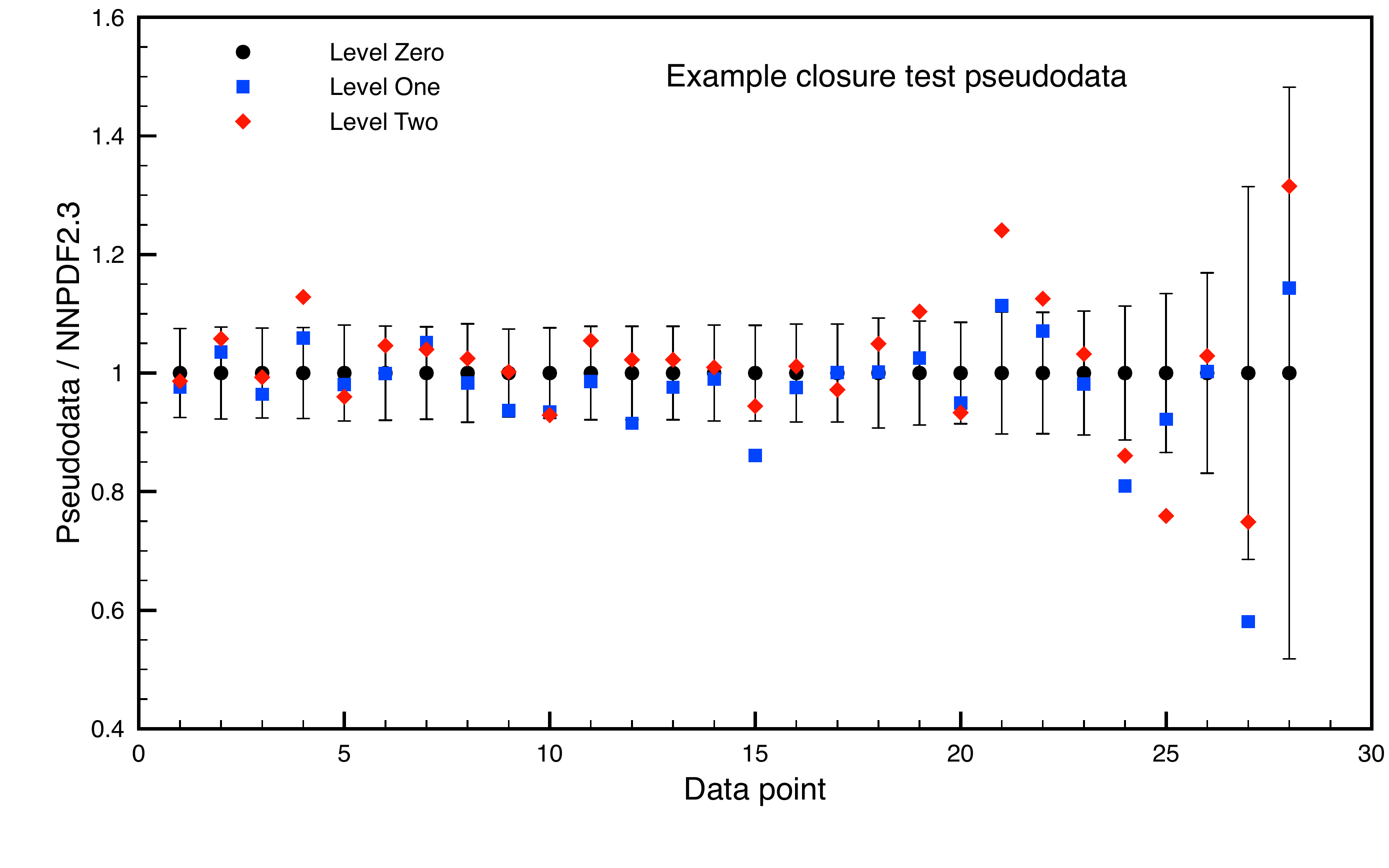}
\caption[Closure test pseudodata example]{Examples of pseudodata used in a closure test for all three levels. The black circles show the level zero pseudodata, and the experimental error bars. The blue squares show the pseudodata after experimental noise has been simulated (level one) and the red diamonds after both statistical noise simulation and Monte Carlo replica generation (level two). All points are normalised to the generating PDF set (NNPDF2.3).}
\label{fig:closurepseudodata}
\end{figure}

Finally one can introduce two levels of noise to the data. The first; applied directly to the pseudo-data, simulates the experimental noise in the distributions. The second level is introduced through the normal Monte Carlo generation of artificial data replicas. This is denoted a \emph{level two} closure test and is the closest to a full fledged PDF fit. The main exception here being the lack of any inconsistency between datasets, as they have all been generated from the same initial distribution. In the case of a level two fit the PDF fitter wishes to reproduce the underlying law to within their quoted PDF uncertainties, the exact reproduction available at level zero is now unavailable due to the introduced pseudo-experimental noise. The level two fit is therefore the most stringent test of a fitting procedure in that it tests the central claim of a fitting group; that the underlying law should lie within the quoted PDF uncertainty band at the quoted confidence level. The settings used in the different closure tests are summarised in Table~\ref{tab:closurelevel}. As a direct comparison of some example pseudodata, Figure~\ref{fig:closurepseudodata} shows example data at closure test levels zero, one and two.

The new structure present in the {\tt nnpdf++} code, particularly the modular treatment of experimental data and theoretical predictions, allows for the straightforward use of predictions in the place of experimental data while keeping the experimental covariance matrices intact. The closure testing method has therefore been extensively applied to the development of the NNPDF methodology, with the procedure used for the NNPDF3.0 determination being guided largely by results from closure testing. Here we shall outline some general results, before demonstrating the application of the procedure to methodological development in the subsequent sections.

\begin{table}[h!]
\label{tab:closurelevel}
\begin{center}
\begin{tabular}{|c|c|c|}
\hline
C.~Level & Exp.~Noise & Art.~Data\\ \hline
0  &  \text{\sffamily X} & \text{\sffamily X} \\
1a  & \checkmark & \text{\sffamily X} \\
1b  & \text{\sffamily X} & \checkmark \\
2  & \checkmark & \checkmark \\
\hline
\end{tabular}
\caption[Levels available in a closure test fit]{Levels available in a closure test fit (C.~Level), Exp.~Noise corresponds to simulating experimental noise in the pseudodata sample. Art.~Data refers to the generation of artificial data replicas in the Monte Carlo uncertainty procedure.}
\end{center}
\end{table}%

\clearpage

\subsubsection{Early closure tests}
The earliest NNPDF closure tests were conducted to assess the usefulness of the procedure, and performed with the full NNPDF2.3 procedure. As an initial test, a fit was performed to the toy PDF parametrisation as used in the Les Houches evolution benchmarks~\cite{Giele:2002hx}, a parametrisation based upon the CTEQ5M determination~\cite{Lai:1999wy}. In this set, the initial state distributions are given as

\begin{eqnarray}
\label{gsav-eq9}
  xu_v(x,\mu_{\rm f,0}^2)       &\! =\! & 5.107200\: x^{0.8}\: (1-x)^3,  
    \nonumber \\
  xd_v(x,\mu_{\rm f,0}^2)       &\! =\! & 3.064320\: x^{0.8}\: (1-x)^4,  
    \nonumber \\
  xg\,(x,\mu_{\rm f,0}^2)       &\! =\! & 1.700000\, x^{-0.1} (1-x)^5, 
    \nonumber \\
  x\bar{d}\,(x,\mu_{\rm f,0}^2) &\! =\! & .1939875\, x^{-0.1} (1-x)^6,
    \nonumber\\
  x\bar{u}\,(x,\mu_{\rm f,0}^2) &\! =\! & (1-x)\: x\bar{d}\,(x,\mu_{\rm f,0}^2),
    \nonumber\\
  xs\,(x,\mu_{\rm f,0}^2)       &\! =\! & x\bar{s}\,(x,\mu_{\rm f,0}^2) 
    \: = \: 0.2\, x(\bar{u}+\bar{d}\,)(x,\mu_{\rm f,0}^2),
\end{eqnarray}
where $u_v$ and $d_v$ refer to the up and down valence distributions respectively. Predictions for the NNPDF2.3 dataset were made according to these distributions, and used in the place of experimental data. Experimental noise was simulated in the pseudodata by application of the same procedure used to provide artificial data replicas. The full NNPDF2.3 procedure including Monte Carlo artificial replicas was then applied to the dataset, the resulting PDF set therefore being a level two type closure test where the generating PDF set should be recovered by the fit within the estimated uncertainties.

\begin{figure}[h!]
\centering
\includegraphics[width=0.48\textwidth]{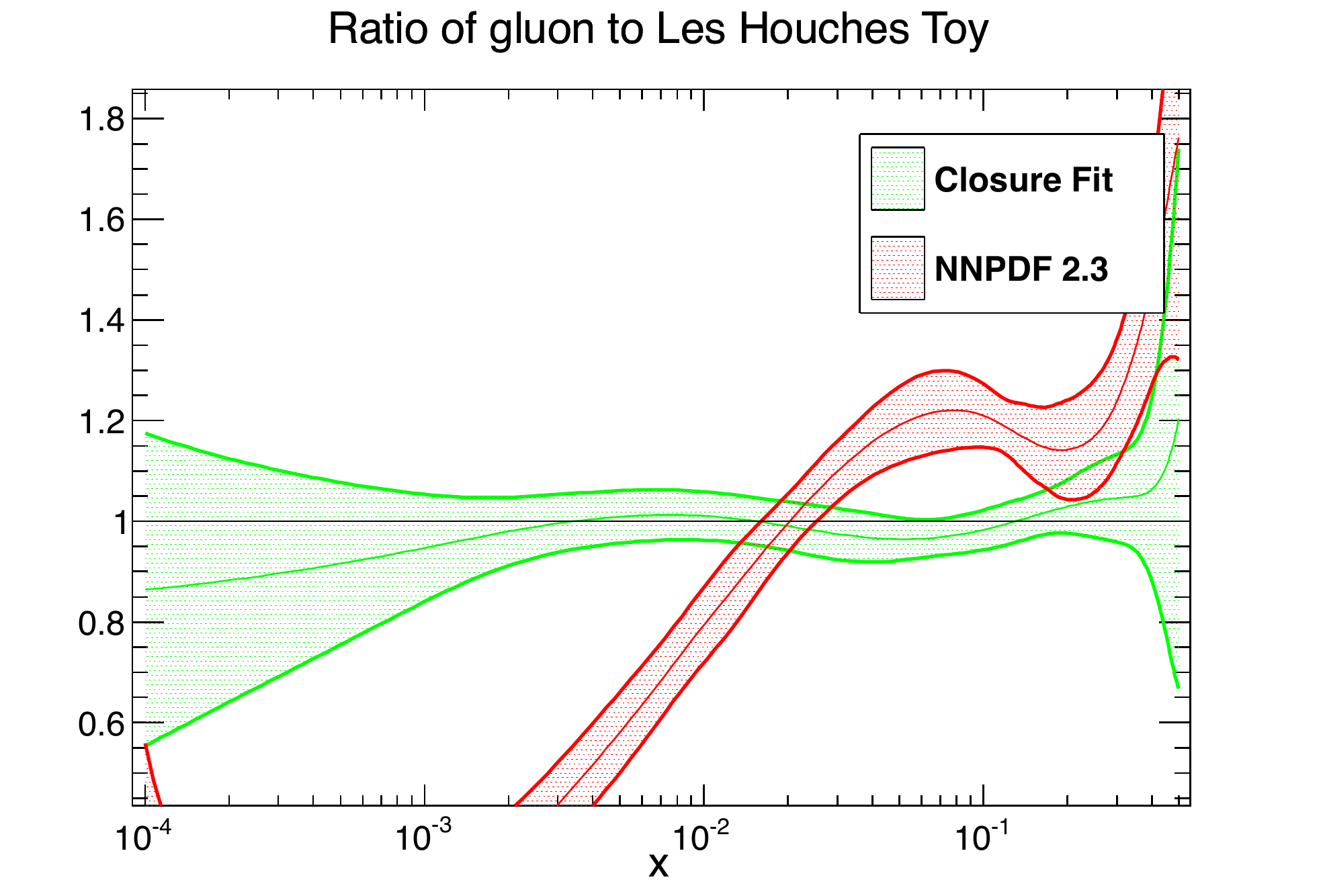}
\includegraphics[width=0.48\textwidth]{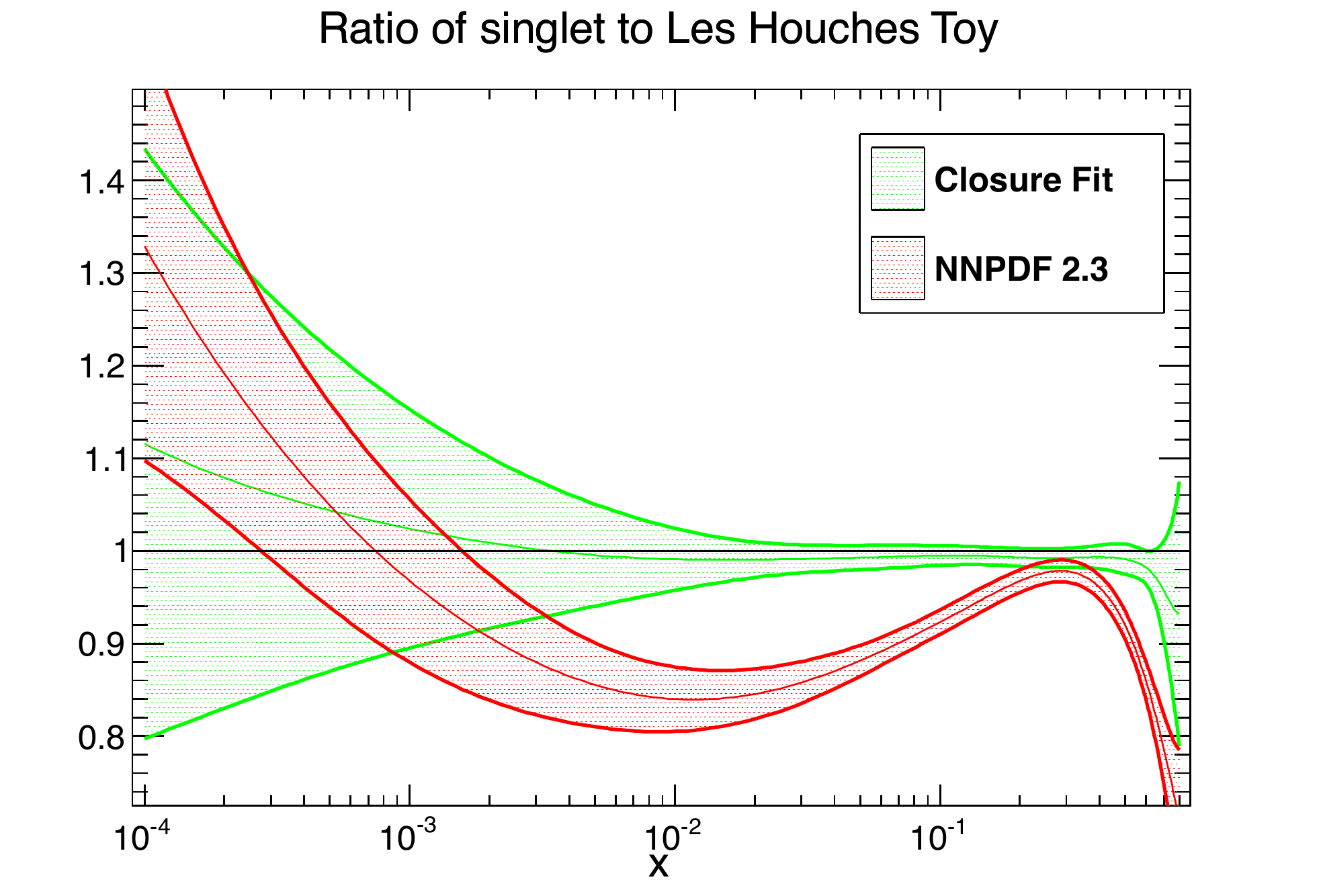}\\
\includegraphics[width=0.48\textwidth]{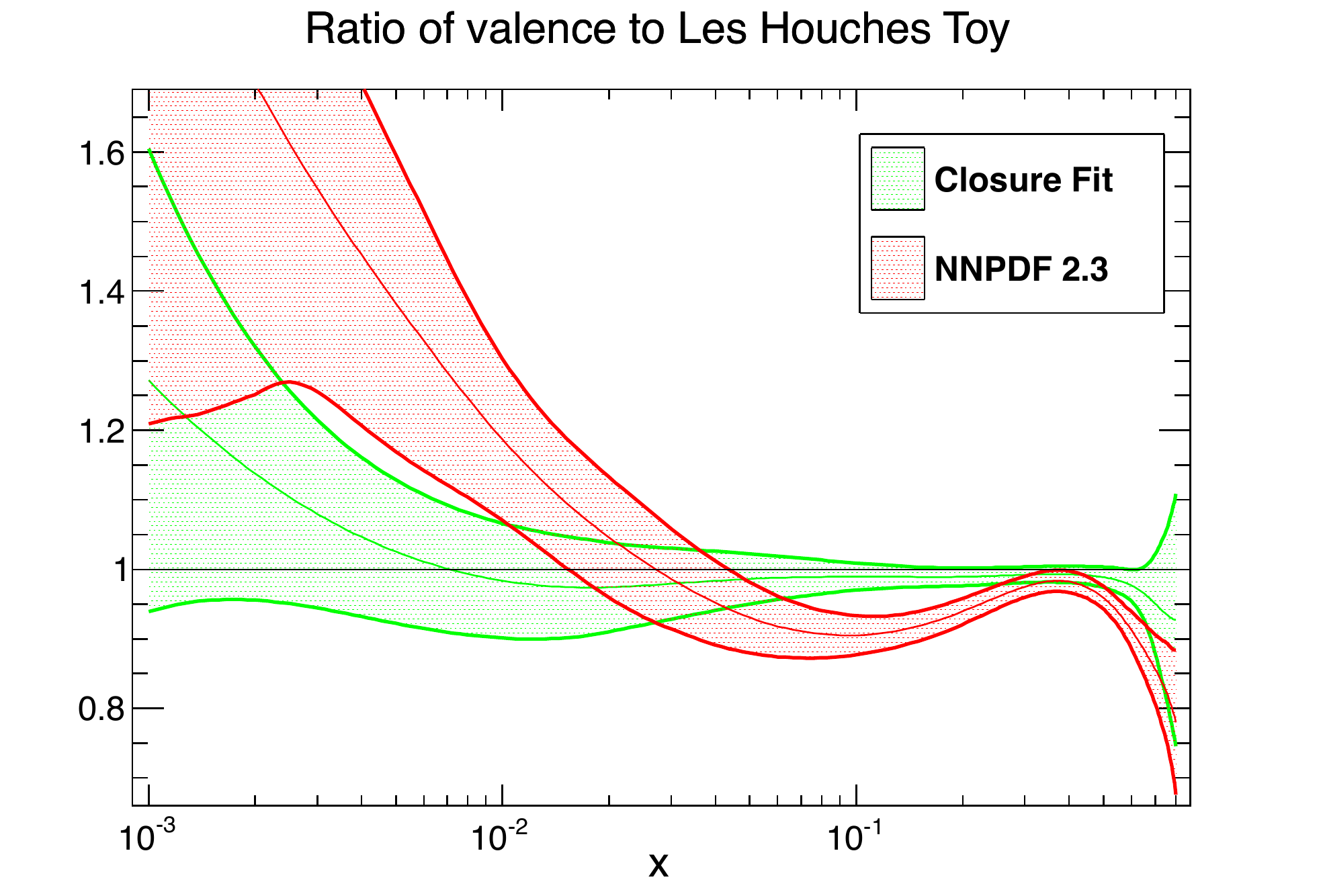}
\includegraphics[width=0.48\textwidth]{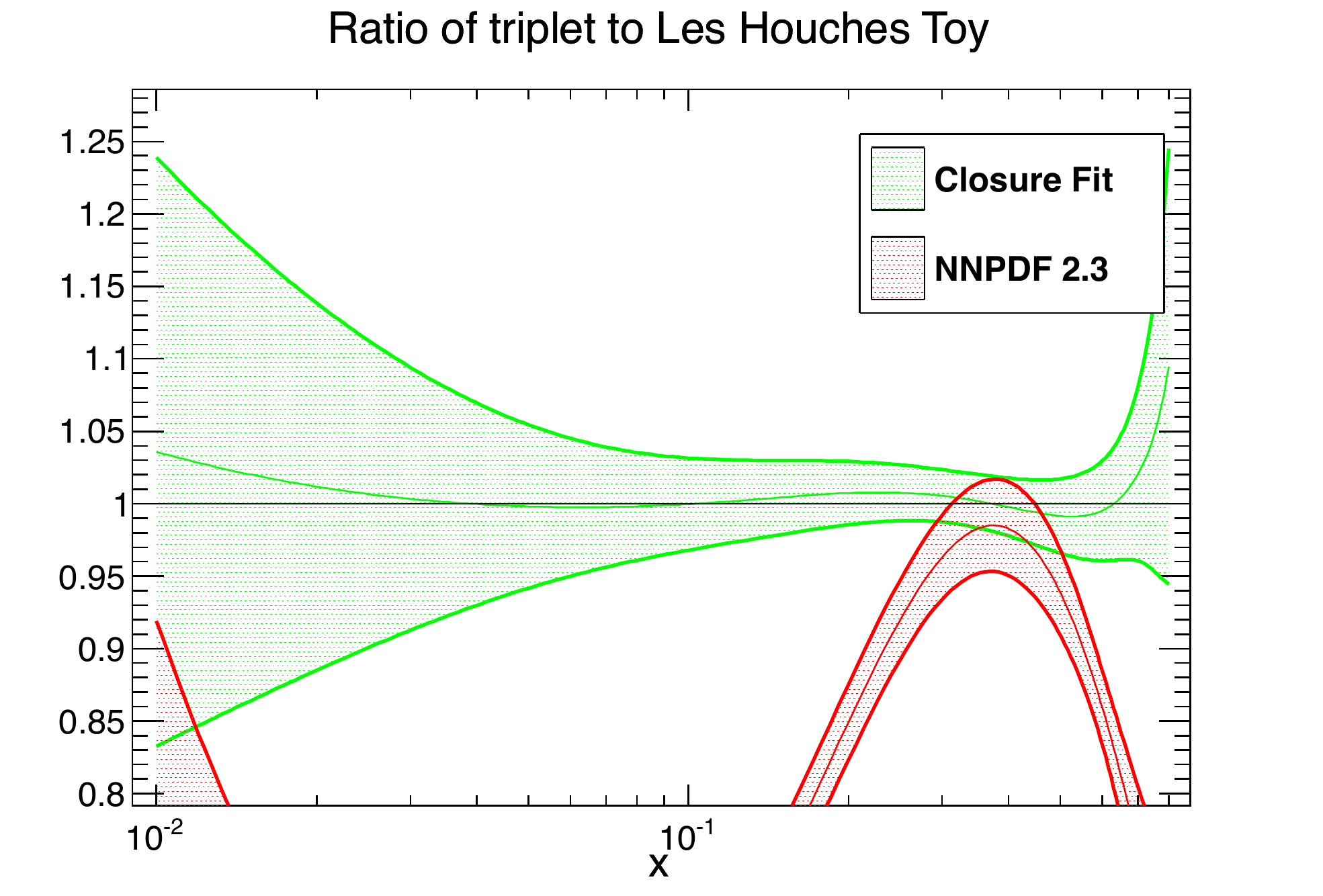}
\caption[PDFs obtained through a Closure test fit with toy PDFs as a generating function]{PDFs obtained through a closure test fit with Les Houches toy PDFs as a generating function, displayed as a ratio to the generating function. Shown are the distributions for the gluon, singlet, valence and triplet PDFs. In green are the results obtained through the closure test, and the red curves show the standard NNPDF2.3 result.}
\label{fig:LHtoyclosure1}
\end{figure}

\begin{figure}[h!]
\centering
\includegraphics[width=0.48\textwidth]{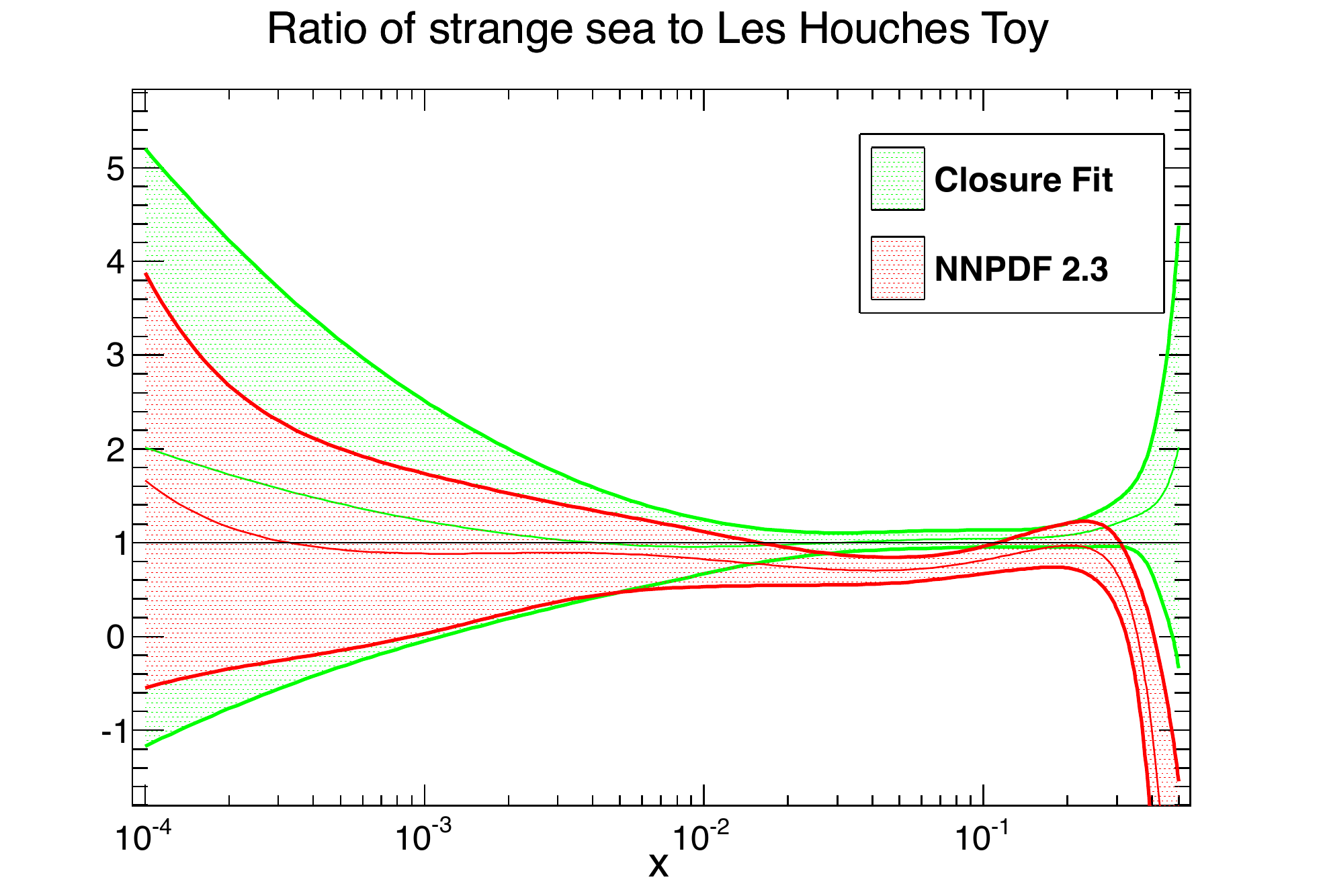}
\includegraphics[width=0.48\textwidth]{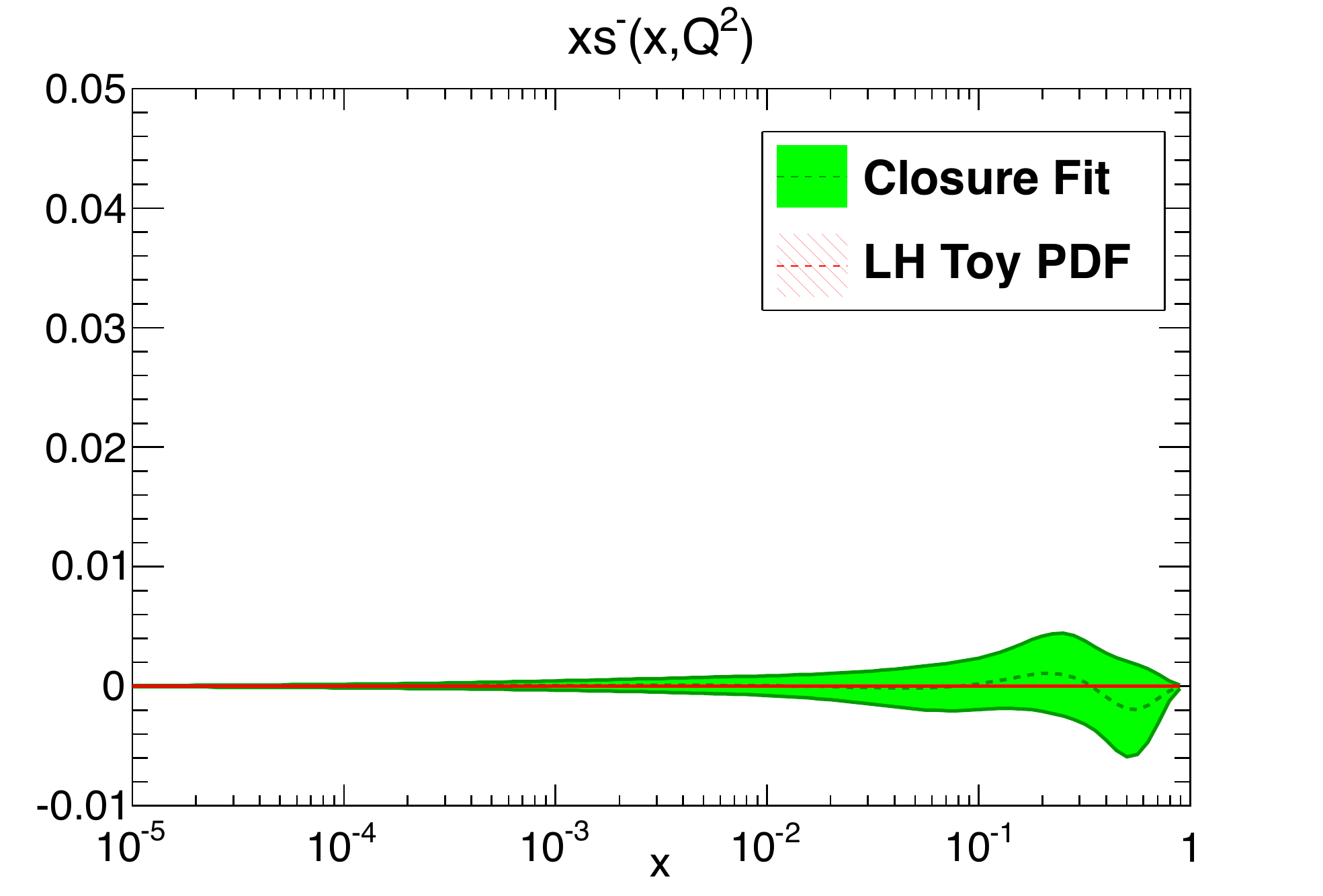}
\caption[Strange PDFs obtained through a Closure test fit with toy PDFs as a generating function]{Strange sea (left) and valence (right) PDFs obtained through a closure test fit with Les Houches toy PDFs as a generating function. The strange sea is presented as a ratio to the LH toy PDF, and the strange valence is presented directly as the PDF, with the (zero) LH toy line shown.}
\label{fig:LHtoyclosure2}
\end{figure}

Figure~\ref{fig:LHtoyclosure1} displays the results of the level two closure test fit with the Les Houches toy PDFs used as a generating function. The result demonstrates impressive agreement, with the NNPDF2.3 methodology able to accommodate the predictions of the Les Houches toy generating function despite it deviating significantly from the standard NNPDF2.3 result. For all four PDF combinations shown, the results of the closure test maintain distances of less than one standard deviation to the generating function across a wide kinematic range. Of additional interest are the strange distributions, relatively poorly constrainted by the data included in the pseudo-dataset. The strange valence in particular is set to zero in the Les Houches toy. Figure~\ref{fig:LHtoyclosure2} shows the results from the closure test for both the total strangeness and strange valence distributions, the NNPDF methodology is able to clearly reproduce the underlying law within uncertainties in both cases, and is able to comfortably resolve a zero strange valence contribution.

The results are particularly impressive considering that this is a test of a methodology that has not been previously verified by closure test. The example case of a pseudo-dataset generated according to the Les Houches toy PDF is however a rather simplified case, and methodological refinements can be made by examining closure tests with greater structure in the generating function.

\begin{figure}[!]
\centering
\includegraphics[width=0.48\textwidth]{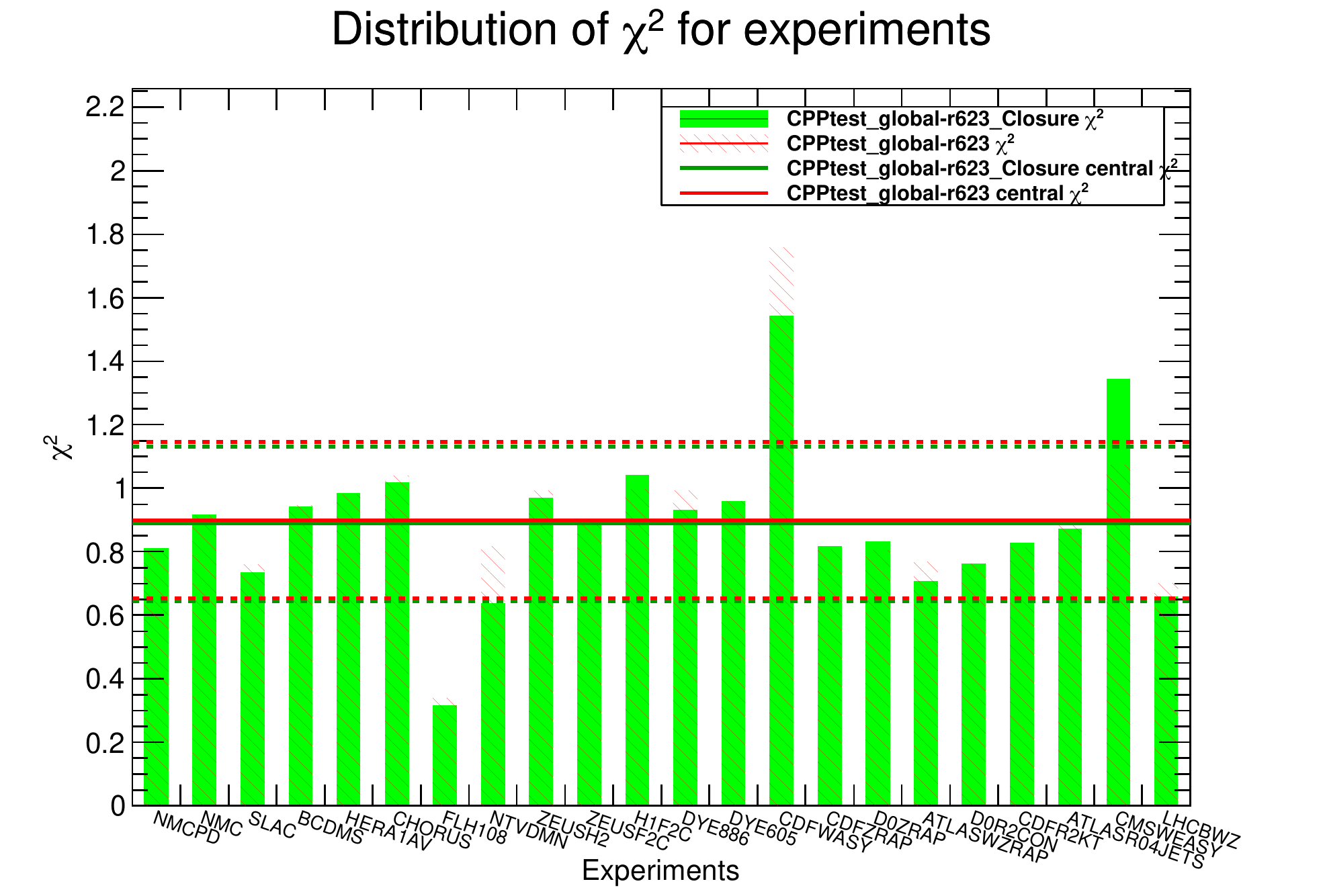}
\includegraphics[width=0.48\textwidth]{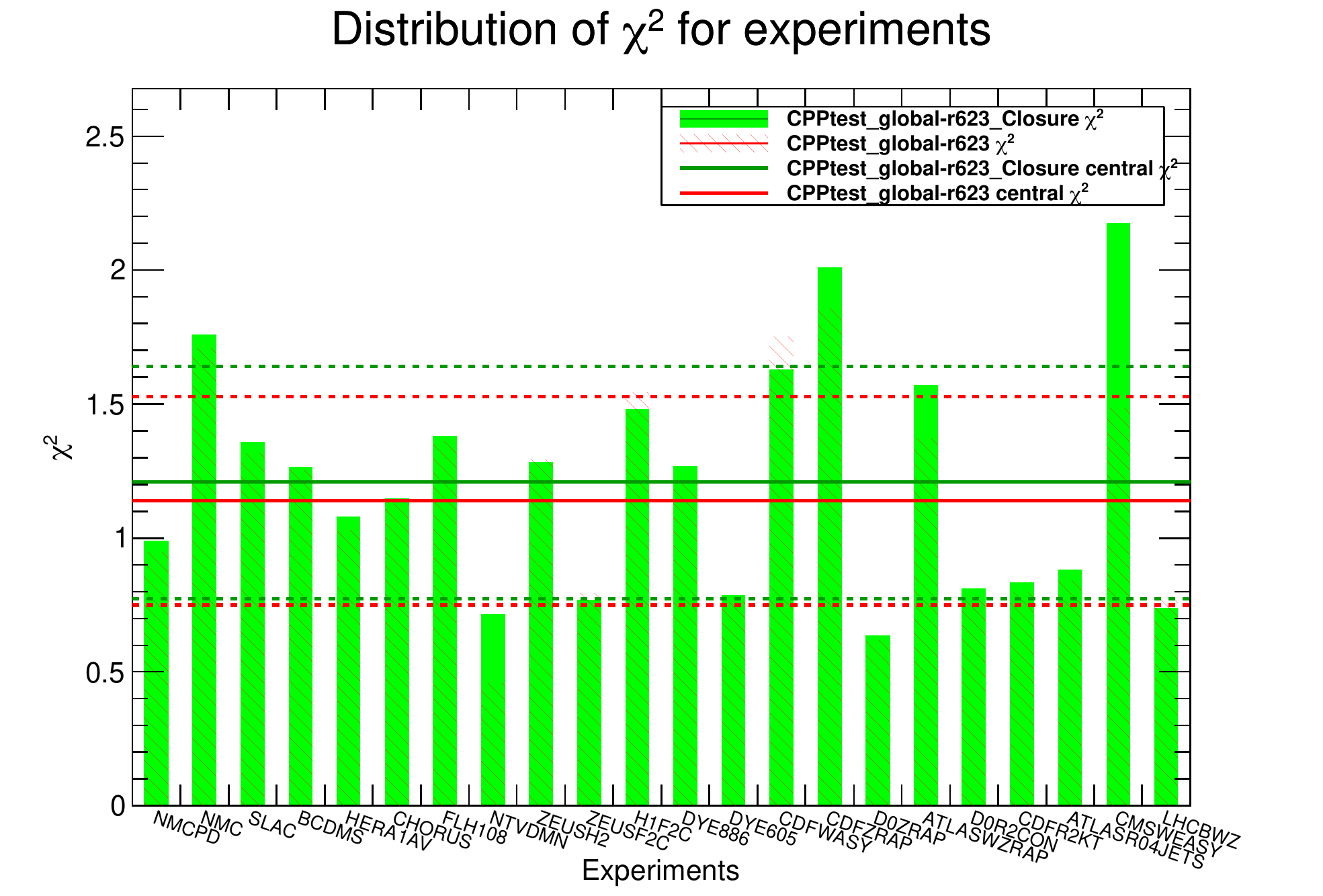}
\caption[$\chi^2$ values to the pseudo- and experimental-datasets of a closure fit and the generating PDF]{Example $\chi^2$ values to the pseudo- (left) and experimental- (right) datasets of a closure fit and the generating PDF from early {\tt nnpdf++} test fits. The red bars show the fit quality of the generating PDF set while the green bars demonstrate the $\chi^2$ for the closure test set. The horizontal lines indicate the average and $1\sigma$ of the fit qualities in their associated colours.}
\label{fig:CPPclosurechi2}
\end{figure}

A good level of agreement can also be found at the level of the $\chi^2$ to both the pseudodata sample, and the real experimental data. In Figure~\ref{fig:CPPclosurechi2} we compare the fit quality of a closure test and its generating PDF dataset by dataset by presenting the $\chi^2$ to each measurement from both the closure test result and the generating PDF. In this case the generating function has considerably greater complexity, being an early {\tt nnpdf++} test fit with most of the NNPDF methodology in place. While agreement is generally very good, especially on the level of total $\chi^2$; we begin to see some elements of discrepancy in datasets sensitive to flavour separation and strangeness such as the NuTeV dataset and electroweak vector boson production data. Such discrepancies can help in pinpointing areas where further development is needed.

\clearpage
\clearpage

\section{Preprocessing}

Early closure tests performed with the NNPDF2.3 methodology showed generally very good agreement between the produced PDFs and the underlying functions used to generate the pseudo-dataset. However some PDF combinations demonstrated rather poorer agreement than others, particularly distributions sensitive to flavour separation. Such disagreements became more apparent when considering closure tests to underlying functions with more structure than available in the Les Houches toy set. The disagreements were found to originate in the choice of the preprocessing exponents used in the definition of the NNPDF parametrisation. Recalling Eqn.~\ref{eq:NNPDF23param}, the structure of the basic NNPDF parametrisation follows
\be f(x) \propto x^{-\alpha} (1-x)^{\beta} \text{NN}(x),\ee
where NN represents the neural network itself, and $\alpha$ and $\beta$ are the preprocessing exponents randomised on a replica-by-replica basis at the start of a fit. The range in which the exponents were randomised has been fixed in the fits up to and including NNPDF2.3, set to a span large enough such that the dependence of the results upon the choice of range was minimised. In such a way the preprocessing was considered to provide a backbone for the neural-network fit and, aside from improving fitting efficiency, to have a minimal impact upon the results.

To study the effect of different preprocessing ranges we can look at estimators for the \emph{effective} asymptotic exponents,
\be \alpha_{\text{eff}} = -\frac{\log{(|f(x)|)}}{\log(x)},\quad\quad \beta_{\text{eff}} = \frac{\log{(|f(x)|)}}{\log(1-x)}, \ee
such that in the limits of $x\to0,1$ the exponents $\alpha,\beta$ are recovered. By examining these effective exponents in the high- and low-$x$ regions, we can ascertain if there is a data preference for a different preprocessing range than was used in a fit, and if the preprocessing range used was too restrictive.

\begin{figure}[hp!]
\centering
\includegraphics[width=0.48\textwidth]{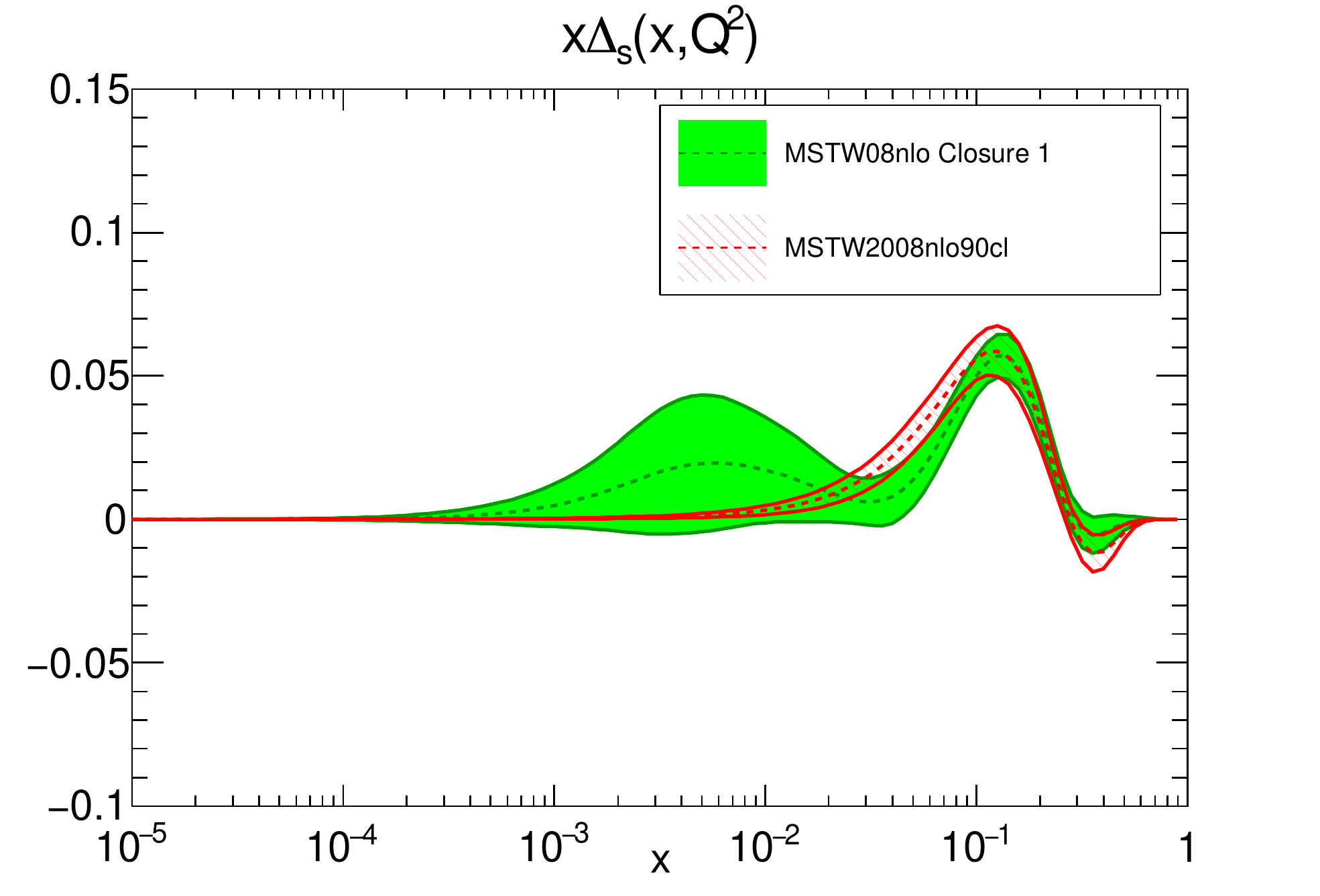}
\includegraphics[width=0.48\textwidth]{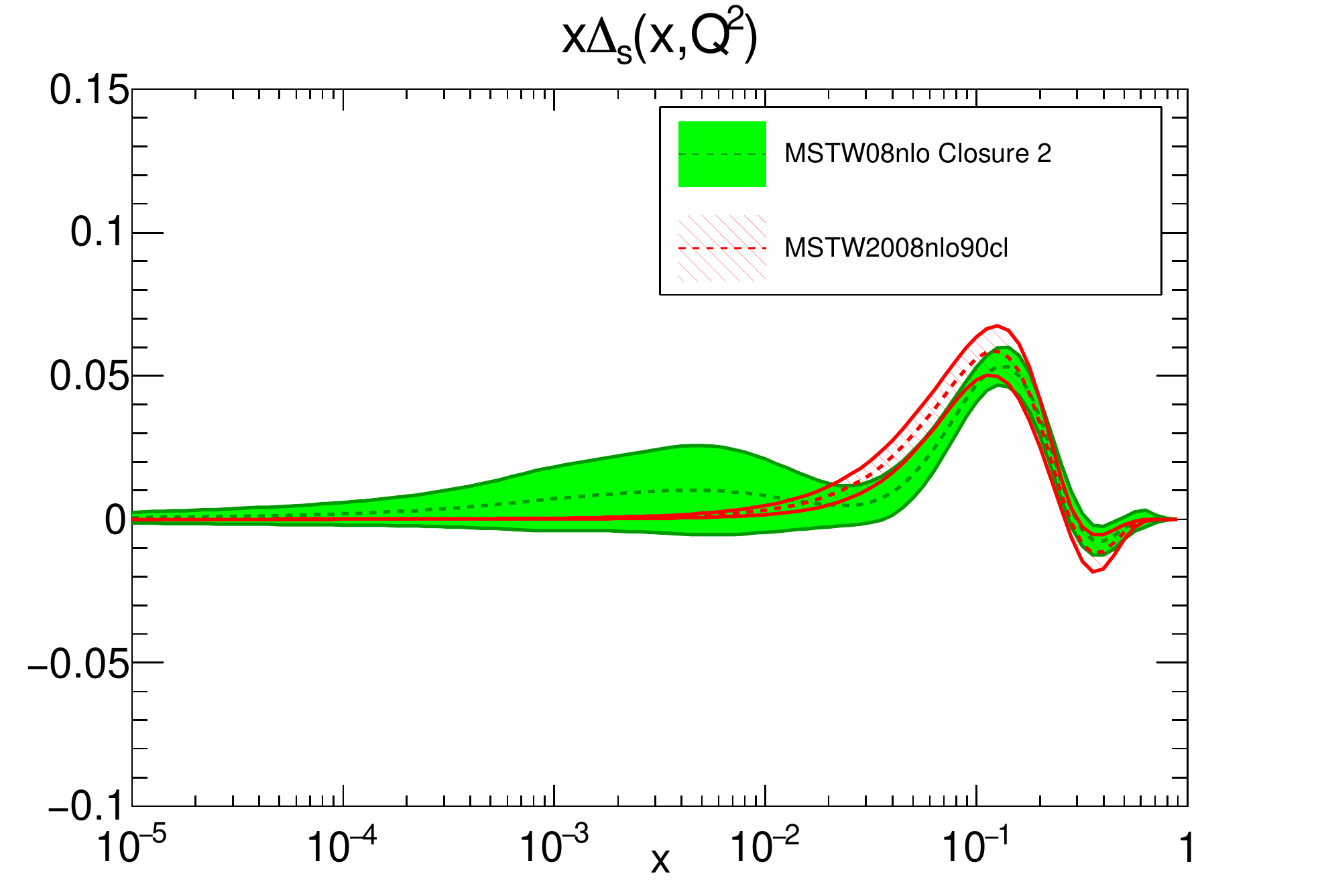}
\includegraphics[width=0.48\textwidth]{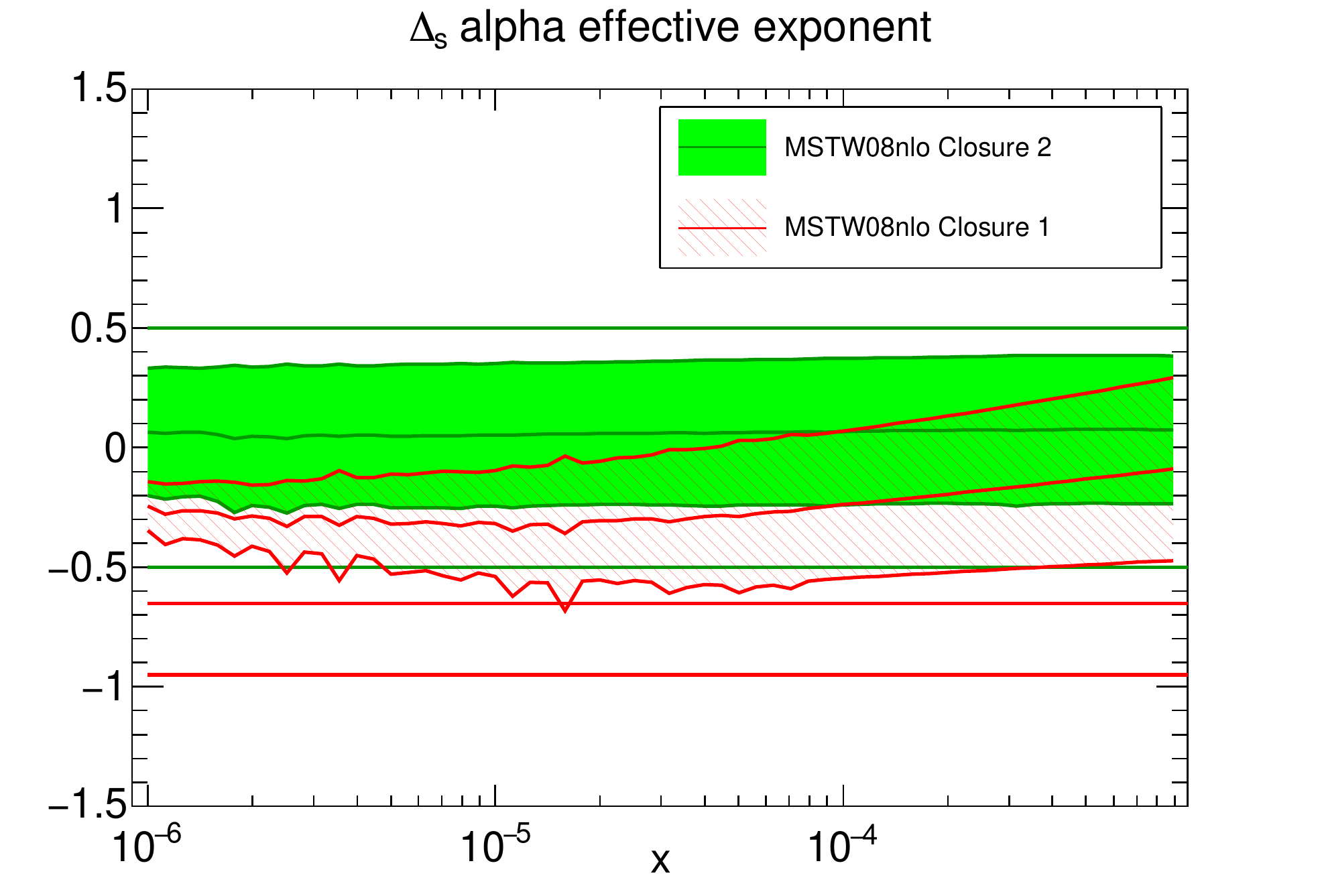}
\includegraphics[width=0.48\textwidth]{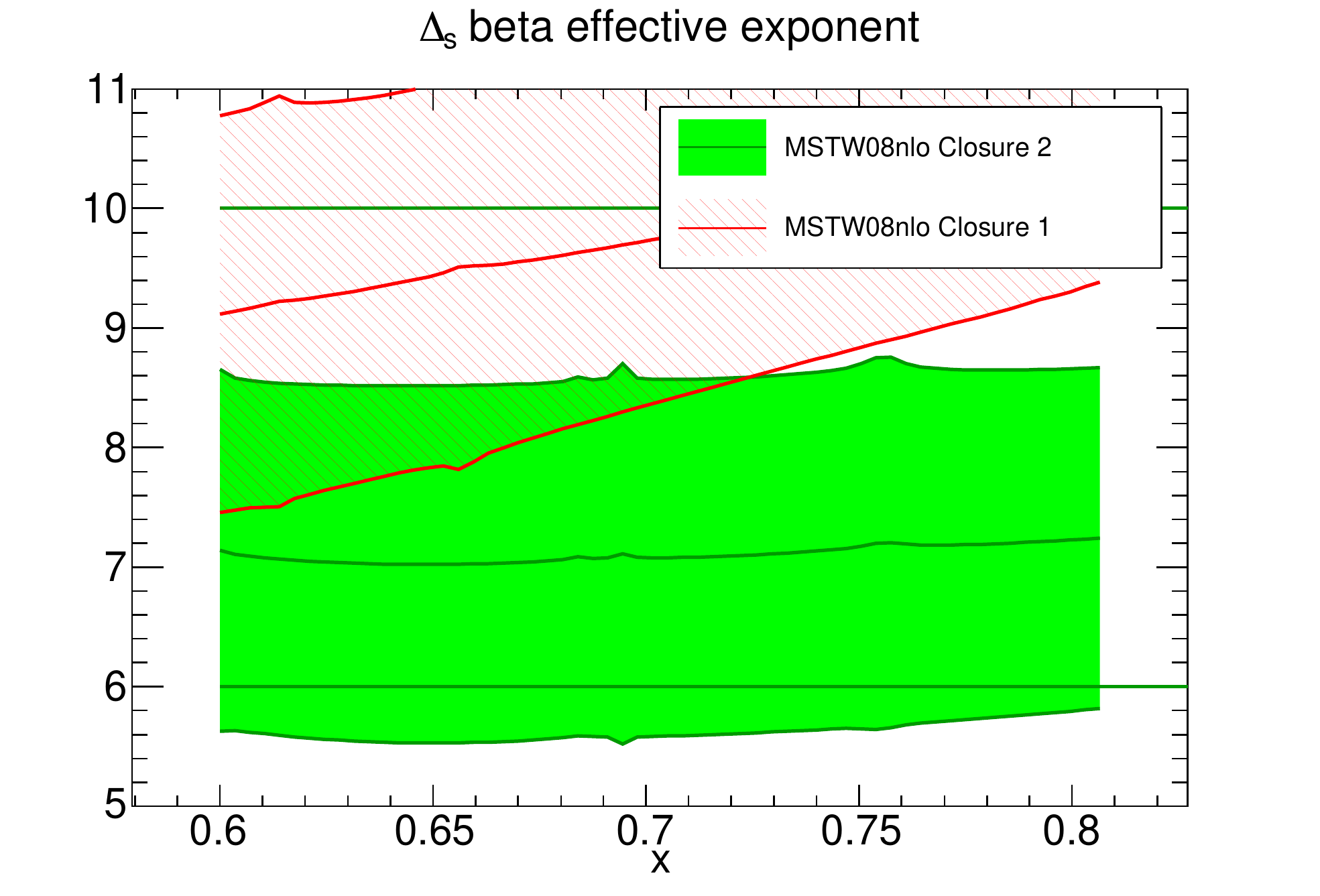}
\caption[Demonstration of the impact made by changes in preprocessing to the sea asymmetry PDF in a closure test fit]{Demonstration of the impact made by changes in preprocessing to the sea asymmetry PDF in a closure test fit to an MSTW08 NLO underlying law. The top two figures demonstrate the results for the $\Delta_s$ distribution for two choices of preprocessing ranges, with the left figure using NNPDF2.3 standard preprocessing. In both cases, the red curve shows the underlying law used in the Closure test. The right figure demonstrates slightly improved agreement, particularly at low-$x$. The lower figures show the low and high $x$ effective exponent plots for the two ranges. The solid horizontal lines delineate the regions in which the preprocessing exponents were initialised, and the bands show the $1\sigma$ contours of the effective exponents.}
\label{fig:preproc1}
\end{figure}

\begin{figure}[!]
\centering
\includegraphics[width=0.42\textwidth]{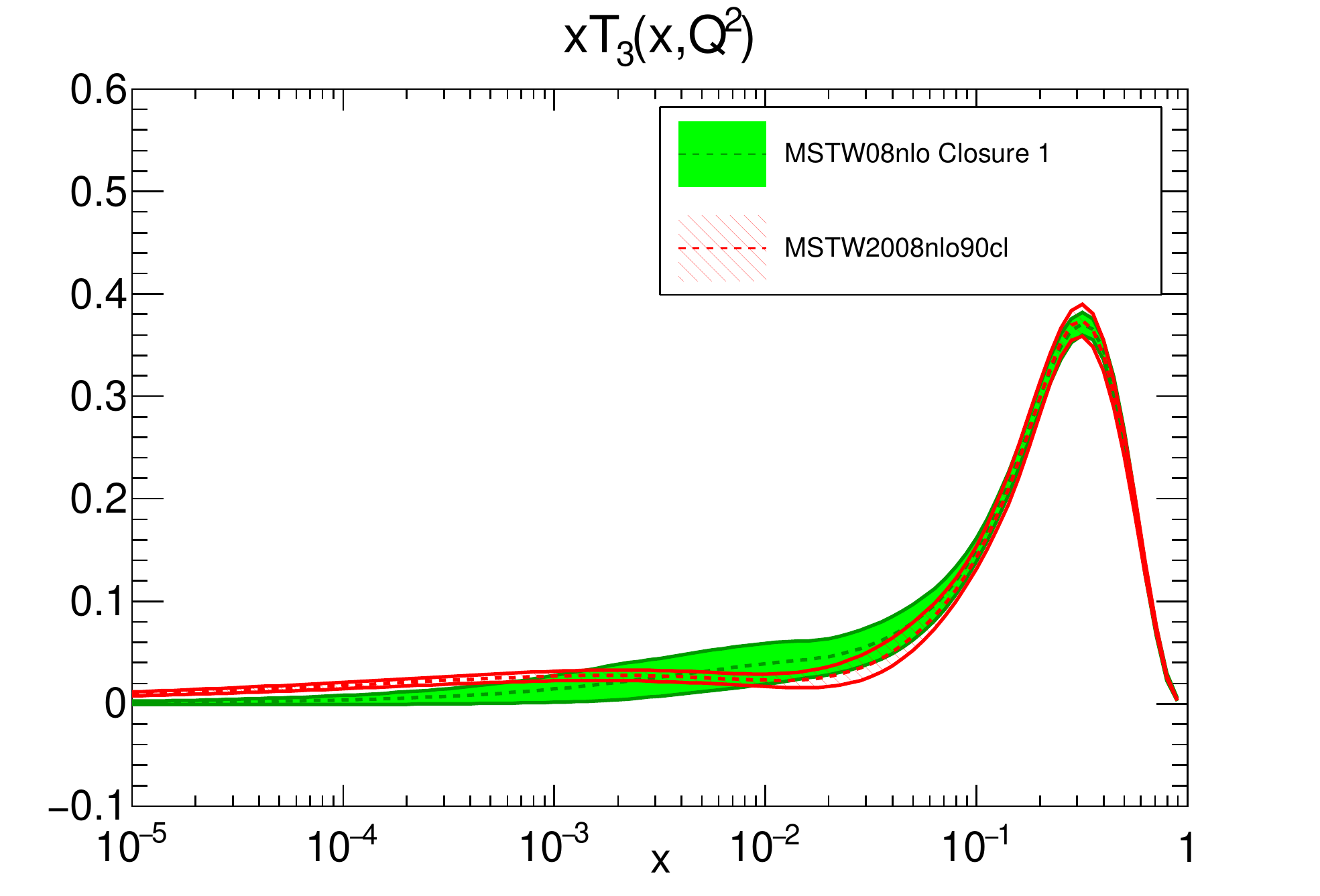}
\includegraphics[width=0.42\textwidth]{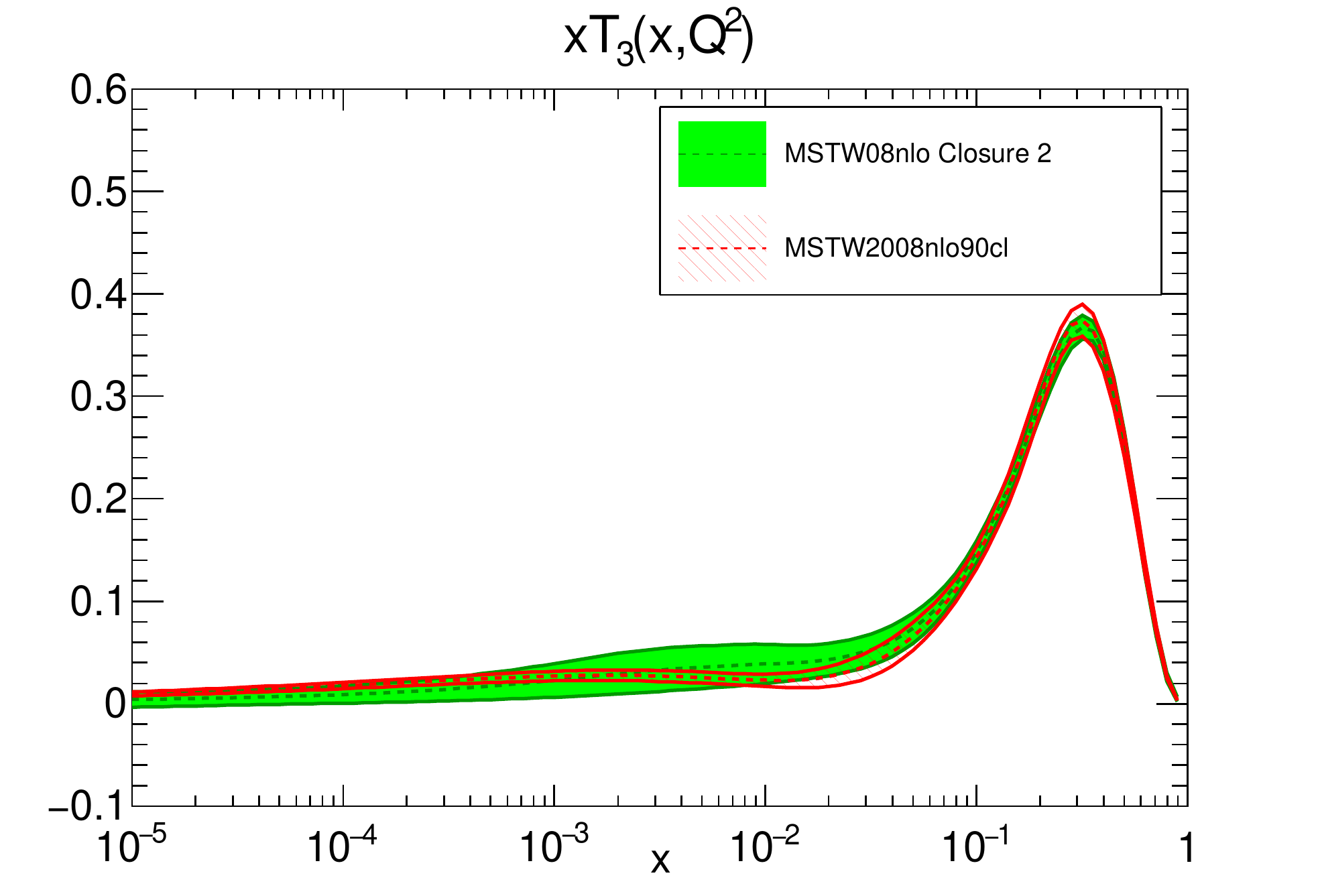}
\includegraphics[width=0.42\textwidth]{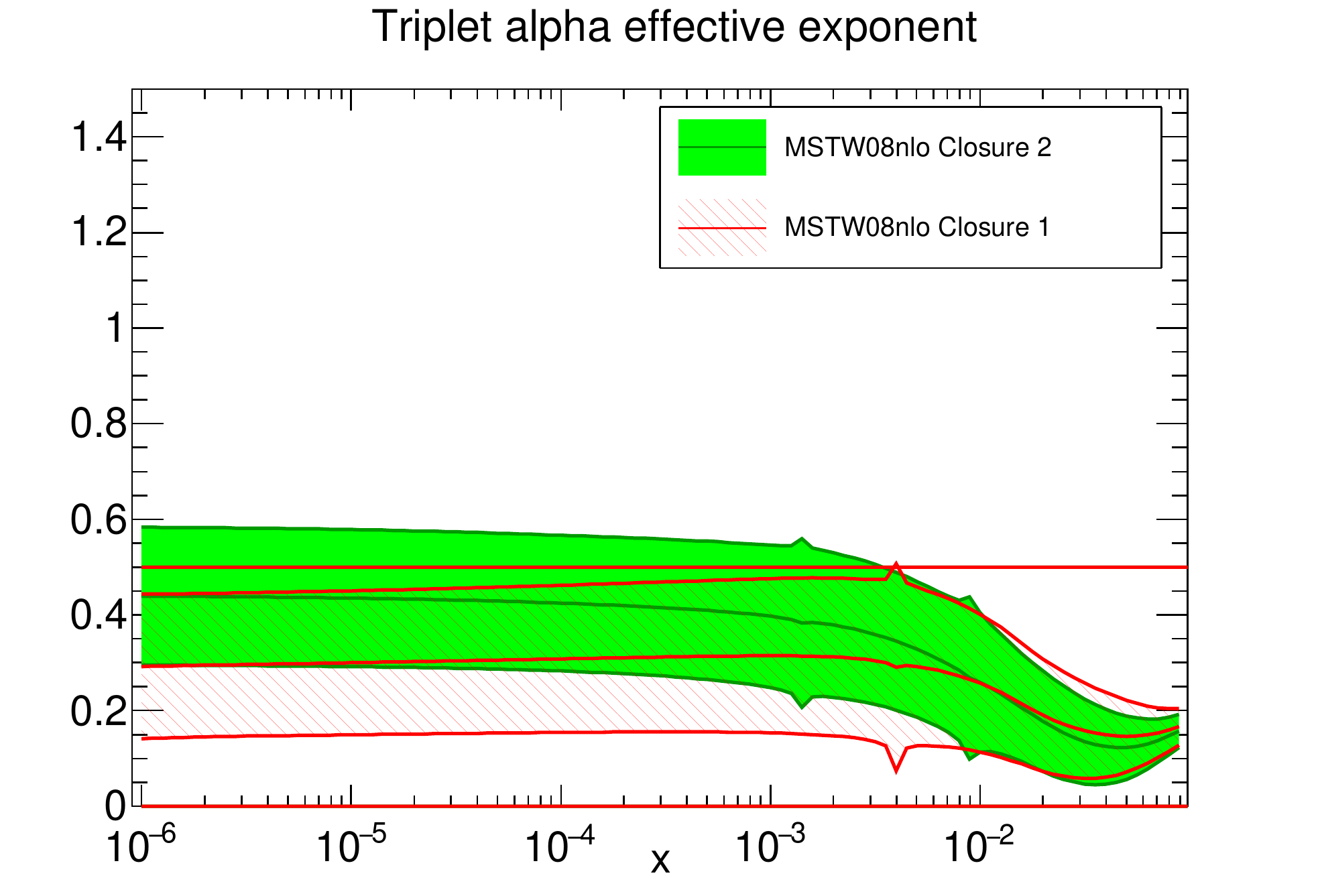}
\includegraphics[width=0.42\textwidth]{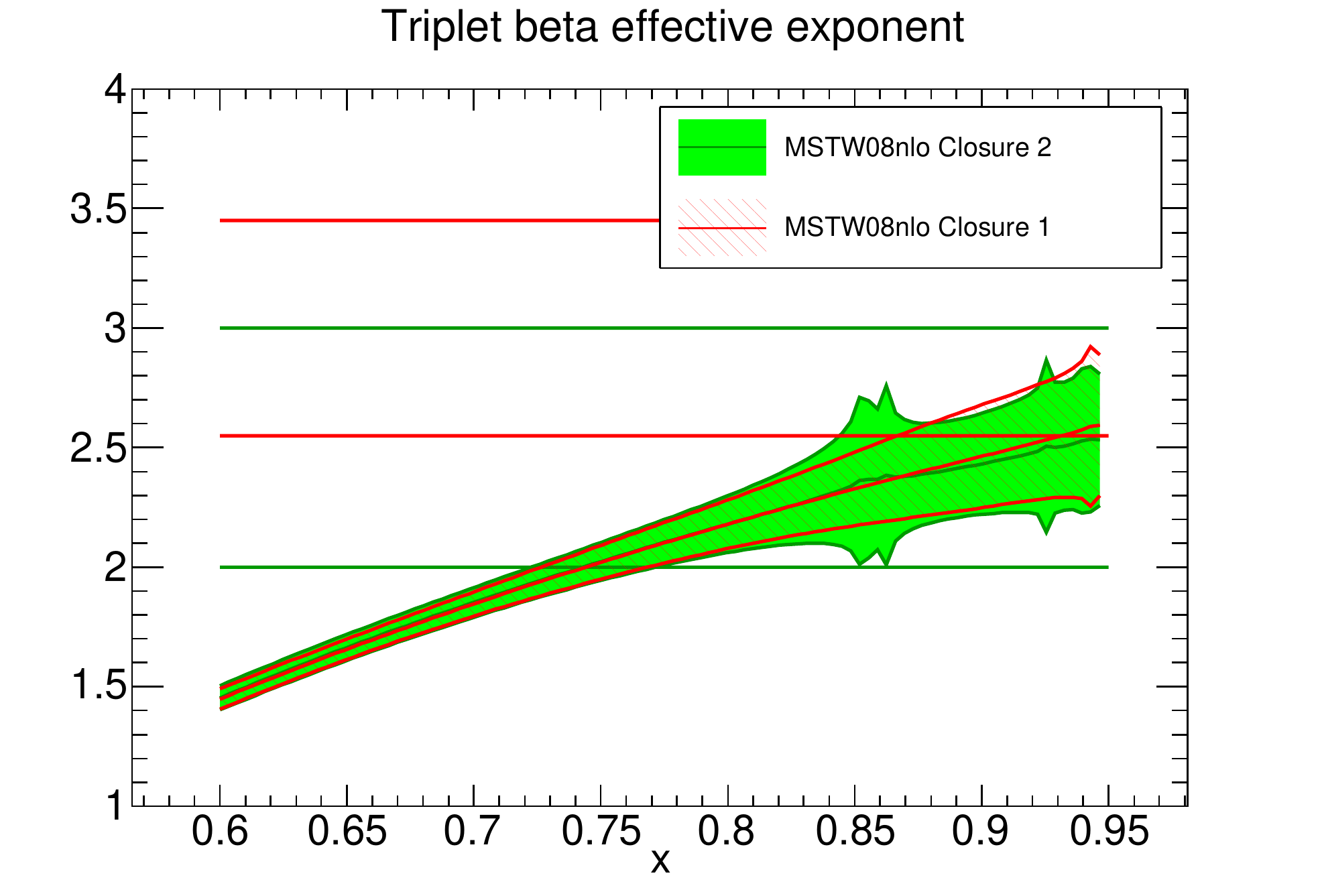}
\caption[Demonstration of the impact made by changes in preprocessing to the triplet PDF in a closure test fit]{A further preprocessing analysis as in Figure~\ref{fig:preproc1}, performed upon the Triplet PDF combination for the same two closure test fits.}
\label{fig:preproc2}
\end{figure}

In Figure~\ref{fig:preproc1} an example preprocessing analysis is shown for a closure test based upon an MSTW08 underlying law at NLO. The sea asymmetry $\bar{u} -\bar{d}$ is shown for two choices of preprocessing range, the NNPDF2.3 standard and a range modified to better accommodate the data preference visible in the effective exponents. From the figure we can see that the choice of exponent randomisation range has a significant effect on the resulting distributions, and that the effective exponents can show a clear data preference for a different range. In Figure~\ref{fig:preproc2} we can see the same analysis applied to the triplet PDF where similar conclusions may be drawn.

These analyses demonstrate that the sensitivity to the preprocessing exponent randomisation ranges is somewhat larger than suspected previously, and needs to be studied in detail in order to avoid minimisation difficulties in a fit where the preprocessing ranges are ill-suited to the dataset. Furthermore, the uncertainty bulges visible in both the triplet and sea asymmetry distributions in Figures~\ref{fig:preproc1} and~\ref{fig:preproc2} are generated by the preprocessing suppressing genuine data uncertainty in the asymptotic regions. These problems may be alleviated by lifting the requirement that such distributions should be preprocessed to zero at low-$x$, and implementing a procedure for the iterative and data-driven determination of preprocessing exponents.

To improve the minimisation performance, hampered by ill-suited preprocessing, NNPDF fits have now adopted the following iterative procedure for the determination of both high and low-$x$ randomisation ranges:

\begin{itemize}
\item \textbf{Singlet and gluon PDFs}\\
Exponent randomisation ranges are set to be twice the $1\sigma$ interval of the previous iteration's effective exponents
at the asymptotic points.
\item \textbf{Nonsinglet PDF combinations}\\
The low-$x$ randomisation interval is set to be the maximal extent of two effective exponent ranges; twice the $1\sigma$ interval at the asymptotic point and twice the $1\sigma$ interval at the point $x=1\times 10^{-3}$. The high-$x$ interval is set identically as with the singlet and gluon.
\end{itemize} 

\begin{figure}[!]
\centering
\includegraphics[width=0.42\textwidth]{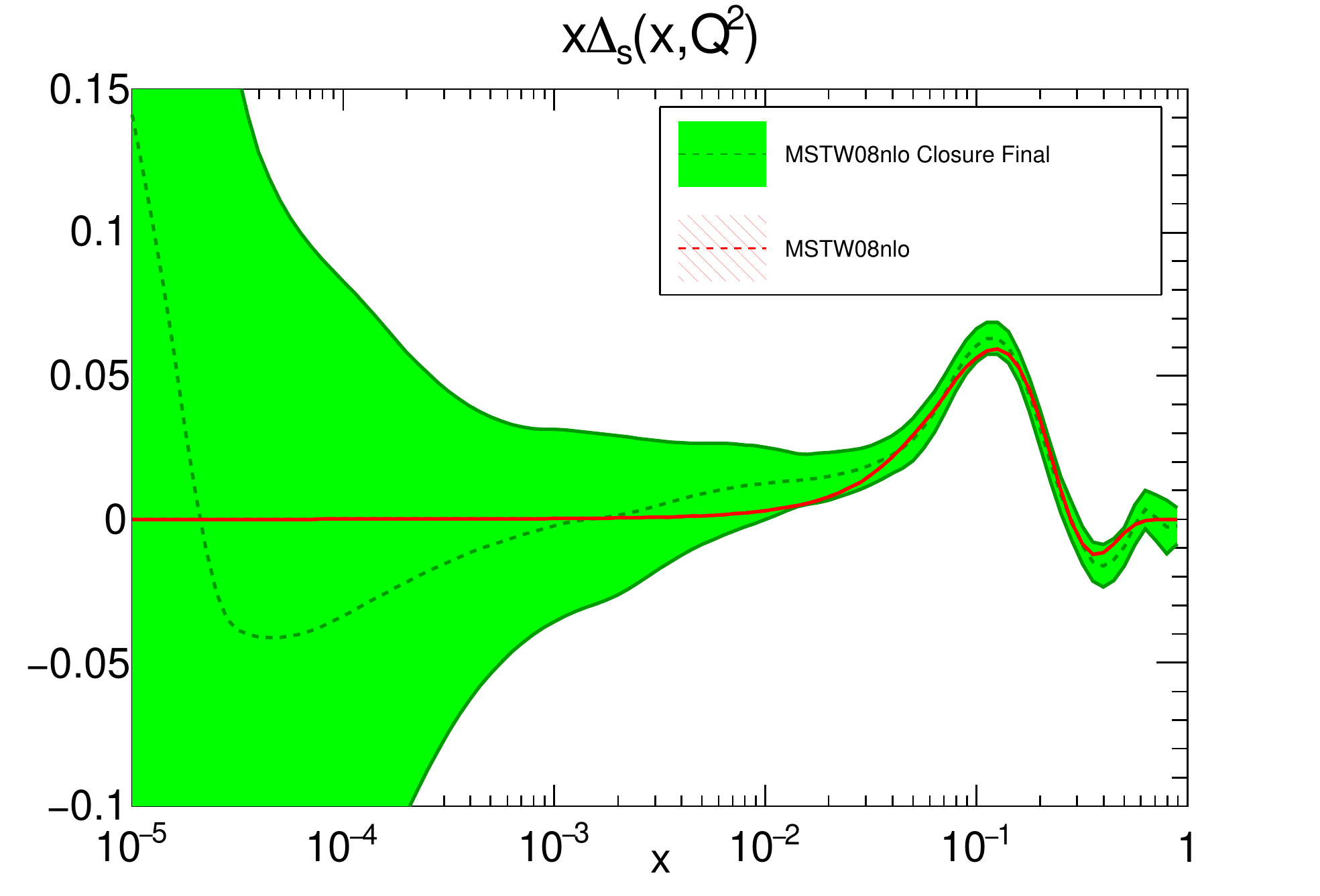}
\includegraphics[width=0.42\textwidth]{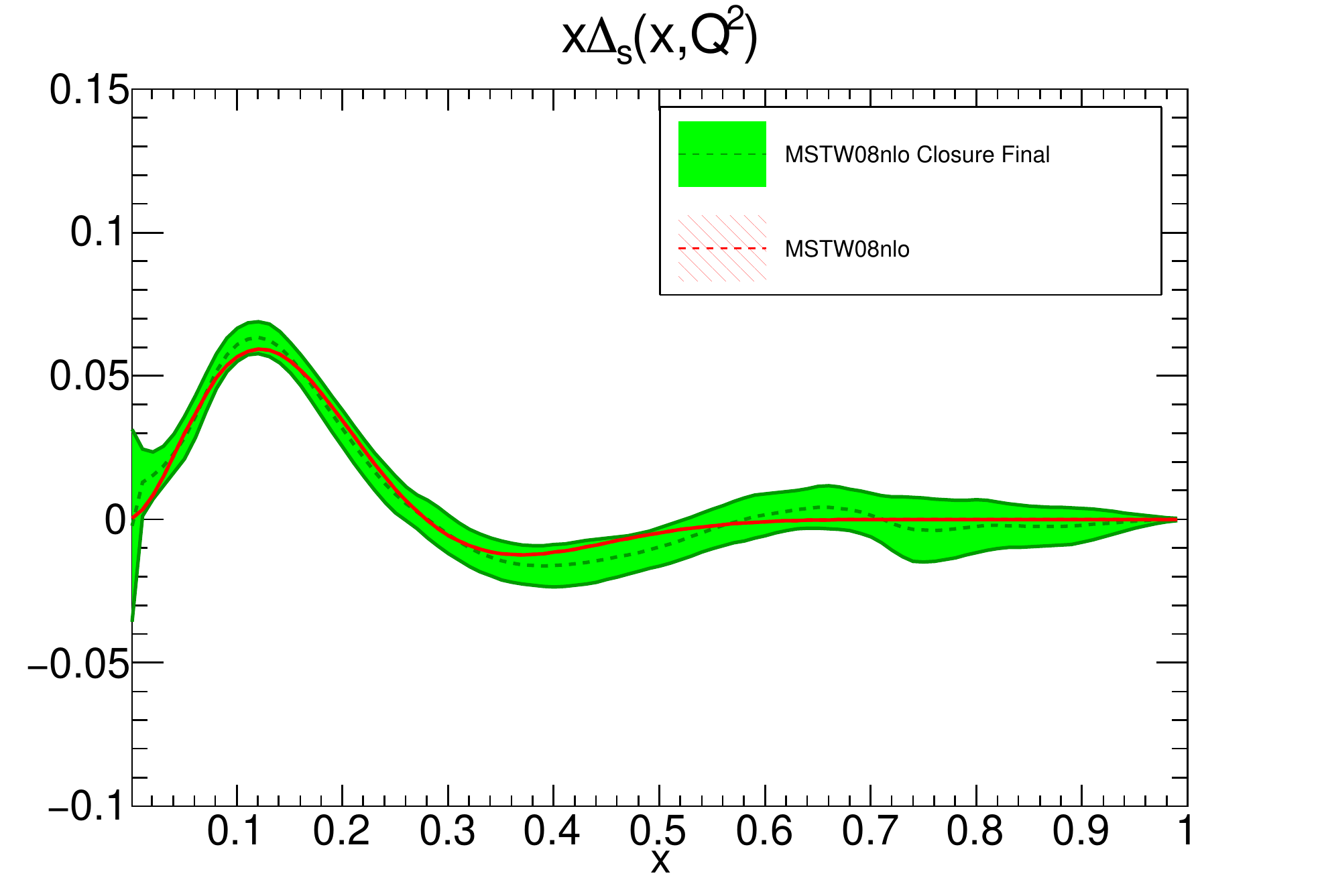}
\includegraphics[width=0.42\textwidth]{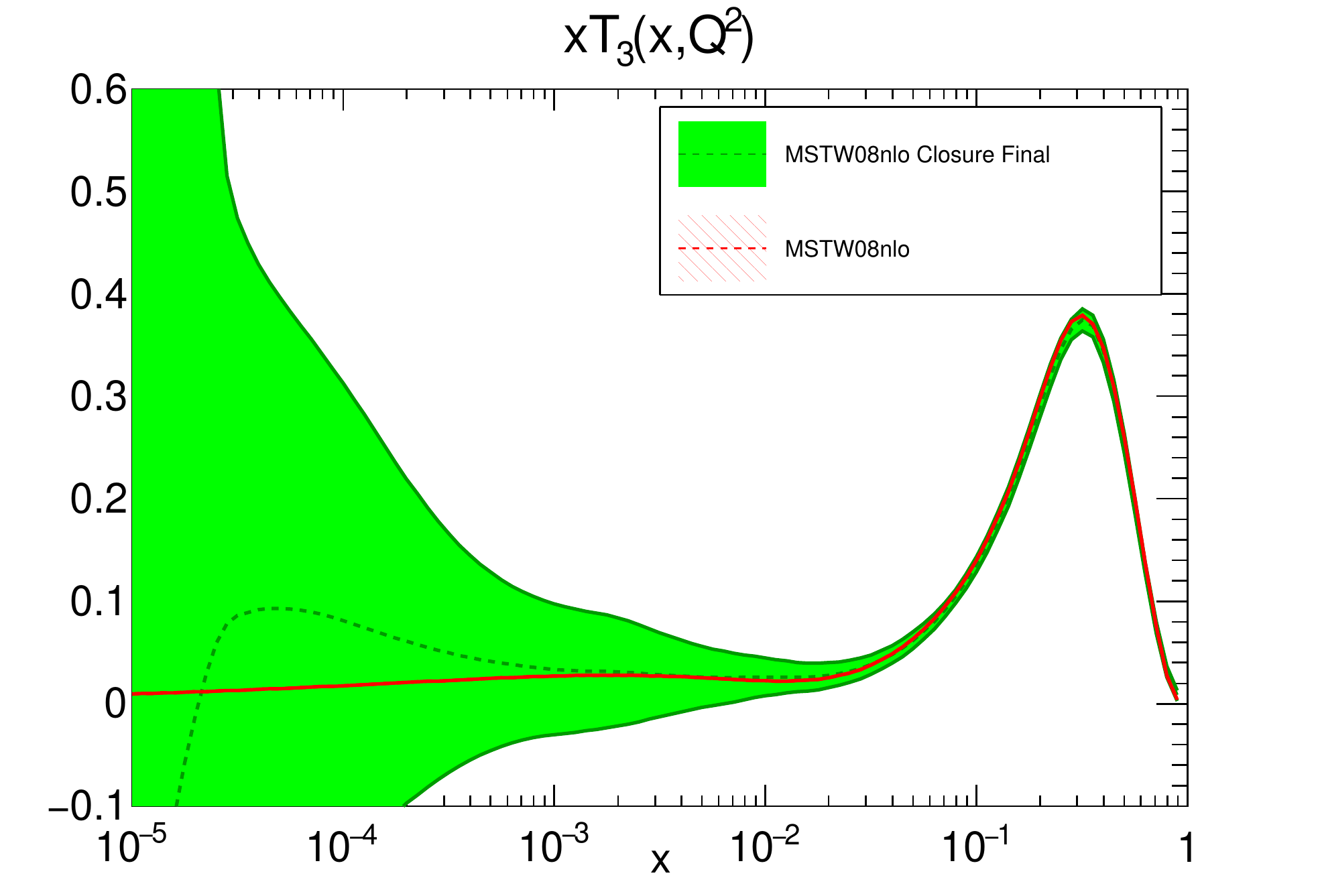}
\includegraphics[width=0.42\textwidth]{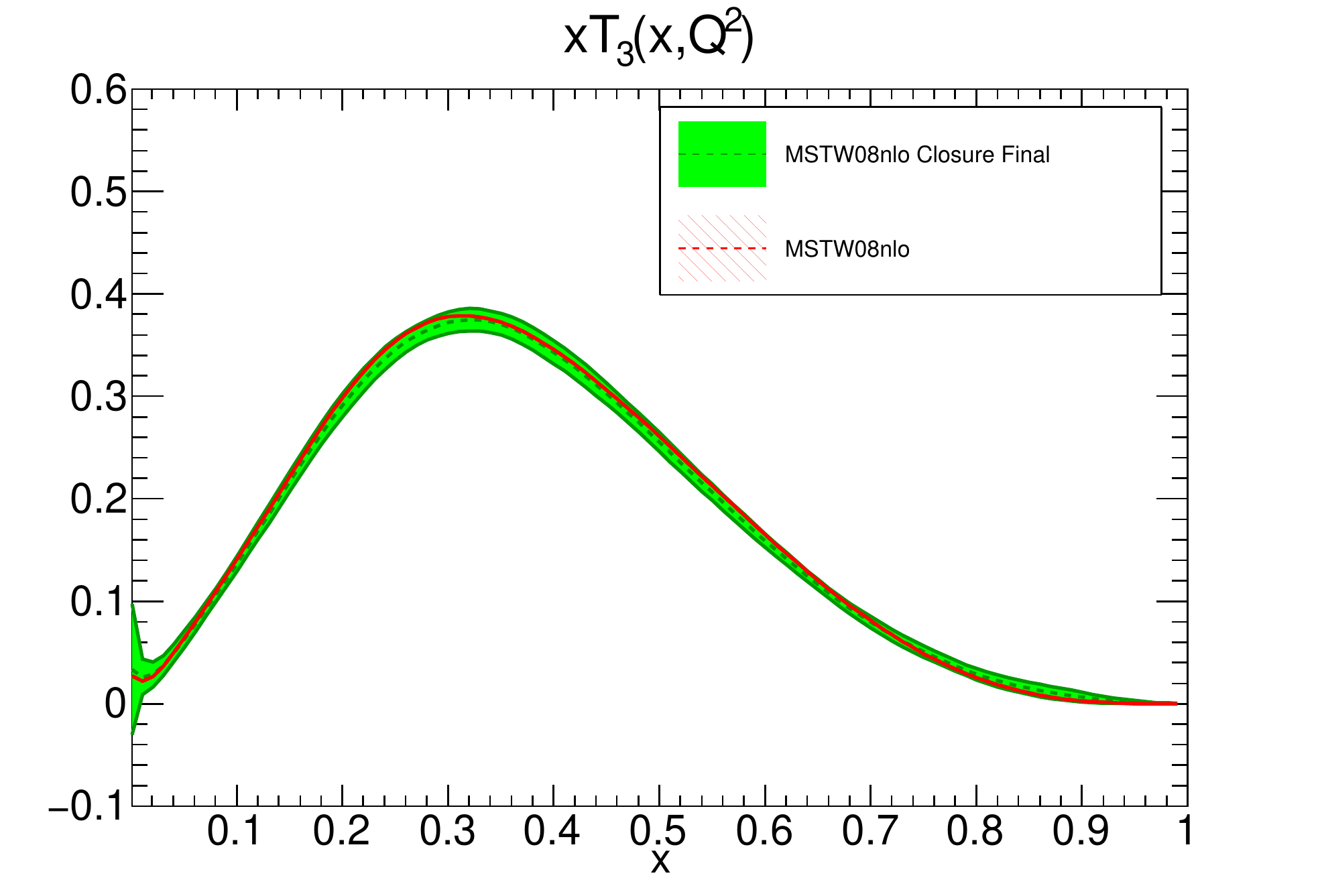}
\caption[Impact of improved preprocessing range selection in the sea asymmetry and triplet PDFs]{Impact of improved preprocessing range selection in the sea asymmetry and triplet PDFs. The top figures demonstrate the $\Delta_s$ PDF obtained via a closure test to MSTW08 using the improved preprocessing procedure in green, with the underlying law shown in red. The figures below show the equivalent plots for the triplet PDF with the improved preprocessing ranges.}
\label{fig:preproc3}
\end{figure}

In such a way convergence of the randomisation interval can be established typically in two or three fit iterations, and the preprocessing exponents are obtained from the preference of the experimental dataset. As an example of a fit generated from such an iterative procedure consider Figure~\ref{fig:preproc3} which demonstrates the preprocessing analysis for the $\Delta_s$ and Triplet distributions resulting from the new procedure. In comparison to Figure~\ref{fig:preproc1} where the old settings are used, the low-$x$ preprocessing ranges have relaxed considerably and are no longer constrained by the chosen exponent range but driven by the experimental data. Furthermore the agreement with the underlying law is noticeably improved over the previous result shown in Figure~\ref{fig:preproc1}. 

\clearpage

\clearpage

\subsection{Strange valence preprocessing}
A special case when considering the preprocessing of the neural networks is that of the strange valence distribution. As specified in Equation~\ref{eq:NNPDF23param}, the strange valence PDF in the NNPDF2.3 determination had an auxiliary term to encourage the PDF to perform its required sign change in the valence region. Such an additional term has been previously needed due to the lack of specific data constraints upon the strange valence distribution before the LHC, introducing a bias, albeit a physically motivated one. Additionally the auxiliary term provides a mechanism by which the strange valence sum rule may be imposed. In the NNPDF3.0 determination and beyond this auxiliary term has been removed given the enlarged dataset and it's improved sensitivity to the strange PDF.

\begin{figure}[h!]
\centering
\includegraphics[width=0.42\textwidth]{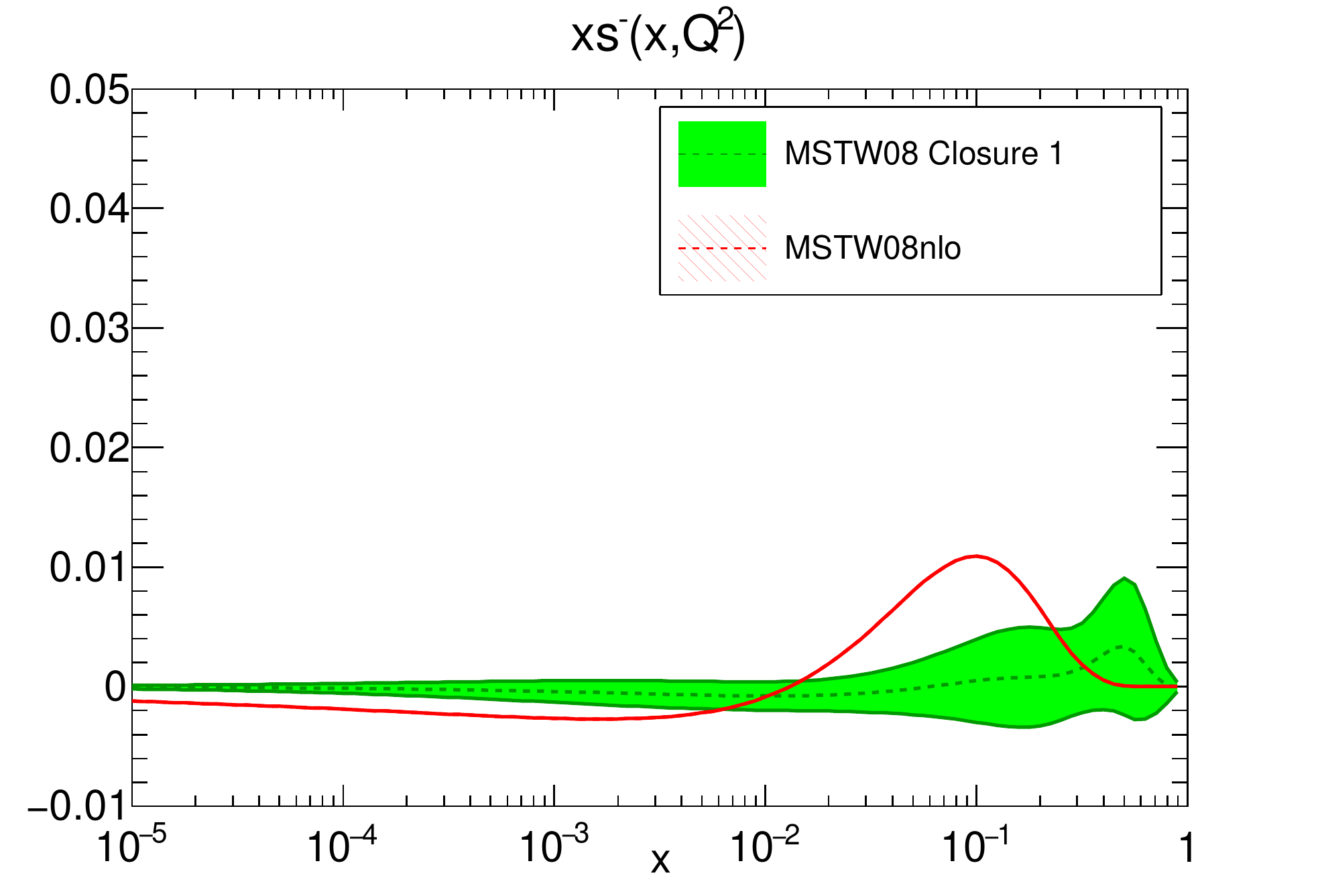}
\includegraphics[width=0.42\textwidth]{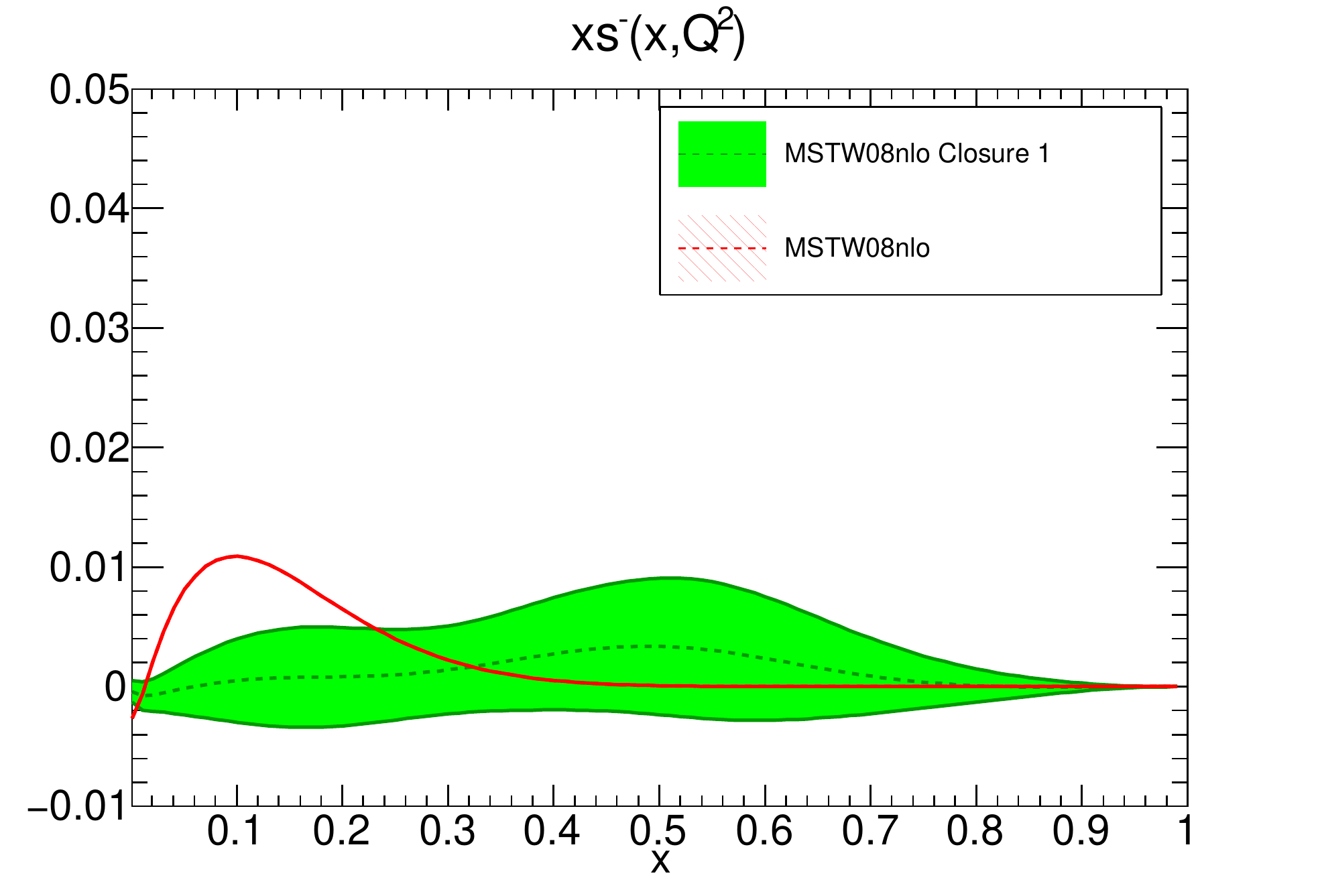}
\includegraphics[width=0.42\textwidth]{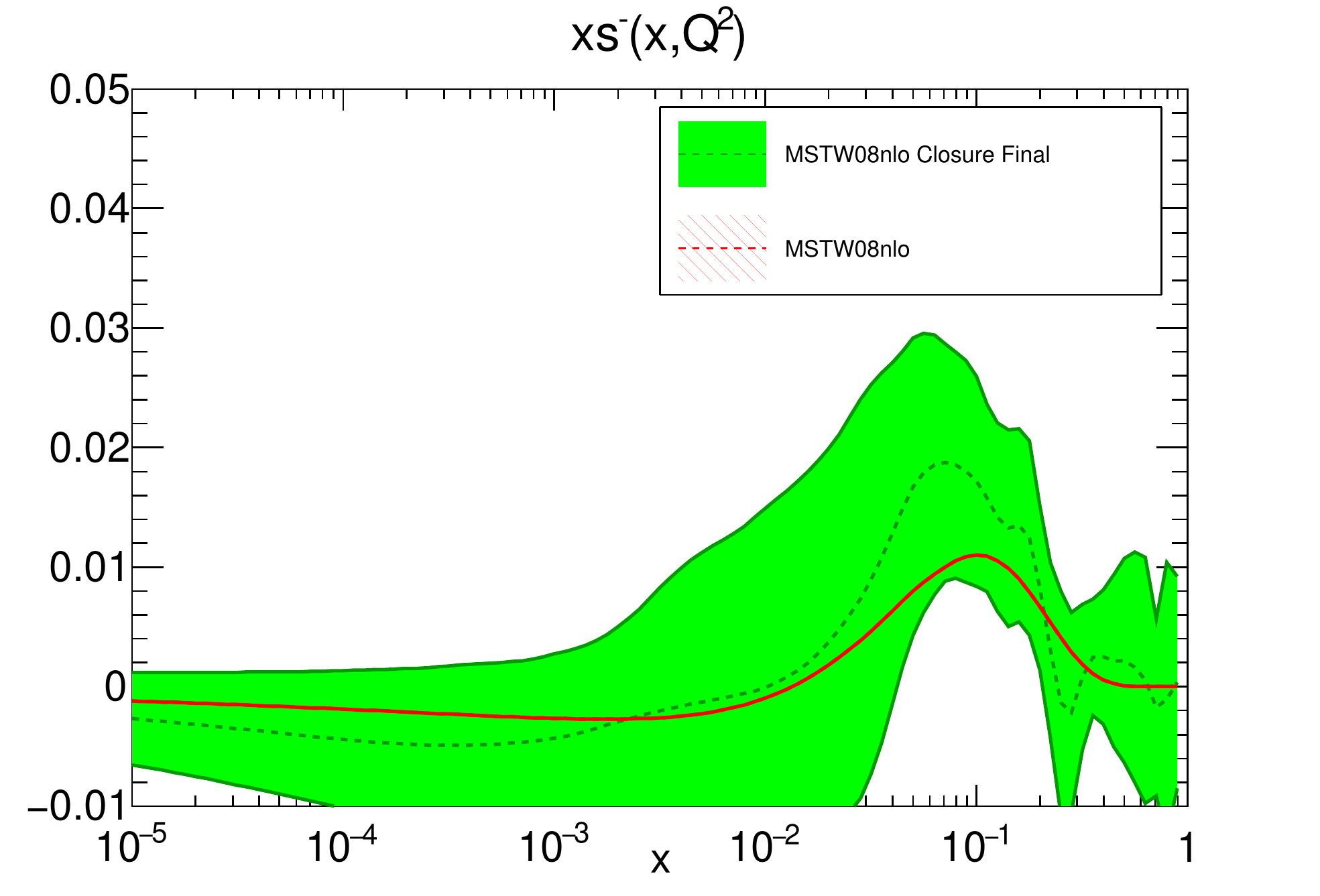}
\includegraphics[width=0.42\textwidth]{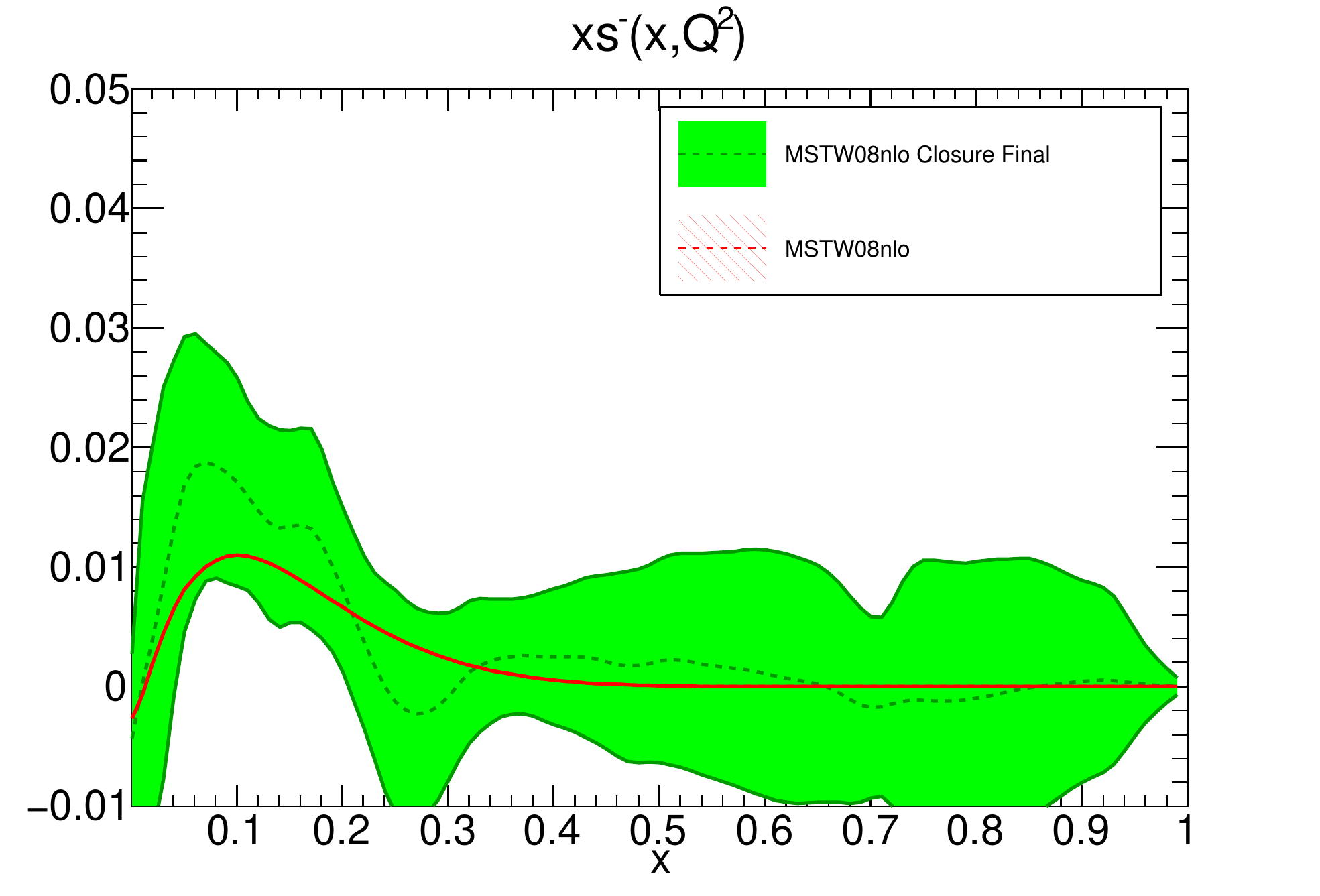}
\caption[Impact of more flexible treatment of strange valence PDF in fits post NNPDF2.3]{Impact of more flexible treatment of strange valence PDF in fits post NNPDF2.3. The top two figures show a comparison of a closure test performed with the NNPDF2.3 preprocessing, and the figures below show the results using the more flexible parametrisation.}
\label{fig:preproc5}
\end{figure}

Figure~\ref{fig:preproc5} demonstrates the effect of the removal of the strange auxiliary term upon a closure test fit to the MSTW08 set. While the NNPDF2.3 methodology closure fit struggles to accommodate the MSTW08 strange valence distribution, the updated methodology is able to reproduce the underlying law well, within enlarged uncertainties.

\clearpage

\section{PDF parametrisation}
The choice of PDF parameterisation and basis used in the fitting procedure has been reassessed with the help of the closure test procedure. In particular, a modification to the choice of fitting basis has been made necessary by the removal of the strange valence sum rule enforcing auxiliary term in the strange valence parametrisation. The most direct choice of fitting basis is the same basis as is used in PDF evolution, and therefore the basis required for PDFs in the {\tt FK} product. In this basis, the required quantum number sum rules may be applied as normalisations to the total valence, $V_3$ and $V_8$ distributions,
\ba V(x,Q^2_0) &=& N_V \left( u^- + d^-+ s^- \right)(x,Q^2_0),\nonumber \\
V_3(x,Q^2_0) &=& N_{V3} \left( u^- - d^- \right)(x,Q^2_0), \nonumber \\
V_8(x,Q^2_0) &=& N_{V8} \left( u^- + d^- - 2s^-\right)(x,Q^2_0),
\ea
where the normalisations $N$ are set such that
\ba \int_0^1 dx\, V(x,Q^2_0) &=& 3,\\
 \int_0^1 dx\, V_3(x,Q^2_0) &=& 1,\\
 \int_0^1 dx\, V_8(x,Q^2_0) &=& 3.\ea

\begin{figure}[!]
\centering
\includegraphics[width=0.9\textwidth]{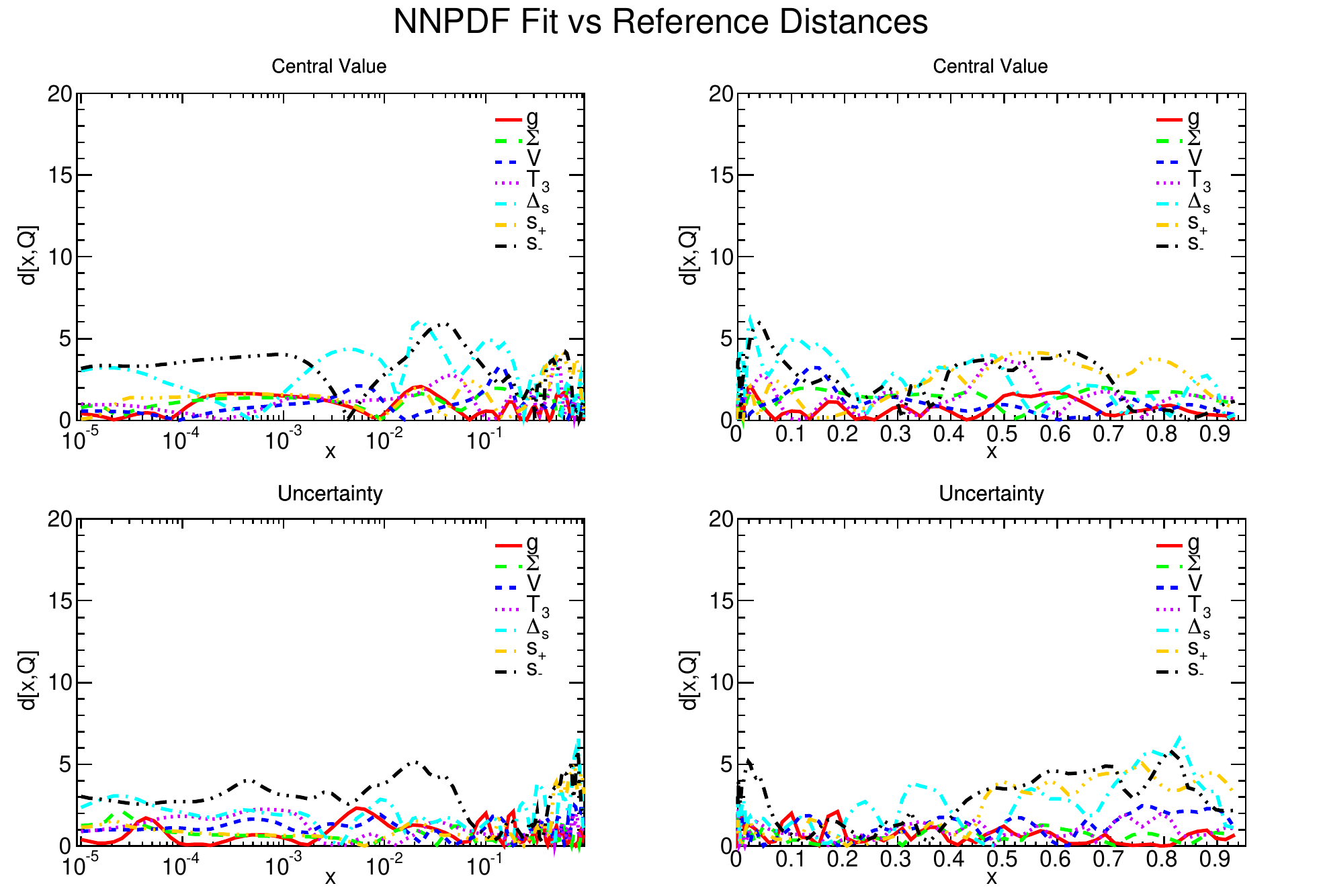}
\caption[Distance comparison of two closure test fits with differing parametrisation bases]{Distance comparison of two closure test fits with differing parametrisation bases. Distances are defined through the measure in Appendix~\ref{app:distances}, whereby a distance of 10 corresponds to $1\sigma$.}
\label{fig:EVOLvs23BASIS}
\end{figure}

In such a way, the total valence quantum number is fixed, along with the up, down and strange valence quantum numbers. The evolution basis also has the advantage of being
particularly efficient, not requiring any transformation before combination with {\tt FK} tables to calculate physical observables. We have shown, based upon closure test results, that the fit results
show a good degree of stability under such a change in parametrisation basis. While the previous strategy was designed to construct PDF combinations with specific data constraints, the flexibility of the fit
means that the results do not suffer when moving away from such a basis. Figure~\ref{fig:EVOLvs23BASIS} shows the statistical distance between a fit with the full evolution basis with a fit based upon the standard NNPDF2.3 parametrisation basis.
As expected, any differences are isolated to those PDFs whose parametrisation (and therefore preprocessing) has substantially changed e.g $\Delta_S$ and the strange PDFs. Even in these PDFs the differences are typically
less than half a standard deviation.

\section{Minimisation and stopping}
In addition to examining areas where the choice of parametrisation may lead to some degree of bias, the closure test procedure is particularly useful for assessing the efficacy of a fitting methodology. Furthermore, the substantial gains in computational efficiency made in the transition to the {\tt nnpdf++} code mean that far more aggressive genetic minimisation strategies may be implemented.

The entirety of the NNPDF minimisation procedure has therefore been re-examined to ensure that it is the most effective methodology in the light of additional constraints coming from the LHC. Here we shall summarise some of the major modifications made since the NNPDF2.3 determination.

\subsection{Target weighted training}
Target Weighted Training (TWT) was a central feature of previous NNPDF determinations. TWT was developed in early NNPDF fits as a method of obtaining a balanced training across datasets, solving a problem with early neural network fits whereby some smaller datasets
were largely ignored by the minimisation in favour of larger, more constraining sets. This typically led to a very uneven fit quality profile over the complete experimental dataset. The TWT procedure solved this problem by introducing a training epoch at the beginning of a fit where each dataset had a target $\chi^2$. In the event where a fit iteration reached a $\chi^2$ value higher than the target, a large weight in fit quality was applied to that dataset in order to bring its fit quality down. 

\begin{figure}[!]
\centering
\includegraphics[width=1\textwidth]{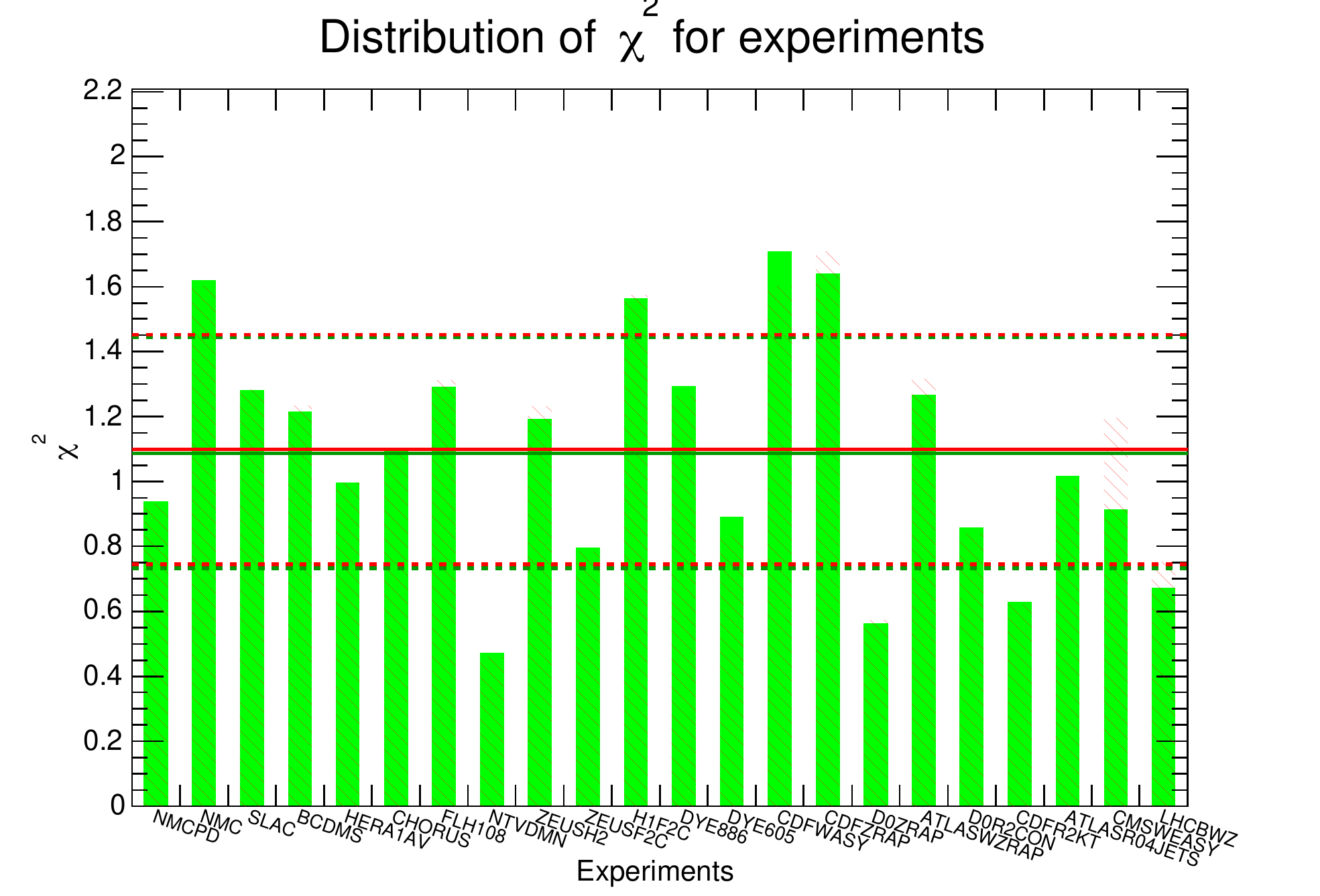}
\caption[Comparison of $\chi^2$ by dataset between real data fits with and without Target Weighted Training]{Comparison of $\chi^2$ by dataset between real data fits with (green bars) and without (red bars) Target Weighted Training.}
\label{fig:twtvsnotwt}
\end{figure}

While ensuring a relatively even training profile, the TWT procedure had a number of difficulties. The most important being the restriction of the early fit to a $\chi^2$ fit quality measure applied on a dataset-by-dataset basis, ignoring experimental uncertainty cross-correlations such as luminosity uncertainties, between datasets. Furthermore the TWT procedure introduced a considerable amount of complexity in the fitting procedure. With this in mind, real data fits with target weighted training were compared to fits without in the {\tt nnpdf++} framework with the large experimental dataset of NNPDF2.3 and updated genetic algorithm parameters. Figure~\ref{fig:twtvsnotwt} compares the dataset-by-dataset fit quality of two such example fits. With these fits we can see clearly that with a larger dataset and more efficient GA procedure, no large training imbalance can be seen in the fits even without the TWT procedure applied. Future NNPDF fits will therefore be performed without target weights, allowing for the consistent application of the experimental correlations across datasets throughout the fitting procedure.

\subsection{Genetic algorithm}
A number of changes have been made to the GA procedure used in NNPDF fits in order to improve fitting efficiency and provide more precise PDF determinations. In the analysis of the efficacy of a GA, level zero closure tests are particularly helpful in that they
directly test the ability of a minimisation procedure to reproduce a given function precisely. In these fits the closure test fit should be able to effectively draw a line between datapoints, leading to an ideal $\chi^2$ of zero to the pseudo-data.
%
%
%140430-r1753-001-sc.ini - DynStop
%140512-r1765-003-cd.ini  - LookBack 30k
%140512-r1765-004-cd.ini  - LookBack 60k
%
%
A number of modifications to the procedure have been tested, the most effective of which is the implementation of \emph{Nodal} mutations in the GA~\cite{Montana:1989:TFN:1623755.1623876}. In previous versions of the NNPDF GA, mutations were performed upon individual parameters of each neural network with no consideration as to their position in the network. 

\begin{figure}[!]
\centering
\includegraphics[width=0.48\textwidth]{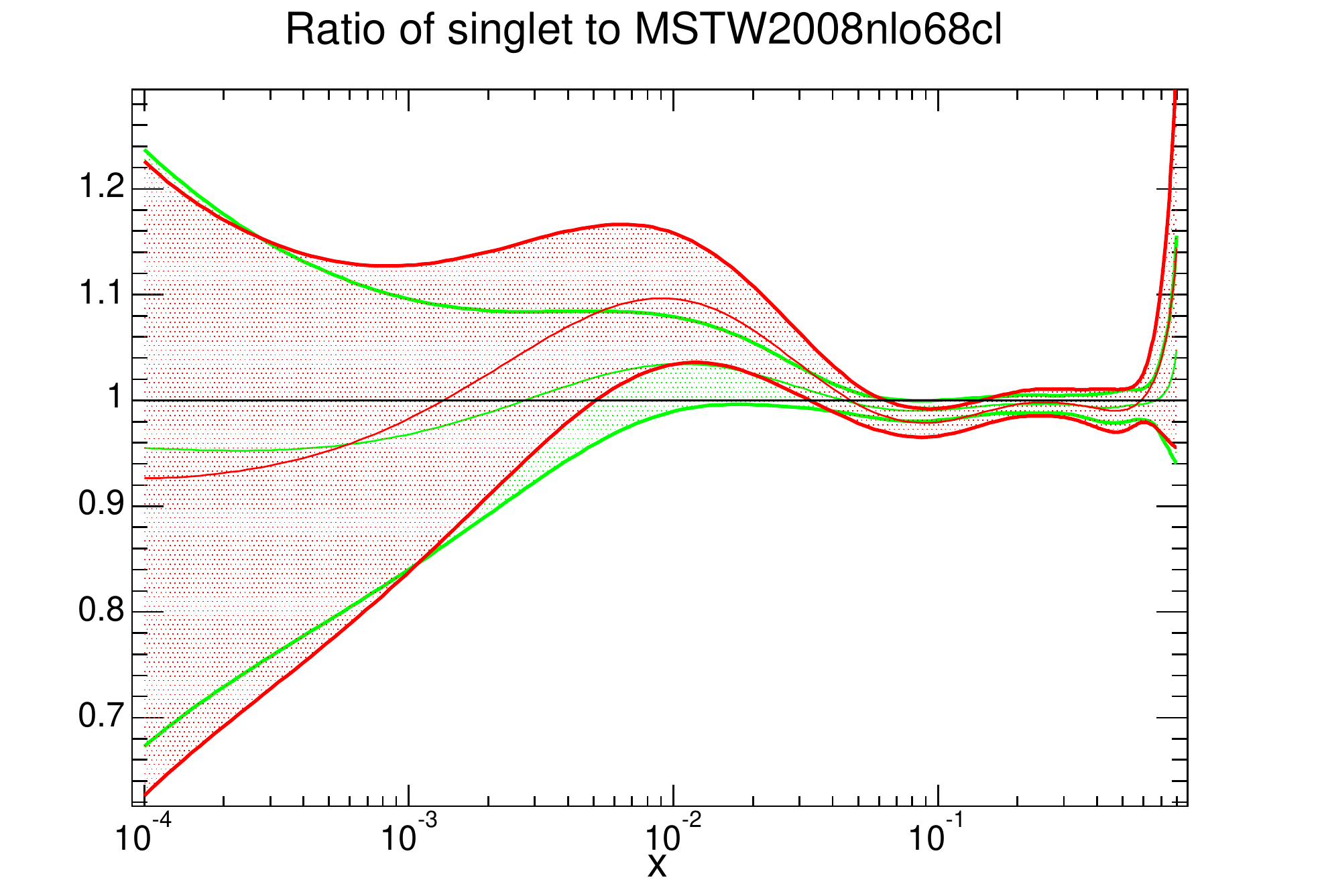}
\includegraphics[width=0.48\textwidth]{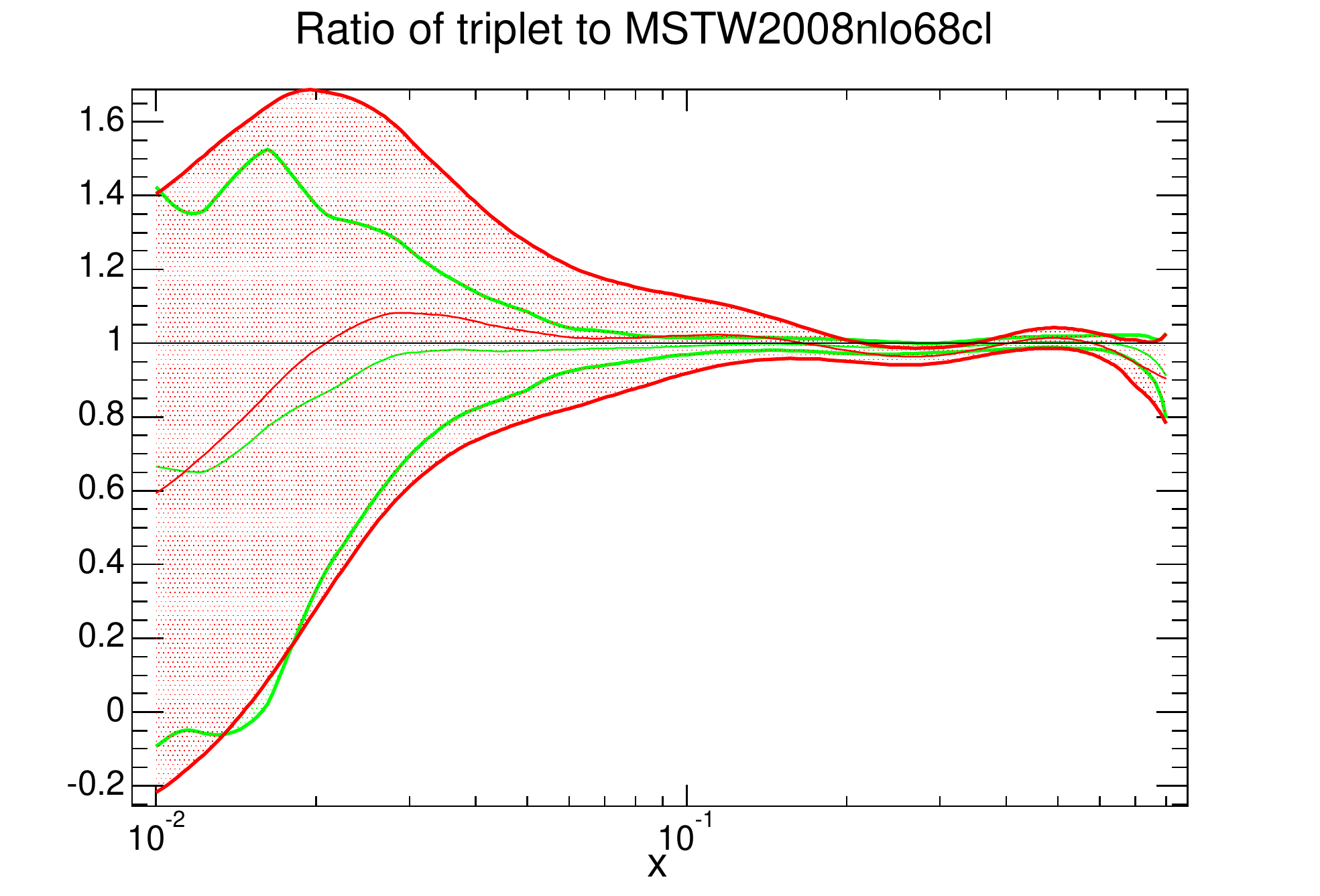}
\caption[Comparison of a conventional NNPDF GA fit with a Nodal GA fit in a closure test to MSTW2008]{Comparison of a conventional NNPDF GA fit (red bands) with a Nodal GA fit (green bands) in a closure test to MSTW08. PDFs are given as a ratio to the generating PDF set for the singlet (left) and triplet (right) distributions.}
\label{fig:nodalvsnonnodal}
\end{figure}

The concept of nodal mutations introduces the strategy of mutating all parameters associated with a particular neural network \emph{node} at once. In this procedure a node of the network is chosen at random, then all of its associated weights connected to the earlier layer are mutated along with its threshold parameter. Doing so yields a much more effective genetic algorithm as demonstrated in the comparison in Figure~\ref{fig:nodalvsnonnodal}, where a standard GA is compared to a nodal mutation GA in their reproduction of the MSTW underlying law. The nodal GA is able to better resolve the underlying law, and to a greater precision. The comparison in Figure~\ref{fig:nodalvsnonnodal} is corroborated by the $\chi^2$ values of the two fits to the perfect pseudo-data in the level zero fit. The standard GA fit shown in the figure obtained a final $\chi^2$ of 0.0279 compared to 0.0043 for the nodal GA. The nodal GA strategy has therefore been adopted for future NNPDF determinations.

\subsection{Dynamical stopping}
The cross-validation dynamical stopping procedure utilised in previous {\tt FORTRAN} based NNPDF fits was triggered by a slope-detection algorithm applied to the fit quality profiles of each replica to the validation dataset. While providing a reasonable stopping criteria and preventing excessive overfitting, the relative balance between the degree of under- and over-learning was governed by the parameters of the slope-detection algorithm. Such sensitivity to the stopping parameters meant that a re-tune was often necessary upon large modifications to the dataset or minimisation algorithm. 

The modular nature of the stopping criteria implemented in the {\tt nnpdf++} framework means that alternative stopping procedures may be quickly and safely implemented to investigate their impact. One such stopping criterion that has demonstrated greater stability than the previous slope-detection based procedure is that of \emph{look-back} cross-validation. 

\begin{figure}[!]
\centering
\includegraphics[width=0.7\textwidth]{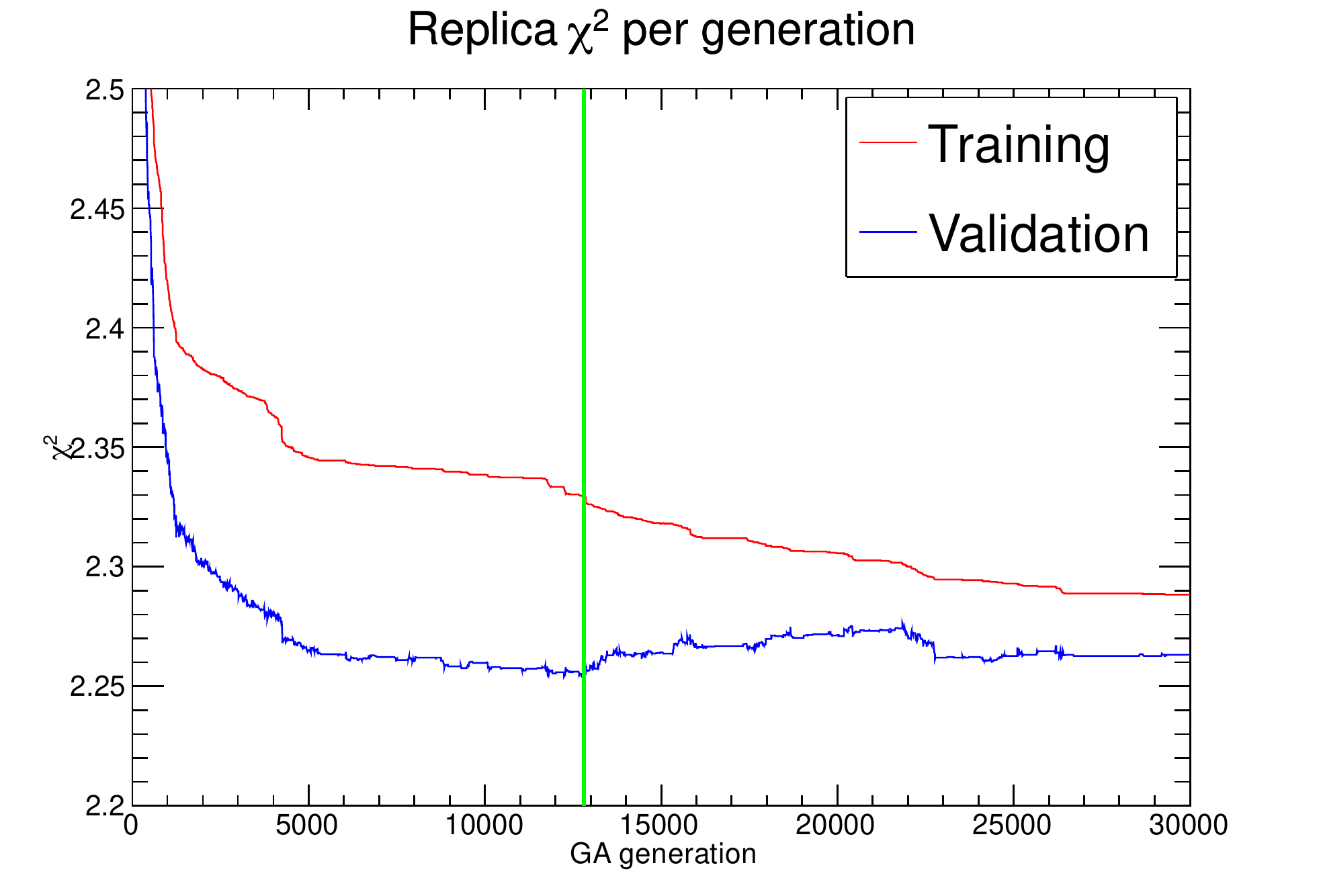}
\caption[Fit quality profiles for the training and validation sets in look-back cross validation]{Fit quality profiles for the training and validation sets in look-back cross validation. The red curve shows the fit quality to the training set, and the blue curve to the validation set as the number of fit generations goes on. The green line indicates the stopping point selected by the look-back criterion, generation 12813 having the minimum validation $\chi^2$.}
\label{fig:LBCVchi2prof}
\end{figure}

In this procedure all replicas are run for the maximum number of generations $N_{\text{gen}}^{\text{max}}$, all the while storing the GA generation that best described the validation dataset. At the end of the fit, the GA generation that minimised the $\chi^2$ to the validation set is selected as the best-fit stopping point, and that replica is used as a member of the Monte Carlo ensemble. This method yields an extremely clean stopping criterion, having no tuneable parameters aside from the maximum number of generations, and offers a very faithful implementation of the cross-validation method. Furthermore, the look-back procedure is not practically more time-consuming to implement despite running each replica to the maximum number of generations, as even in the previous dynamical stopping procedure the time taken to run a fit is typically given by the time taken by the slowest replica. In Figure~\ref{fig:LBCVchi2prof} the fit quality profile for a single PDF replica can be seen for the training and validation sets alongside the look-back stopping point. In this case, the look-back method can clearly discern an overlearning signal, as the fit quality to the validation set worsens while the training set $\chi^2$ improves.

\begin{figure}[h!]
\centering
\includegraphics[width=0.48\textwidth]{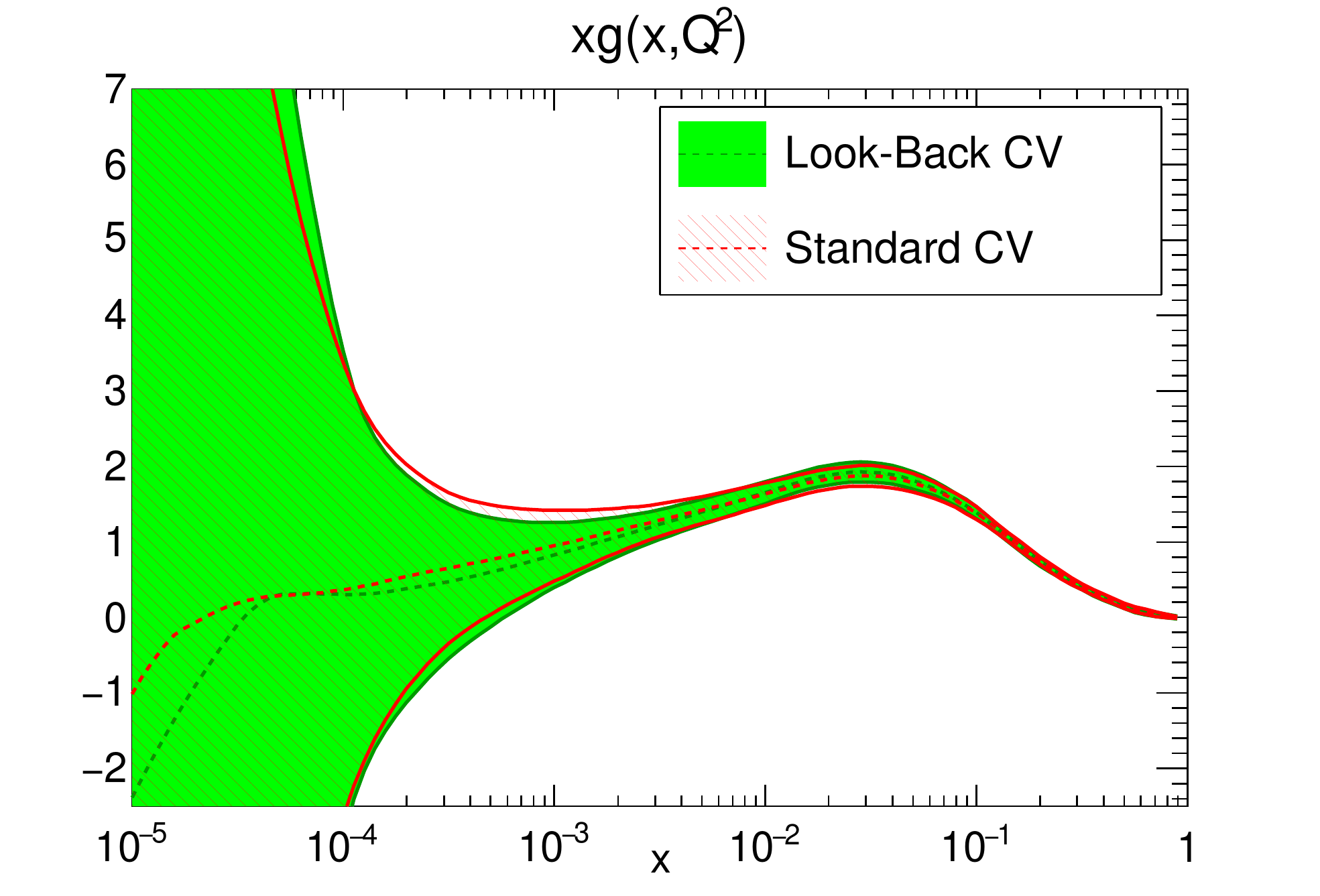}
\includegraphics[width=0.48\textwidth]{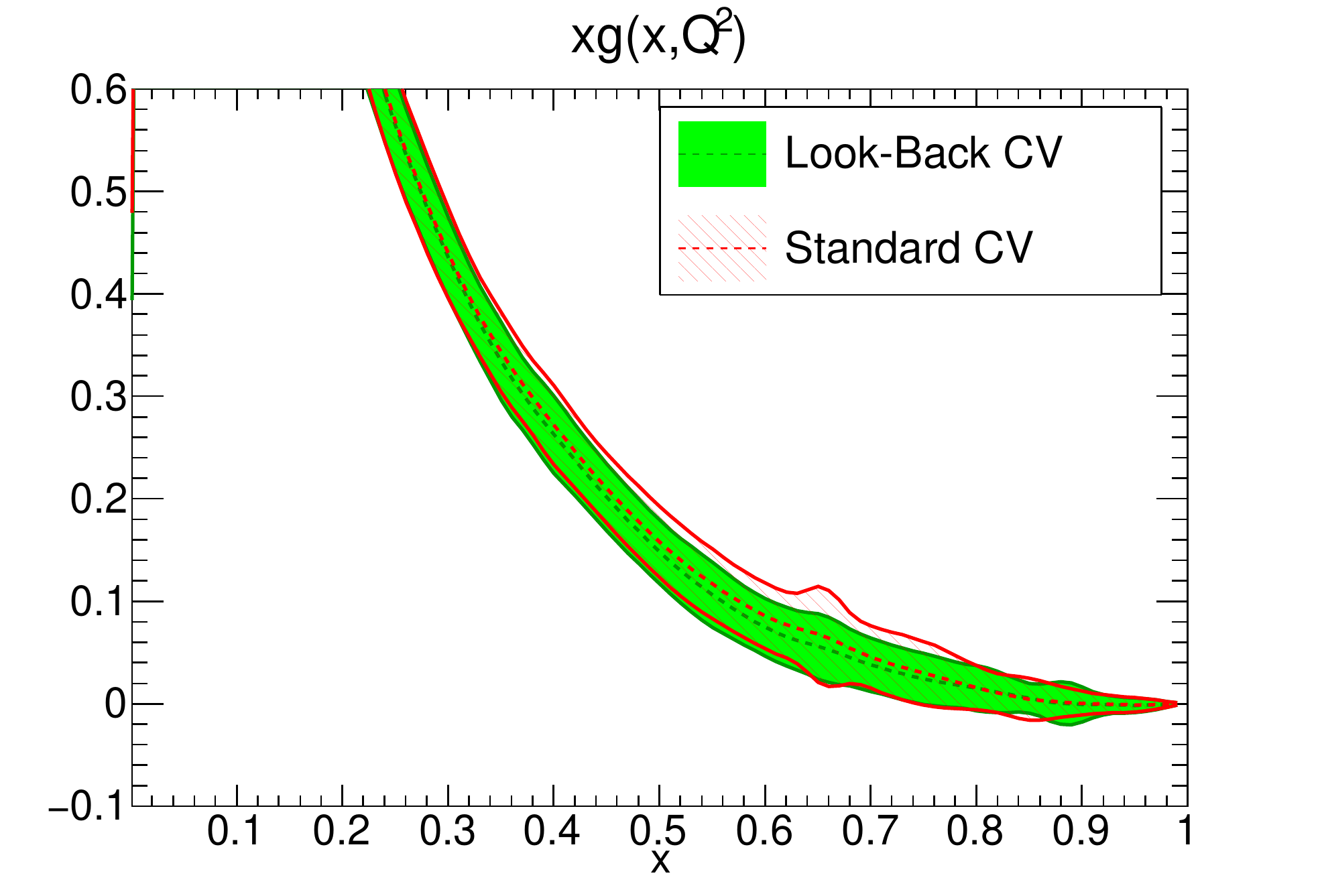}\\
\includegraphics[width=0.48\textwidth]{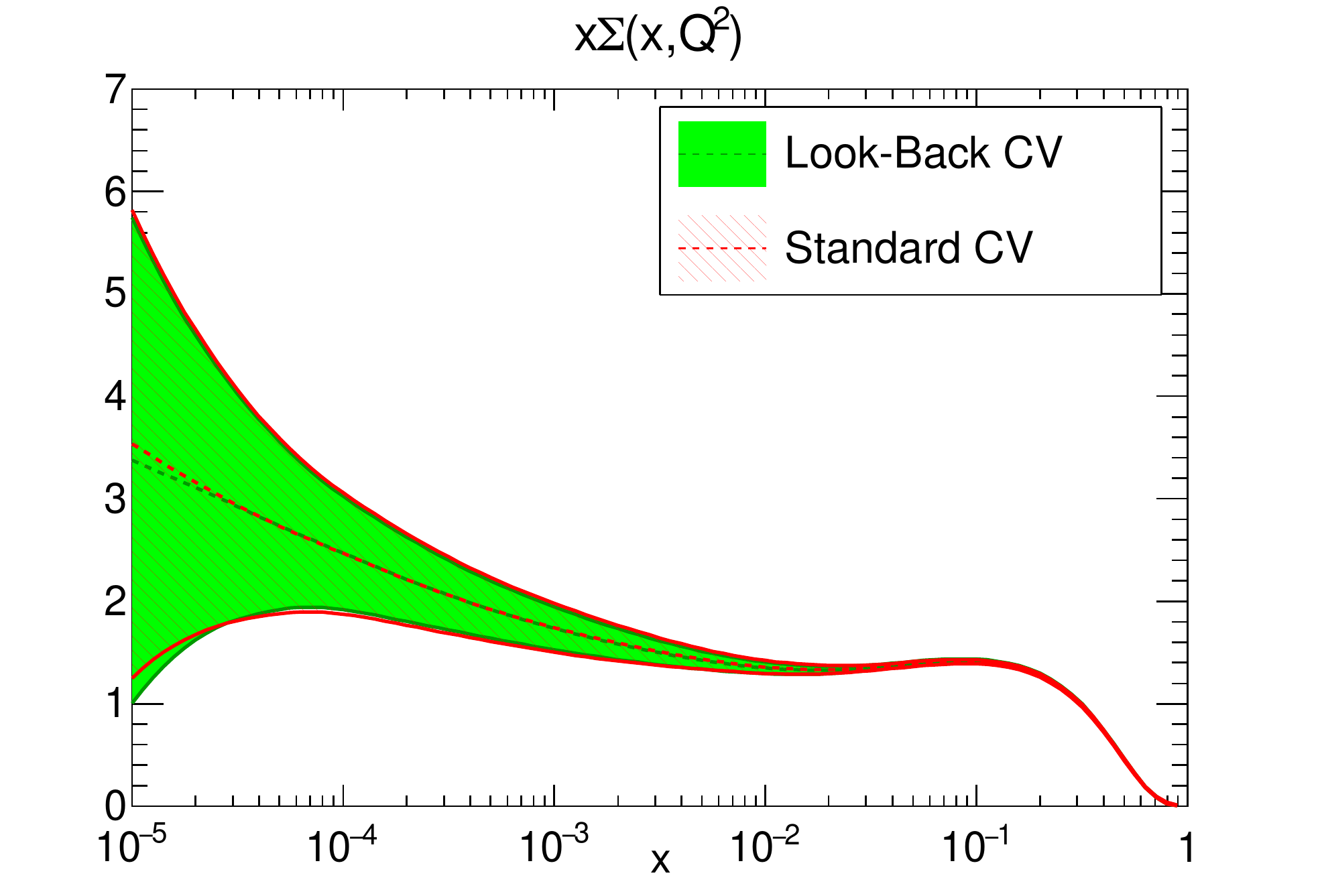}
\includegraphics[width=0.48\textwidth]{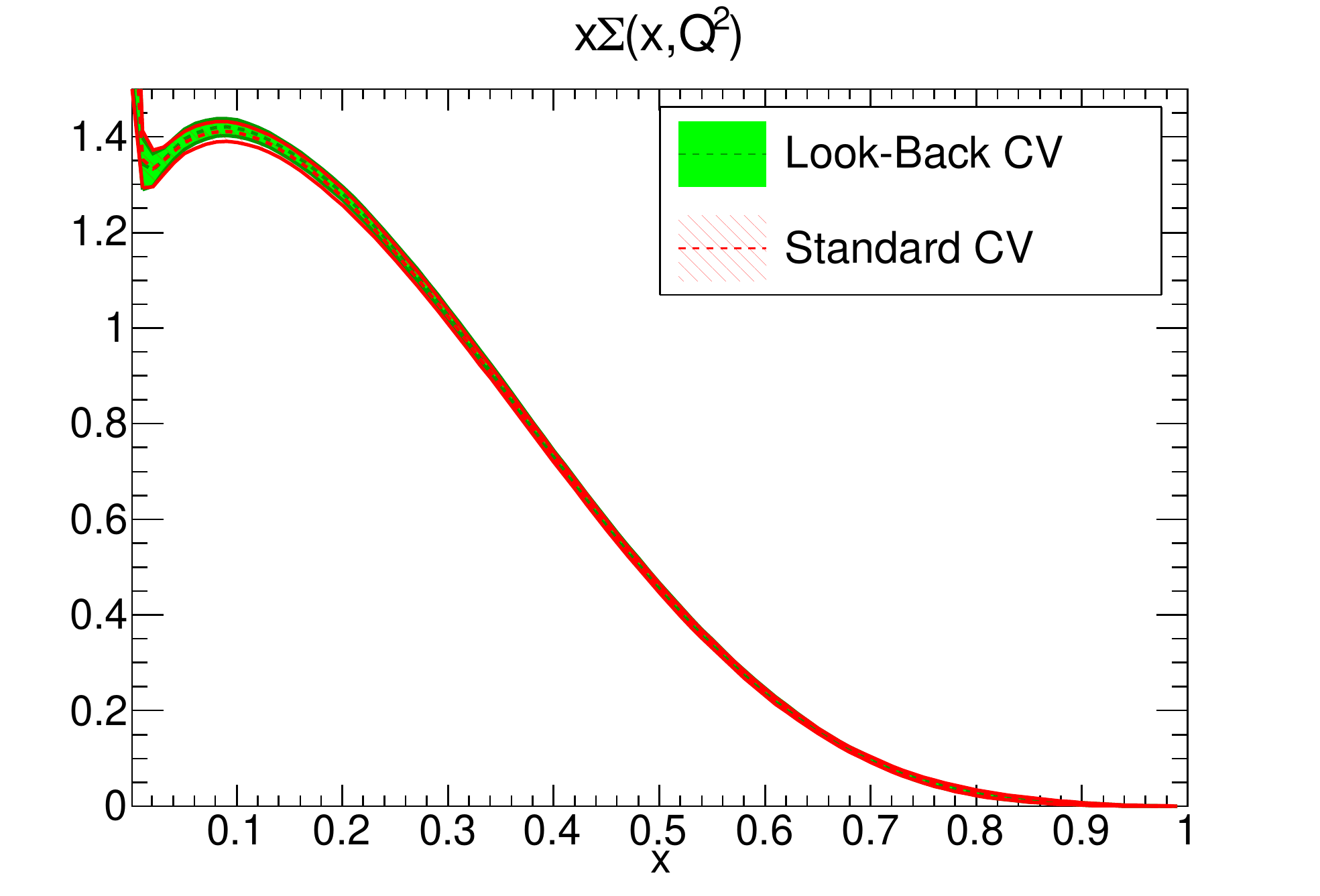}
\caption[Comparison of PDFs obtained through look-back cross validation and NNPDF2.3 standard dynamical stopping]{Comparison of PDFs obtained through look-back cross validation and NNPDF2.3 standard dynamical stopping. PDFs for the singlet and gluon are shown, with green bands representing fits using the look-back method and red demonstrating those with the slope-detection algorithm used in NNPDF2.3 and earlier.}
\label{fig:30kLBvsDYN}
\end{figure}

\begin{figure}[!]
\centering
\includegraphics[width=0.48\textwidth]{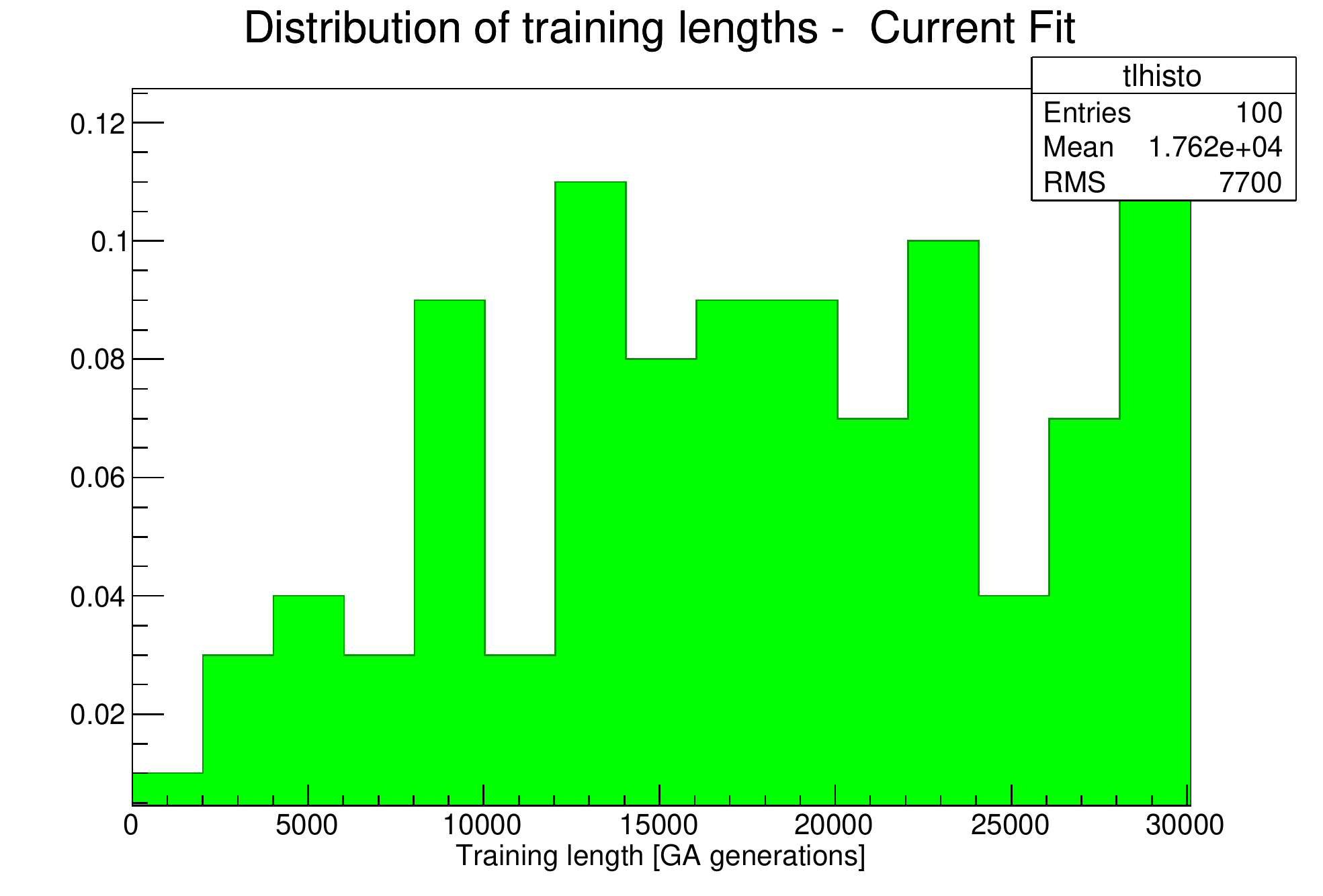}
\includegraphics[width=0.48\textwidth]{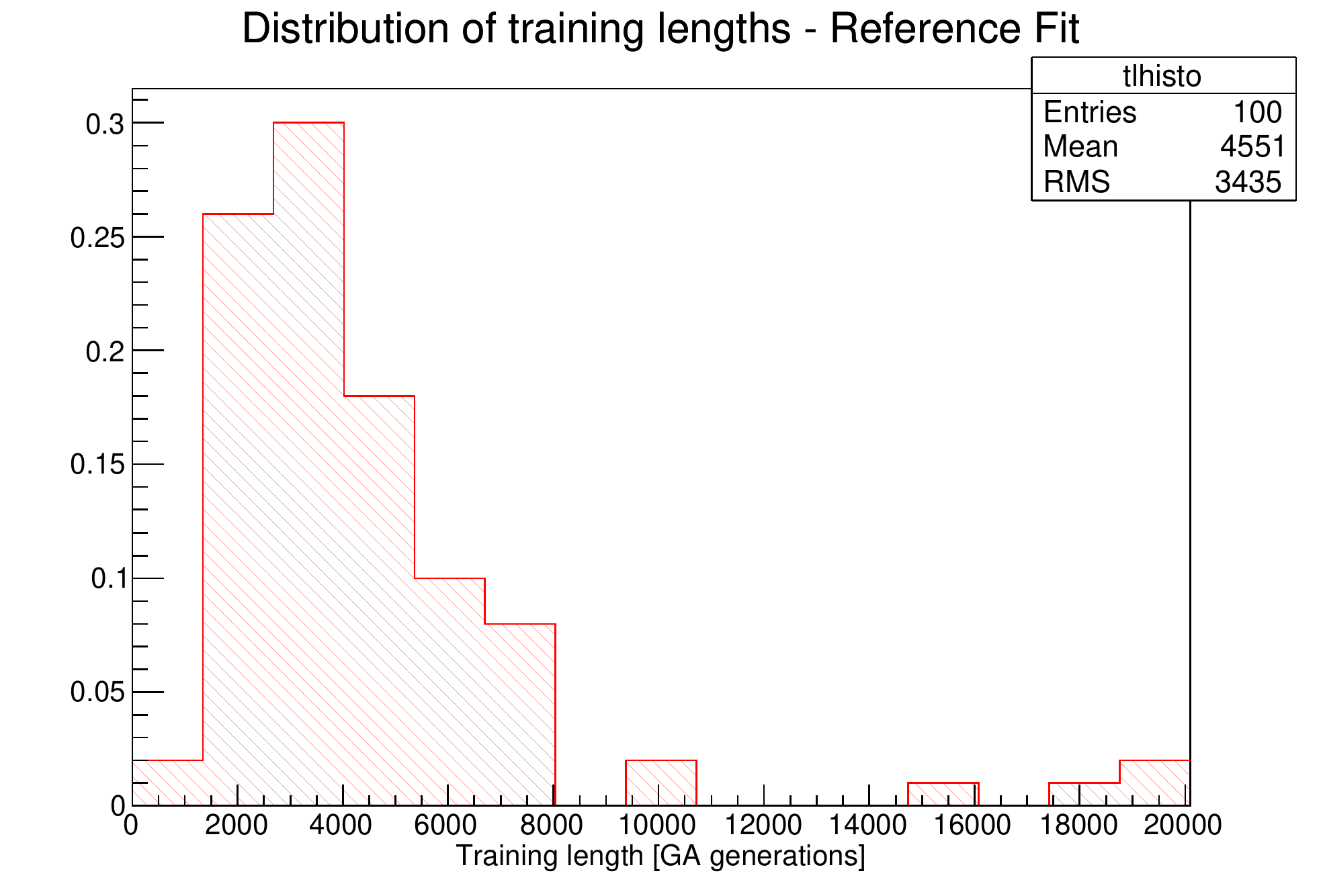}
\caption[Comparison of training lengths in look-back cross-validation and NNPDF2.3 standard dynamical stopping]{Comparison of training lengths in look-back cross-validation and NNPDF2.3 standard dynamical stopping. The left figure demonstrates the 'optimal point' determined by looking back over the while GA interval for the minimum validation $\chi^2$. The right figure shows the stopping point based upon the slope-detection algorithm.}
\label{fig:30kLBvsDYNtl}
\end{figure}

In Figure~\ref{fig:30kLBvsDYN} we compare the results for the singlet and gluon PDFs in the case of a look-back fit with $N_{\text{gen}}^\text{max}=$ 30,000 generations, and a fit with the NNPDF2.3 standard dynamical stopping. In both instances, the fit performed was a level two closure test using MSTW2008 as the underlying law. While differences are small the look-back fit demonstrates slightly smaller uncertainties, implying a marginal underlearning present in the NNPDF2.3 dynamical stopping procedure. The fits yield essentially equivalent results, although the optimal point determined in the look-back method is typically somewhat later than in the dynamical stopping as can be seen in the comparison of training length histograms in Figure~\ref{fig:30kLBvsDYNtl}. In this figure it is clear also that several PDF replicas in the look-back method stop close to the maximum number of generations available, implying that no significant overlearning can been resolved in their cases over the given GA interval.

In order to examine the effect of increasing the length of the look-back period, we compare the 30,000 generation fit to an extended 60,000 generation fit in Figure~\ref{fig:30kLBvs60kLB} where we use the PDF distance definition in Appendix~\ref{app:distances}. Distances of effectively zero throughout the PDF combinations and $x$-range mean that no change is observed between the two fits, demonstrating the stability of the method once a sufficiently large look-back length is used. The look-back cross-validation method as discussed here will therefore be implemented as the default stopping criterion for the NNPDF3.0 family of fits.

\begin{figure}[!]
\centering
\includegraphics[width=0.9\textwidth]{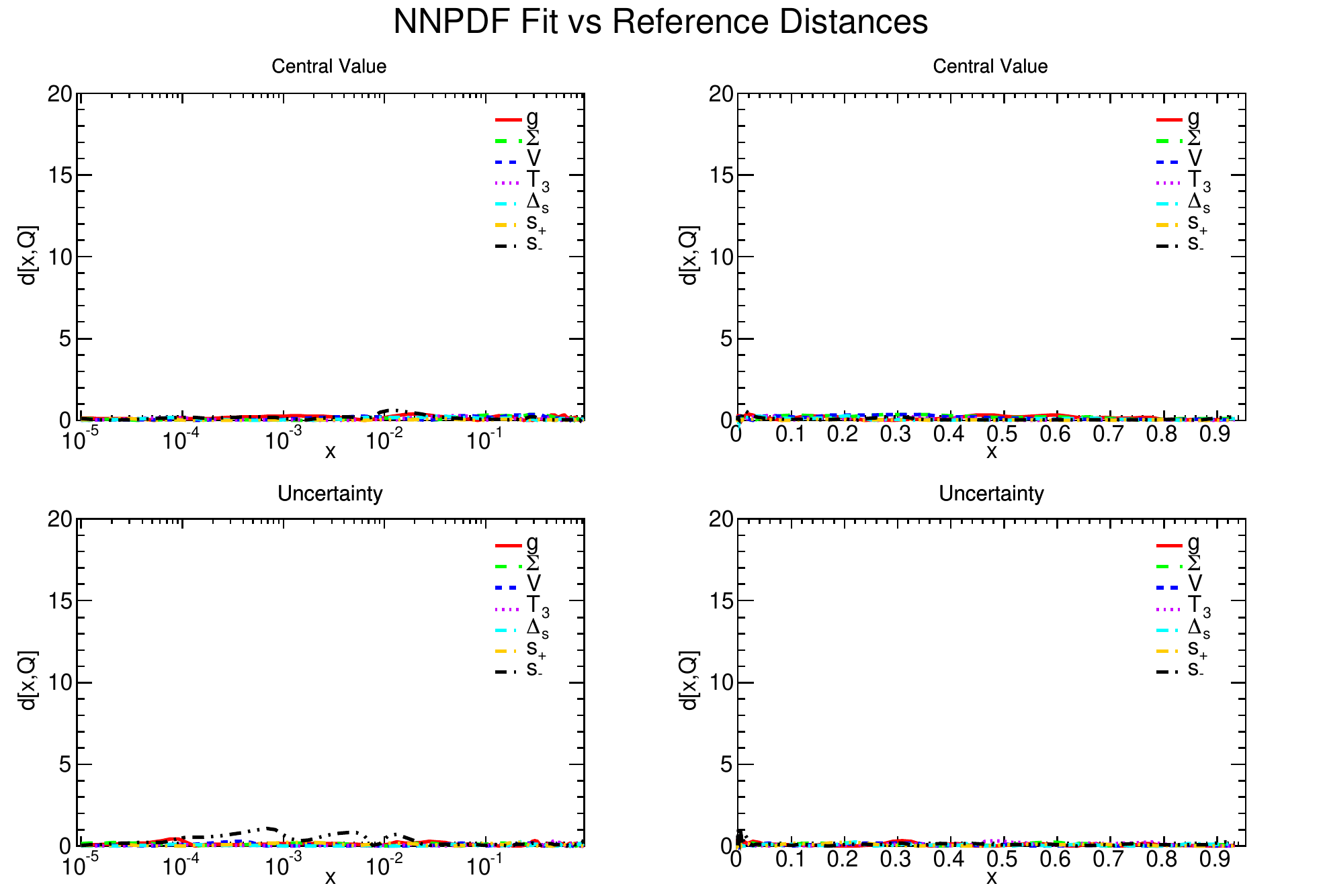}
\caption[Distance comparison of two closure test fits with look-back stopping enabled and different maximum training lengths]{Distance comparison of two closure test fits with look-back stopping enabled and different maximum training lengths. Distances are computed between all evolution basis PDFs at the initial scale between $N_{\text{gen}}^\text{max}=$ 30,000 and $N_{\text{gen}}^\text{max}=$ 60,000 generation look-back fits.}
\label{fig:30kLBvs60kLB}
\end{figure}

\section{Methodology for NNPDF3.0}
We have performed an overview of the methodological developments made since the release of the NNPDF2.3 PDF set, with an aim to outline the procedure to be used in the forthcoming NNPDF3.0 set. To provide a stringent verification of the combined procedure, we shall now examine a set of closure test fits performed at various levels to differing generating PDF sets. In this section we present fits based upon a nodal genetic algorithm minimisation with look-back cross-validation stopping as detailed previously, with the iterative preprocessing procedure and new PDF fitting basis. Therefore the fits represent preliminary closure test results for the NNPDF3.0 methodology, upon a global pseudo-dataset of hadronic and DIS data. Results in this section will be presented using NLO calculations for the observables in the fit, although the conclusions will be very similar for an identical analysis at NNLO, as the closure test procedure is relatively insensitive to theory choices.

\subsection{Closure tests for NNPDF3.0}

\begin{figure}[!]
\centering
\includegraphics[width=0.45\textwidth]{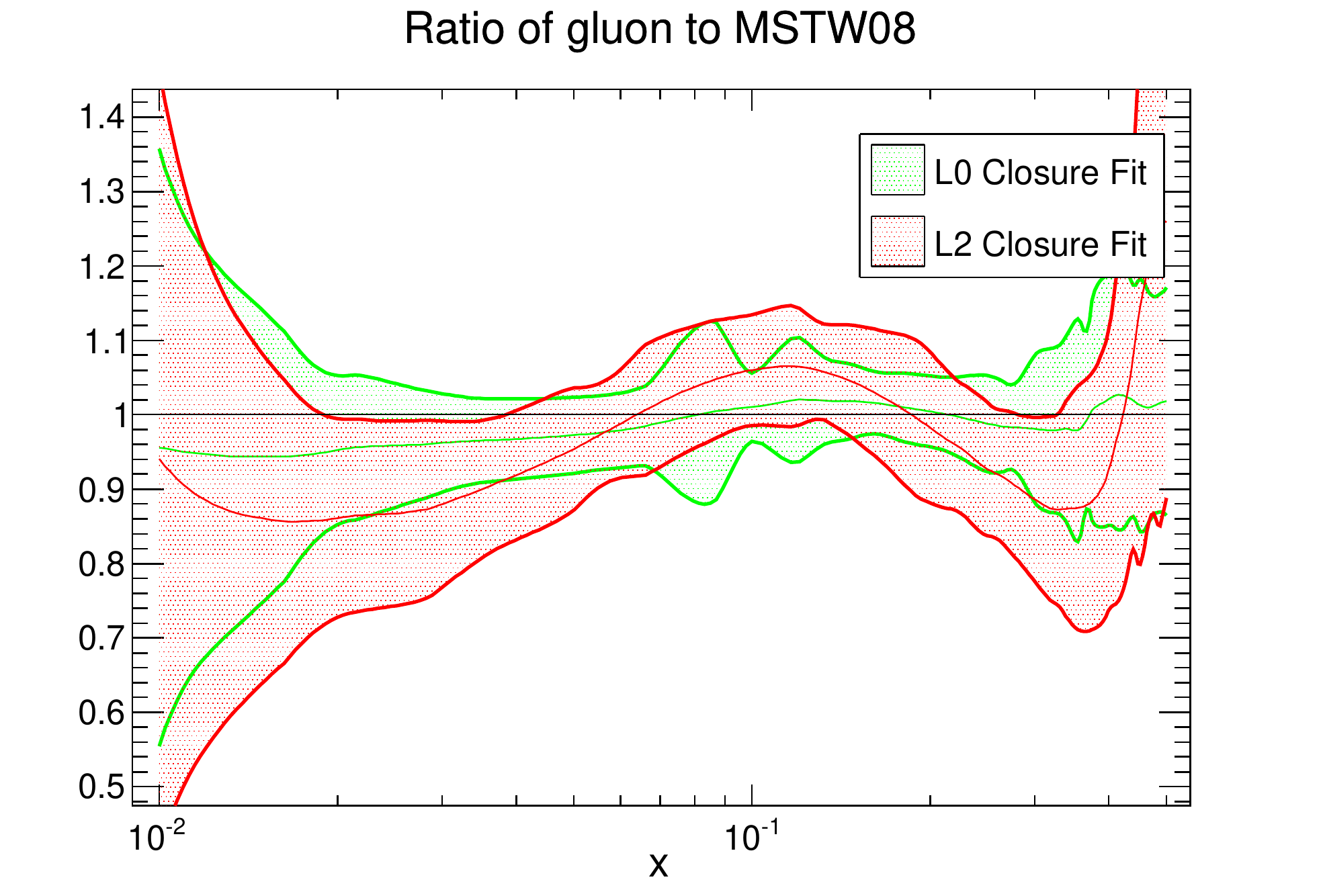}
\includegraphics[width=0.45\textwidth]{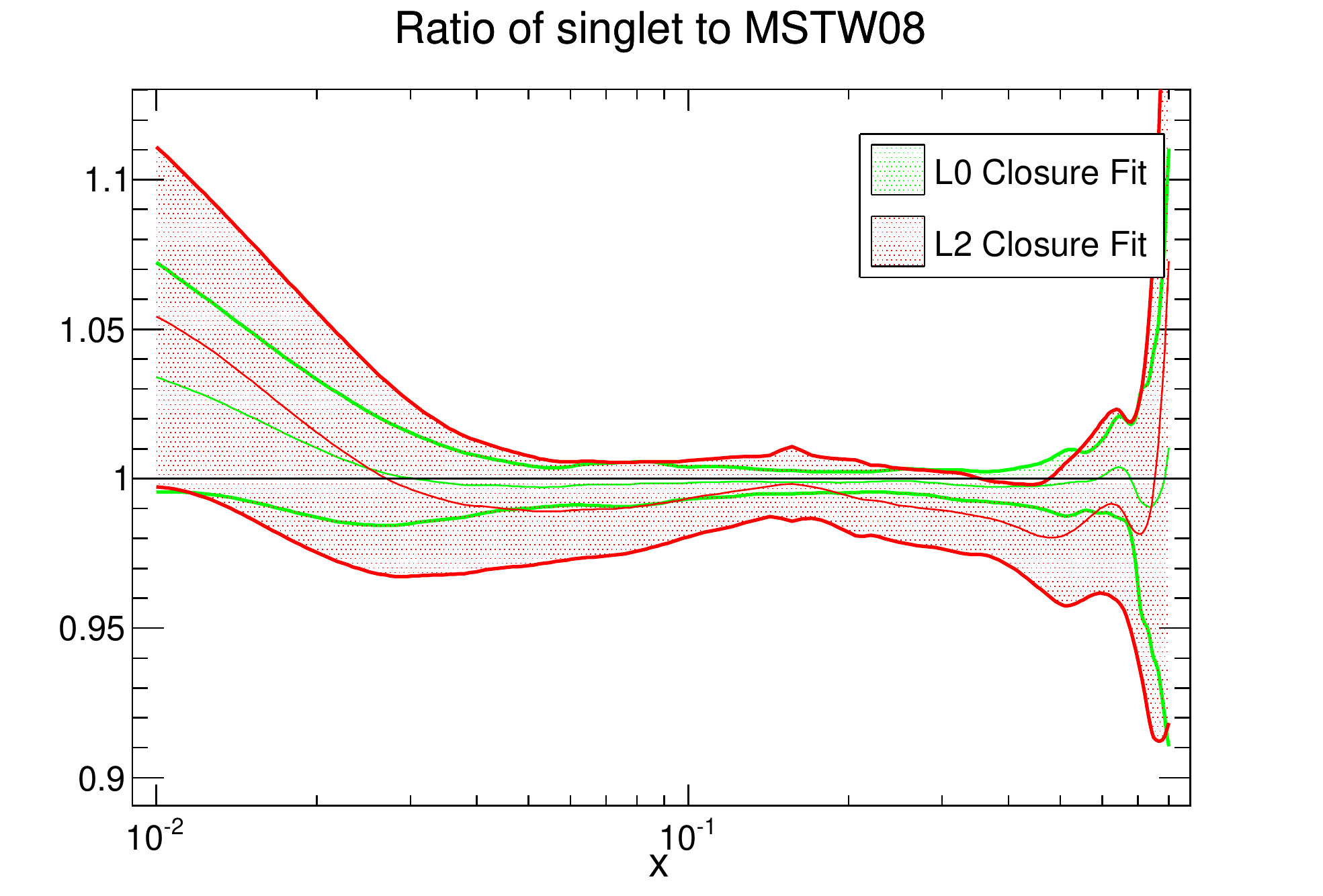}
\includegraphics[width=0.45\textwidth]{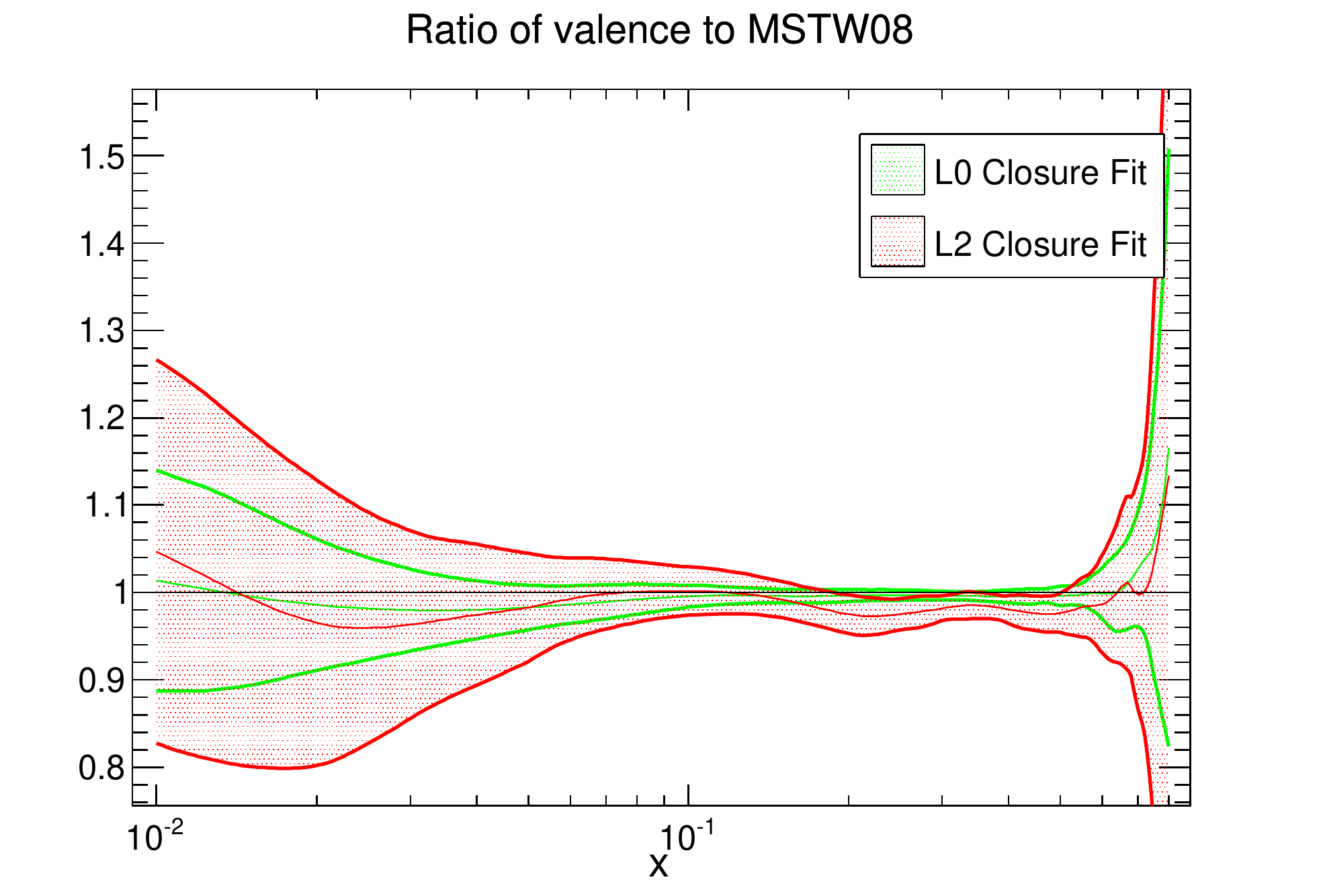}
\includegraphics[width=0.45\textwidth]{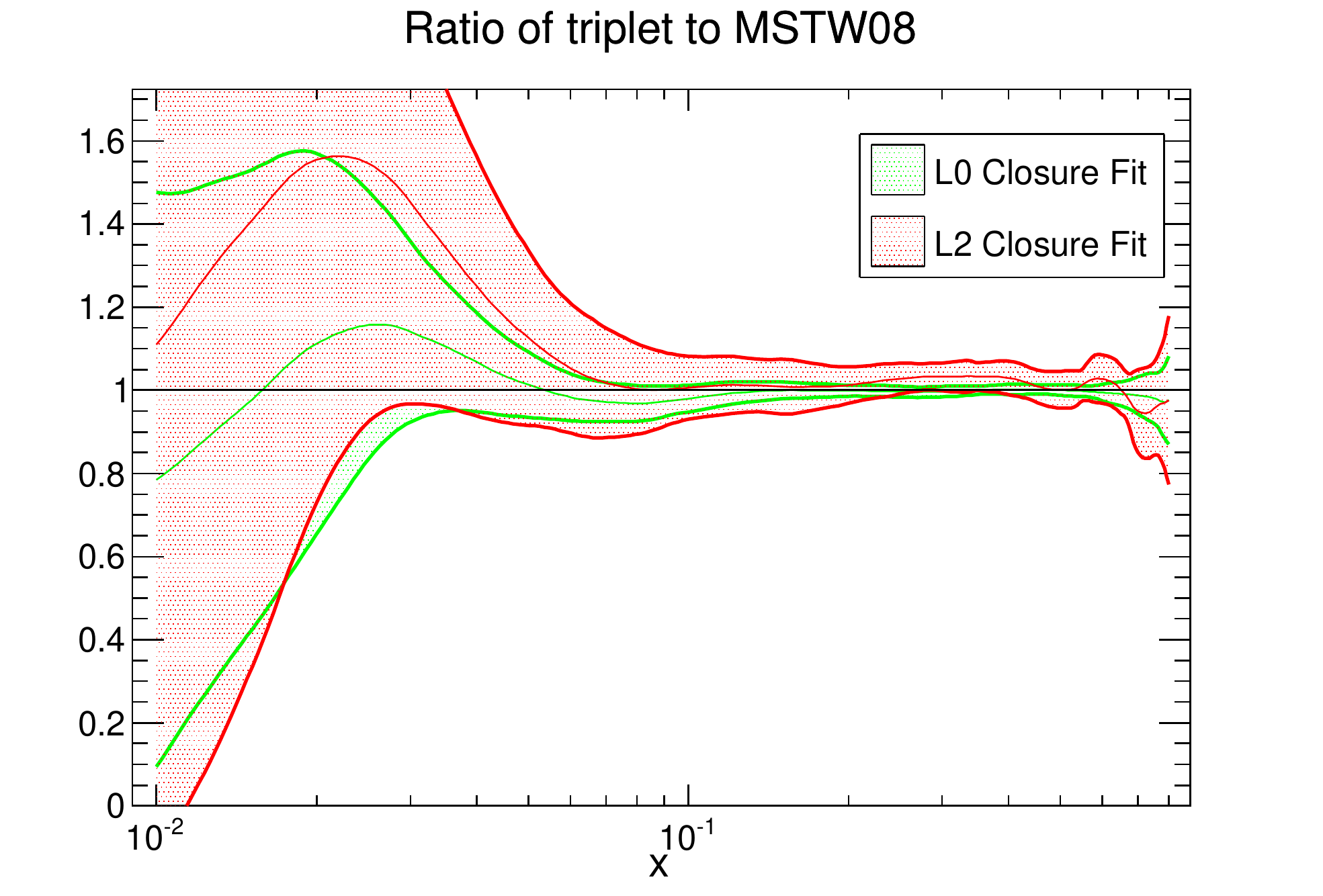}
\caption[NNPDF3.0 methodology closure test fit to MSTW2008 NLO]{NNPDF3.0 methodology closure test fit to MSTW2008 NLO. Curves are shown normalised to the generating PDF for the gluon, singlet, triplet and valence distributions. The green curves show the results of a level zero closure test, while the red curves show the results of a level two test.}
\label{fig:finalClosure_MSTW}
\end{figure}

Firstly let's consider the results obtained when fitting to an MSTW2008 generating PDF, the closure test guiding the methodological choices made so far in this section. In Figure~\ref{fig:finalClosure_MSTW} the ratio of the resulting closure test PDFs to the generating MSTW08 distributions are shown for some of the evolution basis PDFs. Here we show results for the kinematic region most constrained by the experimental pseudo-dataset: $10^{-2} \le x \le 1$. The level zero curves in Figure~\ref{fig:finalClosure_MSTW} closely reproduce the MSTW central values, achieving a final total $\chi^2/N_{\text{dat}} = 0.00182$. The uncertainty band in the case of the level zero result corresponds directly to the functional freedom available within the fitted pseudo-dataset. The level two fit clearly demonstrates the variations introduced by the simulated experimental noise, with the expected level of deviation clearly visible in the resulting PDFs. Given the simulated noise in the pseudo-dataset, the closure test still tracks the central value to an excellent level of accuracy, achieving an almost statistically ideal fit quality of $\chi^2/N_{\text{dat}} = 1.00021$.

\begin{figure}[!]
\centering
\includegraphics[width=0.45\textwidth]{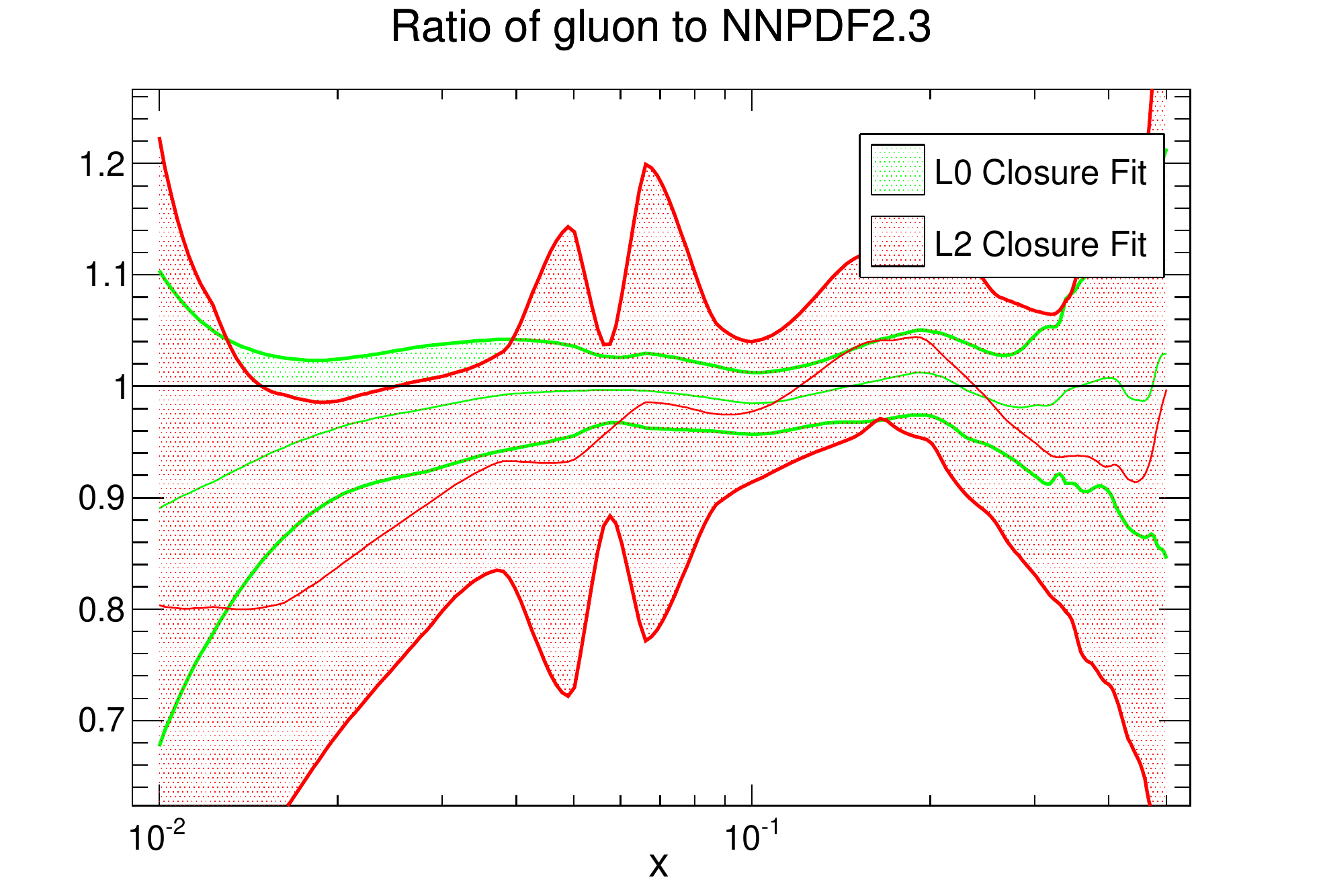}
\includegraphics[width=0.45\textwidth]{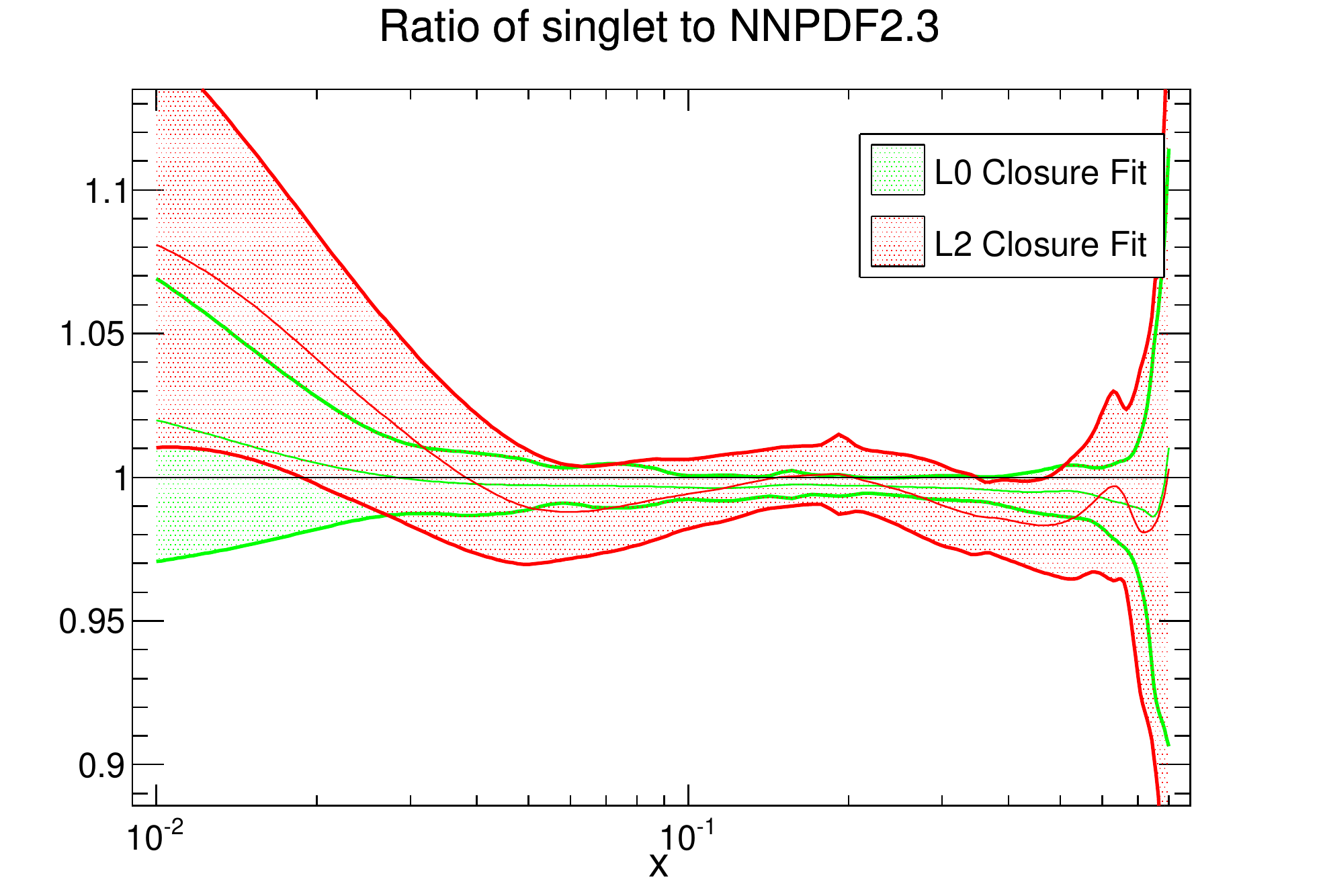}
\includegraphics[width=0.45\textwidth]{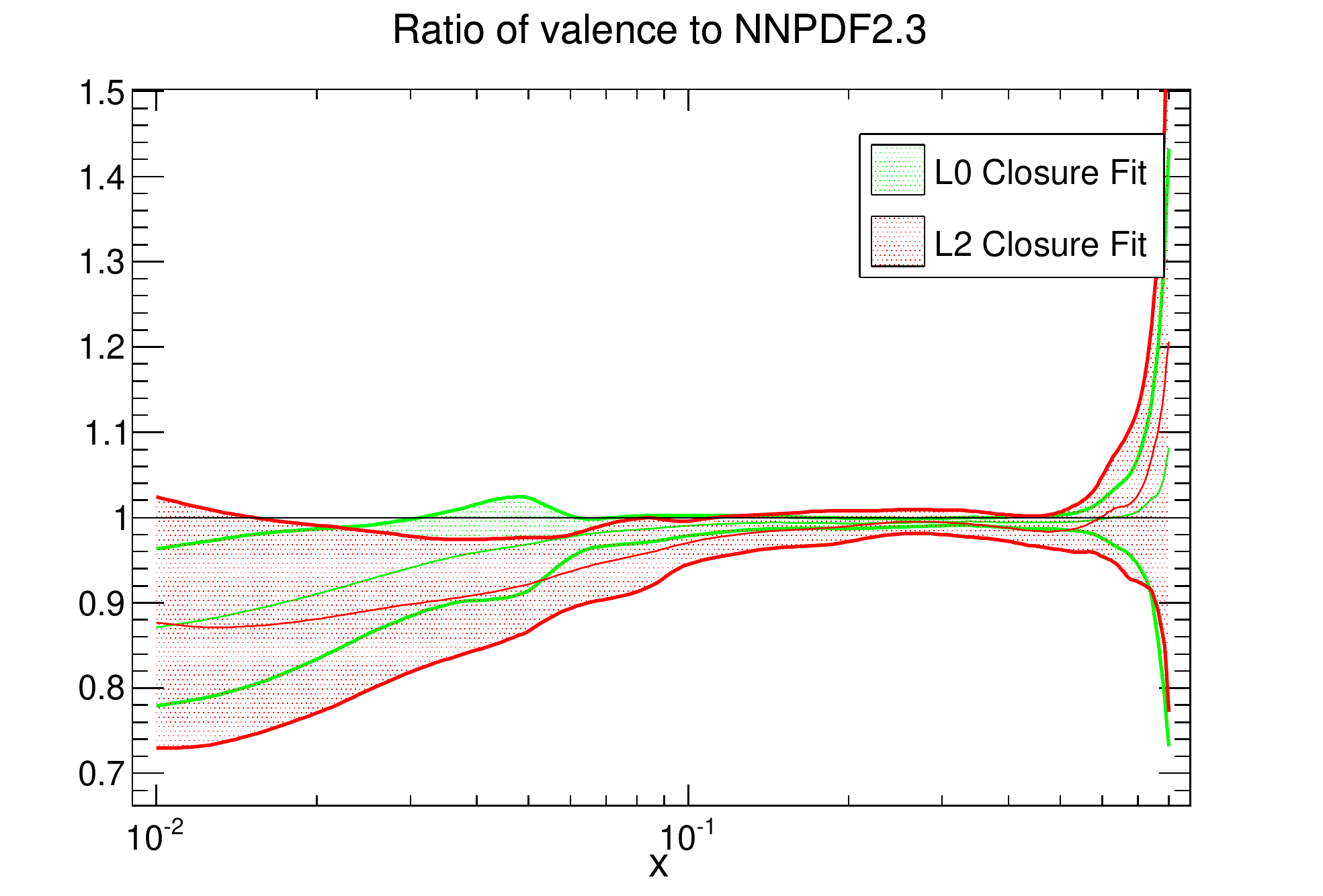}
\includegraphics[width=0.45\textwidth]{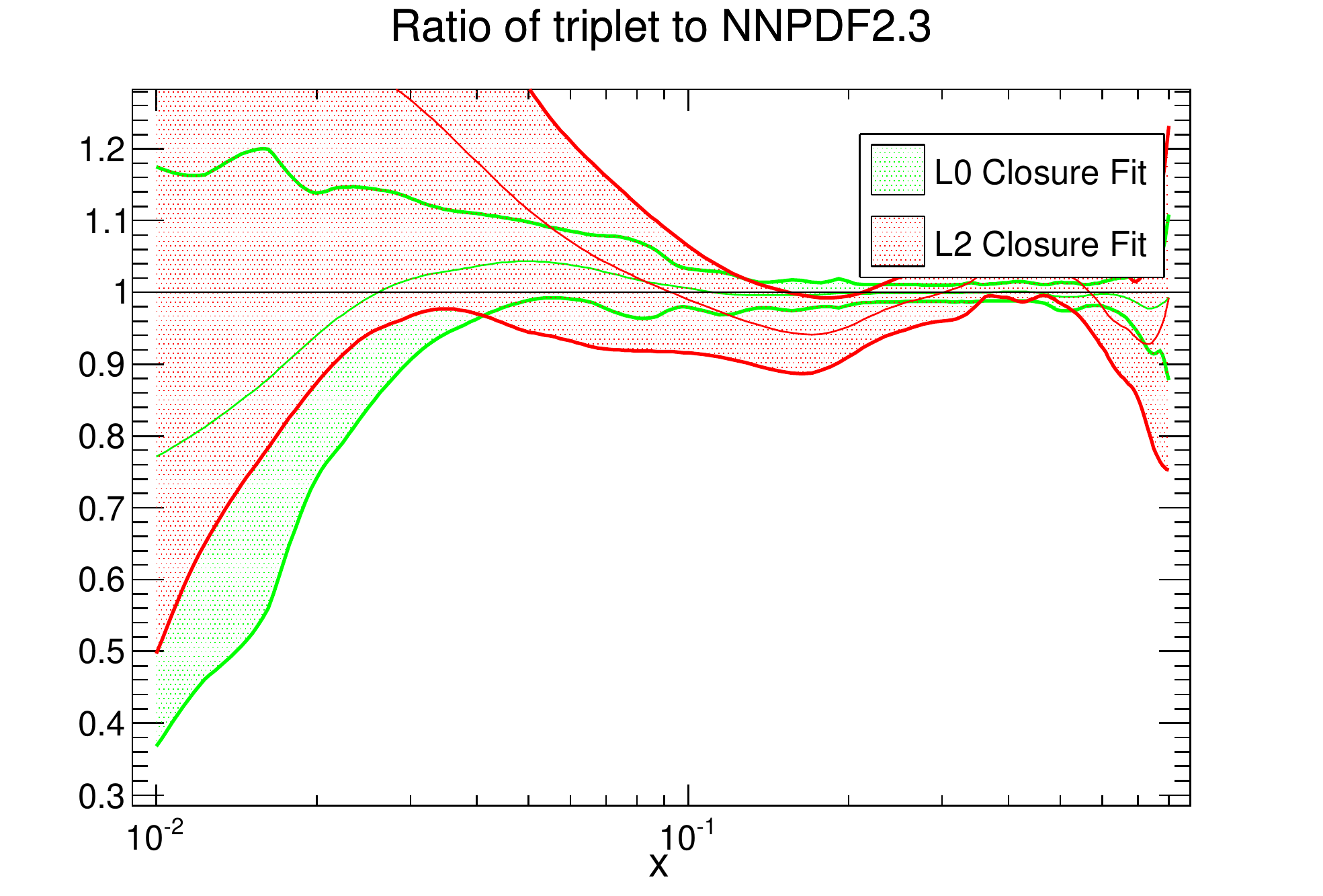}
\caption[NNPDF3.0 methodology closure test fit to NNPDF2.3 NLO]{NNPDF3.0 methodology closure test fit to NNPDF2.3 NLO. Plots as in Figure~\ref{fig:finalClosure_MSTW}. }
\label{fig:finalClosure_NNPDF}
\end{figure}

As the preliminary NNPDF3.0 methodology has been validated against closure test fits to the MSTW2008 set, it is important to test the procedure's ability to reproduce a generating PDF with greater functional complexity. To verify the preliminary methodology in this case we now consider a closure test fit to the NNPDF2.3 PDF set. Figure~\ref{fig:finalClosure_NNPDF} demonstrates once more the level zero and two closure test fits to NNPDF2.3. Even given the greater functional freedom present in the previous NNPDF determination, the 3.0 closure test provides an excellent reproduction of the generating functions, with fit qualities of  $\chi^2/N_{\text{dat}} = 0.00287$ and $1.01356$ respectively. Once again the uncertainty due to parametrisation flexibility is demonstrated in the level zero fit, while the level two fit provides a closer simulation of a full fledged experimental data fit. These figures therefore suggest that the preliminary NNPDF3.0 methodological choices can accurately determine complex functional forms without any modification with respect to fits to much simpler parametrisations.

\begin{figure}[!]
\centering
\includegraphics[width=0.45\textwidth]{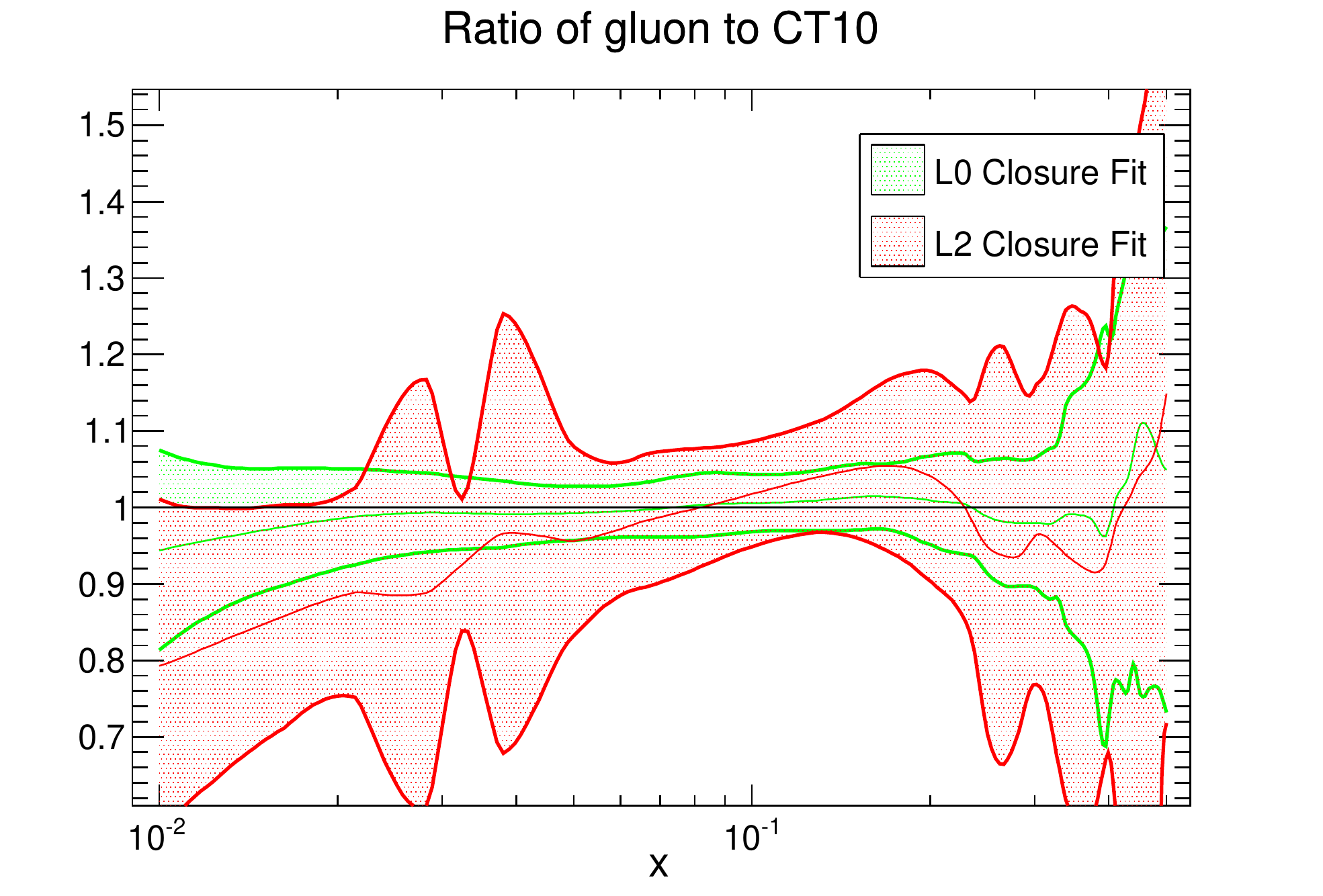}
\includegraphics[width=0.45\textwidth]{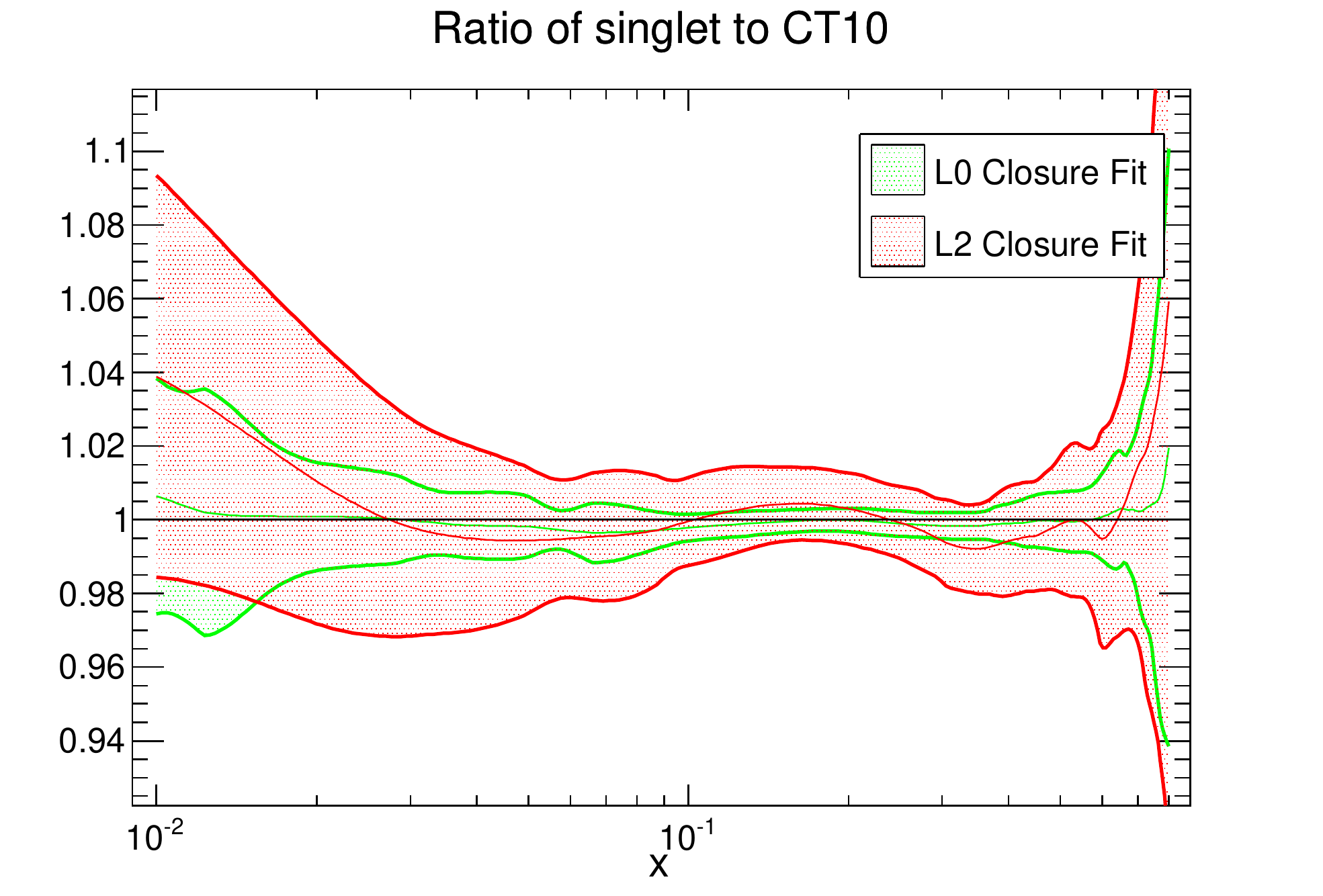}
\includegraphics[width=0.45\textwidth]{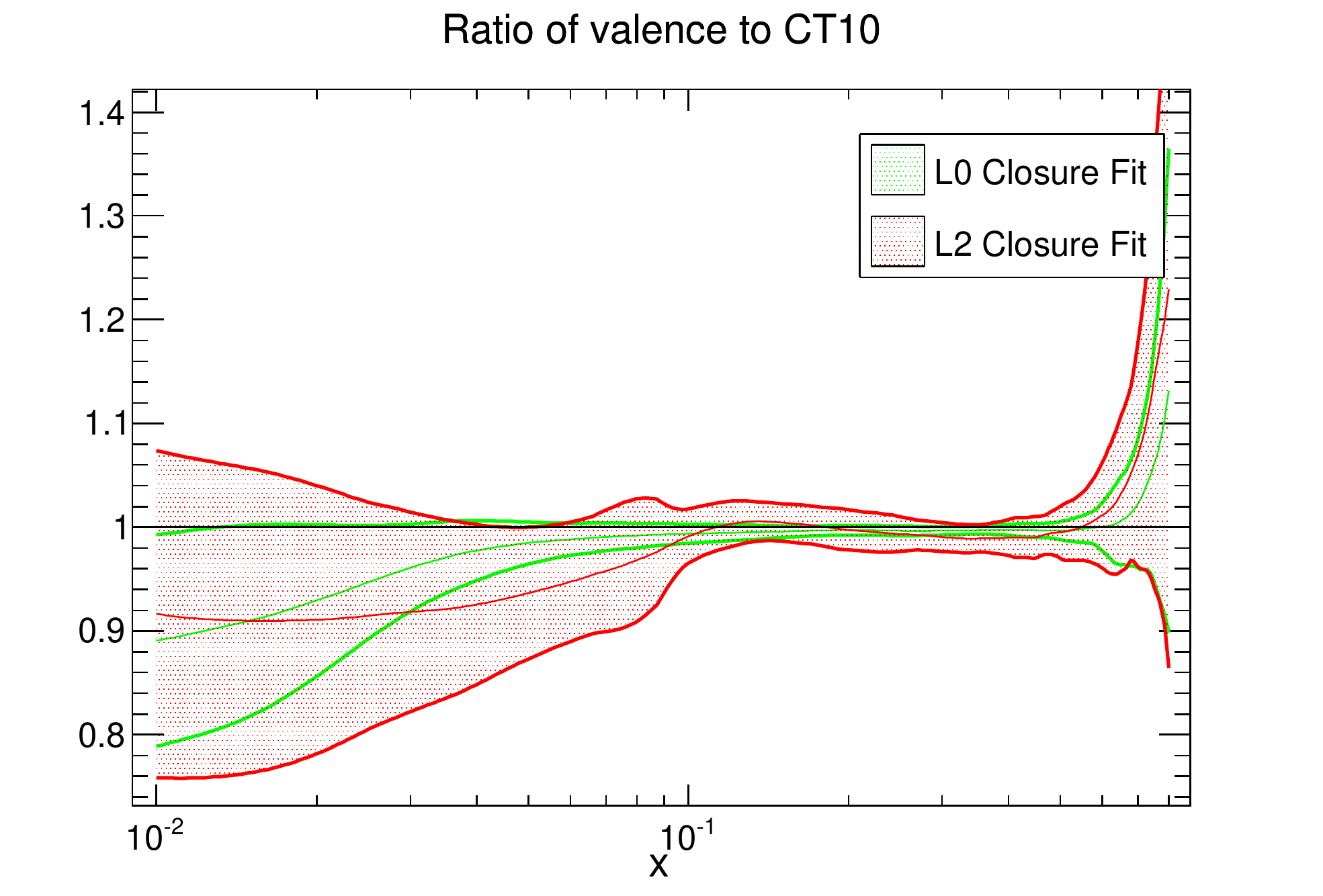}
\includegraphics[width=0.45\textwidth]{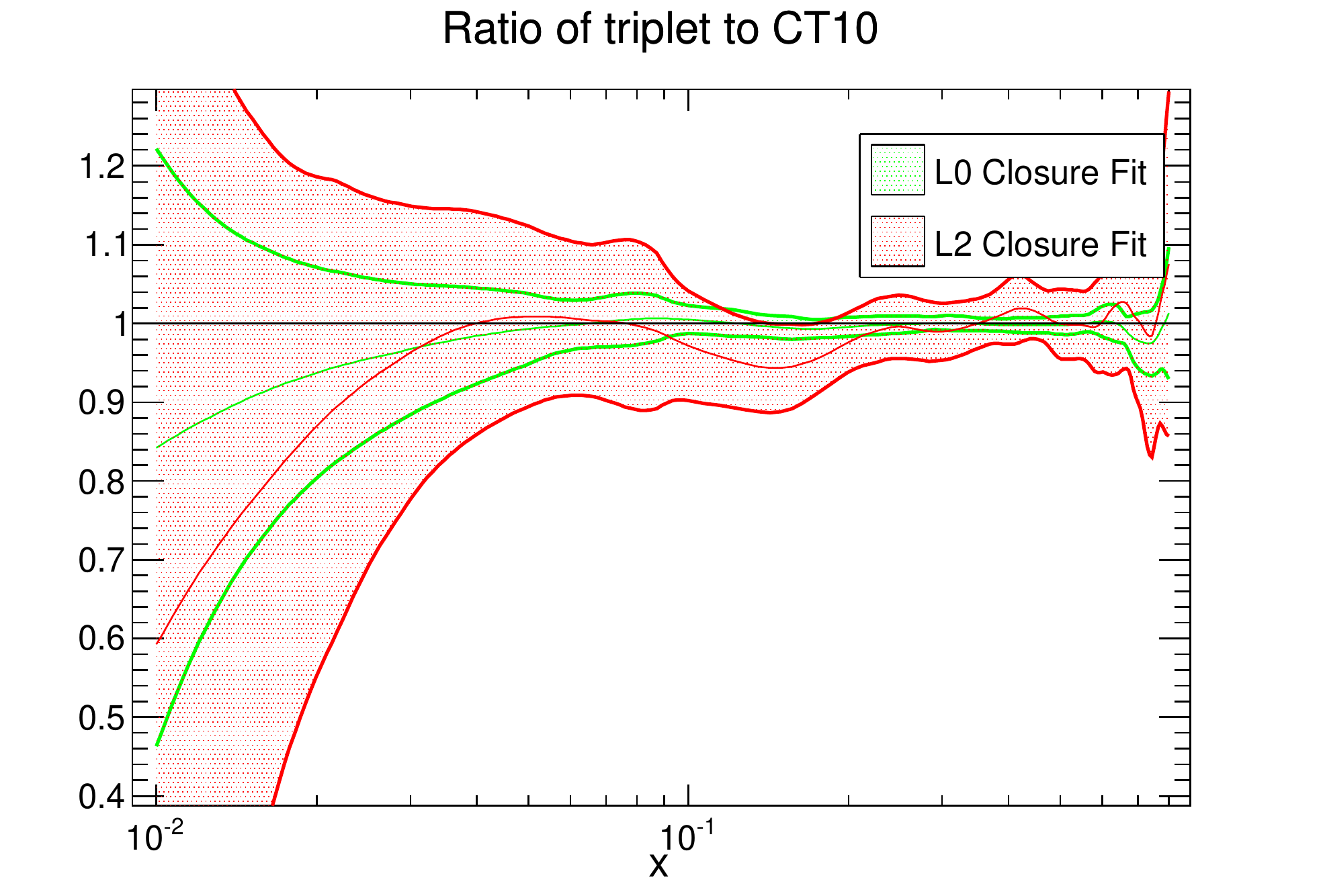}
\caption[NNPDF3.0 methodology closure test fit to CT10 NLO]{NNPDF3.0 methodology closure test fit to CT10 NLO. Plots as in Figure~\ref{fig:finalClosure_MSTW}. }
\label{fig:finalClosure_CT10}
\end{figure}

For a final closure test, we shall now consider a fit using the CT10 PDF set as a set of generating functions. In this way we can verify the NNPDF3.0 method in a way that is independent of the closure PDF set guiding the methodological development (MSTW2008) and previous NNPDF determinations. The results of the test, once more at level zero and two, are shown in Figure~\ref{fig:finalClosure_CT10}. The closure test fit provides once again an excellent description of data, with $\chi^2/N_{\text{dat}} = 0.00130$ for the level zero fit and $1.01324$ for the level one. The procedure detailed here has now been validated against three different generating PDFs in a closure test and is able to convincingly reproduce the generating sets in each of them. We can therefore be confident that when applied to real experimental data the procedure will yield an accurate result up to theoretical uncertainties.


\subsection{Improvements in data fits for NNPDF3.0}
While we have now validated much of the methodology to be used in the NNPDF3.0 determination, we shall now finally investigate some of the expected differences arising with respect to the NNPDF2.3 results in the case of a fit to experimental data. In order to directly assess the changes arising purely from the methodological differences in the two approaches, we shall perform two fits to a small common dataset, one with the full NNPDF2.3 machinery and the second with the improvements implemented in the NNPDF3.0 procedure. It should be noted that these results are of an extremely preliminary nature and as so should only be taken as roughly indicative of the final results. Furthermore, the full NNPDF3.0 set will benefit from a considerably expanded dataset with respect to the NNPDF2.3 determination. 

For these test fits, we use a collider-only dataset to ensure a maximally consistent set of experimental data, including the full NNPDF2.3 LHC and Tevatron datasets, and the HERA-1 combined DIS results. Once more, the fits were run with a maximum number of generations of $N_{\text{gen}} = 30,000$. The NNPDF2.3-like fit was otherwise performed according to the settings of the central NNPDF2.3 fit. The NNPDF3.0 fits were performed with identical settings to the closure test fits described in the previous section. 

\begin{figure}[!]
\centering
\includegraphics[width=0.45\textwidth]{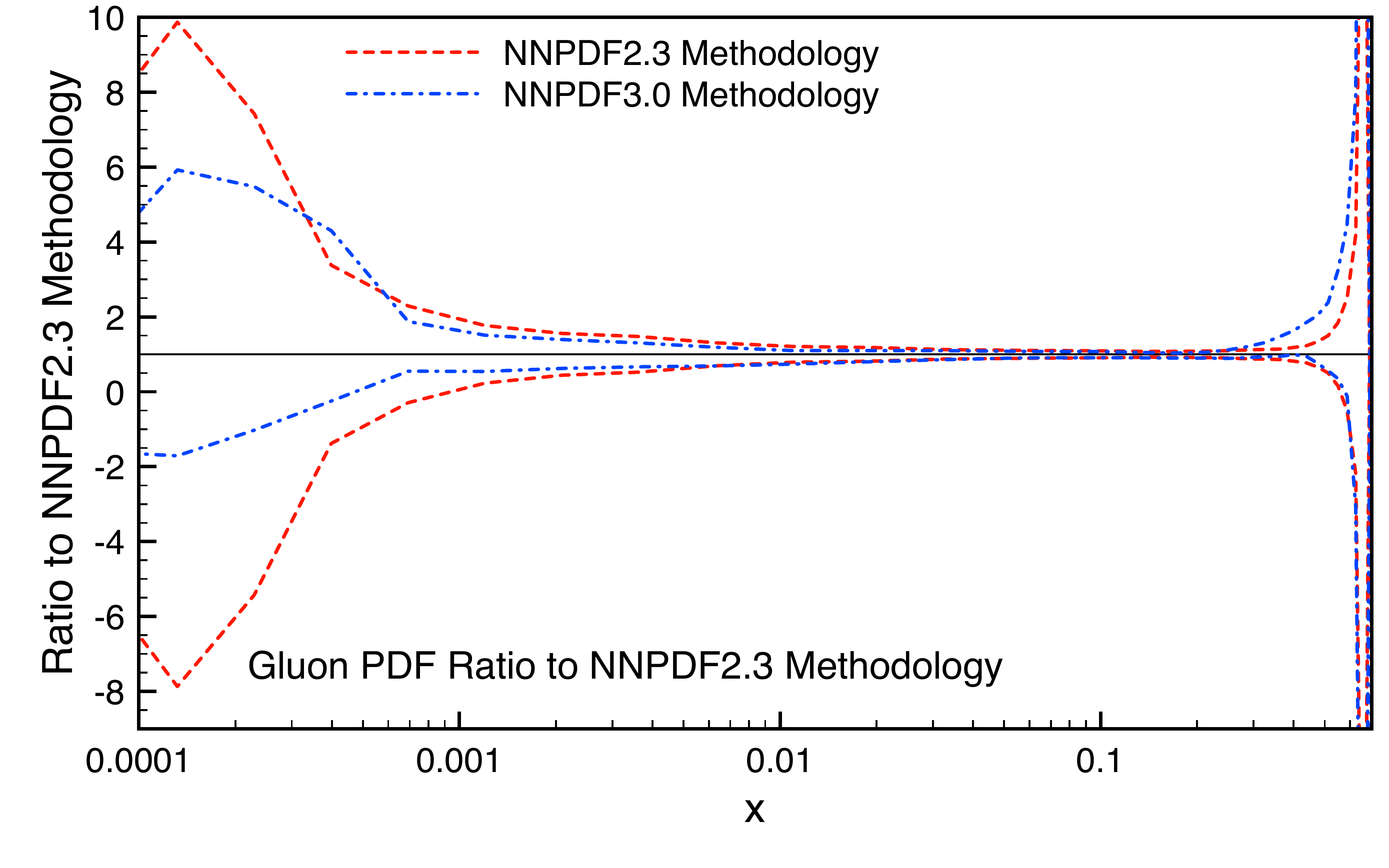}
\includegraphics[width=0.45\textwidth]{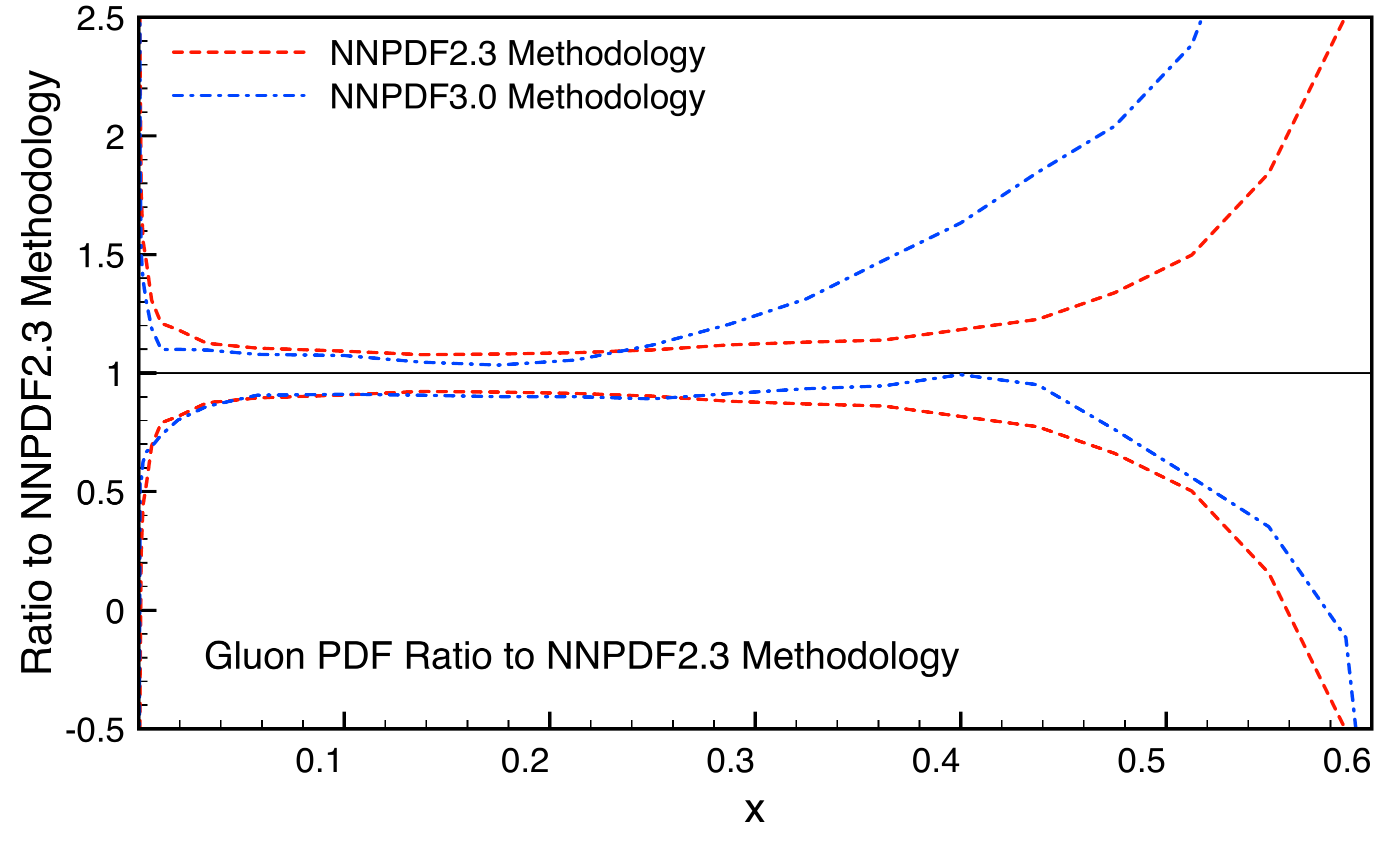}\\
\includegraphics[width=0.45\textwidth]{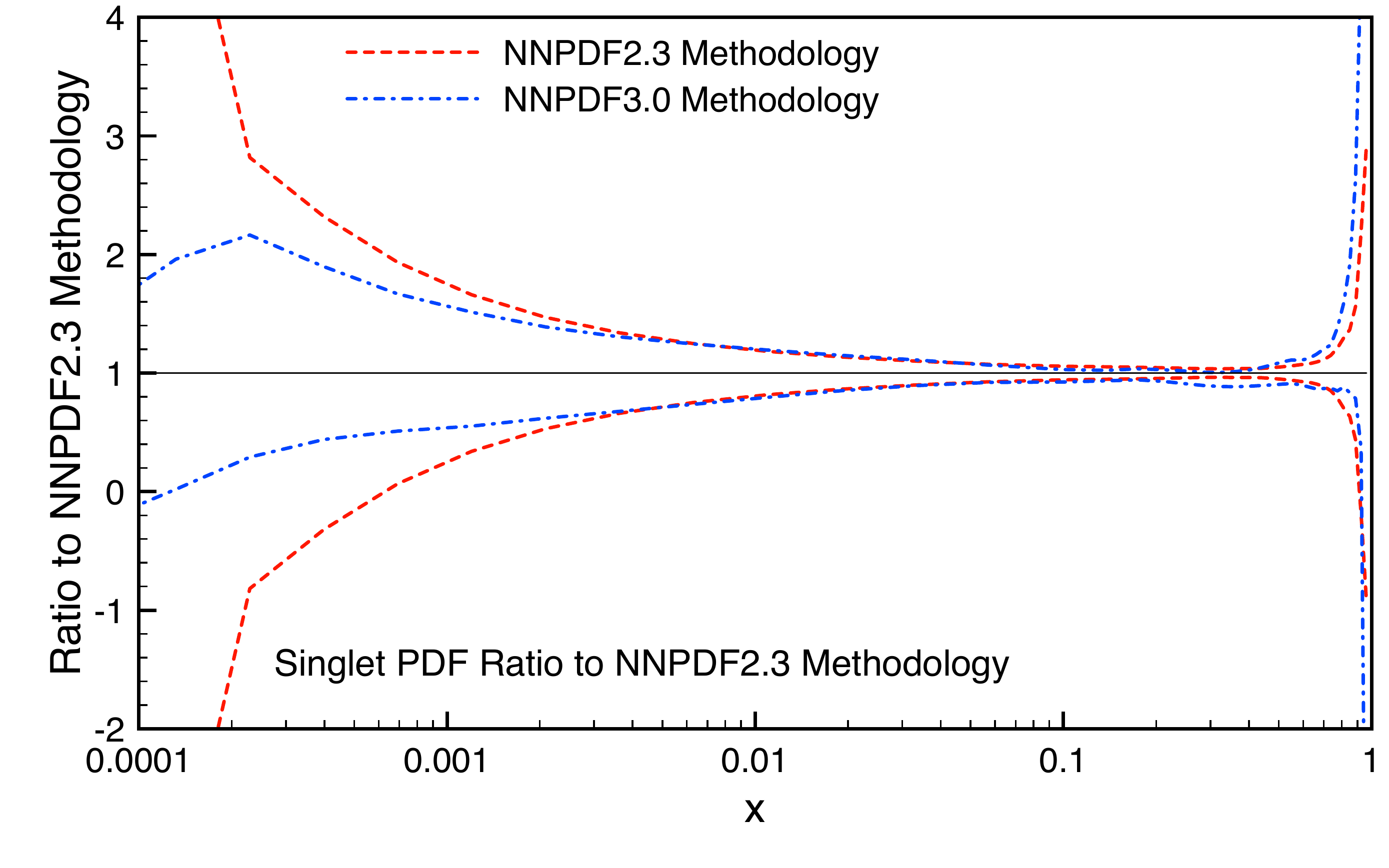}
\includegraphics[width=0.45\textwidth]{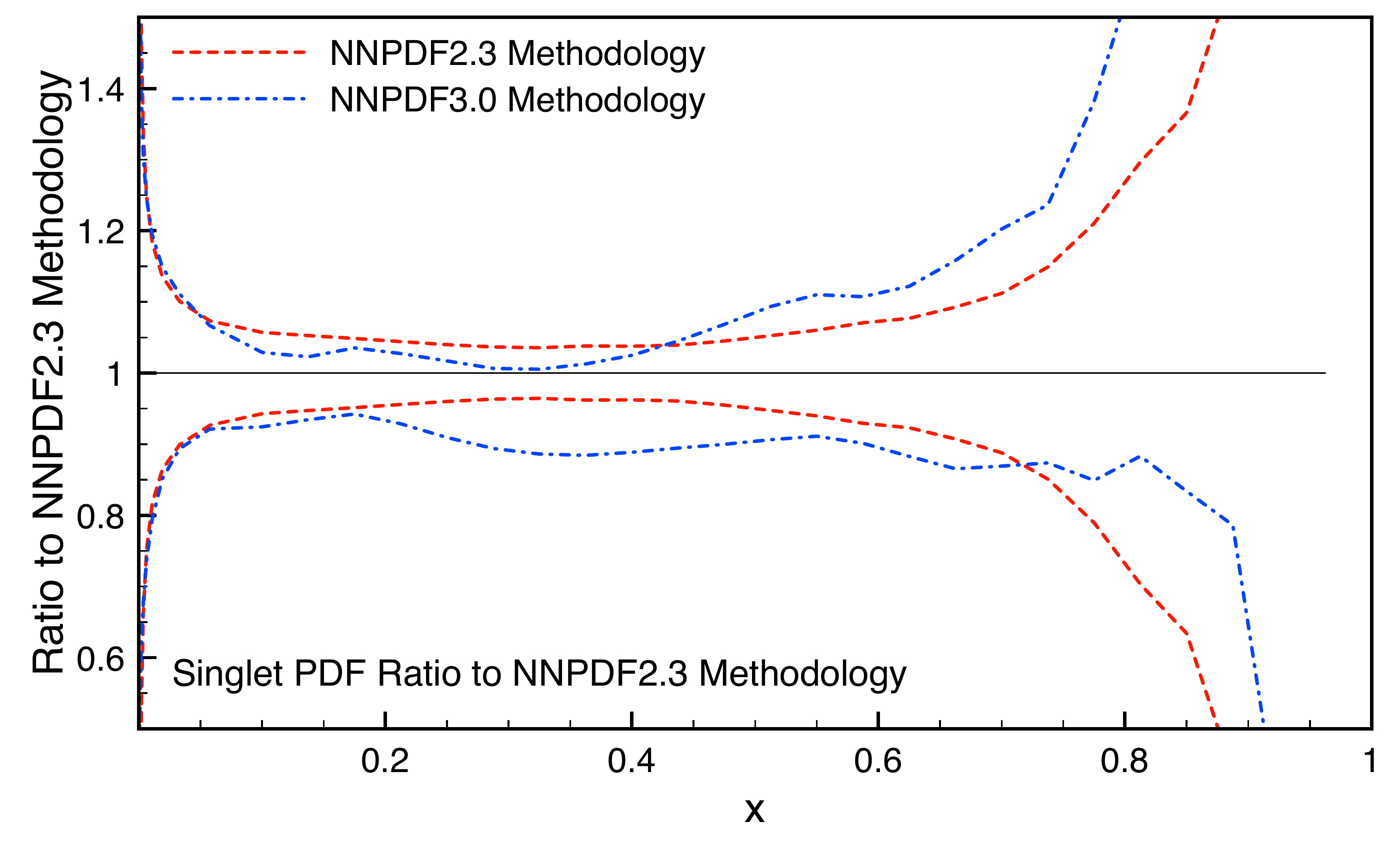}

\caption[Comparison of NNPDF2.3 and NNPDF3.0 fitting methodologies when applied to a common experimental dataset. Gluon and singlet PDF combinations]{Comparison of NNPDF2.3 and NNPDF3.0 fitting methodologies when applied to a common experimental dataset. Here the gluon (top) and singlet (bottom) PDFs are shown, with all values normalised to the result of the NNPDF2.3 methodology fit.}
\label{fig:23vs30methodology_1}
\end{figure}

Looking firstly at the gluon and singlet sectors, in Figure~\ref{fig:23vs30methodology_1} we see the results of the two methodology test fits compared as a ratio to the NNPDF2.3 methodology fit's central value. The first feature to note is that in the region where data constraints in this test fit are largest, the two methodologies remain very consistent in their results, with the most significant changes occurring in the extrapolation regions and for the large-$x$ singlet. At small-$x$ the NNPDF3.0 methodology fit is more confident in the extrapolation for both singlet and gluon PDFs, resulting in a systematically smaller uncertainty. At large-$x$ there is a moderate shift in the gluon central value in the NNPDF3.0 result, and a broadening of uncertainties. The same pattern can be found in the large-$x$ gluon, where once again uncertainties are slightly larger and there is some change in central value. However both distributions remain in agreement within their uncertainties, validating that the two methodologies remain compatible within the experimental uncertainty present in the test dataset.

\begin{figure}[!]
\centering
\includegraphics[width=0.45\textwidth]{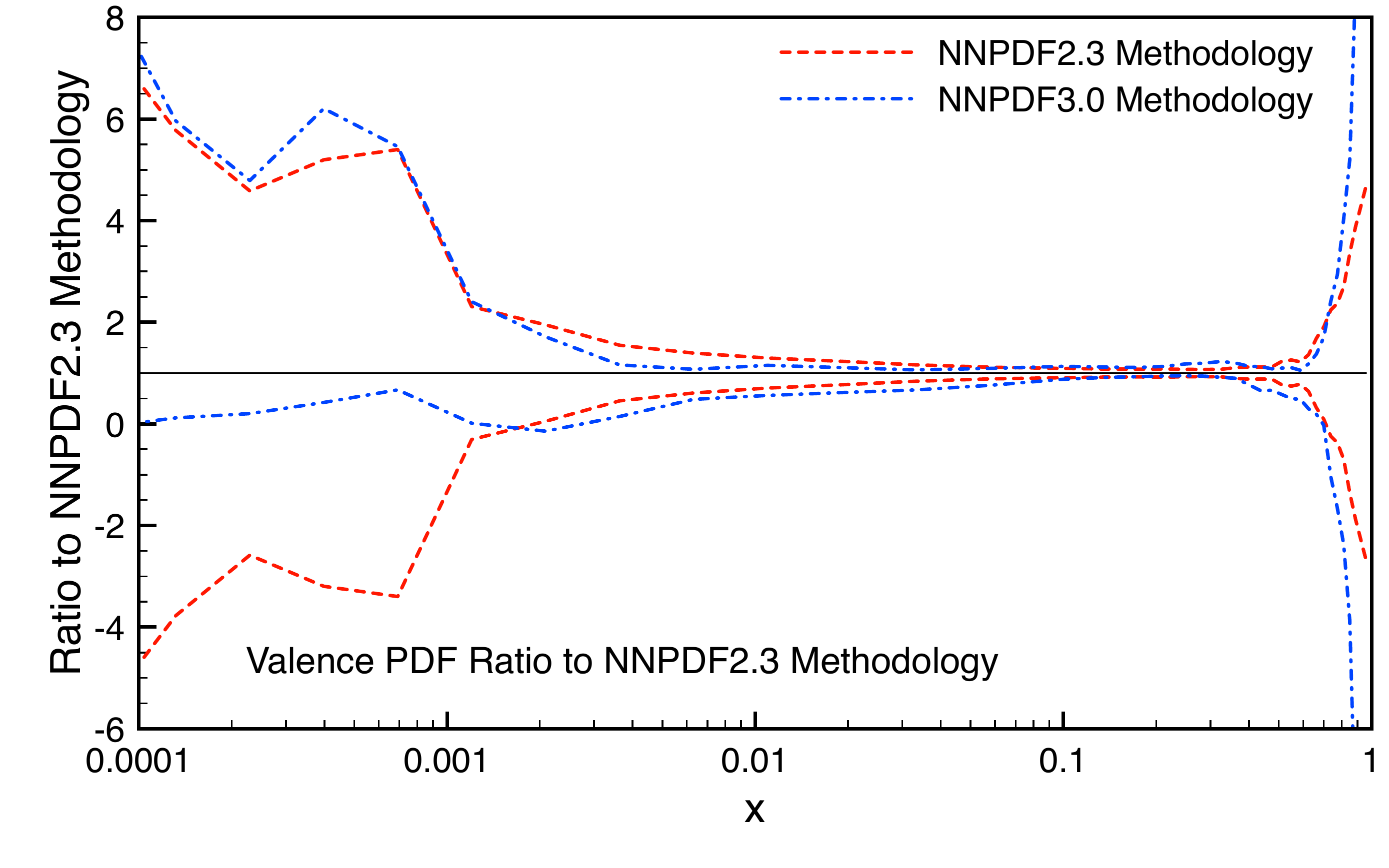}
\includegraphics[width=0.45\textwidth]{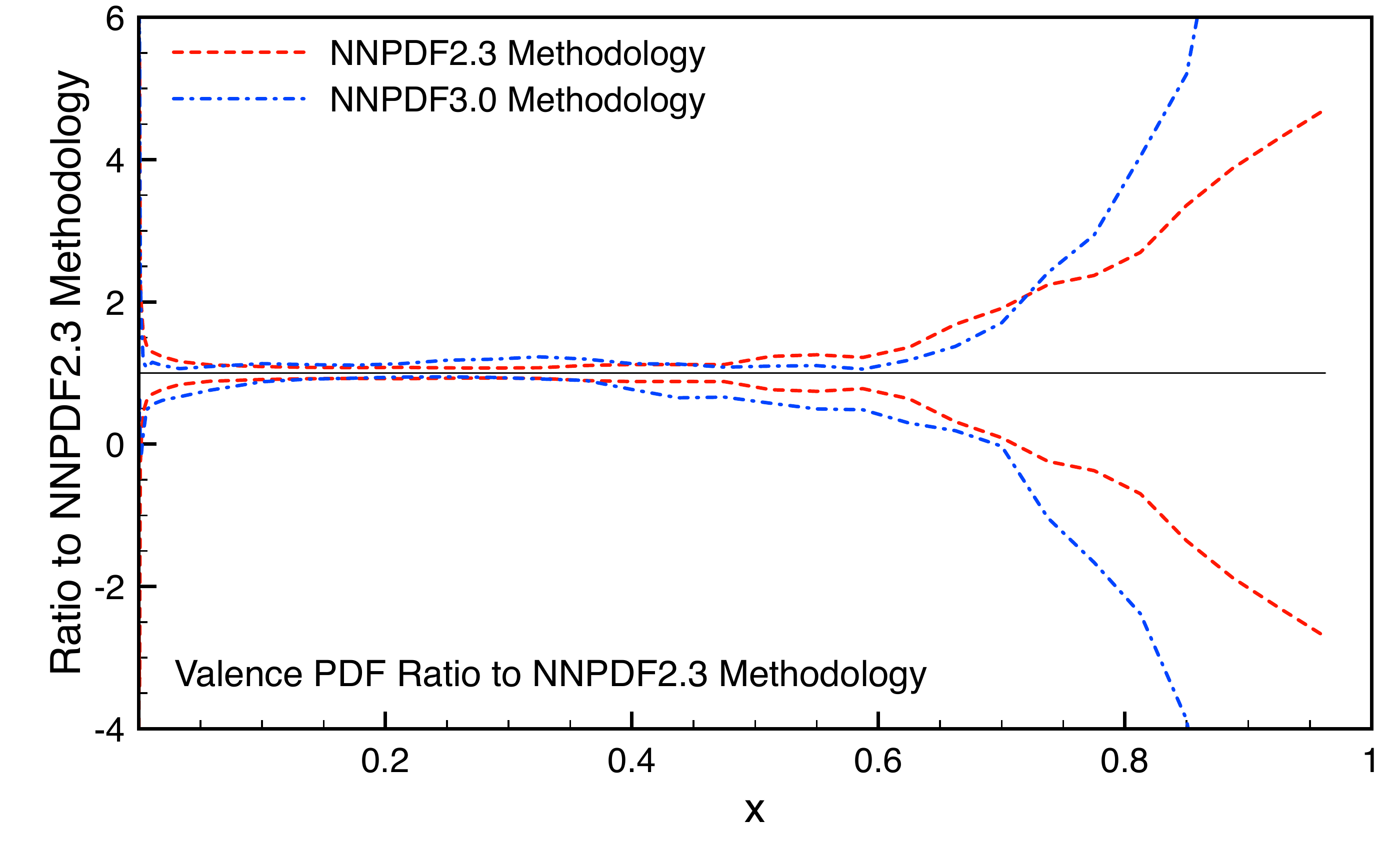}\\
\includegraphics[width=0.45\textwidth]{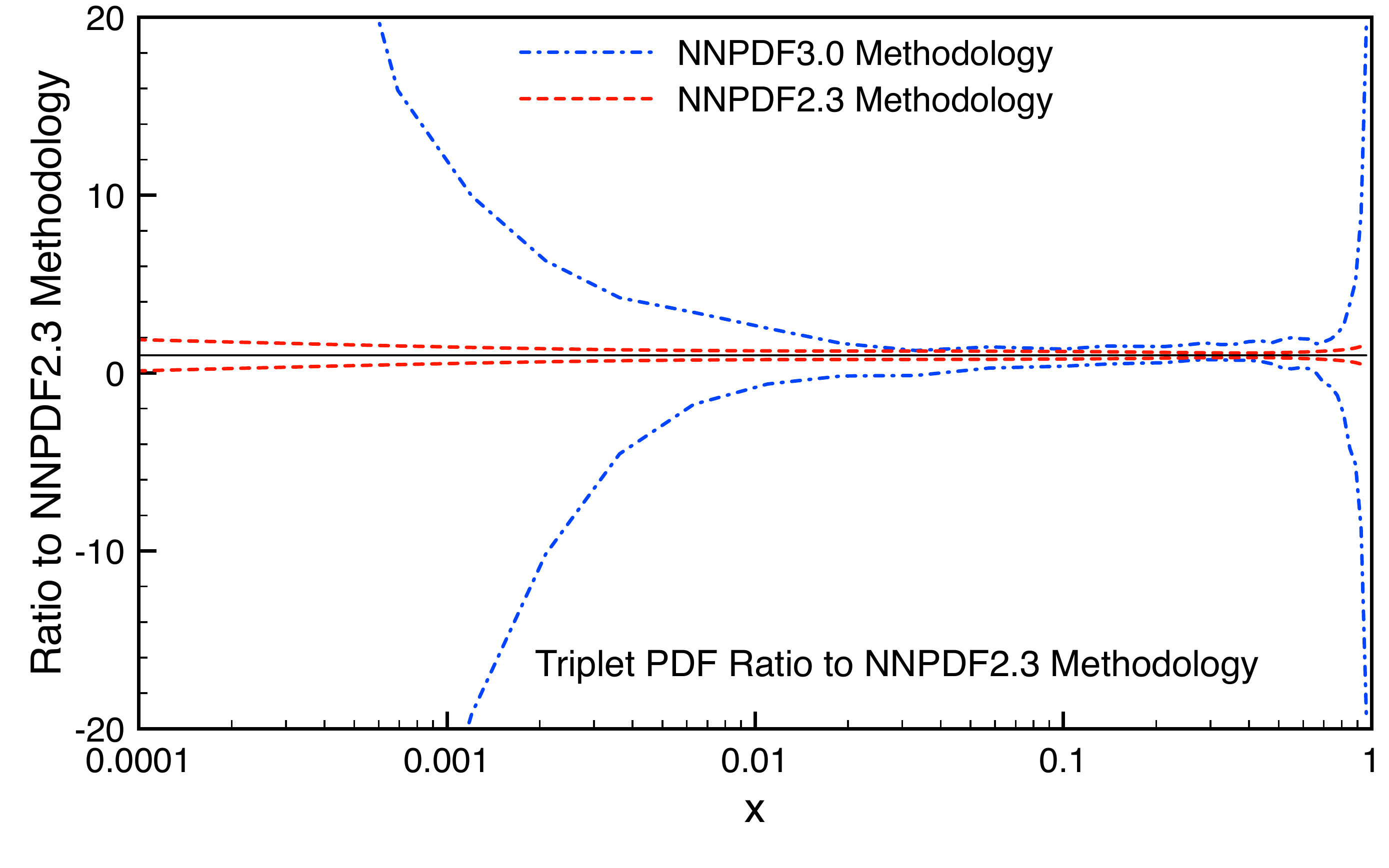}
\includegraphics[width=0.45\textwidth]{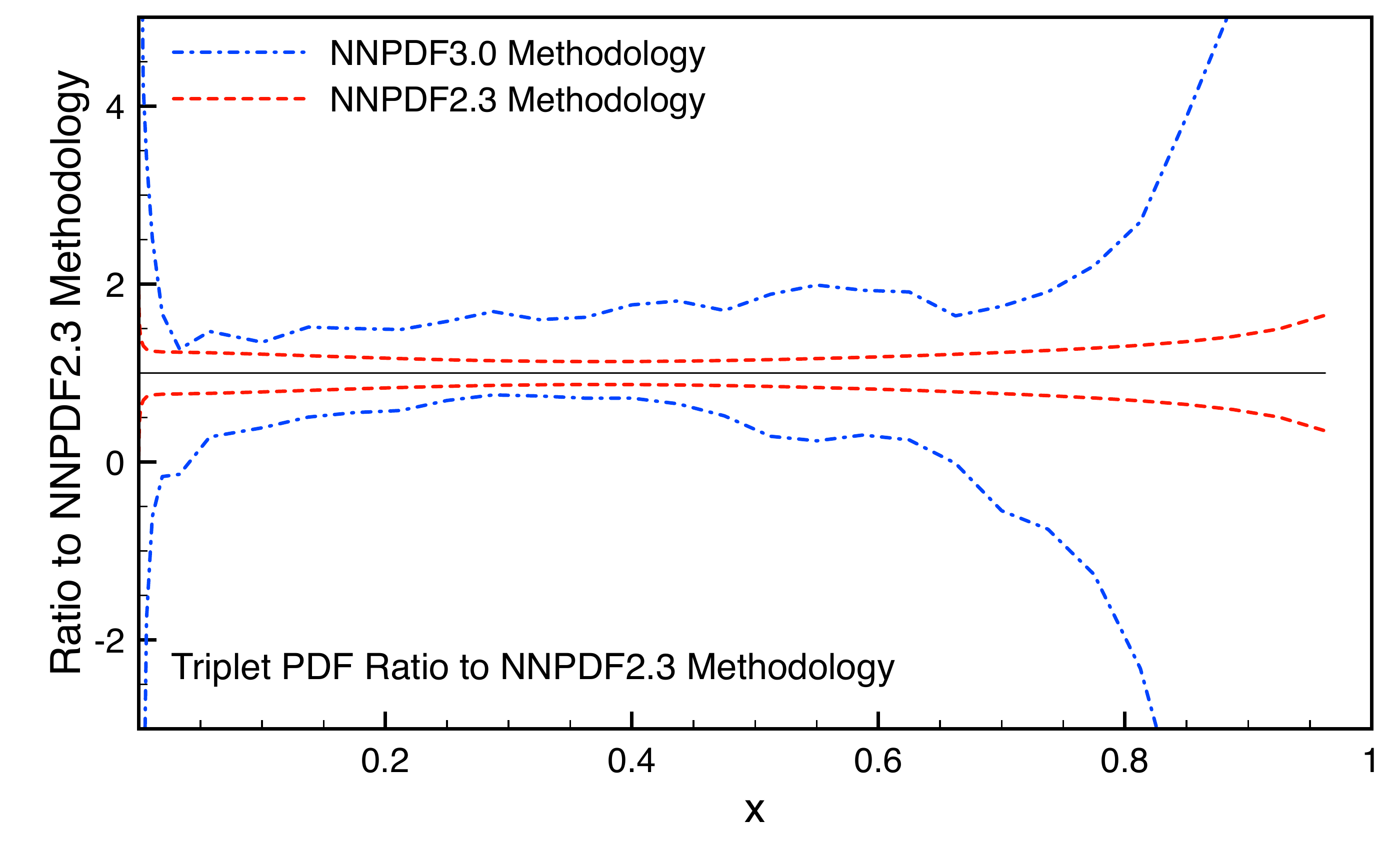}

\caption[Comparison of NNPDF2.3 and NNPDF3.0 fitting methodologies when applied to a common experimental dataset. Valence and triplet PDF combinations]{Comparison of NNPDF2.3 and NNPDF3.0 fitting methodologies when applied to a common experimental dataset, for the valence (top) and triplet (bottom) PDF combinations. Plots as in Figure~\ref{fig:23vs30methodology_1}.}
\label{fig:23vs30methodology_2}
\end{figure}

To investigate the impact of the methodological changes to PDFs sensitive to the valence distributions and quark flavour separation, we plot the valence and triplet PDF combinations in Figure~\ref{fig:23vs30methodology_2}. In the valence PDF comparison, we see a similar pattern as for the singlet and gluon PDFs, where the low-$x$ result from the NNPDF3.0 methodology fit obtained a narrower distribution, and at high-$x$ the uncertainties are systematically larger. The triplet PDF shows by some way the largest differences between the two methodologies, with PDF uncertainties being significantly larger across the whole range of $x$. This effect is largely due to the much more flexible preprocessing used for the triplet PDF, where now there is no requirement that the PDF be preprocessed to zero at low-$x$, the constraint now being entirely based on experimental data. Such a treatment leads to a rather conservative determination of the low-$x$ triplet, however this effect should be at least partially offset by increased data constraints in the full NNPDF3.0 determination.

Here we have seen the two methodologies provide consistent results when applied to the same experimental dataset. However there are significant changes in the fit results due to methodological improvements, particularly important in the PDF extrapolation regions at large and small values of parton-$x$, and for PDF combinations sensitive to light flavour separation. As has been shown in the validation with closure tests, the methodological modifications, particularly in allowing for greater preprocessing flexibility, result in an improved reproduction of a test PDF distribution. The upgraded methodology should therefore provide a more reliable estimate of the parton densities of the proton.

\clearpage

\chapter{Conclusion}
\label{ch:con}

In this thesis we have discussed the impact of LHC measurements upon the extraction of parton distribution functions, and methods for the inclusion of such data into parton fits. With the LHC due to begin collisions at centre of mass energies of $13$ TeV in 2015, the need for precise determinations of proton structure is as great as ever. Here we have described the efforts undertaken to provide the particle physics community with PDFs extracted via a methodologically sound procedure to an extensive experimental dataset including measurements from the first data runs of the LHC.

A number of tools have been developed that enable the study of a large collider dataset in the context of the NNPDF procedure. The Bayesian reweighting procedure, first implemented in a PDF determination by the NNPDF collaboration, allows for a rapid assessment of data impact. Bayesian reweighting also provides a deep check of the consistency of the NNPDF methodology, as a verification of the statistical behaviour of the Monte Carlo PDF ensemble. The method is however unsuitable for the inclusion of a large dataset, due to the need for an unpractically large prior Monte Carlo distribution. Therefore in order to enable the inclusion of a large LHC dataset in an NNPDF fit, further tools are required.

The {\tt FK} method was introduced as a method by which the PDF evolution may be combined with the theoretical predictions for experimental observables in a process independent way. The resulting matrices, or {\tt FK} tables provide an extremely efficient method of computing theoretical predictions based upon a varying input PDF set, reducing the task to a simple scalar product which can be efficiently optimised. This method enabled the inclusion of LHC data into a full NNPDF fit for the first time, without having to compromise on the accuracy of the calculation. While fast, the {\tt FK} procedure is rather specialised to the task of PDF fitting as it only permits for the variation of input PDF set. For wider studies of the dependence of theoretical predictions upon parameters such as the strong coupling or factorisation and renormalisation scales, more flexible approaches as implemented in packages such as {\tt FastNLO} or {\tt APPLgrid} are more relevant.

A package for the interfacing of automated NLO calculational tools to such fast interpolating codes has been developed. The {\tt MCgrid} package allows for the use of Monte Carlo event generators such as {\tt SHERPA} along with a suitable one-loop generator, for the efficient variation of QCD parameters in theoretical predictions. Such an interface opens up the possibility of using such codes in applications such as $\alpha_S$ determinations or PDF fits, alongside making PDF and scale variation more accessible in computationally challenging processes. The interface between event generators and interpolating tools remains however at the level of fixed order perturbation theory. In principle an extension for the fast computation of observables with parton shower effects included would be particularly desirable, making this an important avenue for future research.

These tools have been applied to the study of LHC measurements and their impact upon PDF distributions. To date two NNPDF results including LHC constraints have been published. The NNPDF2.2 determination included a set of $W$ boson production asymmetry measurements at both the LHC and the Tevatron, by the method of Bayesian reweighting. In such a way the reweighting method was extensively validated, and new constraints were placed upon PDFs from LHC data for the first time. As the dataset available from the LHC expanded, the need for a comprehensive refit including all appropriate measurements increased. The resulting PDF set, NNPDF2.3, utilised the {\tt FK} procedure for all of the included processes and so was able to include all available LHC measurements of interest to PDF fits at the time. The NNPDF2.3 set provided a precise determination ideally suited for further applications at the LHC.

Following the NNPDF2.3 set, the development of the {\tt nnpdf++} project has allowed for a greater scope in investigating methodological elements, permitting a large scale re-evaluation of the procedure used in the NNPDF2.3 family of fits. The closure testing procedure introduced in Chapter~\ref{ch:LHClight} now forming the basis for the development in methodology post-NNPDF2.3. Insights provided by a detailed study of the NNPDF procedure when applied to closure tests have informed a number of new approaches in minimisation and stopping for the next global PDF set produced by the collaboration. The next release, NNPDF3.0, being validated using the closure testing procedure and including an expanded LHC dataset will provide the most precise and methodologically sound determination of parton distribution functions.

%%%%%%%%%%%%%%%%%%%%%%%%%%%%%%%%%%%%%%%%%%%%%%%%%%%%%%%%%%%%%%%%%%%%%

\newpage
\cleardoublepage
\phantomsection
\addcontentsline{toc}{chapter}{Bibliography}
\bibliographystyle{hunsrt.bst}
\bibliography{bibliography}

%%%%%%%%%%%%%%%%%%%%%%%%%%%%%%%%%%%%%%%%%%%%%%%%%%%%%%%%%%%%%%%%%%%%%%

\appendix
\chapter{Summary of experimental data}
\label{app:summary_tables}
In this appendix, the experimental measurements discussed in this thesis are summarised. For each experiment, the underlying process and physical observable measured are specified, along with a brief summary of the PDF flavours and combinations targeted by the data.

\clearpage

\begin{table}
\label{tab:FTDIS_summary}
\begin{center}
\begin{tabular}{|c|c|c|c|c|}
\hline
\multicolumn{5}{|c|}{ \textbf{Fixed-Target Deep Inelastic Scattering} }\\
\hline
\textbf{Process} & \textbf{Experiment} & \textbf{Obs.}& \textbf{Ref.} &  \textbf{Target} \\
\hline\hline
$\mu p/d \to \mu X$ & BCDMS& $F_2^p$, $F_2^d$ &\cite{Benvenuti:1989rh,Benvenuti:1989fm} & $q$, $\bar{q}$\\
$\mu p/d \to \mu X$ & NMC &$F_2^p$, $F_2^d/F_2^p$ & \cite{Arneodo:1996kd,Arneodo:1996qe} & $q$, $\bar{q}$, $d/u$\\
$\mu p/d \to \mu X$ & Fermilab E665 & $F_2^p$, $F_2^d$ & \cite{Adams:1996gu} & $q$, $\bar{q}$\\
\hline
$\mu p \to \mu X$ & BCDMS &$F_L$ &\cite{Benvenuti:1989rh} & $g$ \\
$\mu p/d \to \mu X$ & NMC &$F_L$ & \cite{Arneodo:1996qe} & $g$\\
$e p/d \to e X$ & SLAC &$F_L$& \cite{Whitlow:1990gk} & $g$\\
\hline
\end{tabular}
\end{center}
\caption{Summary of discussed Fixed-Target DIS experiments, arranged as in Table~\ref{tab:pp_summary}. Here deuteron and proton structure function data is summarised.}
\end{table}%

\begin{table}
\begin{center}
\begin{tabular}{|c|c|c|c|c|}
\hline
\multicolumn{5}{|c|}{\textbf{HERA Deep Inelastic Scattering}}\\
\hline
\textbf{Process} & \textbf{Experiment} & \textbf{Obs.} &\textbf{Ref.} &  \textbf{Target} \\
\hline\hline
$e p \to e X$ & H1 & NC $\sigma$ & \cite{Adloff:2000qk,Adloff:2000qj,Adloff:2003uh} & $g$, $q$, $\bar{q}$\\
$e p \to e X$ & ZEUS & NC $\sigma$& \cite{Breitweg:1998dz,Chekanov:2001qu,Chekanov:2002ej,Chekanov:2003yv}& $g$, $q$, $\bar{q}$\\
\hline
$e^+ p \to \bar{\nu} X$ & H1 & CC $\sigma$& \cite{Chekanov:2003vw} & $d + s$ , $\bar{u}$ \\
$e^+ p \to \bar{\nu} X$ & ZEUS & CC $\sigma$ & \cite{Adloff:2003uh} & $d + s$ , $\bar{u}$ \\
\hline
$e p \to e X+c$ & H1 & $F^2_c$ & \cite{Chekanov:2003rb,Chekanov:2007ch,Aktas:2005iw,Aktas:2004az} & $g$\\
$e p \to e X+c$ & ZEUS & $F^2_c$ & \cite{Adloff:1996xq,Adloff:2001zj,Breitweg:1999ad} & $g$\\
\hline
$e p \to e X$ & H1 & $F_L$ & \cite{Andreev:2013vha} & $g$ \\
$e p \to e X$ & ZEUS & $F_L$ & \cite{Chekanov:2009na} & $g$\\
\hline\hline
$e p \to e X$ & HERA-I & NC $\sigma$&\cite{aaron:2009wt}& $g$, $q$, $\bar{q}$ \\
$e p \to \nu X$ & HERA-I & CC $\sigma$&\cite{aaron:2009wt}& $q, \bar{q}$\\
\hline
$e p \to e X+c$ & HERA-I & $F_2^c$ & \cite{Abramowicz:1900rp} & $g$\\
\hline
\end{tabular}
\end{center}
\label{tab:HERA_summary}
\caption{Summary of discussed HERA DIS measurements, arranged as in Table~\ref{tab:pp_summary}.}
\end{table}%

\begin{table}
\begin{center}
\begin{tabular}{|c|c|c|c|c|}
\hline
\multicolumn{5}{|c|}{\textbf{Neutrino Deep Inelastic Scattering}}\\
\hline
\textbf{Process} & \textbf{Experiment} & \textbf{Obs.} &\textbf{Ref.} &  \textbf{Target} \\
\hline\hline
$\nu(\bar{\nu})\; \mathrm{ Fe} \to \mu X$ & NuTeV & $F_2/F_3$ & \cite{Tzanov:2005kr} & $q, \bar{q}$\\
$\nu(\bar{\nu})\; \mathrm{ Pb} \to \mu X$ & CHORUS & $F_2/F_3$ & \cite{Onengut:2005kv} & $q, \bar{q}$ \\
\hline
$\nu(\bar\nu)\; \mathrm{ Fe}  \to \mu^+\mu^- X$ & NuTeV/CCFR & Dimuon $\sigma$ & \cite{Goncharov:2001qe} & $s$, $\bar{s}$ \\
\hline
\end{tabular}
\end{center}
\label{tab:NUDIS_summary}
\caption{Summary of discussed Neutrino DIS measurements, arranged as in Table~\ref{tab:pp_summary}.}
\end{table}%

\begin{table}
\begin{center}
\begin{tabular}{|c|c|c|c|c|}
\hline
\multicolumn{5}{|c|}{\textbf{Drell-Yan}}\\
\hline
\textbf{Process} & \textbf{Experiment} & \textbf{Obs.} &\textbf{Ref.} &  \textbf{Target} \\
\hline\hline
$p\; \mathrm{ Cu} \to \mu^+ \mu^- $ & Fermilab E605 & $\sigma^{pp}$ & \cite{Moreno:1990sf} & $q+\bar{q}$\\
\hline
$p\; \mathrm{ H} \to \mu^+ \mu^- $ & NuSea/E866 & $\sigma^{pp}$ & \cite{Webb:2003bj}&  $q+\bar{q}$\\
$p\; \mathrm{ D/H} \to \mu^+ \mu^- $ & NuSea/E866 & $\sigma^{pd}/\sigma^{pp}$ & \cite{Towell:2001nh}& $\bar{d}/\bar{u}$ \\
\hline
\hline
$p\bar{p} \to e^+ e^- $ & D0 & $Z/\gamma$ $y$ & \cite{Abazov:2007jy}& $u$, $d$ \\
$p\bar{p} \to e \nu $ & CDF & $W$ asym. & \cite{Acosta:2005ud}& $u-d$  \\
$p\bar{p} \to \mu \nu $ & D0 & $W$ asym. & \cite{Abazov:2007pm}&  $u-d$ \\
$p\bar{p} \to e \nu $ & D0 & $W$ asym. & \cite{Abazov:2008qv}& $u-d$  \\
\hline
\hline
$pp \to l^+ l^- $ & CMS & $Z$ $p_T/y$ & \cite{Chatrchyan:2011wt}& $u + \bar{u}$, $d + \bar{d}$ \\
$pp \to \mu^+ \mu^- $ & CMS & $Z$ $p_T/y$ & \cite{CMS-PAS-SMP-12-025,CMS-PAS-SMP-13-013}&  $u + \bar{u}$, $d + \bar{d}$\\
$pp \to l \nu $ & CMS & $W$ asym. & \cite{Chatrchyan:2011jz}&  $u-\bar{d}$\\
$pp \to \mu \nu $ & CMS & $W$ asym. & \cite{Chatrchyan:2012xt,Chatrchyan:2013mza} & $u-\bar{d}$ \\
\hline
$pp \to \mu  \nu $ & ATLAS & $W$ asym. & \cite{Aad:2011yna} &  $u-\bar{d}$\\
$pp \to \mu  \mu $ & ATLAS & $Z/\gamma$ $p_T$. & \cite{Aad:2011gj}&  $u + \bar{u}$, $d + \bar{d}$\\
$pp \to l  \nu $ & ATLAS & $W$ $p_T$. & \cite{Aad:2011fp} & $u+\bar{d}$, $\bar{u} + d$  \\
\hline
$pp \to \mu  \nu $ & LHCb & $W$ $p_T$. & \cite{Aaij:2012vn} &  $u+\bar{d}$, $\bar{u} + d$\\
$pp \to \mu  \mu $ & LHCb & $Z/\gamma$ $p_T$. & \cite{Aaij:2012vn}& $u + \bar{u}$, $d + \bar{d}$ \\
\hline
\end{tabular}
\end{center}
\caption{Summary of discussed Drell-Yan measurements, arranged as in Table~\ref{tab:pp_summary}. Here Fixed-Target experiments are shown in the higher segment, and collider experiments in the lower two.}
\label{tab:FTDY_summary}
\end{table}%

\begin{table}
\begin{center}
\begin{tabular}{|c|c|c|c|c|}
\hline
\multicolumn{5}{|c|}{\textbf{Jet Production}}\\
\hline
\textbf{Process} & \textbf{Experiment} & \textbf{Obs.} &\textbf{Ref.} &  \textbf{Target} \\
\hline\hline
$p\bar{p} \to j + X $ & CDF & Inclusive Jets & \cite{Abulencia:2007ez,Aaltonen:2008eq} & $g$\\
$p\bar{p} \to j + X $ & D0 & Inclusive Jets & \cite{Abazov:2008ae} & $g$\\
\hline
$p\bar{p} \to jj + X $ & CDF & Dijets & \cite{Aaltonen:2008dn} & $g$\\
$p\bar{p} \to jj + X $ & D0 & Dijets & \cite{Abazov:2010fr} & $g$ \\
\hline
\hline
$pp \to j + X $ & LHCb & Inclusive Jets &\cite{LHCb:2011xqa} & $g$\\
$pp \to j + X $ & ATLAS & Inclusive Jets &\cite{Aad:2010ad,Aad:2011fc,Aad:2013lpa} & $g$ \\
$pp \to j + X $ & CMS & Inclusive Jets &\cite{CMS:2011ab,Chatrchyan:2012gwa,Chatrchyan:2012bja} & $g$ \\
\hline
$pp \to jj + X $ & LHCb & Dijets & \cite{LHCb:2011xqa} & $g$\\
$pp \to jj + X $ & ATLAS & Dijets &\cite{Aad:2010ad,Aad:2011fc} & $g$\\
$pp \to jj + X $ & CMS & Dijets & \cite{Chatrchyan:2012gwa,Chatrchyan:2012bja} & $g$\\
\hline
\end{tabular}
\end{center}
\label{tab:JET_summary}
\caption{Summary of discussed inclusive jet and dijet measurements, arranged as in Table~\ref{tab:pp_summary}.}
\end{table}%

\begin{table}[h]
\begin{center}
\begin{tabular}{|c|c|c|c|c|}
\hline
\multicolumn{5}{|c|}{\textbf{Prompt Photon}}\\
\hline
\textbf{Process} & \textbf{Experiment} & \textbf{Obs.} &\textbf{Ref.} &  \textbf{Target} \\
\hline\hline
$p\bar{p} \to \gamma X$ & UA1/UA2 & Photon $E_T$ &\cite{Albajar:1988im,Alitti:1992hn,Ansari:1988te} &$q$, $g$ \\
$p\bar{p} \to \gamma X$ & CDF & Photon $p_T$ &\cite{Aaltonen:2009ty,Abazov:2005wc,Abe:1994rra,Acosta:2002ya,Acosta:2004bg}& $q$, $g$\\
$p\bar{p} \to \gamma X$ & D0 & Photon $E_T$ &\cite{Abachi:1996qz,Abazov:2001af,Abbott:1999kd}& $q$, $g$\\
\hline
$pp \to \gamma X$ & PHENIX & Photon $p_T$ &\cite{Adler:2006yt} & $g$, $q+\bar{q}$ \\
\hline\hline
$pp \to \gamma X$ & ATLAS & Photon $E_T$, $\eta$ &\cite{Aad:2011tw} & $g$, $q+\bar{q}$  \\
$pp \to \gamma X + j$ & ATLAS & Photon $E_T$, $\eta$ &\cite{Aad:2011tw} & $g$, $q + \bar{q}$ \\
\hline
$pp \to \gamma X$ & CMS & Photon $E_T$, $\eta$ &\cite{Chatrchyan:2011ue} & $g$, $q + \bar{q}$ \\
\hline
\end{tabular}
\end{center}
\caption{Summary of discussed isolated prompt photon measurements. The process column denotes the reaction observed in each experiment, Obs. refers to the physical observable measured and Target illustrates the most relevant partonic channels for the process and observable in question.}
\label{tab:pp_summary}
\end{table}%

\begin{table}[h]
\begin{center}
\begin{tabular}{|c|c|c|c|c|}
\hline
\multicolumn{5}{|c|}{\textbf{Top production}}\\
\hline
\textbf{Process} & \textbf{Experiment} & \textbf{Obs.} &\textbf{Ref.} &  \textbf{Target} \\
\hline\hline
$p\bar{p} \to t\bar{t}$ & CDF + D0 & $\sigma_{t\bar{t}}$&\cite{Aaltonen:2012ttbar} & $q$ + $\bar{q}$ \\
\hline\hline
$pp \to t\bar{t}$ & ATLAS & $\sigma_{t\bar{t}}$&\cite{ATLAS:2012jyc,ATLAS:2012fja} & $g$ \\
$pp \to t\bar{t}$ & CMS & $\sigma_{t\bar{t}}$&\cite{Chatrchyan:2012bra,CMS:2012iba} & $g$ \\
\hline
\end{tabular}
\end{center}
\caption{Summary of discussed top production measurements, arranged as in Table~\ref{tab:pp_summary}.}
\label{tab:ttbar_summary}
\end{table}%

\chapter{Distance Estimators}
\label{app:distances}
Here we define a set of useful measures in determining the statistical differences between two sets of parton distributions in a Monte Carlo representation, first introduced in Ref.~\cite{Ball:2010de}. Recalling the standard
definitions of the central value of a Monte Carlo PDF with $N_{\text{rep}}$ replicas,

\be \left< f (x,Q^2) \right> = \frac{1}{N_{\text{rep}}}\sum_i^{N_{\text{rep}}} f_k(x,Q^2), \ee
and its associated uncertainty,
\be \sigma^2\left[ f (x,Q^2) \right] = \frac{1}{N_{\text{rep}}-1}\sum_i^{N_{\text{rep}}} \left( f_k(x,Q^2) - \left<f(x,Q^2)\right>\right)^2.  \ee
Further estimators are available~\cite{Beringer:1900zz} for the uncertainty upon the central value,
\be \sigma^2\left[\left<f(x,Q^2)\right>\right] = \frac{1}{N_{\text{rep}}} \sigma^2\left[ f (x,Q^2) \right], \ee
and the uncertainty upon the uncertainty,
\be \sigma^2\left[\sigma^2\left[f(x,Q^2)\right]\right] = \frac{1}{N_{\text{rep}}} \left[ m_4\left[f(x,Q^2)\right] - \frac{N_{\text{rep}} - 3}{N_{\text{rep}} - 1} \left( \sigma^2\left[f(x,Q^2)\right] \right)^2 \right],\ee
where $m_4\left[f(x,Q^2)\right]$ refers to the fourth moment of the distribution $f(x,Q^2)$.

\begin{figure}[ht]
\centering
\includegraphics[width=0.9\textwidth]{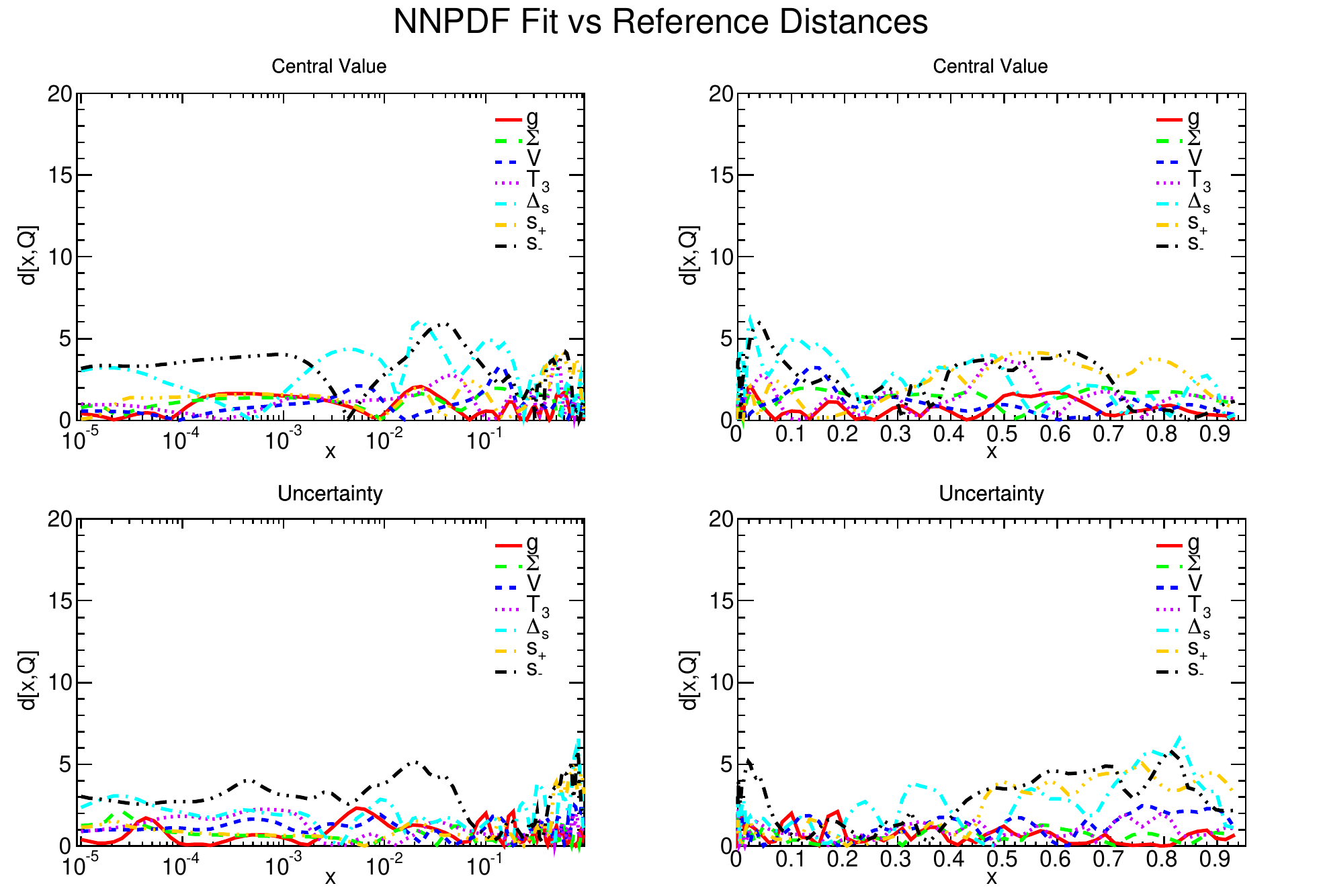}
\caption[Example plot of PDF distances.]{Example of both PDF central value and uncertainty distances for the seven PDFs parametrised at the NNPDF initial scale.}
\label{fig:distance example}
\end{figure}

Given these quantities we can define a distance between the representation of the PDF $f$ in two PDF sets as the square difference of the PDF central values in units of the uncertainty of the mean,
\be d^2_{\text{CV}}\left[f^{(1)},f^{(2)}\right] = \frac{\left(\left<f^{(1)}\right> - \left<f^{(2)}\right> \right)^2}
{\sigma^2 \left[\left<f^{(1)}\right>\right] + \sigma^2 \left[\left<f^{(2)}\right>\right]},
\label{eq:CVdistance}
\ee
where the PDF superscripts enumerate the PDF sets being compared and the dependence upon the kinematical variables $x,Q^2$ is implicit. With this definition of PDF distance, a value of $d^2=1$ corresponds
to a discrepancy between PDF sets consistent with one standard deviation of the central values. A similar measure can be defined for the uncertainties of the distribution,
\be d^2_{\sigma}\left[f^{(1)},f^{(2)}\right] = \frac{\left(\sigma^2 \left[f^{(1)}\right] - \sigma^2 \left[f^{(2)}\right] \right)^2}
{\sigma^2 \left[\sigma^2 \left[f^{(1)}\right]\right] + \sigma^2 \left[\sigma^2 \left[f^{(2)}\right]\right]}.
\label{eq:Vardistance}
\ee
These distances quantities are particularly useful in the systematic comparison of all partons in two PDF sets in order to evaluate the size and statistical significance of differences between the two sets. As an example,
consider Figure \ref{fig:distance example} where distances are shown for both estimators $d_{\text{CV}}$ and $d_\sigma$ (i.e the square-root of Eqns.~\ref{eq:CVdistance}, \ref{eq:Vardistance}) between two PDF sets, for seven PDF combinations.

A distinction should be noted between the distances presented in this work and those defined in Ref.~\cite{Ball:2010de}, where an additional bootstrap sampling of the distributions was used. In this work all distances are presented exactly as in Eqn. \ref{eq:CVdistance} and \ref{eq:Vardistance}.

%%%%%%%%%%%%%%%%%%%%%%%%%%%%%%%%%%%%%%%%%%%%%%%%%%%%%%%%%%%%%%%%%%%

%Environment for own publications list using bibitem (Involves a global renaming of the word Bibliography).
\cleardoublepage
\phantomsection
\renewcommand{\bibname}{Publications}

  \cleardoublepage
\phantomsection
\renewcommand{\bibname}{Proceedings}

\end{spacing}

\end{document}